\documentclass[useAMS,usenatbib]{mn2e}
\pdfoutput=1
\usepackage{graphicx}

\usepackage{amsmath}
\usepackage{amssymb}

\newcommand{\apj}{ApJ}
\newcommand{\apjs}{ApJS}
\newcommand{\aj}{AJ}
\newcommand{\aap}{A\&A}
\newcommand{\aaps}{A\&AS}

\newcommand{\mnras}{MNRAS}
\newcommand{\pasp}{PASP}
\newcommand{\an}{Astr. Nachr.}

\newcommand{\im}{i_\mathrm{m}}

\usepackage{color}

\title[Comprehensive Study of Kepler Triples via ETV]{A Comprehensive Study of the {\em Kepler} Triples via Eclipse Timing}
\author[T. Borkovits et al.]{T. Borkovits$^{1,2}$\thanks{E-mail:
borko@electra.bajaobs.hu (TB)}, T. Hajdu$^3$, J. Sztakovics$^3$, S. Rappaport$^4$, A. Levine$^5$, \\
\newauthor I. B. B\'{\i}r\'o$^1$, and P. Klagyivik$^{6,7}$\\
$^1$Baja Astronomical Observatory of Szeged University, H-6500 Baja, Szegedi \'ut, Kt. 766, Hungary\\
$^2$ELTE Gothard-Lend\"ulet Research Group, H-9700 Szombathely, Szent Imre herceg \'ut 112, Hungary \\ 
$^3$Astronomical Department of E\"otv\"os University, H-1118 P\'azm\'any P\'eter stny. 1/A, Budapest, Hungary\\
$^4$M.I.T. Department of Physics and Kavli Institute for Astrophysics and Space Research, 70 Vassar St.,Cambridge, MA, 02139\\
$^5$M.I.T. Kavli Institute for Astrophysics and Space Research, 70 Vassar St.,Cambridge, MA, 02139\\
$^6$Instituto de Astrofísica de Canarias, 38205, La Laguna, Tenerife, Spain\\
$^7$Dept. Astrofísica, Universidad de la Laguna, 38206, La Laguna, Tenerife, Spain}

\begin{document}

\date{Accepted ??? Received ???; in original form ???}


\maketitle

\label{firstpage}

\begin{abstract}

We produce and analyze eclipse time variation (ETV) curves for some 2600 targeted main-field {\em Kepler} binaries. We find good to excellent evidence for a third body in 222 systems via either the light-travel-time (LTTE) or dynamical effect delays.  Approximately half of these systems have been discussed in previous work, while the rest are newly reported here.  Via detailed analysis of the ETV curves using high-level analytic approximations, we are able to extract system masses and information about the three-dimensional characteristics of the triple for 62 systems which exhibit both LTTE and dynamical delays. For the remaining 160 systems whose ETV curves are dominated by LTTE delays we are able to extract the outer orbital period, eccentricity, and longitude of periastron as well as the mass function of the triple. In general, our solutions improve upon those published earlier.  New techniques of preprocessing the flux time series are applied to eliminate false positive triples and to enhance the ETV curves.  The set of triples with outer orbital periods shorter than $\sim$2000 days is now sufficiently numerous for meaningful statistical analysis.  We find that (i) as predicted, there is a peak near $\im \simeq 40^\circ$ in the distribution of the triple vs. inner binary mutual inclination angles  that provides strong confirmation of the operation of Kozai-Lidov cycles with tidal friction; (ii) the median eccentricity of the third-body orbits is $e_2=0.35$;  (iii) there is a deficit of triple systems with binary periods $\lesssim 1$ day and outer periods between $\sim$50 and 200 days which might help guide the refinement of theories of the formation and evolution of close binaries; and (iv) the substantial fraction of {\em Kepler} binaries which have third-body companions is consistent with a very large fraction of all binaries being part of triples.  

\end{abstract}


\section{Introduction}

The analysis of eclipse time variations (ETVs) via $O-C$ (observed minus calculated) diagrams is a powerful tool for the investigation of period variations in eclipsing binary (EB) systems, and, therefore, has been used in many EB studies over more than a century.   ETVs may arise from different causes that act on various timescales with various amplitudes.  It follows that $O-C$ diagrams may show a wide range of variational forms.  The causes may be either physical, i.e., connected to a real variation of the orbital period, or merely apparent.

Long-term physical ETVs mainly occur as a result of evolutionary effects such as mass exchange between the binary components, wind driven mass loss, magnetic braking, tidal dissipation, or even gravitational radiation. Often the characteristic time scale of the phenomenon substantially exceeds the entire period of human EB observations.  Generally in each such case, the ETVs are manifest as a slow, constant-rate variation of the orbital period which results in a quadratic $O-C$ pattern \citep[for analytic descriptions of the ETVs induced by some of the listed effects see][]{nanourisetal11,nanourisetal15}. Shorter time-scale physical ETVs can arise, e.g., from magnetic activity \citep[see, e.g.][]{hall89,applegate92,lanzarodono02} or from the dynamical effects of a close companion star on a binary orbit \citep{soderhjelm75}. These shorter time-scale effects tend to produce periodic or, at least, quasi-periodic ETV behaviour.

The two most well-known classes of apparent orbital period changes leading to ETVs are the light-travel time effect (LTTE) caused by the changing distance of a binary in a hierarchical multiple-star system, and the apsidal motion effect (AME) which may be seen in eccentric EBs. Apart from the extremely compact triples which were investigated by \citet{borkovitsetal15}, these two phenomena often result in quasi-sinusoidal mono-periodic $O-C$ diagrams. In the case of AME the $O-C$ curve formed from the secondary minima anticorrelates with the curve formed from the primary minima, while in the case of LTTE the two kinds of minima {\em must} vary in the same manner. Additional apparent orbital period changes inducing ETVs may arise, in theory, from the precession of the orbital plane of the EB due to the perturbations induced by either a third-star companion revolving in an inclined orbit or the non-aligned rotation of each or both stars.  Such ETVs are not yet known to have been observed.

In addition to the above effects, erratic variations have been observed as well.  They may indicate physical effects such as variable mass transfer rates or currently unidentified apparent timing effects.
 
Finally, when a light curve is distorted by the effects of, e.g., stellar spots or pulsations, the measurement process tends to yield spurious ETVs that may include periodic or quasiperiodic components (\citealp[see e.g][for spots]{kalimerisetal02,tranetal13,balajietal15} and \citealp[][for stellar oscillations]{borkovitsetal14}). 

The almost continuous four-year-long set of high-precision photometric observations from the {\em Kepler} mission \citep{boruckietal10} offers an unprecedented opportunity to study ETVs in thousands of EBs and ellipsoidal variables (ELVs).  Among a wide range of possibilities, these data are especially suited for searches for short-period third-star companions of these binaries. Third-star companions to binaries are interesting from several perspectives. Third stars may be particularly significant in the formation of close binaries; this has been discussed and investigated intensively over the past two decades \citep[for a short summary, see][]{fabryckytremaine07}. The statistically significant lack of short ($P_2<1000$\,d) outer period ternaries amongst solar or lower-mass binaries \citep{tokovinin14b} makes such investigations especially important.

The first, preliminary, systematic search of {\em Kepler} ETV data for hierarchical triples was carried out by \citet{giesetal12}, who identified possible long-term ETVs in 14 of 41 EBs but did not find any evidence of short period companions ($P_2<700$\,d). Later \citet{rappaportetal13} surveyed the whole available $Q0-Q13$ dataset for some 2100 EBs. They found 39 candidate triple systems in the short outer period domain (48 d $<P_2<$ 960 d), for which they presented combined LTTE+dynamical-effect solutions. This was the first systematic study of the dynamical effect in EBs using the {\em Kepler} data. Nearly contemporaneously, \citet{conroyetal14}, determined eclipse times for all the short period EBs, most of which are overcontact systems, and ELVs, and identified 236 systems for which the ETVs could be compatible with the LTTE. However, the majority of these were observed for less than one complete outer (third body) orbital period.  More recently \citet{borkovitsetal15} investigated 26 {\em Kepler}-field eccentric EBs which feature ETVs that are dominated by dynamical perturbations rather than LTTE. This work featured the simultaneous analysis of both the primary and secondary eclipses so as to break a number of degeneracies in the solutions. In a report published in 2015 June,  \citet{zascheetal15} present lightcurve and ETV analyses of ten detached or semi-detached {\em Kepler}-field EBs. The durations of the flux time series of most of these ten systems were extended by including ground-based timing measurements. Most recently, \citet{giesetal15} report improved analyses of the 41 EBs which were previously investigated in their earlier work \citep{giesetal12}. They now provide third body LTTE solutions for seven EBs. Additional studies of a possible third body affecting the ETVs of individual EBs in the {\em Kepler}-field have also been reported in \citet[][for KOI-928(=KIC~09140402)]{steffenetal11} \citet[][for HD~181068(=KIC~05952403)]{borkovitsetal13}, \citet[][for KIC~02856960]{leeetal13}, \citet[][for V404~Lyr(=KIC~03228863)]{leeetal14}, \citet[][for KIC~05621294]{leeetal15}. Most recently, \citet{baranetal15} reported the detection of a planet-mass companion in the sdB+dM EB 2M1938+4603(=KIC~09472174).

In the present paper we regenerate and reanalyze the ETVs of all the previously investigated triple-body candidate EBs, with the exception of the 26 systems investigated in \citet{borkovitsetal15}, and we extend our analysis to longer period systems which were excluded from the study of \citet{conroyetal14}. While the new study of the previously investigated systems is natural because of the significantly longer time span of the $Q0-Q17$ {\em Kepler} observations, there are additional reasons to further investigate the EBs listed in \citet{conroyetal14}. First, our method for determination of times of minima gives results for semi-detached or detached systems, i.e., systems with relatively sharp and deep minima, that are significantly more accurate than the times for these systems used in \citet{conroyetal14}. Second, for overcontact EBs and ELVs we also analyze quadrature timing variations (QTV), i.e., $O-C$ times of maxima. Third, we checked the individual LTTE solutions in detail with particular attention to whether the inferred masses could be reliable, and, in the cases where further treatment was indicated, we modelled the effects of dynamical perturbations of the binaries. Finally, for the minority of the investigated EBs for which pre-{\em Kepler} ground-based times of minima were available, we also included these data in our analysis. In such a way we were also able to improve the reliability of the LTTE solutions for previously investigated systems.

In Section~\ref{Sect:ThirdbodyETVs} we briefly describe the LTTE and dynamical perturbation effects. Then, in Section~\ref{Sect:dataprep} we outline the method of calculating accurate times of eclipse and non-eclipse minima as well as our method for searching for ETV solutions. We introduce the idea of determining times of light curve maxima, and utilize these so-called `quadrature timing curves' as diagnostics to weed out false positives. Section~\ref{Sect:Ground-based minima} discusses the use of supplementary ground-based timing data for extending the overall span of the observations for a small subset of our triples. Section~\ref{Sect:Overview} gives an overview of the 230 systems that we investigated.  This includes a plot of each ETV curve with fitted solution as well as an extensive set of tables listing fitted system parameters.  In Section~\ref{Sect:Results} we discuss our findings from a number of different perspectives, and we draw some conclusions from this substantial statistical collection of triple star systems. Finally, we summarize our work in Section~\ref{Sect:Summary}.

\section{Effects of a third body on the ETV}
\label{Sect:ThirdbodyETVs}

We define ETV by the $O-C$ time difference:
\begin{equation}
\Delta=T(E)-T_0-P_\mathrm{s}E
\label{Eq:ETVdef}
\end{equation}
where $T(E)$ denotes the observed time of the $E$-th eclipse, $T_0=T(0)$ indicates the reference epoch, i.e., the observed time of the ``zeroth'' eclipse, while the constant $P_\mathrm{s}$ denotes the eclipse period. Our basic model for this time difference is given by
\begin{equation}
\Delta=\sum_{i=0}^{3}c_iE^i+\left[\Delta_\mathrm{LTTE}+\Delta_\mathrm{dyn}+\Delta_\mathrm{apse}\right]_0^E.
\label{Eq:ETVmod}
\end{equation}
The constant and linear terms of the polynomial in $E$ give corrections to the calculated eclipse times in the above definition of ETV, while the quadratic term models any constant-rate period variation ($\Delta P_1/2$), independent of its origin.\footnote{Here we define $\Delta P_1$ in terms of the quadratic coefficient  as $\Delta P_1 = 2 c_2$ which is the change in binary orbital period per orbital cycle (units of [d/c]).  The usual orbital period derivative is given by $\dot P_1 \simeq 2 c_2/P_1$.  Similarly, $c_3$ is related to $\ddot  P_1$ as $\approx 6 c_3/P_1^2$.} The cubic term allows for better approximation of some visible, seemingly non-quadratic, additional long-term ETVs in a small number of the investigated EBs; this term was not used in \citet{borkovitsetal15}. Finally, $\Delta_\mathrm{LTTE}$, $\Delta_\mathrm{dyn}$ and $\Delta_\mathrm{apse}$ refer to the contributions of light-travel time effect (LTTE), short period dynamical perturbations, and apsidal motion effect (AME, including longer time-scale dynamical perturbations) to the ETVs, respectively.

The coefficients $c_0$ and $c_1$ were adjusted {\em simultaneously} with the physical terms in all analyses. The quadratic coefficient $c_2$ was allowed to be nonzero only for originally parabolic shaped ETVs or when the LTTE fitting yielded parabolic residuals; in these cases the quadratic term was determined simultaneously with all other included terms. The coefficient of the cubic term was set to zero except in five cases wherein at least three full outer periods were observed; this yielded reduced-size $O-C$ residuals without substantially altering the orbital parameters. The parameters of the LTTE-term (see below) were adjusted in all cases. Dynamical ETV contributions were considered for a subset of our sample where there was some indication that a pure LTTE solution would not be adequate. Finally, the apsidal motion contributions were also taken into account for a few eccentric EBs.

As will be discussed in Sect.~\ref{Sect:dataprep}, for systems with significant ellipsoidal light variations we also measure and analyze the times of the ellipsoidal maxima. As most of the systems with well-measured ellipsoidal variations -- being overcontact or semi-detached systems -- revolve in circular orbits, the maximum brightnesses occur near quadrature phases (i.e. $\phi=0.25$ and $\phi=0.75$) and, therefore we refer to the $O-C$ times of the maxima as quadrature time variations (QTV). The QTV curves for LTTE and quadratic variations must have the same form as given by Eq.~(\ref{Eq:ETVdef}). The dynamical contribution and the AME-term would, however, be different for quadratures, but, practically speaking, these effects would have needed considerable extra care only for `heart-beat' binaries \citep{thompsonetal12}, of which only one, KIC~03766353, is covered in this paper. Note also, that generalizing the natural convention that the epoch or cycle number ($E$) is integer for primary and half-integer for secondary minima, we calculate it as $E+0.25$ for the first quadrature (at $\phi \sim 0.25$, i.e. after the primary minima) and $E+0.75$ for the second quadrature.

The mathematical form and other properties of the different ETV contributions were discussed comprehensively in \citet{borkovitsetal15}. Here we discuss briefly, and from a bit different point of view, only the two main effects which were applied in this work.

 
\subsection{The Light-Travel Time Effect}
\label{Subsect:LTTEdiscuss}

General criteria for the plausibility of an LTTE model of ETVs have been given by \citet{frieboescondeherczeg73}.  The criteria may be summarized as follows. (1) The shape of the ETV curve must follow the analytical form of an LTTE solution. (2) The ETV of the secondary minima must be consistent in both phase and amplitude with the primary ETV. (3) The estimated mass or lower limit to the mass of the third component, derived from the amplitude of the hypothetical LTTE solution via the mass function -- see below, must be in accord with photometric measurements or limits on third light in the system. (4) Variation of the system radial velocity should be in accord with the LTTE solution. While these criteria do not look very restrictive, none of their candidate systems fulfilled all of them. More than 50 years after the first mathematical description of the problem, there was only one system, Algol itself, where the LTTE was identified clearly via its ETV curve. Even over the ensuing decades, the numbers of confirmed LTTE cases has grown very slowly. The reason is as follows.

\begin{table}
 \caption{Meaning the symbols used in the paper}
 \label{Tab:symbols}
 \begin{tabular}{@{}lll}
  \hline
  Parameter & symbol & explanation \\
\hline
Mass & & \\
~~EB members & $m_\mathrm{A,B}$ &\\
~~total mass of EB & $m_\mathrm{AB}$ & $m_\mathrm{A}+m_\mathrm{B}$ \\
~~ternary's mass & $m_\mathrm{C}$ & \\
~~total mass & $m_\mathrm{ABC}$ & $m_\mathrm{A}+m_\mathrm{B}+m_\mathrm{C}$ \\
\hline
Period &  &\\
~~sidereal/eclipsing & $P_{1,2}$ & \\
~~anomalistic & ${P_\mathrm{a}}_{1,2}$ & \\
\hline
Semi-major axis & & \\
~~relative orbit & $a_{1,2}$ & \\
~~absolute orbit of EB & $a_\mathrm{AB}$ & $m_\mathrm{C}/m_\mathrm{ABC}\cdot a_2$ \\
\hline
eccentricity  & $e_{1,2}$ & \\
\hline
Anomaly      & & \\
~~true       & $v_{1,2}$ & \\
~~eccentric  & $E_{1,2}$ & $E=2\tan^{-1}\left(\sqrt{\frac{1-e}{1+e}}\tan\frac{v}{2}\right)$ \\
~~mean       & $l_{1,2}$ & $l=E-e\sin{E}$\\
\hline
argument of periastron &  & see Fig.~1 and App.~D of\\
                       &  & \citet{borkovitsetal15}\\
~~observable & $\omega_{1,2}$ &  \\
~~dynamical  & $g_{1,2}$      &  \\
\hline
inclination &  & see Fig.~1 and App.~D of\\
            &  & \citet{borkovitsetal15}\\
~~observable & $i_{1,2}$ & \\
~~mutual (relative) & $\im$ & \\
\hline
ascending node & & see Fig.~1 and App.~D of\\
               &  & \citet{borkovitsetal15}\\
~~observational & $\Omega_{1,2}$ & \\
& $\Delta\Omega$ & $\Omega_2-\Omega_1$ \\
\hline
speed of light & $c$ & \\ 
Gravity constant & $G$ & \\
\hline
\end{tabular}
\end{table}

The mathematical form of LTTE can be written as\footnote{In his seminal work \citet{irwin52} shifts the reference plane of the light-time orbit from the center of mass of the triple, i.e. the focal point of the ellipse, to the center of the orbit and, therefore, the extra term of $a_\mathrm{AB}e_2\sin\omega_2\sin i_2/c$ belongs in his equation (compare his Eqs.~[1] and [2]). This extra term has been also given in many of the recent papers dealing with LTTE. Note, however, that this shift was made by the author only for practical reasons, as his graphical solution had a more comfortable form with this formalism. No purpose is served by the use of the shifted form in the era of numeric fitting procedures. An additional caveat is also necessary, as this step can be justified only as far as the elements of the light-time orbit remain constant. In situations where the orbital elements vary, the proper extra term would not be constant and the use of a constant term would not be correct.  Therefore, we recommend omission of this additional $e_2\sin\omega_2$ term in future studies.}
\begin{equation}
\Delta_\mathrm{LTTE}=-\frac{a_\mathrm{AB}\sin{i_2}}{c}\frac{\left(1-e_2^2\right)\sin(v_2+\omega_2)}{1+e_2\cos{v_2}},
\label{Eq:LTTE(v)}
\end{equation}
or changing to eccentric anomaly:
\begin{eqnarray}
\Delta_\mathrm{LTTE}&=&-\frac{a_\mathrm{AB}\sin{i_2}}{c}\left[\sqrt{1-e_2^2}\sin E_2\cos\omega_2\right. \nonumber \\
&&\left.+\left(\cos E_2-e_2\right)\sin\omega_2\right] \nonumber \\
&=&-\frac{a_\mathrm{AB}\sin{i_2}}{c}\left[\sqrt{1-e_2^2\cos^2\omega_2}\sin(E_2+\phi)\right. \nonumber \\
&&\left.-e_2\sin\omega_2\right],
\label{Eq:LTTE(E)}
\end{eqnarray}
and, therefore, the amplitude of the LTTE becomes
\begin{equation}
{\cal{A}}_\mathrm{LTTE}=\frac{a_\mathrm{AB}\sin{i_2}}{c}\sqrt{1-e_2^2\cos^2\omega_2},
\label{Eq:A_LTTE1}
\end{equation}
while its phase is
\begin{equation}
\phi=\tan^{-1}\left(\frac{\sin\omega_2}{\sqrt{1-e_2^2}\cos\omega_2}\right).
\label{Eq:phi}
\end{equation}
By introducing the mass function
\begin{equation}
f(m_\mathrm{C})=\frac{m_\mathrm{C}^3\sin^3i_2}{m_\mathrm{ABC}^2}=\frac{4\pi^2a_\mathrm{AB}^3\sin^3{i_2}}{GP_2^2}
\label{Eq:f(mC)def}
\end{equation}
we obtain that
\begin{equation}
{\cal{A}}_\mathrm{LTTE}\approx1.1\times10^{-4}f(m_\mathrm{C})^{1/3}P_2^{2/3}\sqrt{1-e_2^2\cos^2\omega_2}.
\label{Eq:A_LTTE}
\end{equation}
The meaning of each of the symbols in the above equations, as well as other symbols to be used later, are tabulated in Table~\ref{Tab:symbols}.  In regard to units, masses should be expressed in terms of $M_\odot$, and the period and amplitude in days. For a hierarchical triple composed of three solar-mass stars, the equations above result in ${\cal{A}}_\mathrm{LTTE}\leq0.0027$\,d for $P_2=1$\,yr, and ${\cal{A}}_\mathrm{LTTE}\leq0.0125$\,d for $P_2=10$\,yr. Since, in the first 60-70 years of the last century most of the eclipse timing observations were done with visual brightness estimates having accuracies not better than a few hundredths of a day, and only a very limited number of photographic and photoelectric observations were available, it was almost hopeless to identify light-time orbits with periods shorter than a few decades. Furthermore, in the case of possibly longer period outer orbits, another problem occurs. The ETV of several EBs were found to manifest quite complex and sometimes erratic behaviour over time scales of a few decades; many examples may be found in \citet{kreineretal01}\footnote{http://www.as.up.krakow.pl/o-c/}. These poorly understood variations may act to hide long period LTTEs.

Over the last several decades, the advent of CCD detectors and other advances has led to the acquisition of much new and relatively accurate EB timing data that has, in turn, made it possible to tentatively or definitely detect LTTE in hundreds of EBs. Most of these LTTE solutions reveal companions with orbital periods longer than a decade. Third stars were found in shorter period orbits only for a very limited number of EBs. Before the era of the {\em Kepler} space mission, IU~Aurigae was the only EB system in which there was a detection of LTTE due to a third-star companion with a period shorter than one year \citep{mayer83}. All the other tertiaries with periods less than one year had been discovered spectroscopically in accord with the fact that spectroscopic detection is much more effective for short period outer orbits \citep[see, e.g.][]{mayer90,tokovinin14a}. However, spectroscopy requires much more light and, therefore, larger instruments as well as exposure time for a given system than photometry. Thus, the majority of the EBs are too faint to be suited for spectroscopic third body detection.  

In such a way the {\em Kepler} mission offers an unprecedented opportunity for the discovery of short-period companions orbiting EBs, including also lower mass systems, such as, e.g., the majority of overcontact binaries which are usually too faint for spectroscopic investigations. Furthermore, in contrast to earlier, ground-based observations which were inhomogeneous and generally restricted to small portions of the lightcurves around the eclipses, {\em Kepler} observations provide almost continuous and highly homogeneous lightcurves over intervals as long as four years. As a consequence, we are now in a position to extend our timing investigations to the out-of-eclipse parts of the lightcurves. Accordingly, an additional criterion of reliable LTTE solution can be introduced, as (5) the times of the maxima of the ellipsoidal variations, at least in EBs that have circular orbits, should be in accord both in phase and amplitude with the ETVs.

Another never seen before feature is the presence, in a small number of {\em Kepler} light curves, of outer eclipses. For such systems a further natural criterion for identifying the outer eclipsing body with the source of the observed LTTE is that (6) the LTTE should exhibit the same period as the extra eclipses, and these latter should occur around the extrema of the LTTE. In Sects.~\ref{Sect:dataprep} and \ref{Subsect:extraeclipses} we illustrate the applications of these new criteria. 

\subsection{Dynamical perturbations of a third body}

If an EB has a more or less distant companion, its binary motion no longer remains purely Keplerian since time-dependent perturbations affect all six orbital elements. Naturally, the occurrence times of the eclipses are also affected. The manifestation of the perturbations in the ETVs was first studied in this context by \citet{soderhjelm75} and \citet{mayer90}. Later the third-body effects were elaborated in full in a series of papers by \citet{borkovitsetal03,borkovitsetal11,borkovitsetal15}, and, in the context of transit timing variations (TTV) of exoplanets, by \citet{agoletal05}.

A thorough discussion of the dynamical perturbations may be found in \citet{borkovitsetal15}; here we restrict ourselves to some fundamental notes. The perturbations mostly act on three different timescales, from which we consider those which have a period equal, or related, to the $P_2$ period of the third component. If the inner orbit is circular, which is the case for the majority of the investigated systems, the dominant terms of the ETVs due to the perturbations take the following form:
\begin{eqnarray}
\Delta_\mathrm{dyn}&=&\frac{3}{4\pi}\frac{m_\mathrm{C}}{m_\mathrm{ABC}}\frac{P_1^2}{P_2}\left(1-e_2^2\right)^{-3/2}\nonumber\\
&&\times\left[\left(\frac{2}{3}-\sin^2\im\right){\cal{M}}+\frac{1}{2}\sin^2\im{\cal{S}}\right],
\end{eqnarray}
where
\begin{eqnarray}
{\cal{M}}&=&v_2-l_2+e_2\sin{v_2} \nonumber \\
&=&3e_2\sin{v_2}-\frac{3}{4}e_2^2\sin2v_2+\frac{1}{3}e_2^3\sin3v_2+{\cal{O}}\left(e_2^4\right),
\end{eqnarray}
and
\begin{equation}
{\cal{S}}=\sin(2v_2+2g_2)+e_2\left[\sin(v_2+2g_2)+\frac{1}{3}\sin(3v_2+2g_2)\right].
\end{equation}
These are essentially the same as Eqns. (8)-(10) used in \citet{rappaportetal13}.

Expressions for the dynamical-perturbation ETVs for EBs with elliptical inner orbits ($e_1>0$) are much more complicated \citep[see][]{borkovitsetal11,borkovitsetal15}. In particular, the amplitude of the ETVs depends sensitively on the eccentricities of the binary and third-star orbits and on the mutual inclination of the two orbits.  Therefore, even for a given mass and period ratio, the amplitude may take a value within a wide range, as was illustrated, e.g. in Fig.~3 of \citet{borkovitsetal11}.

There are some dozen eccentric EBs in the {\em Kepler} sample where characteristic shapes of the ETVs, e.g., definite spikes around outer orbit periastron passages or differences between the primary and secondary ETV curves, etc., clearly reveal the dominance of dynamical perturbations \citep[see][for details]{borkovitsetal11}. Dynamical ETV contributions, however, may also be significant when the form of the ETVs is more or less sinusoidal. Therefore, we check each LTTE solution, as follows, in order to determine whether it should be supplemented by the effects of dynamical perturbations. The amplitude of the dynamical ETV contribution\footnote{In \citet{borkovitsetal15} an analogous dynamical amplitude was defined as ${\cal{A}}_\mathrm{dyn}=\frac{15}{16\pi}\frac{m_\mathrm{C}}{m_\mathrm{ABC}}\frac{P_1^2}{P_2}\left(1-e_2^2\right)^{-3/2}$, here, however, we have chosen a different definition, because most of the presently investigated systems have low or zero inner eccentricity ($e_1$) in which case the present definition is more realistic.} is given approximately by
\begin{equation}
{\cal{A}}_\mathrm{dyn}=\frac{1}{2\pi}\frac{m_\mathrm{C}}{m_\mathrm{ABC}}\frac{P_1^2}{P_2}\left(1-e_2^2\right)^{-3/2},
\label{Eq:A_dyn}
\end{equation}   
where the periods and the outer eccentricity are known from the LTTE solution. While the mass ratio is not known, it may be estimated from the mass function of the LTTE solution for different values of the EB's total mass ($m_\mathrm{AB}$) and the inclination of the outer orbit ($i_2$). Then, comparison of the ratio ${\cal{A}}_\mathrm{dyn}/{\cal{A}}_\mathrm{LTTE}$ to unity indicates whether a pure LTTE solution of a given ETV would be satisfactory, or whether a more complex solution is necessary.

The analysis itself was carried out in the same manner and with the same code that was described in detail in \citet{borkovitsetal15}. 

\section{System selection and data preparation}
\label{Sect:dataprep}

We use the present version of the {\em Kepler} EB catalog and lightcurve files available at the Villanova web site\footnote{http://keplerebs.villanova.edu/} \citep{slawsonetal11,matijevicetal12,conroyetal14,lacourseetal15}. All the lightcurve files for the sources in the original {\em Kepler} field were downloaded, and, using the first (BJD), seventh (detrended relative flux), and eighth (uncertainty of the latter) columns of these files, $O-C$ diagrams were formed in an automated way. For a significant portion of the systems to be investigated, some quarters of the observed datasets ($Q4$ and/or $Q12-13$) were not available at the Villanova site; in most of those cases we downloaded the missing data directly from the MAST database operated by the Space Telescope Science Institute\footnote{http://archive.stsci.edu/}, converted it into the proper format, and merged it with the Villanova-derived dataset. We then selected those systems which either were mentioned in the context of having third components in previous literature or had interesting preliminary $O-C$ curves.  For the selected systems we calculated more accurate eclipse times in a somewhat more sophisticated, semi-automated manner. Our method, which is based on forming folded, binned, and averaged light curves for the whole dataset of each EB, then constructing polynomial templates for intervals around the minima of these averaged lightcurves, and finally using these templates for fitting individual minima, was described in detail in Sect.~4 of \citet{borkovitsetal15}. Therefore, here we note only some subtleties and variations specific to the present work.

This procedure yielded $O - C$ diagrams for some 400 systems in which ETVs appear at levels indicating a need for further analysis. The majority of these systems definitely show ellipsoidal variations, which makes it possible to calculate not only times of the eclipses, but also of the maxima in the lightcurves. This latter set of quadrature times (QTVs) was produced in the same manner as the ETVs. We found that in the case of overcontact EBs and most of the ELV binaries, with the exception of a few eccentric ELVs, it was satisfactory to set the phase limits for building up minima and maxima templates to $\phi_\mathrm{p}=[-0.15;0.15]$, $\phi_\mathrm{s}=[0.35;0.65]$, for primary and secondary minima, and $\phi_\mathrm{q1}=[0.10;0.40]$ and $\phi_\mathrm{q2}=[0.60;0.90]$, for the first and second quadratures (maxima), respectively. For semi-detached and detached systems with definite and sharper eclipses, narrower phase limits were set for the minima. We also calculated quadrature (or maxima) templates, applying mostly the same phase constraints, for those systems where the out-of-eclipse sections of the folded, binned lightcurve exhibit ellipsoidal light variations and are not subject to cycle-to-cycle variations (see below). In the cases of a few eccentric systems we departed from the above phase constraints in the calculation of quadrature templates in accordance with the properties of the each lightcurve.)  

\begin{figure*}
\includegraphics[width=84mm]{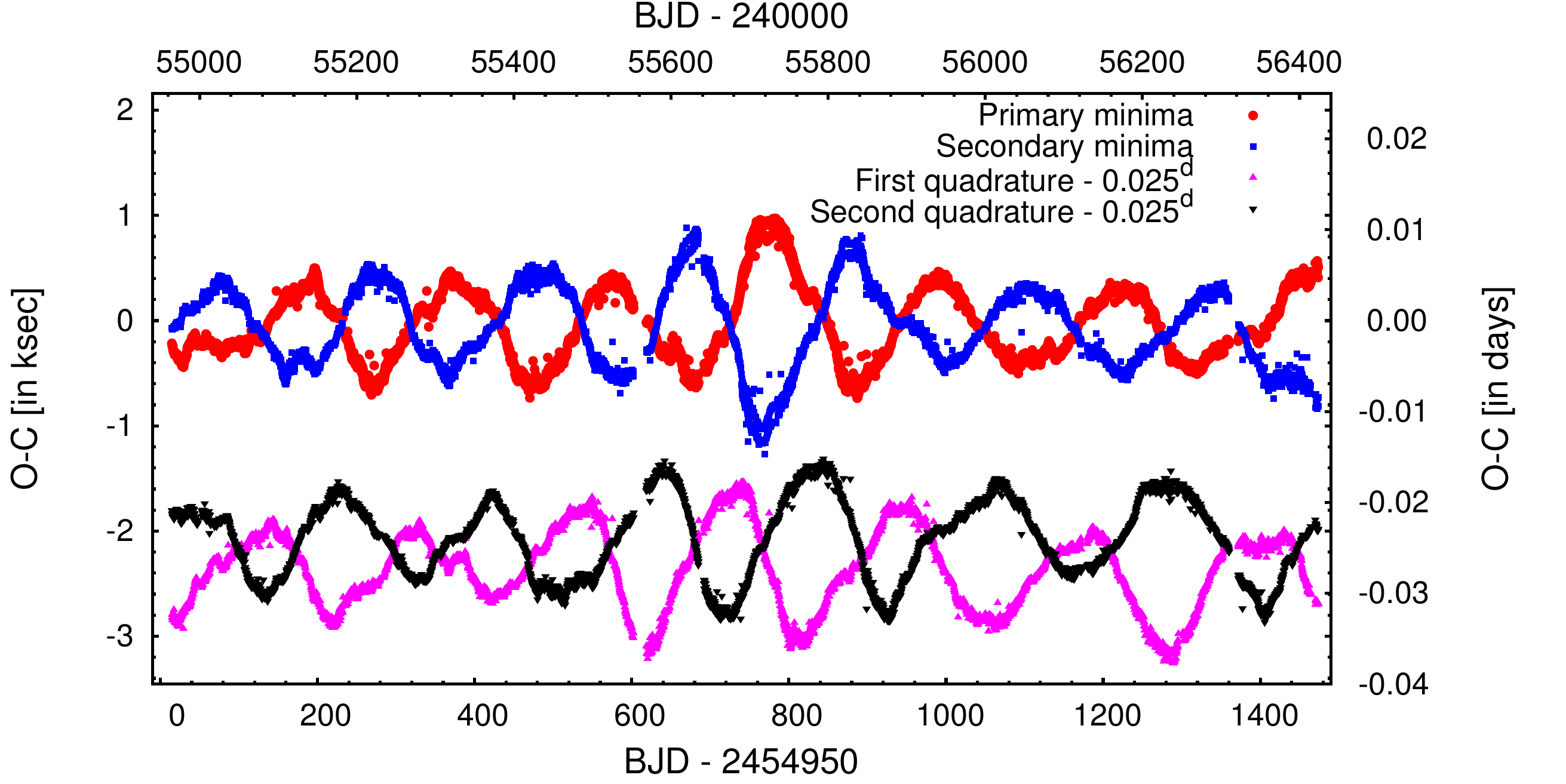}\includegraphics[width=84mm]{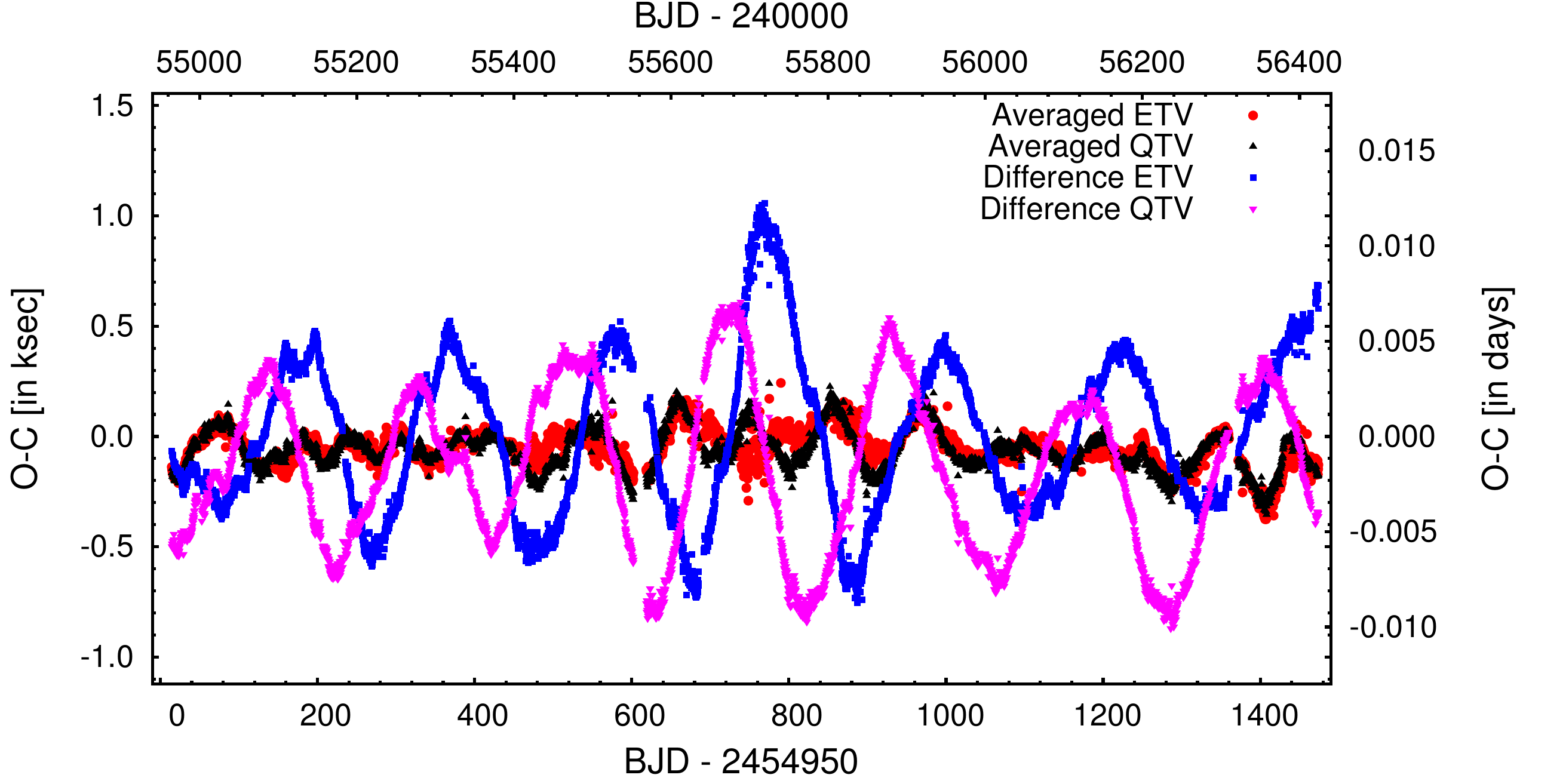}
 \caption{The highly anticorrelated, quasi-periodic ETV and QTV curves of the overcontact EB KIC~06431545. This type of ETV variations, which is likely attributable to large spotted areas on the stellar surface(s), was first reported in \citet{tranetal13}, and was also investigated by \citet{balajietal15}. {\it Left panel:} The individual primary (red circles) and secondary (blue boxes) ETV, and first quadrature (directly after the primary eclipses; magenta upward triangles) and second quadrature (black downward triangles) QTV curves. {\it Right panel:} The averaged ETV (red) and QTV (black) curves show only some low-amplitude residuals, while the difference curves between the two ETVs (blue) and QTVs (magenta) exhibit a phase-shift of one-fourth of a period between the two sets.}
 \label{Fig:EQTVavexample1}
\end{figure*}

Then, having obtained templates, times of individual minima and maxima were determined in exactly the same manner as described in \citet{borkovitsetal15}. In such a way we have obtained 1--2 ETV and 0--2 QTV curves for each system. In several cases, these curves are obviously distorted by the effects of stellar spots, pulsations, or oscillations. Fortunately, as was shown in \citet{tranetal13}, stellar spots in general distort the primary and secondary ETVs in an anticorrelated way. Similarly, the distortions in the two QTVs due to stellar spots also anticorrelate with each other and, furthermore, they are shifted by $\pm90\degr$ in phase from the respective ETVs (see Fig.~\ref{Fig:EQTVavexample1}). Therefore, the effects of starspots can be significantly reduced by averaging the primary and secondary ETVs, and the two QTVs as well. Thus, we also calculated averaged ETVs and QTVs. This process was carried out by interpolating the times of the primary ETVs to the times of the corresponding secondary eclipses with the help of a cubic spline. The same was done for the QTV curves.

\begin{figure*}
\includegraphics[width=84mm]{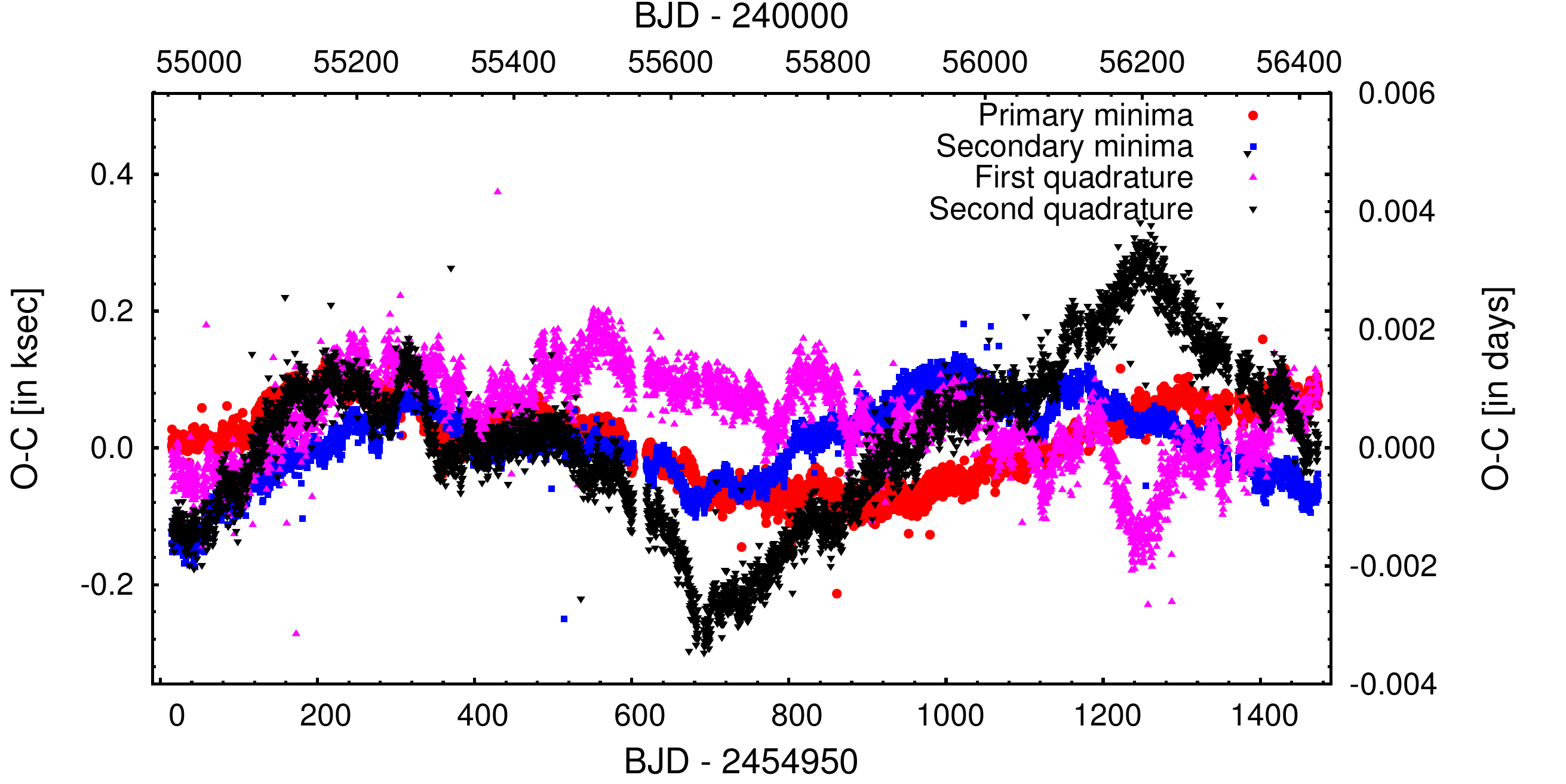}\includegraphics[width=84mm]{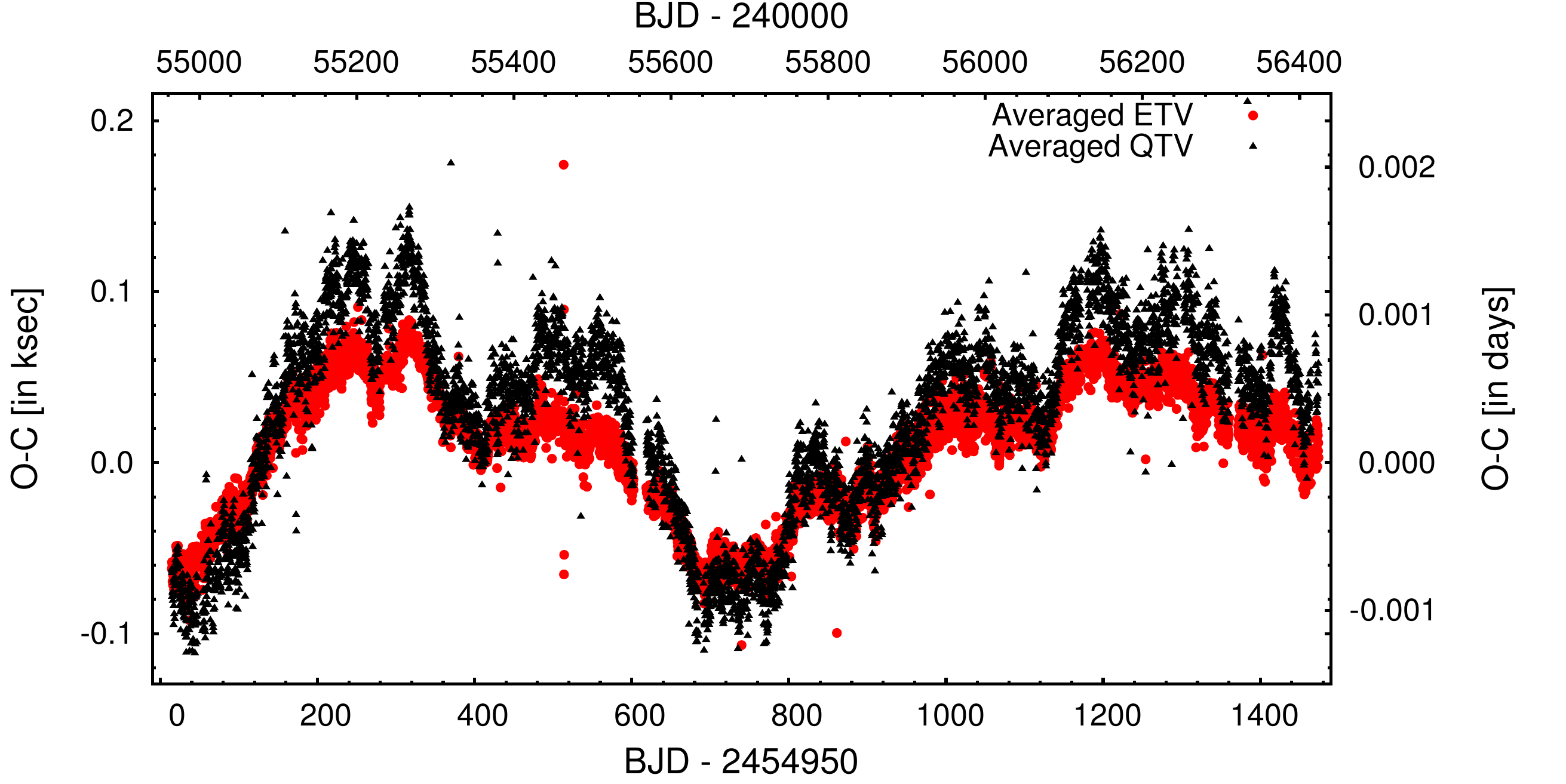}
 \caption{The highly irregular $O-C$ curves of the low-amplitude, short-period, possibly overcontact EB KIC~02715417. {\it Left panel:} The individual primary (red circles) and secondary (blue boxes) ETV, and first quadrature (directly after the primary eclipses; magenta upward triangles) and second quadrature (black downward triangles) QTV curves. {\it Right panel:} The averaged ETV (red) and QTV (black) curves reveal some (quasi-)periodic variations similar both in magnitude and phase for the two curves; this indicates that the LTTE curve could be due a low-mass (or very low inclination) third companion.}
 \label{Fig:EQTVavexample2}
\end{figure*}

In our experience, this averaging process is most effective for overcontact EBs and low-eccentricity or circular orbit ELVs where the two minima, and also the two maxima, are comparable in both amplitude and duration.  Therefore, on the one hand, the times of mid-minima and mid-maxima can be determined with approximately the same accuracy, while on the other hand, they are affected by the distortions more or less at the same level (see Fig.~\ref{Fig:EQTVavexample2}). Another benefit of forming averaged ETVs (and QTVs) is reduced scatter in the $O-C$ curves with respect to the original ones for several systems and, therefore, in these cases we used the averaged curves instead of the individual ETV curves for LTTE fitting. 

\begin{figure*}
\includegraphics[width=84mm]{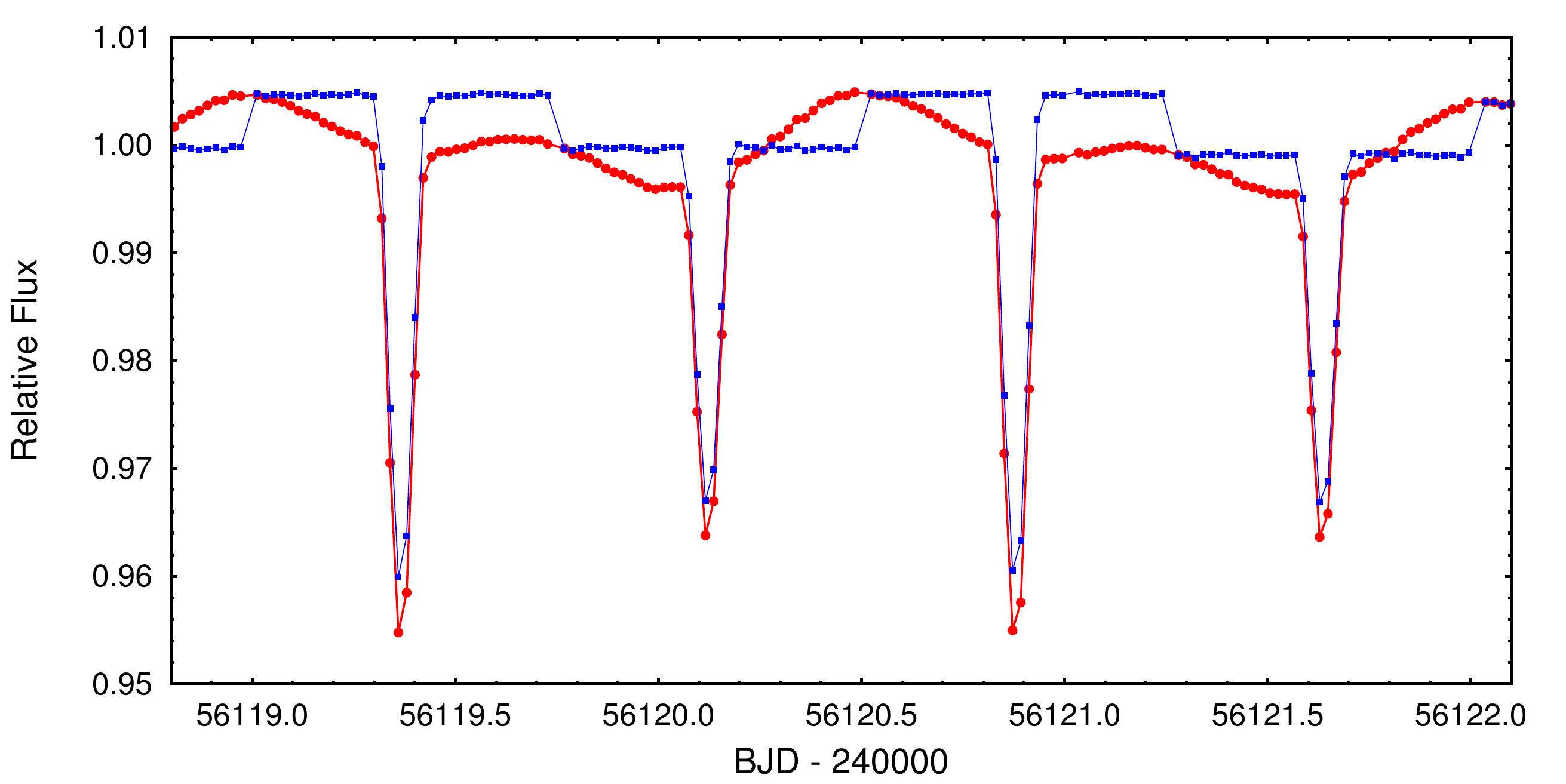}\includegraphics[width=84mm]{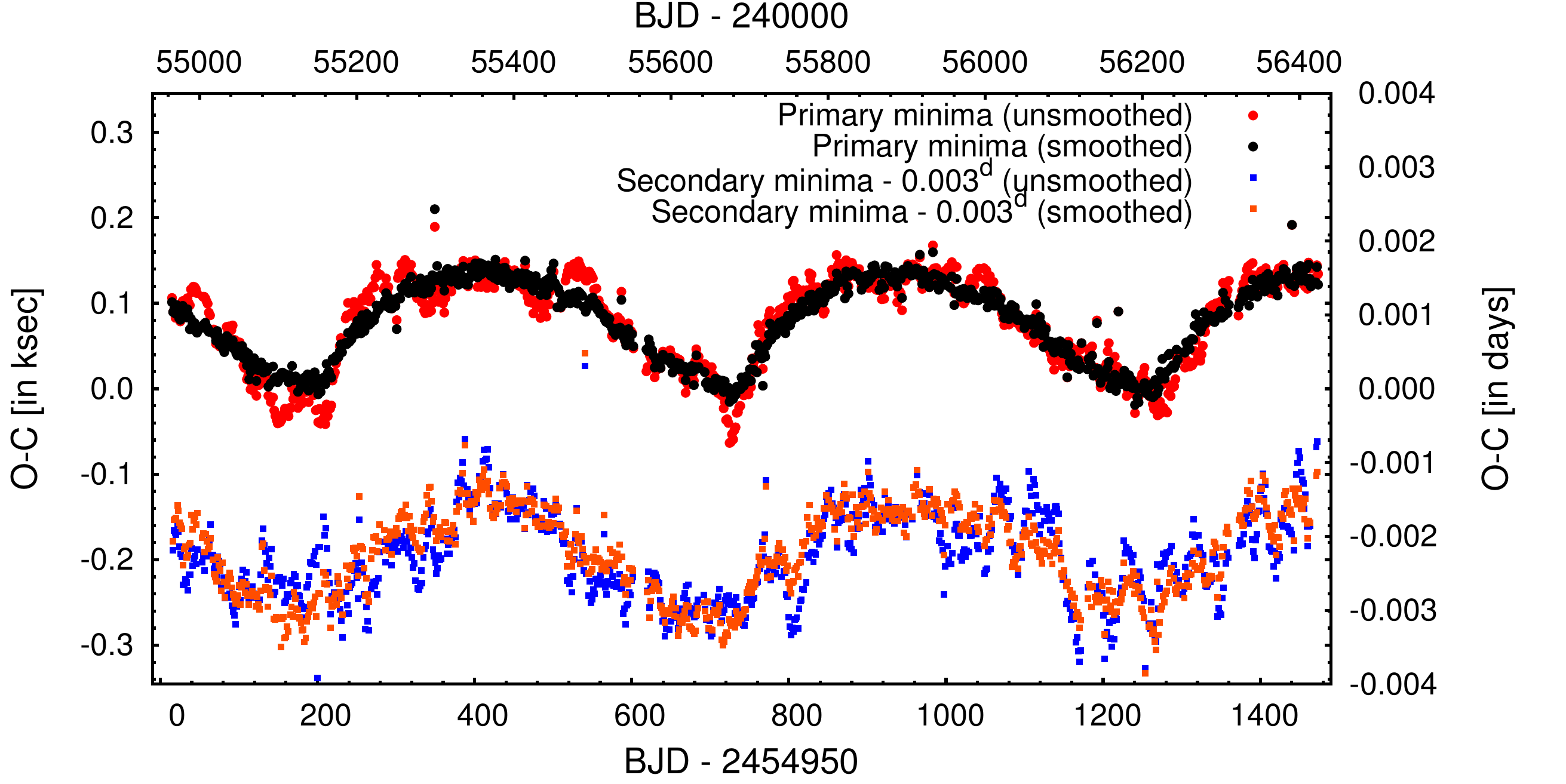}
\includegraphics[width=84mm]{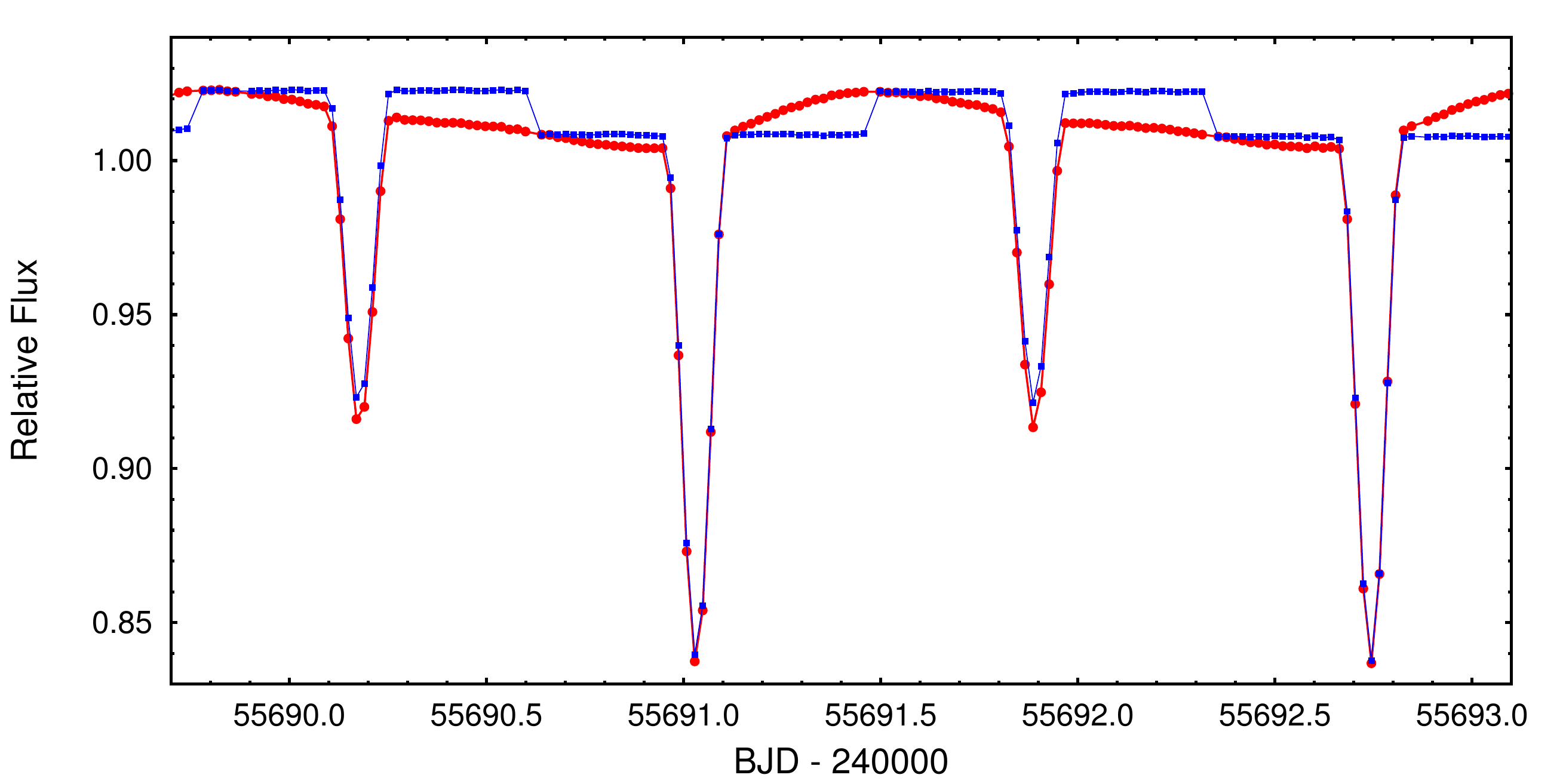}\includegraphics[width=84mm]{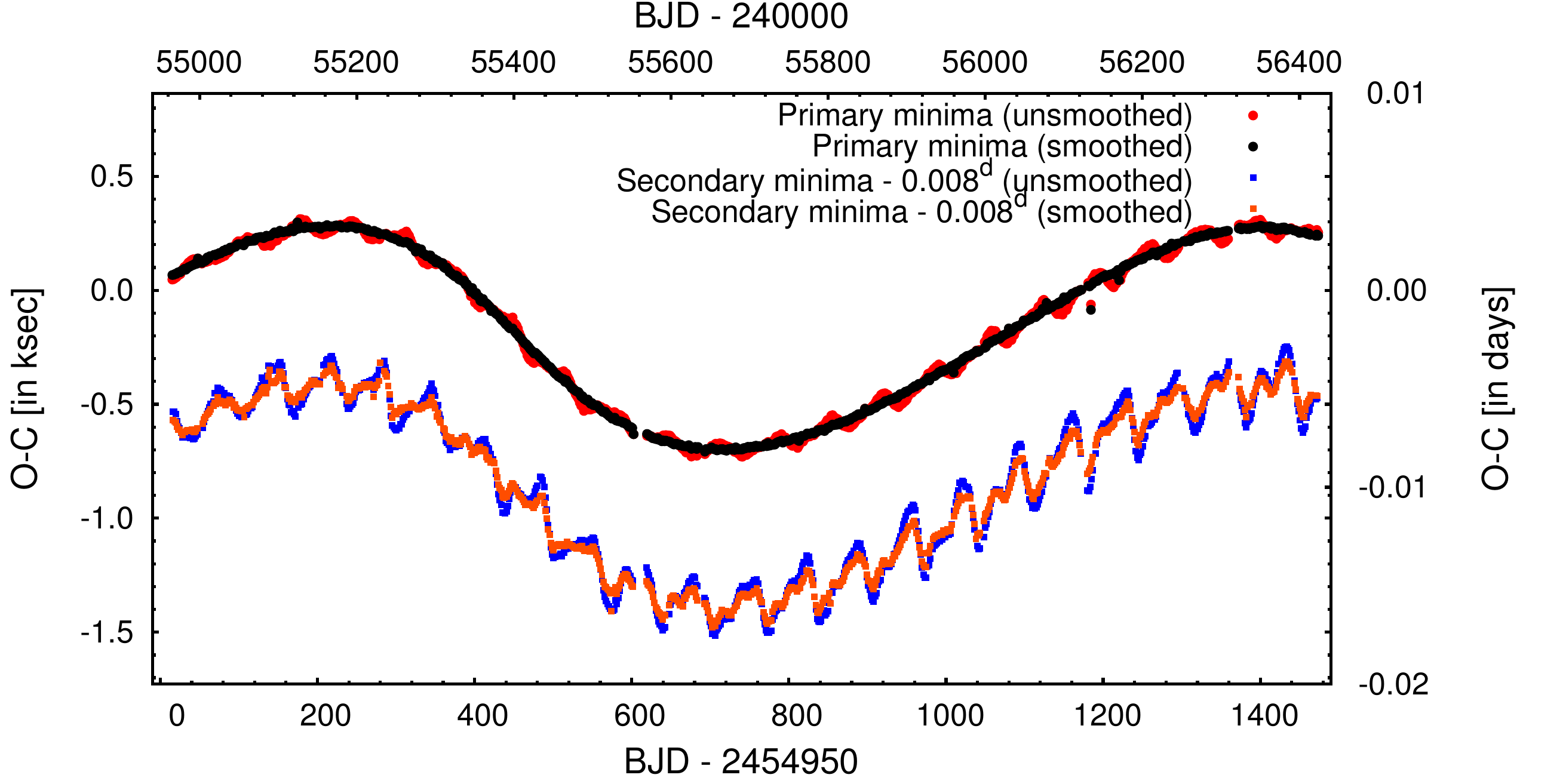}
 \caption{Two examples of the workings and efficiency of local smoothing with $4^{th}$-order polynomial fits. Both KIC~05216727 (upper row) and KIC~09711751 (bottom row) are Algol-type EBs, and exhibit likely rotational variations due to starspots. Left panels show small segments of their detrended {\em Kepler} long-cadence lightcurves (red) and the corresponding locally smoothed curves (blue). Right panels give the ETV curves obtained from both the original unsmoothed and the locally smoothed lightcurves. One can see that the method is more effective, and eliminates nearly perfectly the effects of the (rotational) distortions for the deeper primary minima. As to the shallower secondary minima, some residual distortions survive, but the magnitude has been substantially reduced. (For better visibility, the ETV curves for the secondary minima are shifted down from the primary ETV curves.)}
 \label{Fig:minsmoothing}
\end{figure*}

Another method useful for reducing or eliminating the influences of intrinsic brightness variations on the times of minima is local smoothing of the lightcurves. This method was applied by fitting a low-order (typically $4^\mathrm{th}$ order) polynomial to a portion of each light curve centered on each minimum but excluding the minimum itself, i.e., usually in the intervals [$-0.25$;${\phi_\mathrm{p,s}}_\mathrm{left}-0.02$] and [${\phi_\mathrm{p,s}}_\mathrm{right}+0.02$;$+0.25$]. This polynomial was then subtracted from the entire interval ($[-0.25;0.25]$;
see left panels of Fig.~\ref{Fig:minsmoothing}). This method yielded excellent results for several systems affected by star spots and even for systems affected by stellar oscillations. Some examples are shown in the right panels of Fig.~\ref{Fig:minsmoothing}. An additional example of the oscillating EB system KIC~08560861 can be found in Fig.~1 of \citet{borkovitsetal14}. Local smoothing was found to be effective mainly for detached systems with definite and sharp eclipses, but we could also use it even for some semi-detached binaries. For overcontact EBs and ELVs, however, this algorithm cannot be used.

\begin{figure*}
\includegraphics[width=84mm]{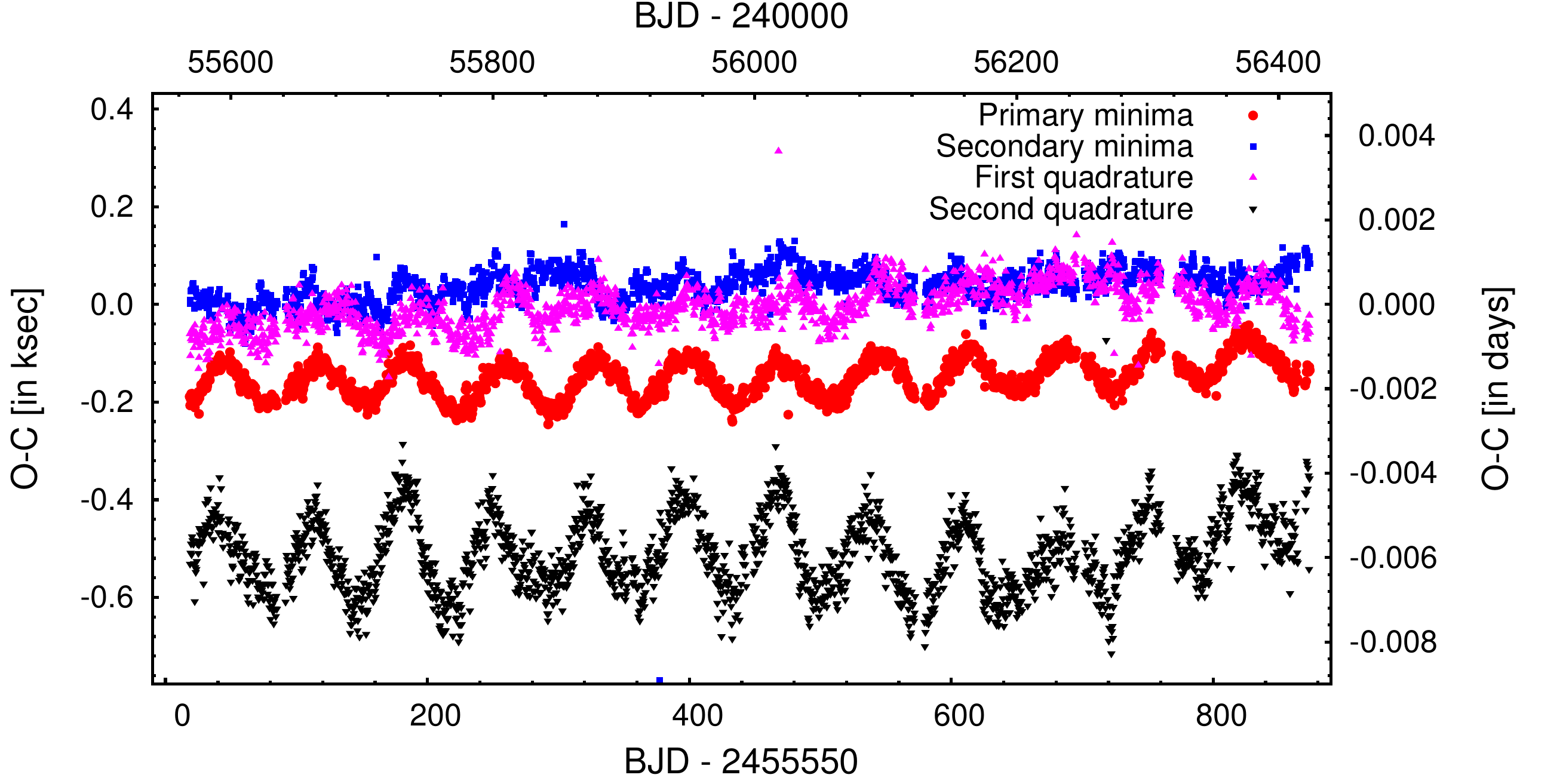}\includegraphics[width=84mm]{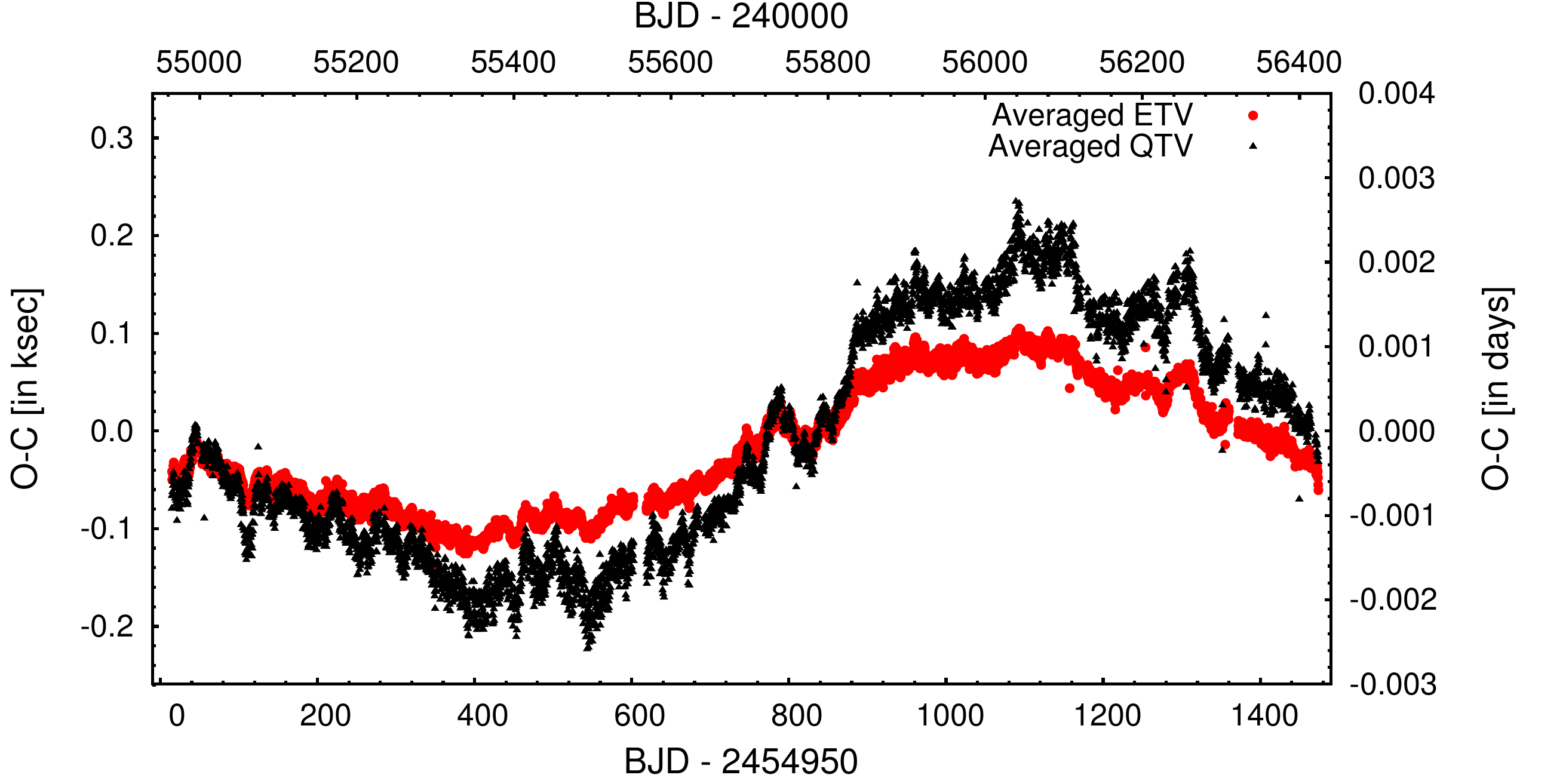}
\caption{Examples of discrepant ETVs and QTVs. {\it Left panel:} A clearly false positive case: the ETV and QTV curves of KIC~11247386. The $P_2=71.2\pm0.1$\,day-periodicity attributed to LTTE by \citet{conroyetal14} is readily visible. The different amplitudes for the different curves clearly show, however, that the origin of this feature cannot be LTTE due to a third body. Therefore, we categorized this system as a false positive in the sense that there is no evidence for this being a triple star system. (Note, careful inspection also reveals discrepancies in the relative phases of the variations in the different curves.) {\it Right panel:} The case of a marginally acceptable LTTE solution: the averaged ETV and QTV curves of KIC~11246163. Although the amplitudes of the two curves differ slightly, we do not rule out a possible LTTE origin.}
\label{Fig:FPornot}
\end{figure*}

The use of QTVs and the averaging and smoothing techniques made it possible not only to find and determine lower amplitude LTTE solutions, but also to apply more stringent criteria for filtering out false positive systems. An example is KIC~11247386,  a possible overcontact EB ($P_1=0.394$\,days) with a remarkable O'Connell-effect\footnote{Unequal brightness levels in the two quadratures, see, e.~g., \citet{oconnell51,milone68}}. \citet{conroyetal14} give an LTTE solution with a period $P_2=71.2\pm0.1$\,days, which would be the shortest outer period in their sample. As one can see in the left panel of Fig.~\ref{Fig:FPornot} this periodicity is definitely present in the primary ETV and the second QTV with quite different amplitudes, is hardly visible in the first QTV, and is even weaker in the secondary ETV. Such amplitude ratios are not typical of LTTE induced by a third body. There are a few additional systems where the averaged QTVs behave similarly to the averaged ETVs, but there are minor discrepancies in the amplitudes. A typical example is shown in the right panel of Fig.~\ref{Fig:FPornot}. In these latter cases we accept the LTTE solution, and note the amplitude discrepancy in the tabulated results.  

Another group of false positives comprises those objects where, although the ETV may suggest an LTTE solution, the {\em Kepler} target was erroneously classified as either an EB or as an ELV binary. For example, $\delta$ Scuti-type oscillating variables can easily be misclassified as ELVs or even low amplitude overcontact EBs. This is quite likely especially in systems with $P \lesssim 0.15$ days. In the absence of radial velocity measurements which would be able to confirm or reject the binary hypothesis for candidate ELV binaries \citep[see, e.g.][]{tal-oretal15}, we could make decisions on the nature of such systems based on temperature or color information when available. Instead, our decisions are based on the characteristics of the folded, binned lightcurves and the ETV and QTV data. 

\begin{figure*}
\includegraphics[width=60mm]{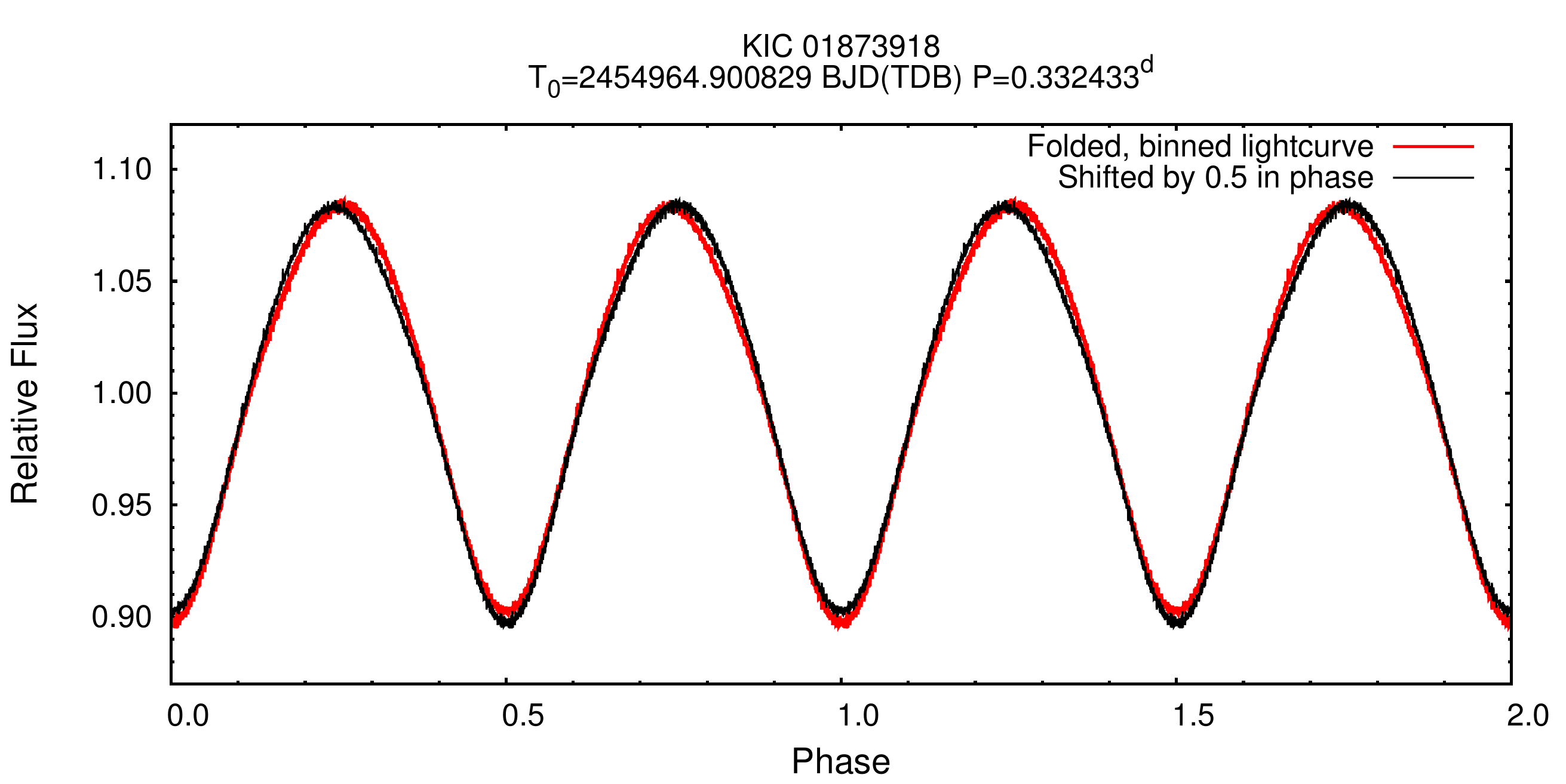}\includegraphics[width=60mm]{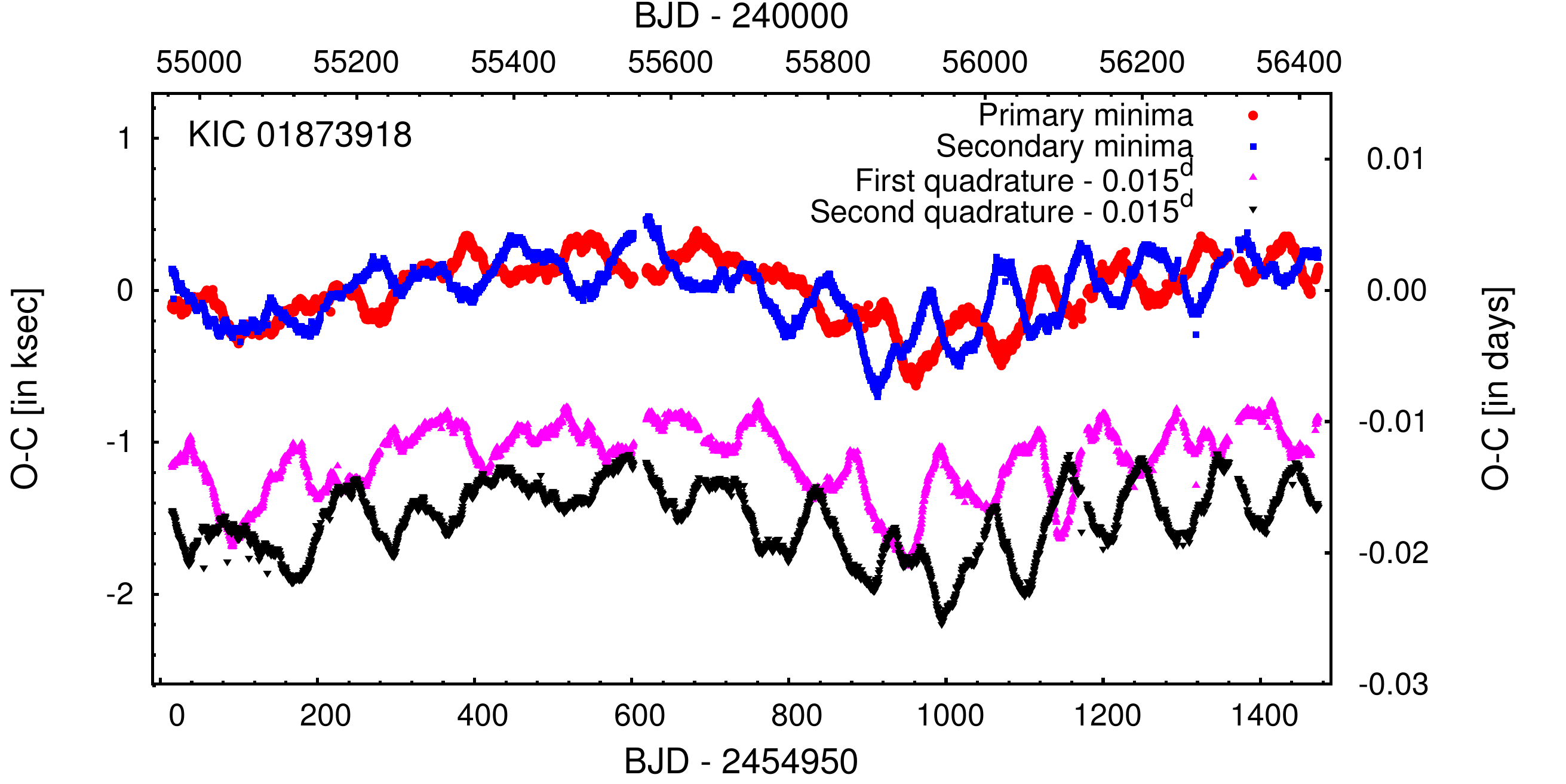}\includegraphics[width=60mm]{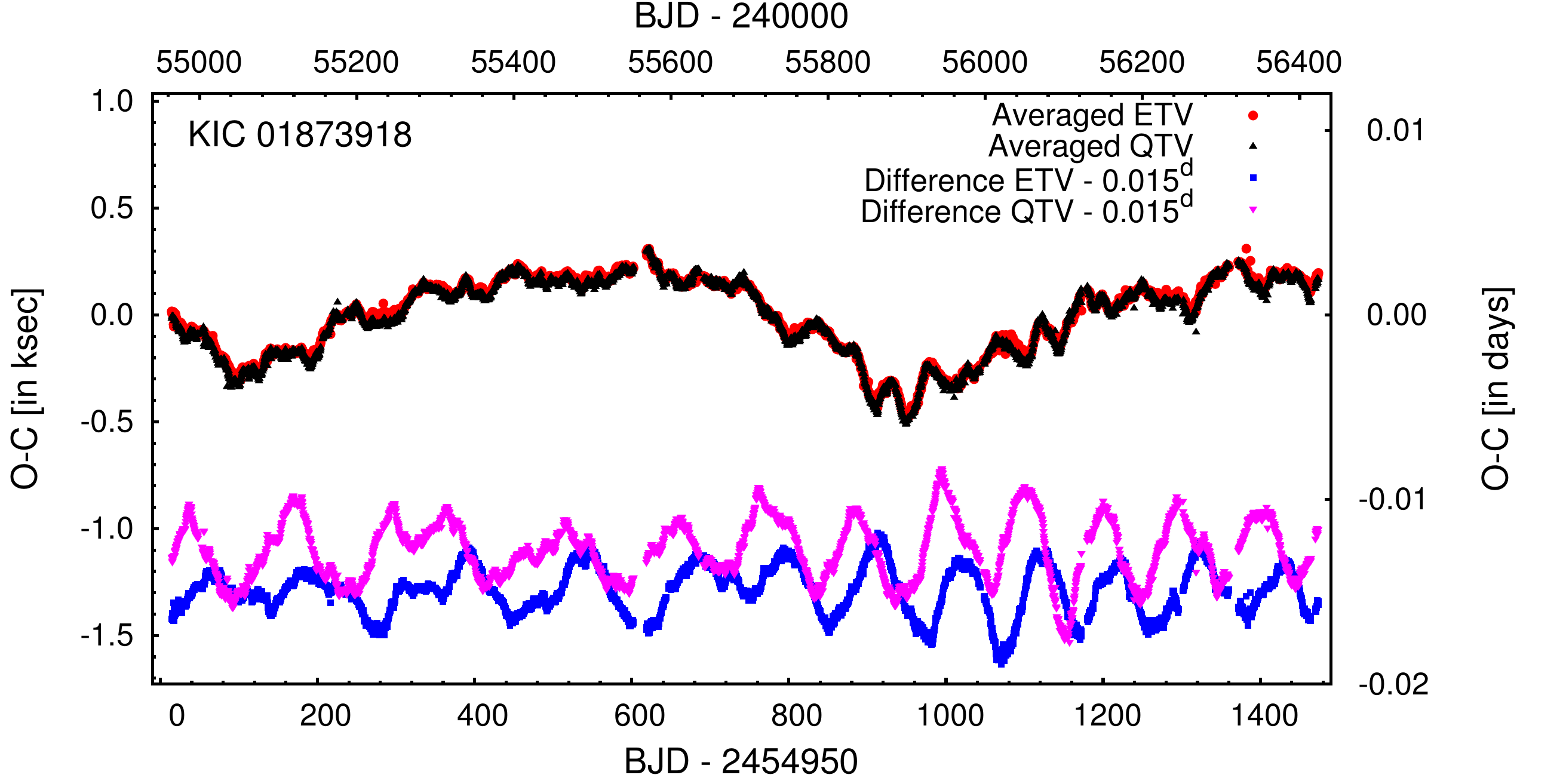}
\includegraphics[width=60mm]{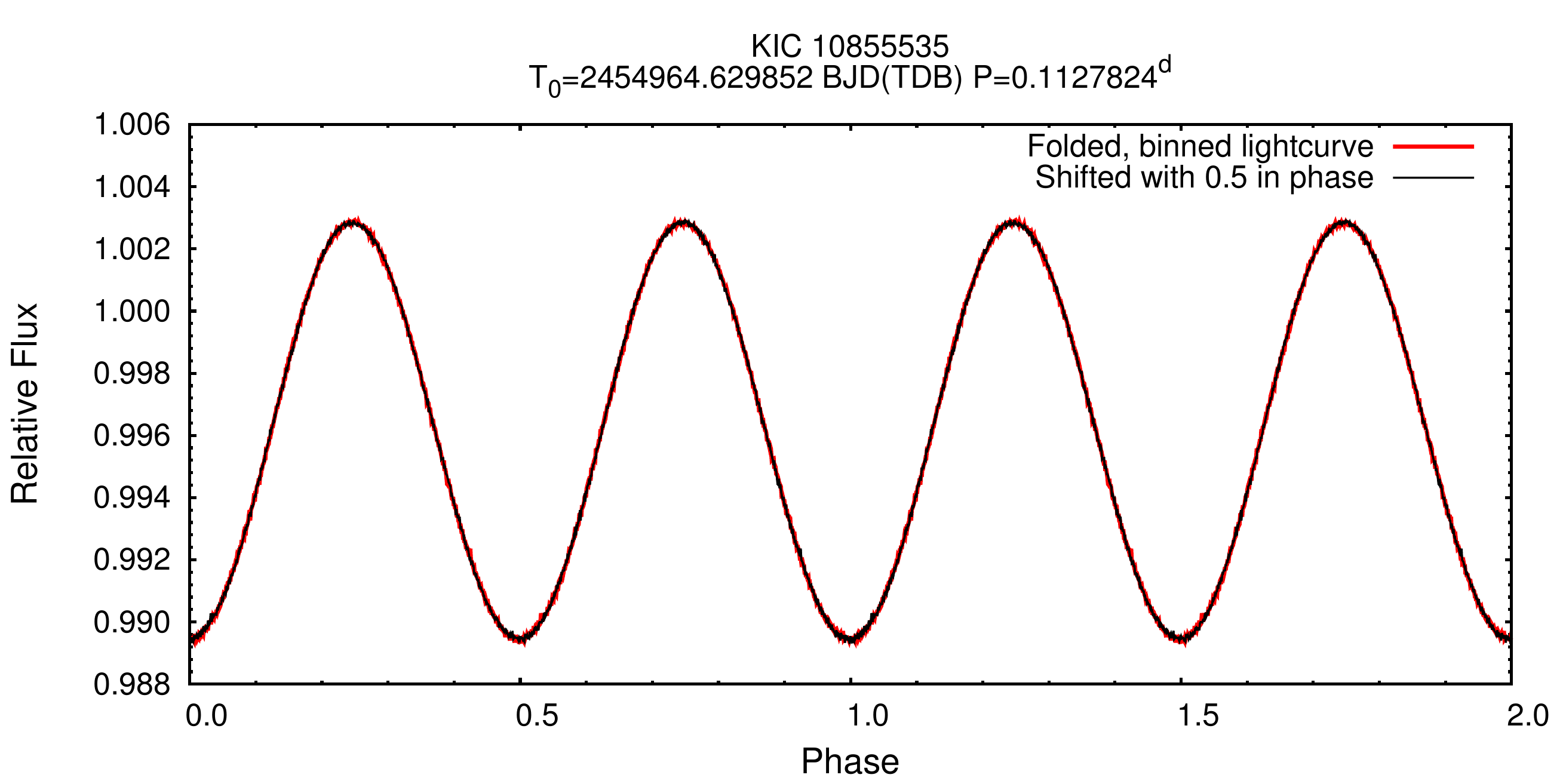}\includegraphics[width=60mm]{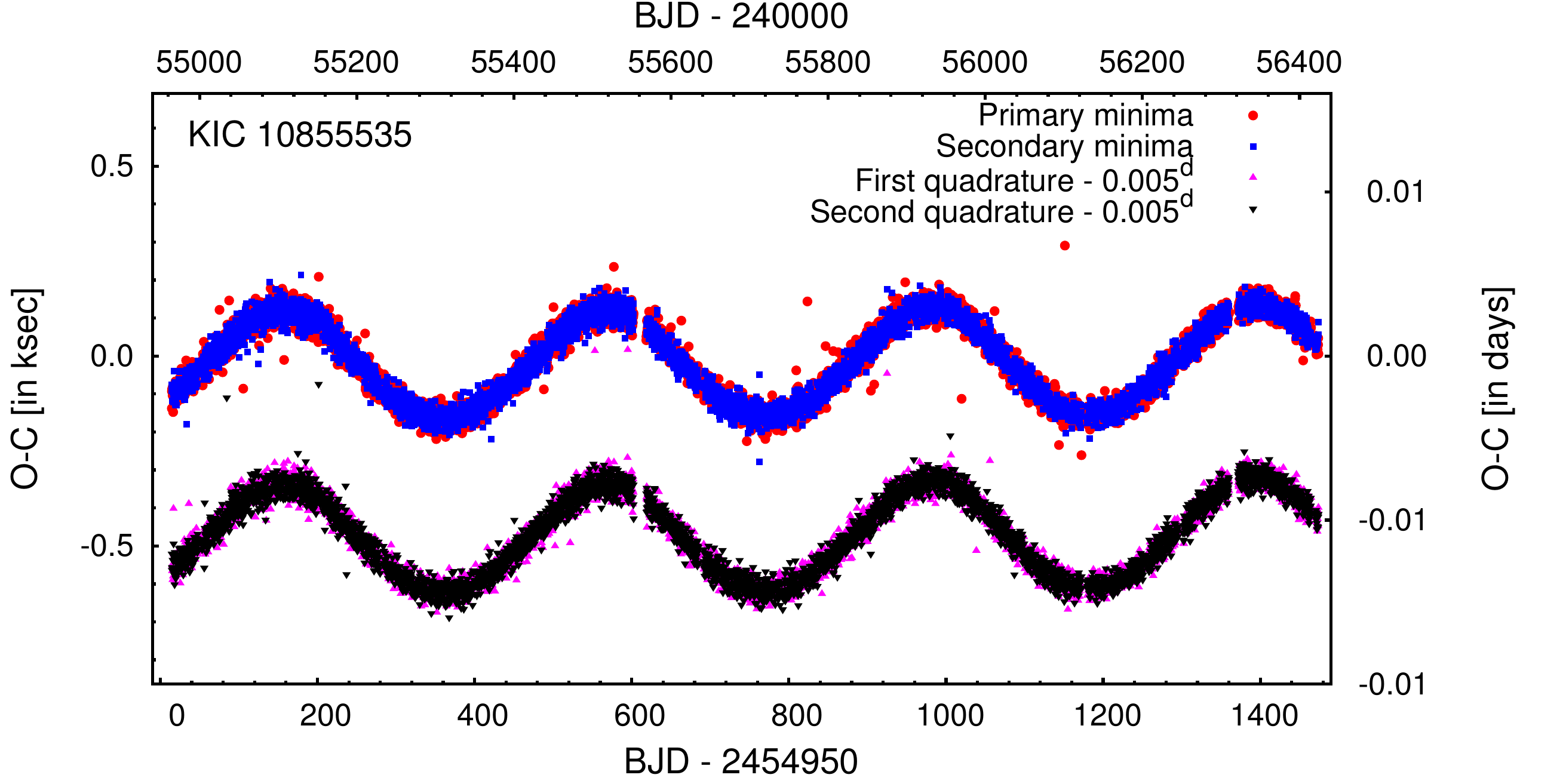}\includegraphics[width=60mm]{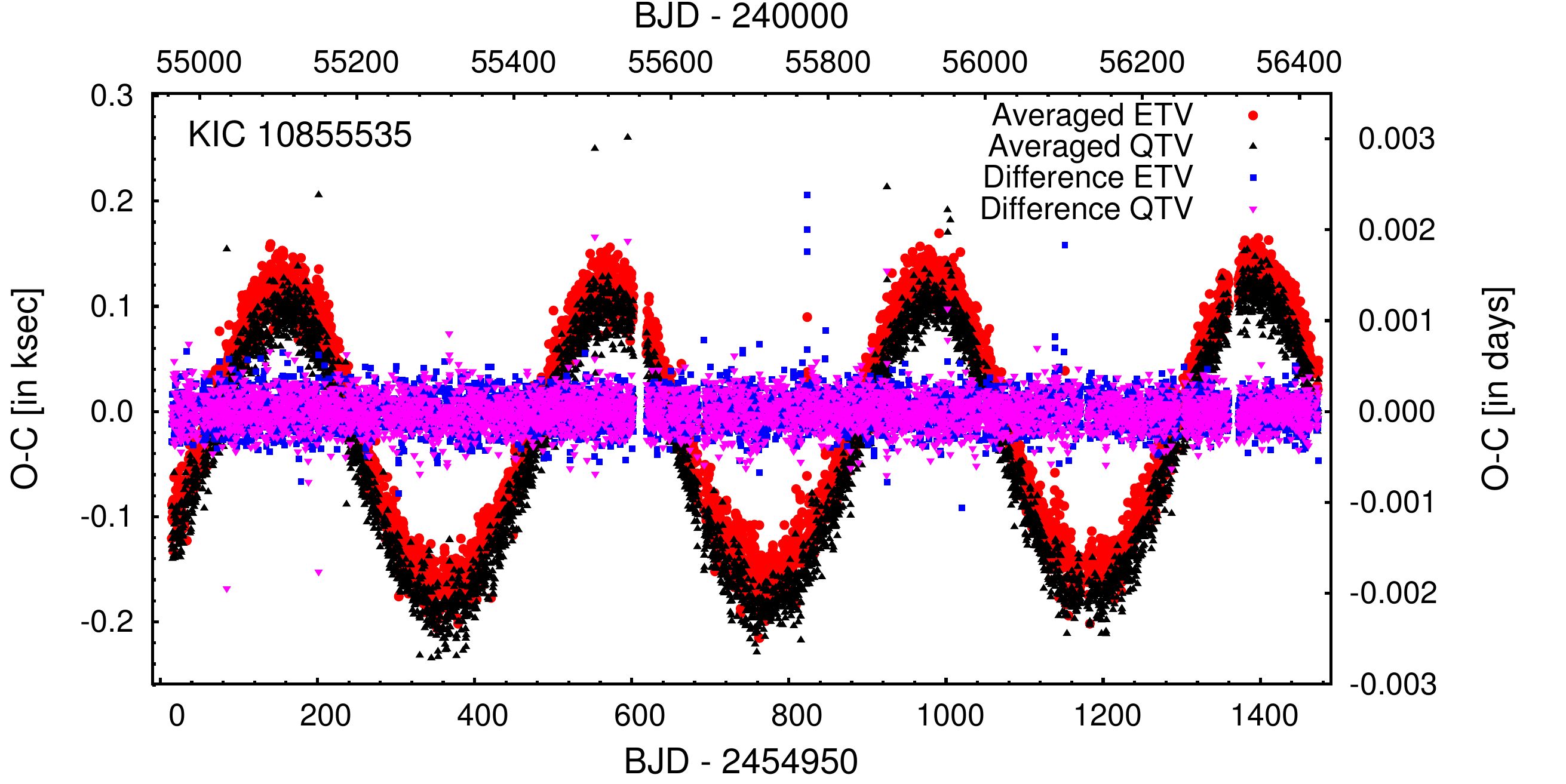}
\includegraphics[width=60mm]{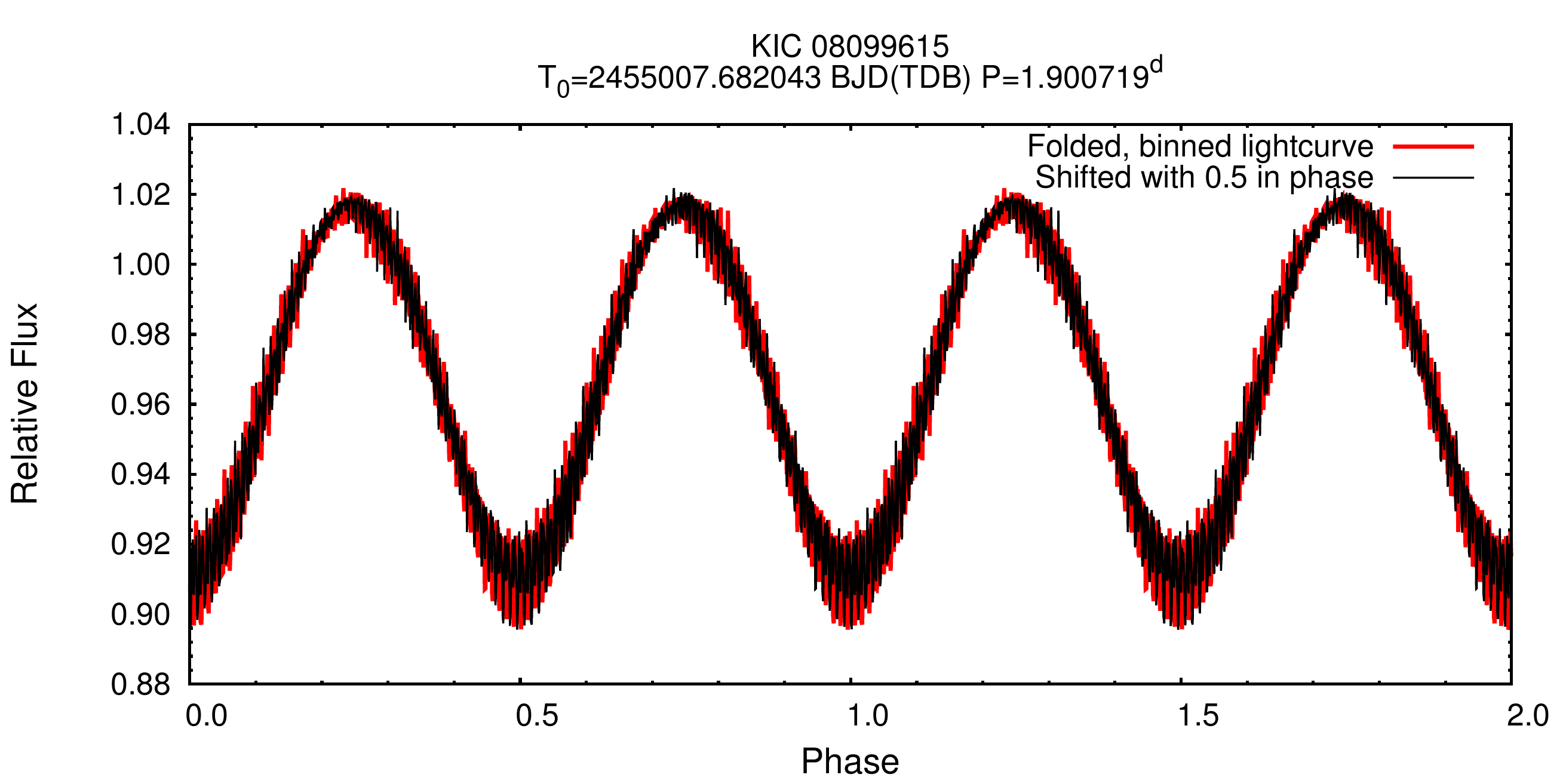}\includegraphics[width=60mm]{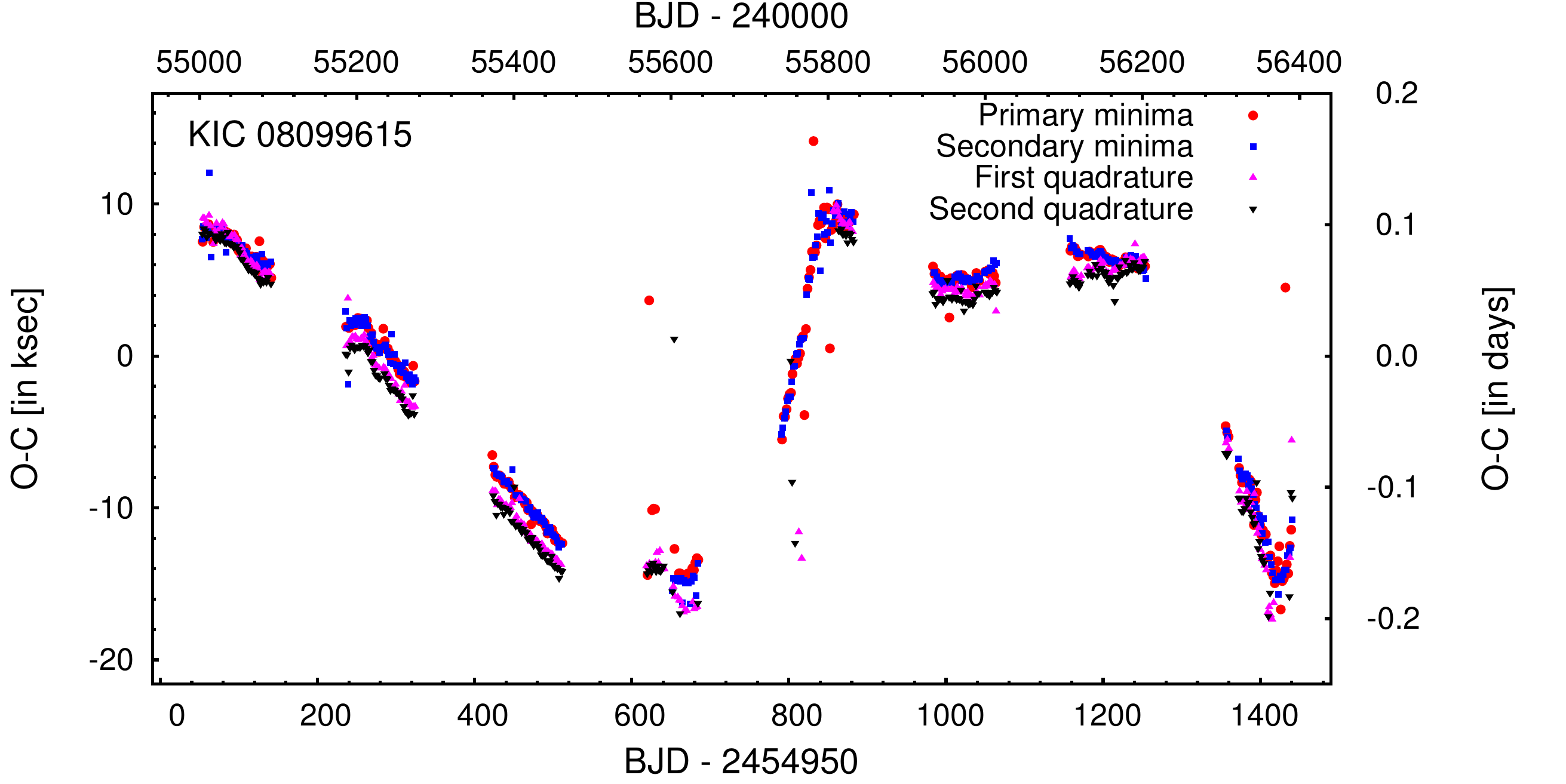}\includegraphics[width=60mm]{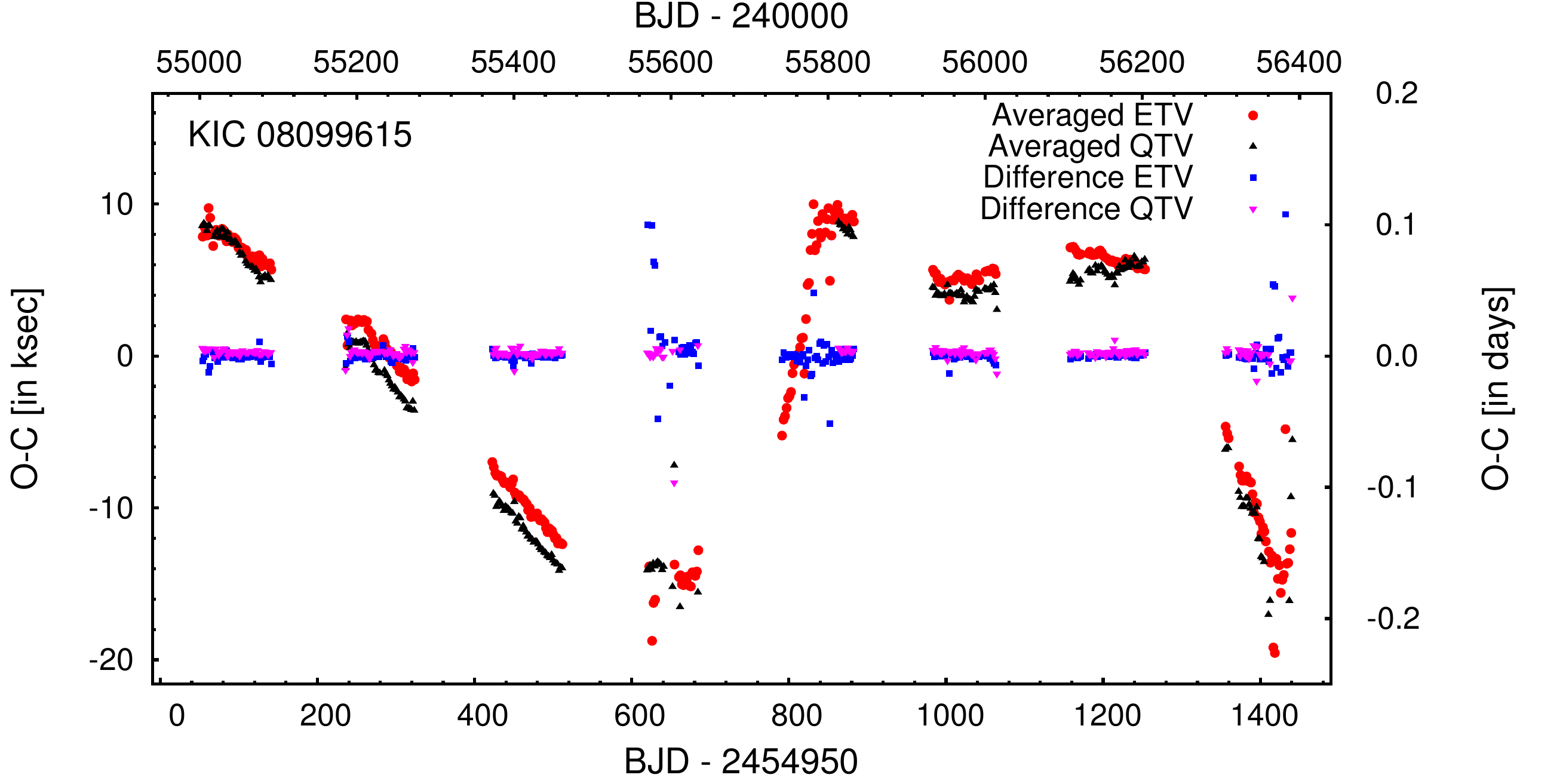}
\caption{One verification and two rejections. Three systems where, at first, the classifications as binaries are ambiguous. {\em Left panels:} The folded and binned long-cadence lightcurves and their $0\fp5$ phase-shifted versions (red and black, respectively). {\it Middle panels} show the individual $O-C$ curves belonging to the purported primary and secondary ETVs as well as the first and second QTVs. {\it Right panels:} The average and the difference of the two ETVs and QTVs are plotted. In the case of KIC~01873918, which has been flagged as a false positive in the {\em Kepler} EB Catalog ({\it first row}), the lightcurve shows alternating maxima and minima that are slightly different in amplitude, thereby indicating that this is not a sinusoidal pulsator. The quasi-anticorrelated behaviour in the ETV curves, and also in the QTV curves, adds confidence to this being a binary. Therefore, we conclude, that this system is indeed a binary within a triple system. In the case of the ultrashort period KIC~10855535 ({\it middle row}) and, the longer period KIC~08099615 ({\it bottom row}) the alternating maxima and minima of the lightcurves look completely the same, suggesting another type of variability with half of the given period. Furthermore, the two ETVs and also the QTVs track each other, which further strengthens the false binary hypothesis. Independent of this fact, the presence of the LTTE effect  in the ETV and QTV curves of KIC~10855535 seems very clear, and thus we may conclude that this system is actually a wide binary (instead of being a triple) with a period of $P_\mathrm{LTTE}=411.9\pm0.2$\,d. For KIC~08099615 the large amplitude, peculiar ETV (and QTV) might have a different origin.}
\label{Fig:FPornot2}
\end{figure*}

There are light curve based checks that may reveal whether a target is actually a physical binary. For example, the light curve of an ELV binary is dominated by a sinusoidal component with a period that is half of the orbital period. The two sections of the folded and binned light curve in the phase ranges [0.0; 0.5] and  [0.5;1.0] typically differ noticeably from each other due to Doppler boosting \citep[see, e.g.][]{vanKerketal10} and reflection effects \citep[see, e.g.][]{zuckeretal07}, not fully averaged-out spot effects, or even, in the case of detached ELVs, orbital eccentricity. On the other hand, if the variations originate in stellar oscillations or pulsations, the underlying oscillation period will be half of the inferred orbital period and the light curve will be more or less identical in the two phase intervals. Furthermore, in such cases the times of consecutive minima (or maxima) tend to consistently follow one ETV (or QTV) curve. By contrast, as was shown in \citet{tranetal13} and \citet{balajietal15}, in the case of ELVs and overcontact systems, especially those which are formed by spotted stars, consecutive minima and maxima timings may show different $O-C$ patterns. Therefore, in accord with the suggestion of \citet{tranetal13}, all the sources which produce significantly different pairs of ETVs (and/or pairs of QTVs) are not likely to be due to oscillations or pulsations. In summary, if the two sections of a light curve differ, the system may be a binary. If the two parts of the light curve happen to be identical, and ETVs and QTVs also look very similar, we take the source to be a false positive binary with a high likelihood. Examples of these checks may be found in Fig.~\ref{Fig:FPornot2}.

After obtaining preprocessed ETV and QTV curves and weeding out likely false positives in the above manner, the next task was to decide whether a pure LTTE solution, or a combined dynamical and LTTE solution, for a given system should be sought. In most cases the decision was evident as, on one hand, some of the ETVs had shown features typical of dynamical perturbations \citep[for a detailed discussion see][]{borkovitsetal15}, or, on the other hand, a large $P_2/P_1$ ratio indicated that dynamical contributions would be negligible. Extra care was necessary, however, for systems with relatively sinusoidal ETVs and moderate $P_2/P_1$ ratios. Therefore, in all cases, when a pure LTTE solution was obtained, we also estimated the possible relative contribution of the dynamical perturbations. For this, the binary mass was approximated by 2M$_\odot$ and then by the use of the mass function $f(m_\mathrm{C})$ obtained from the LTTE solution, a minimum mass of the third body was calculated, as well as the minimum value of the dynamical amplitude (Eqn.~\ref{Eq:A_dyn}). Then, when the estimated ratio $\mathcal{A}_\mathrm{dyn}/\mathcal{A}_\mathrm{LTTE}$ exceeded $\sim$25\% we also calculated a combined LTTE+dynamical solution.

\section{Supplemental ground-based eclipse timing}
\label{Sect:Ground-based minima}

Before giving an overview of our results, we briefly discuss the use of pre- or post-{\em Kepler} ground-based eclipse measurements which are available for a small number of our triples.  As was discussed in Section~\ref{Sect:ThirdbodyETVs}, because of their limited accuracy, ground-based timing measurements of eclipses and light curve minima are generally not suitable for third-body ETV studies in the period range of $P_2\lesssim1-2$\,years. Even for systems where the outer period is comparable to the length of the {\em Kepler} dataset, supplementary ground-based times of minima collected over a somewhat wider time span may serve mainly to confirm or reject a possible solution rather than to quantitatively improve it.  

Supplementary ground-based times of minima are most useful for those systems that were discovered as EBs well before the {\em Kepler} era when times of minima are available over a time span that is much longer, even by orders of magnitude, than the {\em Kepler} data train itself. However, our sample includes only seven EBs for which there are times of minima taken over a time span longer than a decade. Some dozens of our sample EBs, however, were observed a few years before the beginning of the original {\em Kepler} mission in the photometric surveys of ASAS \citep{pigulskietal09}, HATNET \citep{hartmanetal04}, TRESS \citep{devoretal08} and SuperWASP \citep{pollaccoetal06}. Unfortunately, the times of minima obtained from these data bases often offer only lesser benefits because of the restricted extension of the data span and, in several cases, the sampling rate of the observations of each EB was so infrequent that the data do not yield times of individual eclipses with useful accuracy.  For these reasons we did not determine and utilize the times of minima from the observations of the above listed surveys for all systems, we make use of the data only for those EBs for which the eclipse times were determined by \citet{leeetal14} and \citet{zascheetal15}.

Ground-based times of minima obtained from targeted eclipse observations of individual binaries are particularly helpful. In most cases these were collected from the Lichtenknecker-Database of the Bundesdeutsche Arbeitsgemeinschaft f\"ur Ver\"anderliche Sterne e. V. (BAV)\footnote{http://bav-astro.eu/LkDB/index.php?lang=en}, rather than from the journal literature. The sole exceptions are a few recently observed post-{\em Kepler} times of minima published in \citet{zascheetal15}. Some of the oldest times of minima in the extended ground-based datasets are based on visual brightness estimations which have a highly limited accuracy of 5--10 minutes. Despite this, we decided to keep these observations, with the exception of the evident outliers, in cases where they substantially extend the overall span of the data. 

Given the available data and the above considerations, we were able to extend our timing datasets with ground-based measurements for about a dozen systems. In some cases, however, the ground-based minima evidently contradict the {\em Kepler} observations.   In the case of KIC~092883826 (=V2366~Cyg), we found two ground-based times of minimum which were obtained from observations during the {\em Kepler}-era but were inconsistent with the {\em Kepler} times. In the case of KIC~09101279 (=V1580~Cyg), three ground-based times of minimum would extend the data span by a factor of three, but they do not match our ETV-solution from the {\em Kepler} data and, therefore, were not considered further. In a case yielding an opposite conclusion, for KIC~010581918 (=WX~Dra) we rejected the {\em Kepler} LTTE solution, and therefore deleted the EB from our sample because of the contradictory characteristics of the relatively numerous ground-based data. 

In summary we kept all or a part of the ground-based times of minima for eight EBs. In the 221 panels of Fig.~\ref{Fig:ETVs} we plot the $O-C$ curves of almost all of the investigated EBs (see Section~\ref{Sect:Overview}). For the eight systems where ground based minima were also incorporated for the third-body solution, these ground-based minima are plotted together with their uncertainties. As one can see, these uncertainties in some cases are larger than the full amplitude of the third-body ETV feature for these systems. In other cases, however, the extended dataset was found to be suitable for confirming, or even improving, the third-body solutions as will be discussed in the next section. 

\section{Overview of the investigated systems}
\label{Sect:Overview}

We give LTTE and/or dynamical solutions for 230 {\em Kepler} systems. Some parameters of these systems are given in Table~\ref{Tab:Systemproperties}, where the basic properties of the generated ETV and QTV curves, the types of our solutions, and relevant references are also listed. 

Table~\ref{Tab:Systemproperties} contains a column for the system type and a column for "morphology". The correct classification of each binary as an ELV or as a one of the subtypes of real EBs (EA, EB, EW)\footnote{For the definitions of these lightcurve morphology classes, see e.~g., \citet{kallrathmilone09}} generally is difficult except when the binary is well-detached. It is particularly difficult to distinguish low-amplitude overcontact systems (EW) from ELVs, especially when there is a significant amount of third light due to either a bound third star in the system or an unresolved background or foreground light source. Among true EBs, it is difficult to distinguish EWs having low filling factors from tight semi-detached systems; AW UMa is a good example \citep{pribullarucinski08}. Similarly, among ELVs, it may be problematic to separate low inclination overcontact systems from not so low inclination, semi-detached binaries. Finally, the possible misidentification of some kinds of pulsating variables as ELVs or EBs was already noted above. The classifications given in the second column of Table~\ref{Tab:Systemproperties} should be considered with these caveats in mind. The reader may compare our classification results with the automated light curve morphology classifications of \citet{matijevicetal12} that are given in the third column of Table~\ref{Tab:Systemproperties}.  The classifications in the two columns are more or less consistent at least apart from the ambiguities between ELVs and EWs which remain unresolvable by either method.  Note also that our sample contains ten EBs which exhibit extra eclipse event(s) which most probably can be attributed to third bodies which are the subject of our investigations. Amongst these systems, the triply eclipsing nature of KIC~09007918, according to our knowledge, is reported here for the first time. These systems are marked in the second column of Table~\ref{Tab:Systemproperties} with an additional `E3' sign, and will be discussed shortly in Subsect.~\ref{Subsect:extraeclipses}.

The fourth and fifth columns give the epochs and periods that were used for initial light curve folding and binning, template calculations, determinations of times of minima, and for calculating the ETV and QTV $O-C$ diagrams. As corrections of the epoch and period were always obtained during our fitting process via the polynomial coefficients $c_0$ and $c_1$, the final epoch and period values differ from the values in this table. The other columns of Table~\ref{Tab:Systemproperties} are either self-explanatory or are explained in the table notes.


\setlength{\tabcolsep}{4.0pt}
\begin{table*}
\begin{center}
\caption{Properties of the investigated systems} 
\label{Tab:Systemproperties}  
\begin{tabular}{lccccccccccc} 
\hline
KIC No. & Type & Morph. & $T_0$ & $P_1$ & $K_\mathrm{p}$&data length&$\frac{\mathrm{ETV}}{\mathrm{QTV}}$&Fitted&Fit&Tab&Refs. \\ 
        &      &        & (MBJD)& (days)&     (mag)     & (days)&&curves&type&&\\
\hline
1873918 &ELV(EW)&0.86&54964.900829&0.332433& 13.7   & 1459      & 2/2   & a &l+q&L2-13& 7\\
2302092 & EW    &0.89&54964.694441&0.294673& 14.4   & 1459      & 2/2   & a & l &L2-27& 3\\
2305372 & EA    &0.58&54965.956227&1.4046920&13.8   & 1458(4216)& 2s4/2 &p(+e)&l(+q)&L3-25&6,23\\
2450566 & ELV   &0.98&55001.560102&1.8445871&11.7   & 1468      & 2/2   & a & l &L2-24& 3\\
2576692 & EA    &0.04&55027.103323&87.8782329&12.7  & 1406      & 2/0   &p+s&l+d&D3-08&  \\
2708156 & EA    &0.57&54954.336095&1.8912671&10.7   & 33912     & 2s8/0 &p+e&l+c&L1-38& 1\\ 
2715007 & ELV   &0.87&54964.783119&0.2971105&14.7   & 1459      & 2/2   & a & l &L3-17& \\
2715417 &ELV(EW)&0.76&54964.667658&0.2364399&14.1   & 1459      & 2/2   & a &l(+q)&L2-15& \\ 
2835289 &ELV+E3 &0.92&55000.444609&0.857762& 13.0   & 1469      & 2/2   & a & l &L1-35& 3,8\\
2856960 &EA+E3  &0.60&54964.661805&0.258507& 15.6   & 1458      & 2/0   & a &l+q&L1-03& 3,9,10,11\\
2983113 & EW    &0.89&55001.969640&0.3951601&15.2   & 1238      & 2/2   & a & l &L3-04& 3\\
3114667 & EA    &0.52&54999.758222&0.8885832&17.4   &  683      & 2s4/0 & a & l &L2-02 & 3\\
3228863 & EB    &0.65&54954.26185 &0.730944& 11.8   & 6636      & 2/2   &a+e&l+q&L1-29& 2,3,12,22\\
3245776 & ELV   &0.96&55001.663004&1.4920589&14.4   & 1458      & 2/2   & a & l &L1-30& 3\\
3248019 & EA    &0.37&55098.778000&2.6682057&15.4   & 1329      & 2/0   & a & l &L3-24& \\
3335816 & EA    &0.16&54954.355631&7.4220263&12.1   & 1462      & 2/0   & a & l &L3-38& \\
3338660 & EA    &0.60&55002.262623&1.8733806&14.8   &  852      & 2s4/0 & p & l &L2-07& \\
3345675 & EA    &0.00&55083.146716&120.0040103&15.6 & 1320      & 1/0   & p &l+d&D3-03& \\
3440230 & EA    &0.54&54967.238413&2.8811010&13.6   & 1455      & 2s4/2 & p &l+q&L2-35& 1,6\\
3544694 & EA    &0.29&55740.65102 &3.845728& 15.9   &  683      & 2s8/0 &p+s&l+d&D1-05& \\
3766353 & EA(HB)&$-1.00$&54966.722264&2.666966&14.0 & 1456      & 2/2   & p & l &L3-12& 3\\
3839964 &ELV(EW)&0.78&54964.792432&0.2561499&14.6   & 1459      & 2/2   & a & l+q&L3-40& 3\\
3853259 &ELV(EW)&0.98&54964.781808&0.2766478&13.9   & 1459      & 2/2   & a & l+q&L1-10&  \\
4037163 & EA    &0.58&55000.227976&0.6354447&16.7   & 684       & 2/0   & a &l(+q)&L1-07&3\\
4055092 & EA    &0.01&54966.932772&76.464989&15.3   & 1404      & 2/0   &p+s&l+d&D3-16& \\
4069063 & EA    &0.55&54964.906342&0.5042953&13.3   & 1452      & 2s4/0 & a & l &L2-18& 3\\
4074708 & EW    &0.73&54964.856673&0.3021166&15.4   & 1459      & 2/2   & a & l &L2-37& 3\\
4078157 & EA    &0.08&54960.300077&16.025671&15.5   & 1202      & 2/0   &p+s&l+d&D3-02& \\
4079530 & EA    &0.07&54994.805374&17.7271000&15.6  &  579      & 2/0   &p+s&l+d&D1-12& \\
4138301 & ELV   &0.90&54964.685221&0.253379& 14.7   & 1459      & 2/2   & a &l(+q)&L2-14&3\\
4174507 & EA    &0.24&54966.041640&3.891825& 15.4   & 1456      & 2/0   & a &l(+d)&L3-31& \\
4244929 & EW    &0.91&54964.747256&0.341403& 15.1   & 1459      & 2/2   & a & l &L2-55& 3\\
4451148 & EW    &0.82&54954.385233&0.7359815&11.2   & 1470      & 2/2   & a & l &L2-06& 3\\
4547308 & ELV   &0.88&54953.635293&0.5769278&12.5   & 1470      & 2/2   & a & l &L2-17& 3\\
4574310 & EA    &0.56&54954.662614&1.3062201&13.2   & 1468      & 2s4/2 & p & l &L2-56& 23\\
4647652 & EB    &0.68&54953.945894&1.06482495&11.8  & 1470      & 2/2   & p & l &L2-08& 2,3\\
4670267 & EA    &0.60&54966.375624&2.0060974&15.1   & 1456      & 2s4/2 & a &l(+d)&L2-09&3\\
4681152 & EA    &0.55&54954.060778&1.835930& 13.1   & 1456      & 2s4/2 & p & l &L2-43& 3\\
4753988 & EA    &0.16&54968.025737&7.304476& 15.0   & 1454      & 2/0   &p+s&l+d&D3-14& \\
4758368 & EA    &0.57&54958.206761&3.749935& 10.8   & 1468      & 2s4/2 &p+s&l+a&L3-45& 3,13\\
4762887 & ELV   &0.95&54964.771668&0.7365737&14.4   & 1458      & 2/2   & a & l &L2-47& 3\\
4769799 & EA    &0.12&54968.515532&21.928614&10.9   & 1438      & 2/0   &p+s&l+d&D2-11& 4,5\\
4848423 & EA    &0.48&55000.595941&3.003613& 11.8   &  922      & 2/2   & a & l &L3-03& 1,23\\
4859432 & EW    &0.76&54949.996305&0.3854799&15.5   & 1421      & 2/2   & a &l(+q)&L2-05&3\\
4909707 & EB    &0.72&54953.913193&2.3023675&10.7   & 1470      & 2/2   &p+s&l+d&D1-28& 2,3\\
4937217 & EW    &0.82&54964.627330&0.4293416&15.4   & 1459      & 2/2   & a &l+q&L3-42& 3\\
\hline	
\end{tabular}
\end{center}

{\bf Notes.} {(1) E3 refers to tertiary eclipse(s) in the lightcurve. (2) In columns 2 and 3 we give the lightcurve classifications according to both the classical eclipsing binary typology \citep[see, e.~g.,][]{kallrathmilone09} and the recently introduced morphology of \citet{matijevicetal12}. (3) Sidereal period ($P_1$) and epoch ($T_0$) were used for plotting $O-C$ curves. (4) {\em Kepler} magnitudes were taken from the Kepler Input Catalog \citep{batalha10}. (5) In column ETV/QTV the number of calculated ETV and QTV curves are given. If both ETVs and/or QTVs were obtained, their average and (half-difference) curves were also determined. In cases where we used local smoothing polynomials on the lightcurves, this is denoted by putting s$n$ after the ETV number, where $n$ gives the order of the smoothing polynomial. (6) Abbreviations in ``Fitted curves'' column: `p' -- primary, `s' -- secondary, `a' -- averaged ETV curves, `e' -- ground-based times of minima were also included; (7) Abbreviations in ``Fit type'' column: `l' -- LTTE; `a' -- AME (noted separately only for non-`d'-type solutions); `d' -- dynamical; `q' -- quadratic; `c' -- cubic. Parentheses in this column indicate that two types of fits were performed; the unparenthesized terms were included in both fits while the term(s) in parentheses were included in only the less preferred fit. (8) Column ``Tab'' is the location of the solution of the given system in one of the Tables~\ref{Tab:Orbelem}--\ref{Tab:Orbelem3}, \ref{Tab:Orbelemdyn1}--\ref{Tab:Orbelemdyn3} and \ref{Tab:OrbelemFP} (`L1'--`L3' for pure LTTE, `D1'--`D3' for combined LTTE and dynamical, and `F' for false positive systems, respectively).  \\
{\small References: 1: \citet{giesetal12}; 2:\citet{rappaportetal13}; 3: \citet{conroyetal14}; 4: \citet{borkovitsetal15}; 5: \cite{orosz15}; 6: \citet{zascheetal15}; 7: \citet{tranetal13}; 8: \citet{conroyetal15}; 9: \citet{armstrongetal12}; 10: \citet{leeetal13}; 11: \citet{marshetal14}; 12: \citet{leeetal14}; 13: \citet{gaulmeetal13}; 14: \citet{leeetal15}; 15: \citet{carteretal11}; 16: \citet{borkovitsetal13}; 17: \citet{masudaetal15}; 18: Fabrycky~et~al., in prep.; 19: \citet{steffenetal11}; 20: \citet{baranetal15}; 21: \citet{liska14}; 22: \citet{csizmadiasandor01}; 23: \citet{giesetal15}}}
\end{table*}

\addtocounter{table}{-1}
\setlength{\tabcolsep}{4.0pt}

\begin{table*}
\begin{center}
\caption{({\em continued})} 
\begin{tabular}{lccccccccccc} 
\hline
KIC No. & Type & Morph. & $T_0$ & $P_1$ & $K_\mathrm{p}$&data length&$\frac{\mathrm{ETV}}{\mathrm{QTV}}$&Fitted&Fit&Tab&Refs. \\ 
        &      &        & (MBJD)& (days)&     (mag)	& (days)&&curves&type&&\\
\hline
4940201 & EA    &0.15&54967.276926&8.816578& 15.0   & 1455      & 2/0   &p+s&l+d&D1-23& 2,4,5\\
4945857 & EW    &0.74&54964.830222&0.335416& 14.0   & 1459      & 2/2   & a & l &L2-59& 3\\
4948863 & EA    &0.10&54972.831420&8.6435903&15.4   & 1452      & 2/0   &p+s&l+d&D2-09& \\
5003117 & EA    &0.37&54986.095638&37.610001&14.0   & 1429      & 2/0   &p+s&l+d&D3-06& 4,5\\
5039441 & EA    &0.39&54955.351360&2.151383& 12.9   & 1469      & 2s4/0 &p+s&l+a&L1-33& 2\\
5080652 & EA    &0.30&54968.166959&4.144357& 15.1   & 1422      & 2/0   & p &l+d&D1-15& \\
5095269 & EA    &0.05&54966.865286&18.6119616&13.5  & 1433      & 1/0   & p &l+d&D1-16& 5\\
5128972 & EW    &0.74&54965.047601&0.505323& 13.2   & 1459      & 2/2   & a & l &L1-16& 2,3\\
5216727 & EA    &0.48&54964.929149&1.513023& 13.4   & 1459      & 2s4/2 & p & l &L1-22& \\
5255552 & EA+E3 &0.17&54970.636491&32.458635&15.2   & 1414      & 2/0   &p+s&l+d&D1-31& 4,5\\
5264818 & ELV   &0.92&54955.241047&1.905050&  8.9   & 1469      & 2/0   & a &l+d(+q)&D1-20&2,3\\
5269407 & EA    &0.53&54965.651124&0.9588631&14.2   & 1458      & 2s4/0 & a & l &L3-30& 3\\
5307780 & EW    &0.88&54964.977524&0.308851& 14.9   & 1459      & 2/2   & a &l+q&L2-38& \\
5310387 & EW    &0.96&54953.664664&0.441669& 12.7   & 1470      & 2/2   & a &l+q&L1-04& 2,3\\ 
5353374 & EW    &0.78&54964.661848&0.3933205&14.1   & 1459      & 2/2   & a & l &L3-11& 3\\
5376552 & EW    &0.82&54954.083210&0.5038188&12.9   & 1470      & 2/2   & a &l(+q)&L1-11&2,3\\ 
5384802 & EA    &0.17&54966.988768&6.083093& 13.7   & 1454      & 2/0   & a &l+d&D1-19& 2,5\\
5459373 & ELV   &0.97&54964.670887&0.2866088&15.1   & 1459      & 2/2   & a & l &L1-14& 3\\
5478466 & EW    &0.97&54964.859645&0.4825005&14.2   & 1459      & 2/2   & a & l &L2-04& 3\\
5513861 & EA    &0.57&54954.995935&1.5102117&11.6   & 3010      & 2/2   &a+e& l &L2-63& 1,3,6,23\\
5611561 &ELV(EW)&0.74&55000.011420&0.25869465&14.0  & 1421      & 2/2   & a & l &L2-33& 3\\
5621294 & EA    &0.60&54954.510518&0.938905& 13.6   & 1470      & 2s4/2 & p &l+q&L2-36& 1,6,14\\
5653126 & EA    &0.09&54985.913152&38.493382&13.2   & 1424      & 2/0   &p+s&l+d&D2-06& 4,5\\
5731312 & EA    &0.08&54968.093163&7.946382& 13.8   & 1456      & 2/0   &p+s&l+d&D2-05& 4,5\\
5771589 & EA    &0.12&54962.130765&10.738342&11.8   & 1434      & 2/0   &p+s&l+d&D1-10& 2,4,5\\
5897826 & EA+E3 & .. &55069.313   &1.76713  &13.1   &  ..       & ..    & ..& ..&D1-01& 15\\
5903301 & EA    &0.41&55003.431007&2.320302& 15.1   & 1330      & 2/2   & a & l &L2-49& \\
5952403 & EA+E3 &0.52&55051.237191&0.9056774& 7.0   & 1426      & 2s4/0 & a &l+d(+q)&D1-02&16\\
5956776 & EA    &0.61&55000.305505&0.5691150&16.7   &  855      & 2s4/2 & p & l &L3-21& \\
5962716 & EA    &0.47&54965.398009&1.804586& 13.9   & 1458      & 2s4/0 & p & l &L3-32& \\
5975712 & ELV   &0.87&54953.924190&1.136083& 11.5   & 1469      & 2/2   & a &l(+q)&L3-39&3\\
6103049 & EA    &0.59&54964.888912&0.6431712&15.1   & 1426      & 2s4/0 & a & l &L3-09& \\
6144827 & ELV   &0.79&54964.642040&0.234650& 15.0   & 1459      & 2/2   & a &l+q&L1-05& 3\\
6233903 & EA    &0.36&55001.719115&5.9908477&16.5   &  851      & 2s4/2 &p+s&l+a&L3-54& \\
6265720 & EW    &0.93&54964.729666&0.3124277&14.8   & 1426      & 2/2   & a & l &L3-06& 3\\ 
6281103 &ELV(EW)&0.98&54964.870642&0.3632811&14.9   & 1459      & 2/2   & a &l+q&L2-50& 3\\
6287172 & FP?   &0.95&54953.651911&0.2038732(/2)&12.7&1469      &2/2(1/1)&a & l &F-06& \\
6370665 & EW    &0.96&54965.405240&0.9323155&14.0   & 1458      & 2/2   & a &l+q&L1-08& 2,3\\
6516874 & EA    &0.60&55001.4643225&0.9163260&15.9  & 1237      & 2s4/0 & a & l &L2-20& 3\\
6525196 & EA    &0.36&54954.353139&3.420598& 10.2   & 1467      & 2s4/0 & a &l+d&D1-26& 2\\
6531485 & EA    &0.53&54964.801481&0.676990& 15.6   & 1459      & 2/0   &p+s&l+d&D1-03& 2\\
6543674 & EA+E3 &0.53&54965.303847&2.391030& 13.5   & 1456      & 2s4/2 & a & l &L1-36& 3,17\\
6545018 & EA    &0.42&54965.835642&3.991460& 13.7   & 1457      & 2/2   &p+s&l+d&D1-07& 2,4,5\\
6546508 & EA    &0.20&55189.798579&6.107057& 15.7   & 1237      & 2/0   &p+s&l+d&D2-10& \\
6606282 & EA    &0.31&54965.433543&2.107130& 13.0   & 1456      & 2/0   & a & l &L3-22& \\
6615041 & EW    &0.75&54964.807732&0.3400856&13.9   & 1459      & 2/2   & a &l(+q)&L3-49&3\\
6669809 & EB    &0.64&54953.997571&0.7337388&10.8   & 1437      & 2s4/2 & p &l+c&L1-02& \\
6671698 & EW    &0.73&54954.077303&0.471525& 13.5   & 1437      & 2/2   & a &l+q&L2-52& 3\\
6766325 &ELV(EW)&0.92&54964.713835&0.4399657&13.8   & 1459      & 2/2   & a & l &L3-26& 3\\
6794131 & ELV?  &0.81&54954.298318&1.613328& 12.5   & 1455      & 2/2   & p &l(+q)&L3-52&3\\
6877673 & EA    &0.11&54989.092003&36.7587372&13.7  & 1454      & 2/0   &p+s&l+d&D3-07& \\
6964043 &EA+E3  &0.35&55190.170   &10.725518&15.6   & 1233      & 2/0   &p+s&l+d&D1-17& 4\\
6965293 & EA    &0.18&54957.473848&5.077746& 12.8   & 1468      & 2/0   &p+s&l+a(+d)&L2-39&\\
7119757 & EA    &0.64&54965.304131&0.7429217&15.6   & 1459      & 2s4/2 & a & l &L2-57& 3\\
7177553 & EA    &0.06&54954.545842&17.996467&11.5   & 1458      & 2/0   &p+s&l+d&D1-29& \\
7272739 & EW    &0.75&54964.853794&0.2811644&13.0   & 1459      & 2/2   & a &l(+q)&L3-58&3\\
7289157 & EA+E3 &0.37&54969.966600&5.266525& 12.9   & 1459      & 2/0   &p+s&l+d&D1-18& 2,4,5\\
7339345 & EW    &0.74&54964.6478878&0.2596643&15.2  & 1459      & 2/2   & a &l+q&L2-19& 3\\ 
7362751 &ELV(EW)&0.73&54964.744494&0.338249& 15.8   & 1459      & 2/2   & a &l+q&L1-25& 3\\
7375612 & FP?   &0.98&54953.639904&0.1600728(/2)&12.0&1470      &2/2(1/1)&a &l(+q)&F-07&3\\ 
7385478 & EA    &0.54&54954.534784&1.655478& 11.5   & 1468      & 2s4/2 & p & l &L2-31& 3\\ 
7440742 &EW(ELV)&0.71&54949.930411&0.2839922&11.8   & 1388      & 2/2   & a & l &L2-45& \\
7518816 & EB    &0.65&54953.692277&0.4665805&12.8   & 1470      & 2s4/2 & a & l &L3-13& 3\\
\hline	 
\end{tabular}
\end{center}
\end{table*}

\addtocounter{table}{-1}
\setlength{\tabcolsep}{4.0pt}

\begin{table*}
\begin{center}
\caption{({\em continued})} 
\begin{tabular}{lccccccccccc} 
\hline
KIC No. & Type & Morph. & $T_0$ & $P_1$ & $K_\mathrm{p}$&data length&$\frac{\mathrm{ETV}}{\mathrm{QTV}}$&Fitted&Fit&Tab&Refs. \\ 
        &      &        & (MBJD)& (days)&     (mag)	& (days)&&curves&type&&\\
\hline
7552344 & EA    &0.24&54964.948438&2.001491& 15.4   & 1457      & 2/0   & a & l &L2-25& \\
7593110 & EA    &0.17&54999.192999&3.549384& 15.9   & 1235      & 2/0   & p &l+d&D1-22& \\
7630658 & EA    &0.47&55003.279035&2.151155& 13.9   & 1418      & 2s4/2 & a & l &L2-22& 6\\
7668648 & EA+E3 &0.08&54963.315401&27.825590&15.3   & 1433      & 2/0   &p+s&l+d&D1-13& 2,4,5\\
7670617 & EA    &0.07&54969.139216&24.703160&15.5   & 1433      & 2/0   &p+s&l+d&D3-09& 4,5\\
7680593 &ELV(EW)&0.97&54964.639100&0.2763915&15.4   & 1459      & 2/2   & a &l+q&L2-32& 3\\
7685689 & EW    &0.77&55001.994674&0.3251596&15.5   & 1238      & 2/2   & a &l(+q)&L1-21&3\\
7690843 & EB    &0.69&54954.158345&0.786260& 11.1   & 1470      & 2s4/2 & a &l+d+c&D1-04&2,3,13\\
7811211 & EA    &0.49&54964.825947&0.9024037&14.6   & 1458      & 2s4/0 & p &l(+q)&L1-19&\\
7812175 & EA    &0.06&55002.612666&17.793925&16.3   &  658      & 2/0   &p+s&l+d&D2-01& 4\\
7821010 & EA    &0.03&54969.615845&24.2382426&10.8  & 1454      & 2/0   &p+s&l+d&D2-07& 18\\
7837302 & EA    &0.06&54982.935571&23.837136&13.7   & 1430      & 1/0   & p &l+d&D2-12& 2\\
7877062 & EW    &0.81&54964.779743&0.3036520&13.8   & 1459      & 2/2   & a & l &L2-54& 3\\
7955301 & EA    &0.14&54967.950750&15.32784& 12.7   & 1448      & 2/0   &p+s&l+d&D1-14& 2,4,5,13\\
8016214 & EA    &0.53&54966.725645&3.1749714&14.4   & 1454      & 2s4/2 & p &l(+q)&L3-57&3\\
8023317 & EA    &0.13&54979.733478&16.579002&12.9   & 1465      & 2/2   &p+s&l+d&D1-30& 2,4\\
8043961 & EA    &0.63&54954.555903&1.5592127&10.7   & 1469      & 2/2   & a &l(+d)&L1-20&2,3\\
8045121 & FP?   &1.00&54953.761839&0.2631774(/2)&12.0&1470      &2/2(1/1)&a &l(+q)&F-02&3\\
8081389 & EA    &0.56&54965.003801&1.4894435&14.0   & 1458      & 2s4/2 & p &l(+q)&L2-58& \\
8094140 & EA    &0.49&54965.145553&0.7064292&15.2   & 1459      & 2s4/0 & a & l &L1-32&  \\
8143170 & EA    &0.15&54970.113064&28.785943&12.9   & 1455      & 2/0   &p+s&l+d&D3-04& 4\\
8145477 & EW    &0.89&54965.076077&0.5657843&14.8   &  497      & 2/2   & a & l &L2-01& 3\\
8190491 & ELV   &0.95&54965.198125&0.7778768&14.3   & 1459      & 2/2   & a &l(+q)&L1-27&3\\
8192840 &ELV(EW)&0.95&54965.013933&0.43354925&13.5  & 1459      & 2/2   & a &l(+q)&L2-40&2,3\\
8210721 & EA    &0.08&54971.157082&22.672816&14.3   & 1451      & 2/0   &p+s&l+d&D2-03& 4,5\\
8242493 & EW    &0.73&54964.621844&0.2832856&14.7   & 1459      & 2/2   & a &l(+q)&L2-29&3\\
8265951 & EW    &0.81&54954.246763&0.7799575&12.7   & 1469      & 2/2   & a & l &L3-48& 3\\
8330092 &ELV(EW)&0.79&54964.940576&0.32172355&13.5  & 1459      & 2/2   & a & l &L1-26& 3\\
8386865 & ELV   &0.99&54953.942556&1.258041& 12.0   & 1466      & 2/2   & a &l(+d)&L1-09&2,3\\
8394040 &ELV(EW)&0.77&54964.878453&0.3021262&14.5   & 1459      & 2/2   & a & l &L1-12& 2,3\\
8429450 & EA    &0.47&54954.217684&2.7051516&13.1   & 1466      & 2/2   & a & l &L3-46& 5\\
8444552 & EA    &0.49&54964.595346&1.178090& 13.6   & 1459      & 2s4/0 & a & l &L3-41& \\
8553788 & EA    &0.54&54954.997634&1.6061632&12.7   & 2771      & 2s4/2 &p+e& l &L3-51& 1,5,6,23\\
8563964 & FP?   &1.00&54953.846748&0.338436(/2)&12.9& 1470      &2/2(1/1)&a & l &F-03& 3\\
8690104 & EW    &0.77&54964.834110&0.4087744&14.9   & 1459      & 2/2   & a & l &L3-27& 3\\
8719897 & EA    &0.50&54955.237444&3.151420& 12.4   & 1469      & 2s4/0 & a &l+d&D1-21&2,13\\
8739802 & ELV   &0.93&55001.999865&0.2745129&14.9   & 1238      & 2/2   & a & l &L2-21& 3\\
8758161$^a$& EA & .. &54953.834107&1.9964352&12.5   & 1467      & 2s4/2 & a & l &L3-43& \\
8868650 & EA    &0.62&54957.940589&4.447430& 11.9   & 1463      & 2s4/2 & p &l(+q)&L3-36&\\
8904448 & EW    &0.74&54965.059034&0.865983& 13.9   & 1458      & 2s2/2 & p &l+c&L1-23& 2,3\\
8938628 & EA    &0.14&54966.603088&6.862216& 13.7   & 1455      & 2/0   &p+s&l+d&D1-25& 2,4\\
8957887 & EW    &0.76&54964.884185&0.3473543&15.4   & 1459      & 2/2   & a & l &L2-11& 3\\
8982514 & EW    &0.83&54953.930563&0.4144906&13.2   & 1470      & 2/2   & a & l &L3-28& 3\\
9007918 & EA+E3 &0.52&54954.748782&1.3872066&11.7   & 1469      & 2s4/2 & p &l(+d)&L1-18&6\\
9028474 & EA    &0.00&55010.672516&124.9365792&12.3 & 1374      & 2/0   &p+s&l+d&D3-11& \\
9075704 & EB    &0.68&54999.891435&0.5131516&16.2   &  855      & 2/2   & a &l+q&L1-13& 3\\
9083523 & EB    &0.65&54954.484907&0.9184208&12.7   & 1470      & 2s4/2 & p & l &L3-16& 3\\
9084778 & EA    &0.49&54964.654261&0.5922444&15.7   & 1459      & 1/0   & p & l &L1-34& \\
9091810 & EB    &0.69&54953.600339&0.4797214&12.8   & 1470      & 2s4/2 & a & l &L2-53& 3\\
9101279 & EA    &0.58&54965.932213&1.8114606&13.9   & 1456(4987)& 2s6/2 &p(+e)&l+q&L2-46&3\\ 
9110346 & EA    &0.43&55002.222003&1.7905531&16.4   & 1330      & 2s4/2 & a & l &L3-47& \\
9140402 & EA    &0.27&54966.441095&4.9883312&15.3   & 1457      & 2/0   &p+s&l+d&D1-11& 19\\
9159301 & EA    &0.55&54956.304393&3.0447712&12.1   & 1468      & 2s4/0 & p & l &L2-34& 1\\
9181877 & EW    &0.74&54953.797919&0.3210098&11.7   & 1470      & 2/2   & a & l &L3-55& 3,13\\
9272276 & EW    &0.78&54953.693247&0.280615& 13.2   & 1470      & 2/2   & a & l &L2-61& 3\\
9283826 & EW    &0.84&54953.801153&0.3565232&13.1   & 1470      & 2/2   &a(+e)&l&L3-08& 3\\
9353234 & ELV   &0.86&54965.446983&1.4865274&13.7   & 1458      & 2/2   & a & l &L2-28& 3\\
9392702 & EA    &0.37&54964.893911&3.9093245&14.6   &  868      & 2s4/0 & p & l &L3-02& \\
9402652$^b$& EA &0.65&54954.290416&1.073106& 11.8   & 2048(5723)& 2/2   & p+s+e&l&L2-62& 1,6,23\\
9412114 & ELV   &0.85&55001.895873&0.2502532&15.2   & 1147      & 2/2   & a &l(+q)&L3-56&3\\
9451096 & EA    &0.53&54954.729422&1.2503906&12.6   & 1470      & 2s4/2 &p+s&l+d&D1-09& 2,3,5\\
\hline	 
\end{tabular}
\end{center}

{\bf Notes.} {$^a$: True period is twice of the given in the Villanova Catalog; $^b$: HAT, ASAS, SWASP minima were omitted}
\end{table*}

\addtocounter{table}{-1}
\setlength{\tabcolsep}{4.0pt}
\begin{table*}
\begin{center}
\caption{({\em continued})} 
\begin{tabular}{lccccccccccc} 
\hline
KIC No. & Type & Morph. & $T_0$ & $P_1$ & $K_\mathrm{p}$&data length&$\frac{\mathrm{ETV}}{\mathrm{QTV}}$&Fitted&Fit&Tab&Refs. \\ 
        &      &	& (MBJD)& (days)&     (mag)	& (days)&&curves&type&&\\
\hline
9472174$^a$&oEA,sdB+dM&0.78&54953.643197&0.12576528&12.3&1437   & 2/0   & p &l+c&L1-15& 20\\
9532219 & EW    &0.74&55001.947386&0.1981551&16.1   & 1330      & 2/2   & a & l &L3-50& 3\\
9574614 & EA    &0.40&54965.687069&0.9820954&15.9   & 1458      & 1/0   & p & l &L2-48& \\
9592145 & EB    &0.65&54965.015451&0.4888674&14.0   & 1459      & 2/2   & p &l+q&L2-03& 3\\
9596187 & EA    &0.47&54964.705879&0.9532917&14.5   & 1459      & 2/0   & p & l &L3-18& \\
9612468 & FP?   &1.00&54953.604225&0.1334715(/2)&11.5&1470      &2/2(1/1)&a & l &F-08& 3\\
9664215 & EA    &0.27&54964.925032&3.3194959&15.1   & 1459      & 2s4/0 &p+s&l+d&D2-04& \\
9665086 & EB    &0.67&55000.087903&0.296536& 13.9   & 1421      & 2/2   & a &l(+q)&L2-16&3\\
9706078 & EA    &0.56&54954.140288&0.6135606&12.8   & 1470      & 2s4/0 & a & l &L3-20& 3\\
9711751 & EA    &0.49&54965.352420&1.7115283&13.8   & 1458      & 2s4/0 & p & l &L2-44& \\
9714358 & EA    &0.13&54967.395501&6.474177& 15.0   & 1454      & 2/0   &p+s&l+d&D1-08& 2,4,5\\
9715925 & EA    &0.10&54998.920053&6.308299& 16.5   &  830      & 2/0   &p+s&l+d&D2-02& 4\\
9722737 & EW    &0.78&54964.973629&0.4185284&14.9   & 1459      & 2/2   & a & l &L1-17& 2,3\\
9777987 & EW    &0.74&55000.068578&0.2585001&16.3   &  684      & 2/2   & a & l &L1-01& \\
9788457 & EA    &0.60&54965.186856&0.9633378&13.0   & 1459      & 2s2/2 & p & l+q&L3-33& 3\\
9821923 & EW    &0.95&54964.814614&0.3495329&14.2   & 1459      & 2/2   & a & l &L3-10& 3\\
9838047 & EW    &0.84&54953.713063&0.436162& 13.5   & 1470      & 2/2   & a & l &L2-41& 3\\
9850387 & EA    &0.47&54956.416799&2.7484986&13.5   & 1468      & 2s4/0 & a & l(+d)&L1-31&\\
9912977 & EA    &0.59&54966.709125&1.887874& 13.7   & 1457      & 2s4/2 & a & l &L2-10& 2,3\\
9963009 & EA    &0.06&54986.018248&40.069657&14.5   & 1443      & 2/0   &p+s&l+d&D3-12& 4\\
9994475 & EW    &0.76&54964.733082&0.3184064&14.3   & 1459      & 2/2   & a &l+q&L1-28& 3\\
10095469& EA    &0.60&54999.865835&0.6777625&14.7   &  855      & 2s4/0 & p & l &L3-01& 3\\
10095512& EA    &0.24&54953.888455&6.017207 &13.1   & 1468      & 2/0   &p+s&l+d&D1-27& 2\\
10226388& EW    &0.77&54954.120530&0.6606583&10.8   & 1470      & 2/2   & a & l &L2-26& 2,3\\
10268809& EA    &0.05&54971.999951&24.708999&13.7   & 1450      & 2/0   &p+s&l+d&D3-15& 4\\
10268903& EA    &0.39&54999.901602&1.1039788&17.4   &  683      & 2/0   & a & l &L3-05& \\
10275197& EW    &0.79&54953.707304&0.3908377&12.9   & 1470      & 2/2   & a & l &L3-37& 3\\
10296163& EA    &0.17&54959.387400&9.2967444&13.2   & 1463      & 2/0   &p+s&l+d&D3-17& \\
10319590& EA    &0.09&54965.716743&21.320459&13.7   &  405      & 2/0   &p+s&l+d&D3-01& 2,4,5\\
10383620& EA    &0.64&54954.123817&0.7345658&12.8   & 1470      & 2/2   & a & l &L3-14& 3\\
10483644& EA    &0.12&54966.314610&5.1107711&14.0   & 1457      & 2/0   & p &l+d&D1-24& \\
10549576& EA    &0.20&54972.078799&9.0894658&13.0   & 1454      & 2/0   &p+s&l+d&D2-13& \\
10557008& EW    &0.77&54964.639092&0.2654186&14.7   & 1459      & 2/2   & a & l &L3-15& \\
10583181& EA    &0.47&54955.206895&2.696353& 11.0   & 1467      & 2s4/2 & p & l &L2-42& \\
10613718& EA    &0.39&54953.886226&1.175878& 12.7   & 1469      & 2s4/0 & a &l+d&D1-06& 2\\
10686876& EA    &0.45&54953.951815&2.6184153&11.7   & 3820      & 2s4/2 &p+e& l &L3-53& 6,23\\
10724533& EB    &0.75&54954.395189&0.7450918& 9.0   & 1470      & 2s4/2 & p & l &L3-35& 3\\
10727655& EW    &0.74&54953.910817&0.3533652&13.4   & 4374      & 2/2   &a+e&l+q&L1-37& 3\\
10848807& EW    &0.74&54999.987867&0.3462467&15.8   & 1421      & 2/2   & a & l &L2-12& 3\\
10855535& FP?   &0.99&54964.629852&0.1127824(/2)&13.9&1459      &2/2(1/1)&a & l &F-01& 3\\
10916675& EW    &0.86&54953.700609&0.4188675&13.4   & 1470      & 2/2   & a & l &L3-19& 3\\
10934755& EB    &0.68&54964.840450&0.786486& 14.4   & 1459      & 2/2   & p & l &L3-07& 3\\
10979716& EA    &0.10&54967.081259&10.684056&15.8   & 1453      & 2/0   &p+s&l+d&D2-08& 4\\
10991989& EA    &0.54&54954.650910&0.9744775&10.3   & 1470      & 2s4/0 & a & l &L1-24& 2,3,13\\
11042923& EW    &0.76&54964.970492&0.390162& 14.4   & 1459      & 2/2   & a & l &L2-30& 2,3\\
11234677& EA    &0.42&54953.872607&1.587425& 13.3   & 1470      & 2s4/2 & p & l &L3-23& \\
11246163& EW    &0.77&54964.565448&0.2792271&14.5   & 1459      & 2/2   & a &l(+q)&L3-29&3\\
11502172& EA    &0.05&54968.617081&25.4319585&14.2  & 1435      & 2/0   &p+s&l+d&D3-10& \\
11519226& EA    &0.03&54972.990000&22.161715&13.0   & 1463      & 2/0   &p+s&l+d&D2-14& 4\\
11558882& EA    &0.01&54987.716793&73.914770&15.4   & 1384      & 2/0   &p+s&l+d&D3-13& \\
11604958& EW    &0.72&54964.653176&0.2989297&13.9   & 1459      & 2/2   & a & l &L2-51& 3\\
11825204& FP?   &0.98&54964.751093&0.2096356(/2)&13.8&1458      &2/2(1/1)&a &l+q&F-05& 3\\
11968490& EA    &0.49&54965.437249&1.078890& 13.7   & 1458      & 2s4/0 & a &l+q(+d)&L1-06&2\\
12019674& EW    &0.76&53363.5350  &0.3544975&13.0   & 5188      & 2/2   &a+e& l &L2-64& 3,21\\
12055014& EW    &0.85&54965.041294&0.4999046&13.5   & 1459      & 2/2   & a & l &L3-34& \\
12055255&ELV(EW)&0.90&54964.528184&0.2209404&15.9   & 1459      & 2/2   & a & l &L3-44& 3\\
12071741&ELV(EW)&0.94&54964.820555&0.3142642&14.8   & 1459      & 2/2   & a & l &L2-23& 3\\
12356914& EA    &0.03&54976.492322&27.308455&15.5   & 1459      & 2/0   &p+s&l+d&D3-05& 4\\
12508348& FP?   &0.97&54951.682693&0.255596(/2)&13.4& 1457      &2/2(1/1)&a &l+q&F-04& \\
12554536& EB    &0.63&54953.964623&0.6844956&12.8   & 1470      & 2s4/2 & p & l &L2-60& 3\\
\hline	 
\end{tabular}
\end{center}

{\bf Notes.} {$^a$: short-cadence (SC) data only}
\end{table*}

The results of our analyses are tabulated in Tables~\ref{Tab:Orbelem}--\ref{Tab:AMEparam}. Of our 230 EBs, pure LTTE ETV solutions, some of which are supplemented by quadratic or cubic terms, were calculated for 160 systems, while combined LTTE and dynamical solutions were obtained for another 62 EBs.  The remaining 8 `systems' were found to be false positives in the sense that was previously discussed. Despite this, we give LTTE solutions for these cases as well (Table~\ref{Tab:OrbelemFP}) but do not plot them in Fig.~\ref{Fig:ETVs}.

\subsection{EBs with LTTE solution}
\label{Subsect:LTTEgeneral}

In this section we consider the systems with pure LTTE solutions. We divide them into three groups approximately following \citet{conroyetal14}. Broadly speaking, for groups one, two, and three, the data span more than two, more than one, and less than one outer orbital periods.  The motivation for this grouping is that the more outer orbital periods that are covered, the more secure are the solutions.  In what follows we give more specifics on the systems included in each group.

The {\em first group} consists of 38 EBs, with outer periods of $95\,\mathrm{d}\lesssim P_2\lesssim5532\,\mathrm{d}$. Generally our data on each of these cover more than two outer orbital cycles. The only exception is KIC~06543674 where even though the {\em Kepler} observations cover only $\sim1.32$ outer orbits, outer eclipses at the expected times evidently verify the third-body solution. For KIC~10727655(=V2280~Cyg) and KIC~02708156(=UZ~Lyr), ground-based observations extend the observing interval sufficiently to justify their inclusion.  The {\em second group}, the most numerous subgroup with its 64 members, contains EBs whose outer periods are shorter than the length of the time series, i.e., at least one full outer orbital cycle was observed. The period range is $364\,\mathrm{d}\lesssim P_2\lesssim2800\,\mathrm{d}$. KIC~05513861, KIC~09402652 (=V2281~Cyg), and KIC~12019674 (=V2294~Cyg), are included in this group on the basis of both {\em Kepler} and ground-based observations.  Finally, 58 triples are included in the {\em third group} wherein each system was observed over less than one complete outer orbit. The outer period domain for this group is $932\,\mathrm{d}\lesssim P_2\lesssim9256\,\mathrm{d}$.

For these three groups the orbital elements and their uncertainties are given in Tables~\ref{Tab:Orbelem}--\ref{Tab:Orbelem3}. In each table the systems are ordered by increasing outer orbital period ($P_2$). As is well-known, similar to single-line radial velocity observations, the LTTE solution does not allow either the inclination ($i_2$) of the wide orbit or the mass of the third companion ($m_\mathrm{C}$) to be uniquely deduced.  Nevertheless, a crude estimate can readily be found for $m_\mathrm{C}$ by use of the reasonable approximation that the mass of the EB is likely to be about 2\,M$_\odot$. Then, solving Eq.~(\ref{Eq:f(mC)def}) which is third order in $m_\mathrm{C}$, the mass of the third object can be estimated for different $i_2$ inclinations. We list these approximate minimum values of $m_\mathrm{C}$, i.e., for $i_2=90\degr$, in our tables. Naturally, if the inner binary mass, $m_\mathrm{AB}$, of any of our investigated systems is more accurately known, a better estimate for $(m_\mathrm{C})_\mathrm{min}$ can be obtained. In most cases our rough estimate is likely to be fairly satisfactory for judging the nature of the third object, and can also be used to forecast the expected amount of third light either for any future photometric lightcurve solutions, or spectroscopic follow-up observations. We use the same approximation to estimate the ratio of the amplitudes of the dynamical and LTTE contributions (${\cal{A}}_\mathrm{dyn}/{\cal{A}}_\mathrm{LTTE}$) of the ETVs.
 
The vast majority of the EBs in these three groups have inner periods in the range of $0.23\,\mathrm{d}\leq P_1\leq3\,\mathrm{d}$. The shortest period in our sample, $P_1\sim0\fd13$, belongs to the low-mass sdB+dM binary KIC~09472174, while the longest two periods, $P_1\sim5\fd08$ and $5\fd99$, belong to, respectively, the slightly eccentric detached systems KIC~06965293 and KIC~06233903. While the lower end of the inner period distribution is in accord with the short period limit of contact binaries, the low upper limit requires a brief explanation. For this, one can see that in our approximation 
\begin{equation}
\frac{{\cal{A}}_\mathrm{dyn}}{{\cal{A}}_\mathrm{LTTE}}=\frac{c}{\left(2\pi G m_\mathrm{ABC}\right)^{1/3}\sin i_2}{\cal{E}}(e_2,\omega_2)\left(\frac{P_1}{P_2}\right)^2P_2^{1/3},
\end{equation}
where
\begin{equation}
{\cal{E}}(e_2,\omega_2)=\left(1-e_2^2\right)^{-3/2}\left(1-e_2^2\cos^2\omega_2\right)^{-1/2}
\end{equation}
and therefore, for a given total mass
\begin{eqnarray}
\frac{{\cal{A}}_\mathrm{dyn}}{{\cal{A}}_\mathrm{LTTE}}&\geq&\frac{c}{\left(2\pi G m_\mathrm{ABC}\right)^{1/3}}\left(\frac{P_1}{P_2}\right)^2P_2^{1/3} \nonumber \\
&\geq&1.45\times10^3m_\mathrm{ABC}^{-1/3}\,\frac{P_1^2}{P_2^{5/3}}.
\end{eqnarray}
where $P$'s are expressed in days and $m_\mathrm{ABC}$ in solar units. 
Since in our sample, $P_2$ has a strong upper limit, i.e., practically the duration of the {\em Kepler} observations, substituting this limit, i.e. $P_2=1470\,\mathrm{d}$, into the equation above, we obtain
\begin{equation}
\frac{{\cal{A}}_\mathrm{dyn}}{{\cal{A}}_\mathrm{LTTE}}\geq m_\mathrm{ABC}^{-1/3}\left(\frac{P_1}{11.46}\right)^2\left(\frac{1470}{P_2}\right)^{5/3},
\end{equation}
which illustrates that if $P_1$ exceeds 5\,days, the dynamical contribution is likely to be comparable or larger than the LTTE contribution. Therefore, all the triples with longer inner binary periods are included in the LTTE plus dynamical effect groups listed in Tables~\ref{Tab:Orbelemdyn1}--\ref{Tab:Orbelemdyn3}.

\subsection{EBs with combined dynamical and LTTE solution \label{Subsect:LTTEdyngeneral}}

We list 62 triples with combined dynamical and LTTE solutions. With the exception of the two shortest outer period systems discussed below, our fitting process was practically identical in great detail with that described in \citet{borkovitsetal15}. These systems allow, in principle, the determination of all the system masses, though in principle, there are some degeneracies in the parameters \citep{rappaportetal13} unless the inner binary is eccentric and the ETV curves for both the primary and secondary eclipses can be measured and fit simultaneously \citep{borkovitsetal15}.

The two exceptional systems in this group are KIC~05897826 and KIC~05952403. The inner binary in KIC~05897826 is just barely an eclipsing binary; the two stars do actually eclipse each other at favorable phases of the rapid precession of the binary orbital plane.  Consequently, we cannot find an ETV solution for this triple. Therefore, we borrow the orbital elements and masses from the photodynamical solution of \citet{carteretal11}. KIC~05952403 (HD 181068) is a triply eclipsing system where the inner and outer orbits are both circular and coplanar. Hence the usually dominant quadrupole dynamical term disappears. In our analysis, we have obtained a pure LTTE solution for this triple, and the other parameters, which can usually be deduced from the dynamical part of the combined solution, were taken from \citet{borkovitsetal13}. This is also a prime example for emphasizing that the theoretical ratio ${\cal{A}}_\mathrm{dyn}/{\cal{A}}_\mathrm{LTTE}$ is merely a rough estimate (for this system ${\cal{A}}_\mathrm{dyn}/{\cal{A}}_\mathrm{LTTE}=1.22$). 

As noted above, a combined solution offers several parameters which cannot be obtained from a pure LTTE solution. Therefore, Tables~\ref{Tab:Orbelemdyn1}--\ref{Tab:Orbelemdyn3} contain information somewhat different from that in Tables~\ref{Tab:Orbelem}--\ref{Tab:Orbelem3}. The masses of the two components of the wide binary, $m_\mathrm{AB}$ and $m_\mathrm{C}$, are calculated from the mass function, $f(m_\mathrm{C})$, and the outer mass ratio, $m_\mathrm{C}/m_\mathrm{ABC}$, which are direct outputs of the combined solution. All four of these quantities are listed in the present tables. We are also able to give the full semimajor axis of the outer orbit ($a_2$) instead of the projected semimajor axis of the LTTE orbit of the binary ($a_\mathrm{AB}\sin i_2$). Lastly, instead of the theoretically calculated ratio ${\cal{A}}_\mathrm{dyn}/{\cal{A}}_\mathrm{LTTE}$, we give the actual `measured' value. In regard to this latter point, we note that in the case of an eccentric EB, the true amplitude of the dynamical term may differ by as much as a factor of two for the primary and secondary minima of a given system.  In all cases, we tabulate the larger of the two dynamical amplitudes. See \citet{borkovitsetal11} for a discussion, and, as examples, the ETV curves of KICs~05255512, 07670617, 08143170, and 10258809 in the appropriate panels of Fig.~\ref{Fig:ETVs}.

Additional parameters, none of which can be obtained from a pure LTTE solution, are tabulated in Table~\ref{Tab:AMEparam}. These refer to the orbital elements of the inner binary orbit and spatial orientations. The mutual inclination ($\im$), has extraordinary importance in connection to the dynamical evolution of a triple system. Its determination, and more generally the complete three-dimensional orientation of a triple, were discussed in great detail in \citet[][especially in Appendix~D]{borkovitsetal15}, but some additional remarks are in order here. 

During the first step in our analysis, the mutual inclination, $\im$, and one of the additional node-like angles were taken as adjustable parameters. For many of the systems the result is a low, but definitely non-zero mutual inclination (typically $\im<10\degr$). For such low values of $\im$ the EB's orbital plane should precess very rapidly with a low amplitude. KIC~05897826, discussed above, is a good example of this.  Orbital inclination ($i_1$) variations should then be visible as eclipse depth variations. In most cases, variation of the eclipse depths is not seen. In these cases we fixed $\im$ at $0\degr$ manually, and then reran our parameter solver. We believe this is reasonable because the lack of variation of the eclipse depths rules out a small non-zero misalignment of the orbits, and, because the ETV solution fundamentally rules out the possibility of a larger mutual inclination, which would result in a larger amplitude, substantially slower precession. Furthermore, from a dynamical and/or evolutionary point of view, there is no fundamental difference between strictly and nearly coplanar configurations; therefore, we believe it is justified to use these systems with $\im=0\degr$ in our statistical studies.

As was done for the LTTE systems, we divided the set of 62 triples into three subgroups according to coverage of their outer orbits. The first group contains 31 members with outer periods $34\,\mathrm{d}\lesssim P_2\lesssim862\,\mathrm{d}$. The middle group has 14 triples in the period range $583\,\mathrm{d}\lesssim P_2\lesssim1437\,\mathrm{d}$, while the systems for which less than one outer cycle was observed include 17 potential triples with $452\,\mathrm{d}\lesssim P_2\lesssim15271\,\mathrm{d}$.

\setlength{\tabcolsep}{3.3pt}
\begin{table*}
\begin{center}
\caption{Orbital Elements from LTTE solutions for systems, where more than two outer periods are covered, or/and triply eclipsing systems} 
\label{Tab:Orbelem}  
\begin{tabular}{lccccccccccc} 
\hline
KIC No. & $P_1$ & $\Delta P_1$ & $P_2$ & $a_\mathrm{AB}\sin i_2$ & $e_2$ & $\omega_2$ & $\tau_2$ & $f(m_\mathrm{C})$ & $(m_\mathrm{C})_\mathrm{min}$ & $\frac{{\cal{A}}_\mathrm{dyn}}{{\cal{A}}_\mathrm{LTTE}}$ & $m_\mathrm{AB}$\\
        & (day) &$\times10^{-10}$ (d/c)&(day)&(R$_\odot$)  &       &   (deg)    &   (MBJD) & (M$_\odot$)       & (M$_\odot$)            & &  (M$_\odot$)    \\
\hline
9777987 & 0.25850259(5)&$-6.3$(1)&95.28(6)&20.6(2)&0.19(2)& 245(5) & 54979(2) & 0.0129(3)& 0.42 & 0.04 & 2: \\
6669809$^a$&0.73374152(7)&$-101$(2)&193.8(1)&25.9(2)&0.12(2)&74(8) & 54958(4) & 0.0062(2)& 0.32 & 0.09 & 2: \\
2856960 & 0.25850790(6)&$-2.2$(2)&204.8(2)&94(3)& 0.55(3) & 164(3) & 55007(2) & 0.27(2)  & 1.48 & 0.02 & 2: \\
5310387 & 0.44166866(1)&2.80(4)&214.0(1)&13.5(1)& 0.23(1) & 210(4) & 55051(2) &0.00072(2)& 0.15 & 0.03 & 2: \\
6144827 & 0.23465160(1)&$-5.80$(4)&228.0(2)&32.8(3)&0.15(2)& 52(6) & 54921(4) & 0.0091(2)& 0.37 & 0.008& 2: \\
11968490& 1.07889066(9)&$-3$(1)&253.9(1)&111.9(5)&0.374(8)& 284(1) & 54862(1) & 0.291(4) & 1.54 & 0.13 & 2: \\ 
4037163 & 0.63544461(3)& $-$  & 268(2) &  20(2) & 0.66(7) & 356(4) & 54901(10)& 0.0015(4)& 0.19 & 0.12 & 2: \\
6370665 & 0.93231431(7)&14.8(8)&286.4(5)&27.7(4)& 0.08(3) & 354(19)& 55027(15)& 0.0035(1)& 0.26 & 0.08 & 2: \\
8386865 & 1.25804169(7)& $-$  &294.0(5)&  84(2) & 0.49(3) & 314(4) & 55028(3) & 0.092(6) & 0.92 & 0.19 & 2: \\ 
3853259:& 0.27664714(1)&2.48(4)&325.7(6)&17.2(4)& 0.60(3) & 121(3) & 54900(3) &0.00064(4)& 0.01 & 0.14 & 2: \\
5376552 & 0.50381878(1)& $-$  &334.8(1)& 41.9(2)&0.349(6) &355.2(9)& 54874(1) & 0.0088(1)& 0.37 & 0.02 & 2: \\
8394040 & 0.30212624(1)& $-$  &388.9(1)&124.6(4)&0.520(5) & 295(1) & 54809(1) & 0.171(2) & 1.21 & 0.007& 2: \\
9075704 & 0.5131488(1) &15.2(7)&402.0(5)&63.4(3)&0.160(9) & 252(3) & 55084(4) & 0.0212(3)& 0.51 & 0.01 & 2: \\ 
5459373 & 0.28660872(1)& $-$  &412.7(2)& 98.6(3)&0.361(6) & 271(1) & 55051(1) & 0.0754(7)& 0.85 & 0.004& 2: \\
9472174$^b$& 0.12576528(1)&$-0.063$(4)&418(2)&0.63(2)&0.38(4)&124(6)&55118(8) & 20(2)E-9 &0.0019&0.000&0.60(3)\\
5128972 & 0.50532338(1)& $-$  &442.1(2)&114.6(3)&0.285(5) & 285(1) & 54940(1) & 0.1032(9)& 0.97 & 0.01 & 2: \\
9722737 & 0.41852837(1)& $-$  &444.2(1)&103.4(2)&0.174(4) & 223(1) & 54913(2) & 0.0750(5)& 0.85 & 0.007& 2: \\
9007918 & 1.38720655(1)& $-$  &470.9(6)& 17.7(2)& 0.68(2) & 271(1) & 54827(2) &0.00033(1)& 0.11 & 0.19 & 2: \\
7811211 & 0.90240346(9)& $-$  & 477(6) &  35(3) & 0.29(12)& 169(24)& 55168(33)& 0.0026(6)& 0.24 & 0.04 & 2: \\
8043961 & 1.55921280(1)& $-$  &478.6(2)& 82.6(2)&0.245(5) &  13(1) & 54817(2) & 0.0330(3)& 0.61 & 0.10 & 2: \\
7685689 & 0.32515963(1)& $-$  &514.9(5)& 80.2(3)&0.125(8) & 170(4) & 54774(5) & 0.0261(3)& 0.55 & 0.003& 2: \\
5216727 & 1.51302292(1)& $-$  &532.9(6)& 30.1(2)& 0.50(1) & 129(1) & 55158(2) &0.00129(3)& 0.18 & 0.11 & 2: \\
8904448$^c$& 0.8659838(1) &$-99$(3)&543.7(6)&68.5(4)&0.525(9)&307.7(9)&54796(2)&0.0146(3)& 0.44 & 0.04 & 2: \\
10991989& 0.97447759(5)& $-$  & 548(1) & 106(1) & 0.35(2) &  29(4) & 54960(6) & 0.053(2) & 0.73 & 0.03 & 2: \\
7362751 & 0.33825080(4)&$-9.2$(2)&552.0(6)&120.1(6)&0.256(9)&107(2)& 54930(3) & 0.076(1) & 0.85 & 0.004& 2: \\
8330092 & 0.32172365(1)& $-$  & 581(1) & 52.9(4)& 0.18(1) &   2(4) & 55134(7) & 0.0059(1)& 0.32 & 0.003& 2: \\
8190491 & 0.77787699(4)& $-$  & 621(3) &  65(1) & 0.54(3) &  67(4) & 54789(7) & 0.0097(7)& 0.38 & 0.02 & 2: \\
9994475 & 0.31840931(2)&$-13.28$(6)&626.6(5)&85.3(3)&0.288(6)&199(1)&54779(2) & 0.0212(2)& 0.51 & 0.003& 2: \\ 
3228863 & 0.73094352(1)&3.31(7)&642.8(6)&83.8(6)& 0.05(1) & 57(16) & 55052(28)& 0.0191(4)& 0.49 & 0.01 & 2: \\
3245776 & 1.4920589(2) & $-$  & 663(9) &  54(3) & 0.45(12)& 263(16)& 55003(33)& 0.0048(9)& 0.29 & 0.07 & 2: \\
9850387 & 2.7484978(1) & $-$  & 671(2) &  98(1) & 0.46(2) & 121(3) & 54683(6) & 0.028(1) & 0.57 & 0.22 & 2: \\
8094140 & 0.70642857(1)& $-$  & 676(1) & 56.3(4)& 0.35(1) & 171(2) & 54774(4) & 0.0052(1)& 0.30 & 0.01 & 2: \\
5039441 & 2.15138294(6)& $-$  & 678(1) & 87.2(7)& 0.25(1) & 163(3) & 55217(6) & 0.0194(5)& 0.49 & 0.10 & 2: \\
9084778 & 0.59224375(8)& $-$  & 680(9) &  69(3) & 0.20(9) & 176(25)& 55168(49)& 0.009(1) & 0.38 & 0.008& 2: \\
2835289 & 0.8577610(1) & $-$  & 755(5) & 138(7) & 0.74(6) & 294(4) & 54933(11)& 0.06(1)  & 0.78 & 0.04 & 2: \\  
6543674 & 2.39103051(1)& $-$  &1101.4(4)&115.2(1)&0.617(2)&267.1(1)& 55038(1) &0.01689(6)& 0.47 & 0.10 & 2: \\  
10727655& 0.35336509(1)&0.38(4)&1138.1(6)&141.8(2)&0.247(2)&36.4(5)& 55063(2) & 0.0295(1)& 0.58 & 0.001& 2: \\
2708156:$^d$&1.8912615(2)&$-29.4$(7)&5532(26)&137(7)&0.46(3)&242(9)&55955(153)& 0.0011(2)& 0.17 & 0.003& 2: \\
\hline
\end{tabular}
\end{center}
{\bf Notes.} {$a$: Cubic ephemeris -- $c_3=1.84(3)\times10^{-12}$\,d/c$^3$; $b$: Cubic ephemeris -- $c_3=3\times10^{-16}$\,d/c$^3$; $c$: Cubic ephemeris -- $c_3=2.57(6)\times10^{-12}$\,d/c$^3$; $d$: Cubic ephemeris -- $c_3=-0.058(2)\times10^{-12}$\,d/c$^3$}
\end{table*}

\setlength{\tabcolsep}{3.3pt}
\begin{table*}
\begin{center}
\caption{Orbital Elements from LTTE solutions which cover more than one but less than two outer periods} 
\label{Tab:Orbelem2}  
\scalebox{0.95}{
\begin{tabular}{lccccccccccc} 
\hline
KIC No. & $P_1$ & $\Delta P_1$ & $P_2$ & $a_\mathrm{AB}\sin i_2$ & $e_2$ & $\omega_2$ & $\tau_2$ & $f(m_\mathrm{C})$ & $(m_\mathrm{C})_\mathrm{min}$ & $\frac{{\cal{A}}_\mathrm{dyn}}{{\cal{A}}_\mathrm{LTTE}}$ & $m_\mathrm{AB}$\\
        & (day) &$\times10^{-10}$ (d/c)&(day)&(R$_\odot$)  &       &   (deg)    &   (MBJD) & (M$_\odot$)       & (M$_\odot$)            & &  (M$_\odot$)    \\
\hline
8145477 & 0.56578395(8)& $-$  & 364(4) & 59.2(8)& 0.41(2) & 190(2) & 54911(4) & 0.021(1) & 0.51 & 0.03 & 2: \\
3114667:& 0.88858302(3)& $-$  & 573(4) &  21(1) & 0.73(4) & 347(2) & 55275(9) &0.00038(7)& 0.12 & 0.10 & 2: \\
9592145:& 0.48886728(1)&0.75(3)&730(2) & 6.88(6)& 0.28(2) & 260(3) & 55041(7) & 82(2)E-7 &0.03& 0.005& 2: \\
5478466 & 0.48250055(1)& $-$  &739.2(9)& 91.2(8)& 0.505(9)& 17.6(9)& 54961(2) & 0.0186(5)& 0.49 & 0.007& 2: \\
4859432 & 0.38547990(1)& $-$  &747.3(8)& 67.3(3)& 0.565(8)&270.0(8)& 54634(2) & 0.0073(1)& 0.34 & 0.005& 2: \\ 
4451148 & 0.73598174(1)& $-$  &749.7(6)&139.6(5)& 0.288(6)&  44(1) & 55002(3) & 0.0649(7)& 0.80 & 0.01 & 2: \\
3338660:& 1.8733818(1) & $-$  & 752(18)&  7.7(8)& 0.76(8) & 308(5) & 54821(30)&0.000011(3)&0.04 & 0.26 & 2: \\ 
4647652 & 1.06482497(1)& $-$  &754.7(3)&106.2(2)& 0.291(3)& 30.7(5)& 54760(1) & 0.0282(1)& 0.57 & 0.02 & 2: \\
4670267 & 2.00609728(2)& $-$  & 755(2) & 29.4(4)& 0.56(2) &  63(2) & 55052(5) &0.00060(3)& 0.14 & 0.13 & 2: \\
9912977 & 1.88787319(1)& $-$  &756.2(5)& 47.4(1)& 0.282(5)&  76(1) & 55168(2) &0.00250(2)& 0.23 & 0.07 & 2: \\
8957887 & 0.34735418(1)& $-$  &774.2(4)&186.1(4)& 0.468(3)&151.2(4)& 55275(1) & 0.144(1) & 1.12 & 0.003& 2: \\
10848807:&0.34624670(1)& $-$  & 785(3) & 11.3(2)& 0.36(3) & 339(1) & 55013(11)&0.000031(2)&0.05 & 0.003& 2: \\
1873918 & 0.33243183(3)&$-3.6$(1)&840.9(9)&124.4(7)&0.663(7)&64.5(7)&55035(2) & 0.0364(7)& 0.63 & 0.004& 2: \\
4138301 & 0.25337919(1)& $-$  & 844(1) &122.8(7)& 0.423(8)& 221(1) & 55049(3) & 0.0348(6)& 0.62 & 0.001& 2: \\ 
2715417 & 0.23643993(1)& $-$  & 856(2) & 21.9(2)& 0.22(1) & 124(3) & 54939(8) &0.00019(1)& 0.09 & 0.001& 2: \\
9665086 & 0.29653688(2)& $-$  & 856(2) & 253(3) & 0.51(1) &  66(2) & 54659(4) & 0.297(9) & 1.55 & 0.002& 2: \\
4547308 & 0.57692798(1)& $-$  & 871(4) & 105(4) & 0.90(2) &  54(2) & 55078(7) & 0.020(2) & 0.50 & 0.06 & 2: \\
4069063 & 0.50429527(3)& $-$  & 876(1) & 233(2) & 0.64(1) & 132(1) & 55013(3) & 0.220(7) & 1.35 & 0.007& 2: \\
7339345:& 0.25966151(1)&10.01(2)&892(2)& 15.3(1)& 0.40(1) & 286(2) & 55120(5) &0.000060(1)&0.06 & 0.001& 2: \\
6516874 & 0.91632549(7)& $-$  & 905(4) &102.3(8)& 0.24(1) & 205(2) & 54663(9) & 0.0175(4)& 0.48 & 0.01 & 2: \\
8739802 & 0.27451278(1)& $-$  & 907(5) & 41.1(6)& 0.40(2) & 191(3) & 55357(9) &0.00113(5)& 0.17 & 0.001& 2: \\
7630658 & 2.15115567(2)& $-$  &921.6(3)&179.8(5)& 0.673(2)&326.2(2)& 55353(1) &0.0917(7) & 0.92 & 0.16 & 2: \\ 
12071741& 0.31426438(1)& $-$  & 927(2) & 176(2) & 0.64(1) &149.5(8)& 54932(2) & 0.085(3) & 0.89 & 0.003& 2: \\
2450566 & 1.8445840(7) & $-$  & 935(11)& 236(19)& 0.72(8) & 142(6) & 55050(19)& 0.20(5)  & 1.30 & 0.12 & 2: \\
7552344 & 2.0014910(9) & $-$  & 952(14)& 237(9) & 0.26(8) & 261(17)& 55330(47)& 0.20(2)  & 1.29 & 0.05 & 2: \\
10226388& 0.66065835(1)& $-$  &954.6(8)&211.3(5)& 0.276(4)& 108(1) & 54716(3) & 0.139(1) & 1.10 & 0.005& 2: \\
2302092 & 0.29467288(1)& $-$  &986.1(7)&175.2(4)& 0.442(4)&109.9(5)& 55112(2) &0.0741(5) & 0.84 & 0.001& 2: \\
9353234 & 1.4865278(2) & $-$  & 987(20)&  60(4) & 0.18(11)& 121(39)&55240(108)& 0.0029(5)& 0.24 & 0.03 & 2: \\
8242493 & 0.28328569(1)& $-$  &1013(2) & 27.4(1)& 0.182(7)&   1(2) & 55094(7) &0.000268(3)&0.11 & 0.001& 2: \\
11042923& 0.39016214(1)& $-$  &1041.7(8)&120.6(2)&0.274(2)&167.6(5)& 54483(2) &0.02165(9)& 0.52 & 0.002& 2: \\
7385478:& 1.655473(1)  & $-$  &1049(9) &  67(1) & 0.47(4) & 119(4) & 55039(14)& 0.0037(2)& 0.27 & 0.04 & 2: \\
7680593:& 0.27639826(5)&$-25.7$(2)&1051(5)&43.6(8)&0.54(2)& 148(2) & 55315(6) &0.00101(6)& 0.17 & 0.001& 2: \\
5611561 & 0.25869469(1)& $-$  &1052(2) & 44.5(2)& 0.205(9)& 347(2) & 55362(7) &0.00106(2)& 0.17 & 0.001& 2: \\
9159301 & 3.0447717(1) & $-$  &1072(23)& 12.6(3)& 0.40(4) & 263(7) & 54922(27)&0.000023(2)&0.05 & 0.12 & 2: \\
3440230 & 2.8811326(2) &$-1277$(10)&1082(8)&17.6(5)& 0.56(3)&178(3)& 55203(11)&0.000063(5)&0.06 & 0.17 & 2: \\
5621294:& 0.93890979(5)&$-59.2$(8)&1083(15)&6.8(2)&0.43(5)& 323(7) & 55143(22)& 36(4)E-7 & 0.02 & 0.01 & 2: \\
4074708 & 0.30211649(1)& $-$  &1110(3) & 28.1(1)& 0.09(1) &  42(6) & 54843(18)&0.000243(4)&0.10 & 0.001& 2: \\
5307780 & 0.30884972(3)&5.7(1)&1115(2) &  48(2) & 0.86(1) &  15(1) & 55101(5) & 0.0012(1)& 0.18 & 0.01 & 2: \\
6965293 & 5.0777443(1) & $-$  &1119(2) &197.4(6)& 0.204(7)& 312(2) & 54716(6) & 0.0823(9)& 0.88 & 0.23 & 2: \\
8192840 & 0.43354928(1)& $-$  &1145(4) &118.4(7)& 0.655(4)& 3.5(3) & 55486(3) &0.0170(3) & 0.47 & 0.005& 2: \\
9838047 & 0.43616206(3)& $-$  &1154(2) &221.1(8)& 0.267(6)& 174(1) & 55008(4) & 0.109(1) & 0.99 & 0.002& 2: \\
10583181& 2.69635389(2)& $-$  &1169.2(9)&154.0(1)&0.060(2)&  99(2) & 54503(6) & 0.0358(1)& 0.63 & 0.06 & 2: \\
4681152 & 1.8359276(2) & $-$  &1177(21)& 36.3(9)& 0.22(3) & 155(7) & 54998(26)&0.00046(4)& 0.13 & 0.03 & 2: \\
9711751 & 1.71152818(1)& $-$  &1186.1(7)&218.1(2)&0.259(1)&351.0(4)& 55385(1) & 0.0989(3)& 0.95 & 0.02 & 2: \\
7440742 & 0.28399218(1)& $-$  &1200(4) & 29.9(4)& 0.66(3) & 287(2) & 55048(8) &0.000249(9)&0.10 & 0.002& 2: \\
9101279:& 1.81146057(5)& $-$  &1202(8) & 46.8(4)& 0.17(1) & 129(5) & 55342(16)&0.00095(3)& 0.16 & 0.03 & 2: \\
4762887 & 0.73657344(4)& $-$  &1233(37)&  25(1) & 0.25(8) &   5(20)& 55288(72)&0.00013(2)& 0.08 & 0.005& 2: \\
9574614 & 0.982095(1)  & $-$  &1234(43)& 266(12)& 0.02(5) &208(113)&55093(387)& 0.17(2)  & 1.19 & 0.007& 2: \\
5903301 & 2.3203030(4) & $-$  &1255(32)& 153(3) & 0.43(4) &  22(6) & 54977(31)& 0.031(2) & 0.59 & 0.06 & 2: \\
6281103 & 0.36328330(1)&$-9.98$(6)&1254(3)&76.9(4)&0.024(9)&239(23)& 55179(81)&0.00388(7)& 0.27 & 0.001& 2: \\
11604958& 0.29892982(1)& $-$  &1256(8) & 22.7(3)& 0.50(1) &  36(1) & 55122(6) &0.000099(4)&0.08 & 0.001& 2: \\
6671698 & 0.471532(1)  &$-42$(6)&1261(47)&129(9)& 0.316(5)& 131(1) & 54743(39)& 0.018(4) & 0.48 & 0.002& 2: \\
9091810 & 0.47972130(1)& $-$  &1298(33)& 17.1(3)& 0.25(4) & 257(8) & 54488(37)&0.000040(3)&0.06 & 0.002& 2: \\
7877062 & 0.30365194(5)& $-$  &1321(34)&  52(2) & 0.128(9)&  89(9) & 54716(41)& 0.0011(1)& 0.17 & 0.001& 2: \\
4244929 & 0.3414038(1) & $-$  &1342(23)& 129(4) & 0.28(1) & 235(2) & 55116(13)& 0.016(2) & 0.46 & 0.001& 2: \\
4574310 & 1.30622013(1)& $-$  &1347(20)& 14.9(2)& 0.56(2) & 154(1) & 54557(16)&0.000025(1)&0.05 & 0.02 & 2: \\
8081389 & 1.48944301(3)& $-$  &1383(15)& 13.8(2)& 0.27(1) & 217(2) & 55181(12)&0.000018(1)&0.04 & 0.02 & 2: \\ 
7119757 & 0.7429197(2) & $-$  &1402(43)& 179(5) & 0.60(1) &181.8(5)& 54451(32)& 0.039(4) & 0.65 & 0.008& 2: \\
4945857 & 0.33541778(4)& $-$  &1423(6) & 346(2) & 0.402(2)&343.0(2)& 54286(5) & 0.273(4) & 1.49 & 0.001& 2: \\
12554536& 0.68449643(1)& $-$  &1448(7) & 44.8(2)& 0.515(9)&246.9(6)& 54914(5) &0.000574(9)&0.14 & 0.004& 2: \\ 
9272276 & 0.28061416(2)& $-$  &1458(7) & 235(1) & 0.252(3)&325.4(9)& 55291(5) & 0.082(1) & 0.88 & 0.001& 2: \\
9402652 & 1.07310692(2)& $-$  &1506(2) &163.9(3)& 0.805(1)& 86.6(2)& 54838(2) & 0.0260(2)& 0.55 & 0.03 & 2: \\
5513861 & 1.51020953(9)& $-$  &2140(6) & 306(2) & 0.140(6)& 208(4) & 54202(22)& 0.084(1) & 0.89 & 0.007& 2: \\
12019674& 0.35449743(3)& $-$  &2800(13)& 408(2) & 0.216(5)& 145(1) & 52584(14)& 0.116(2) & 1.02 & 0.001& 2: \\
\hline
\end{tabular}}
\end{center}
\end{table*}

\setlength{\tabcolsep}{3.3pt}
\begin{table*}
\begin{center}
\caption{Orbital Elements from LTTE solutions which cover less than a full period} 
\label{Tab:Orbelem3}  
\begin{tabular}{lccccccccccc} 
\hline
KIC No. & $P_1$ & $\Delta P_1$ & $P_2$ & $a_\mathrm{AB}\sin i_2$ & $e_2$ & $\omega_2$ & $\tau_2$ & $f(m_\mathrm{C})$ & $(m_\mathrm{C})_\mathrm{min}$ & $\frac{{\cal{A}}_\mathrm{dyn}}{{\cal{A}}_\mathrm{LTTE}}$ & $m_\mathrm{AB}$\\
        & (day) &$\times10^{-10}$ (d/c) &(day)&(R$_\odot$)  &       &   (deg)    &   (MBJD) & (M$_\odot$)       & (M$_\odot$)            & &  (M$_\odot$)    \\
\hline
10095469& 0.67776245(2)& $-$  &932(15) & 40.9(4)& 0.19(2) &  67(4) & 54958(19)&0.00106(5)& 0.17 & 0.006& 2: \\
9392702 & 3.90933(1)   & $-$  &976(170)& 103(29)& 0.29(3) & 281(14)& 55044(56)& 0.02(1)  & 0.45 & 0.19 & 2: \\
4848423 & 3.003612(6)  & $-$  &1190(68)& 227(22)& 0.14(3) & 358(7) & 55450(68)& 0.11(4)  & 1.00 & 0.07 & 2: \\
2983113 & 0.39515998(2)& $-$  &1249(36)& 36.9(9)& 0.49(4) & 303(3) & 55246(22)&0.00043(4)& 0.12 & 0.002& 2: \\ 
10268903& 1.103978(1)  & $-$  &1286(318)&211(33)& 0.66(8) & 130(4) &54904(293)& 0.08(5)  & 0.85 & 0.02 & 2: \\
6265720 & 0.31242762(2)& $-$  &1447(15)& 220(2) & 0.652(7)&192.4(6)& 55345(6) & 0.068(2) & 0.82 & 0.002& 2: \\
10934755& 0.78648549(8)& $-$  &1466(32)&  51(1) & 0.26(1) &  47(3) & 54354(24)&0.00083(7)& 0.16 & 0.004& 2: \\
9283826 & 0.35652321(5)& $-$  &1475(27)&  98(2) & 0.31(3) &  30(3) & 54850(19)& 0.0057(4)& 0.31 & 0.001& 2: \\
6103049 & 0.6431713(2) & $-$  &1482(62)&  59(5) & 0.49(4) & 243(2) & 54945(20)&0.0013(3) & 0.18 & 0.004& 2: \\
9821923 & 0.3495323(2) & $-$  &1493(63)& 110(7) & 0.48(2) & 293(12)& 54786(52)& 0.008(2) & 0.35 & 0.001& 2: \\
5353374 & 0.39332061(1)& $-$  &1494(33)& 29.1(5)& 0.13(3) & 268(8) & 55009(35)&0.00015(1)& 0.09 & 0.001& 2: \\
3766353 & 2.6669672(8) & $-$  &1522(84)& 152(5) & 0.24(4) & 196(17)& 55267(84)& 0.020(3) & 0.50 & 0.04 & 2: \\
7518816 & 0.46658065(6)& $-$  &1523(35)&  49(2) & 0.27(2) & 171(3) & 55022(22)&0.00067(8)& 0.15 & 0.001& 2: \\
10383620& 0.7345688(1) & $-$  &1541(12)& 277(3) & 0.219(4)& 0.9(3) & 54250(9) & 0.120(4) & 1.03 & 0.003& 2: \\
10557008& 0.26541872(1)& $-$  &1545(15)& 79.2(5)& 0.343(5)& 188(2) & 55243(11)&0.00279(7)& 0.24 & 0.001& 2: \\
9083523 & 0.9184227(3) & $-$  &1573(77)&  59(5) & 0.39(1) &  97(2) & 54664(53)& 0.0011(3)& 0.17 & 0.006& 2: \\
2715007 & 0.29711140(4)& $-$  &1598(21)& 256(3) & 0.623(6)&213.3(7)& 54766(16)& 0.088(4) & 0.90 & 0.001& 2: \\
9596187 & 0.953283(3)  & $-$  &1599(97)& 508(52)& 0.18(4) &  41(7) & 54892(46)& 0.69(22) & 2.35 & 0.004& 2: \\
10916675& 0.41886753(4)& $-$  &1626(77)&  20(1) & 0.31(3) &  36(7) & 55133(47)&0.00004(1)& 0.06 & 0.001& 2: \\
9706078:& 0.613561(2)  & $-$  &1632(287)&109(54)& 0.49(11)&  73(10)& 54973(60)& 0.007(10)& 0.33 & 0.003& 2: \\
5956776 & 0.5691161(6) & $-$  &1655(1122)&33(19)& 0.57(20)&  16(5) &54222(788)& 0.0002(4)& 0.09 & 0.003& 2: \\
6606282 & 2.107135(1)  & $-$  &1681(61)& 317(9) & 0.32(3) & 134(3) & 55781(45)& 0.15(2)  & 1.14 & 0.02 & 2: \\
11234677& 1.587418(2)  & $-$  &1738(171)&135(18)& 0.20(4) & 156(6) & 55586(99)& 0.011(5) & 0.40 & 0.01 & 2: \\
3248019 & 2.668200(5)  & $-$  &1749(331)&130(48)& 0.44(7) &  19(12)&54706(162)& 0.010(11)& 0.38 & 0.04 & 2: \\
2305372 & 1.40469157(8)& $-$  &1772(25)& 140(2) & 0.206(9)& 305(2) & 54966(14)&0.0117(6) & 0.41 & 0.009& 2: \\
6766325:& 0.4399650(4) & $-$  &1801(202)& 78(17)& 0.41(3) & 231(4) &54460(149)& 0.002(1) & 0.21 & 0.001& 2: \\
8690104:& 0.4087740(2) & $-$  &1835(222)& 46(9) & 0.24(6) & 287(7) &54704(133)& 0.0004(2)& 0.12 & 0.001& 2: \\
8982514:& 0.41449027(4)& $-$  &1901(150)& 63(2) & 0.12(1) & 299(9) & 54288(70)& 0.0009(1)& 0.16 & 0.001& 2: \\
11246163& 0.27922679(9)& $-$  &1902(149)& 56(5) & 0.36(3) & 210(3) & 55840(94)& 0.0006(2)& 0.14 & 0.001& 2: \\
5269407 & 0.958860(2)  & $-$  &1905(172)&268(31)& 0.53(2) &  82(6) & 55206(61)& 0.07(3)  & 0.83 & 0.005& 2: \\
4174507 & 3.89179(1)   & $-$  &1922(333)&647(87)& 0.82(3) & 202(3) &55730(181)& 0.98(52) & 2.85 & 0.34 & 2: \\
5962716 & 1.8045827(2) & $-$  &1935(29)& 208(2) & 0.507(8)&253.4(9)& 55804(17)& 0.032(1) & 0.60 & 0.02 & 2: \\ 
9788457:& 0.96333879(1)& $-$  &1960(425)&27.9(2)& 0.46(1) &  17(1) &55737(167)&0.00007(3)& 0.07 & 0.005& 2: \\
12055014& 0.4999043(1) & $-$  &1961(175)& 31(5) & 0.32(4) &  29(5) & 54568(72)&0.00010(6)& 0.08 & 0.001& 2: \\
10724533& 0.7450940(4) & $-$  &2028(198)& 70(10)& 0.499(9)&  77(3) &54178(131)& 0.0011(5)& 0.17 & 0.003& 2: \\
8868650 & 4.4474056(9) & $-$  &2040(88)& 367(9) & 0.62(2) & 234(2) & 55374(32)& 0.16(2)  & 1.17 & 0.13 & 2: \\
10275197& 0.390846(1)  & $-$  &2127(82)& 612(72)& 0.268(4)&210.7(9)& 54851(25)& 0.68(24) & 2.34 & 0.001& 2: \\
3335816 & 7.422028(5)  & $-$  &2250(1234)&66(42)& 0.16(24)& 233(69)&54351(703)& 0.001(2) & 0.15 & 0.16 & 2: \\
5975712 & 1.136080(1)  & $-$  &2308(118)&347(21)& 0.43(1) & 115(4) & 55530(59)& 0.11(2)  & 0.98 & 0.004& 2: \\ 
3839964 & 0.2561427(4) & 29(1)&2404(371)&311(22)& 0.17(1) &   4(5) &53795(184)& 0.07(3)  & 0.82 & 0.002& 2: \\
8444552 & 1.1780785(7) & $-$  &2441(73)&376(13) & 0.492(6)& 104(1) & 55301(22)& 0.12(1)  & 1.03 & 0.005& 2: \\
4937217 & 0.4293407(2) &3.8(6)&2468(1187)&24(12)& 0.49(14)& 176(8) &55622(453)&0.00003(5)& 0.05 & 0.001& 2: \\
8758161 & 1.9964243(2) & $-$  &2501(276)&133(2) & 0.196(7)& 103(1) & 55375(47)& 0.005(1) & 0.30 & 0.01 & 2: \\
12055255& 0.2209449(6) & $-$  &2530(166)&544(57)& 0.416(8)& 280(2) & 55142(33)& 0.34(12) & 1.65 & 0.001& 2: \\ 
4758368 & 3.74998(1)   & $-$  &2876(1289)&357(127)&0.7(1) & 313(10)&55556(429)& 0.07(10) & 0.84 & 0.08 & 2: \\
8429450 & 2.705145(7)  & $-$  &3088(1698)&128(75)&0.38(17)& 185(11)&56136(894)& 0.003(6) & 0.24 & 0.02 & 2: \\
9110346 & 1.790580(3)  & $-$  &3645(695)&350(59)& 0.74(2) & 307(3) &55406(129)& 0.04(3)  & 0.68 & 0.01 & 2: \\
8265951 & 0.7799554(2) & $-$  &3721(247)&423(19)& 0.76(1) &215.9(5)& 55390(41)& 0.07(1)  & 0.84 & 0.003& 2: \\
6615041 & 0.34008660(2)& $-$  &3951(1200)&105(1)& 0.616(7)& 30.3(8)&55747(238)& 0.0010(6)& 0.17 & 0.001& 2: \\
9532219 & 0.19815367(5)& $-$  &4401(900)&217(7) & 0.38(1) & 205(2) &55770(159)& 0.007(3) & 0.34 & 0.001& 2: \\
8553788 & 1.606184(2)  & $-$  &4579(552)&473(66)& 0.75(1) & 56.2(7)&56474(249)& 0.07(3)  & 0.81 & 0.008& 2: \\
6794131 & 1.613324(2)  & $-$  &4743(2105)&446(141)&0.87(5)& 150(2) &55889(604)& 0.05(7)  & 0.73 & 0.03 & 2: \\
10686876& 2.618397(8)  & $-$  &5280(1590)&400(147)&0.33(10)&174(3) & 54912(51)& 0.03(4)  & 0.59 & 0.006& 2: \\
6233903 & 5.99090(3)   & $-$  &5359(2135)&642(223)&0.69(8)&   2(2) &56036(582)& 0.12(16) & 1.05 & 0.08 & 2: \\
9181877 & 0.321019(5)  & $-$  &5497(2957)&963(711)&0.35(15)&332(12)&55078(263)& 0.40(97) & 1.78 & 0.001& 2: \\
9412114 & 0.2502592(2) & $-$  &5596(353)&922(48)& 0.70(1) &   1(1) & 55540(53)& 0.34(7)  & 1.65 & 0.001& 2: \\
8016214 & 3.174930(5)  & $-$  &7350(2008)&484(113)&0.71(5)& 173(3) &55328(162)& 0.03(2)  & 0.57 & 0.02 & 2: \\
7272739 & 0.28116304(6)& $-$  &9256(910)&218(17)& 0.75(2) & 184(1) &55988(144)& 0.0016(5)& 0.20 & 0.001& 2: \\
\hline
\end{tabular}
\end{center}
\end{table*}

\setlength{\tabcolsep}{3.3pt}
\begin{table*}
\begin{center}
\caption{Orbital Elements from combined dynamical and LTTE solutions for systems, where more than two outer periods are covered, or/and triply eclipsing systems} 
\label{Tab:Orbelemdyn1}  
\begin{tabular}{lccccccccccc} 
\hline
KIC No. & $P_1$ & $P_2$ & $a_2$ & $e_2$ & $\omega_2$ & $\tau_2$ & $f(m_\mathrm{C})$ & $\frac{m_\mathrm{C}}{m_\mathrm{ABC}}$ & $m_\mathrm{AB}$ & $m_\mathrm{C}$ & $\frac{{\cal{A}}^\mathrm{meas}_\mathrm{dyn}}{{\cal{A}}_\mathrm{LTTE}}$\\
        & (day) & (day) &(R$_\odot$)  &       &   (deg)    &   (MBJD) & (M$_\odot$)       & (M$_\odot$)            &  (M$_\odot$)  &   \\
\hline
5897826$^a$& 1.76713(19)&33.921(1)&53.6(4)& 0.304(2)& 52.9(3)&55168.63(3)&0.754(1) & 0.748(1)& 0.454(4)& 1.35(3)&10.67\\
5952403 & 0.90567828(5) &45.47(2) &90.3(7)& 0.0     & $-$    &  $-$     & 1.19(1)  & 0.63(1) & 1.79(4) & 3.0(1) & 0.00\\
6531485 & 0.67699050(2) &48.267(6)& 73(4) & 0.57(1) &  22(2) & 54983(1) & 0.198(9) & 0.45(2) & 1.3(2)  & 1.0(2) & 2.89\\
7690843$^c$&0.7862597(1)& 74.25(3)&123(13)& 0.369(2)&   4(2) & 54919(1) & 1.25(5)  & 0.66(7) & 1.6(6)  &3.0(1.0)& 0.60\\
3544694 & 3.8457246(6)  & 80.99(9)&120(7) & 0.109(6)& 334(2) & 55724(7) & 0.11(2)  & 0.32(3) & 2.4(4)  & 1.1(2) & 3.17\\
10613718& 1.17587788(3) & 88.20(4)& 93(10)& 0.10(3) &  49(12)& 54994(3) & 0.21(5)  & 0.53(3) & 0.7(2)  & 0.7(2) & 0.31\\
6545018 & 3.99145688(7) &90.586(5)&118(3) & 0.225(4)& 236(1) & 54971(1) & 0.044(9) & 0.25(2) & 2.0(2)  & 0.7(1) & 6.68\\
9714358 & 6.4742247(8)  &103.77(2)&113(12)& 0.29(1) & 120(2) & 54977(1) & 0.010(3) & 0.178(8)& 1.5(5)  & 0.3(1) &23.98\\ 
9451096 & 1.25039069(1) &106.89(1)&121(12)& 0.093(2)& 159(2) & 54993(1) & 0.045(1) & 0.28(3) & 1.5(5)  & 0.6(2) & 0.21\\
5771589 & 10.738233(3)  &112.97(2)&152(5) &0.1294(8)&290.8(5)& 54978(1) & 0.44(9)  & 0.50(4) & 1.9(2)  & 1.9(2) &21.88\\
9140402 & 4.988351(6)   & 117.0(2)&112(9) & 0.24(1) & 300(32)& 55022(15)& 0.48(14) & 0.71(9) & 0.41(2) & 1.0(3) & 8.89\\
4079530:& 17.72714(2)   & 144(1)  &134(106)&0.06(5) &  89(90)& 54967(51)& 20(1)E-6 &0.02(2) & 1.5(3.6)& 0.04(9)&18.15\\
7668648 & 27.8256(2)    & 204.8(4)&179(17)& 0.33(2) & 341(5) & 54917(4) & 0.006(1) & 0.15(1) & 1.6(5)  & 0.27(8)&11.05\\
7955301 & 15.32775(1)   & 209.1(1)&229(26)& 0.310(7)& 309(1) & 54879(1) & 0.22(7)  & 0.40(1) & 2.2(8)  & 1.5(5) &33.26\\
5080652:& 4.1443558(2)  & 220.9(8)&187(44)& 0.13(3) &  18(4) & 54966(8) & 0.16(10) & 0.45(5) & 1.0(7)  & 0.8(6) & 0.99\\ 
5095269 & 18.611868(5)  &236.26(8)&204(28)& 0.071(3)& 324(3) & 55004(2) & 13(5)E-7 &0.0090(5)&2.0(8) &0.018(8)&30.50\\ 
6964043 & 10.72553(2)   & 239.1(2)&248(25)& 0.52(1) & 311(2) & 55110(2) & 0.27(8)  & 0.42(2) & 2.1(6)  & 1.5(5) &30.59\\
7289157 & 5.2665478(4)  &243.36(8)&215(6) & 0.309(3)&156.5(7)& 54942(1) & 0.14(2)  & 0.39(3) & 1.4(1)  & 0.9(1) & 4.41\\ 
5384802 & 6.0830921(3)  &255.23(5)&244(11)& 0.357(5)&  11(2) & 55000(2) & 0.24(3)  & 0.44(3) & 1.7(2)  & 1.3(2) & 5.06\\
5264818 & 1.9050517(1)  & 299.4(6)&296(40)& 0.44(3) & 214(6) & 54948(6) & 0.029(8) & 0.34(5) & 2.6(1.1)& 1.3(6) & 0.44\\
8719897 & 3.15141994(9) & 333.1(2)&264(12)& 0.265(7)& 128(2) & 54997(2) & 0.12(2)  & 0.38(3) & 1.4(2)  & 0.9(1) & 0.62\\
7593110 & 3.5493857(3)  & 353(1)  &267(140)&0.10(6) & 144(29)& 54997(29)& 0.0248(2)& 0.24(12)& 1.6(2.5)& 0.5(8) & 0.24\\
4940201 & 8.816559(1)   & 364.9(3)&278(24)& 0.24(2) & 247(5) & 54864(7) & 0.0618(1)& 0.31(3) & 1.5(4)  & 0.7(2) & 3.62\\
10483644& 5.1107702(2)  & 371(2)  &287(131)&0.17(4) & 343(5) & 54929(12)& 0.04(2)  & 0.25(10)& 1.7(2.4)& 0.6(8) & 0.77\\
8938628 & 6.8622000(2)  & 388.6(2)&308(27)& 0.21(1) &  63(2) & 54824(4) & 0.17(6)  & 0.41(6) & 1.5(4)  & 1.1(3) & 1.46\\
6525196 & 3.42059733(4) & 418.2(1)&334(27)& 0.295(5)&  94(2) & 55070(3) & 0.066(10)& 0.29(3) & 2.0(5)  & 0.8(2) & 0.51\\
10095512& 6.0172059(1)  & 473.4(2)&324(76)& 0.19(1) & 329(4) & 54865(8) & 0.17(4)  & 0.44(10)& 1.1(8)  & 0.9(7) & 0.81\\
4909707 & 2.3023671(2)  & 514.8(6)&406(14)& 0.60(1) & 176(1) & 54848(2) & 0.276    & 0.43(2) & 1.9(2)  & 1.5(2) & 0.65\\
7177553 & 17.99628(6)   & 529(2)  &339(50)& 0.46(2) & 201(5) & 54701(9) & 41(1)E-9 &0.0028(3)&1 .9(8)  &0.005(2)&47.93\\
8023317 & 16.57907(1)   & 610.6(5)&342(11)& 0.249(4)& 164(1) & 55014(3) & 0.0015(7)& 0.10(2) & 1.3(1)  &0.15(3) & 7.86\\ 
5255552 & 32.465339(2)  & 862.1(2)&510(17)&0.4342(7)& 37.3(1)& 54875(1) & 0.0609(1)& 0.29(1) & 1.7(2)  & 0.7(1) &17.21\\
\hline
\end{tabular}
\end{center}
{\bf Notes.} {$a$: From photodynamical solution of \citet{carteretal11}; $b$: Combination of ETV, radial velocity and lightcurve soolution of \citet{borkovitsetal13}; $c$: Cubic ephemeris: $\Delta P=-30(4)\times10^{-10}$\,d/c, $c_3=1.09(6)\times10^{-12}$\,d/c$^3$. }
\end{table*}

\setlength{\tabcolsep}{3.3pt}
\begin{table*}
\begin{center}
\caption{Orbital Elements from combined dynamical and LTTE solutions which cover more than one but less than two outer periods} 
\label{Tab:Orbelemdyn2}  
\begin{tabular}{lccccccccccc} 
\hline
KIC No. & $P_1$ & $P_2$ & $a_2$ & $e_2$ & $\omega_2$ & $\tau_2$ & $f(m_\mathrm{C})$ & $\frac{m_\mathrm{C}}{m_\mathrm{ABC}}$ & $m_\mathrm{AB}$ & $m_\mathrm{C}$ & $\frac{{\cal{A}}^\mathrm{meas}_\mathrm{dyn}}{{\cal{A}}_\mathrm{LTTE}}$\\
        & (day) & (day) &(R$_\odot$)  &       &   (deg)    &   (MBJD) & (M$_\odot$)       & (M$_\odot$)            &  (M$_\odot$)  &   \\
\hline
7812175 & 17.79359(2)   & 583(2)  &389(50)& 0.030(4)& 207(6) & 54783(11)& 0.07(3)  & 0.31(7) & 1.6(6)  & 0.7(3) & 5.88\\
9715925 & 6.308265(3)   & 736(36) &325(56)& 0.38(2) & 136(7) & 55083(42)& 0.007(2) & 0.21(4) & 0.7(4)  & 0.2(1) & 1.23\\
8210721 & 22.67318(4)   & 789.7(4)&492(19)& 0.259(2)& 212(1) & 54628(4) & 0.10(3)  & 0.34(3) & 1.7(2)  & 0.9(1) & 9.74\\
9664215 & 3.3195345(8)  & 910(7)  &539(68)& 0.536(8)& 190(2) & 54861(7) & 0.161(6) & 0.40(5) & 1.5(6)  & 1.0(4) & 0.42\\
5731312 & 7.9464246(2)  & 911(3)  &423(42)& 0.584(2)& 25.9(4)& 54837(3) & 0.0015(5)& 0.11(2) & 1.1(3)  &0.13(4) & 4.96\\
5653126 & 38.49233(5)   & 968(2)  &586(31)& 0.189(4)& 326(1) & 55469(4) & 0.15(2)  & 0.38(1) & 1.8(3)  & 1.1(2) &26.91\\
7821010 & 24.2382191(1) & 991(3)  &551(23)& 0.372(9)& 126(2) & 55124(6) & 3(1)E-9  &0.00111(4)&2.3(3)  &0.0025(3)&32.60\\  
10979716& 10.684099(2)  & 1047(4) &530(6) & 0.445(5)& 60.3(5)& 54518(4) & 0.099(2) & 0.389(5)& 1.12(4) &0.71(3) & 2.57\\
4948863 & 8.6435529(9)  & 1060(11)& 80(2) & 0.11(2) & 124(7) & 55107(24)& 0.0060(5)& 0.15(3) & 1.7(9)  & 0.3(2) & 0.28\\
6546508 & 6.107118(6)   & 1154(31)&523(77)& 0.34(3) & 321(3) & 55123(19)& 0.26(2)  & 0.56(8) & 0.6(3)  & 0.8(4) & 0.47\\
4769799 & 21.9284(1)    & 1231(8) &653(74)& 0.191(8)& 233(9) & 55542(40)& 0.04(1)  & 0.26(4) & 1.8(6)  & 0.6(2) & 3.29\\
7837302 & 23.83679(6)   & 1382(2) &213(238)&0.260(4)&	3(5) & 54974(26)& 0.07(23) & 0.31(38)& 1.6(2.7)&0.7(1.4)& 5.06\\ 
10549576& 9.08946(3)    & 1411(52)&821(461)&0.54(7) & 139(6) & 55015(52)& 0.05(3)  & 0.24(13)& 2.8(4.8)&0.9(1.6)& 1.31\\
11519226& 22.161767(7)  & 1437(1) &745(8) & 0.332(2)&321.7(5)& 55010(2) & 0.27(1)  & 0.463(8)& 1.44(5) &1.25(4) & 5.21\\ 
\hline
\end{tabular}
\end{center}
\end{table*}

\setlength{\tabcolsep}{3.3pt}
\begin{table*}
\begin{center}
\caption{Orbital Elements from combined dynamical and LTTE solutions which cover less than a full outer period} 
\label{Tab:Orbelemdyn3}  
\begin{tabular}{lccccccccccc} 
\hline
KIC No. & $P_1$ & $P_2$ & $a_2$ & $e_2$ & $\omega_2$ & $\tau_2$ & $f(m_\mathrm{C})$ & $\frac{m_\mathrm{C}}{m_\mathrm{ABC}}$ & $m_\mathrm{AB}$ & $m_\mathrm{C}$ & $\frac{{\cal{A}}^\mathrm{meas}_\mathrm{dyn}}{{\cal{A}}_\mathrm{LTTE}}$\\
        & (day) & (day) &(R$_\odot$)  &       &   (deg)    &   (MBJD) & (M$_\odot$)       & (M$_\odot$)            &  (M$_\odot$)  &   \\
\hline
10319590& 21.32116(6)   &  452(2) & 330(11)&0.146(4)& 316(2) & 54857(3) & 0.10(3)  & 0.35(3) & 1.5(2)  & 0.8(1) &10.02\\
4078157 & 16.02554(2)   & 1377(26)& 736(66)&0.480(8)&  70(3) & 54630(24)& 0.100    & 0.34(3) & 1.9(5)  & 1.0(3) & 5.09\\ 
3345675:& 120.033(2)    & 1662(94)&671(336)& 0.39(2)&  95(65)&54894(428)& 0.001    & 0.10(5) & 1.3(2.0)& 0.1(2) &117.39\\
8143170 & 28.78680      & 1710(35)& 864(21)&0.704(6)&108.7(9)& 54411(26)& 0.005(1) & 0.13(1) & 2.6(2)  &0.37(5) &25.48\\
12356914& 27.3083183(3) & 1804(1) & 807(43)&0.385(1)& 36.5(1)& 55860(1) & 0.0096(1)& 0.19(1) & 1.8(3)  &0.41(7) & 7.03\\
5003117 & 37.6094(2)    & 2128(50)&892(145)& 0.26(1)& 191(6) & 54750(49)& 0.06(1)  & 0.33(6) & 1.4(7)  & 0.7(4) & 6.32\\
6877673:& 36.759691(6)  & 2870(11)&1112(254)&0.468(2)&155.5(5)&54286(11)& 0.03(2)  & 0.27(3) & 1.6(1.1)& 0.6(4) & 7.81\\
2576692 & 87.8797(1)    &2884(173)&936(216)& 0.56(3)& 161(15)&54277(213)&0.00004(2)& 0.032(5)& 1.3(9)  &0.04(3) &33.85\\
7670617 & 24.70317(4)   &3304(108)&1054(30)&0.707(7)& 86.4(9)& 55642(35)& 0.082(9) & 0.38(2) & 0.9(1)  &0.55(6) & 7.81\\ 
11502172:&25.431831(7)  & 3313(58)&1081(140)&0.17(1)&  86(4) & 54359(47)& 0.020(3) & 0.25(3) & 1.2(5)  & 0.4(2) & 0.83\\
9028474 & 124.93573(1)  & 3378(94)&1258(421)&0.09(2)& 242(8) & 54286(78)& 10(5)E-6 &0.0163(4)&2.3(1.2)&0.04(2) &44.23\\
9963009 & 40.0716(1)    & 3770(10)&1447(46)& 0.24(6)& 189(6) & 54074(79)& 0.111(7) & 0.41(1) & 1.7(2)  & 1.2(1) & 2.98\\
11558882:&73.9135(2)    & 4050(50)&1417(301)&0.30(2)& 105(5) & 54919(80)& 0.016(7) & 0.19(3) & 1.9(1.2)& 0.4(3) & 7.16\\
4753988:& 7.30451(1)    &5567(2325)&1597(577)&0.67(8)&349(3) &55359(238)& 0.007(9) & 0.20(3) & 1.4(1.9)& 0.4(5) & 0.13\\
10268809& 24.70843(1)   &7000(1000)&2208(60)&0.737(1)&292.6(6)&56147(169)& 0.32(10)& 0.48(2) & 1.5(5)  & 1.4(4) & 2.66\\
4055092 & 76.464532(9)  &11548(88)&2353(39)&0.533(2)&276.2(4)& 56487(21)& 0.242(2) & 0.65(1) & 0.5(1)  & 0.9(1) & 3.08\\
10296163& 9.296847(4)   &15271(760)&3172(286)&0.73(1)&355(3) &55918(132)& 0.016(4) & 0.26(1) & 1.4(4)  & 0.5(1) & 0.13\\ 
\hline
\end{tabular}
\end{center}
\end{table*}

\setlength{\tabcolsep}{3.3pt}
\begin{table*}
\begin{center}
\caption{Orbital Elements from LTTE solutions for systems which probably are oscillating variables instead of binaries (i.e. false positive EBs).} 
\label{Tab:OrbelemFP}  
\begin{tabular}{lccccccccccc} 
\hline
KIC No. & $P_1$ & $\Delta P_1$ & $P_2$ & $a_\mathrm{AB}\sin i_2$ & $e_2$ & $\omega_2$ & $\tau_2$ & $f(m_\mathrm{C})$ & $(m_\mathrm{C})_\mathrm{min}$ & $\frac{{\cal{A}}_\mathrm{dyn}}{{\cal{A}}_\mathrm{LTTE}}$ & $m_\mathrm{AB}$\\
        & (day) &$\times10^{-10}$ (d/c) &(day)&(R$_\odot$)  &       &   (deg)    &   (MBJD) & (M$_\odot$)       & (M$_\odot$)            & &  (M$_\odot$)    \\
\hline
10855535& 0.11278241(1)& $-$  &411.9(2)& 61.4(2)&0.096(5) & 296(3) & 55135(3) & 0.0183(1)& 0.48 & 0.006& 2: \\
        & 0.05639121(1)& $-$  &411.9(2)& 60.6(2)&0.106(8) & 292(4) & 55131(5) & 0.0176(2)& 0.48 & $-$  & 2: \\
8045121 & 0.26317782(1)& $-$  & 896(2) & 139(1) & 0.37(1) & 342(2) & 55237(6) & 0.045(1) & 0.69 & 0.001& 2: \\
        & 0.13158891(1)& $-$  & 896(3) & 140(2) & 0.37(2) & 342(3) & 55238(7) & 0.045(2) & 0.69 & $-$  & 2: \\ 
8563964 & 0.33843576(2)& $-$  &1183(6) & 98.7(7)&0.199(9) & 345(2) & 55035(7) & 0.0092(2)& 0.37 & 0.001& 2: \\
        & 0.16921788(1)& $-$  &1184(6) & 98.7(7)&0.196(9) & 345(2) & 55034(8) & 0.0092(2)& 0.37 & $-$  & 2: \\
12508348& 0.255619(6)  &$-86$(12)&1839(472)&789(235)&0.36(9)&218(3)&55770(298)&1.95(2.01)& 4.23 & 0.001& 2: \\
        & 0.127810(4)  &$-22$(4)&1814(618)&754(319)&0.30(13)&213(5)&55754(391)&1.75(2.51)& 3.96 & $-$  & 2: \\
11825204& 0.2096193(1) &46.8(4)&2230(236)&107(9)& 0.75(3) & 297(2) &55894(140)& 0.003(1) & 0.26 & 0.001& 2: \\
        & 0.1048096(2) &11.8(2)&2588(966)&112(22)&0.79(5) & 294(2) &55887(488)& 0.003(3) & 0.24 & $-$  & 2: \\
6287172 & 0.2038728(2) & $-$  &3583(1875)&365(159)&0.95(3)& 170(2) &56053(822)& 0.05(9)  & 0.72 & 0.005& 2: \\
        & 0.10193641(9)& $-$  &3320(1216)&345(109)&0.95(2)& 170(1) &56052(575)& 0.05(6)  & 0.72 & $-$  & 2: \\
7375612 & 0.16007308(6)& $-$  &4417(835)&287(46)& 0.41(7) & 306(4) &55957(259)& 0.016(10)& 0.46 & 0.001& 2: \\
        & 0.08003657(5)& $-$  &5859(2075)&365(106)&0.49(11)&302(4) &55938(478)& 0.019(21)& 0.49 & $-$  & 2: \\
9612468 & 0.13347101(9)& $-$  &5307(1624)&162(38)&0.76(5) & 193(4) &55450(225)& 0.002(2) & 0.22 & 0.001& 2: \\
        & 0.06673554(5)& $-$  &4888(1842)&133(38)&0.75(7) & 192(5) &55455(274)& 0.001(1) & 0.18 & $-$  & 2: \\
\hline
\end{tabular}
\end{center}
\end{table*}

\begin{table*}
\begin{center}
\caption{Apsidal motion and/or orientation parameters from AME and dynamical fits} 
\label{Tab:AMEparam}
\scalebox{0.88}{
\begin{tabular}{lccccccccccc} 
\hline
KIC No. & $P_\mathrm{anom}$ & $a_1$      & $e_1$ & $\omega_1$ & $\tau_1$ & $P_\mathrm{apse}$ & $\im$ & $i_1$ & $i_2$ & $\Delta\Omega$ & $P_\mathrm{node}$\\
        & (days)            &(R$_\odot$) &       & (deg)      & (MJD)    &   (years)         & (deg) & (deg) & (deg) &   (deg)   &  (years) \\
\hline
4758368 & 3.750(1)    & $-$     & 0.0043(5)&   4(69) & 54959.2(7) & 123(527)    &&&&&\\
5039441 & 2.151385(8) & $-$     &  0.01(4) & 283(42) & 54955.4(2) & 5286(18159) &&&&&\\   
6233903 & 5.9910(2)   & $-$     & 0.006(15)& 290(48) & 55002.1(8) & 1690(4648)  &&&&&\\
6965293 & 5.077754(6) & $-$     & 0.020(7) & 239(12) & 54957.1(2) & 7095(4286)  &&&&&\\
\hline
2576692 & 87.8770(1)  &  90(21) & 0.15(2)  & 338(19) & 55040(4)   & $-7924$     & 42(2) & 88 &102 & 40(5)  &2118\\
3345675 & 120.054(2)  & 112(57) & 0.11(4)  & 279(97) & 55086(26)  & 1820        & 24(19)& 86 & 62 &$-1$(27)&348\\
3544694 & 3.8478379(6)&13.8(8)  &0.00135(4)& 329(6)  & 55741.28(6)&  19         & 0     & 84 & 84 & 0     & $-$\\
4055092 &76.460808(9) &58.5(1.1)&0.34515(7)&309.53(1)&54970.961(1)&$-4298$      & 54(1) & 88 &119 & 46(1) &6297\\ 
4078157 & 16.02631(2) &32.9(3.0)& 0.198(6) & 205(4)  & 54958.3(1) & 913         & 10(3) & 84 & 75 &  5(14)& 697\\
4079530 & 17.72746(8) &  33(26) & 0.2985(5)& 315(10) & 54996.0(3) & 2642(648)   & 0     & 88 & 88 & 0     & $-$\\ 
4753988 & 7.30451(2)  &17.8(8.1)& 0.020(3) &  75(2)  & 54971.35(6)& 21051(37135)& 47(7) & 84 & 53 & 38(7) &40986\\
4769799 & 21.9300(1)  &40.3(4.6)& 0.10(2)  & 330(21) & 54972(1)   & 805         & 22(2) & 86 & 69 & 14(10)& 826\\
4909707 & 2.3023959(2)& 9.1(3)  & 0.013(3) & 241(6)  & 54953.74(4)& 503         &  6(1) & 88 & 87 &$-6$(1)& 471\\  
4940201 & 8.817798(1) &20.6(1.8)& 0.0014(1)& 194(16) & 54965.4(4) & 172         &  6(2) & 85 & 86 &$-6$(2)& 139\\
4948863 & 8.6436174(9)&21.0(3.7)&0.01810(9)& 255     & 54972.5(3) & 3172        & 0     & 82 & 82 & 0     & $-$\\
5003117 & 37.6141(2)  &53.0(8.8)& 0.14(3)  & 309(10) & 54989.2(8) & 826         & 43(1) & 89 & 66 & 38(4) &1484\\
5080652 & 4.1436823(2)&10.8(2.6)& 0        & $-$     &  $-$       &  $-$        & 0     & 80 & 80 & 0     & $-$\\
5095269 & 18.612758(5)&37.3(5.2)& 0.05(5)  & 270(10) & 54966.9(5) & 1066        & 40(1) & 86 & 73 & 39(1) & 136\\
5255552 & 32.478076(2)&51.0(1.8)&0.30668(6)&105.27(1)& 54956.79(1)& 227         & 6.4(1)&83.8&89.5&$-2.8$(1)&140\\ 
5264818 & 1.9050371(1)& 8.8(2.0)& 0        & $-$     &  $-$       &  $-$        & 39(3) &70(3)&35 & 23(4) & 433\\
5384802 & 6.0812488(3)&16.7(8)  & 0        & $-$     &  $-$       &  $-$        &  5(3) & 83 & 78 &$-0.8$(7)&65\\
5653126 & 38.50848(5) &58.1(3.1)& 0.247(6) & 313(1)  & 54988.65(8)& 251         & 10(1) & 87 & 78 &$-5$(3)& 157\\
5731312 & 7.9463939(2)&17.2(1.7)& 0.4196(1)& 183.9(3)&54967.198(5)& $-5622$     &37.8(4)&88.5&77.3&36.4(4)&1013\\ 
5771589 & 10.7866(1)  &25.3(1.1)&0.01285(8)& 237.7(3)&54961.139(9)& 6.53(2)     & 7.9(8)& 86 & 82 &$-6.9$(8)&7.5\\ 
5897826 & 1.7671(2)   & 4.72(2) & 0.0223(4)& 269.5(4)&55168.754(2)&             & 8     &92.1&96.9&8.01(4)& \\
5952403 & 0.9056768(2)& 4.78(4) & 0        & $-$     &  $-$       &  $-$        & 0     &87.5&87.5& 0     & $-$\\
6525196 & 3.4205160(1)&12.1(1.0)& 0        & $-$     &  $-$       &  $-$        & 0$^a$ & 80 & 80 & 0     & $-$\\
6531485 & 0.6770720(1)& 3.5(2)  & 0.0014(1)&  46(3)  &54965.056(6)&   15        & 0     & 80 & 80 & 0     & $-$\\
6545018 & 3.9914569(1)&13.3(4)  &0.00294(1)& 176.0(4)&54964.796(4)&   27        & 0     & 86 & 86 & 0     & $-$\\
6546508 & 6.107205(6) &12.0(1.9)& 0.002(2) &  65(27) & 55192.4(5) & 1172        & 0     & 86 & 86 & 0     & $-$\\
6877673 & 36.75992(4) &54.8(12.6)&0.18038(3)&57.196(6)&55002.8378(9)&16411(2783)& 35(1) & 88 & 56 & 16(1) &1998\\
6964043 & 10.73721(2) &26.0(2.7)& 0.0548(8)&  77.0(2)&55195.103(6)&   27        & 19(1) &91.2&89.5& 19(1) &  26\\ 
7177553 & 17.9970(4)  &35.5(5.2)&0.39412(1)& 179.7(4)& 54952.23(1)& 1173(676)   & 26(3) & 84 & 81 & 26(3) & 293\\
7289157 & 5.2673864(4)&14.1(4)  & 0.0828(2)& 65.43(4)&54972.1908(8)&  91        & 4.3(3)&85.8&89.5& 2.2(7)&  80\\ 
7593110 & 3.5493317(3)&11.4(6.0)& 0        & $-$     &  $-$       &  $-$        & 30(13)& 82 & 77 & 30(13)& 536\\
7668648 & 27.865(5)   &45.0(4.4)& 0.08(1)  &  85.7(8)& 54976.85(7)&   54(6)     & 42(1) & 84 & 93 &$-41$(2)& 25\\
7670617 & 24.7049(1)  &34.3(1.3)& 0.249(5) & 135(1)  & 54961.5(1) & 965(64)     &147.4(4)&86 & 89 &$-147.8$(4)&$-1678$\\
7690843 & 0.7861873(1)& 4.1(5)  & 0        & $-$     &  $-$       &  $-$        & 0     & 80 & 80 & 0     & $-$\\  
7812175 & 17.79638(2) &33.5(4.4)& 0.169(4) & 321(2)  & 55004.44(7)&  311        & 17(2) & 85 & 79 &$-16$(2)&176\\
7821010 & 24.238246(2)&46.4(2.0)&0.6791$^b$&239.234(1)&54969.3138(1)&60500(5000)& 25(1) & 88 &105 &$-19$(2)&618\\ 
7837302 & 23.83859(6) &40.9(22.7)&0.15(5)  & 314(6)  & 54985.1(4) &  865        & 0     & 86 & 86 & 0     & $-$\\
7955301 & 15.3713(6)  &33.8(3.9)&0.02886(8)& 115.5(7)& 54961.45(3)& 14.8(2)     &18.4(8)& 80 & 79 &$-18.7$(8)&72(34)\\  
8023317 & 16.57780(1) &29.8(1.0)& 0.2511(2)& 177.7(9)& 54976.81(4)& $-595$      &49.5(6)& 88 & 93 &$-49.3$(6)&588\\ 
8143170 & 28.78924(2) &54.3(1.6)& 0.146(4) & 291.3(5)& 54971.38(3)& 929         &38.5(3)& 89 & 114&$-30.5$(3)&890\\
8210721 & 22.67727(4) &40.2(1.7)& 0.140(1) & 158(1)  & 54965.04(8)& 344         & 14(1) &89.5&81.6&$-11$(2)&235\\ 
8719897 & 3.1512989(1)&10.1(5)  & 0        & $-$     &  $-$       &  $-$        & 0$^b$ & 80 & 80 & 0     & $-$\\
8938628 & 6.8628468(2)&17.5(1.6)&0.00271(3)& 345(3)  & 54968.04(6)& 199         & 14(1) & 87 & 80 & 12(1) & 170\\ 
9028474 & 124.93403(2)& 139(24) &0.80575(5)&   2.2(3)& 55013.96(2)& $-25145$    &50.6(9)& 88 & 87 &$-50.7$(9)&1557\\
9140402 & 4.981371(6) & 91.(1.1)& 0        & $-$     &  $-$       &  $-$        & 0     & 85 & 85 & 0     & $-$\\
9451096 & 1.2504286(1)& 5.6(6)  &0.00067(1)& 181(8)  & 54954.42(3)&  113        &  7(1) & 86 & 79 &$-1$(1)& 102\\
9664215 & 3.3195565(8)&10.8(1.4)& 0.02(1)  &  96(3)  & 54963.33(3)& 1371        & 0     & 86 & 86 & 0     & $-$\\
9714358 & 6.4742247(8)&16.6(1.8)&0.01518(4)& 142.1(4)&54965.109(7)&   30        & 0     & 83 & 83 & 0     & $-$\\
9715925 & 6.308231(3) &12.6(2.2)& 0.201(8) & 355(18) & 55000.0(3) & $-3182$     & 37(2) & 83 & 76 &$-37$(2)&1163\\
9963009 & 40.0714(1)  &58.7(1.9)& 0.22(10) & 258(5)  & 54985.2(4) & $-18152$    & 34(3) &89.5&55.7& 0(2)  &2703\\
10095512& 6.0175433(1)&14.6(3.5)&0.00114(5)& 195(9)  & 54952.6(1) &  294        & 0     & 83 & 83 & 0     & $-$\\  
10268809& 24.70935(5) &41.3(4.1)& 0.314(2) & 143.1(3)& 54965.57(3)& 1830(99)    &23.7(4)& 84 & 94 &21.6(4)&3333\\
10296163& 9.296861(7) &20.7(2.0)& 0.354(5) & 45.7(9) & 54962.00(4)& 16784(7355) & 55(5) & 86 &127 &$-40$(3)&121561\\
10319590& 21.33946(6) &37.3(1.4)&0.0256(5) & 247.7(4)& 54964.45(2)&   68        &40.2(4)& 88 &102 &38.0(5)& 110\\
10483644& 5.110517(2) &15.0(6.9)& 0        & $-$     &  $-$       &  $-$        & 0     & 86 & 86 & 0     & $-$\\ 
10549576& 9.08958(3)  &26.0(14.7)&0.00419(7)&355(5)  & 54974.2(1) & 1985        & 0     & 89 & 89 & 0     & $-$\\ 
10613718& 1.1757655(1)& 4.1(4)  & 0        & $-$     &  $-$       &  $-$        & 0     & 86 & 86 & 0     & $-$\\ 
10979716& 10.684099(2)& 21.2(3) & 0.0753(8)& 106.0(2)&54962.300(6)&  755        &  9(1) & 86 & 77 & 0(1)  & 616\\
11502172& 25.431970(7)&38.2(5.0)&0.10074(2)& 334(10) & 54972.4(6) & 12746       & 26(1) & 88 &110 & 15(2) &5700\\
11519226& 22.163175(7)& 37.5(4) &0.18718(4)& 358.4(9)& 54977.11(5)&  955        &17.0(4)& 88 & 89 &17.0(4)& 510\\
11558882& 73.9103(2)  &91.6(19.5)& 0.365(4)& 169(3)  & 54975.8(6) &$-4653$      & 43(3) & 88 & 84 &$-43$(3)&2702\\
12356914& 27.30812(2) &46.0(2.4)& 0.325(1) & 113.2(9)& 54966.0(1) &$-10309$(1210)&40.2(1)&88 & 60 &$-30.4$(1)&1329\\
\hline
\end{tabular}}
\end{center}
{\bf Notes.} {$^a$: Adjusted mutual inclination resulted in $\im=25\degr\pm2\degr$ which would lead to $\Delta i_1\sim1\degr$ during {\em Kepler} observations and consequently, significant eclipse depth variations which is not the case; $^b$: $e_1$ was kept fixed on the radial velocity solution result of Fabrycky~et~al., in prep; $^c$: Adjusted mutual inclination resulted in $\im=23\degr\pm2\degr$ which would lead to $\Delta i_1\sim1.7\degr$ during {\em Kepler} observations and consequently, significant eclipse depth variations which is not the case}
\end{table*}

\section{Results}
\label{Sect:Results}

\subsection{The reliability of the results}
\label{Subsect:Reliability}

In the following subsections we provide a general statistical analysis of our sample, and then discuss the specific properties of selected subsets of our triples. Before this, however, it is crucial to establish the reliability of our third body solutions. Two questions naturally arise. First, do our third-body model and solution really demonstrate the presence of a third body in a given system?  Second, if the third component does exist, how reliable are the parameters we have obtained?  The answers to these questions are somewhat different for the systems with pure LTTE solutions and those with combined LTTE and dynamical solutions, and even for the three different subgroups of each of these two overall categories. For the ten triply eclipsing systems in our sample, the eclipses involving the third star validate not only the existence of the third object, but also the orbital parameters of the wide orbits as well. For the remaining 212 EBs, we can expect that a larger number of complete outer orbits covered generally yields a solution with higher reliability. However, there may be other effects which can mimic short-period periodic or quasi-periodic LTTE-like ETVs, as was discussed previously in Sect.~\ref{Sect:dataprep}. Although, by introducing the QTV analysis and using smoothing polynomials, we are able to filter out the majority of such false positives, we cannot completely eliminate the possibility that a few false positives might remain in our sample. On the other hand, for the subgroup with the shortest outer periods we can say that if an LTTE model turns out to be real, the estimated orbital parameters and mass ratio that are obtained are expected to be reliable and accurate enough for statistical analyses.  With regard to the well-covered systems with combined LTTE and dynamical solutions, there is only a slight chance of misidentifying a non-dynamical ETV curve as having a dynamical third-body origin.  This is due to the fact, that in most cases the dynamically perturbed ETVs have very characteristic shapes. 

For the systems where less than two but more than one outer periods are covered, the differences between pure LTTE and dynamical systems become even more clear. Because of the specific features of the dynamical ETVs, we are convinced that all the systems listed in Tables~\ref{Tab:Orbelemdyn1}--\ref{Tab:Orbelemdyn3} are triples or higher-order multiples, though the reliability of the outer periods and orbital elements depends as usual on the length of the orbital coverage. For pure LTTE systems we cannot offer general rules. When fitting LTTE solutions to systems in the second and third subgroups, we generally tried to avoid incorporating quadratic functions because, in the absence of well-defined and separable short-period ETV modulations, quadratic terms can easily produce spurious LTTE solutions.  For example, \citealp{borkovitsetal05} illustrated that artificial ETVs consisting of two constant period sections with an abrupt period jump between them were nicely fitted with the combination of a quadratic polynomial and an LTTE orbit. However, there were a few cases, e.g., KICs~03440230, 05621294, 07339345 and 07680593, where we were only able to obtain an LTTE solution with the combination of a quadratic fit and an LTTE orbit. In these instances we generally obtained a very low mass third-star companion with a period of about 1000 days. In our opinion, such types of solutions should be considered with considerable caution. Regarding those LTTE solutions where the inferred outer period exceeds the length of the observed data train, the only thing we can say is that, in most cases, the ETVs really signify the presence of a third companion. However, the parameters obtained in most cases are necessarily uncertain and are less suited for statistical analysis.


\begin{figure*}
\includegraphics[width=60mm]{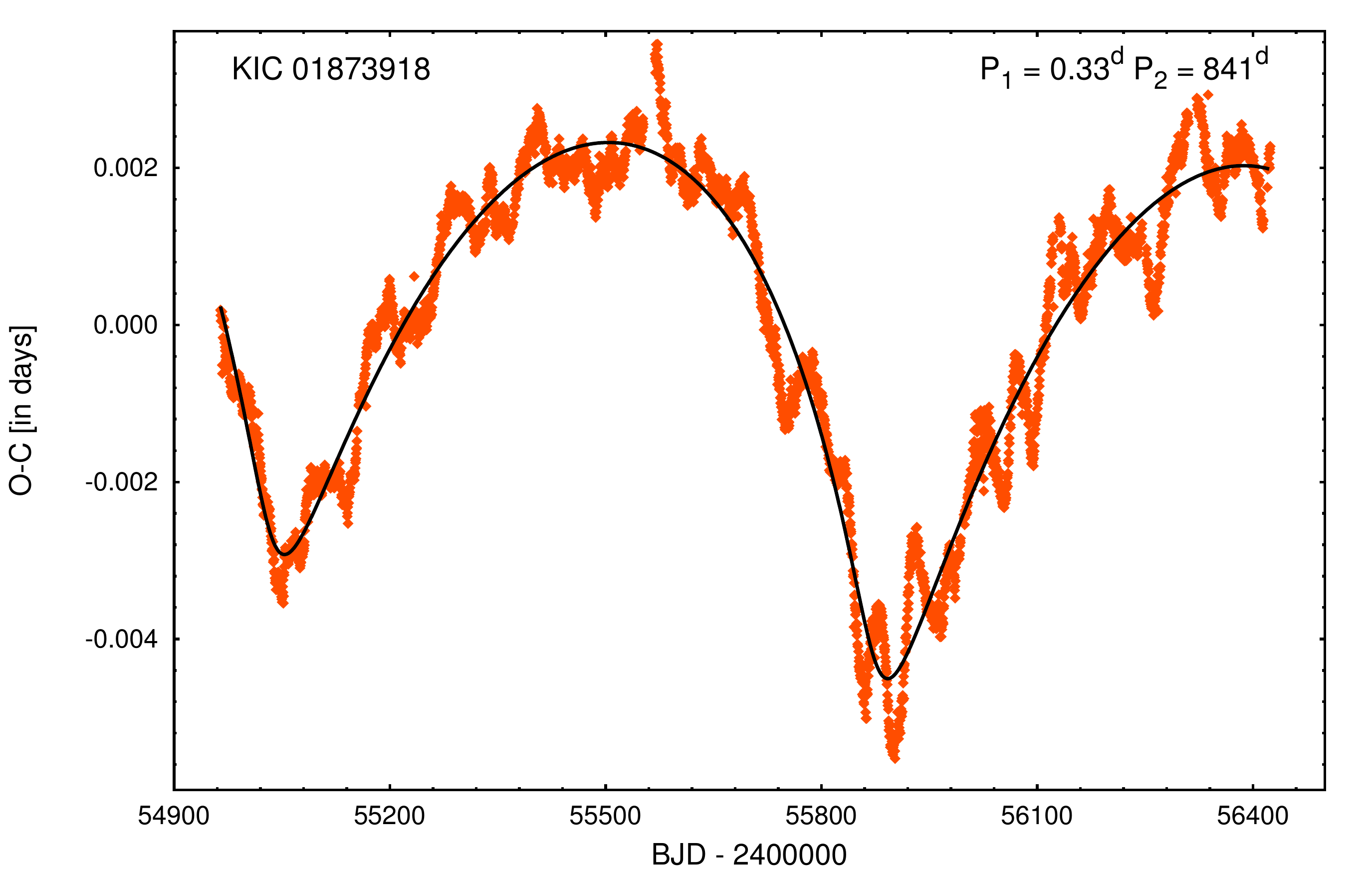}\includegraphics[width=60mm]{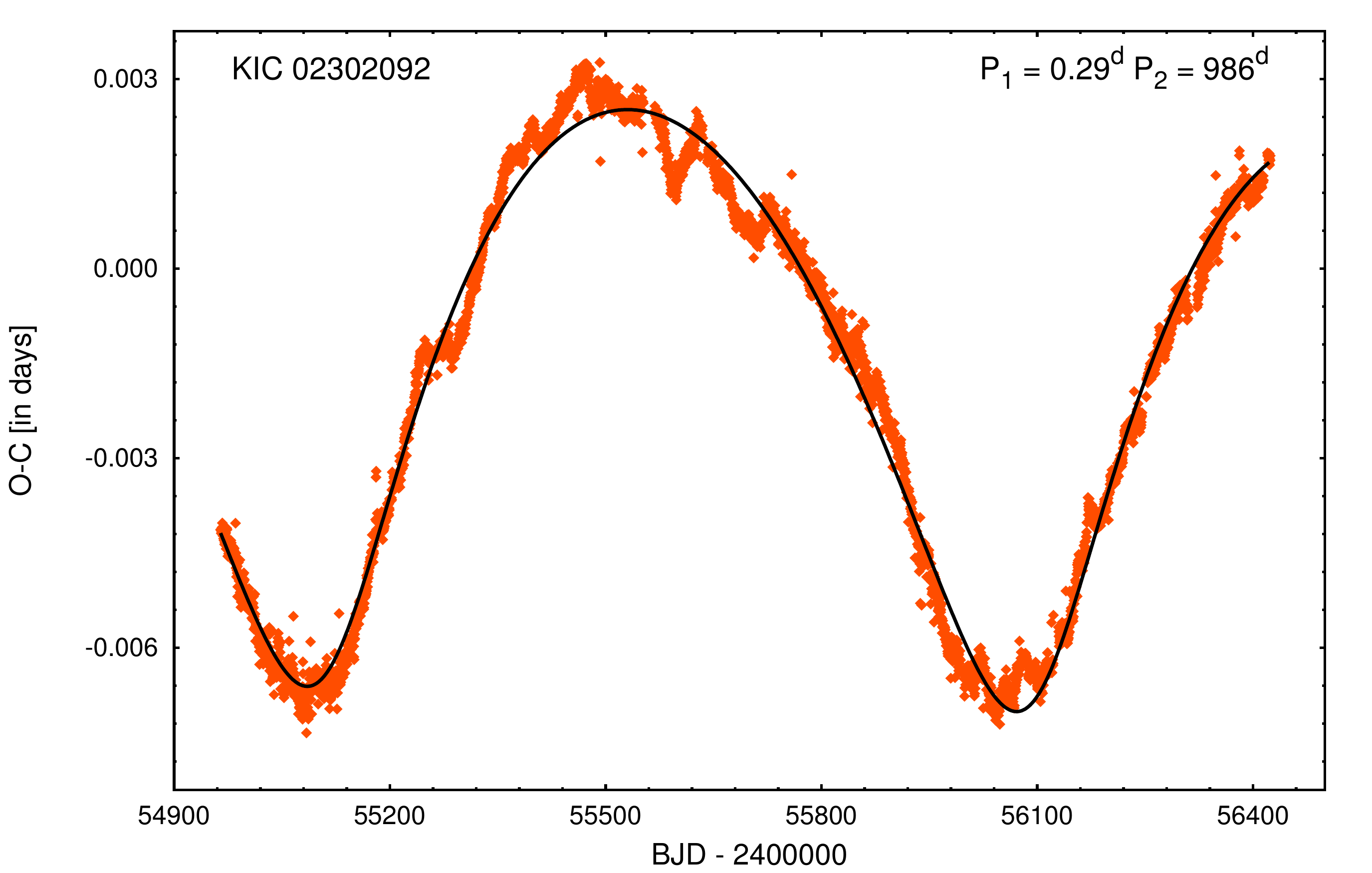}\includegraphics[width=60mm]{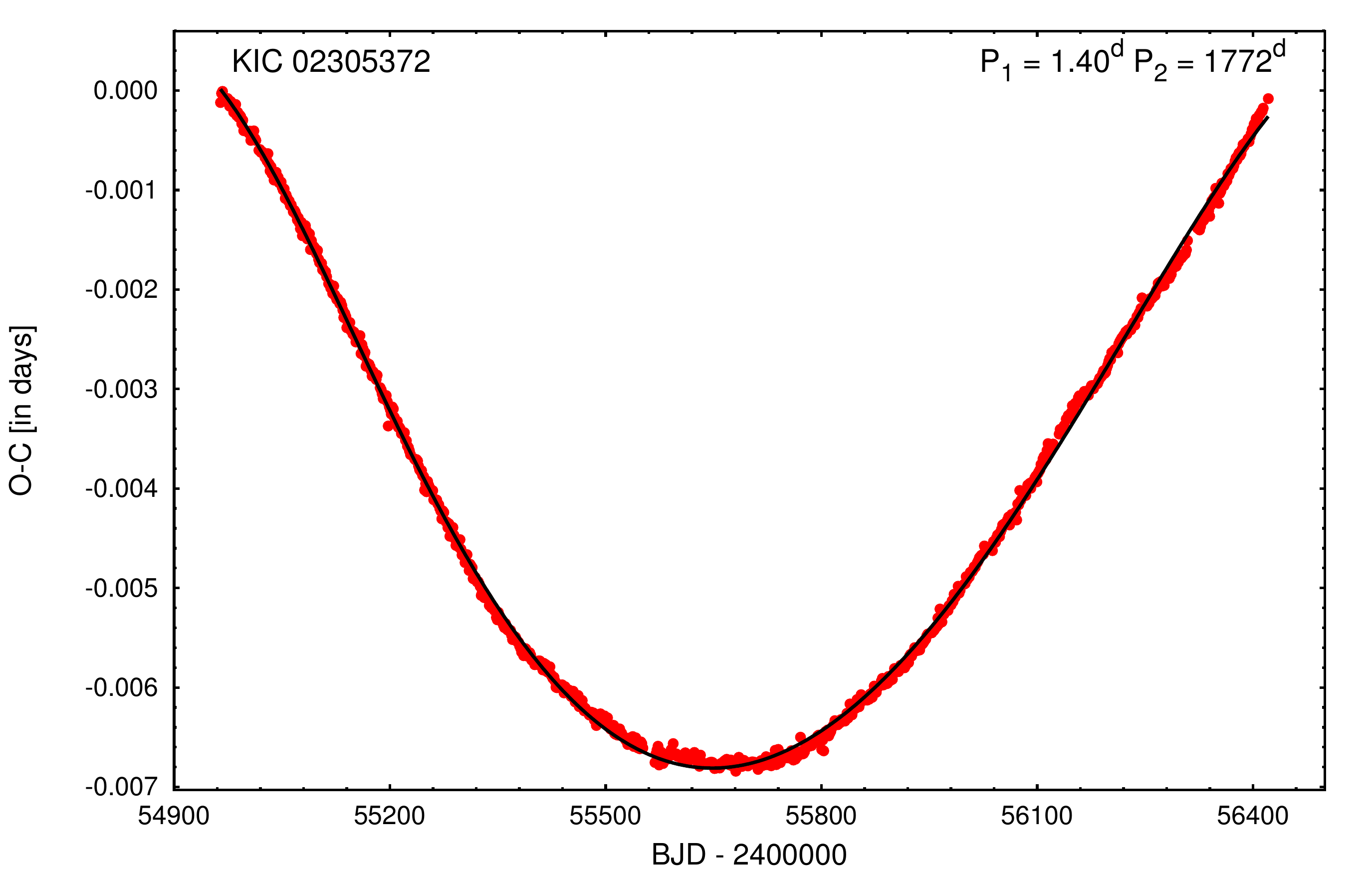}
\includegraphics[width=60mm]{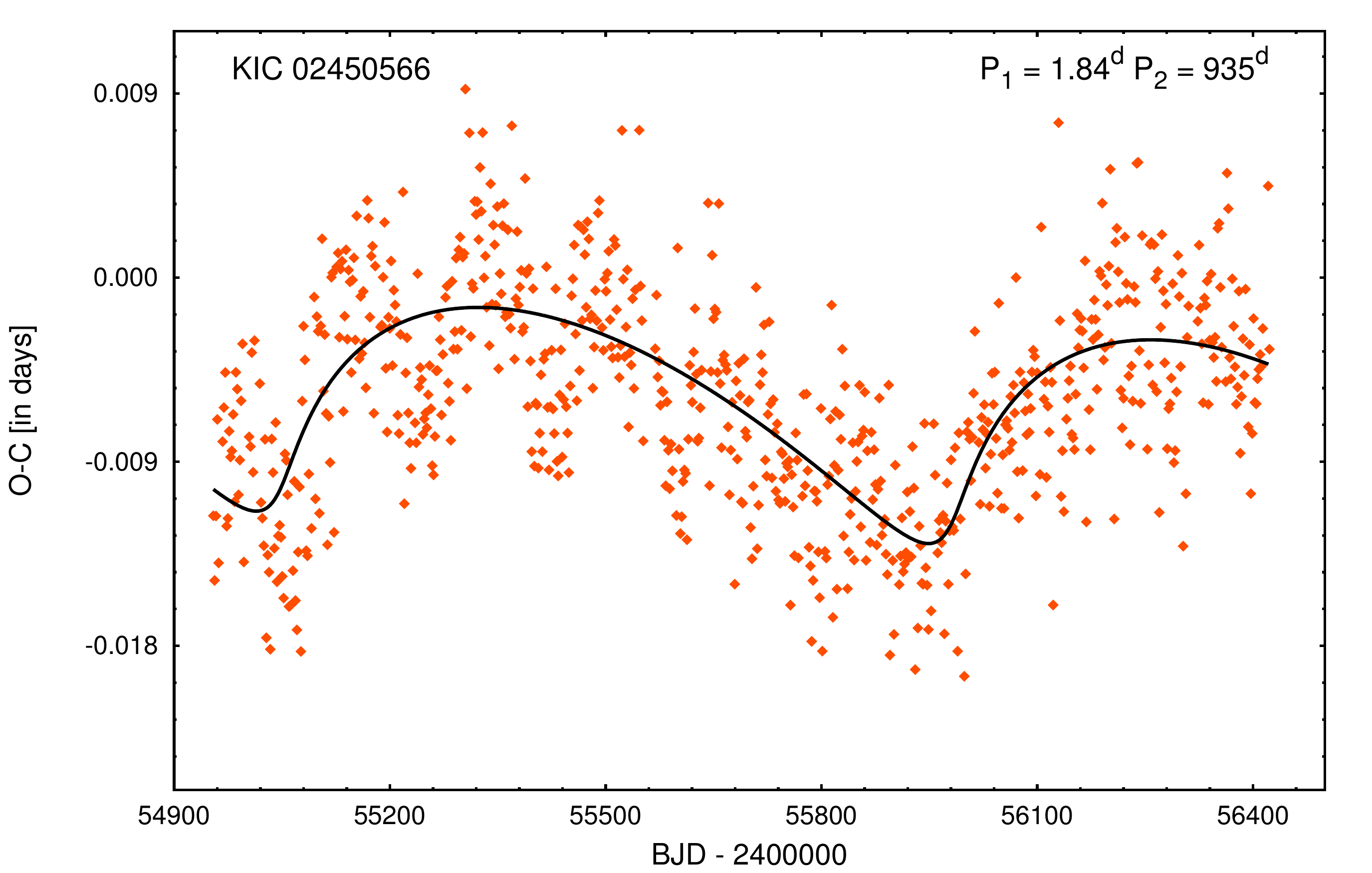}\includegraphics[width=60mm]{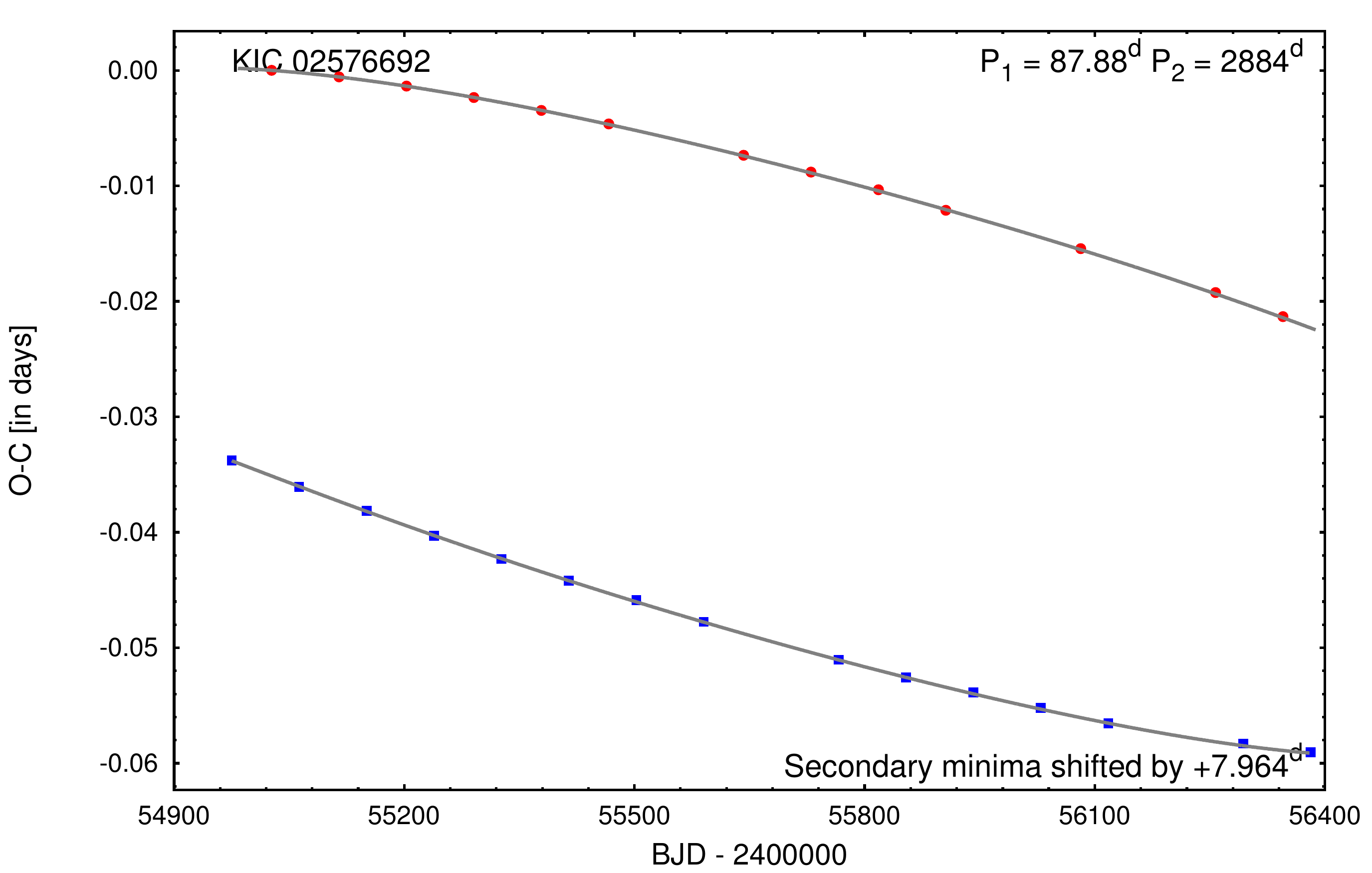}\includegraphics[width=60mm]{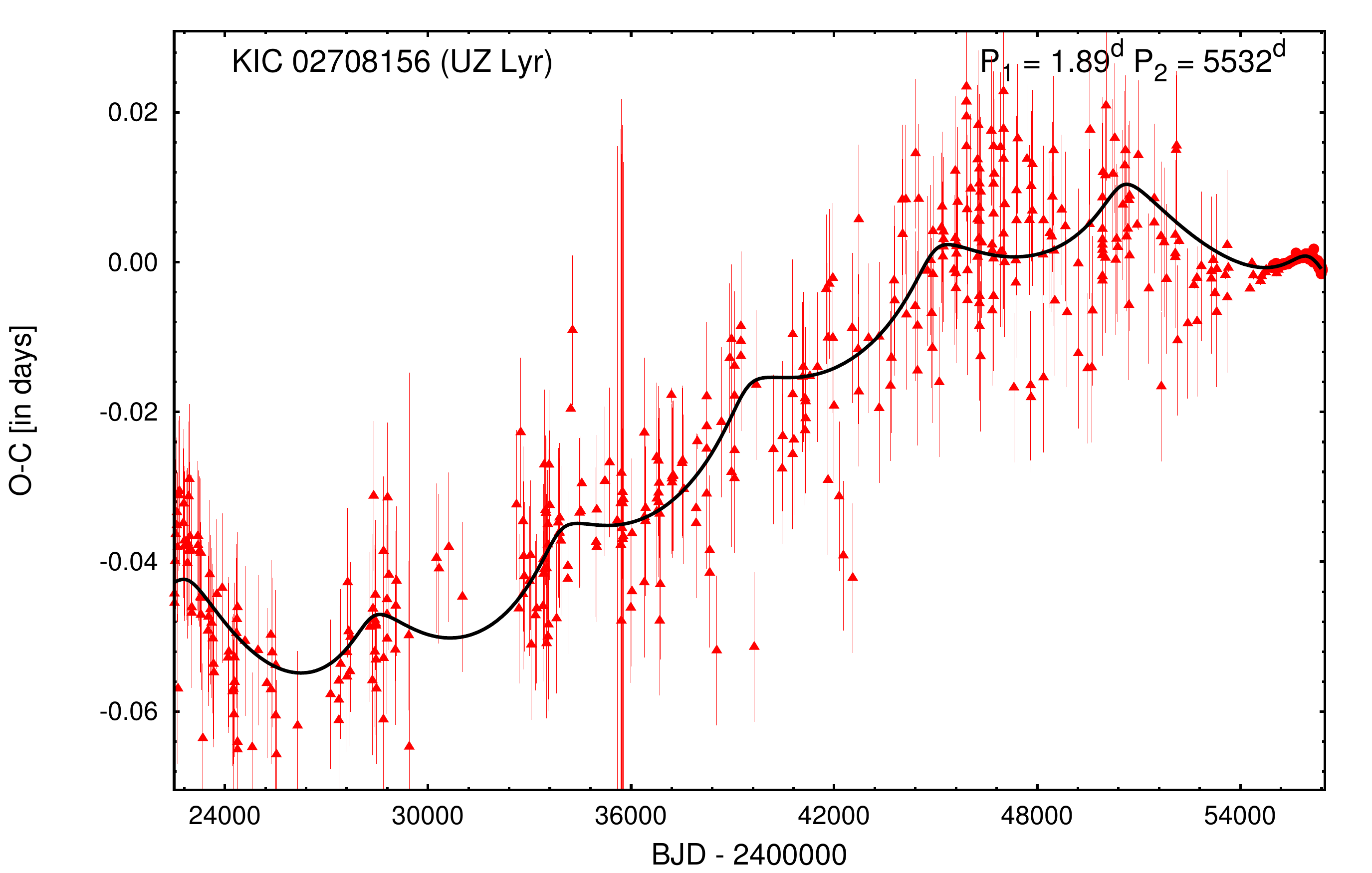}
\includegraphics[width=60mm]{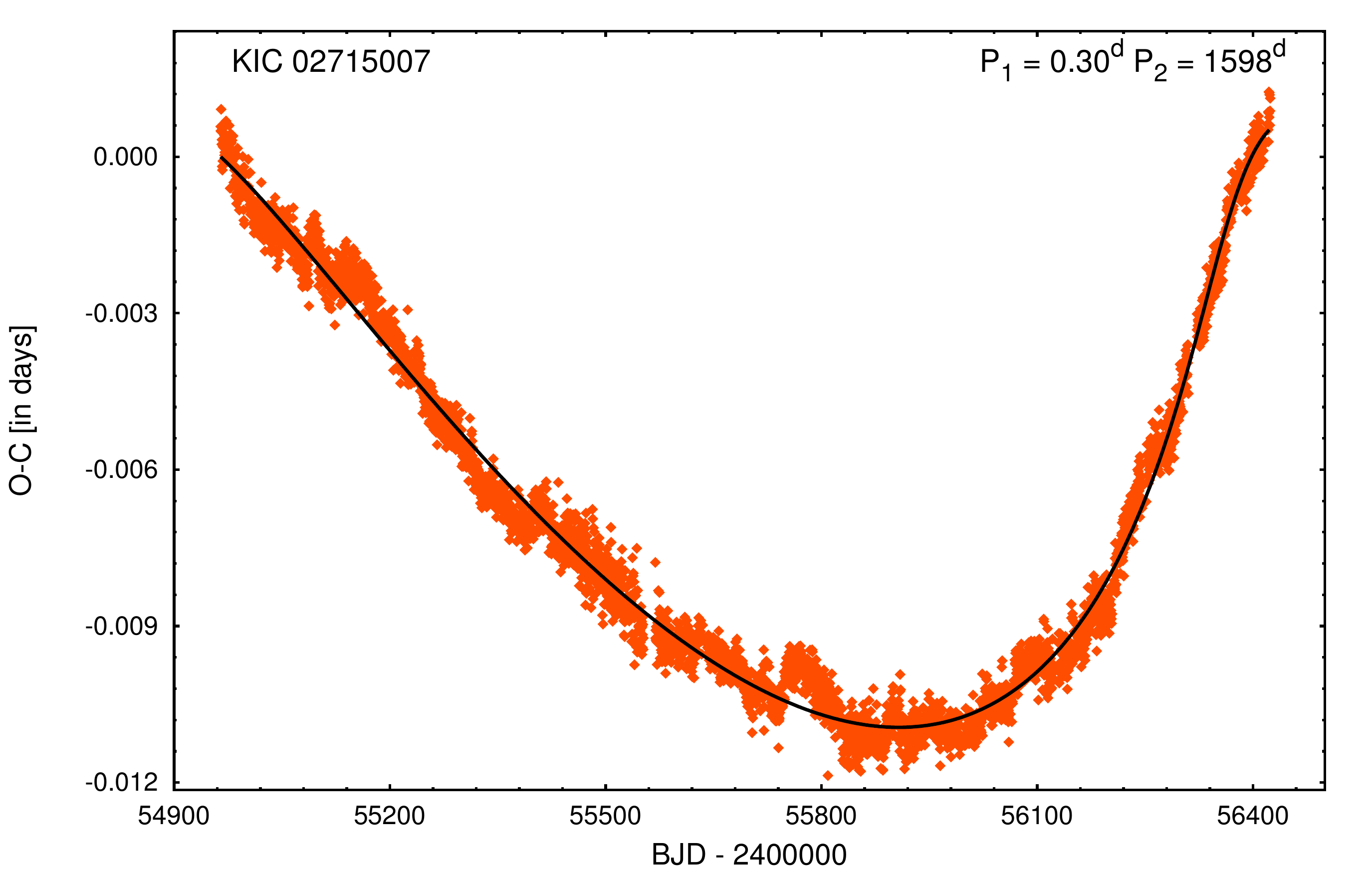}\includegraphics[width=60mm]{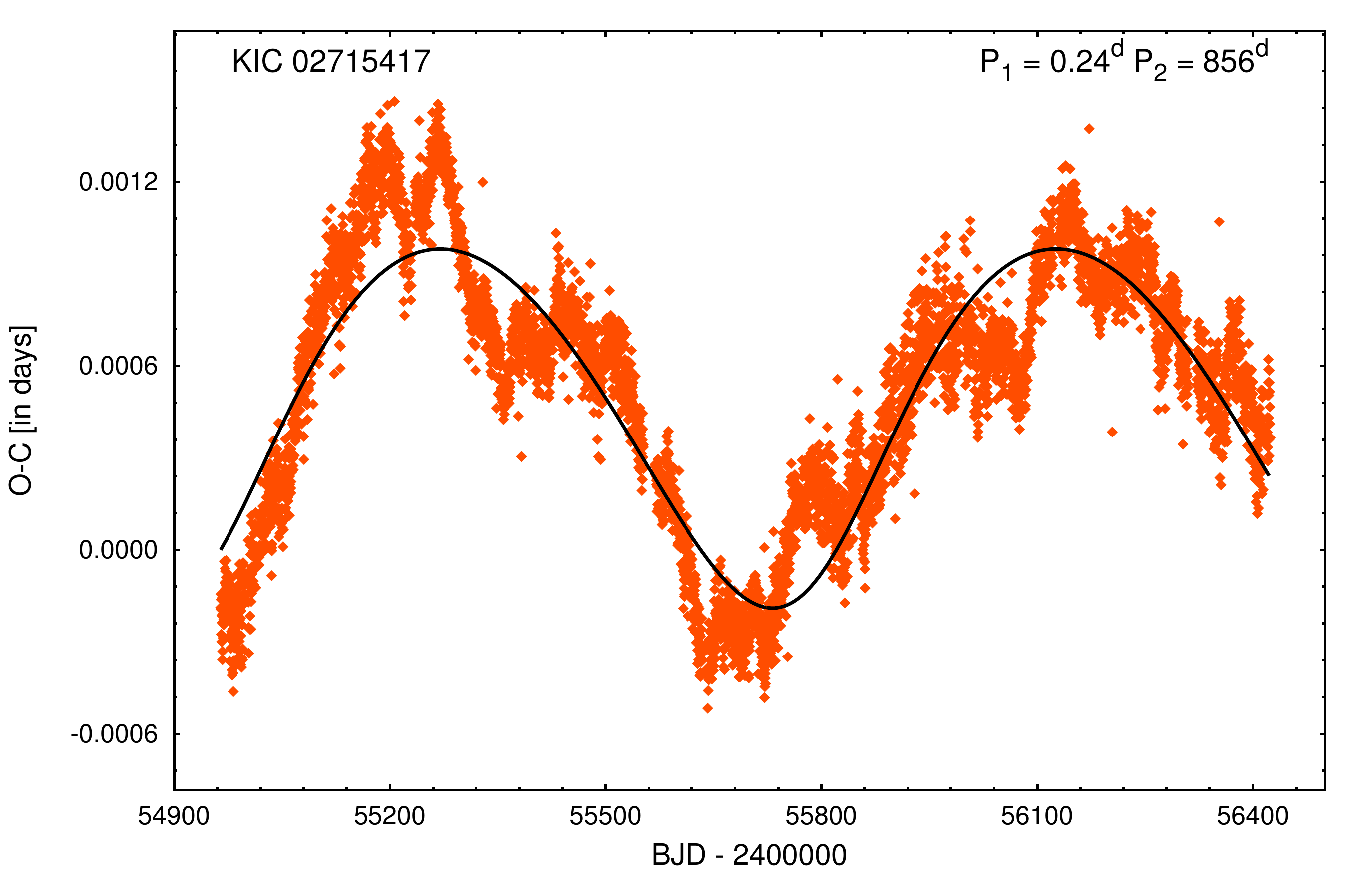}\includegraphics[width=60mm]{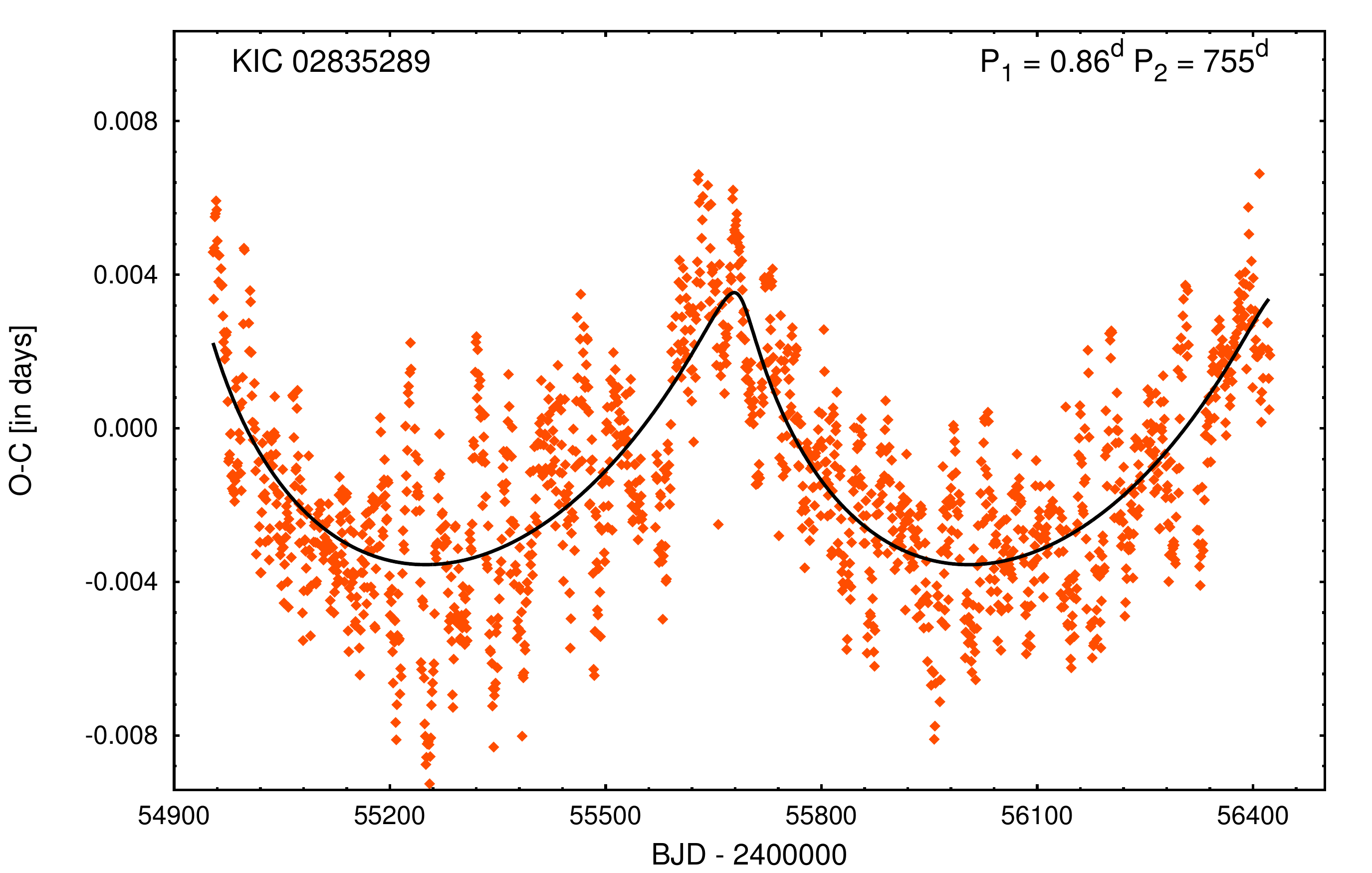}
\includegraphics[width=60mm]{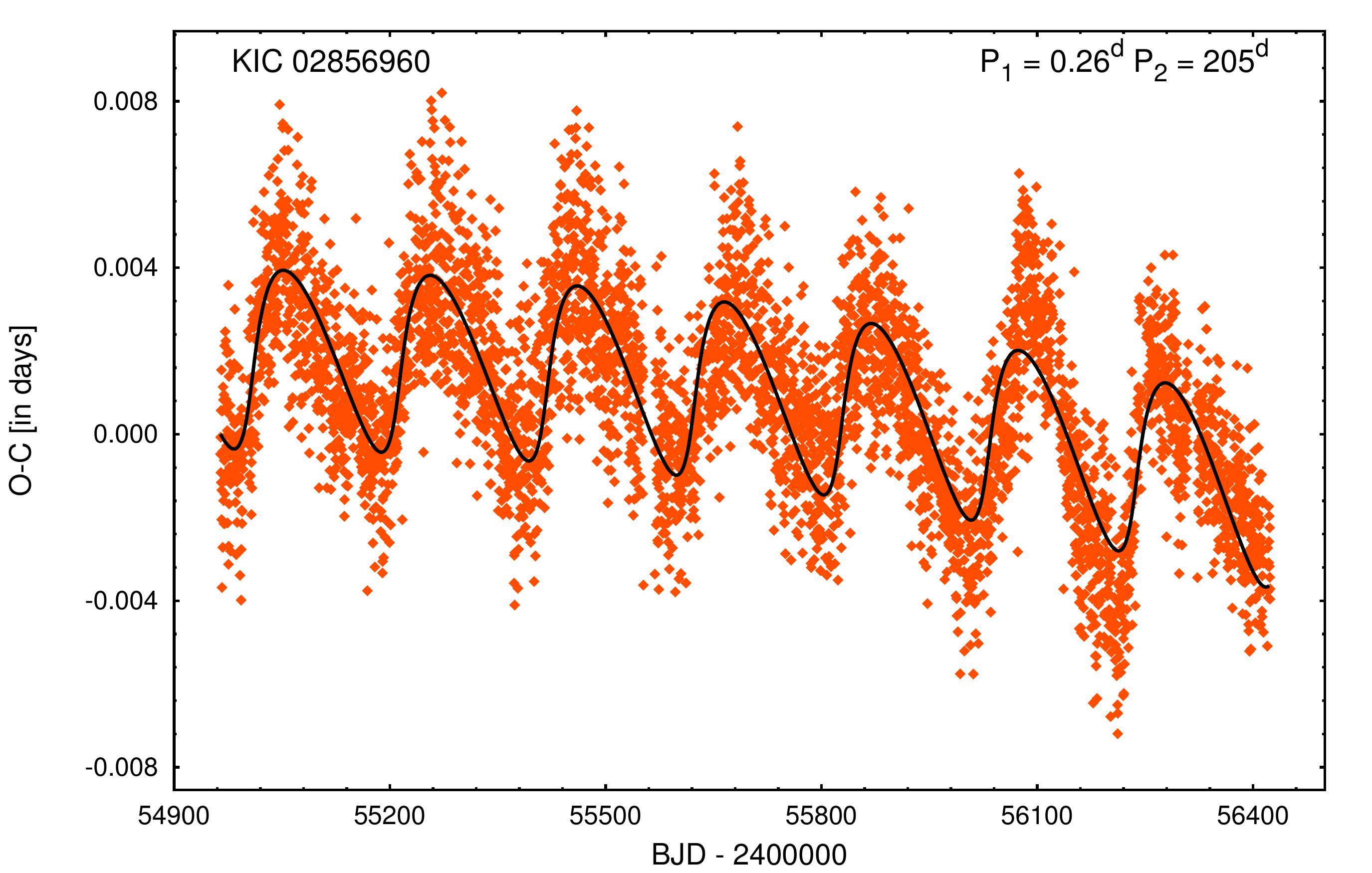}\includegraphics[width=60mm]{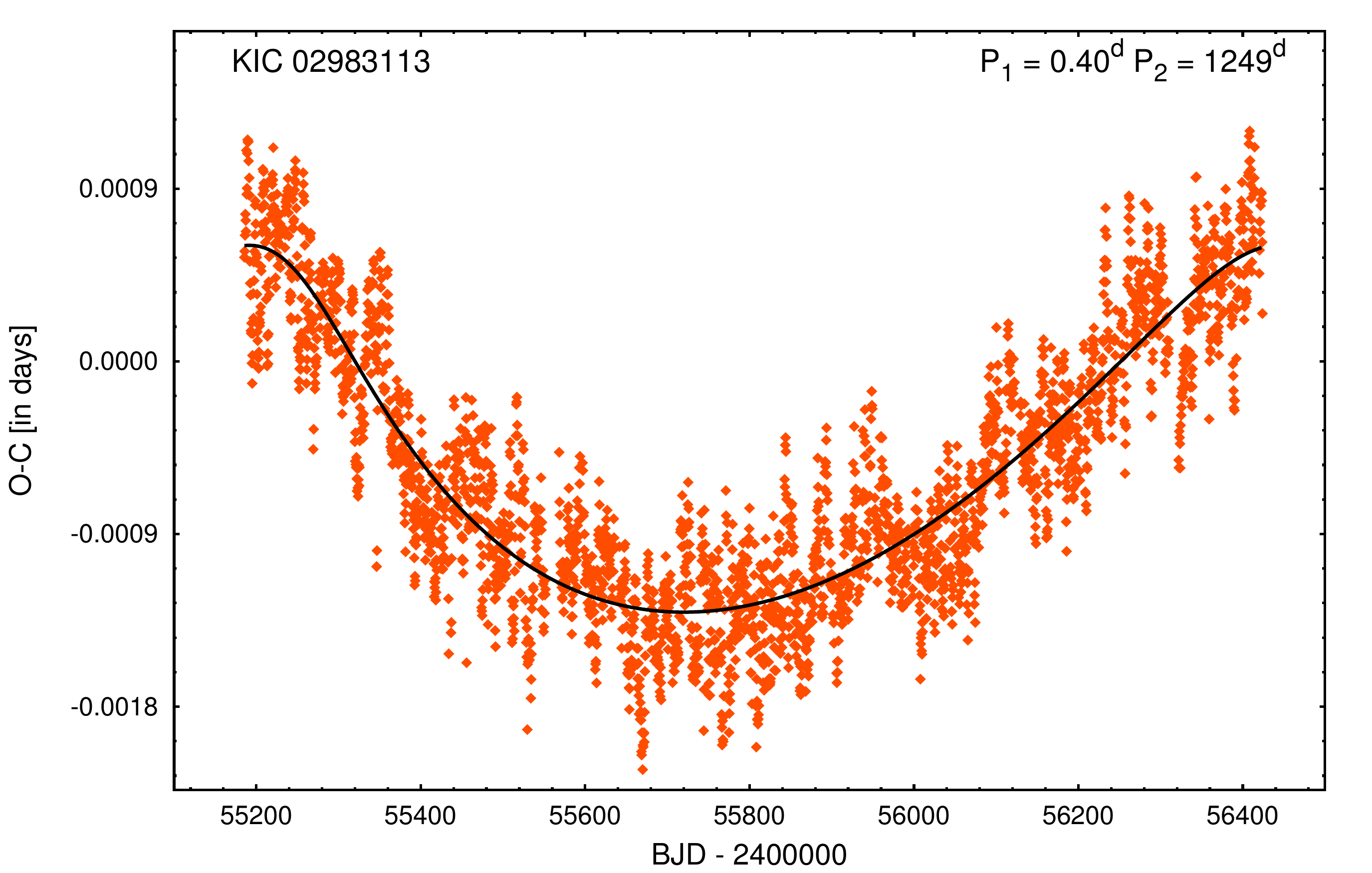}\includegraphics[width=60mm]{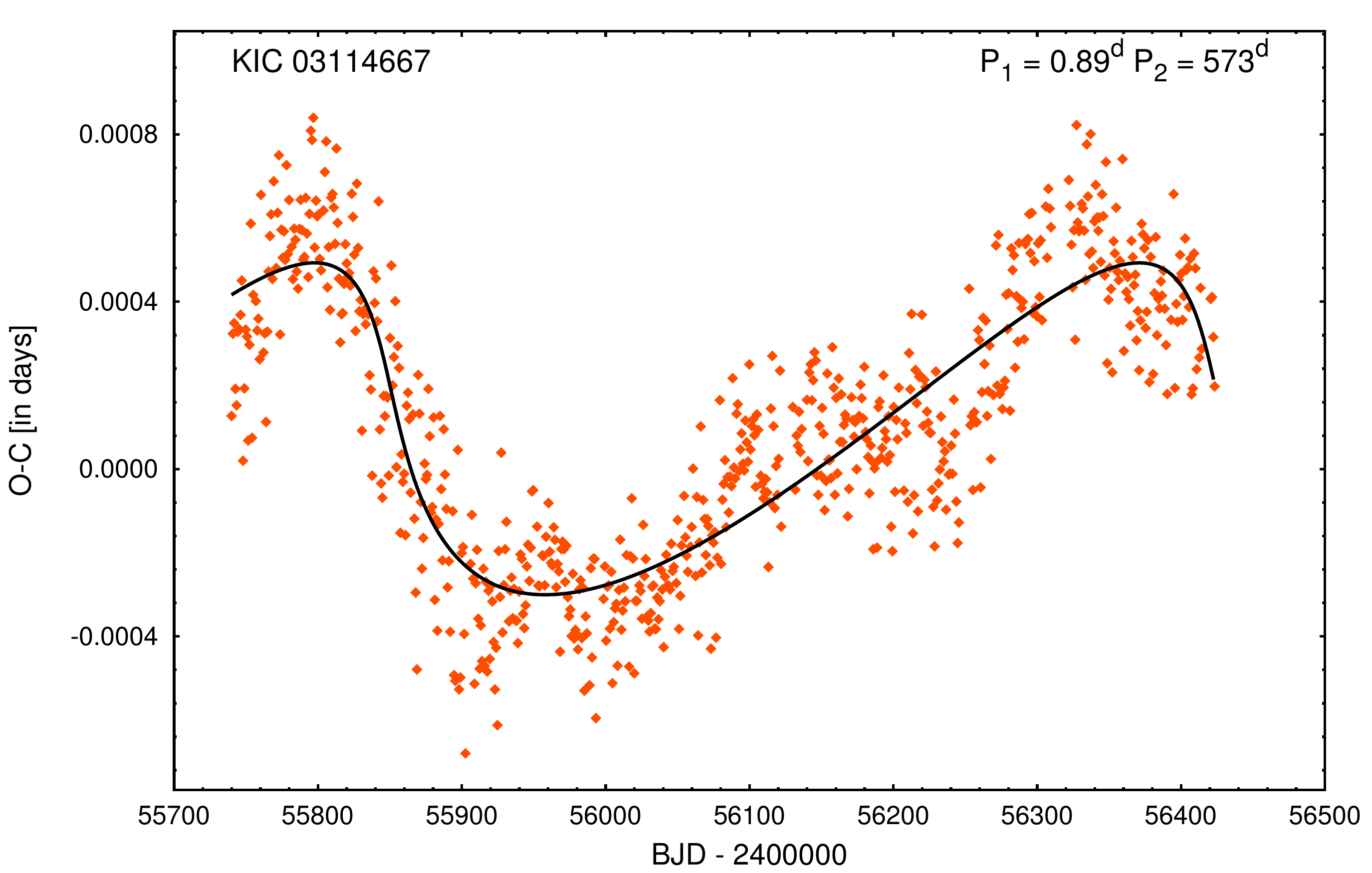}
\includegraphics[width=60mm]{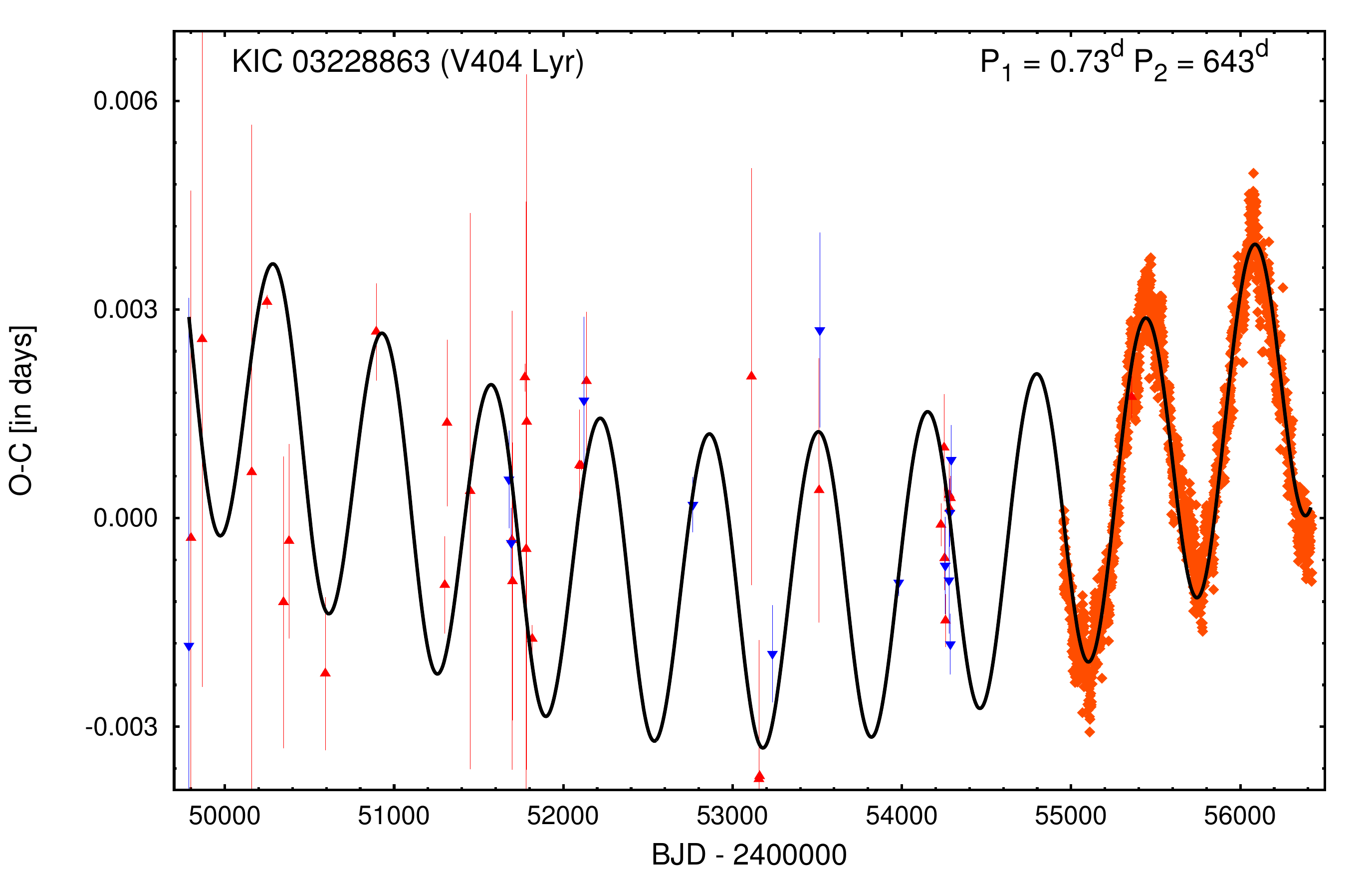}\includegraphics[width=60mm]{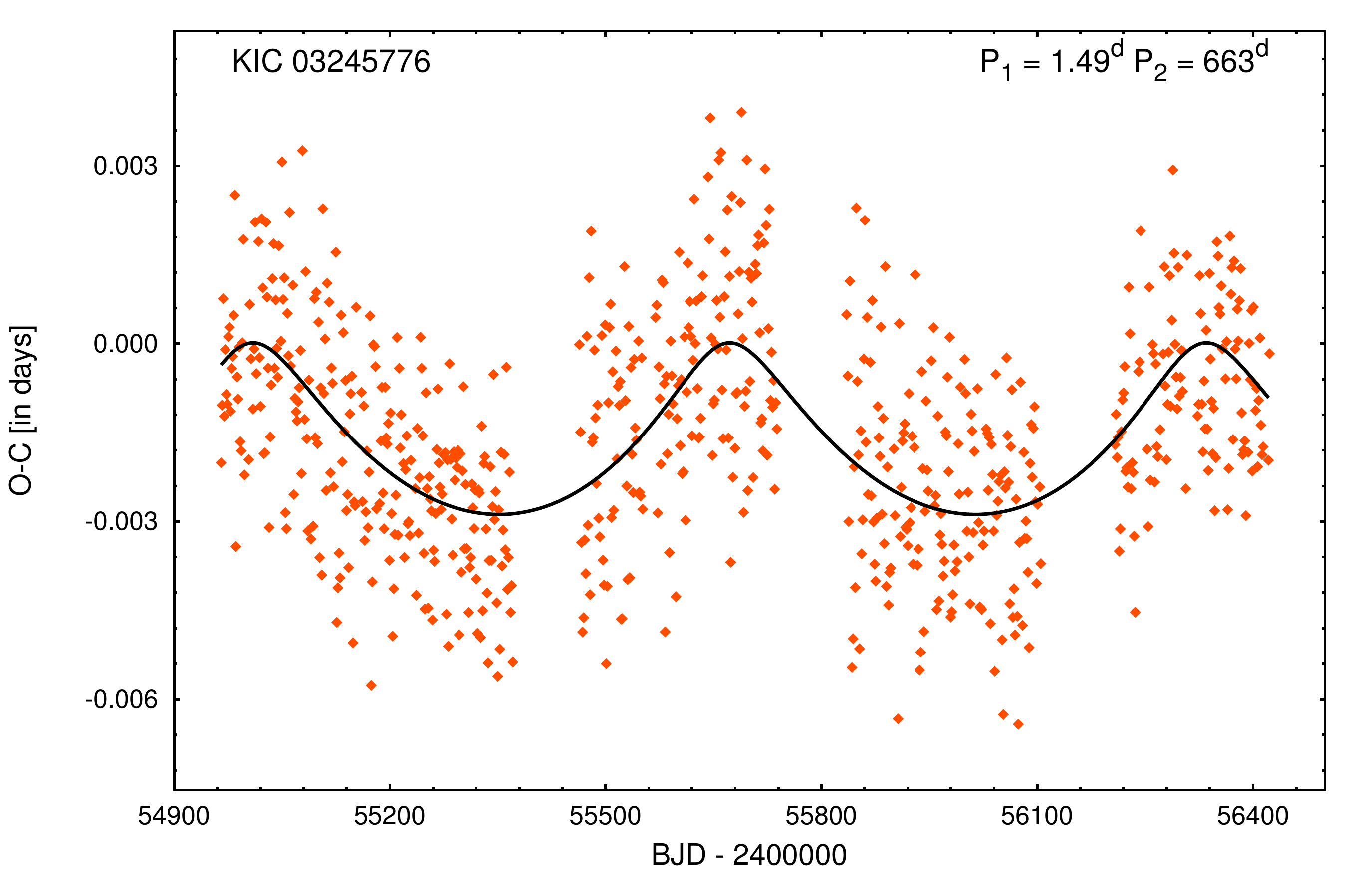}\includegraphics[width=60mm]{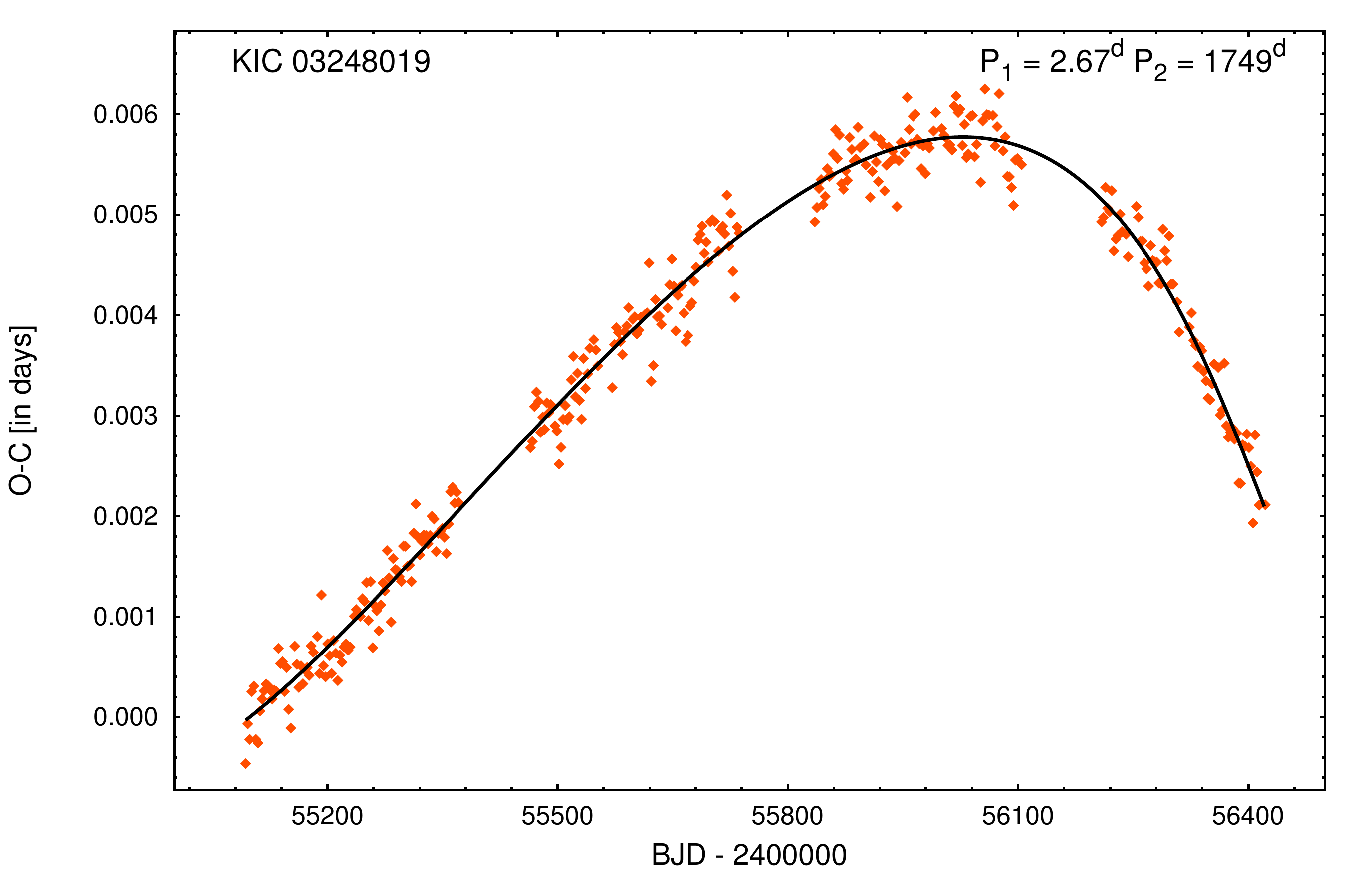}
\caption{ETVs with third body solutions. ETV curves calculated from {\em Kepler} observations of primary and secondary minima, and the average of the two, are denoted by red circles, blue boxes, and orange diamonds, respectively. We display and fit the ETV curves for both the primary and secondary eclipses only when the data quality warrant a joint analysis and the binary is eccentric. If the primary and secondary ETV curves are of comparable quality and the binary eccentricity is nearly zero, we display and fit only the average of the two ETV curves. If the quality of the primary ETV curve is significantly better than that of the secondary curve or, if only primary eclipses are present, we present only the plot and the fit for the primary eclipses. Ground-based minima (taken from the literature, and available only for a few systems) are denoted by upward red triangles (primary) and downward blue triangles (secondary); their estimated uncertainties are also indicated.  Pure LTTE solutions are plotted with black lines, while combined dynamical and LTTE solutions are drawn with grey lines. (Note, the use of quadratic or cubic terms is not indicated; for these and other details, see Table~\ref{Tab:Systemproperties}.) The complete Figure 6 covering 221 ETV curves is available in the on-line version of the Journal. {\it Note, in this arXiv edition the remaining 206 panels are included at the end of the paper.}}
\label{Fig:ETVs}
\end{figure*}

\subsection{Distributions and Statistics}
\label{sec:stats}

Since this is the largest collection of triple systems with known outer orbital periods, $P_2 \lesssim$ few years, it makes sense to examine distributions of several of the system parameters and other statistics.  Certain of these parameters, including $P_1$, $P_2$, $e_2$, $\omega_2$, and $f(m_{\rm C})=m_\mathrm{C}^3 \sin^3 i/m_{\rm ABC}^2$, can be determined using only the LTTE delays.  Hence these parameters are available for 222 systems(see Tables~\ref{Tab:Orbelem}--\ref{Tab:Orbelemdyn3}).  

For many of the 62 LTTE-dynamical combined solution systems listed in Tables~\ref{Tab:Orbelemdyn1}--\ref{Tab:Orbelemdyn3} and \ref{Tab:AMEparam}, there is also information on parameters associated with the three-dimensional structure of the triple, including the mutual orbital inclination angle, $\im$, and with the system masses, i.e., $f(m_{\rm C})$, $m_{\rm C}/m_{\rm ABC}$, $m_{\rm C}$, and $m_{\rm AB}$.

\begin{figure}
\includegraphics[width=\columnwidth]{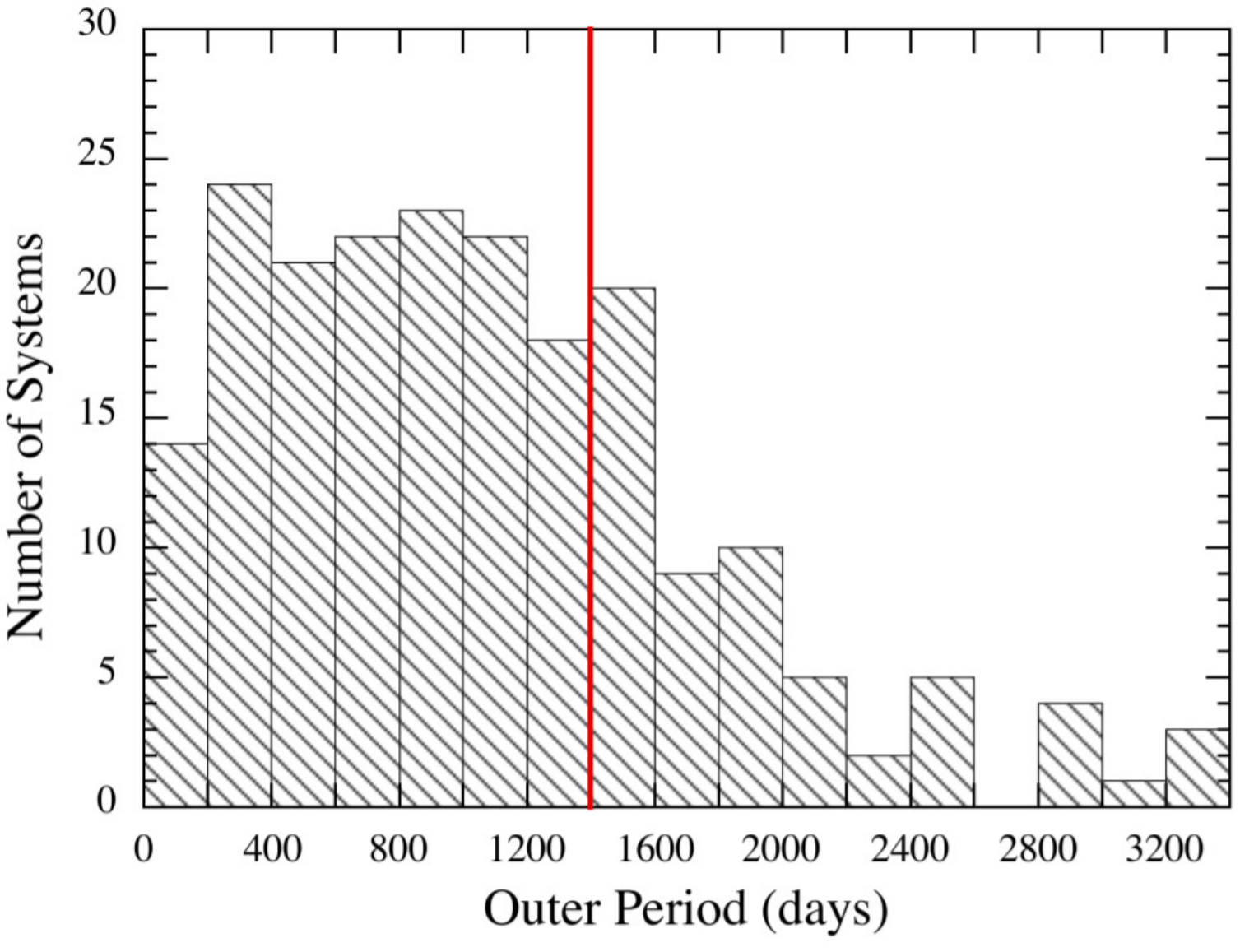}
\caption{Distribution of the outer orbital periods, $P_2$, for 222 triple systems found in the {\em Kepler} field.  The vertical red line denotes the duration of the {\em Kepler} mission.}
\label{Fig:P2_dist}
\end{figure}

In Fig.~\ref{Fig:P2_dist} we show the outer orbital period distribution, $f(P_2)$, for some 200 triple systems. This distribution is flat, at least within the limits of Poisson fluctuations, out to $P_2 \simeq 1600$ days, a value   comparable to the $\sim1400$-day duration of the {\em Kepler} mission. For longer periods the distribution declines rapidly. This may be wholly or partially due to observational selection effects, since longer period ETV curves are more difficult to definitively identify.  At the same time it also suggests a possible $f(P_2) \propto P_2^{-1}$ decrease with increasing $P_2$.  Let us define $F(P_2)$ and $f(P_2)$ as the orbital period distributions per logarithmic and per linear period intervals, respectively.  In that case, $F(P_2) \equiv P_2 \,f(P_2)$.  The possible functional forms of $F(P_2)$ include uniform per logarithmic interval, log-normal with a peak at about $180$\,years which was found to fit the period distribution of a large sample of binary stars \citep{abtlevy78,duquennoymayor91}, or even a form with a peak near $\sim$$3\,000$\,years which was obtained both from observations \citep{tokovinin08} and numerical simulations \citep{naozfabrycky14} of triples containing close binaries.  For any of these three possibilities, $f(P_2)$, would vary roughly like $P_2^{-1}$ in the period range of Fig.~\ref{Fig:P2_dist}.

\begin{figure}
\includegraphics[width=\columnwidth]{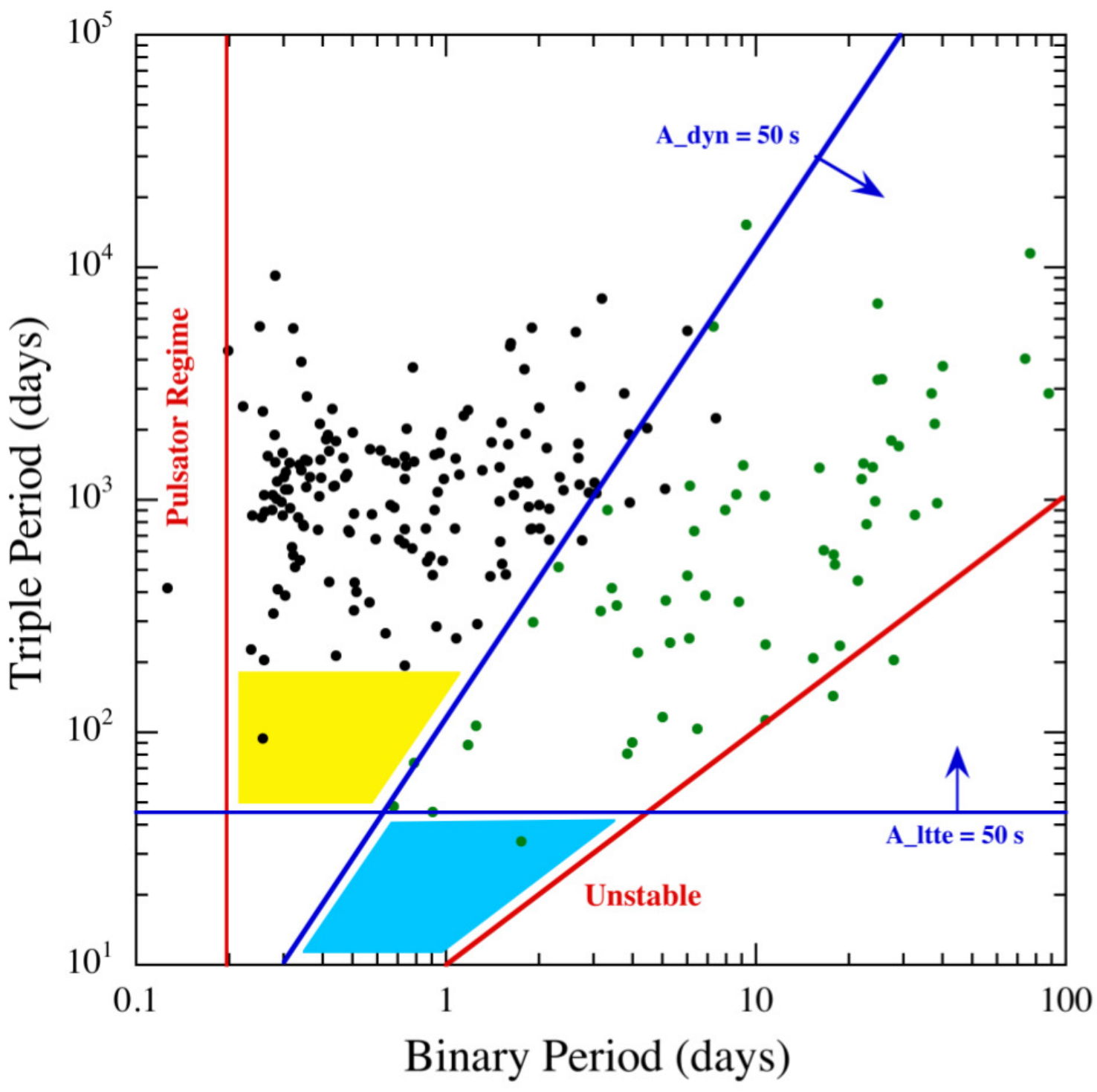}
\caption{Outer triple orbital period, $P_2$ vs.~the inner binary period, $P_1$ for 222 triple systems found in the {\em Kepler} field. The vertical red line denotes the typical minimum orbital period of contact binaries, while the sloped red line roughly separates regions of stability and instability. The horizontal and sloped blue lines are boundaries that roughly separate detectable ETVs from undetectable ETVs assuming that the ETVs must be $\sim$50 seconds or greater in amplitude to be detectable.  These amplitudes were calculated using $m_\mathrm{A}=m_\mathrm{B}=m_\mathrm{C}=1\,\mathrm{M}_\odot$, $e_2=0.35$, $i_2=60\degr$, and $\omega_2=90\degr$.  The arrows indicate the direction of greater detectability as long as $P_2 \lesssim 2000$ days.  The shaded cyan region indicates the period ranges where the dynamical delays are still detectable even though the LTTE delays might not be. There is only one known system in this region (see text for a discussion).  The region shaded in yellow indicates the period ranges where systems should nominally be detectable via the LTTE delays even though the dynamical delays might be undetectable. The fact that there are almost no systems in this region may have interesting physical implications (see text).}
\label{Fig:P1vsP2}
\end{figure}

As for the lower end of the outer period distribution, the question arises as to whether the limit is set by observational selection effects or results from actual dynamical effects.  Fig.~\ref{Fig:P1vsP2}, which shows a correlation plot of $P_2$ vs.~$P_1$ for all 222 systems, provides an answer. In this figure the blue lines denote the limits of the regions where the $\mathcal{A}_\mathrm{LTTE}$ and the $\mathcal{A}_\mathrm{dyn}$ amplitudes are likely to exceed 50\,s, a value which roughly approximates the threshold for likely detection of an ETV.  The shaded cyan region indicates the period ranges where the dynamical delays are still detectable even though the LTTE delays might not be.  There is only one system in this region, which is KIC~05897826 (=KOI-126). This system, however, was discovered via its triply eclipsing nature rather than via an ETV analysis.  The region shaded in yellow indicates part of the lower outer period range where systems should nominally be detectable via the LTTE delays even though the dynamical delays might be undetectable. The fact that there are almost no systems in this region, where detection of the LTTE delays should be straightforward, proves that our sample of triples at the lower edge of the outer period distribution has most probably been shaped by dynamical or evolutionary processes rather than by observational selection effects. For the cyan region one might argue that the combination of tightest binaries and tightest ternary orbits would lead to two circular or, at least low eccentricity, nearly aligned orbits due to tidal effects or other interactions; in that case $\mathcal{A}_\mathrm{dyn}$ is rather small. For the yellow region, however, the ETV amplitude is dominated by $\mathcal{A}_\mathrm{LTTE}$ rather than by $\mathcal{A}_\mathrm{dyn}$, so this objection is not relevant. Therefore, we can surely conclude that the tightest EBs, and especially the contact binaries, do not have very close ternary companions. This result might imply some additional differences between the dynamical processes which lead to the formation of the tightest close binaries, e.g., those with $P_1\lesssim 1/2$\,days, and to the processes which lead to the formation of binaries with longer $P_1$. 

\begin{figure}
\includegraphics[width=\columnwidth]{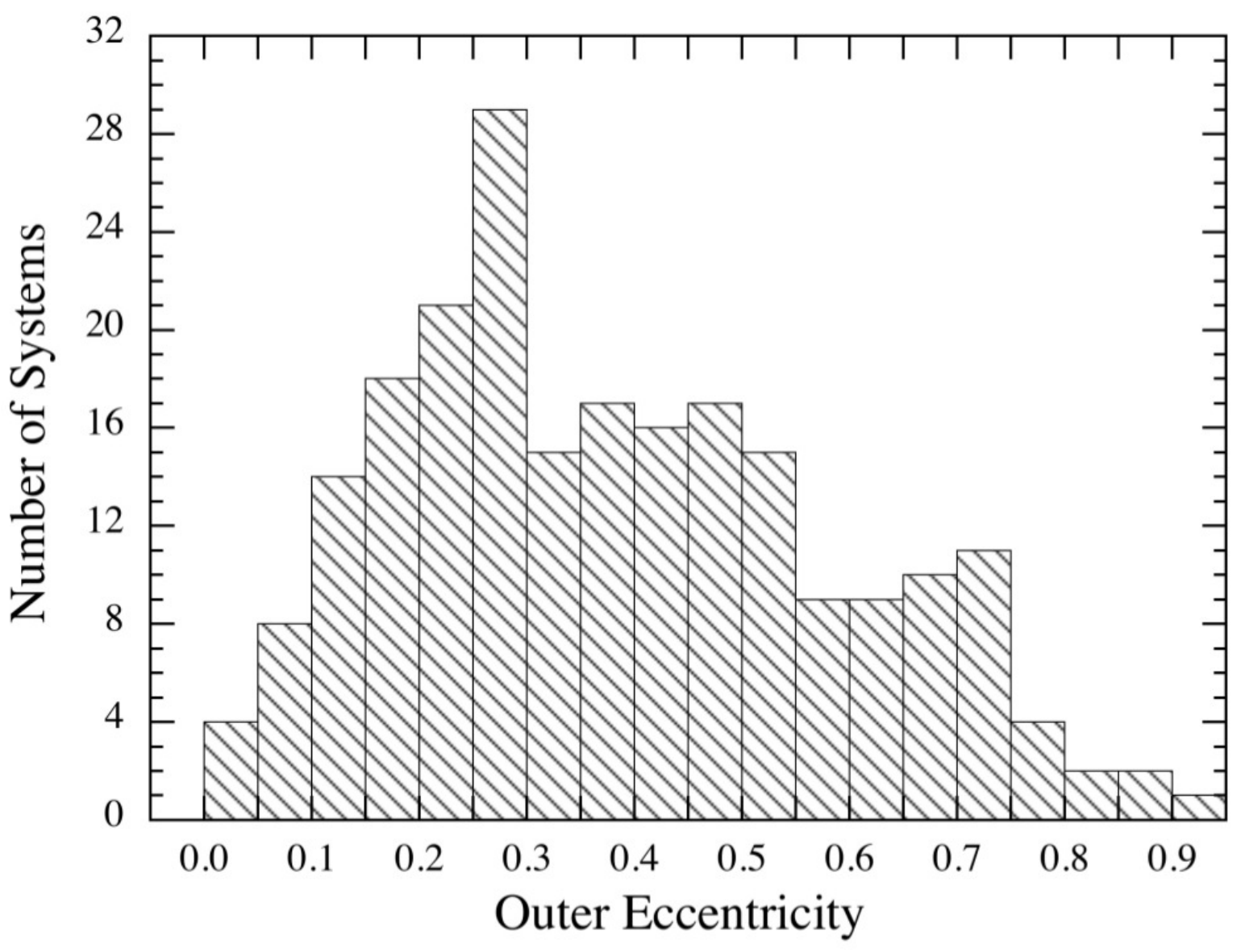}
\caption{Distribution of the eccentricities, $e_2$, of the outer orbits for 222 triple systems found in the {\em Kepler} field.}
\label{Fig:e2_dist}
\end{figure}

In Fig.~\ref{Fig:P1vsP2}, the sloped red line approximately separates dynamically stable systems from unstable systems.  The position of the line is based on an expression for dynamical stability in hierarchical triples in \citeauthor{mardling01} (\citeyear{mardling01}; see Eqn.~(27) of \citealt{borkovitsetal15}).  In applying this expression we assumed that the outer orbital eccentricity $e_2$ is equal to the median value of $0.35$ computed from the eccentricity distribution in Fig.~\ref{Fig:e2_dist}.  The vertical line in this figure indicates a value of $P_2 \simeq 0.2$ days, approximately the shortest orbital period of ordinary contact binaries.  All but 3 of the 222 systems lie between these two limiting curves, and, given the approximate nature of both constraints, this seems entirely satisfactory.

\begin{figure}
\includegraphics[width=\columnwidth]{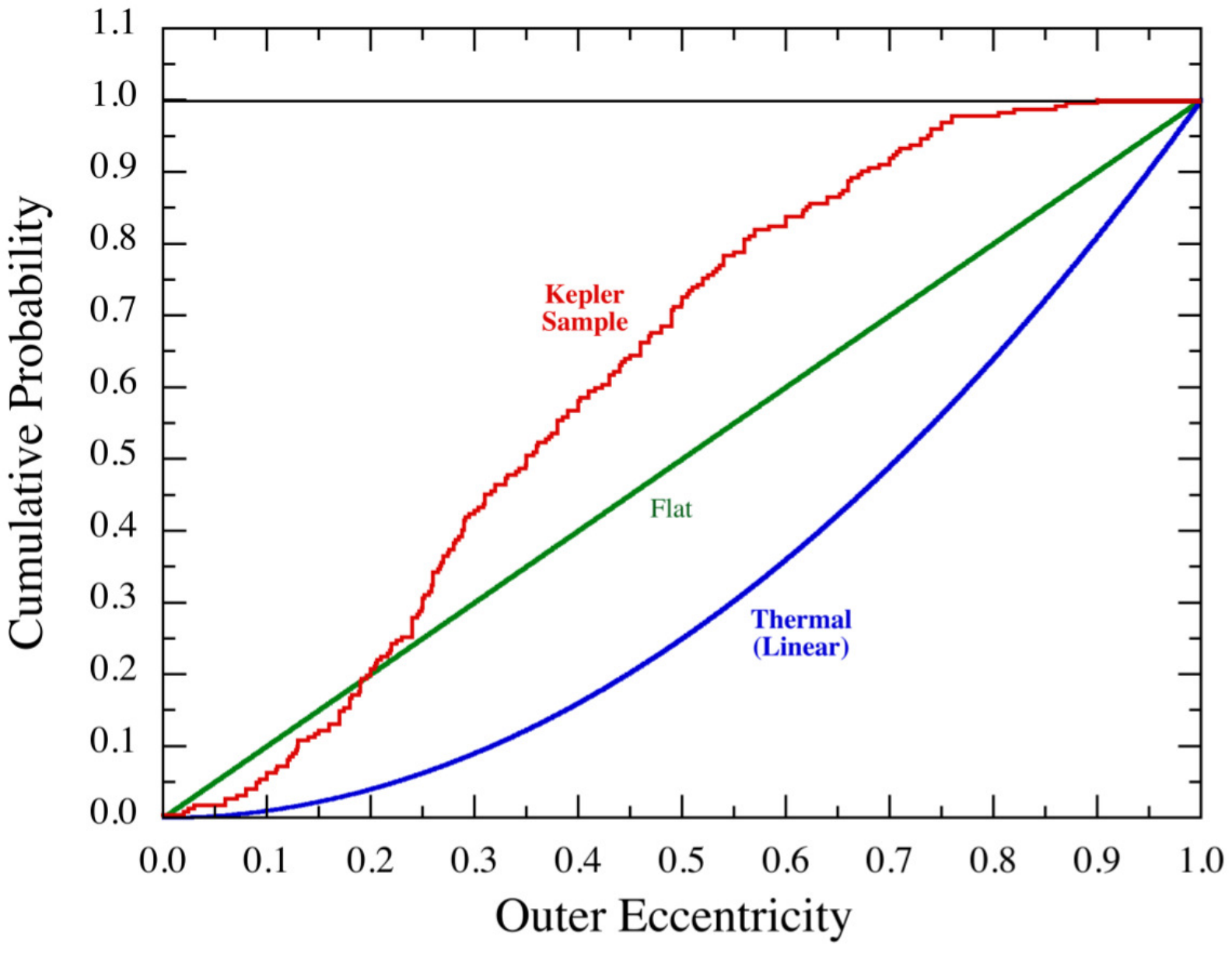}
\caption{Cumulative distribution of the outer eccentricity ($e_2$) for all 222 {\em Kepler} triples in our sample. The green curve, shown for comparison, represents the cumulative distribution expected for a uniformly distributed set of eccentricities between zero and 1. The blue curve is for an eccentricity distribution that increases linearly with $e_2$.  Neither comparison curve is a good match to the observed distribution, which results from the eccentricities tending to peak near $\sim$0.3. For comparison to the eccentricities of unperturbed wide field binaries in the same period regime, see \citet{duchenekraus13}.}
\label{Fig:e2cum}
\end{figure}

The outer orbit eccentricities have a wide range of values (Fig.~\ref{Fig:e2_dist}).  The distribution is characterized by a broad peak together with a narrow peak near $e_2 \simeq 0.28$.  We have no immediate explanation for either feature.  In any event, the distribution is clearly inconsistent with a `thermal' distribution of eccentricities such as that originally posited by \citet{jeans19} which would be linearly rising with $e_2$. In contrast, our finding is in good accord with the eccentricity distributions of different populations of field binaries obtained from recent surveys. This may be seen by comparison of the cumulative distribution of the outer eccentricities of our complete sample (Fig.~\ref{Fig:e2cum}) with the distributions shown in Fig.~3 of \citet{duchenekraus13} for homogenous subgroups of binaries with periods in the regime of $100\leq P\leq10\,000$ days. (Note, here we are treating the triple systems as binaries composed of the outer body and the inner binary.) For further comparison, we also plot the cumulative distributions expected for a uniformly distributed set of eccentricities and for an eccentricity distribution that increases linearly with $e_2$. As is the case for the binaries in \citet{duchenekraus13}, neither comparison curve is a good match to the observed distribution, which results from the eccentricities tending to peak near $\sim$0.3.

\begin{figure}
\includegraphics[width=\columnwidth]{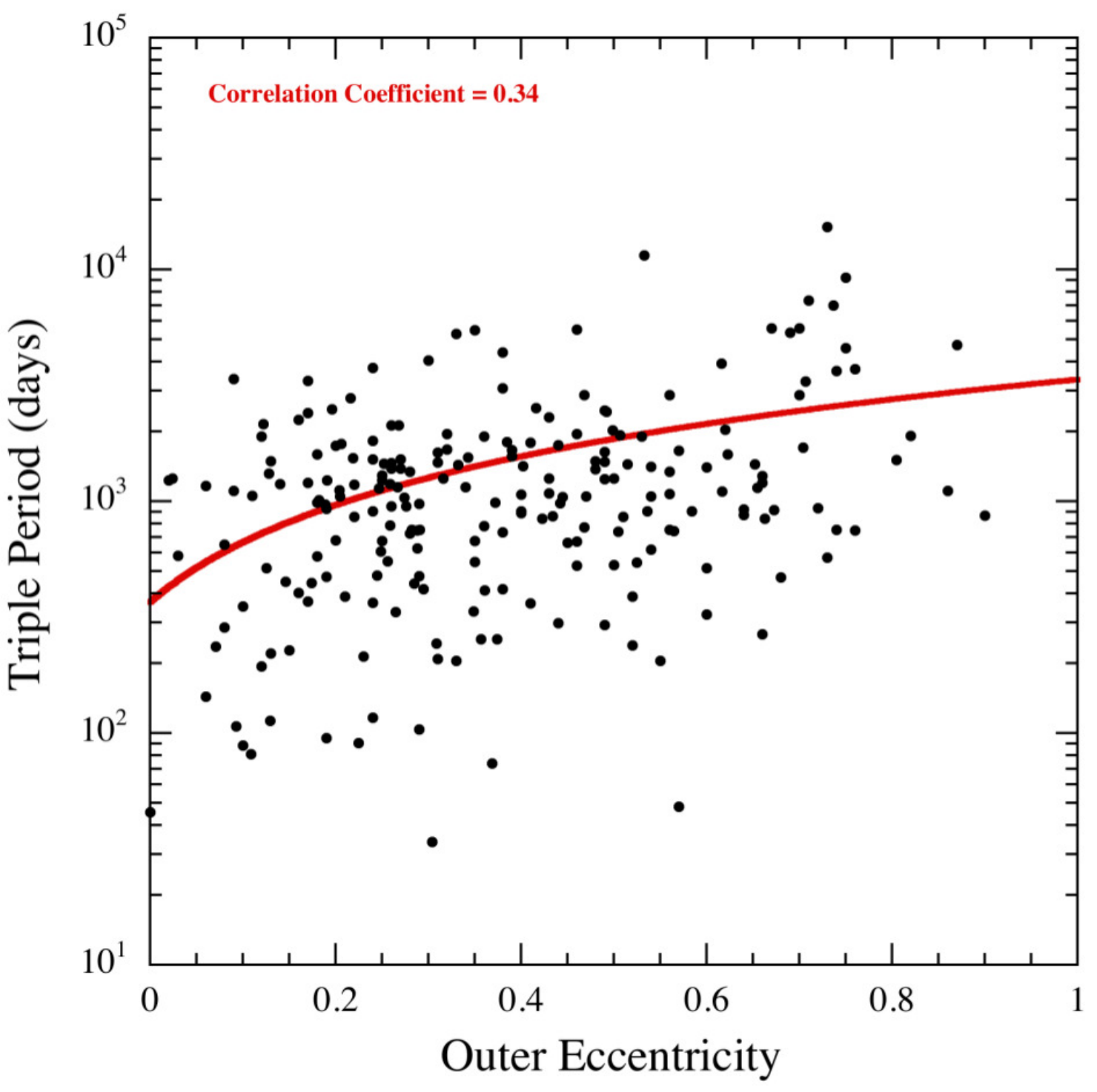}
\caption{Triple period, $P_2$, vs.~eccentricity, $e_2$, for 222 triple systems found in the {\em Kepler} field. The red curve is the best linear fit which has a correlation coefficient of 0.36.}
\label{Fig:P2vse2}
\end{figure}

The relation between the outer orbital period and outer orbital eccentricity is shown in Fig.~\ref{Fig:P2vse2}.  The red curve shows the result of a fit to a linear relation; the correlation coefficient is 0.36.  For a sample of 222 systems, this is significant with a false-alarm probability of only $10^{-6}$.  In spite of this, the correlation is clearly not particularly striking.  \citet{jeans19} showed that for a population of binaries in `thermal equilibrium', the eccentricity would be uncorrelated with the period; this does {\em not} appear to be the case for the currently observed population of binaries \citep{duchenekraus13}.

\begin{figure}
\includegraphics[width=\columnwidth]{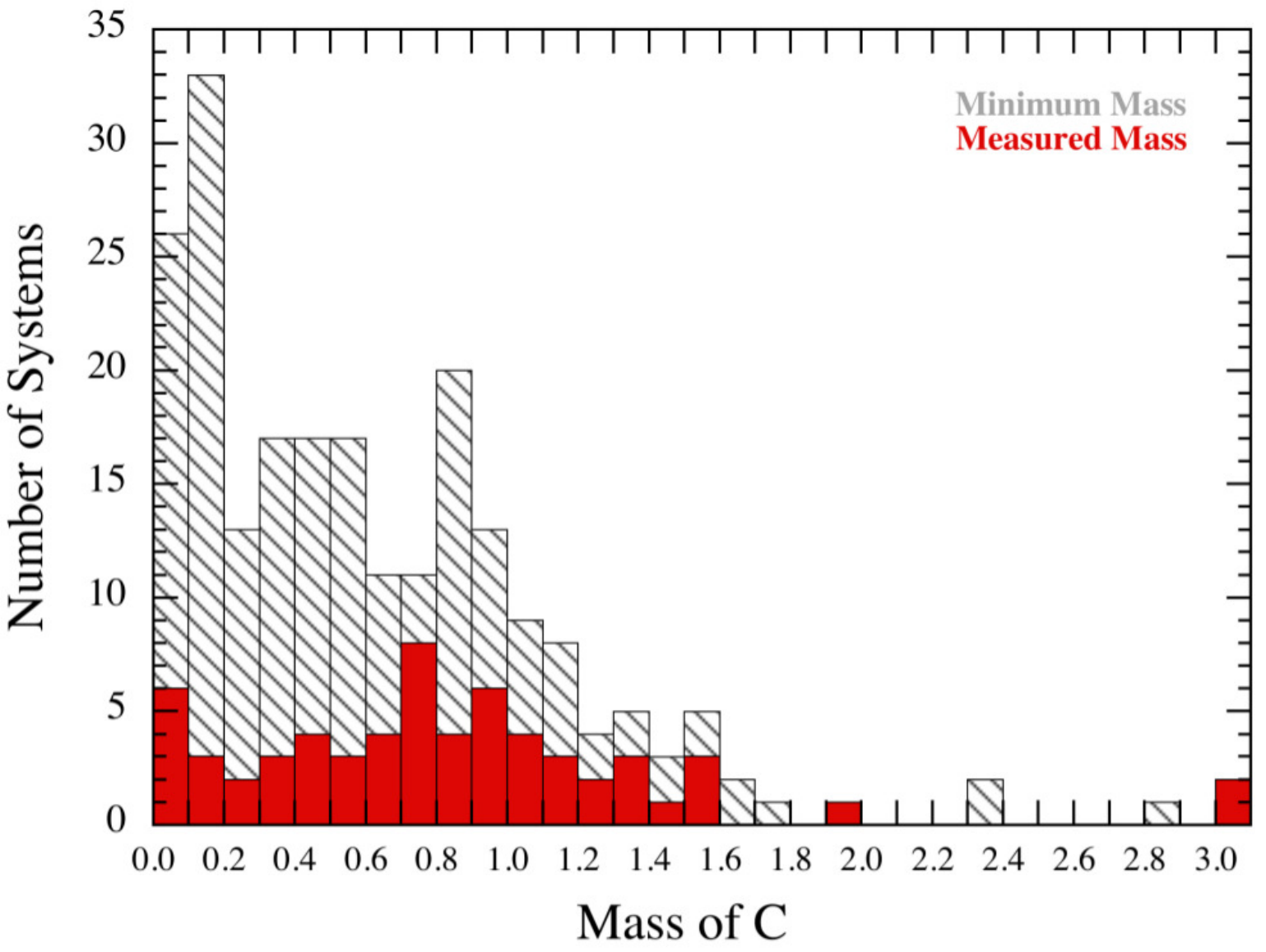}
\caption{Distribution of the tertiary masses, $m_{\rm C}$ (in M$_\odot$), for 222 triple systems found in the {\em Kepler} field.  The 62 systems marked in red are the ones for which there is sufficient information in the ETV curves from both dynamical and LTTE effects so that both the tertiary mass and the inner binary mass,  $m_\mathrm{C}$ and $m_{\rm AB}$, respectively, can be determined. The other tertiary masses are based on the LTTE solutions, and make the assumption that $m_{\rm AB}\simeq 2\,\mathrm{M}_\odot$.}
\label{Fig:mc_dist}
\end{figure}

Fig.~\ref{Fig:mc_dist} presents the distribution of the tertiary masses. For 62 triples in which both the LTTE and dynamical effects are measured, there is sufficient information to determine reasonably accurate masses  $m_{\rm C}$ for the third star.  For the remaining 160 systems, we estimate $m_{\rm C}$ from the mass function, $f(m_{\rm C})$ after adopting the reasonable assumption/approximation that $m_{\rm AB} \simeq 2\,\mathrm{M}_\odot$.  We see from this figure that, overall, the mass distribution is well populated out to $m_{\rm C} \simeq 1 \,\mathrm{M}_\odot$ and then falls off steeply toward higher masses.  We also note that the vast bulk of the systems have $m_{\rm C} \lesssim 1.8 \,\mathrm{M}_\odot$.  This is likely a selection effect since {\em Kepler} targets included relatively few ($\lesssim 1/2\%$) stars with masses greater than this.  The masses of the tertiaries in the LTTE systems tend to be low, at least relative to the tertiary masses for the systems with combined solutions.  This is likely a natural consequence of the fact that the LTTE mass values only represent lower limits. Therefore, for triples with small outer inclinations, $i_2$, the true tertiary masses may be substantially larger. Despite this, however, the modestly enhanced peak at masses between 0.1 and 0.2\,M$_\odot$, suggests that caution should be used before accepting the LTTE interpretations of the lowest amplitude ETVs.  Here we mention again those systems where the combination of low amplitude, $2-3$ year-periodicities plus quadratic terms might have been misinterpreted as LTTE orbits.

\begin{figure}
\includegraphics[width=\columnwidth]{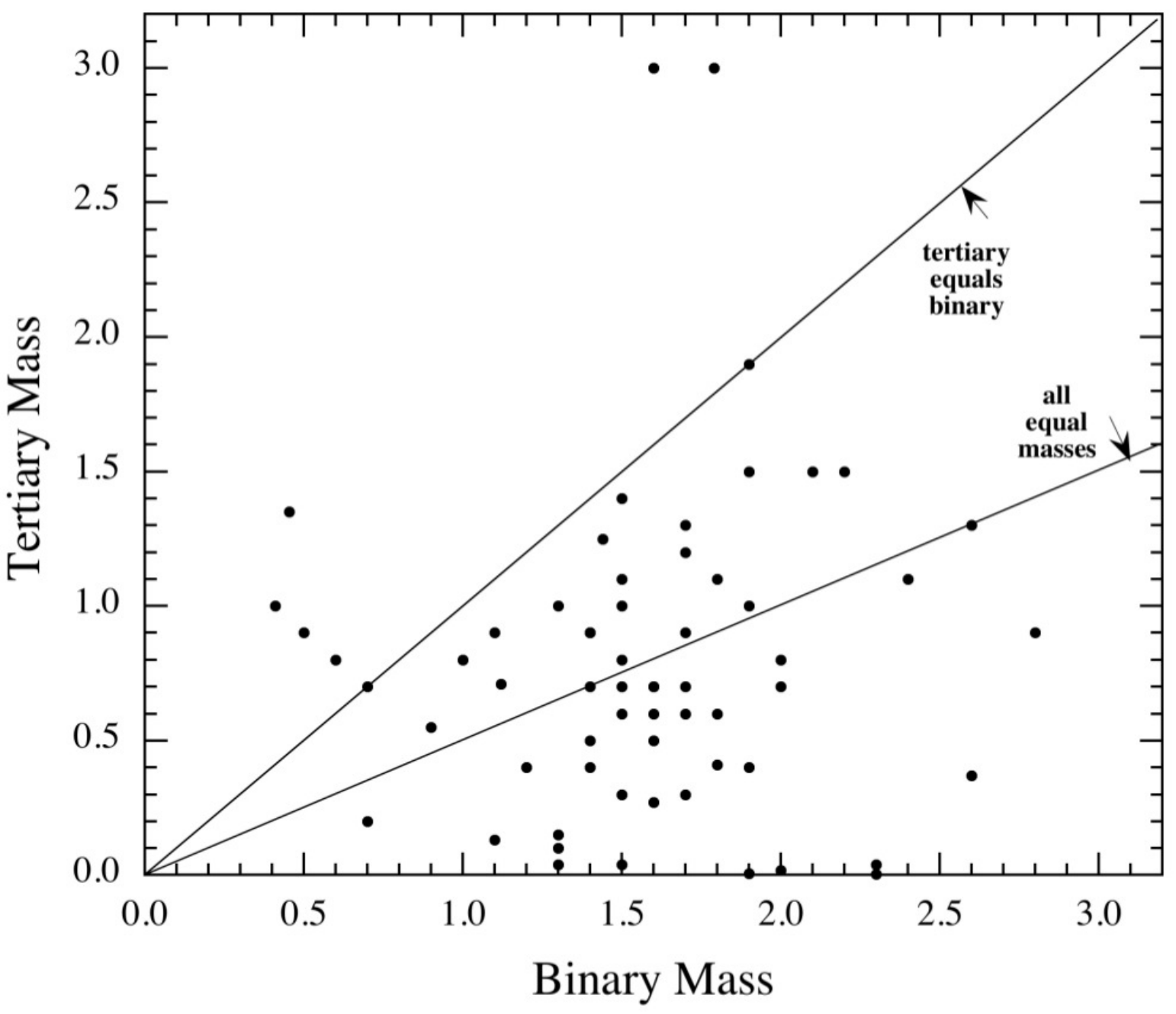}
\caption{Relation between the tertiary mass, $m_{\rm C}$ (in M$_\odot$), and the inner binary mass, $m_{\rm AB}$, for the 62 for which the ETV curves yield combined dynamical and LTTE solutions. }
\label{Fig:mcvsmab}
\end{figure}

Fig.~\ref{Fig:mcvsmab} shows the correlation between $m_{\rm C}$ and $m_{\rm AB}$ for the 62 combined-solution systems.  The straight line with the smaller slope indicates what would be expected for the special case of $m_{\rm A} = m_{\rm B} = m_{\rm C}$, while the line with the larger slope illustrates the locus of points where $m_{\rm C} = m_{\rm AB}$.  Roughly half the systems lie between these two lines, while a nearly equal number lie below the lower line.  Only a few systems lie above the higher line.  Broadly speaking, the tertiary masses range from rivaling that of the binary to being quite low.  The systems with very low tertiary masses (near the very bottom of the plot) are discussed below in Sect.~\ref{Sect:substellar}.

\begin{figure}
\includegraphics[width=\columnwidth]{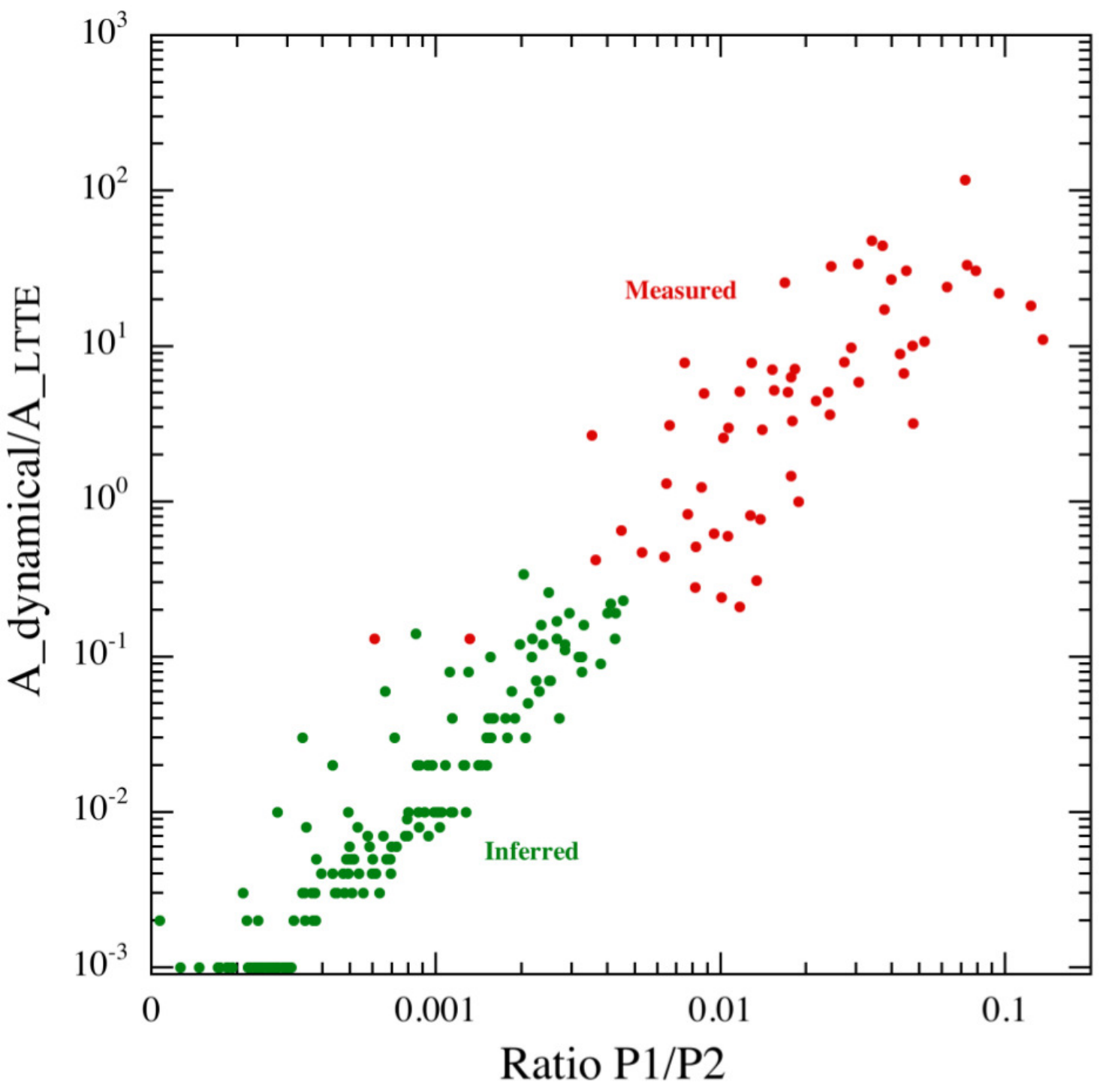}
\caption{The relation between the dynamical and LTTE amplitudes, $\mathcal{A}_{\rm dyn}/\mathcal{A}_{\rm LTTE}$, and the ratio of inner to outer periods, $P_1/P_2$.  The systems marked in red are directly measured from the dynamical plus LTTE solutions to the ETV curves.  By contrast, the green points are estimates based on the assumption that $m_{\rm AB} \simeq 2 \,\mathrm{M}_\odot$. }
\label{Fig:dynoltte}
\end{figure}

In Fig.~\ref{Fig:dynoltte} we plot the ratio of the dynamical to LTTE amplitude vs.~the ratio of periods $P_1/P_2$. For the 62 systems shown in red symbols, the ratio $\mathcal{A}_{\rm dyn}/\mathcal{A}_{\rm LTTE}$ is directly measured from the fits to the ETV curves.  For the remaining systems where the ETV curve is dominated by the LTTE effect, $\mathcal{A}_{\rm dyn}/\mathcal{A}_{\rm LTTE}$ is estimated using the measured periods and outer eccentricity and is also based on the assumption that $m_{\rm AB} \simeq 2\, \mathrm{M}_\odot$.  The quite strong correlation can be understood with the help of the theoretical ratio of the two amplitudes, as was discussed in Subsect.~\ref{Subsect:LTTEgeneral}. It was shown there that, aside from dependencies on masses and eccentricities, the ratio is proportional to $P_1^2/P_2^{5/3}$.  This would give a slope of $\sim2$ in a $\log-\log$ plot, i.e., a value close to the slope exhibited in the figure.

\begin{figure}
\includegraphics[width=\columnwidth]{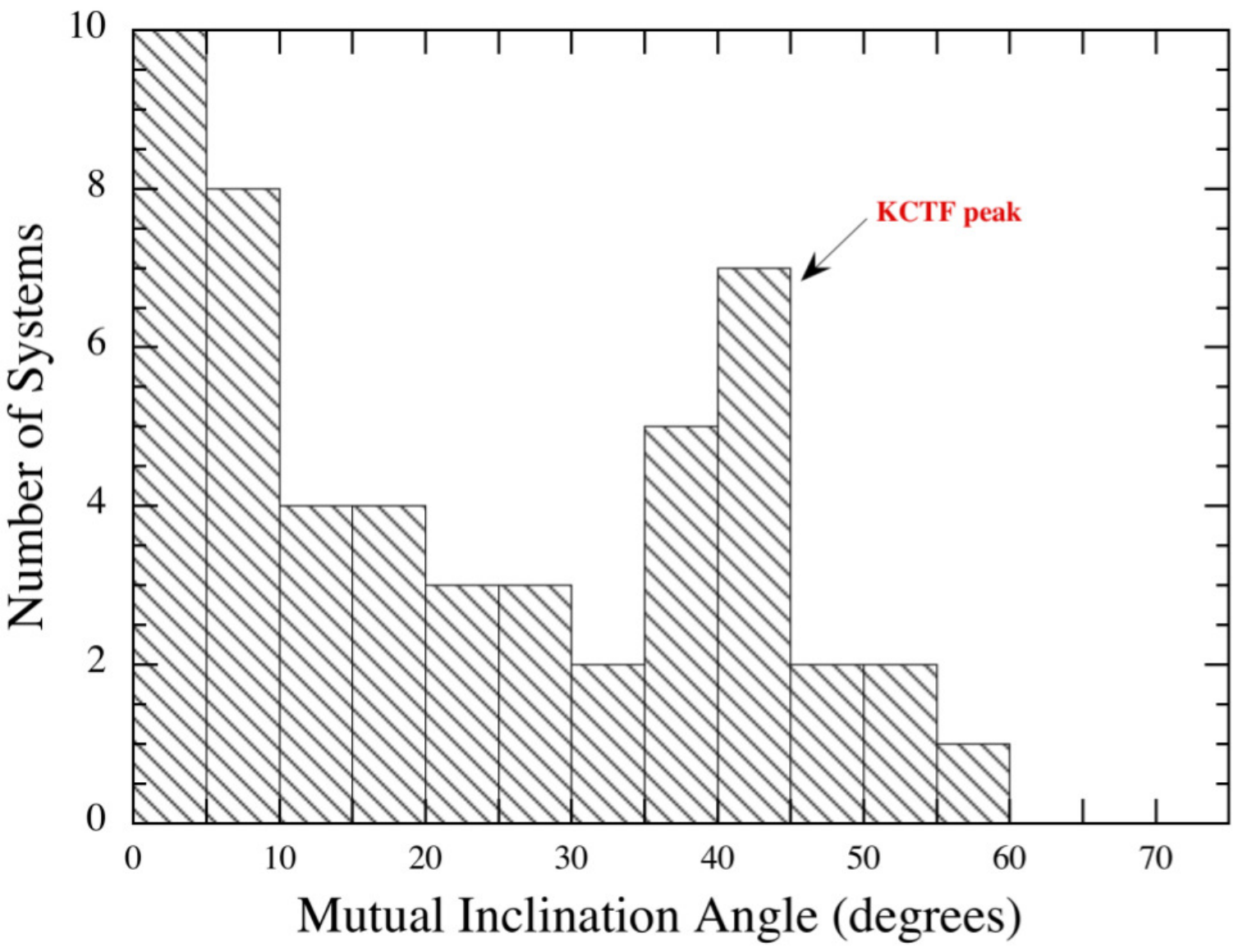}
\caption{Distribution of the mutual orbital inclination angle, $\im$, for 62 systems where there was sufficient information in the ETV curves to allow for its determination.  Note the peak centered around $\im \simeq 40^\circ$ which we associate with Kozai cycles with tidal friction in systems with initial values of $39^\circ \lesssim \im \lesssim 141^\circ$ (see text for a discussion and references). The peak between $\im=0\degr$ and $5\degr$ actually contains 21 systems, but goes off the top of the plot.}
\label{Fig:im}
\end{figure}

\begin{figure}
\includegraphics[width=\columnwidth]{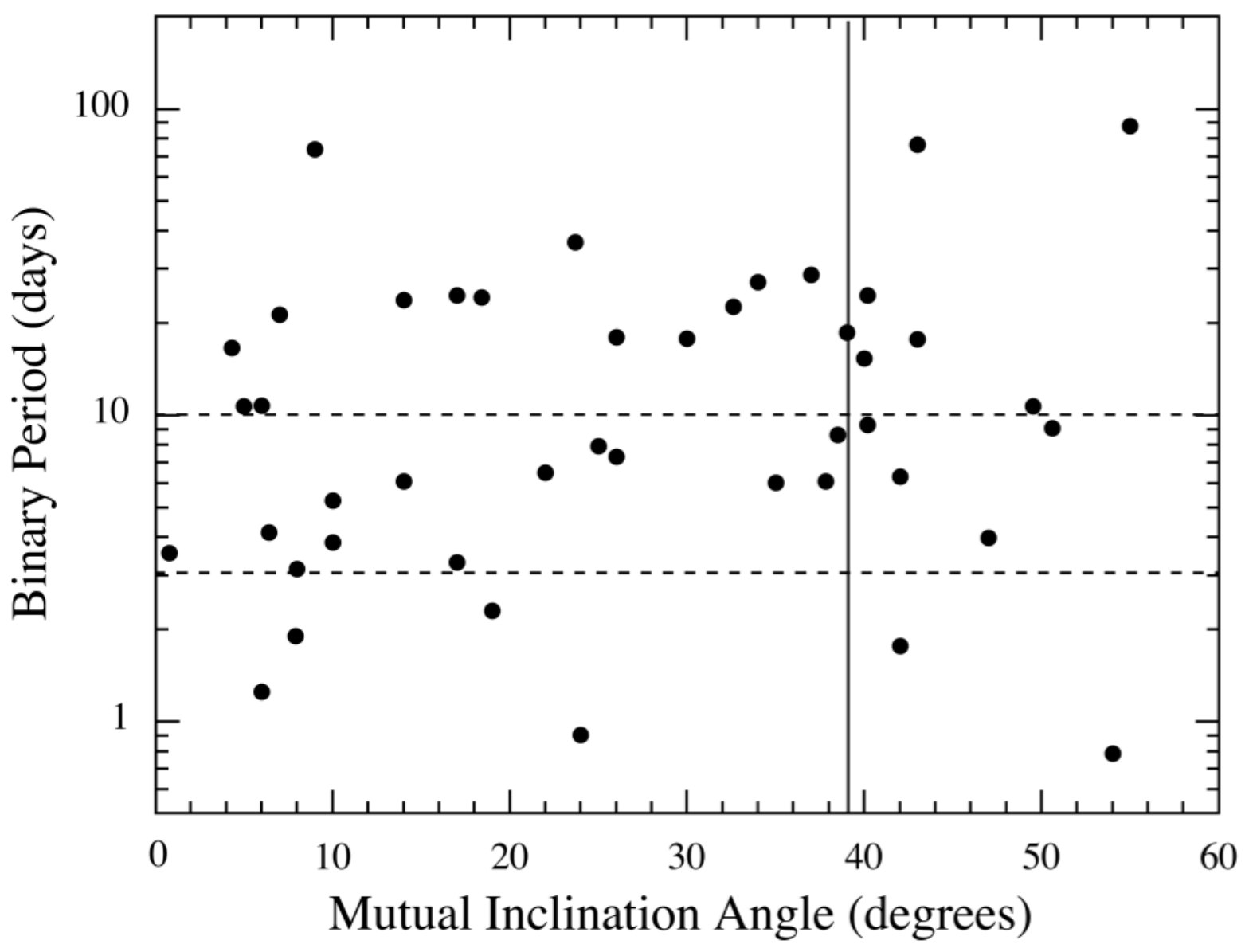}
\caption{The relation between the mutual orbital inclination angle, $\im$, and the inner binary period ($P_1$) for 44 systems where there was sufficient information in the ETV curves to allow for their determination (see text). Only systems with non-zero $\im$ are shown for clarity. The two dashed horizontal lines indicate the expected range of $P_1$ values near $\im \sim 39^\circ$ (vertical line) from the \citet{fabryckytremaine07} model.}
\label{Fig:P1vsim}
\end{figure}

In Figure~\ref{Fig:im}, we show the distribution of mutual inclination angles, $\im$ for the combined-solution systems. Some 32\% of the systems are contained in a peak centered at $\im \simeq 40^\circ$.  For systems where the primordial value of $\im$ lies in the range $39.2^\circ \lesssim \im \lesssim 140.8^\circ$, it has been shown that the tertiary star drives Kozai-Lidov cycles with tidal friction \citep[hereafter KCTF; see, e.g,][]{kozai62,lidov62,kiselevaetal98,fabryckytremaine07} that may involve large-amplitude oscillations of the eccentricity and inclination of the inner binary.  The large eccentricities thereby induced in the binary ultimately lead via tidal friction to shrinkage and circularization of the orbit---with $\im$ `frozen out' near $\im \simeq \sin^{-1}(\sqrt{2/5)}$.  This explains the peak in the $\im$ distribution near 40$^\circ$. In that regard, our results may be taken as confirmation of the KCTF model. On the other hand, however, some caution is necessary because the inner period--mutual inclination relation (Fig.~\ref{Fig:P1vsim}) does not confirm the expected final period distribution of \citet{fabryckytremaine07} which predicts an enhancement of $3\lesssim P_1\lesssim10$ day short-period binaries amongst the $\im\sim40\degr$ mutual inclination triples. Finally, we note that many of the lower mutual inclination systems in our sample are likely to have been originally formed with mutual inclination angles less than $39.2^\circ$.

\begin{figure}
\includegraphics[width=\columnwidth]{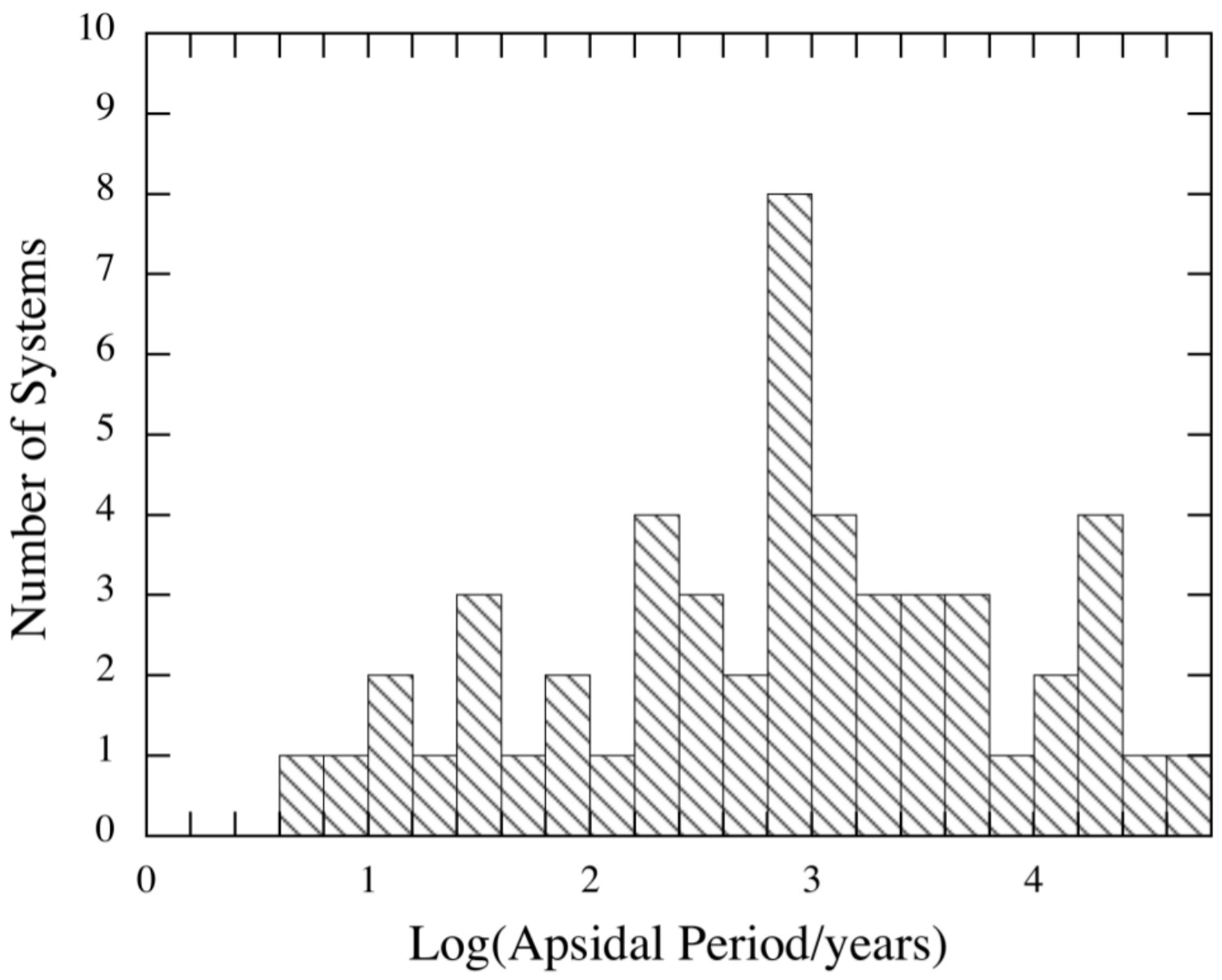}
\caption{Distribution of the apsidal period in the inner eccentric binary driven by the tertiary star.  There was sufficient information in the ETV curves of 45 systems to derive $P_{\rm apse}$.}
\label{Fig:Papse}
\end{figure}

Apsidal motion time scales are discussed extensively in \citet{borkovitsetal15}.  As noted there, in the presence of a close ternary, the dynamically forced apsidal motion of an eccentric EB can substantially exceed, even by several orders of magnitude, the classical and the relativistic apsidal motion contributions. The apsidal advance rates in the present set of EBs are fully constrained by the dynamical ETV solutions and vice versa; these constraints are built into our ETV solution procedures\footnote{Two different approximations which are used by our code for determining the constrained apsidal motion parameters and, furthermore, the difference between the dynamical and the apparent (geometrical) apsidal advance rates are explained in Appendix~C of \citealp{borkovitsetal15}.}.  As a consequence, the dynamical apsidal motion time scale is a derived output of our combined ETV solution. These $P_{\rm apse}$ time scales, which are shown in Fig.~\ref{Fig:Papse}, are distributed widely, e.g., over the range $\sim$$10-10^4$ years, with more than half of them above $500$ years.

We conclude this section by noting that our collection of 222 {\em Kepler} triples constitutes nearly 10\% of the entire {\em Kepler} catalog of  $\sim$2600 binaries.  The outer periods range from approximately 30 to more than 2000 days; the sample is rather incomplete for outer periods longer than 2000 days.  Thus our sample covers only 1.8 dex out of the possible total range of approximately $6-7$ dex.  If the outer periods of triples are roughly uniformly spaced logarithmically, this immediately implies that at least 30\% of all binaries are located in triples or higher-order multiples.  Furthermore, we may have missed some substantial fraction of the triples in the {\em Kepler} data set due to a variety of causes.  We conclude that a very substantial fraction, perhaps approaching unity, of the binaries are likely bound together with one or more other bodies.  

\subsection{Systems with extra eclipse events}
\label{Subsect:extraeclipses}

As mentioned before, ten EBs among our sample exhibit extra eclipsing events which we have attributed to the same third bodies identified as the sources of the ETVs. Nine of these ten systems were recognized earlier. They are: KIC~05897826 (=KOI-126) \citep{carteretal11}, KIC~05952403 (=HD~181068) \citep{derekasetal11}, KICs~06543674, 07289157 \citep{slawsonetal11}, KIC~02856960 \citep{armstrongetal12}, KIC~02835289 \citep{conroyetal14}, and KICs~05255552, 06964043, 07668648 \citep{borkovitsetal15}. The tenth triply eclipsing EB is KIC~09007918 which shows one extra eclipsing event around BJD~2\,456\,326.2 that has not been reported previously (see Fig.~\ref{Fig:K9007918E3}). These extra eclipses have a variety of shapes and, in most cases, large depths.  In a minority of cases these extra events are manifest only as barely discernible short disturbances or shallow transit-like fadings which might even be aperiodic, and their real nature can only be verified with the help of an LTTE or combined ETV solution.  Such events are seen in the light curves of KIC~06543674, KIC~07668648, and, most notably, KIC~09007918. 

The modeling of eclipses involving a third body brings great sensitivity to the determination of the complete configuration of a system and of its dynamical properties.  On the other hand, this great sensitivity implies that it may be extremely difficult to obtain a model that accurately predicts the extra eclipse times and other characteristics. For example, even if the outer orbit is wide enough to nearly eliminate any dynamical perturbations, the lightcurve may be affected not only by the presence of the easily detectable third eclipses but also by small changes in the binary eclipses, e.g., when the two orbits are not perfectly coplanar and there is precession of the orbital planes, or when either of the two orbits is eccentric and undergoes apsidal motion.  In practice, the most accurate interpretation of such a system can be carried out only by simultaneous modeling of its photometric and dynamical properties, as was done for KIC~05897826 (=KOI-126) by \citet{carteretal11}.


Two additional triply eclipsing systems for which lightcurve solutions are available in the literature are KIC~05952403 (=HD~181068) \citep{borkovitsetal13} and KIC~06543674 \citep{masudaetal15}. For neither of them is there a complete photodynamical solution. In the case of KIC~05952403 this can be understood from the fact that this is the only system in our sample which consists of two nearly perfectly coplanar circular orbits and, therefore, cannot show significant perturbations such as orbital plane precession or apsidal motion. In the case of KIC~06543674, only one set of outer eclipse events has been observed.  It is therefore unfit for a complete photodynamical analysis. Note also that the outer period of $P_2=1101\fd4\pm0\fd4$ of this system is the longest period known for any triply eclipsing system. The outer orbit thus represents the `eclipsing binary' with the longest period in the entire {\em Kepler} EB sample.

There are other systems in the {\em Kepler} EB sample which have light curves that exhibit extra eclipsing events or other complex features, but do not turn out to be hierarchical triples or do not show ETVs.  They are not included in our sample.

Amongst these systems, KIC~07670485 shows only one extra fading event around BJD~2\,455\,665 \citep{orosz15}. The primary and secondary $O-C$ curves of this EB, however, do not show any ETVs, but only some scatter with an amplitude of $\sim3\times10^{-4}$\,d. 

For KICs~04247791 and 07622486 the strict periodicity and unaltered shapes of the extra eclipses make it evident that what is seen in these two lightcurves are the blends of two EBs. In the case of KIC~04247791 it has already been reported by \citet{lehmannetal12} that this source consists of two double-lined (SB2) binaries. The question that naturally arises for these blended EBs is whether these form 2+2 hierarchical quadruple systems or not.  In order to investigate this question, we proceeded to disentangle the lightcurve of each of the two targets in the following manner. First, we folded, binned, and averaged the complete {\em Kepler} lightcurve independently according to each of the two periods. Each folded lightcurve allowed the determination of the average phases of the first and last contacts of each eclipse. The folding, binning, and averaging procedures were then repeated for each of the two binaries, but this time excluding those lightcurve sections containing eclipses of the other binary. Then, each of the folded, binned, and averaged lightcurves was subtracted from the original time-series with a three-point local Lagrangian polynomial interpolation. In such a manner we obtained two residual lightcurves, each of which primarily contained only the eclipsing structure of the other binary of the blended source. In a final stage, the times of minima were determined from these residual lightcurves in the same way as was done for all the other systems in this study. 

For KIC~04247791 the four $O-C$ curves (two primary and two secondary, respectively), do not exhibit any significant curvature. This does not eliminate the possibility that the two EBs could be gravitationally bound, but we can conclude that the period of the possible wide (quadruple) orbit most probably exceeds a few decades. 

The situation in KIC~07622486 is a bit complicated. This source consists of a long period ($P_\mathrm{1A}=40\fd25$) eccentric EB with a sharp and relatively deep primary eclipse. A secondary eclipse is not observed (but the disentangled average lightcurve reveals a low amplitude, asymmetric, heartbeat-like feature around the edges of the primary eclipse). The other binary is most probably a semi-detached system ($P_\mathrm{1B}=2\fd28$) with shallow transit-like dips in flux as primary eclipses, and with nearly invisible secondary occultations. Therefore, we used only the $O-C$ diagrams of the primary events for our ETV analysis.  According to this analysis, the longer period binary does not exhibit any interesting ETVs (with an accuracy of $3-4\times10^{-4}$\,days). The primary minima of the shorter period binary show a cyclic feature with a period of $P \simeq 231\pm4$\,days. More in-depth analysis indicates that this periodicity is the consequence of stellar oscillations in the short-period range of some hundredths to tenths of a day; these alter the times of the shallow primary transits in a quasi-periodic manner. Therefore, we conclude that there is neither a periodic signal nor any curvature in the ETVs of the two blended binaries in KIC~07622486. Thus, our assessment of this system is the same as that for KIC 04247791.

Perhaps the most complex EB lightcurve {\em ever} observed is that of KIC~04150611. It exhibits eclipses with three different periods, of which the longest period eclipses exhibit very complex and variable features. Therefore, the multiple nature of this system is beyond question. Instead of the comprehensive analysis of the ETVs, we determined $O-C$ diagrams only for those eclipses which belong to the $\sim8\fd65$-day eccentric binary component.  We were able to obtain its disentangled lightcurve with only a little effort (by the use the above described technique). Neither the primary nor the secondary $O-C$ curves exhibit any curvature or periodicity; therefore, due to the lack of interesting and informative ETVs we have not included this intriguing system in our sample.
  
\begin{figure*}
\includegraphics[width=84mm]{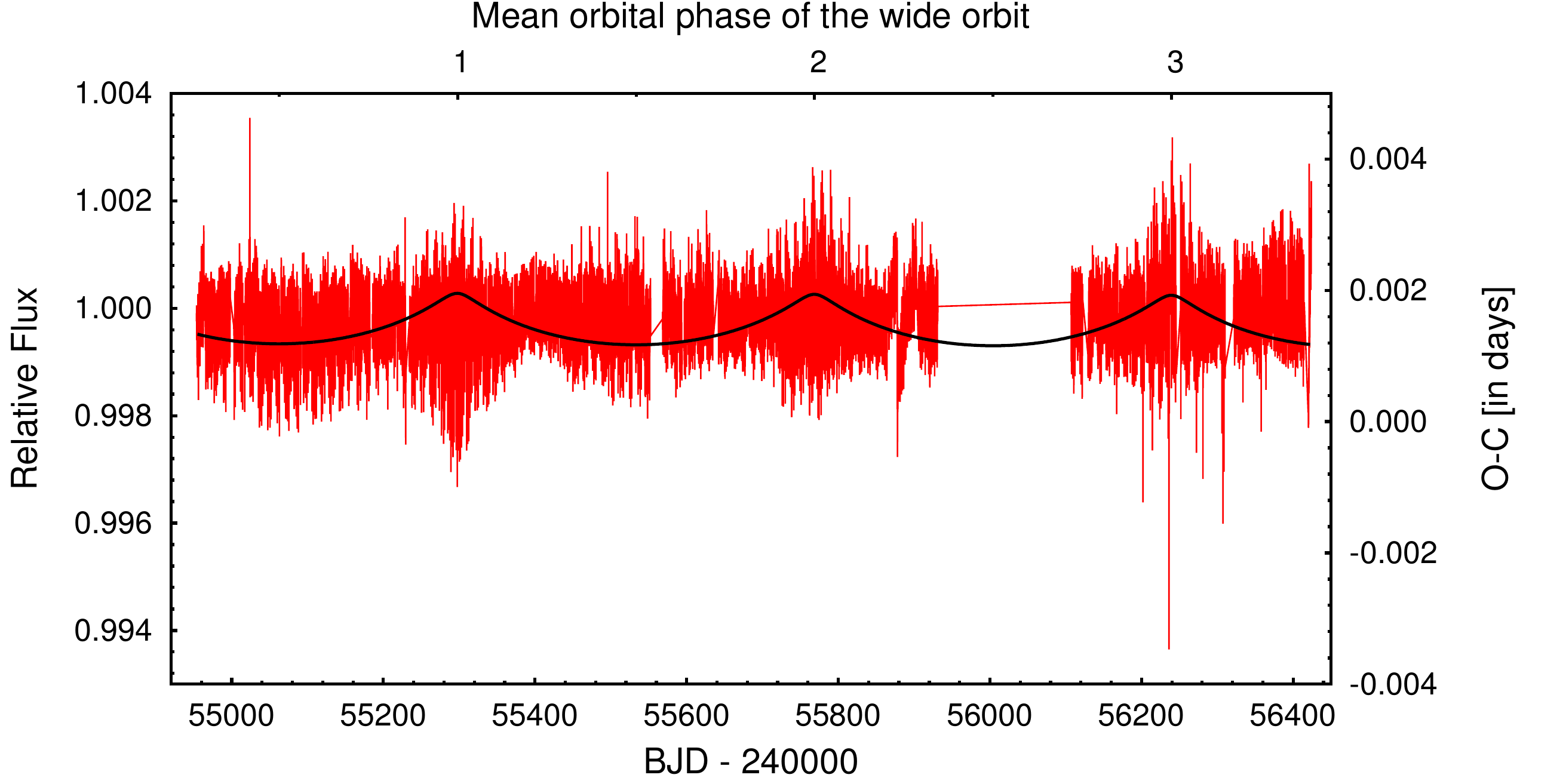}\includegraphics[width=84mm]{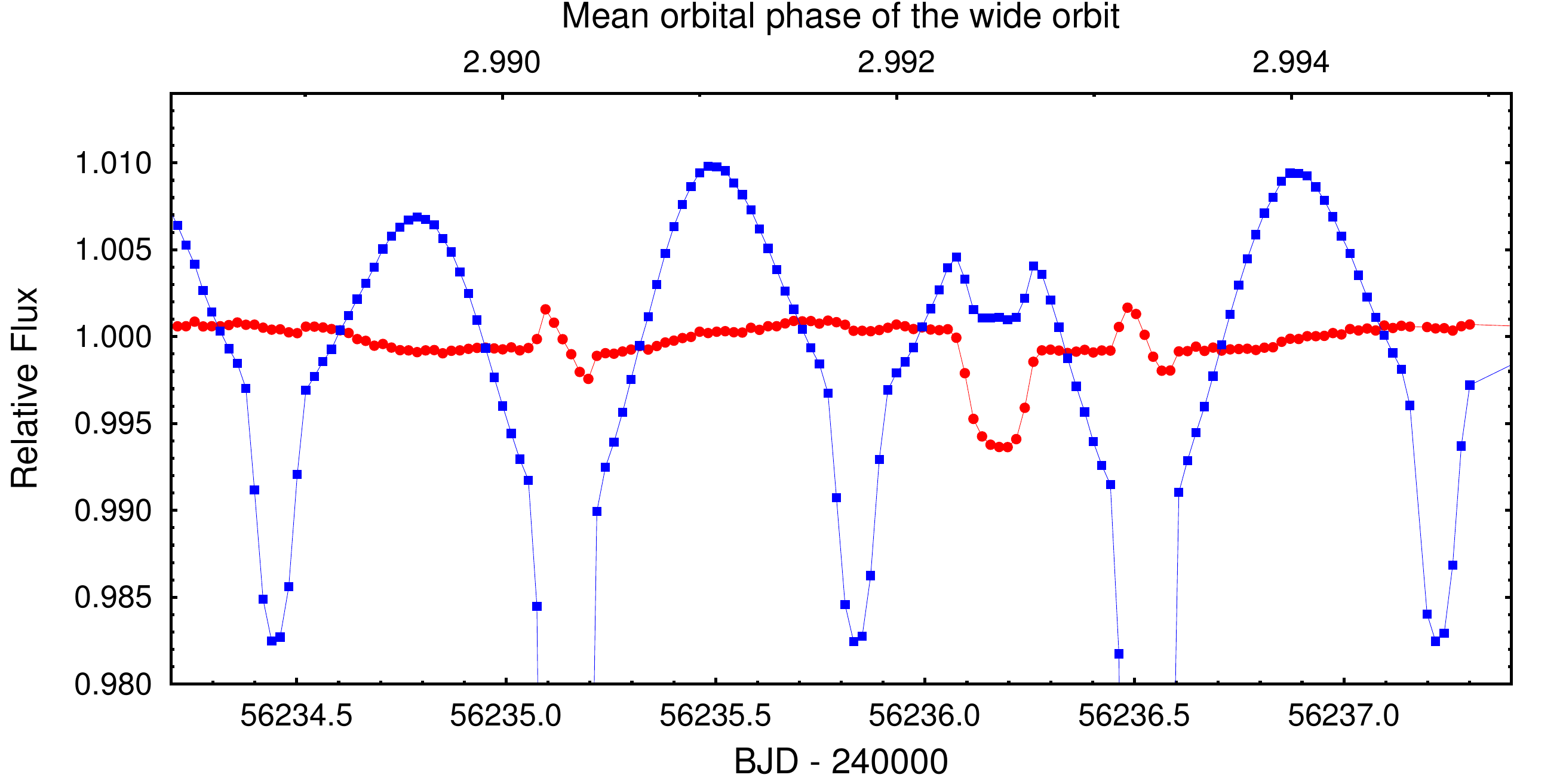}
 \caption{The identification of an extra eclipse event in the {\em Kepler} lightcurve of KIC~09007918. {\it Left panel:} After subtraction of the folded, binned, averaged lightcurve from the detrended full time-series, a definite fading event can be seen around MBJD~56326.2 in the residual lightcurve (red) which coincides with one of the sharply peaked maxima of the ETV curve and, therefore, the of LTTE solution (black curve) as well. {\it Right panel:} A close-up this fading shows a transit-like extra eclipse event, which can nicely be identified not only on the residual lightcurve (red), but also on the original, detrended lightcurve (blue). The event occurred very close to the time of the maximum Roemer-delay of the EB, which, in this triple happens almost at the same moment as periastron passage of the wide orbit. In such a scenario the physical (i.e., spatial) and the projected distances of the EB and the ternary reach their minimum values at the same time, which increases the likelihood of the outer eclipses. The regular, transit-like shape of the fading, and the fact that it happened during the second quadrature of the EB, i.e., when the projected distance of the two binary members is maximal, makes it most likely that only one of the binary members was eclipsed by the ternary.}
 \label{Fig:K9007918E3}
\end{figure*}

\subsection{Non-transiting circumbinary planet candidates}
\label{Sect:substellar}

Our ETV analysis has identified three triples where the third body is most probably a planetary-mass object. These systems are: KICs~07177553, 07821010 and 09472174. 

KIC~09472174 contains the only short-period, low-mass sdB+dM binary in our sample. The periodic ETVs have already been interpreted as being due to the LTTE effect by \citet{baranetal15}. Because our analysis yielded results similar to those of this previous study, we simply report the analyzed ETV curve and the corresponding orbital solution. We also note that if we accept $m_\mathrm{AB}=0.60\pm0.03\,\mathrm{M}_\odot$ for the total mass of the EB \citep{ostensenetal10}, then we obtain $(m_\mathrm{C})_\mathrm{min}=2.0\,\mathrm{M}_\mathrm{J}$. This implies that the third body would exceed the lower mass limit of a brown dwarf only for $i_2 \lesssim15\degr$. If the periodic signal really arises from the LTTE effect, the third object may well have a mass in the planetary range.

Ten {\em transiting} circumbinary planets have been previously reported \citep{doyleetal11,welshetal12,oroszetal12a,oroszetal12b,schwambetal13,kostovetal13,kostovetal14,welshetal15}. Candidate circumbinary exoplanets may also be found in KIC~07177553 and KIC~07821010.  The two candidates revolve around relatively wide eccentric binaries ($P_1=18\fd00$ and $24\fd24$; $e_1=0.39$ and 0.68 for KICs~07177553 and 07821010, respectively), with periods of $P_2=529\pm2$ and $991\pm3$ days. In both cases the ETVs are dominated by dynamical effects  (${\cal{A}}_\mathrm{dyn}/{\cal{A}}_\mathrm{LTTE}\sim48$ and $33$). The possible non-transiting circumbinary planet in the KIC~07177553 system is reported here for the first time, while the circumbinary planet candidate in KIC~07821010 has been recently investigated by D. Fabrycky and his collaborators (Fabrycky et al., in prep.). Their preliminary results have been presented at a conference by W. Welsh\footnote{http://www.astro.up.pt/investigacao/conferencias/toe2014/files/wwelsh.pdf}. Considering our own finding for KIC~07177553, because of the very low contribution of the LTTE term to the ETV solution, instead of the individual masses, our ETV solution yields only the ratio $m_\mathrm{C}/m_\mathrm{ABC}$ with satisfactory accuracy. Therefore, strictly speaking, we can say only that if the total mass of the EB $m_\mathrm{AB}$ is less than $\sim3\,\mathrm{M}_\odot$, then the potential third body is in the mass range of a giant planet instead of a brown dwarf. Spectroscopic follow up to confirm or reject this result is in progress.

\subsection{Comparison with previous surveys}
\label{Sect:prior_survey}

Here we compare our results with those of previous systematic ETV surveys of the {\em Kepler} EB sample. As mentioned in the Introduction, apart from the pioneering investigations of \citet{giesetal12}, which due to its very preliminary nature does not allow for quantitative comparisons, third-body solutions via ETV analyses for {\em Kepler} EBs were first published by \citet{rappaportetal13}. The latter reported combined LTTE and dynamical ETV solutions constrained by the circular-inner-orbit approximation for a sample of 39 EBs. Twenty of these 39 triples, i.e., those where the dynamical terms had yielded a negligible contribution, were also considered in \citet{conroyetal14}. The remaining ten eccentric EBs out of the 39 systems of \citet{rappaportetal13} were reinvestigated by \citet{borkovitsetal15} with the first application of an improved, much more sophisticated approximation for the dynamical contribution of the ETVs. The present sample includes the 39 EBs of \citet{rappaportetal13}. For 38 of 39 triples the present solutions differ only slightly in terms of numerical values, which is in accord with the statement in Section~\ref{Subsect:Reliability} that, for well- and multiply-covered outer orbits, the ETV solutions yield robust and reliable orbital parameters. The one exception of the 39 systems is KIC~07837302 for which, due to insufficient data-length, the ETV behaviour was misinterpreted. For this triple, by the use of the entire, 4-year-long {\em Kepler} dataset, we give a completely different dynamically dominated solution. This latter solution, however, should also be considered with caution, since the outer period we obtain is shorter than the data length only by a small amount.

The largest sample of triple star candidates amongst {\em Kepler} EBs was published by \citet{conroyetal14}. They produced and investigated the $O-C$ diagrams of all the short-period {\em Kepler} EBs and ELVs and identified 236 systems for which they found that the ETV might be compatible with the presence of a third companion. Our compilation contains only 115 of their 236 triple system candidates, mainly as a result of our more stringent selection criteria. 

To be specific, our criteria filtered out seven of the 35 systems in the first group of \citet{conroyetal14}, the most likely of their triples candidates.  Amongst these are KICs~05560831 and 10014830 where the smoothing polynomial killed the cyclic ETV pattern, while in the cases of KICs~03641446, 07657914, 08211618 and 11247386 we found highly discrepant, and/or anticorrelated ETV and QTV curves (see the left panel of Fig.~\ref{Fig:FPornot}).  The seventh rejected system, KIC~06302592, has a morphological classification \citep{matijevicetal12} value of 0.93, indicating this system is an ELV binary, although the folded, averaged lightcurve reveals clear, very low amplitude grazing eclipses.  For this system we were unable to find a third-body solution for the distorted quasi-periodic primary ETV curve.  One of the remaining 28 of the 35 first-group systems in \citet{conroyetal14}, KIC~010855535, proved to be a false positive in the sense that although the ETV signal is quite possibly due to the LTTE effect induced by a third star, the modulations in the {\em Kepler} lightcurve are most probably due to the pulsations of single star instead of an EB or ELV (Fig.~\ref{Fig:FPornot2}). We dropped ten of the 80 members of the middle group of \citet{conroyetal14} for reasons like those used in rejecting the seven systems of the first group, and an additional four of the 80 were found to be false positive EBs. Most of the systems we eliminated belonged to the third group of \citet{conroyetal14}. These are EBs where no complete LTTE solutions were given, but only a possible outer period was listed. In most cases we confirm the claim of \citet{conroyetal14} that these ETVs might arise from long-period LTTEs. Due to insufficient length of the available data, however, we were unable to obtain reasonable LTTE solutions for most of these ETVs and, therefore, they are not included in our sample.

All of the 26 eccentric EBs with strongly dynamically dominated ETVs which were investigated by \citet{borkovitsetal15} are naturally included in the present survey.  We repeated the analysis only for those systems for which additional {\em Kepler} light curve data is now available relative to that used in the previous study. Our results on these do not depart significantly from those in \citet{borkovitsetal15}. The only remarkable difference is that, while in the previous work there was an ambiguity regarding the mutual inclination of KIC~12356914, being either prograde or retrograde, our new solution clearly prefers a prograde configuration.   

During the preparation of this work, an additional study was published by \citet{zascheetal15}. These authors give third-body LTTE solutions for ten {\em Kepler} EBs based on both ground-based and {\em Kepler} eclipse times, and thereby extend the time intervals covered by the available datasets. We confirm the solutions of seven of the ten targets. We attempted to find a solution for KIC~10581918 (=WX~Dra), but were not able to obtain a reliable solution which covered the ground-based eclipse times.  The two other exceptions are KIC~05621294 and KIC~03440230. The very questionable nature of the combined quadratic and low-amplitude LTTE solution given for KIC~05621294 was discussed above in Section~\ref{Subsect:Reliability}. In addition, for KIC~05621294, after the application of the smoothing polynomials, the ETVs were found to be very low in amplitude and even significantly smaller than those of the low-amplitude solutions of both \citet{zascheetal15} and \citet{leeetal15}. For KIC~03440230, our smoothed $O-C$ curves, especially for the primary eclipse (see Fig.~\ref{Fig:ETVs}), do not show the periodic pattern which is visible in Fig.~3 of \citet{zascheetal15}. Therefore, unfortunately, we are not able to confirm their findings of a low-mass third body in a one-year orbit. We also note that if the one-year periodic feature happens to be real, this system most probably would require a combined LTTE plus dynamical solution.  We give instead a parabolic plus low-amplitude LTTE solution. The reliability of this latter LTTE solution is, however, questionable.

In summary, we find that our work (i) is in reasonable agreement with earlier studies, (ii) effectively doubles the sample of well-diagnosed {\em Kepler} triples (iii) substantially improves on many of the earlier solutions, and (iv) adds a significant degree of rigor in selecting valid triples.

 \subsection{Additional interesting ETVs}

There are hundreds of other EBs in the {\em Kepler} sample for which $O-C$ diagrams show a wide variety of ETVs. Unfortunately, however, these cannot be interpreted either qualitatively or quantitatively because of the short length of the data train with respect to the probable timescale(s) of these features. Not counting the simply diverging or converging primary and secondary ETV curves, which are clear markers of the classical and/or relativistic apsidal motions of eccentric EBs,  i.e., apsidal motions not due to third-body forced perturbations, the most typical examples of these ETVs are more or less parabolically shaped.  Because of the large numbers of such systems we do not list them individually in this work. There are a few other systems, however, where the features of the $O-C$ diagrams make it very probable that they indicate the presence of third-body perturbed dynamical ETVs. We list those systems in Table~\ref{Tab:ETVinteresting}.

\begin{table}
\begin{center}
\caption{Additional systems with interesting, potentially dynamically originated ETVs} 
\label{Tab:ETVinteresting}  
\begin{tabular}{lll} 
\hline
KIC No. & $P_1$ & ETV characteristics \\ 
\hline
05393558 & 10.22 & displaced secondary eclipses\\
         &       & with different curvature \\
05553624 & 25.76 & displaced secondary eclipses\\
         &       & with different curvature \\
06146838$^a$&27.47&periastron passage event of an eccentric, \\
         &       & inclined tertiary? (only primary eclipses)\\
09032900 & 67.42 & sine-like curve with enormous amplitude \\
10666242 & 87.24 & section of a large amplitude sine? \\
         &       & (eclipse depth decreases, no sec. eclipses) \\
\hline
\end{tabular}
\end{center}
{\bf Notes.} {$^a$: See also: http://www.exoplanet-science.com/koi-6668.html}
\end{table}

\section{Summary and Conclusions}
\label{Sect:Summary}

We have carried out ETV analyses for the complete EB sample of the original {\em Kepler} mission.  Our precise determinations of times of lightcurve extrema were enhanced by the use of refinements such as averaging primary and secondary ETVs, and fitting and subtracting smoothing polynomials over intervals of the lightcurves around individual eclipses.  For the first time, we extended our analyses to include the portions of the lightcurves around the quadrature-phase brightness maxima of the tidally distorted EBs, most of which are contact systems and ELV binaries, and in such a way produced `QTV curves'. We have thereby obtained ETVs for all and QTVs for many of more than 2500 binary systems.  We then selected systems for further analysis where the ETV curves most probably indicate LTTE delays and/or dynamical perturbations caused by a third body in the system. We selected 230 systems, $\sim$$9\%$ of the entire {\em Kepler} sample, that appear to harbor third-body companions.  According to the results of our investigations we have classified these 230 EBs into three main groups, as follows.
\begin{itemize}
\item[]{{\it Group I:} These are EBs for which the ETVs are dominated by the LTTE delays and the dynamical contributions to the ETVs are likely to be negligible. With 160 systems, this is the most highly populated group. The outer periods fall in the range $95\lesssim P_2\lesssim9256$\,days.   In 25 cases, an additional quadratic term was fitted simultaneously to the ETV curve, while a cubic polynomial was required for 4 of the EBs. Furthermore, for 4 of these 160 EBs the apsidal motion effect was also considered.}
\item[]{{\it Group II:} This group contains 62 EBs that exhibit remarkable dynamical perturbations. Therefore, in each of these cases we fit for a combined LTTE plus dynamical ETV solution including apsidal motion terms for the eccentric EBs.  The fits yield several system parameters beyond those which can be obtained from a pure LTTE solution. The most important such parameters are the masses of the EB ($m_\mathrm{AB}$) and the ternary component ($m_\mathrm{C}$), as well as the mutual inclination angle ($\im$) between the inner and outer orbits.  In most cases, the masses can be obtained only with a limited accuracy not appropriate for deeper astrophysical considerations. In addition, a cubic polynomial was also fitted for one system.  The outer period range for the Group II systems is $34\lesssim P_2\lesssim15271$\,days.}
\item[]{{\it Group III:} Each of the remaining eight systems was categorized as a false positive in the sense that, although the observable ETVs most probably arise from LTTE delays due to a companion body, the modulations of the {\em Kepler} lightcurve are likely due to intrinsic variability of the target star rather than to a binary orbit. For these systems we also give LTTE solutions that are naturally excluded from our statistical analyses.} 
\end{itemize}
Groups I and II were also divided into subgroups according to the lengths of the ETV data sets relative to the outer periods. Those systems for which the observational data, in some cases extended with ground-based eclipse timing observations, cover more than two orbital periods were selected to be in the first subgroup.  These can generally be considered as the most reliable candidates and those for which we can expect the most accurate parameters.  We also placed into this first subgroup all the triply eclipsing systems, irrespective of their outer period/data-length ratio, as long as the locations of the outer eclipse(s) were in accord with the corresponding ETV solutions. This subgroup contains a total of 69 triple candidates, 38 and 31 for pure LTTE and combined ETV solutions, respectively. The second subgroup comprises 78 triples, 64 LTTE and 14 combined solution systems, that have outer periods shorter than the length of the available data, but longer than half the length of the data train. For the remaining 75 EBs, 58 LTTE and 17 combined solution systems, less than one outer period was observed and, therefore, the solutions are generally the least certain.

Among our candidates there are ten systems which exhibit triple eclipses which are consistent with the third-body ETV solutions. From this set, the occurrence of an outer eclipse in KIC~09007918 is reported here for the first time. In the case of four additional EBs, where the lightcurves also reveal extra eclipse event(s), we were not able to confirm the multiplicity via our ETV analysis. This does not refute the possible multiple nature of these systems, but rather provides restrictions on the period(s) of the outer orbit(s).

There are three EBs in our sample where our analysis revealed companions that are probably of planetary mass. These non-transiting circumbinary planet candidates are found in KICs~07177553, 7821010 and 94721714; that in KIC 07177553 is reported here for the first time. For the other two planet-candidates our solutions are in accord with earlier reported findings.

In Sect.~\ref{sec:stats} we have presented a statistical analysis of the system parameters obtained for our sample of 222 triple candidates.  Here we highlight three interesting results. The first concerns the distribution of mutual inclination angles obtained for 51 favorable cases among the systems with combined LTTE and dynamical effect solutions.  Two peaks are seen in the distribution.  The larger of the two peaks is at small values that indicate coplanar or nearly coplanar configurations. A significant portion (some 38\%) of the systems are contained in a second peak centered at $\im \simeq 40\degr$.  The centroid of this peak is in good agreement with the predictions of models of Kozai-Lidov cycles with tidal friction.  Second, our collection, which contains 104 triple candidates with outer period $P_2<1000$\,days and 155 triples with $P_2 < 1500$ days, is the richest sample to date of short-outer-period triples. We find that the outer period distribution is more or less flat in the range $200 \lesssim P_2 \lesssim 1600$\,days. For longer periods, the distribution decreases rapidly. Third, we note the almost complete absence of ternaries with $P_2\lesssim200$\,days among the short period mostly overcontact binaries. This cannot be an observational selection effect since we expect to be able to detect the majority of the shortest outer period companions of the closest EBs down to the limit of $P_2\gtrsim40-50$\,days. This result is in agreement with the findings of \citet{conroyetal14} and might offer additional guidelines for the refinement of theories of the formation and evolution of close binaries.  

Finally, we stress the importance of future follow-up observations of the systems investigated here. In cases where ternary eclipses and dynamical perturbations are absent, spectroscopic observations could yield definitive confirmations of the presence of the third stars.  Such confirmations would be of special importance for the shorter outer period systems, because of their significance in the statistics, formation, dynamics, and evolution of hierarchical triples. Furthermore, it is also possible that in some cases spectroscopy may reveal that the third body is also a binary even though this was not apparent in our ETV solution.  

Extension of the eclipse-time data sets via new photometric observations is also highly desirable.  Many of the Kepler EBs offer ideal targets even for proficient amateur astronomers. While the amplitudes of the LTTE and/or the dynamical effects in the shorter outer period systems remain below the realistically available accuracy of ground-based observations, the long-term follow up of these systems would still be useful for detecting longer time-scale variations in the ETV curves. For EBs with longer outer periods, and therefore, larger amplitude ETVs, the ground-based follow-up may even be critical for the confirmation or rejection of the triple system hypothesis, not to mention the quantitative refinement of orbital parameters. A significant fraction of {\em Kepler} EBs and of our sample have eclipses that are too shallow or too long to be good targets for ground-based eclipse monitoring.  Nonetheless, we are convinced that, for many of the wider {\em Kepler} triple candidates, the triplicity can be confirmed within a few years with the help of ground-based observations.

\section*{Acknowledgements}
This project has been supported by the Hungarian OTKA Grant K113117.
This research has made use of data collected by the {\em Kepler} mission, which is funded by the NASA Science Mission Directorate.  Some of the data presented in this paper were obtained from the Mikulski Archive for Space Telescopes (MAST).  STScI is operated by the Association of Universities for Research in Astronomy, Inc., under NASA contract NAS5-26555.  Support for MAST for non-HST data is provided by the NASA Office of Space Science via grant NNX13AC07G and by other grants and contracts. T.\,B. would like to thank the City of Szombathely for support under Agreement No. S-11-1027.

\clearpage

\setcounter{figure}{5}

\begin{figure*}
\includegraphics[width=60mm]{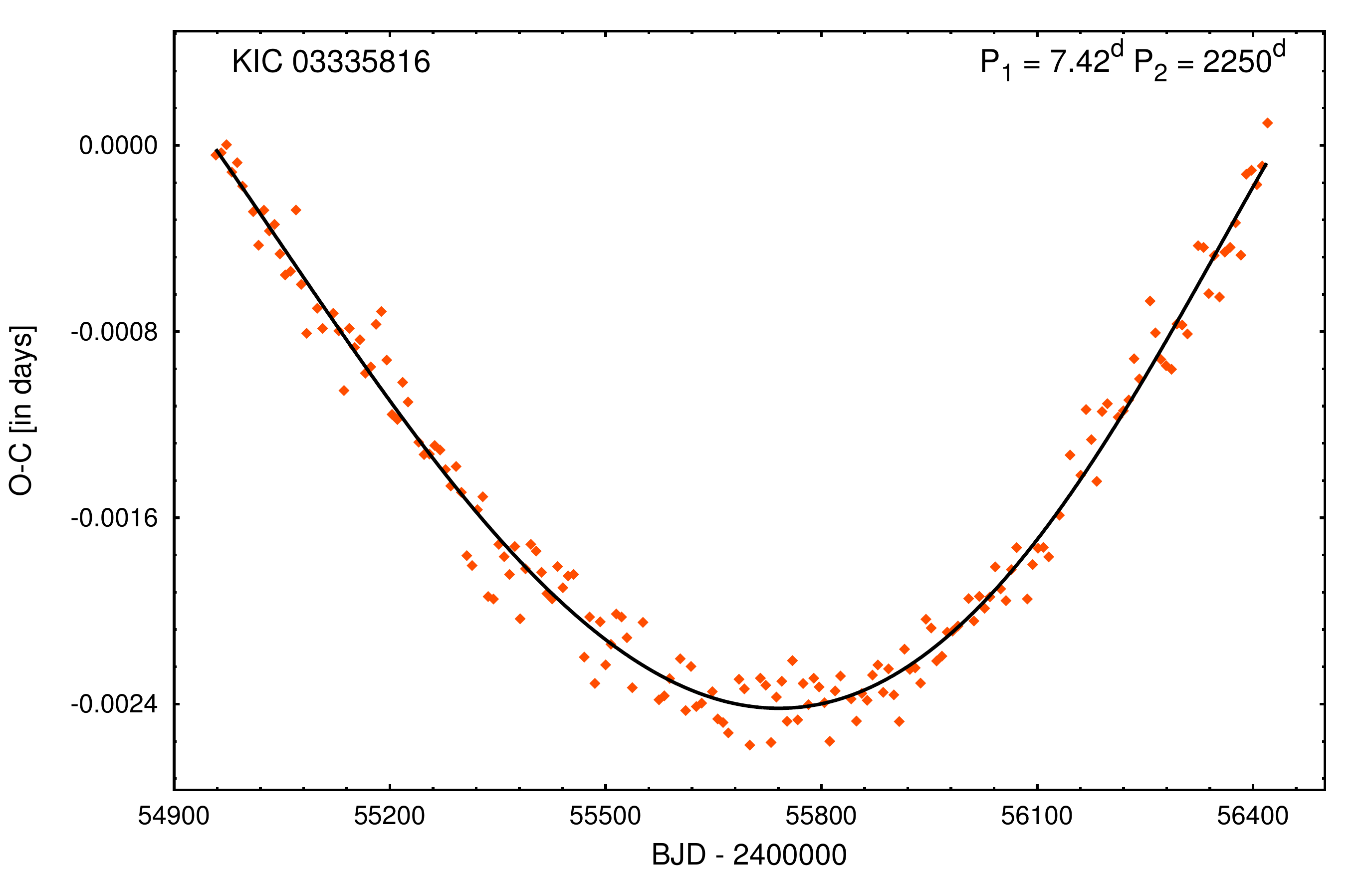}\includegraphics[width=60mm]{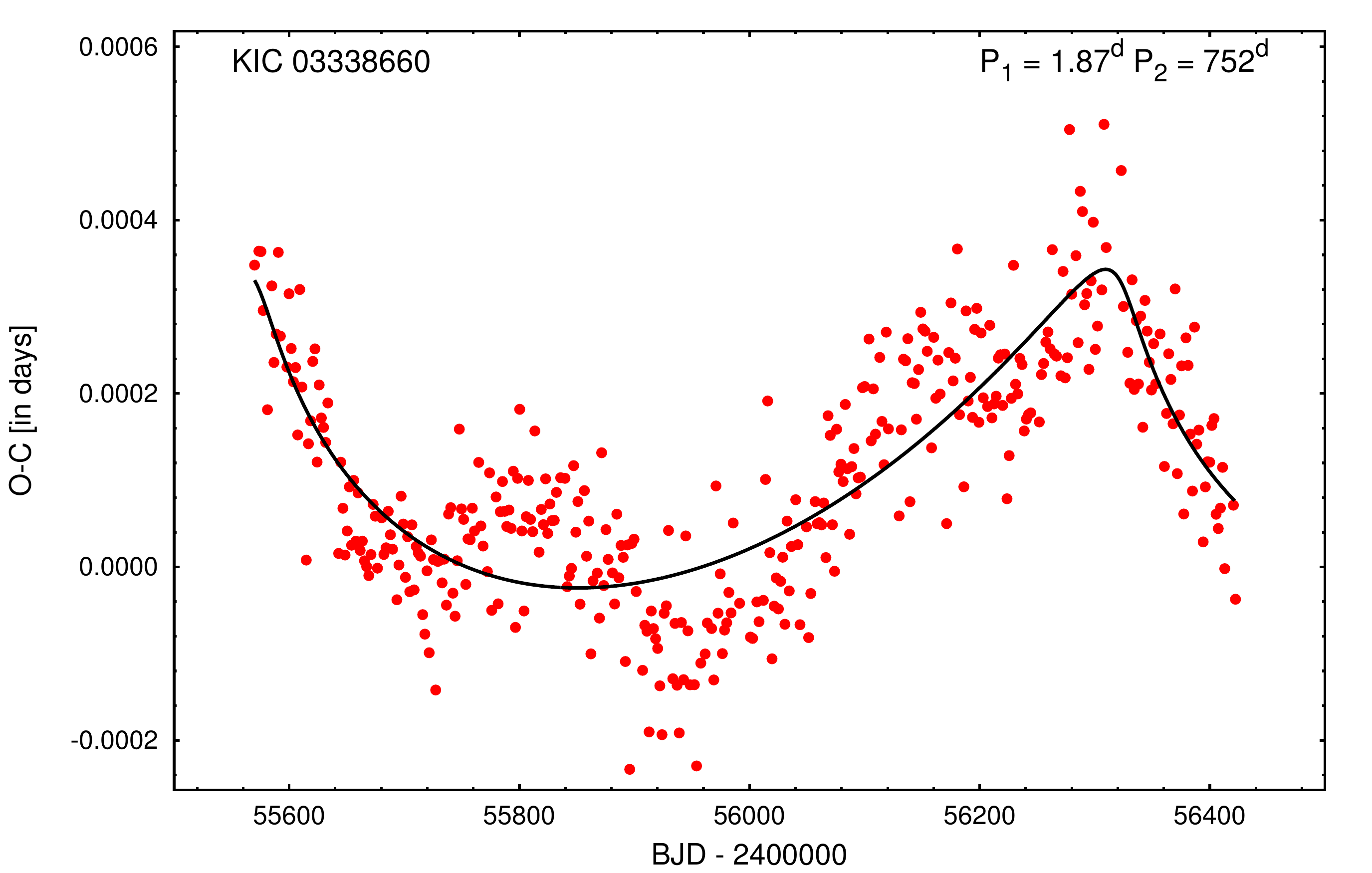}\includegraphics[width=60mm]{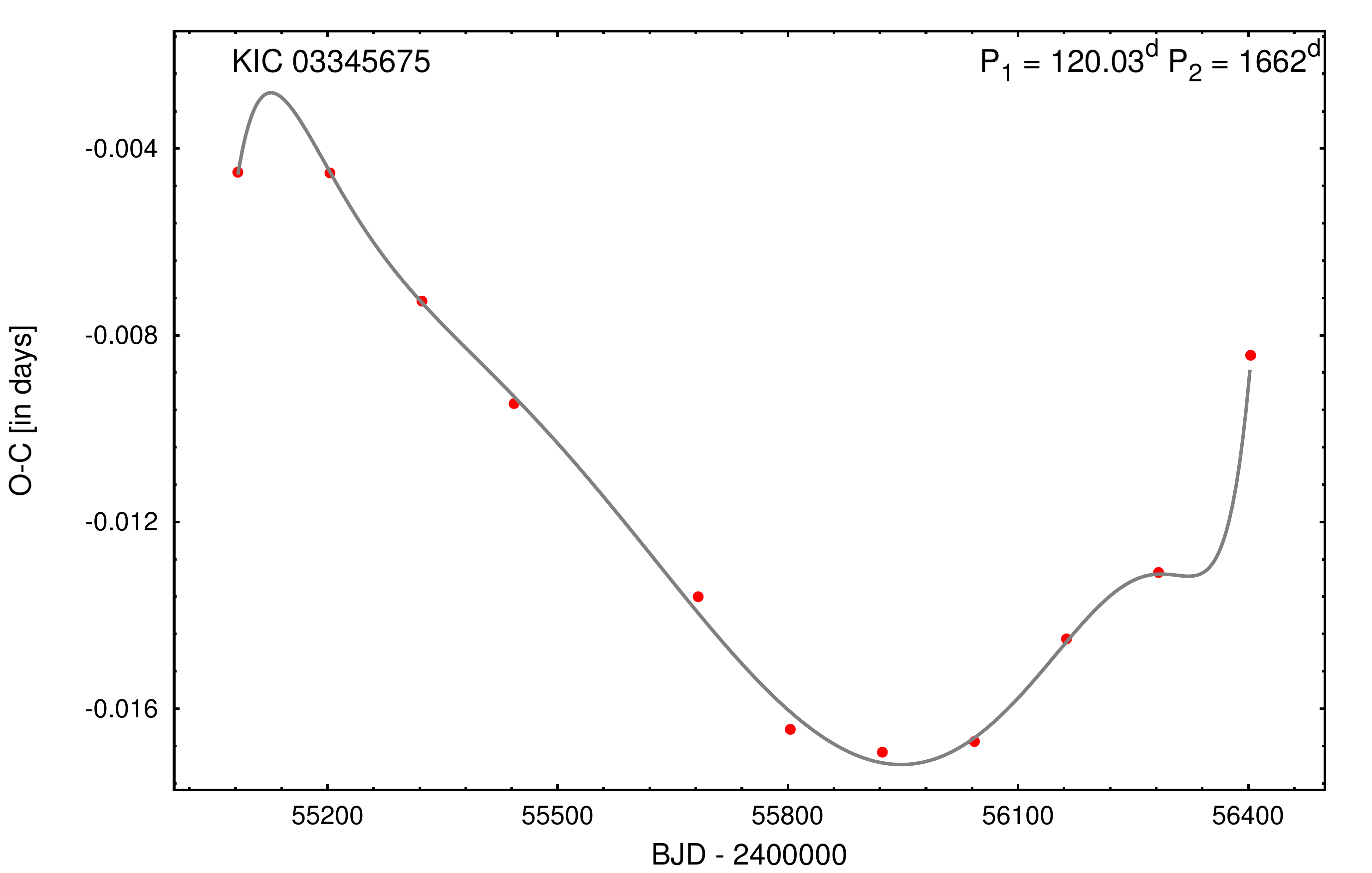}
\includegraphics[width=60mm]{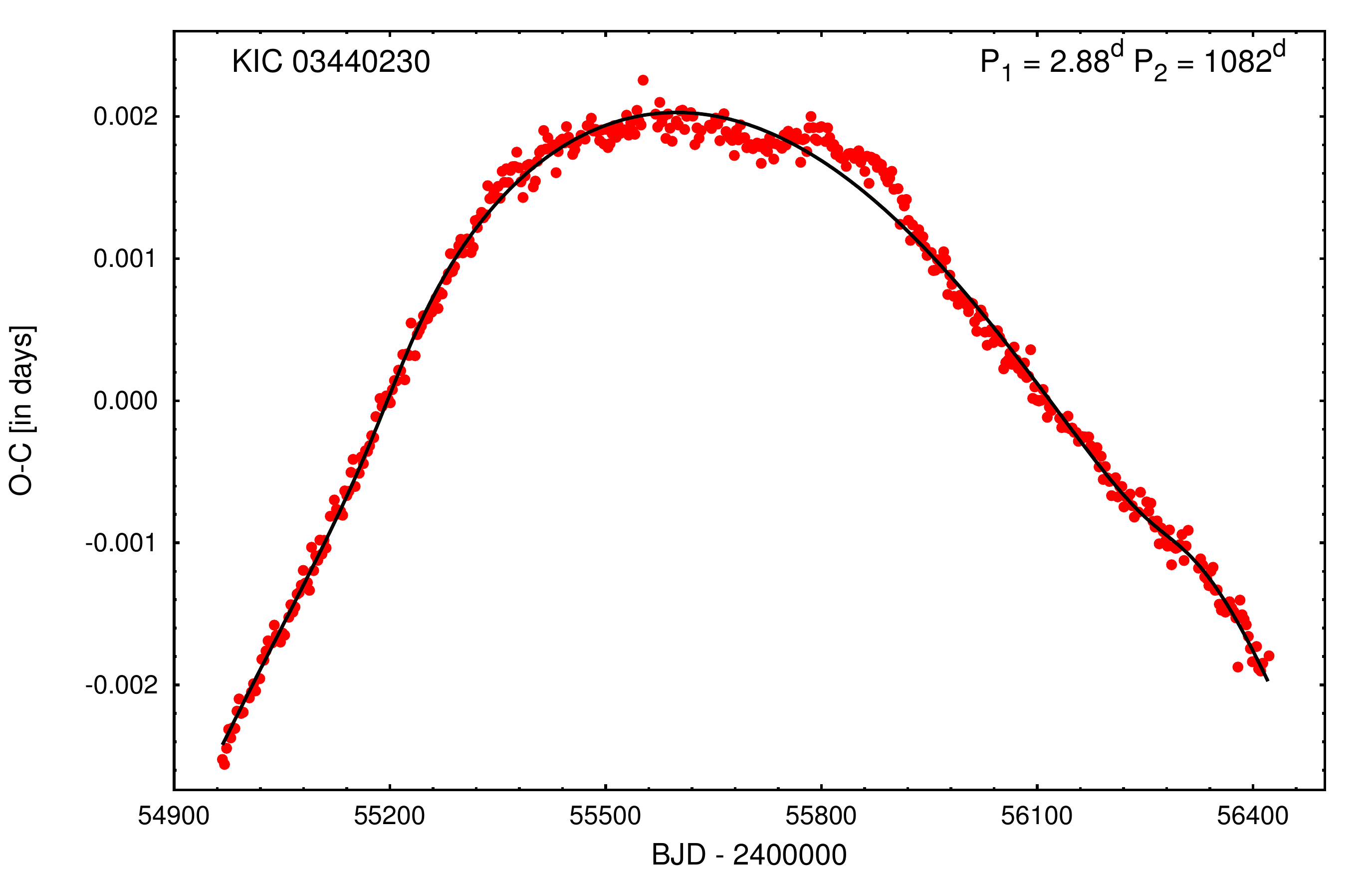}\includegraphics[width=60mm]{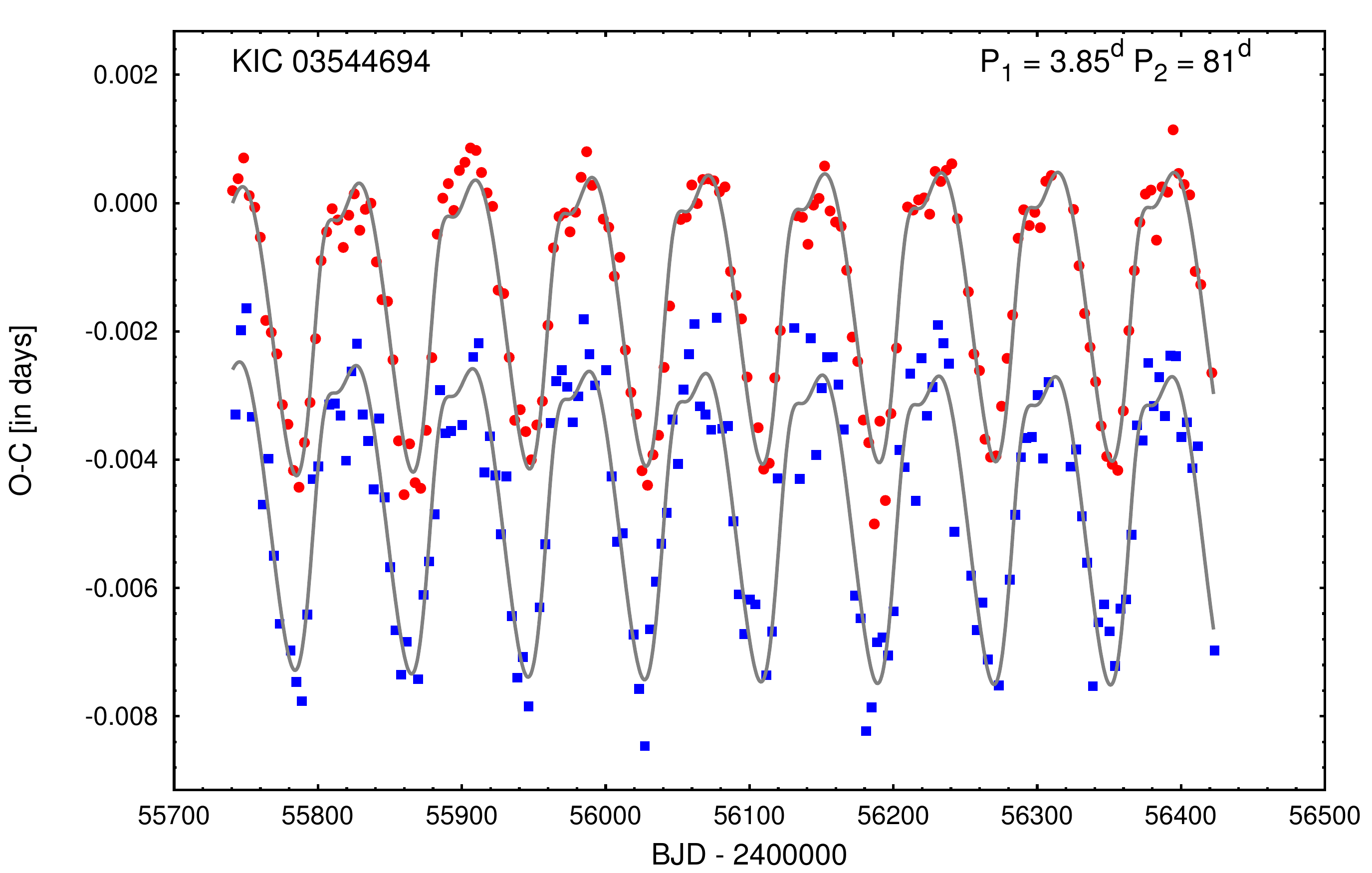}\includegraphics[width=60mm]{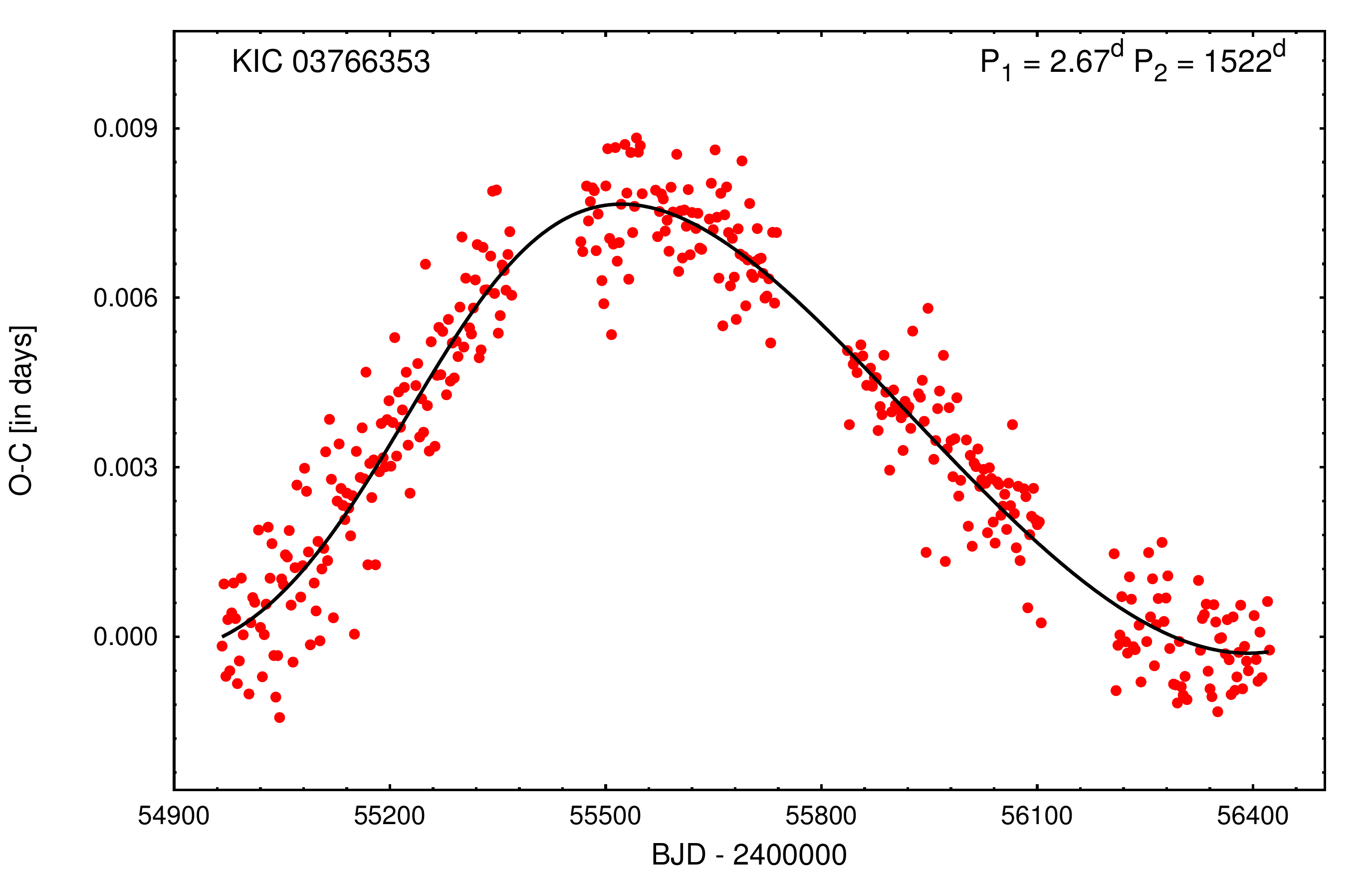}
\includegraphics[width=60mm]{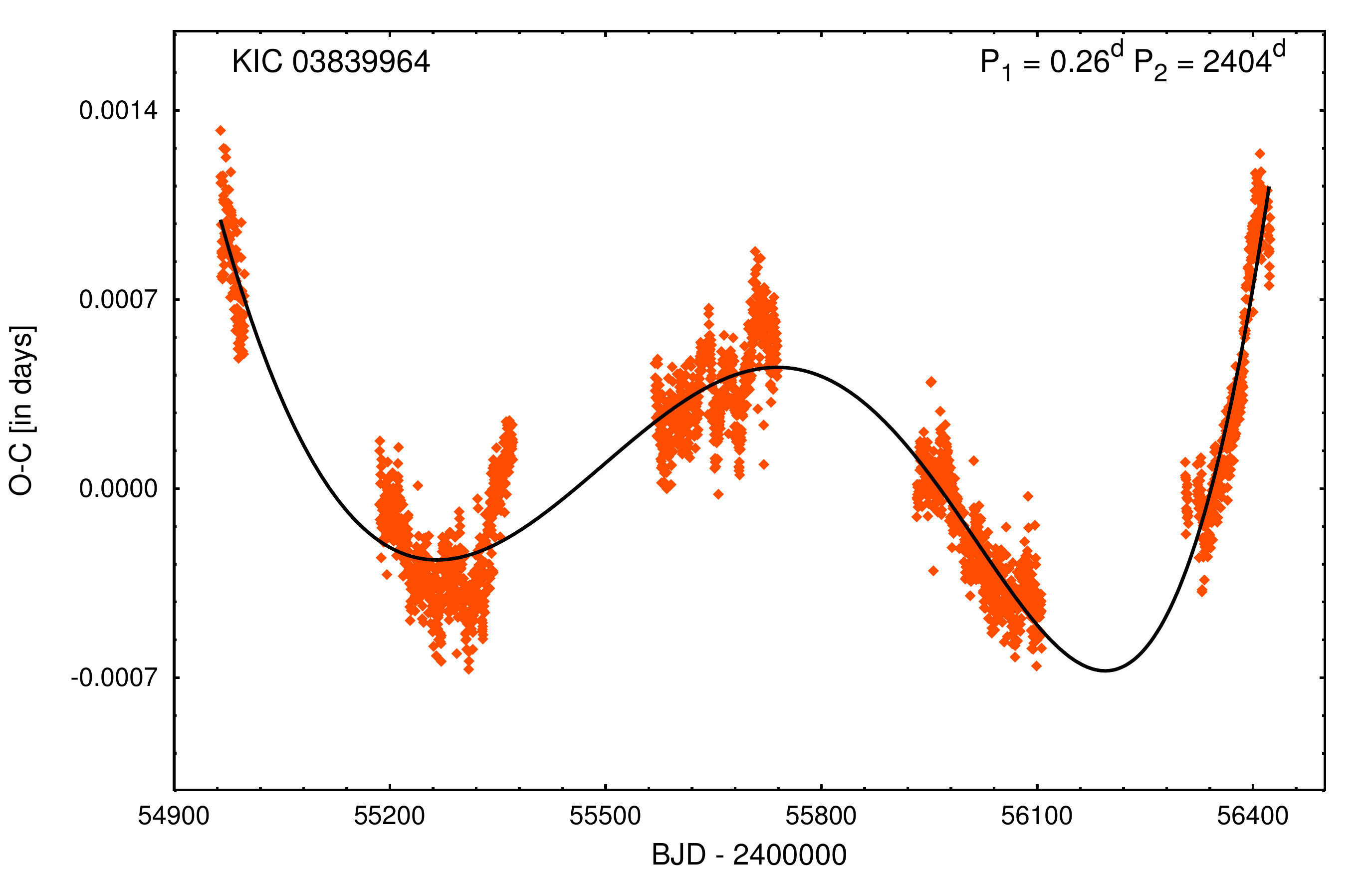}\includegraphics[width=60mm]{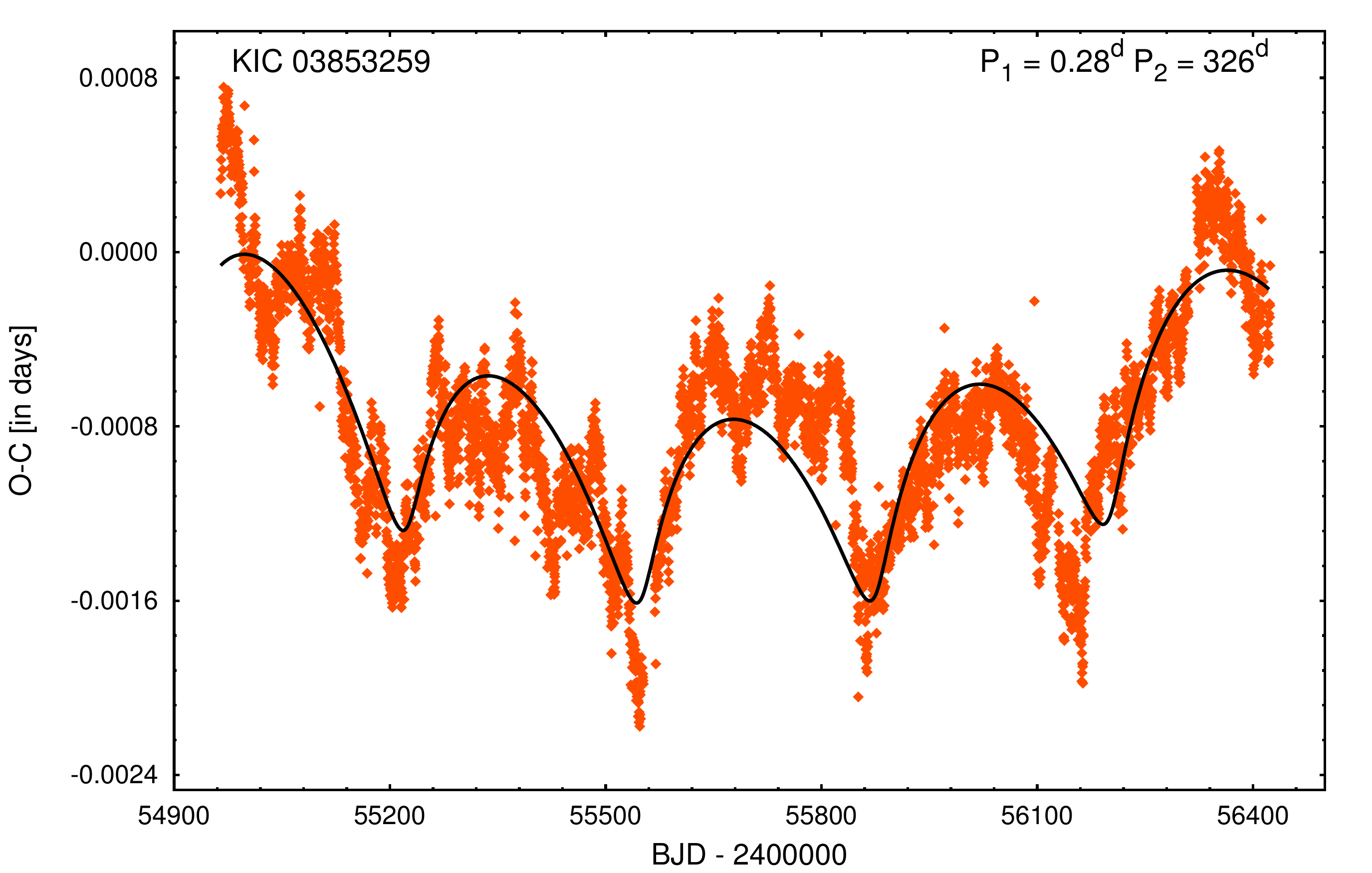}\includegraphics[width=60mm]{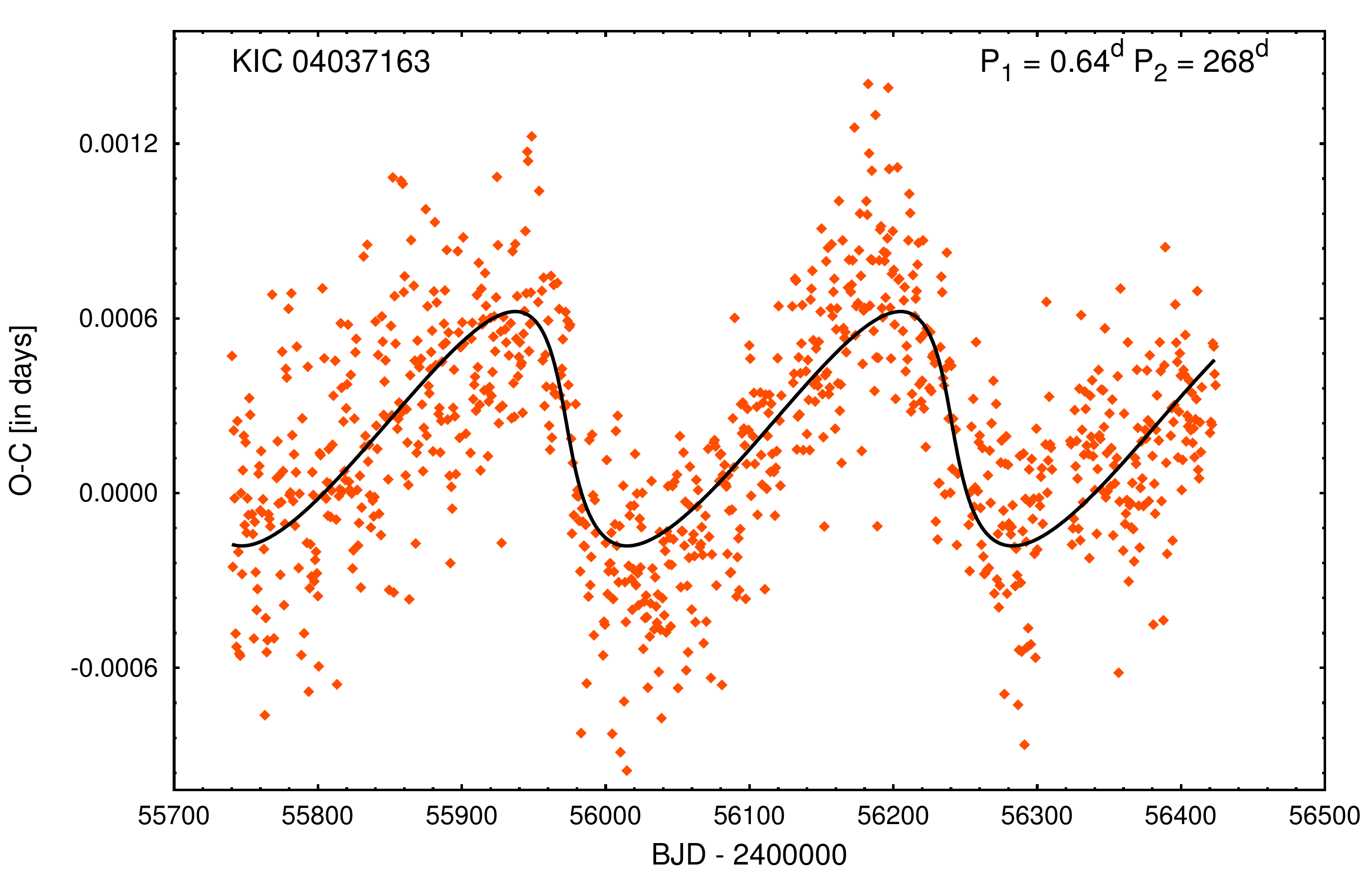}
\includegraphics[width=60mm]{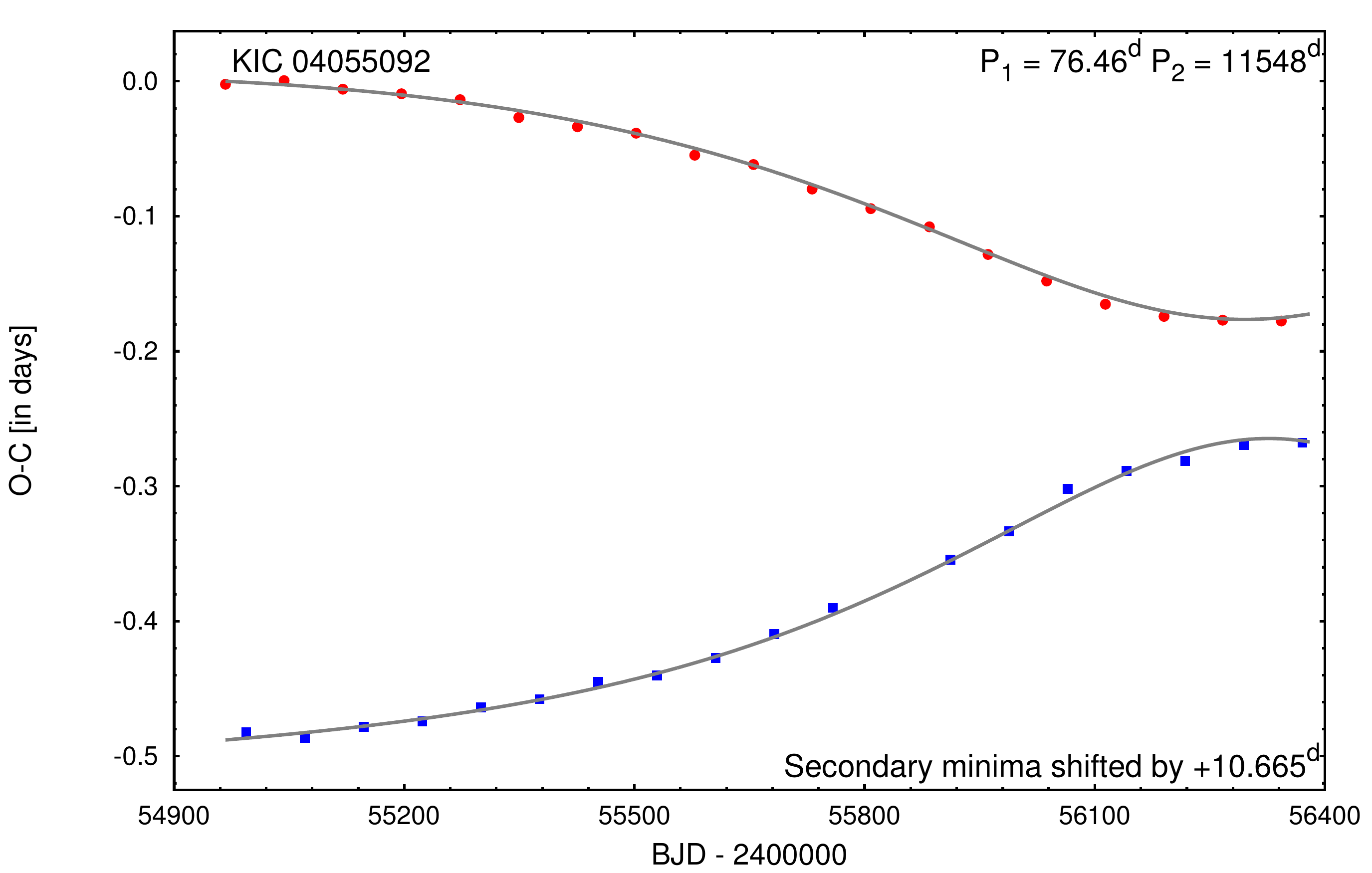}\includegraphics[width=60mm]{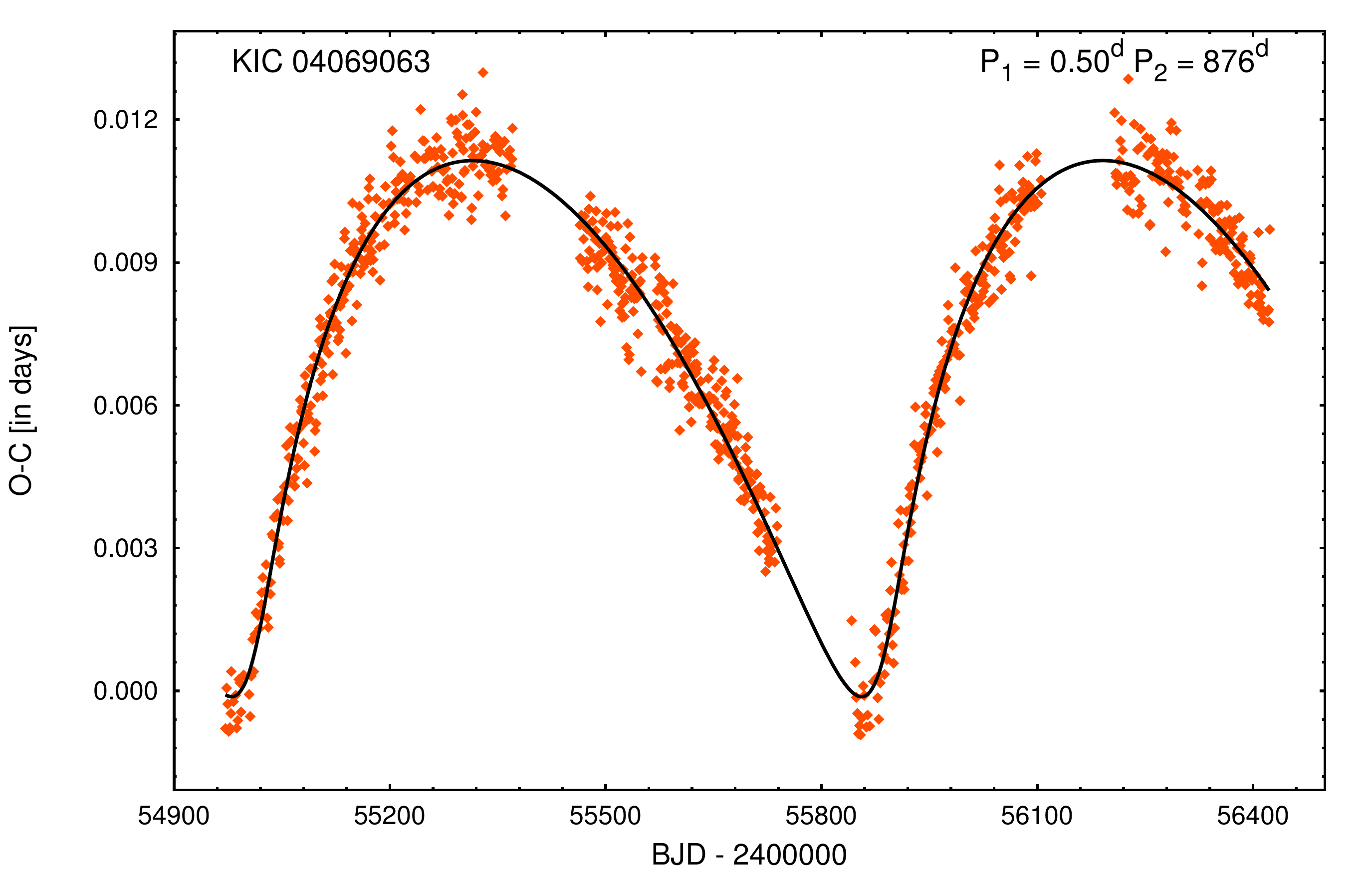}\includegraphics[width=60mm]{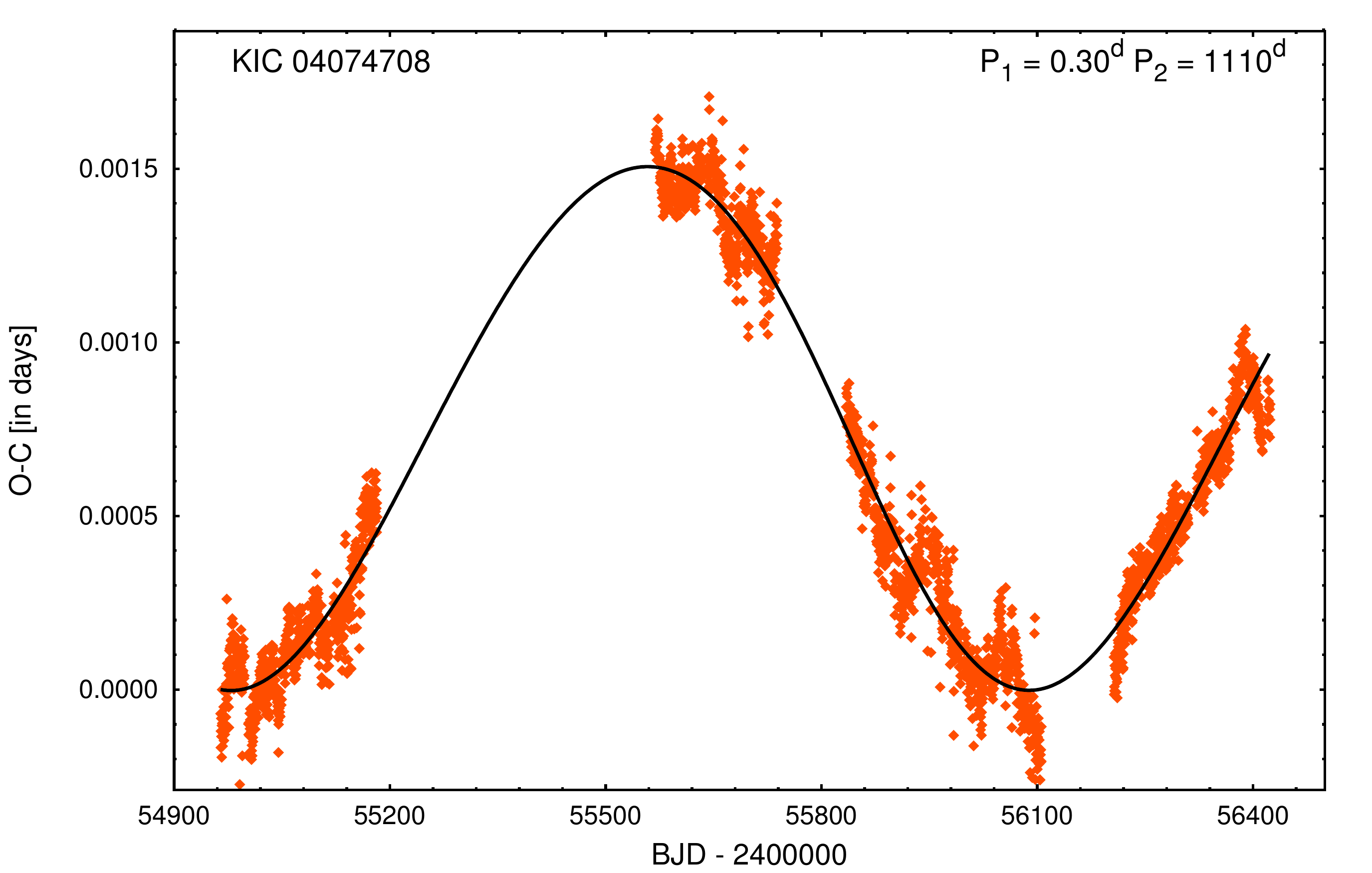}
\includegraphics[width=60mm]{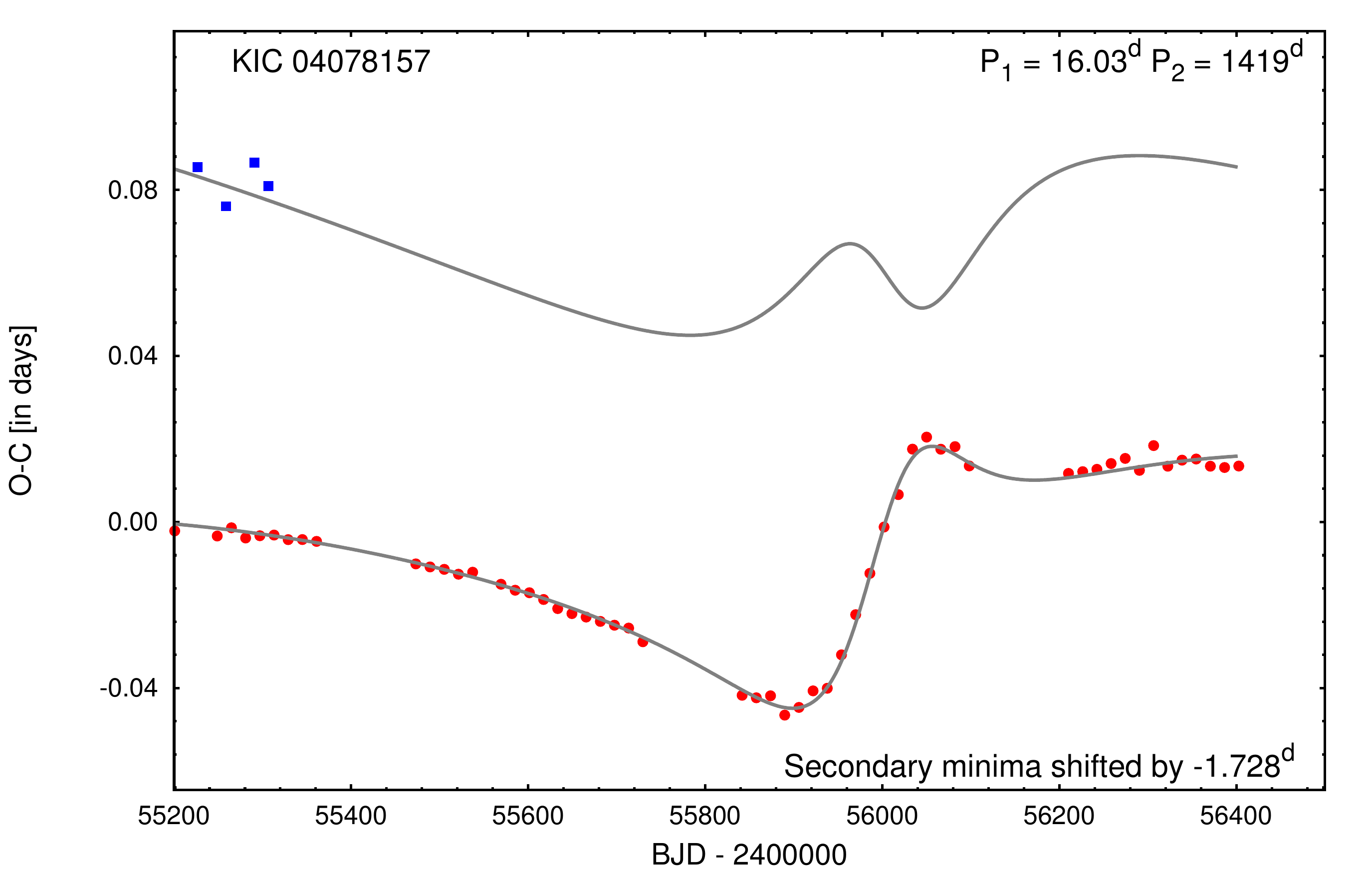}\includegraphics[width=60mm]{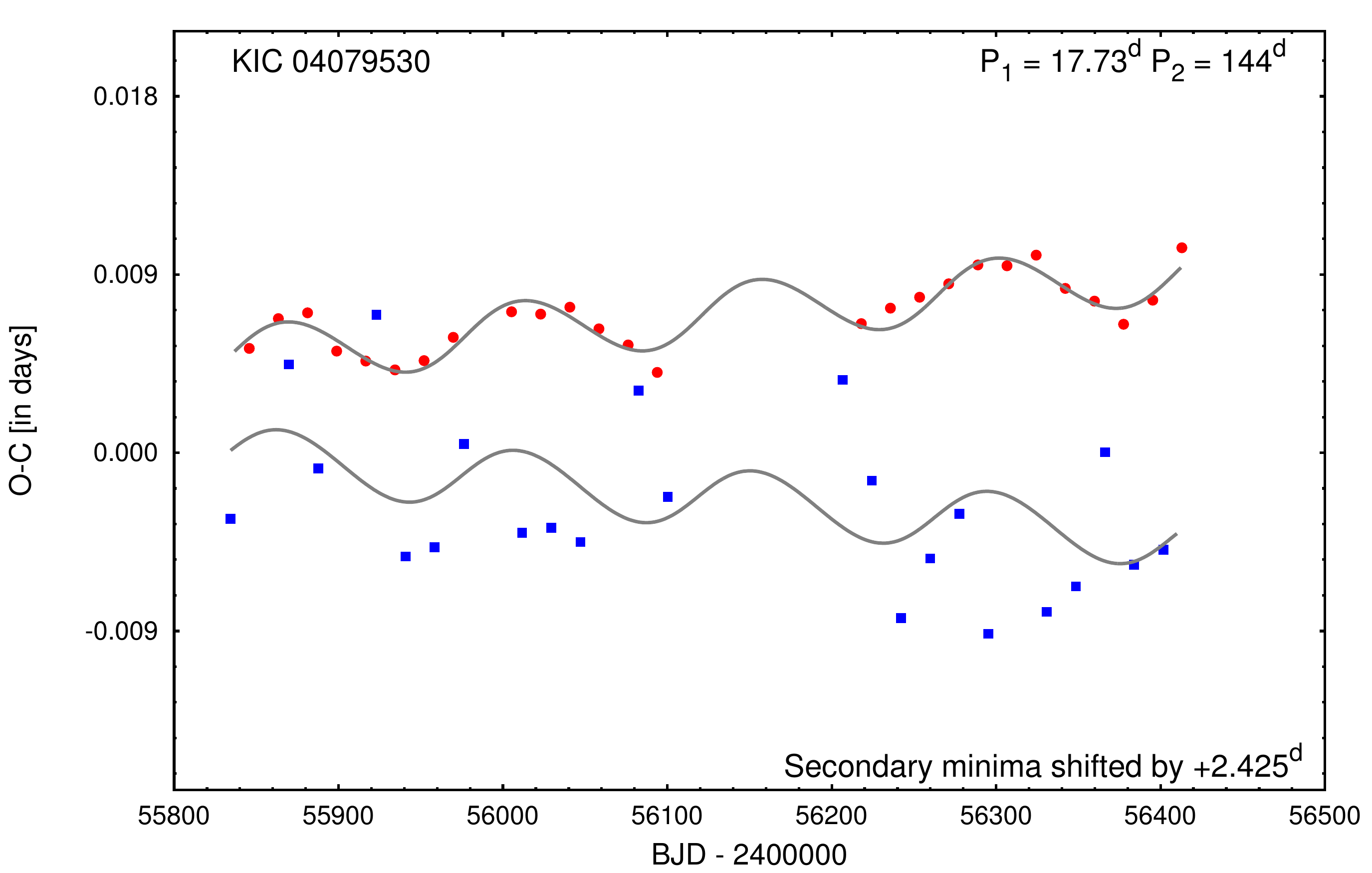}\includegraphics[width=60mm]{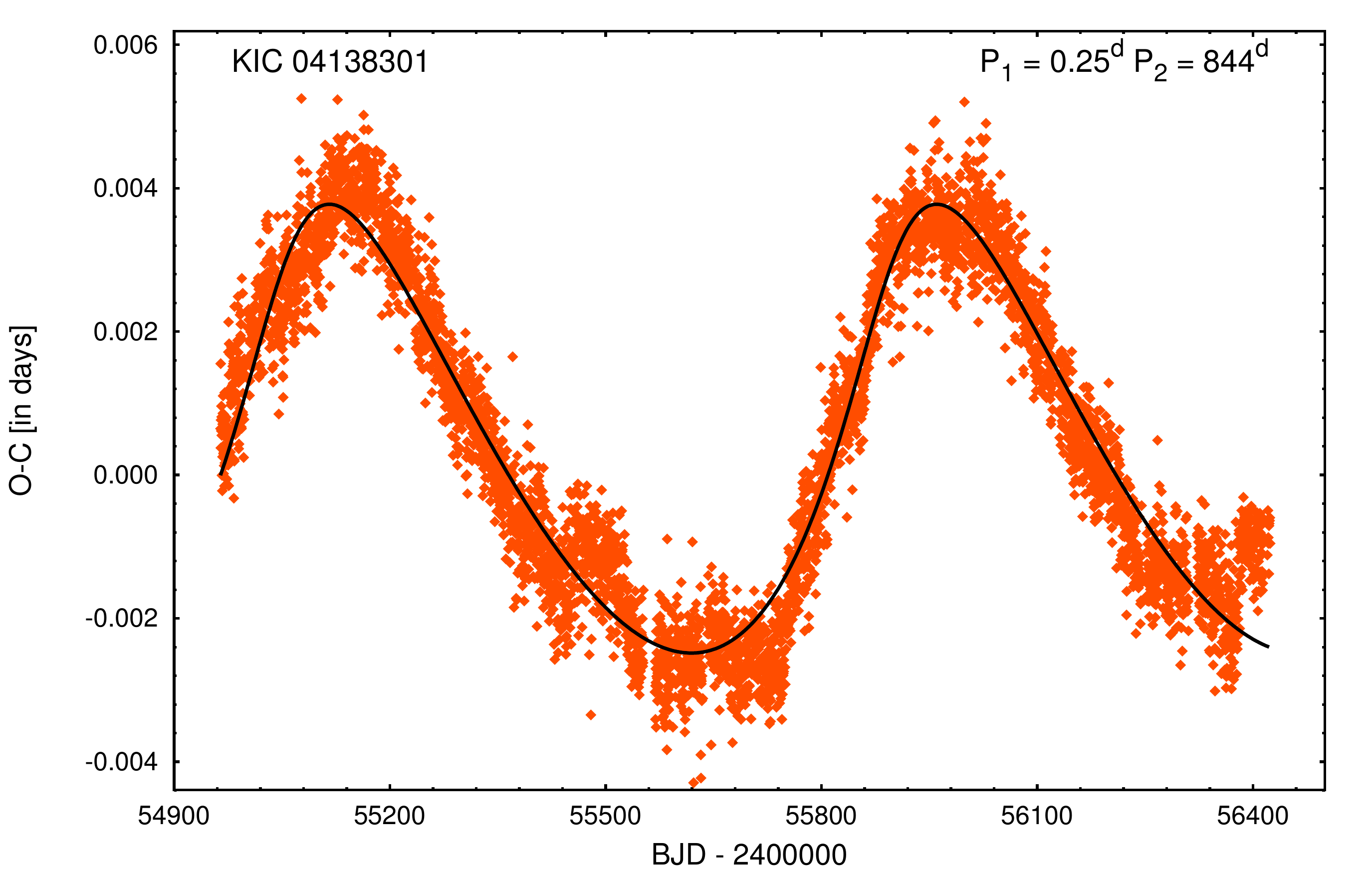}
\includegraphics[width=60mm]{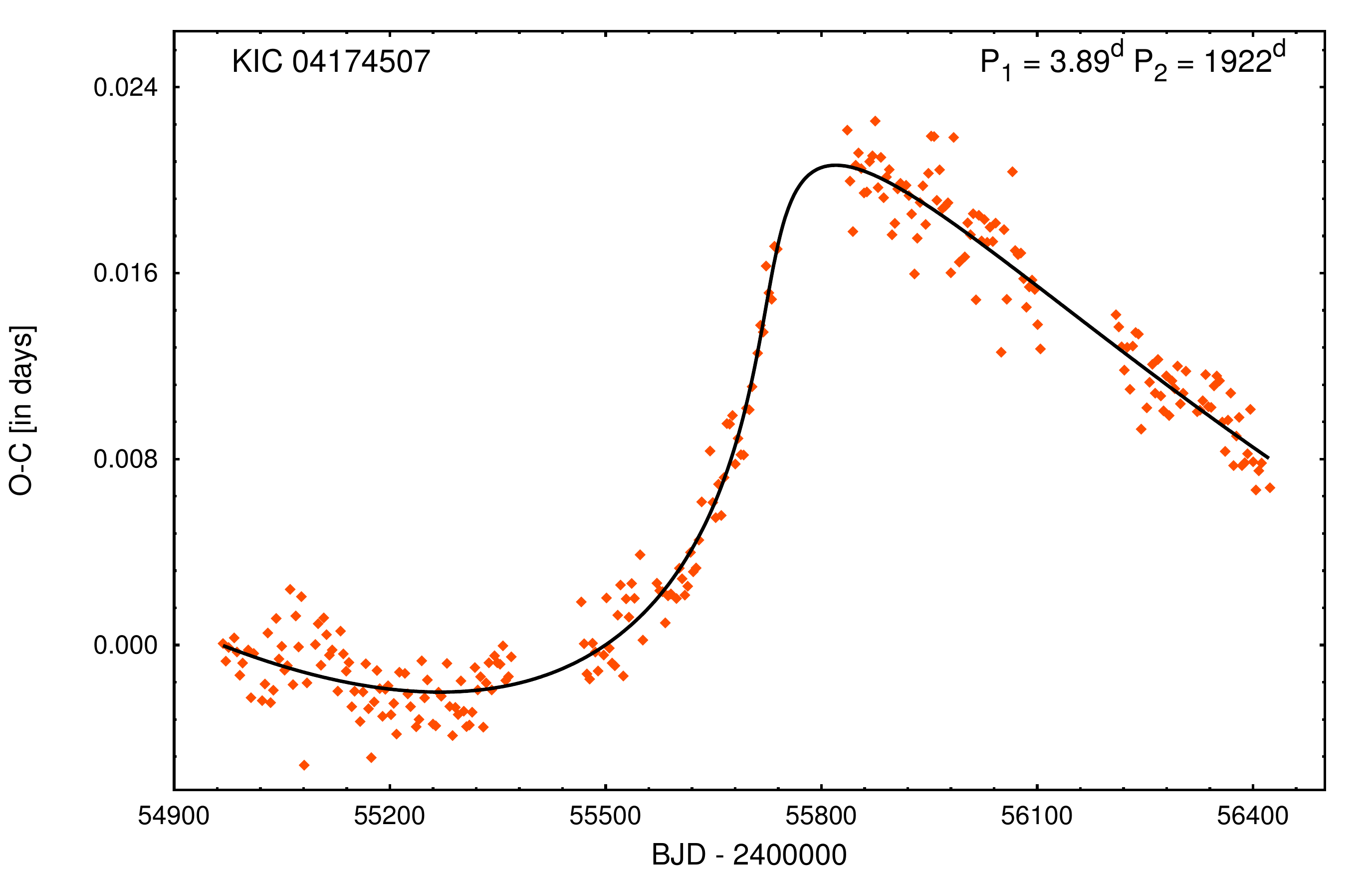}\includegraphics[width=60mm]{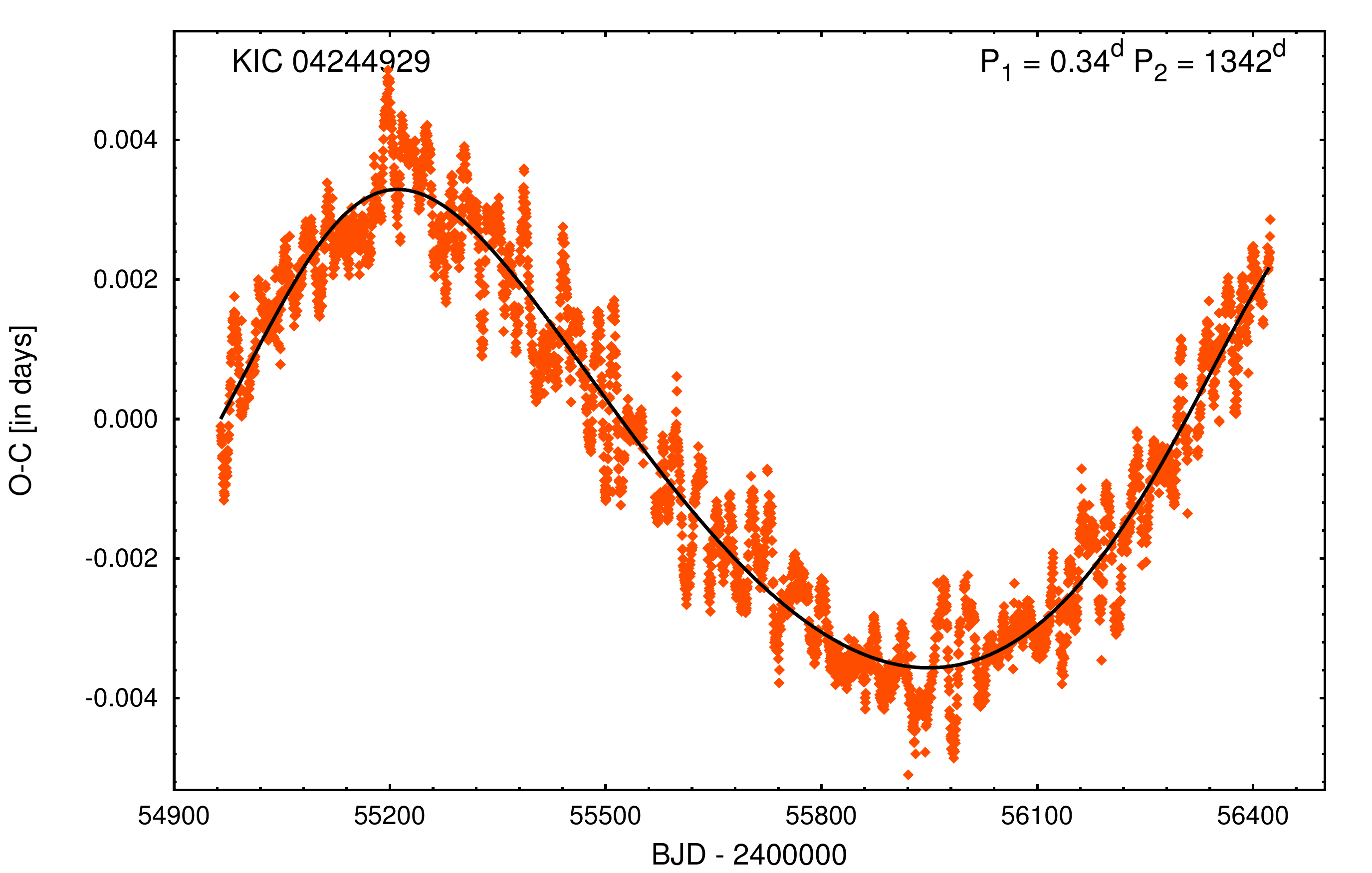}\includegraphics[width=60mm]{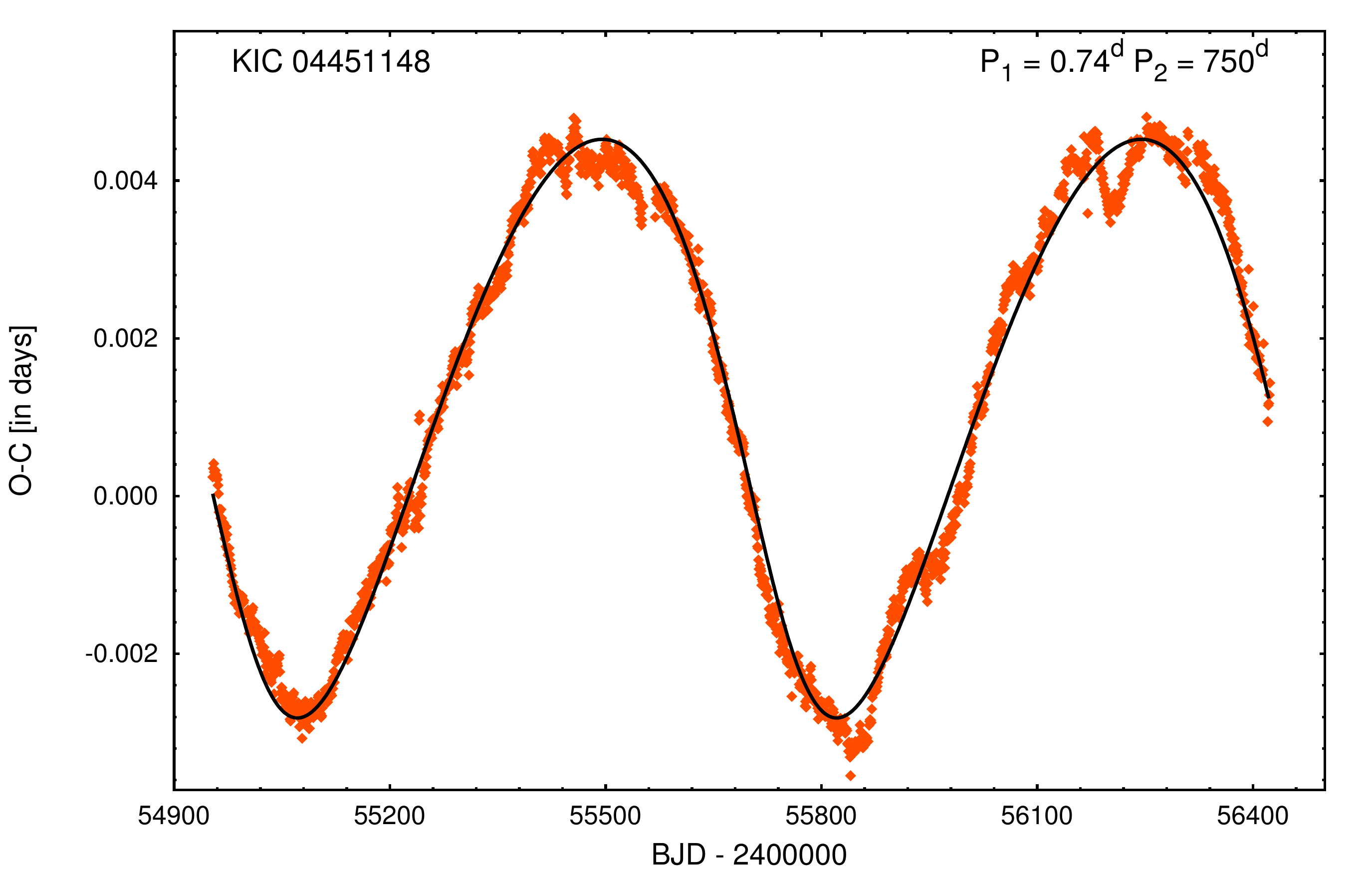}
\caption{(continued)}
\end{figure*}

\addtocounter{figure}{-1}

\begin{figure*}
\includegraphics[width=60mm]{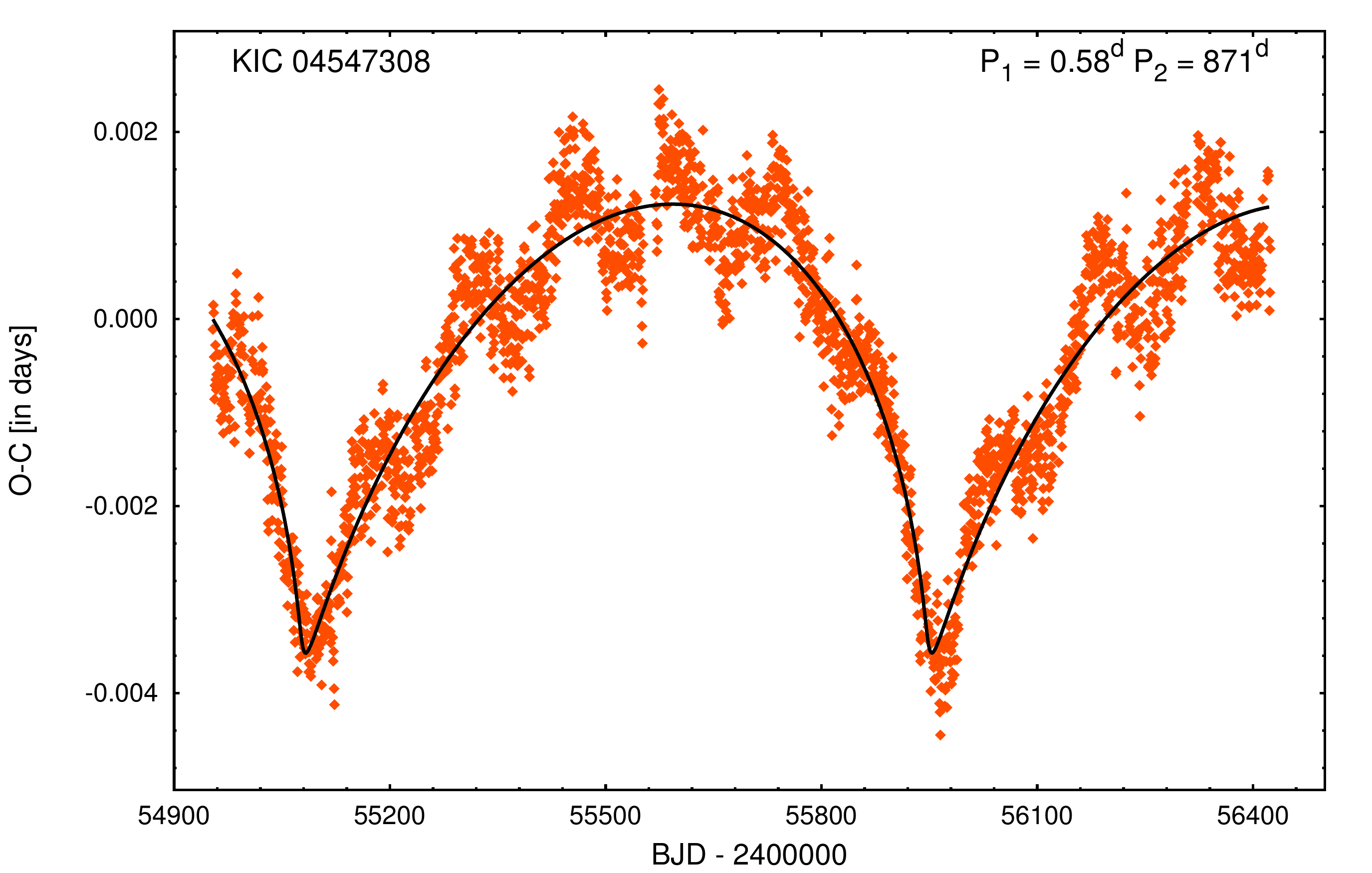}\includegraphics[width=60mm]{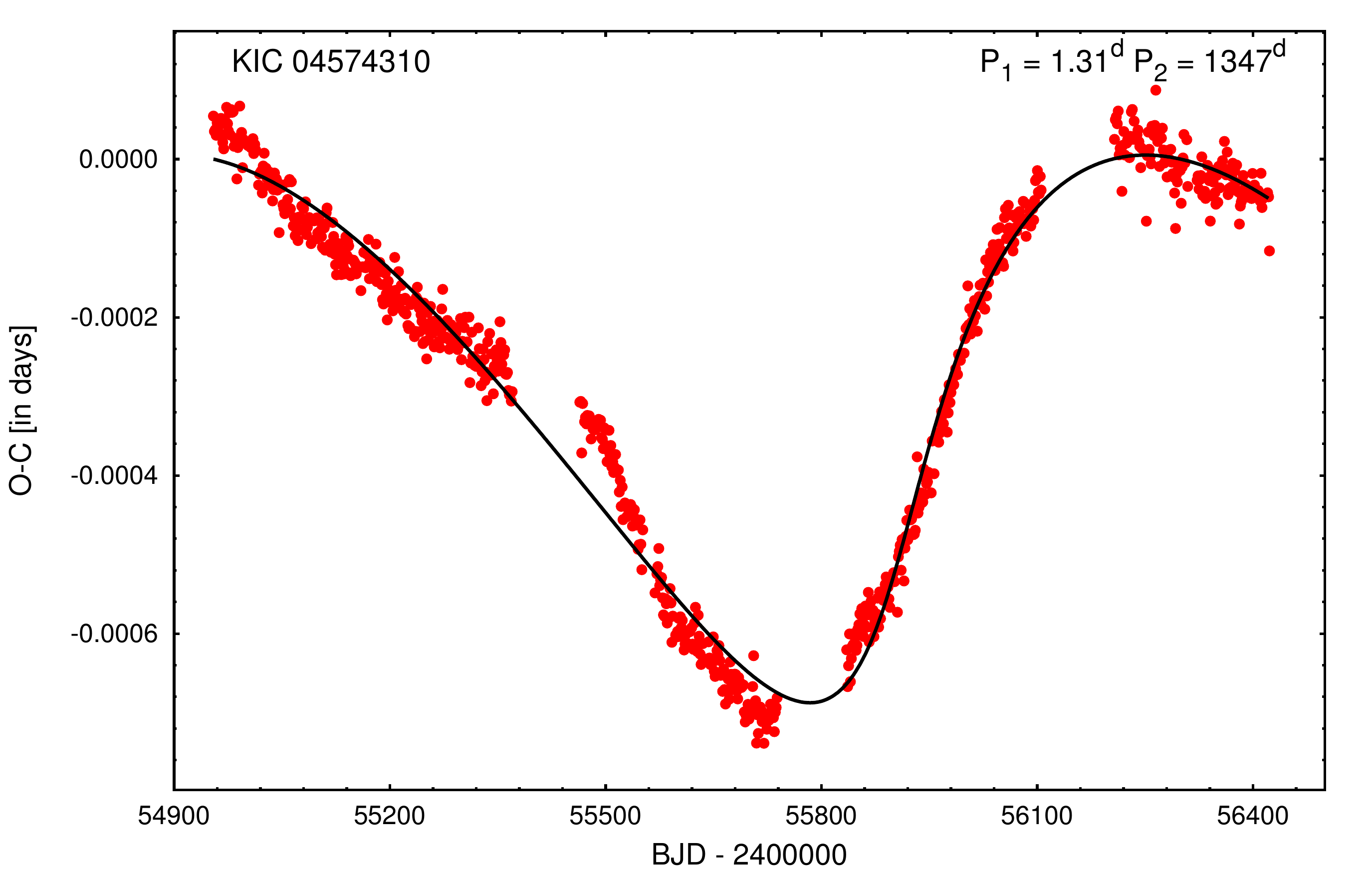}\includegraphics[width=60mm]{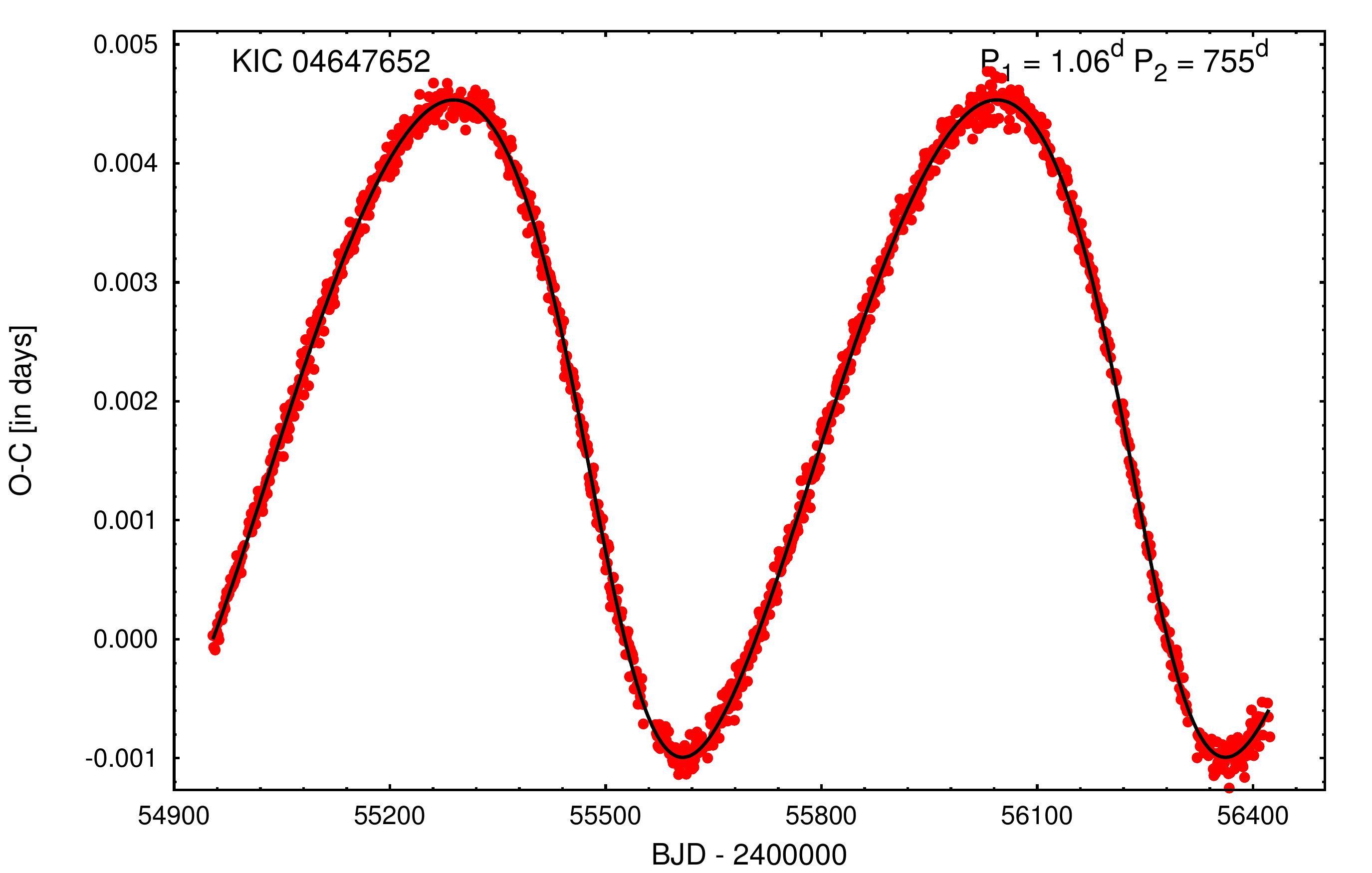}
\includegraphics[width=60mm]{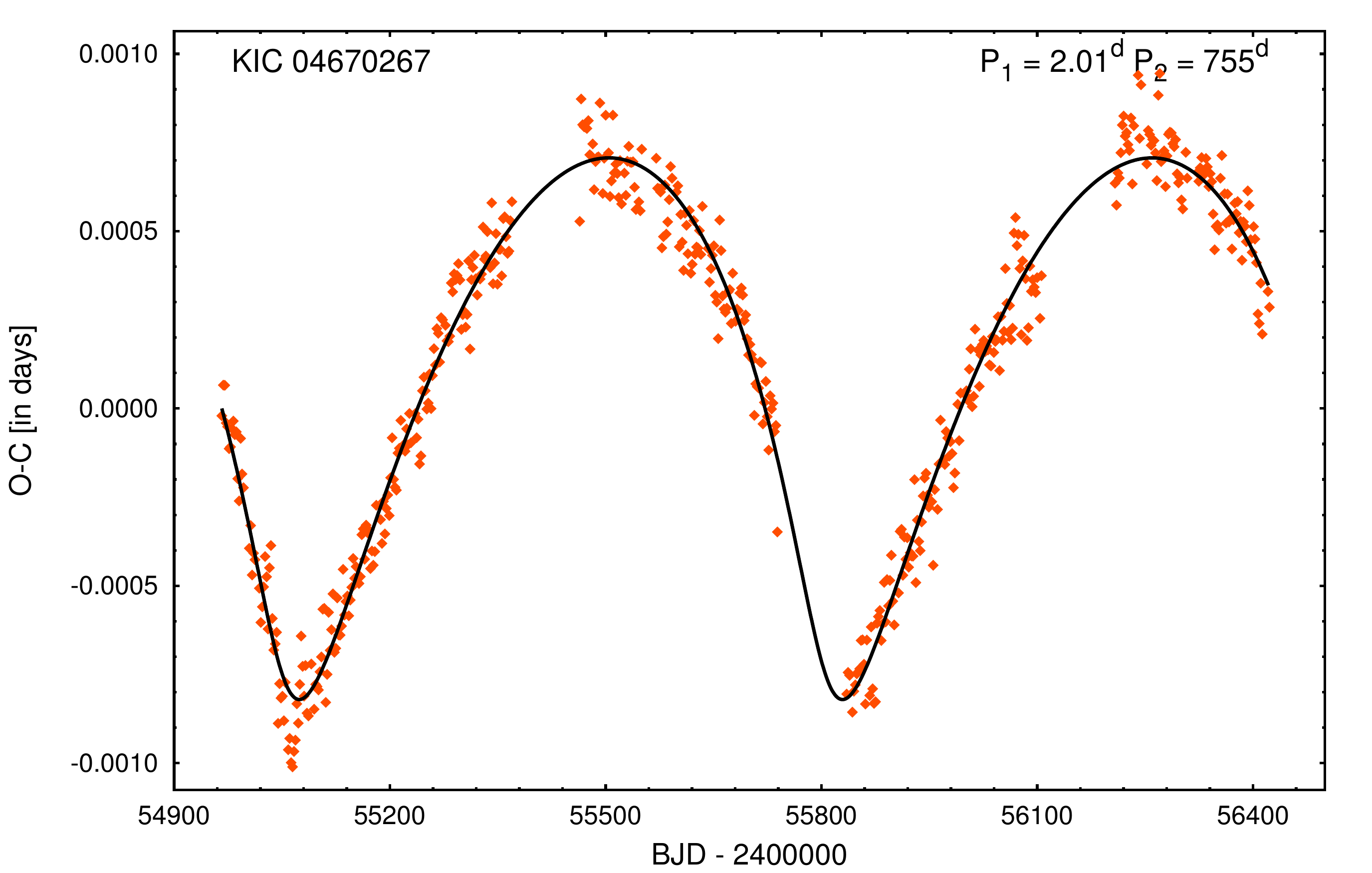}\includegraphics[width=60mm]{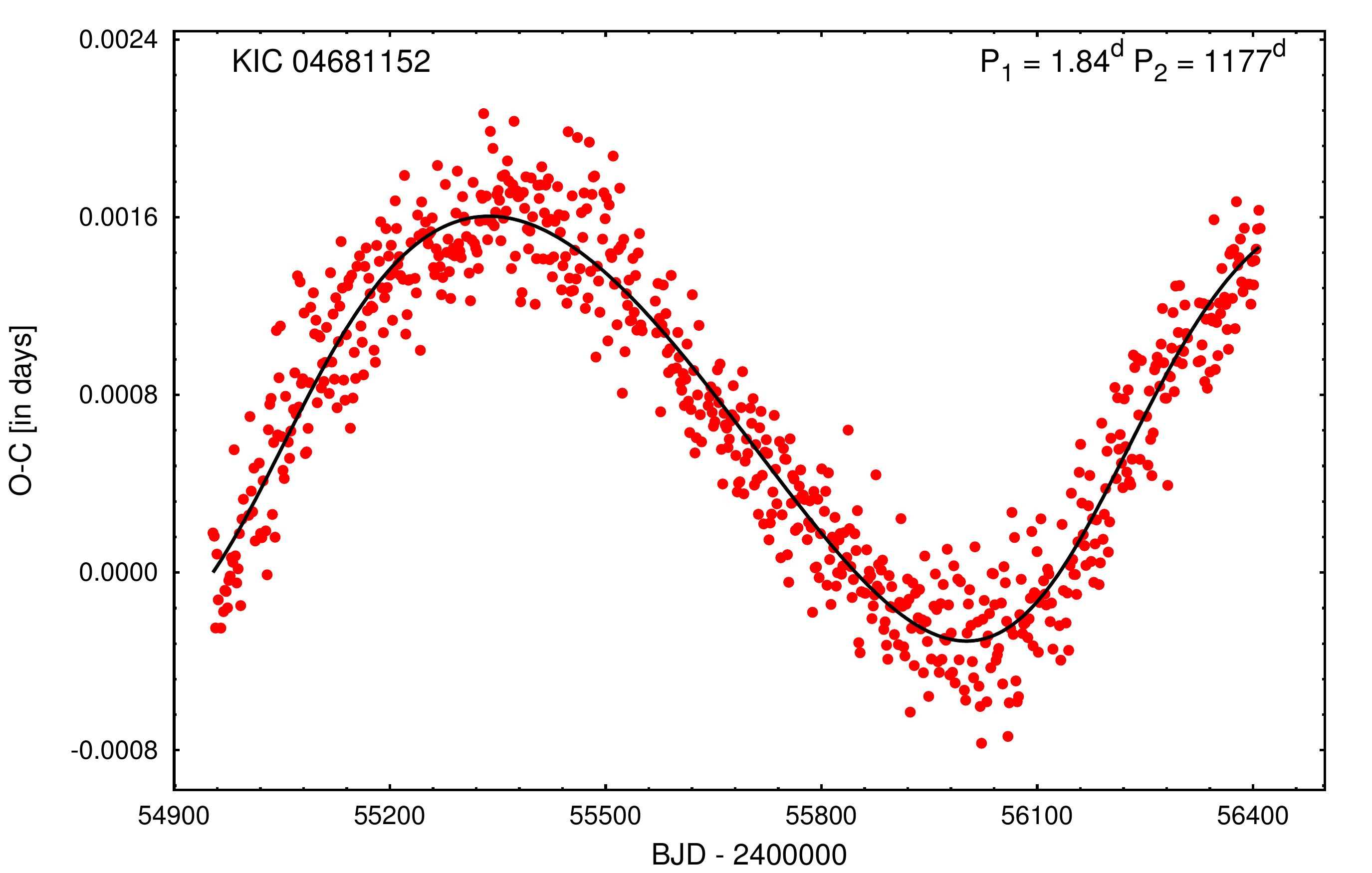}\includegraphics[width=60mm]{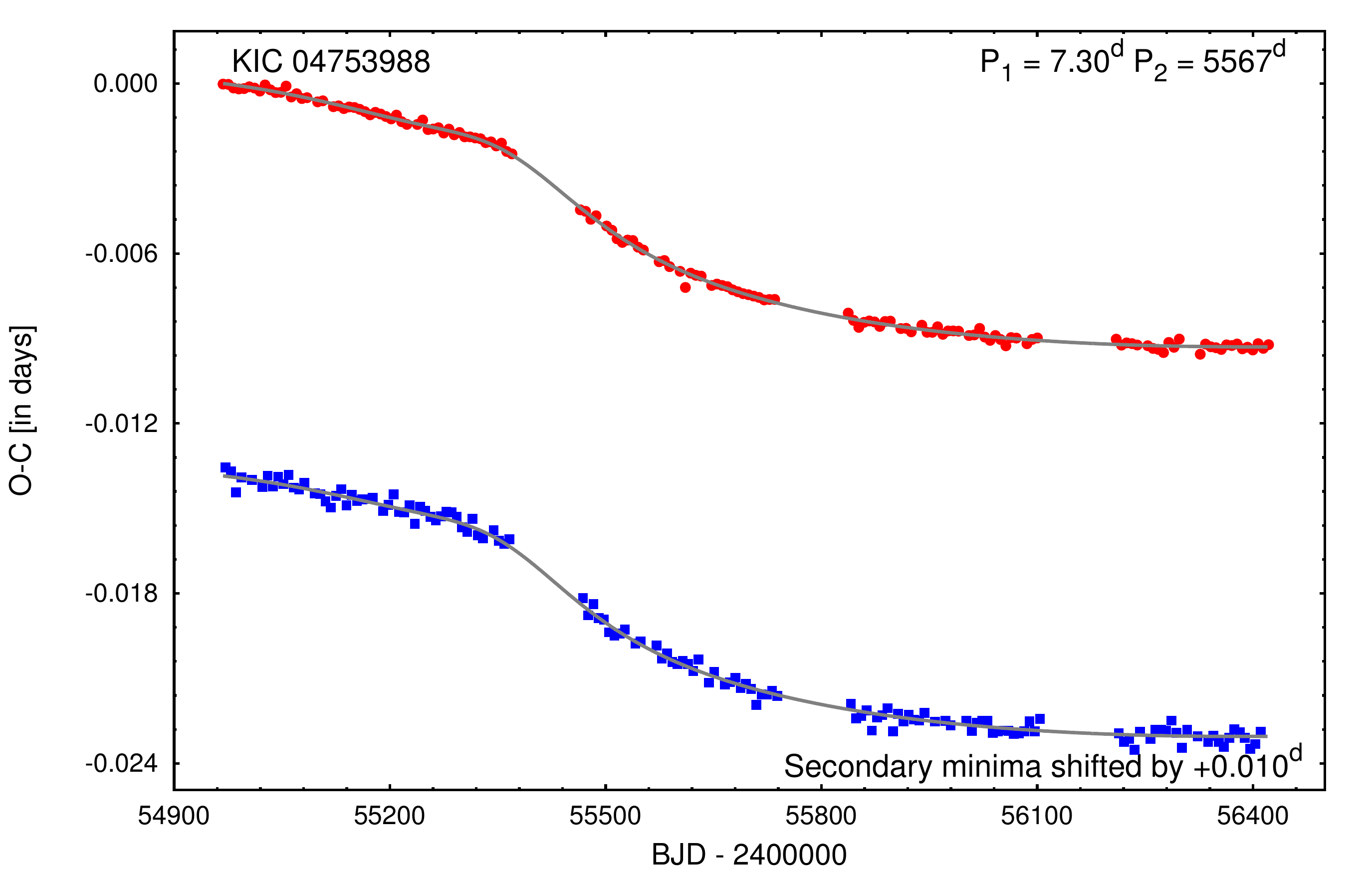}
\includegraphics[width=60mm]{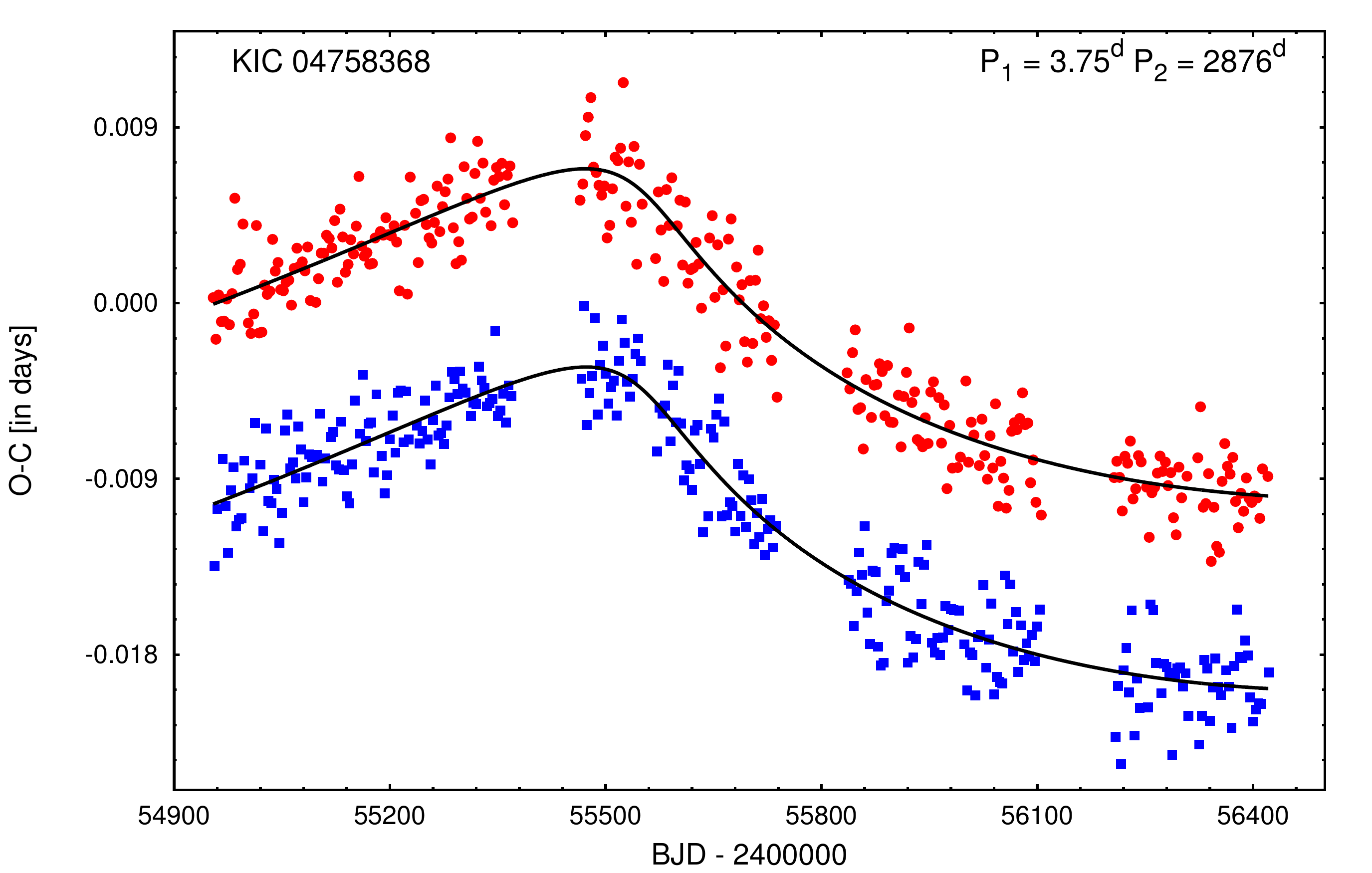}\includegraphics[width=60mm]{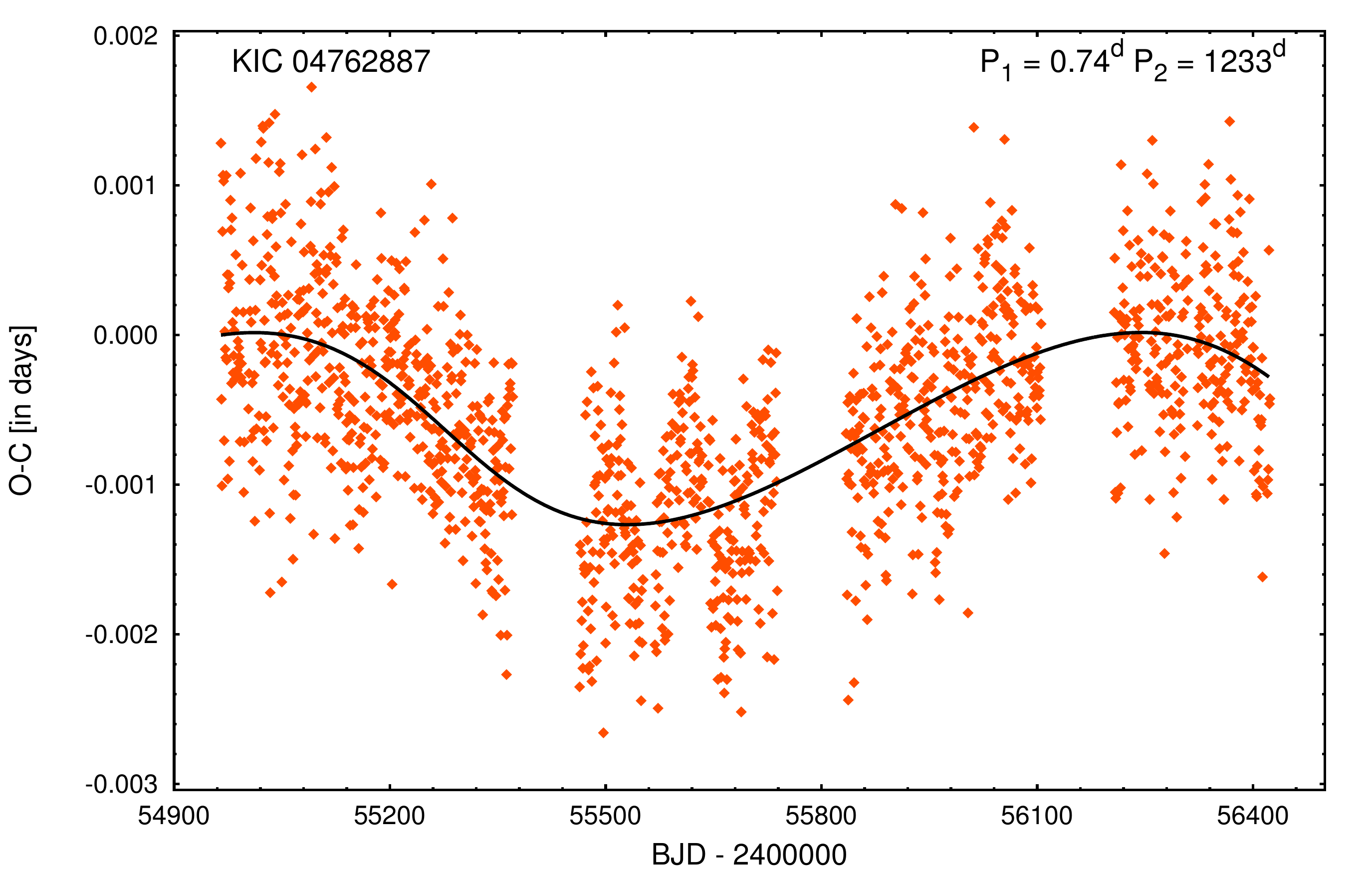}\includegraphics[width=60mm]{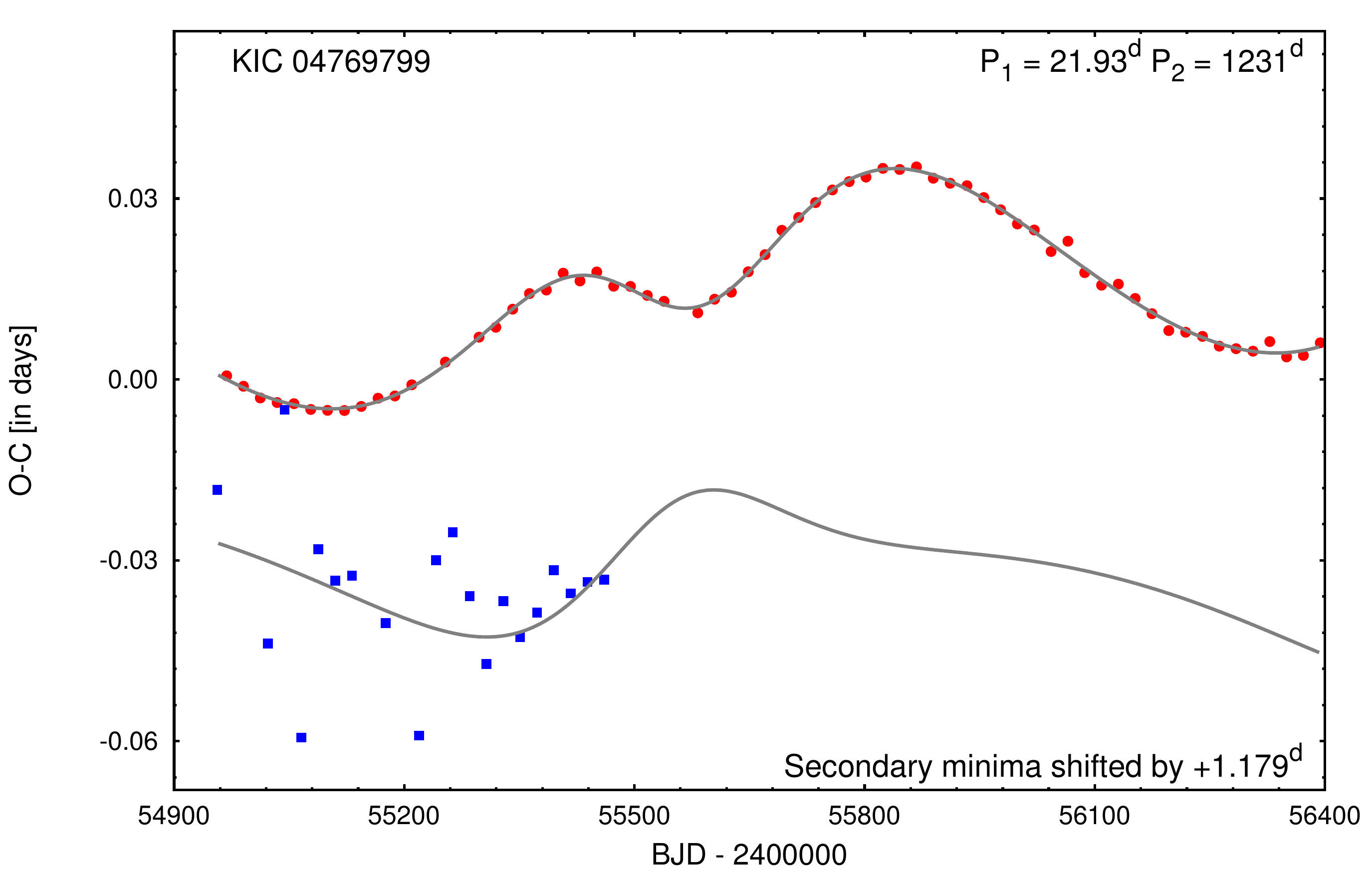}
\includegraphics[width=60mm]{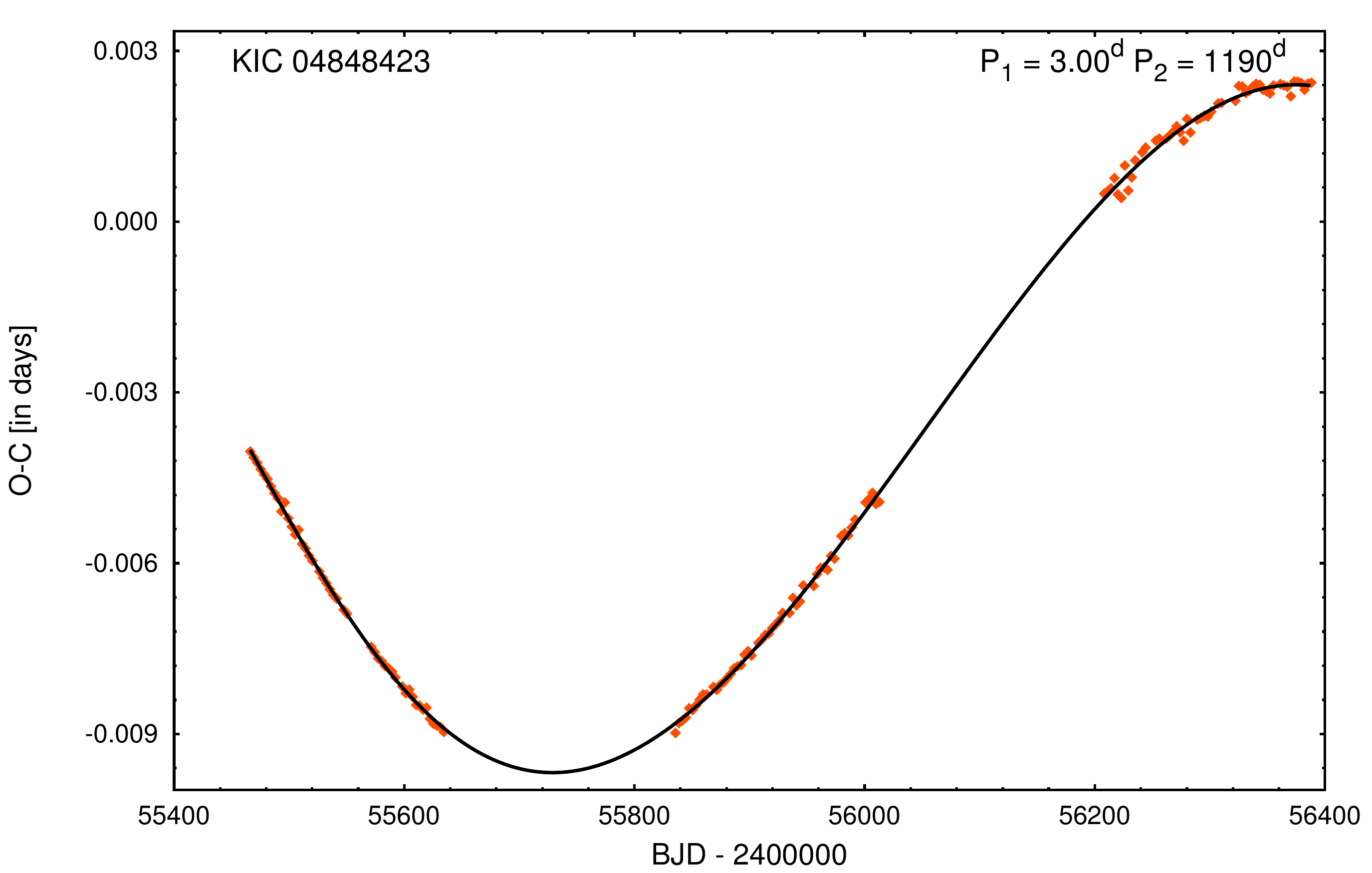}\includegraphics[width=60mm]{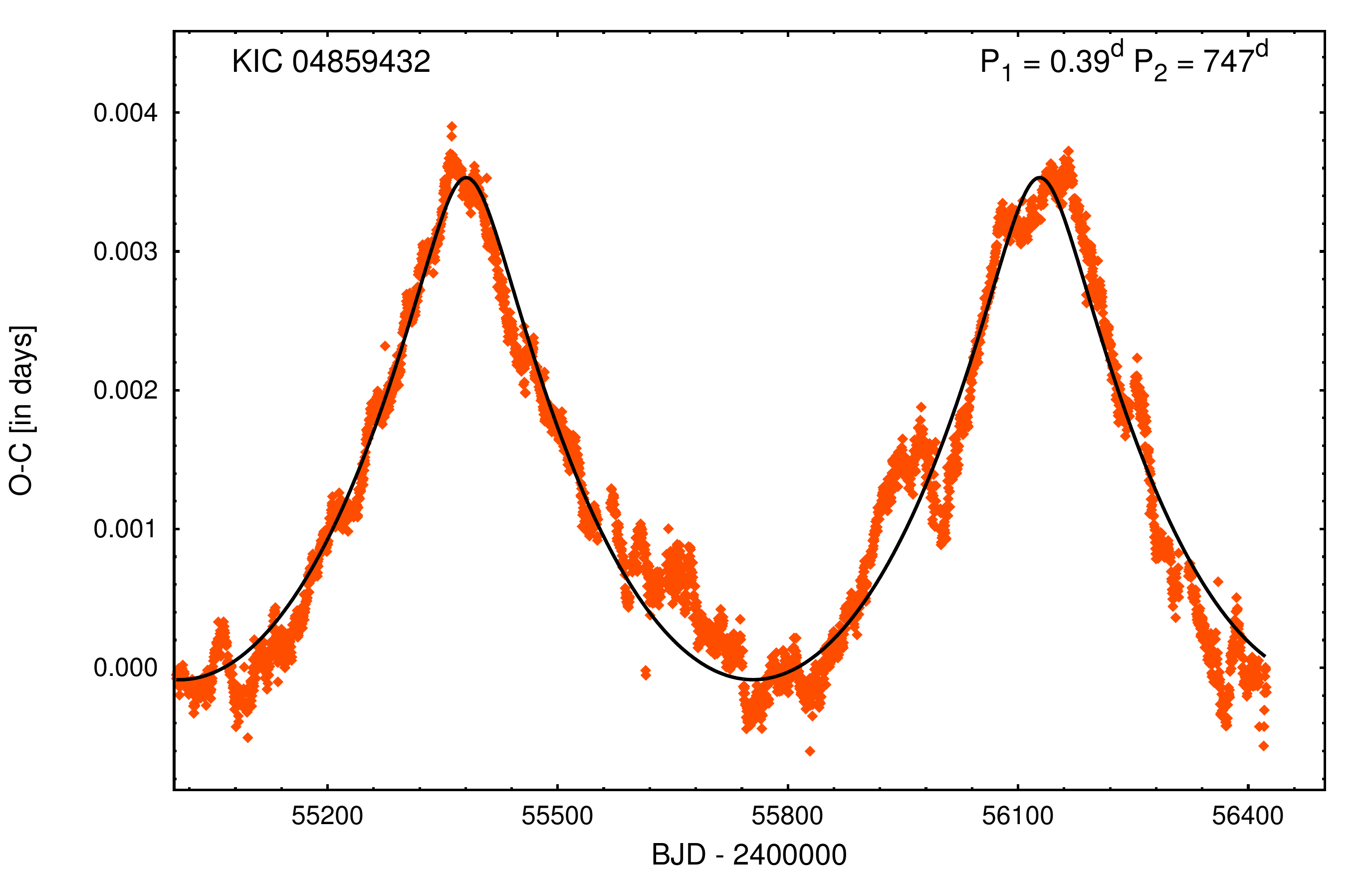}\includegraphics[width=60mm]{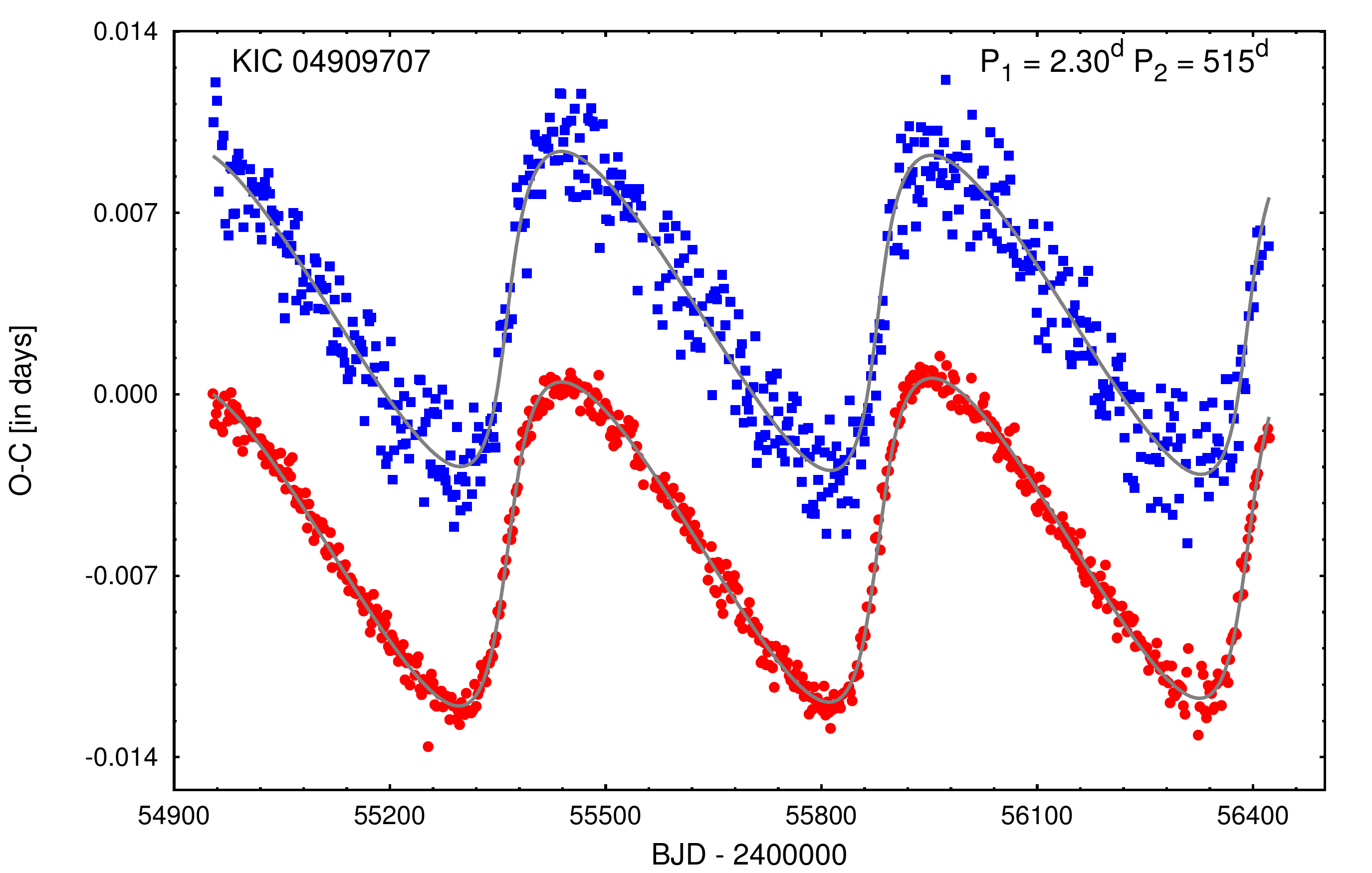}
\includegraphics[width=60mm]{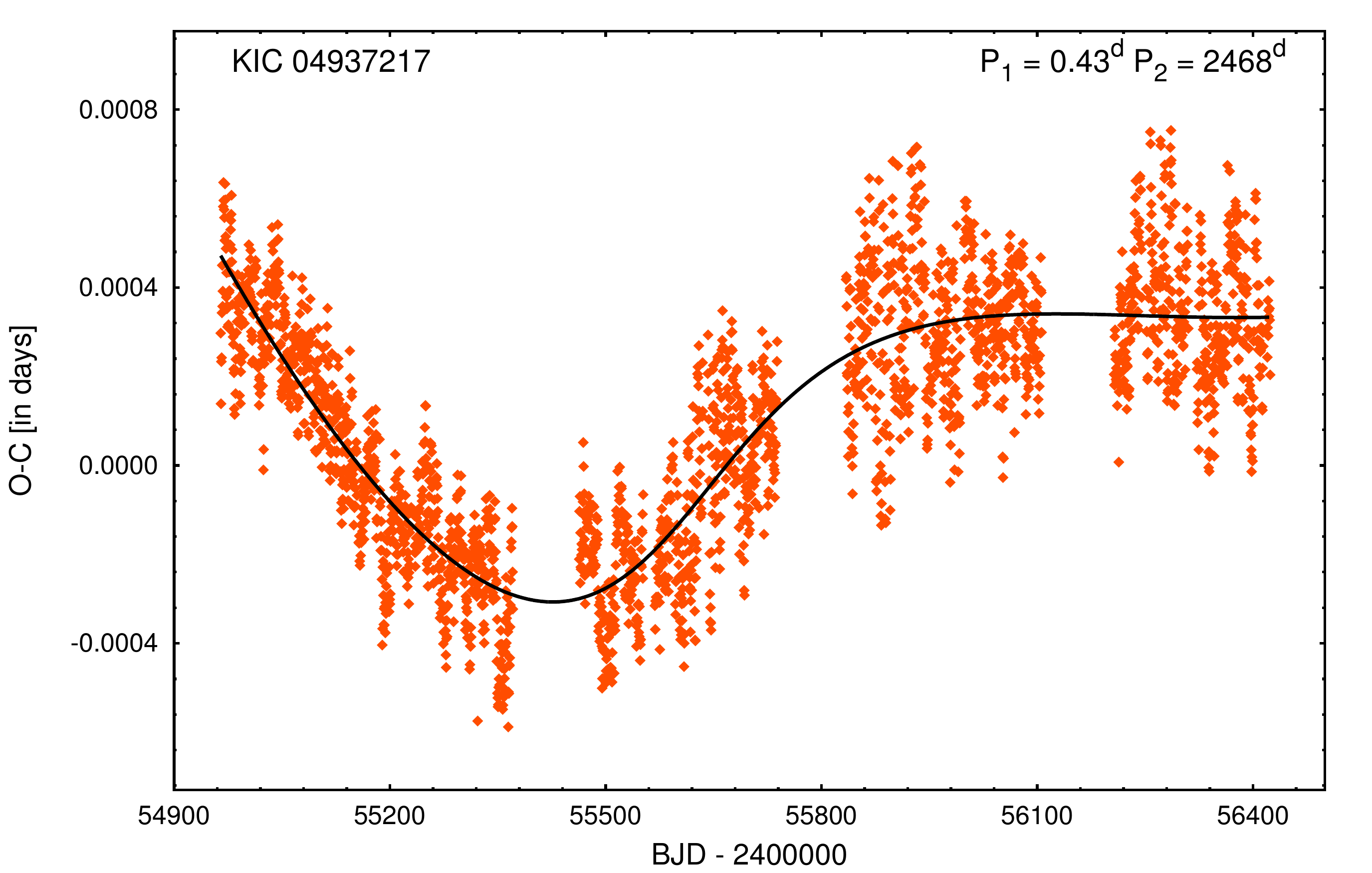}\includegraphics[width=60mm]{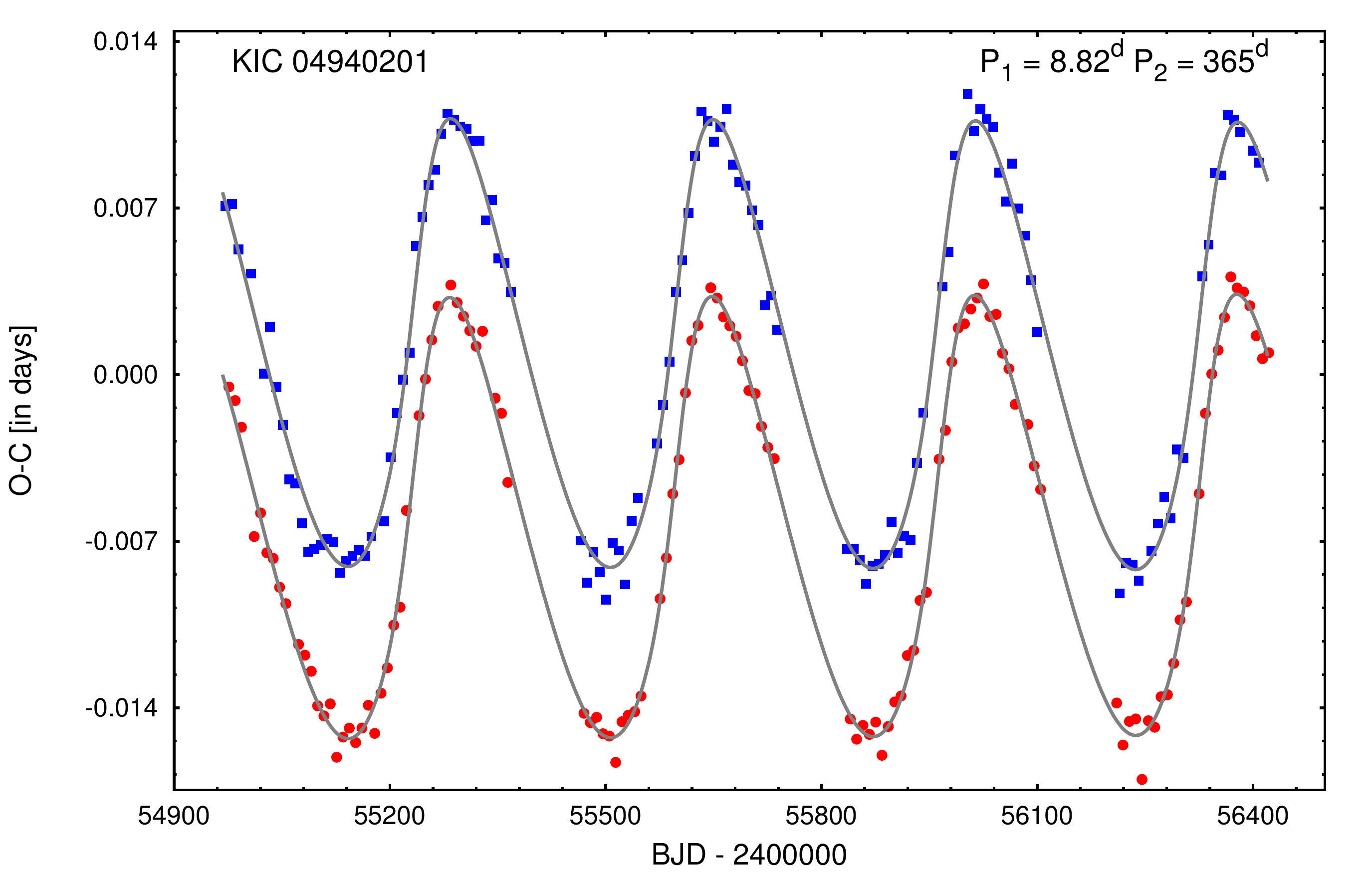}\includegraphics[width=60mm]{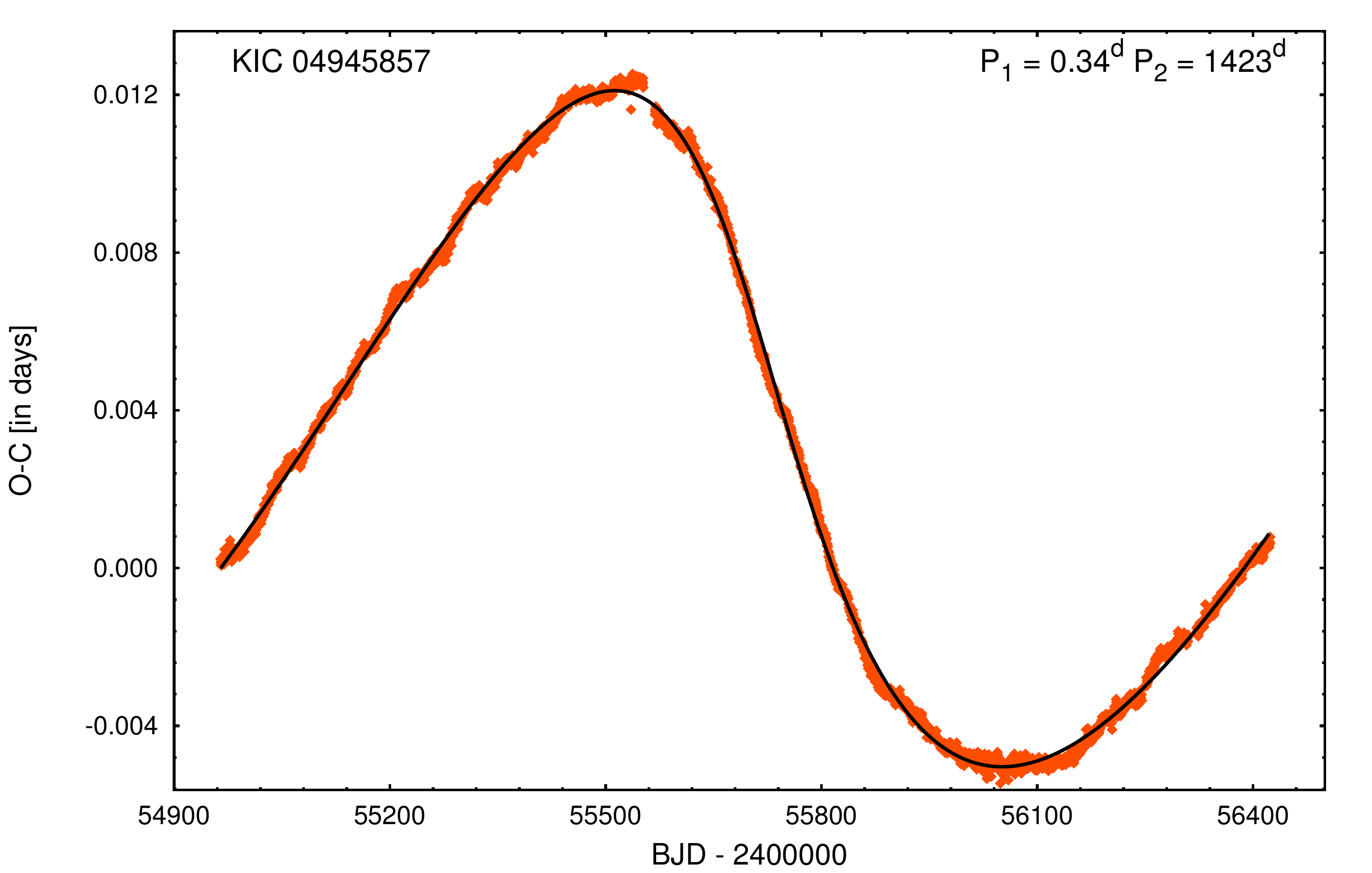}
\includegraphics[width=60mm]{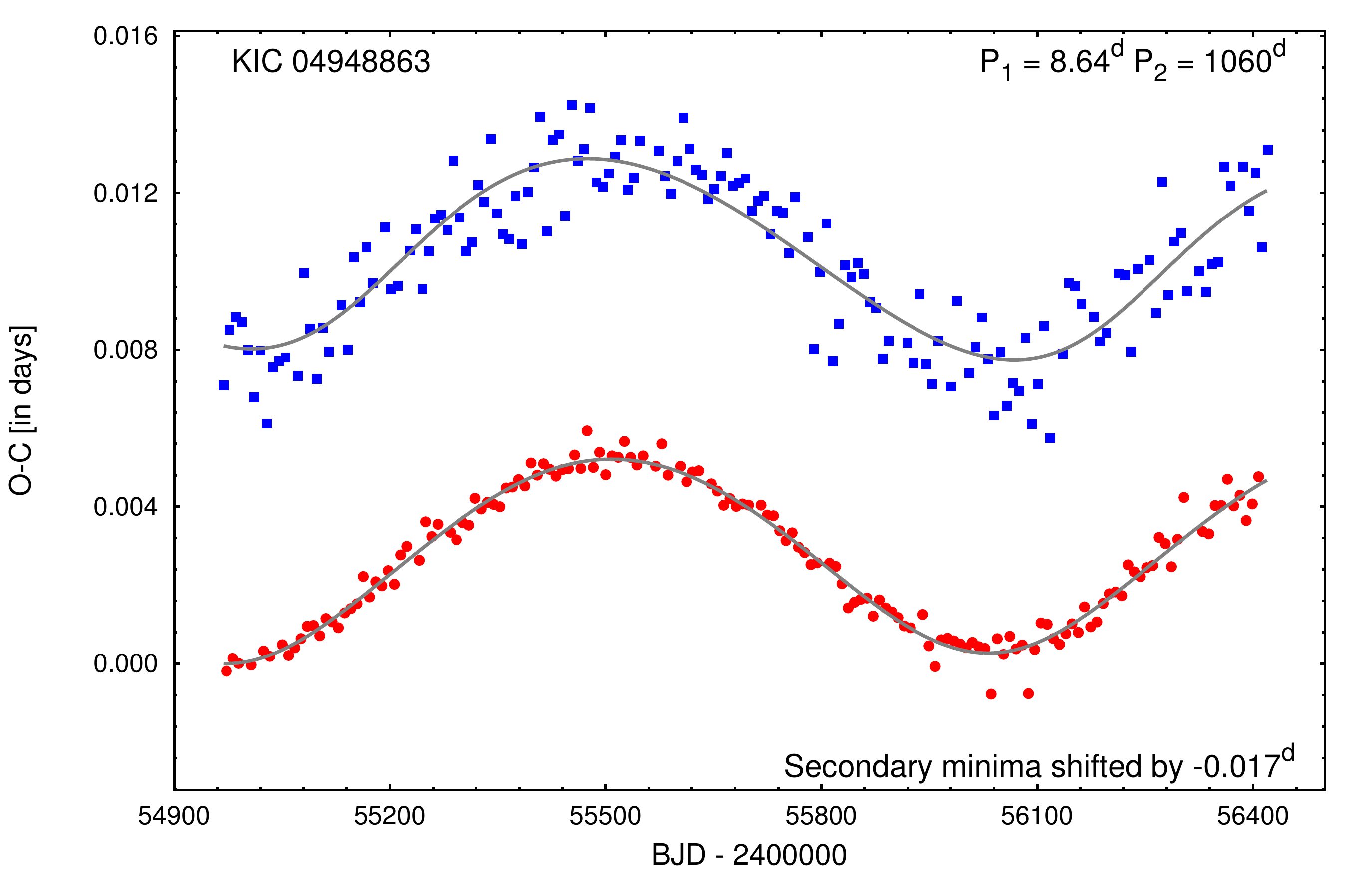}\includegraphics[width=60mm]{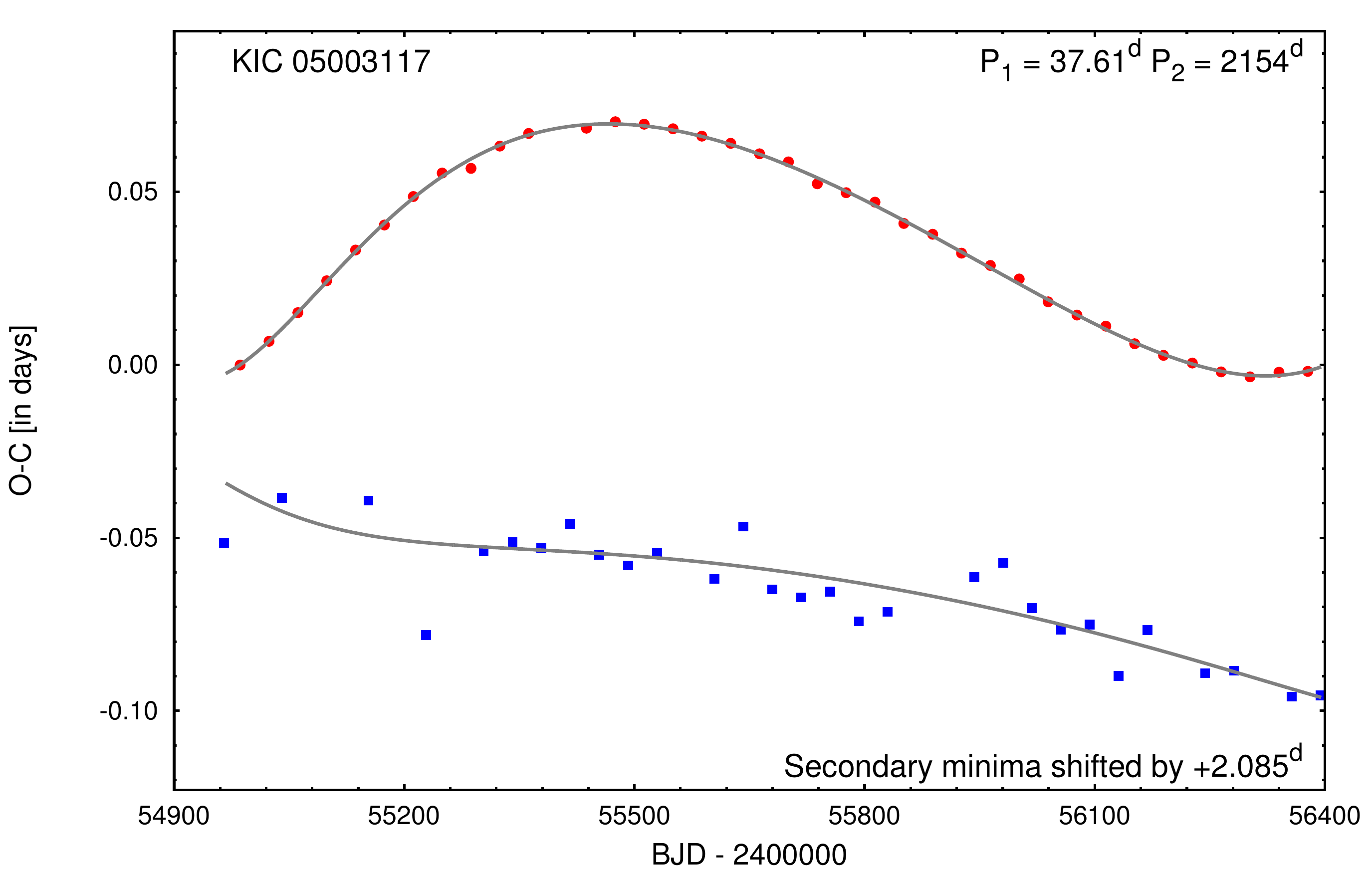}\includegraphics[width=60mm]{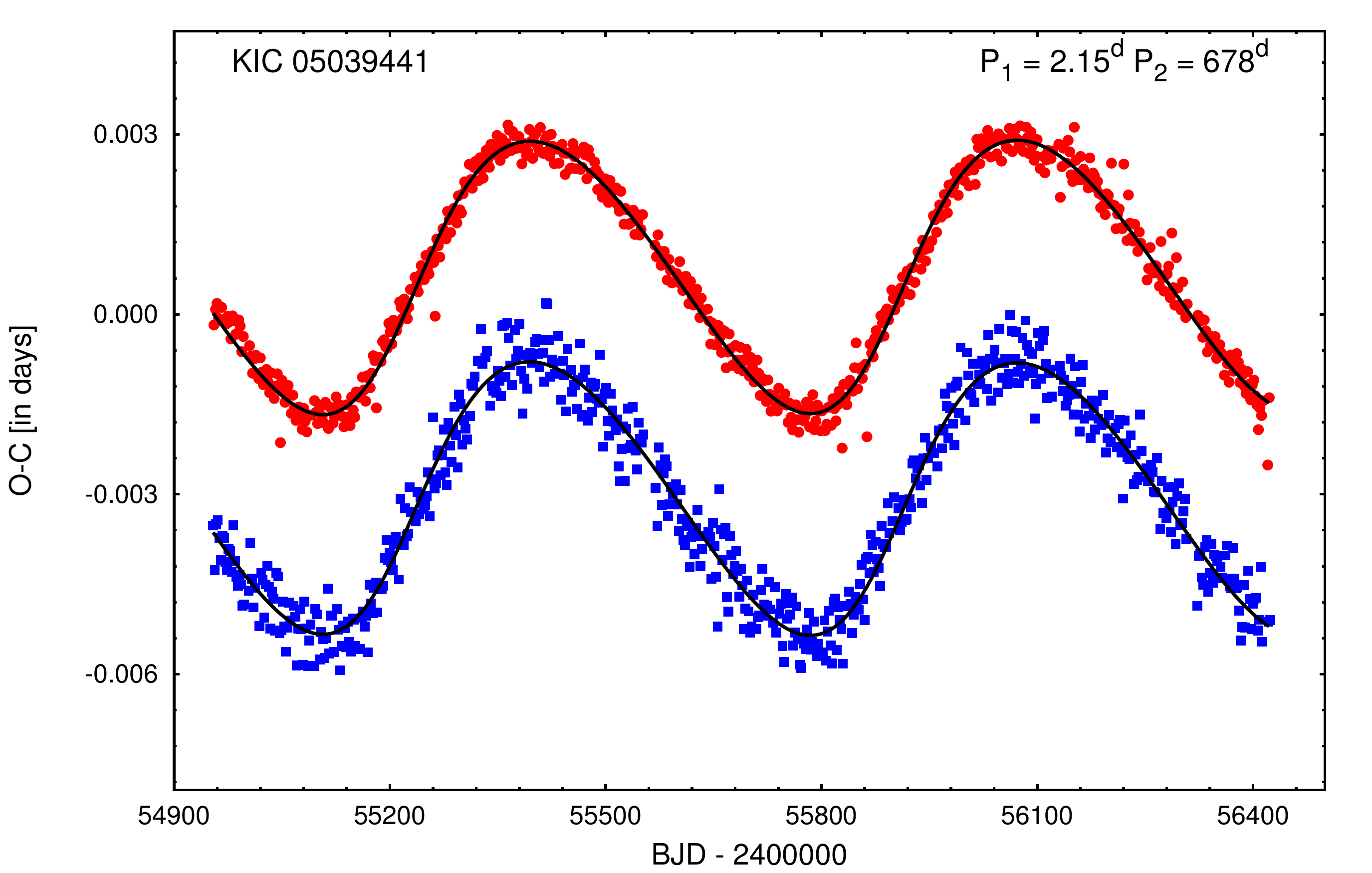}
\caption{(continued)}
\end{figure*}

\addtocounter{figure}{-1}

\begin{figure*}
\includegraphics[width=60mm]{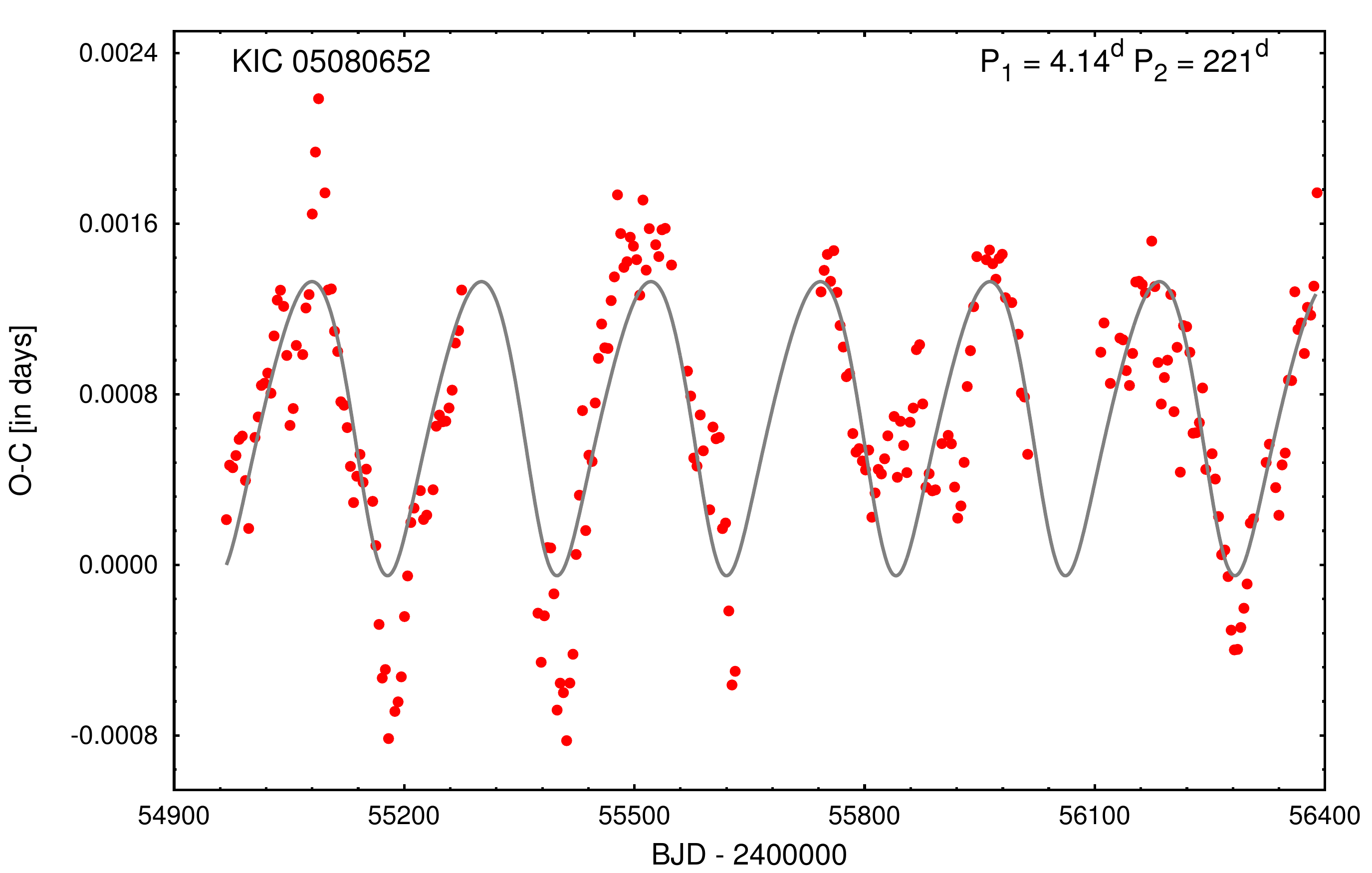}\includegraphics[width=60mm]{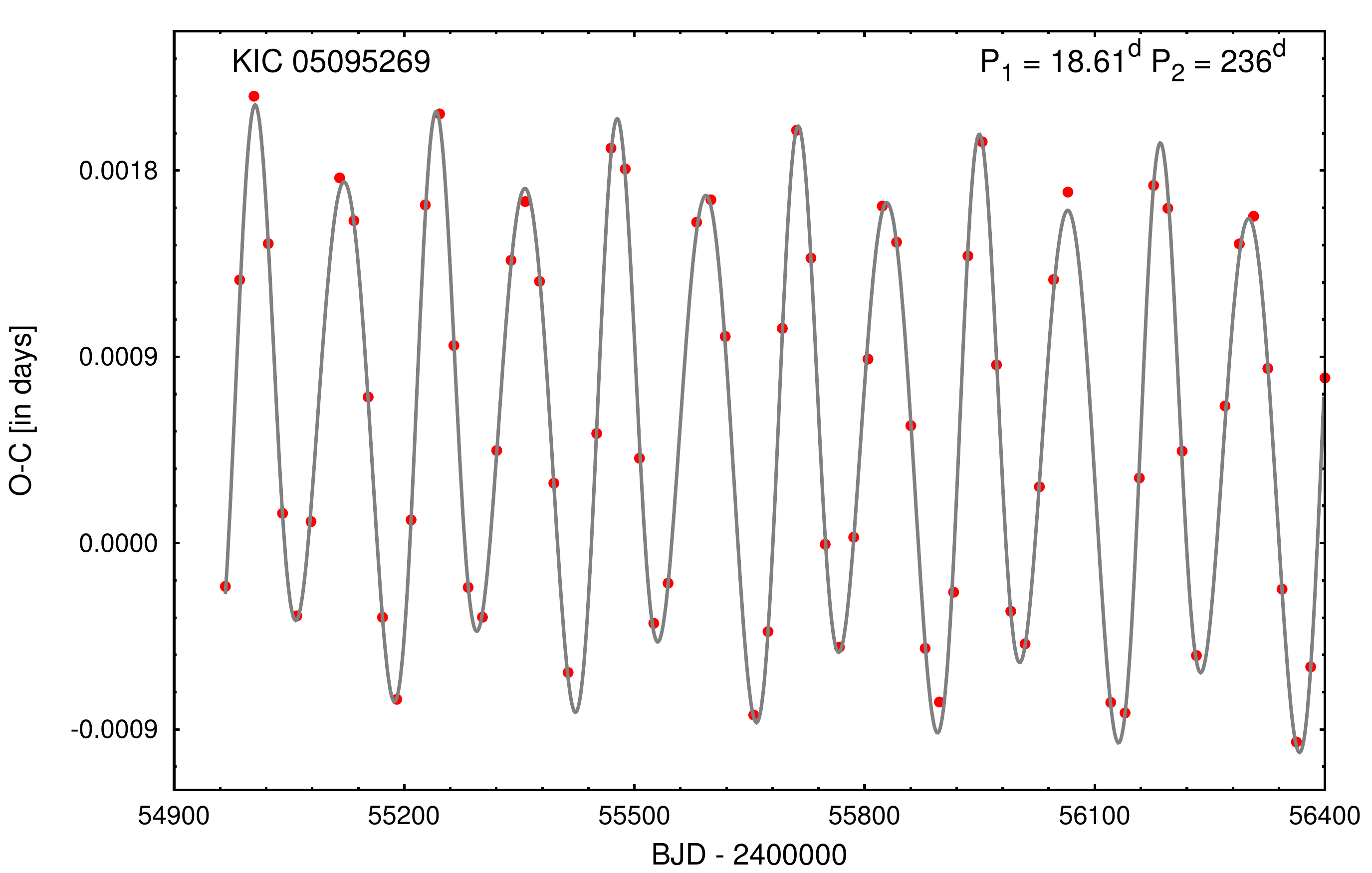}\includegraphics[width=60mm]{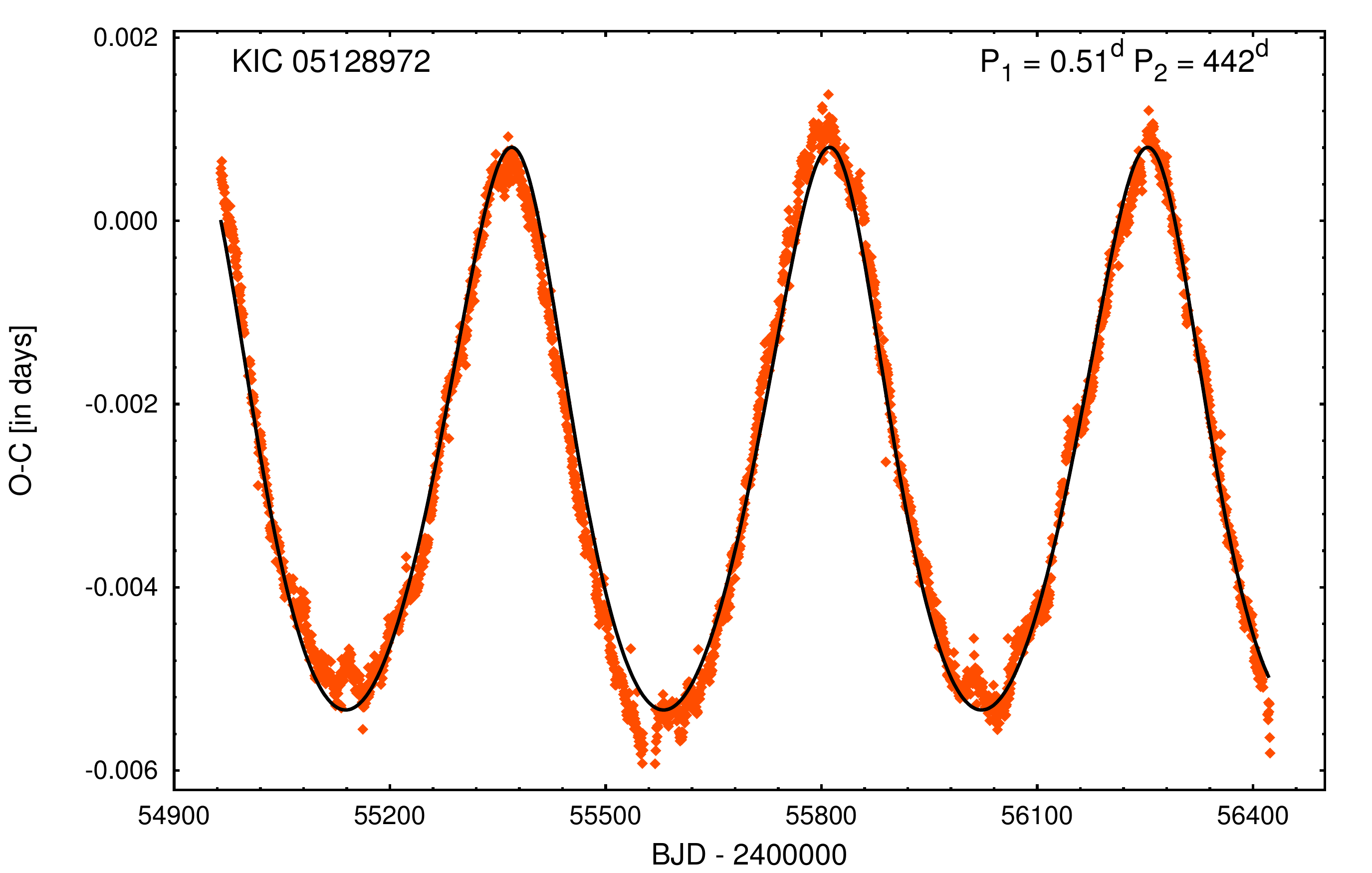}
\includegraphics[width=60mm]{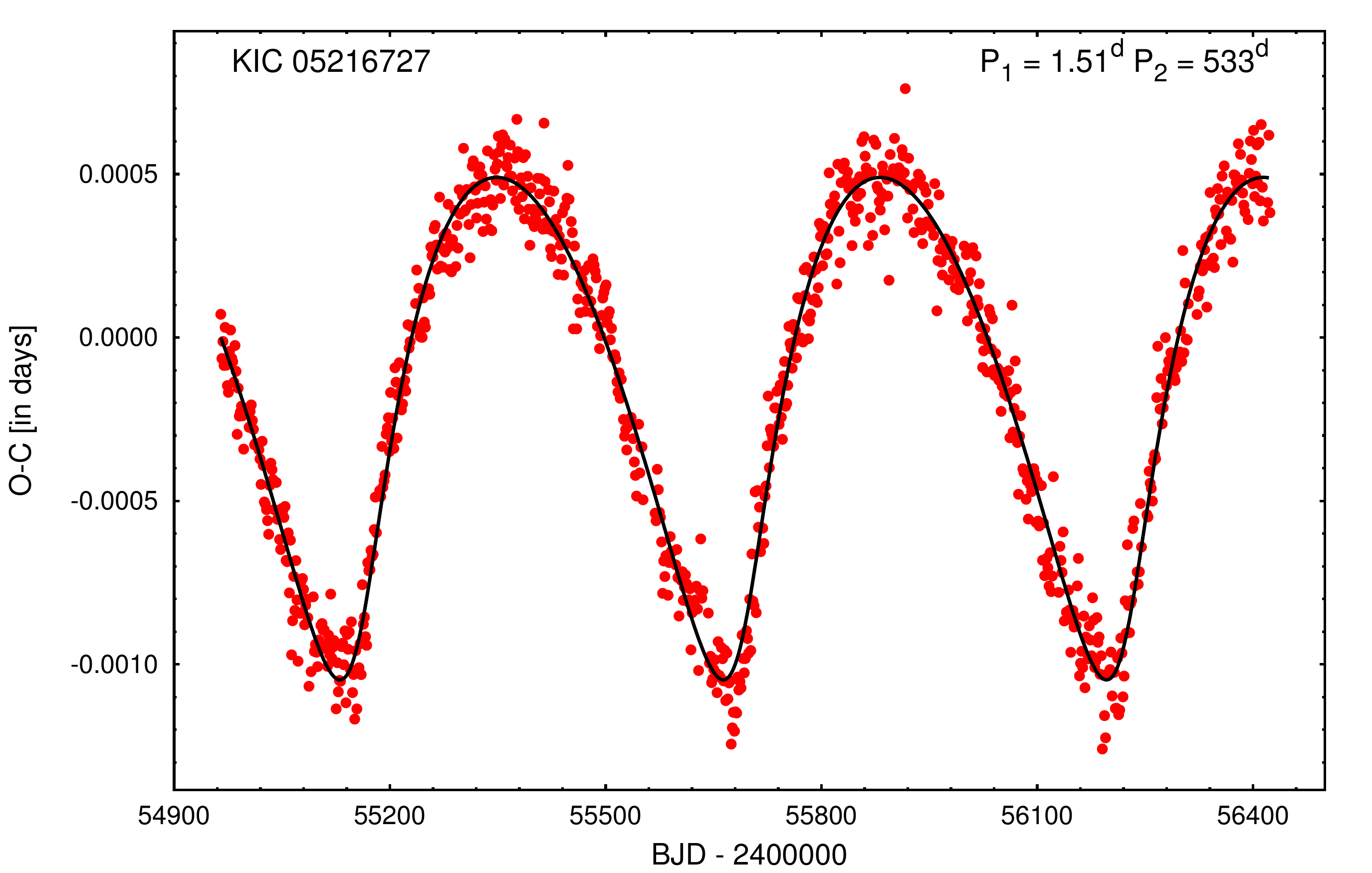}\includegraphics[width=60mm]{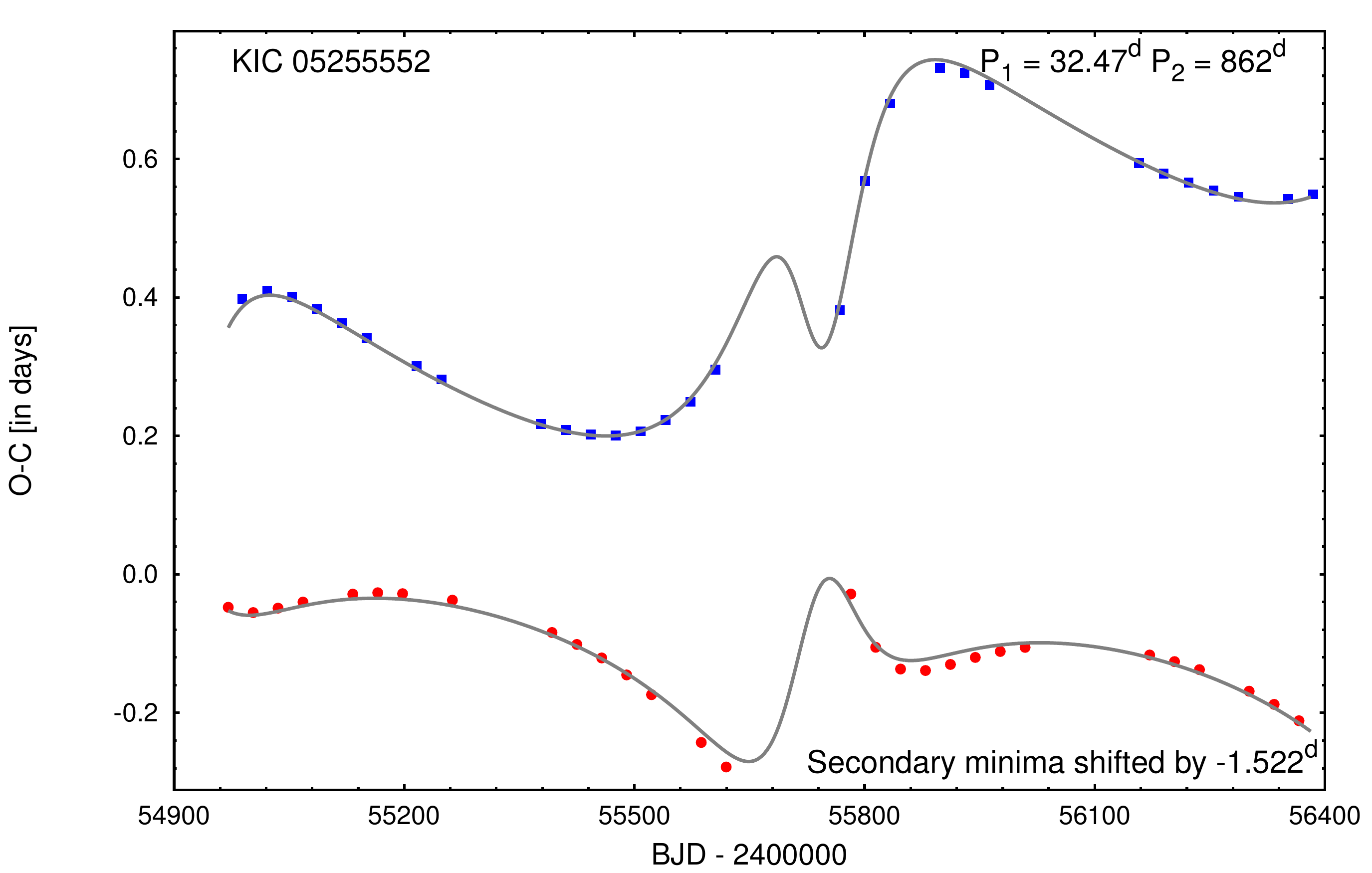}\includegraphics[width=60mm]{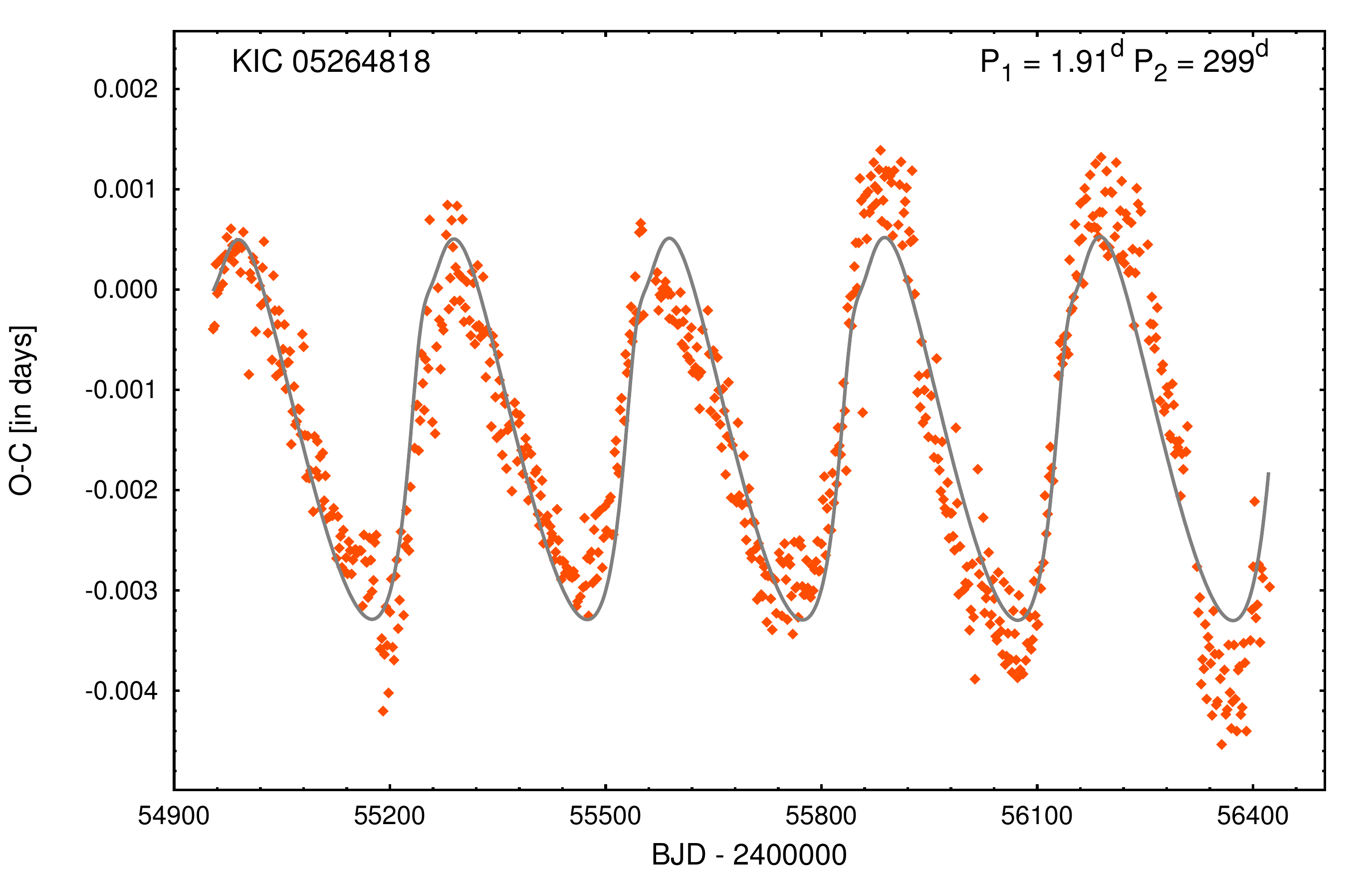}
\includegraphics[width=60mm]{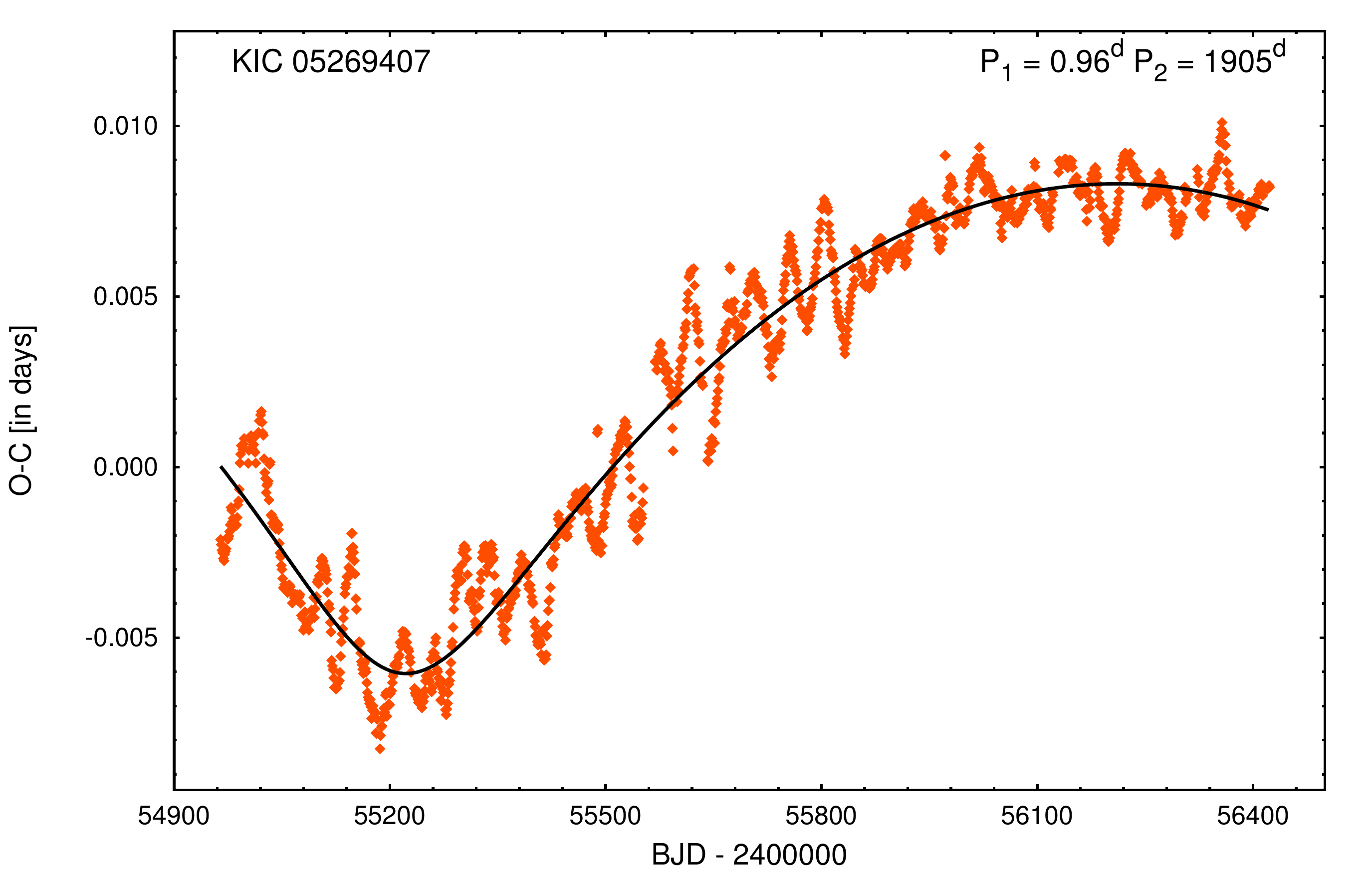}\includegraphics[width=60mm]{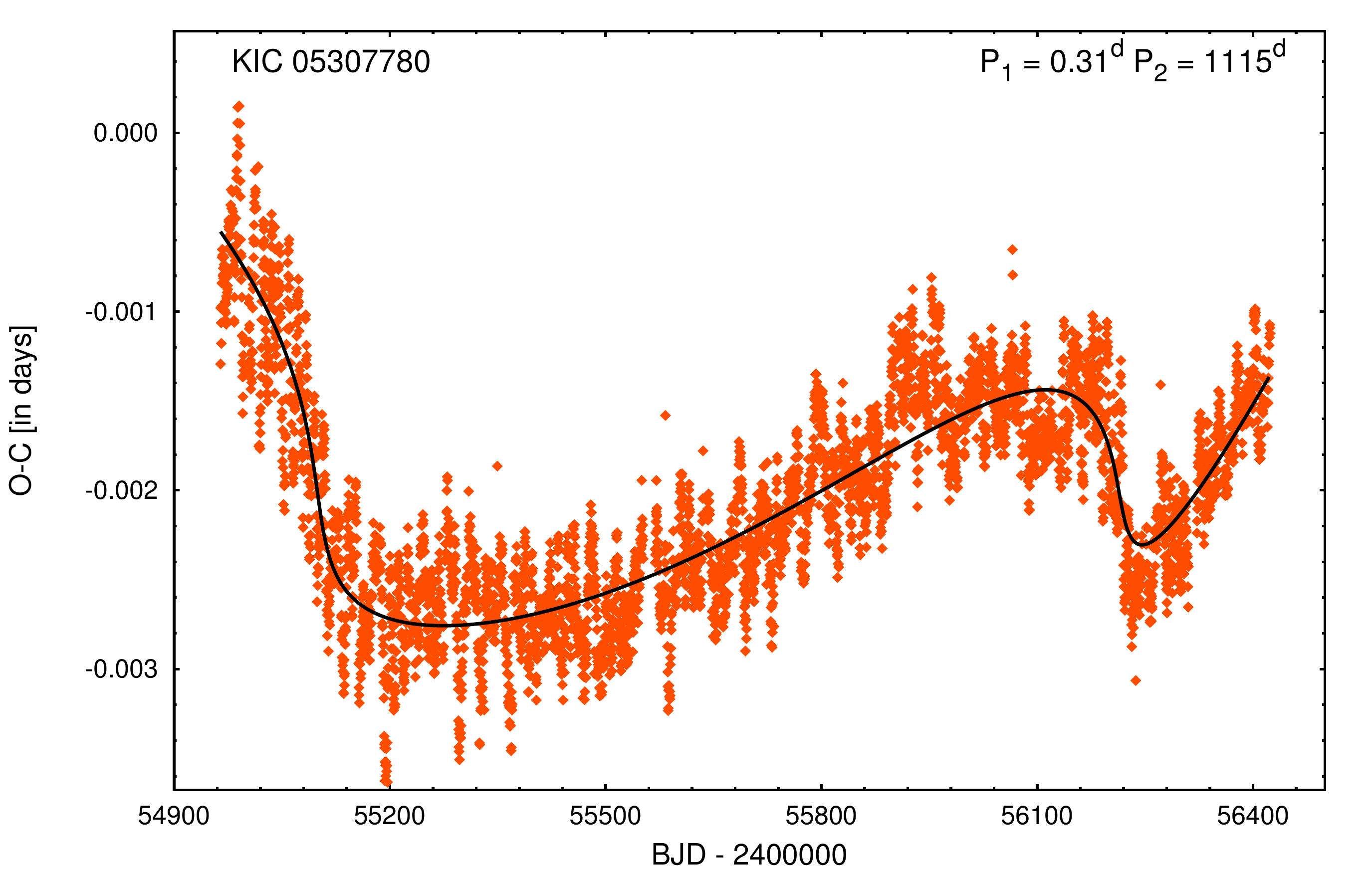}\includegraphics[width=60mm]{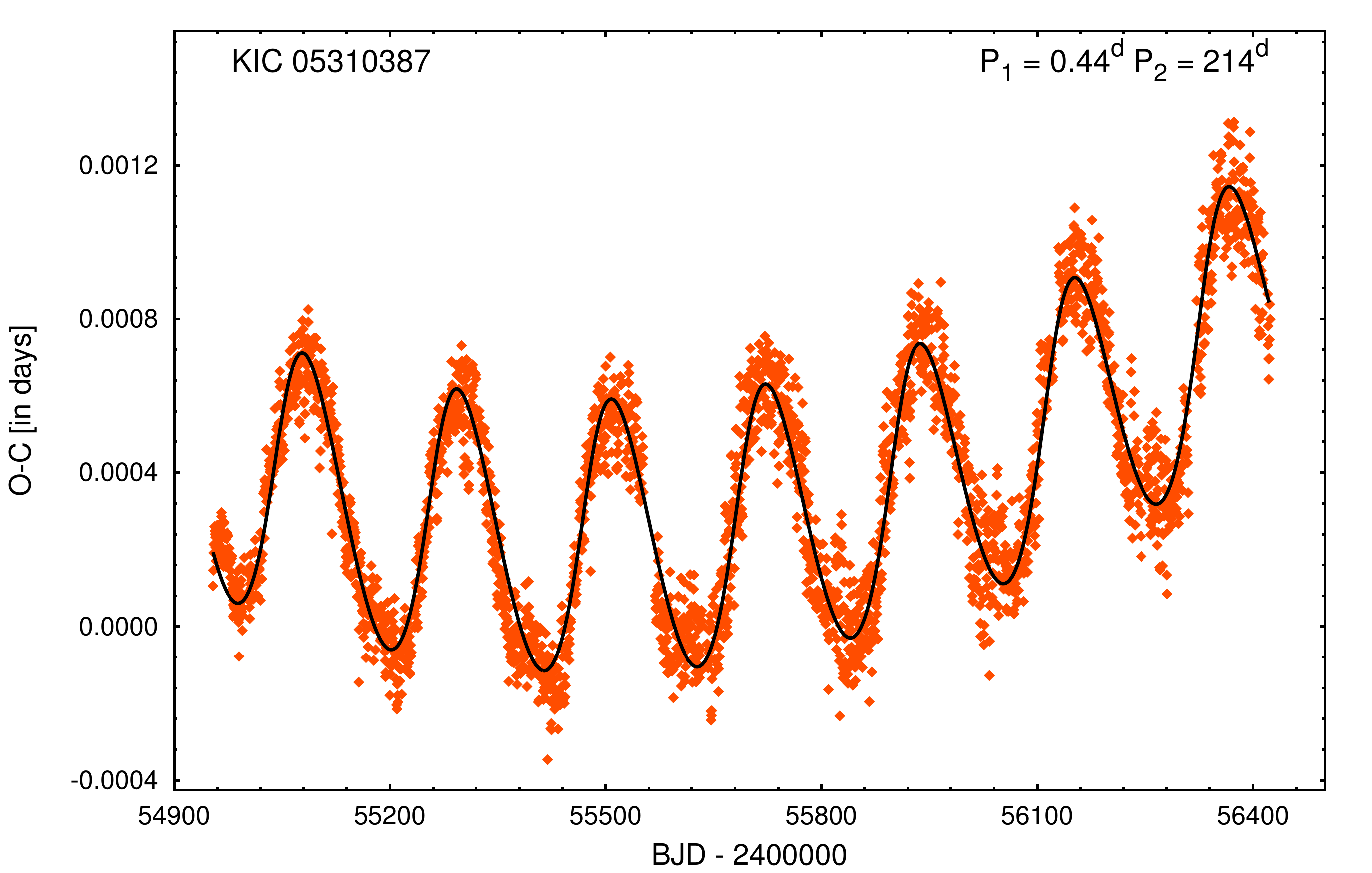}
\includegraphics[width=60mm]{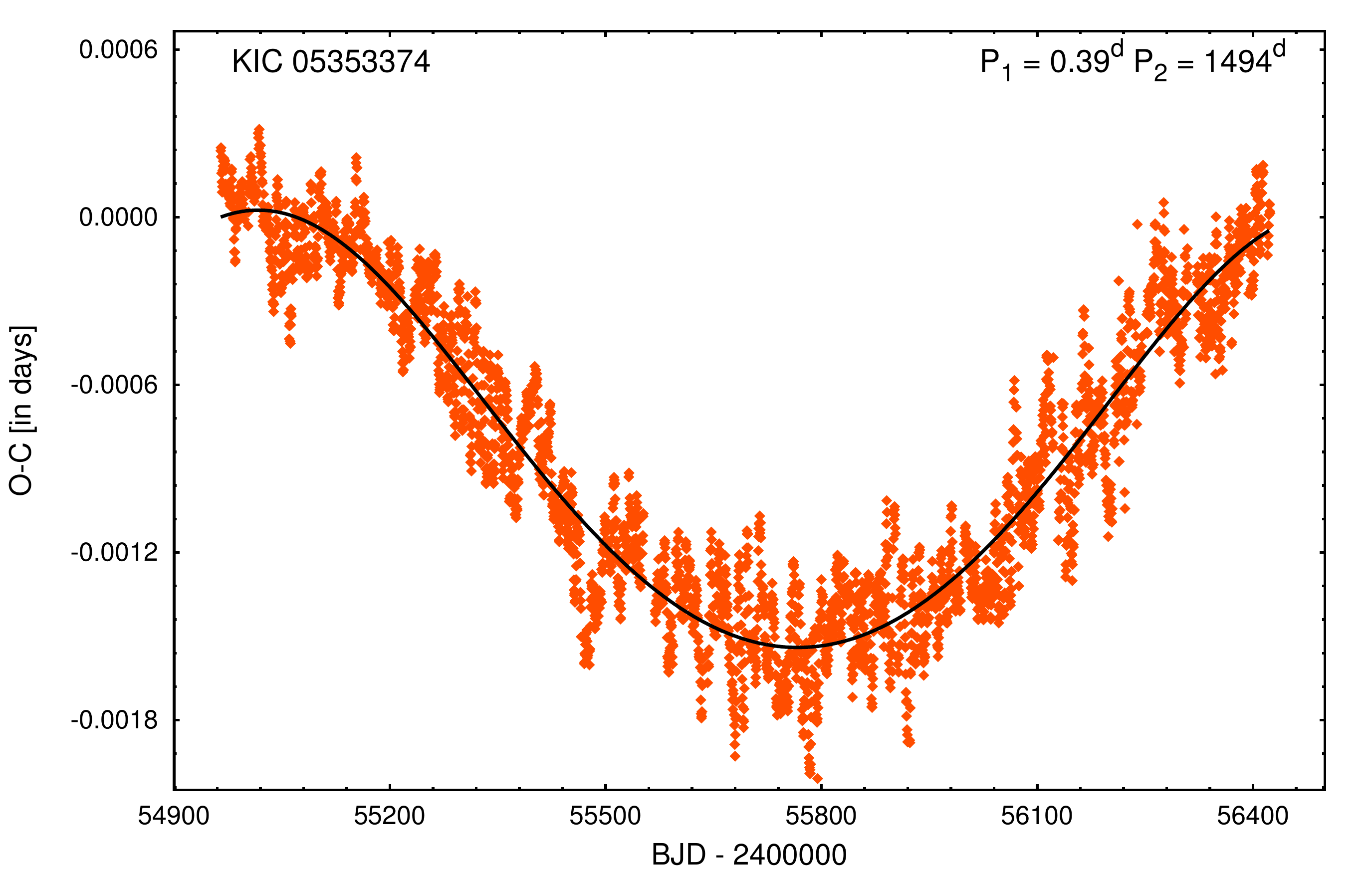}\includegraphics[width=60mm]{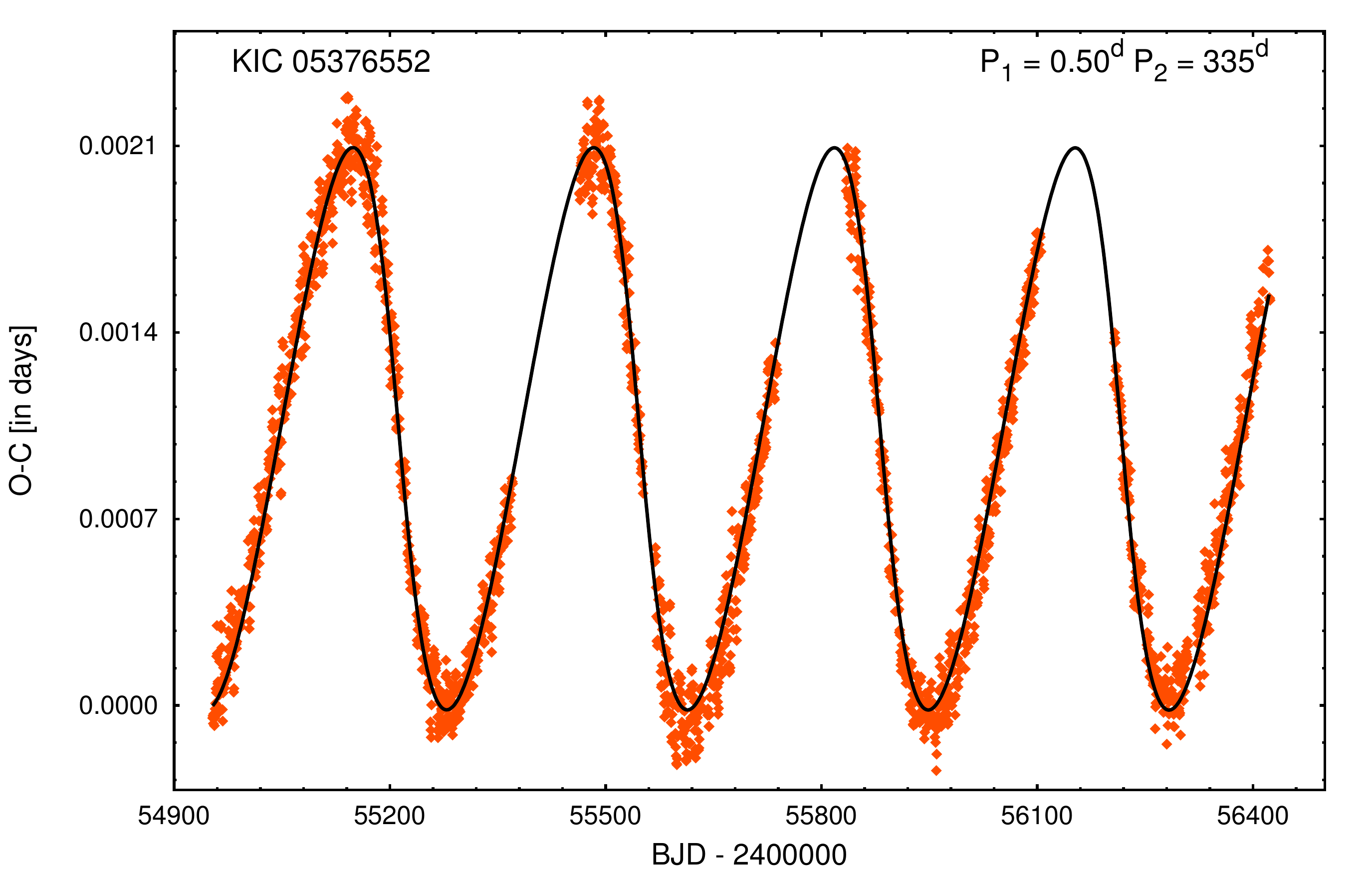}\includegraphics[width=60mm]{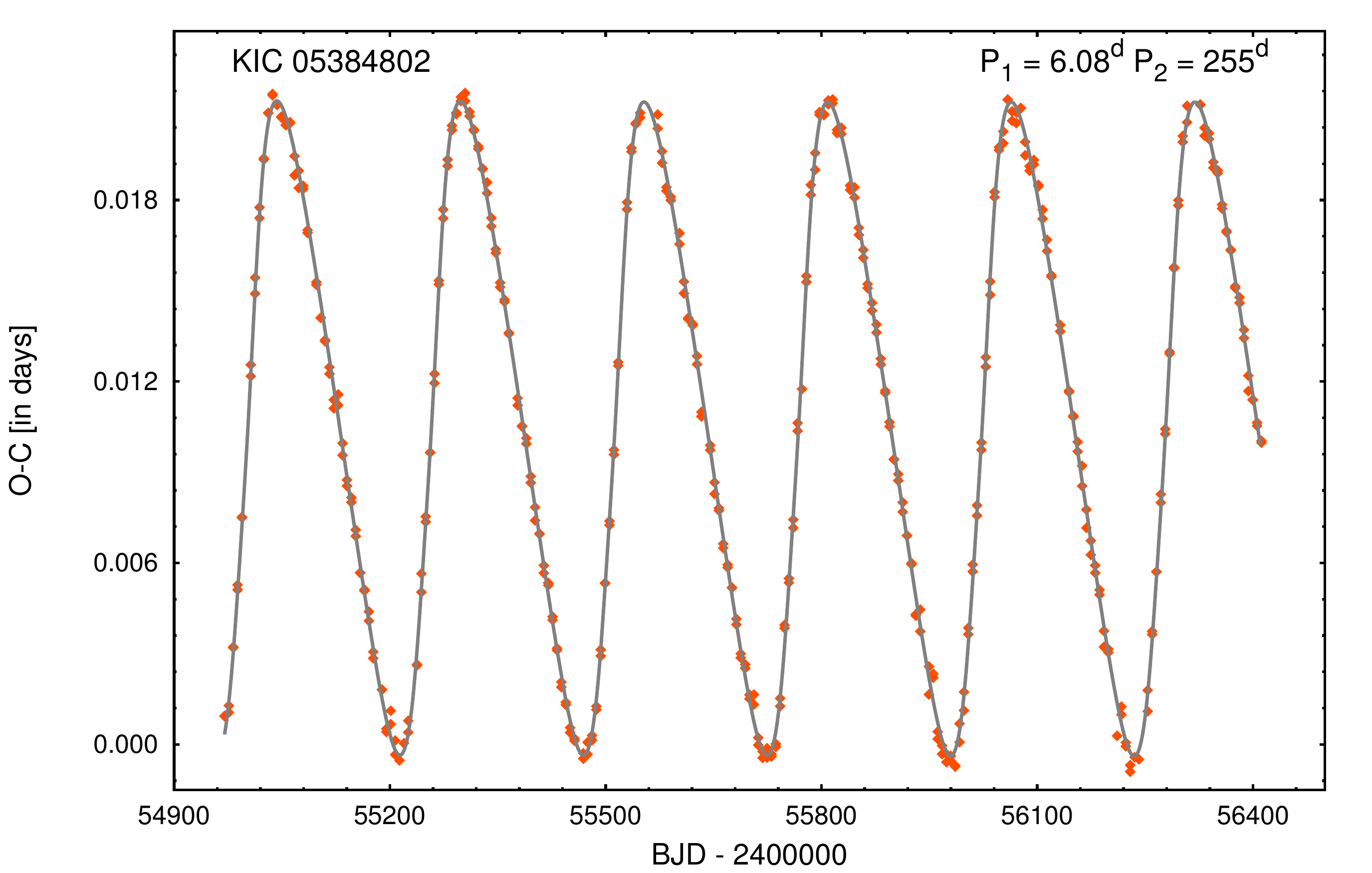}
\includegraphics[width=60mm]{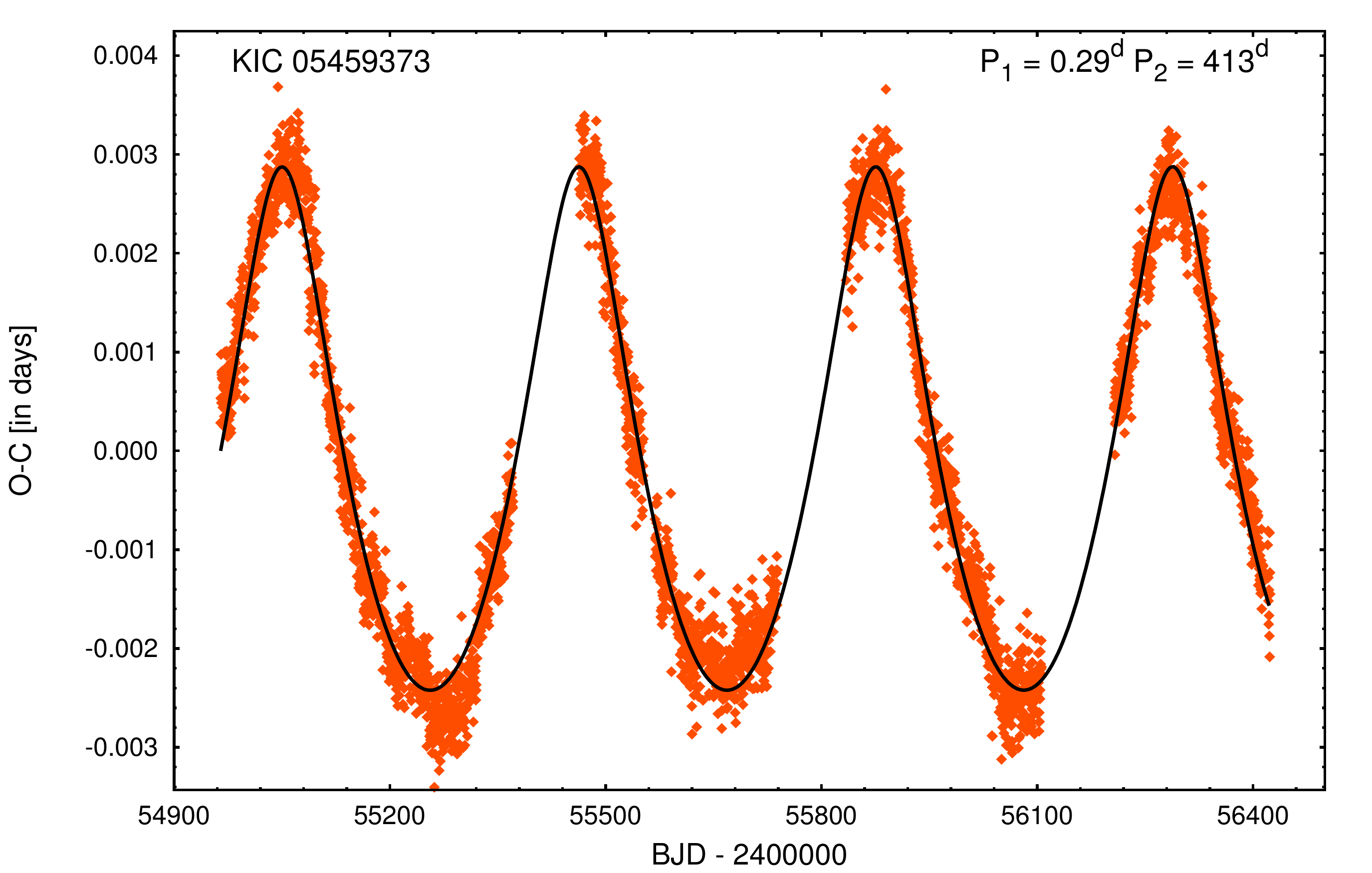}\includegraphics[width=60mm]{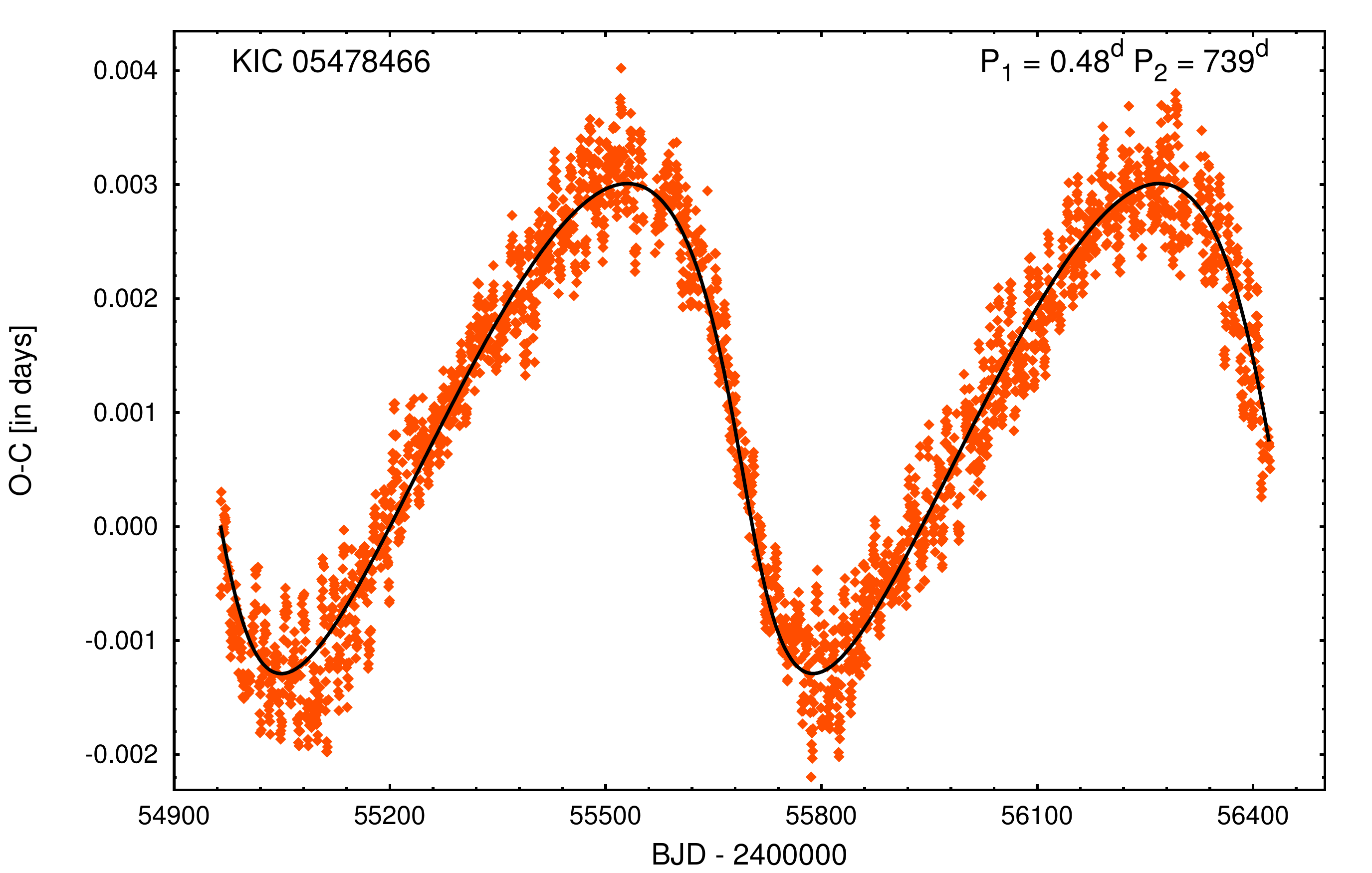}\includegraphics[width=60mm]{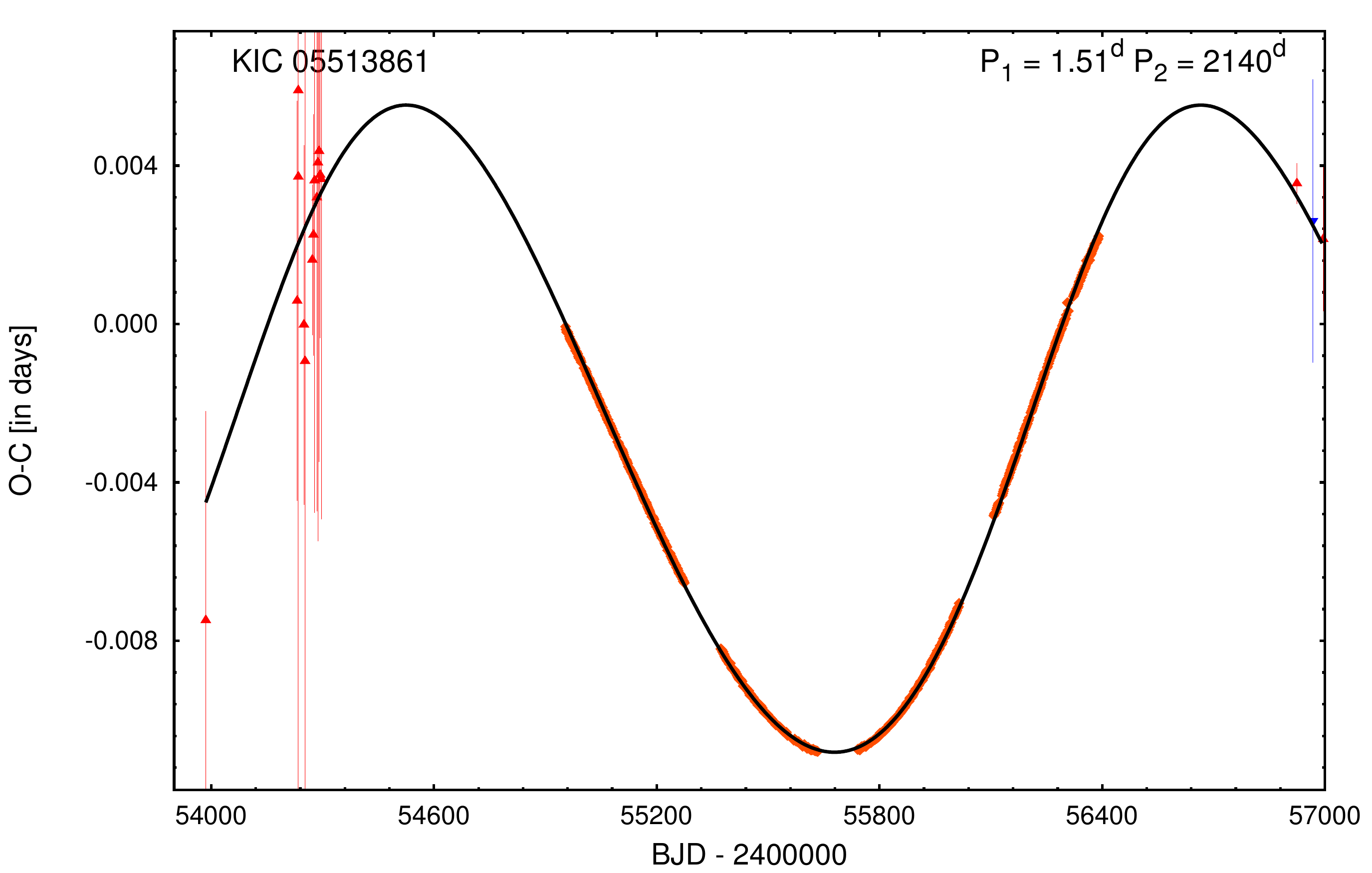}
\includegraphics[width=60mm]{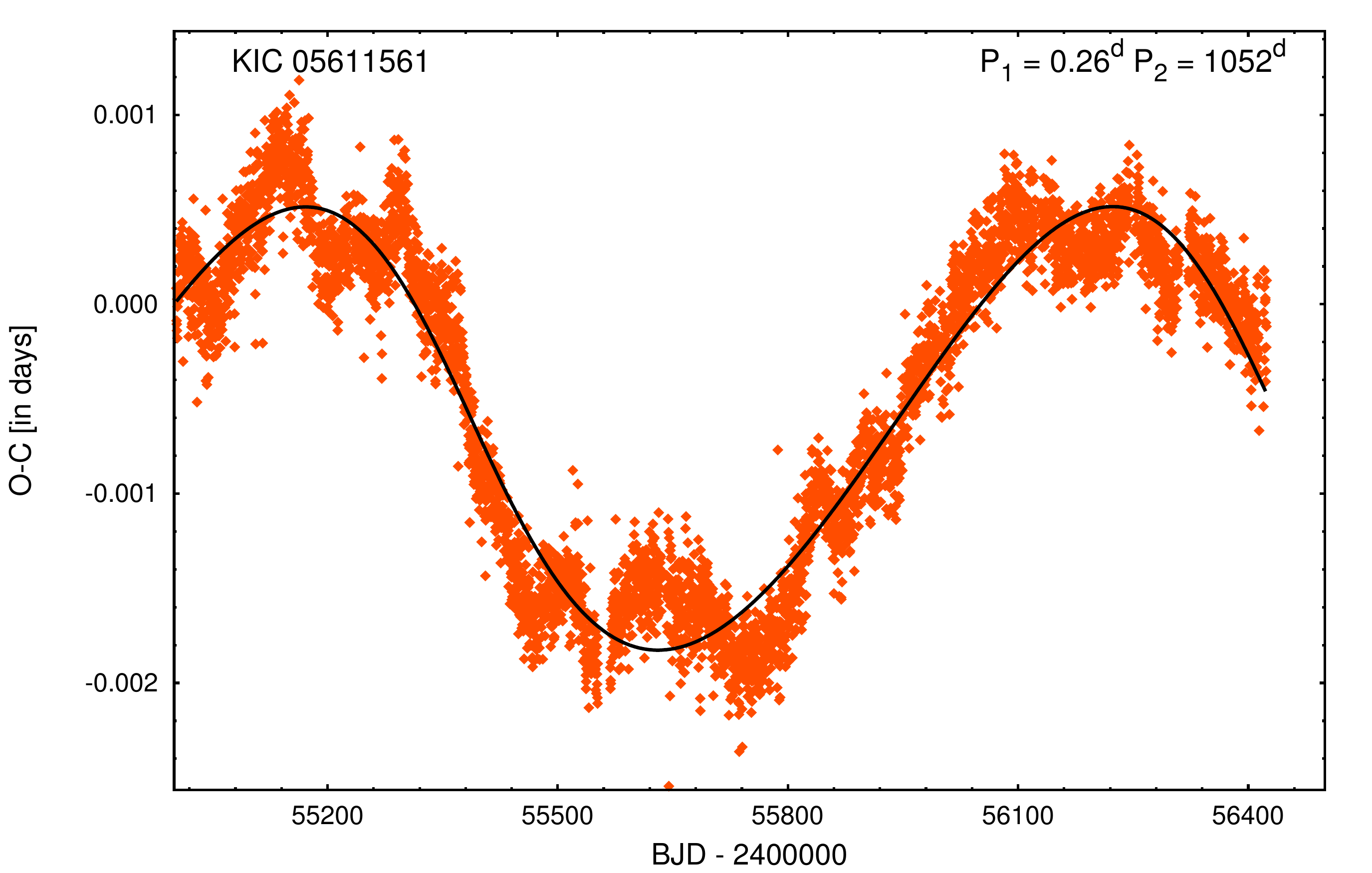}\includegraphics[width=60mm]{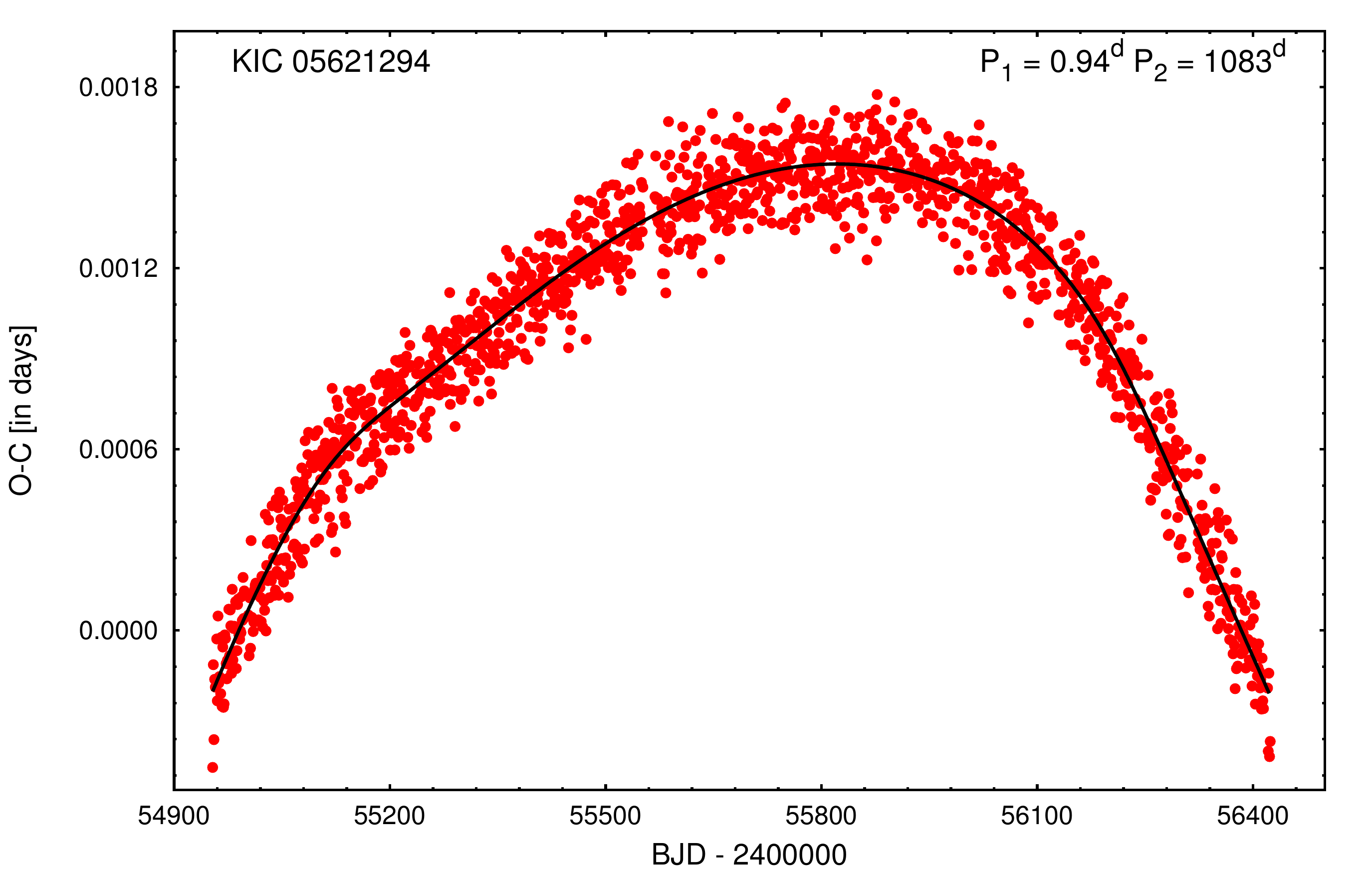}\includegraphics[width=60mm]{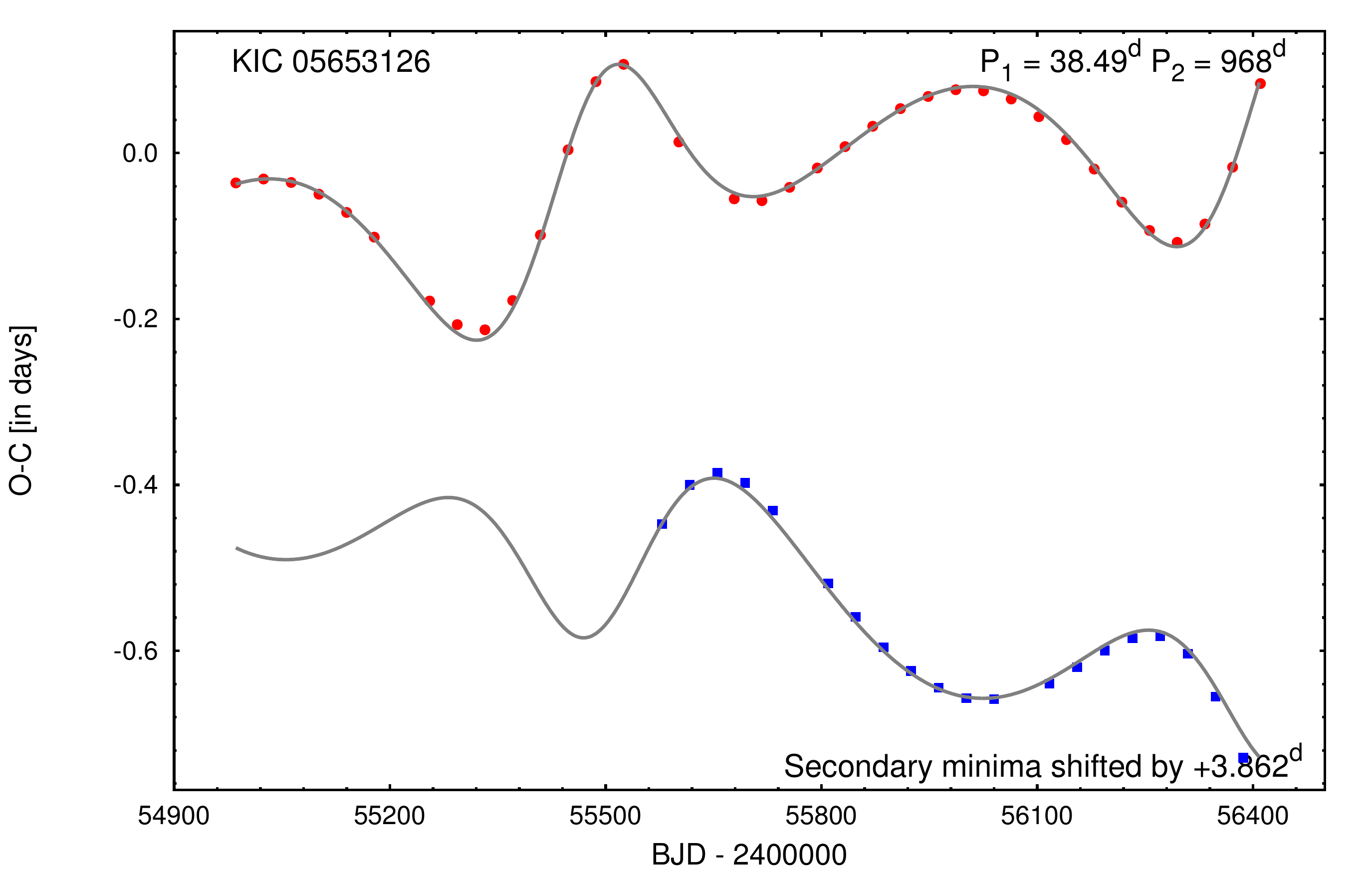}
\caption{(continued)}
\end{figure*}

\addtocounter{figure}{-1}

\begin{figure*}
\includegraphics[width=60mm]{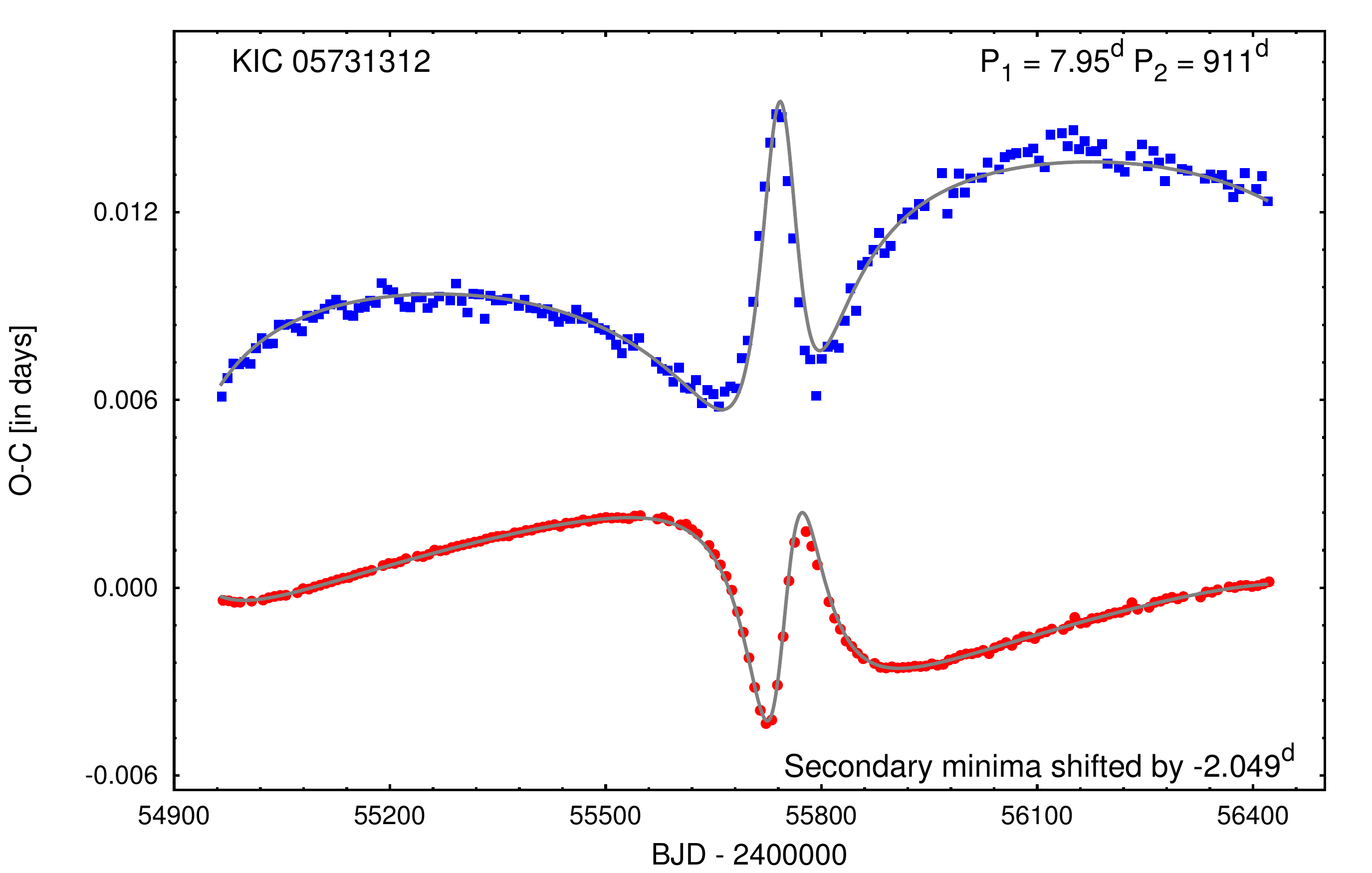}\includegraphics[width=60mm]{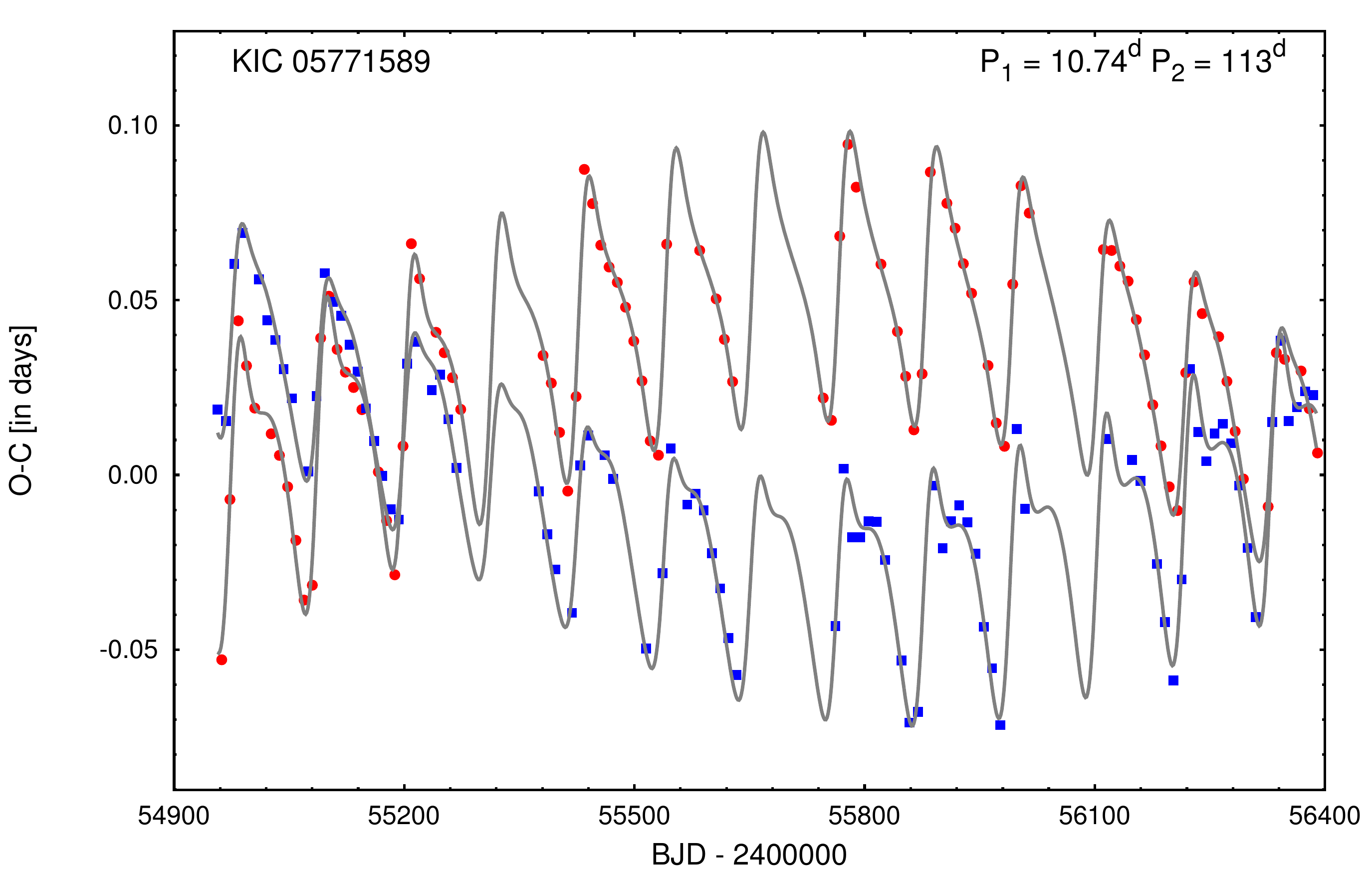}\includegraphics[width=60mm]{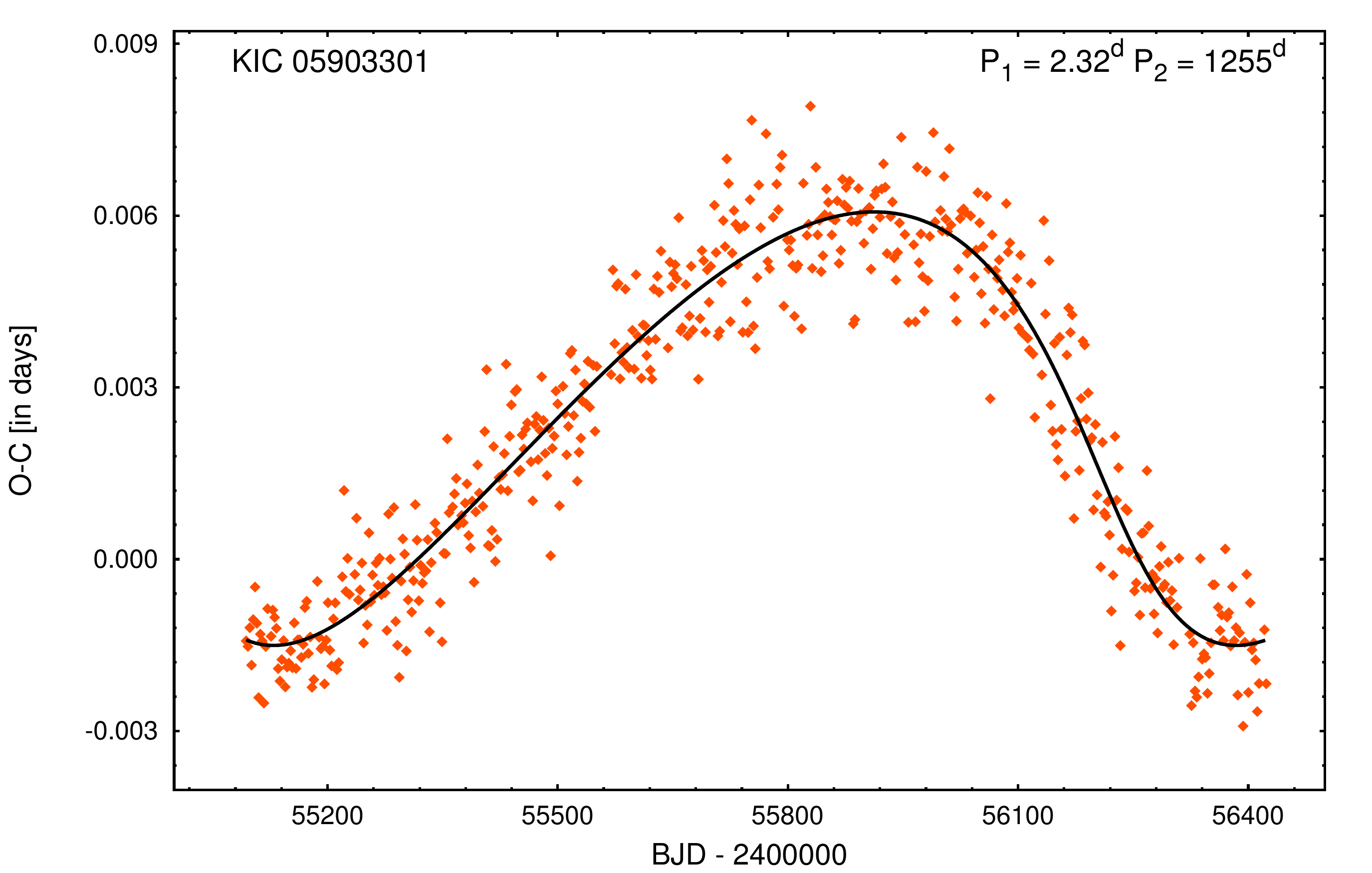}
\includegraphics[width=60mm]{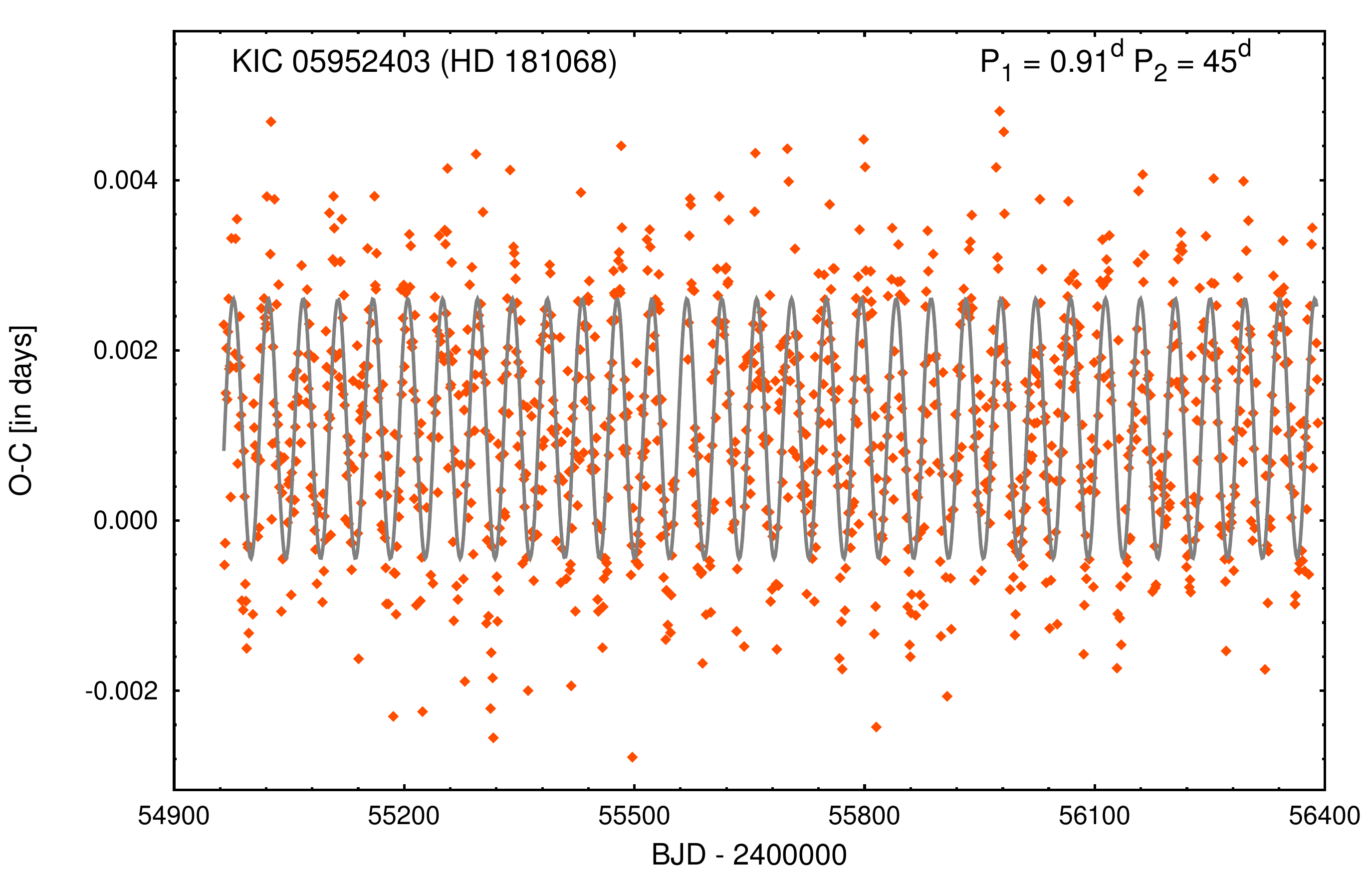}\includegraphics[width=60mm]{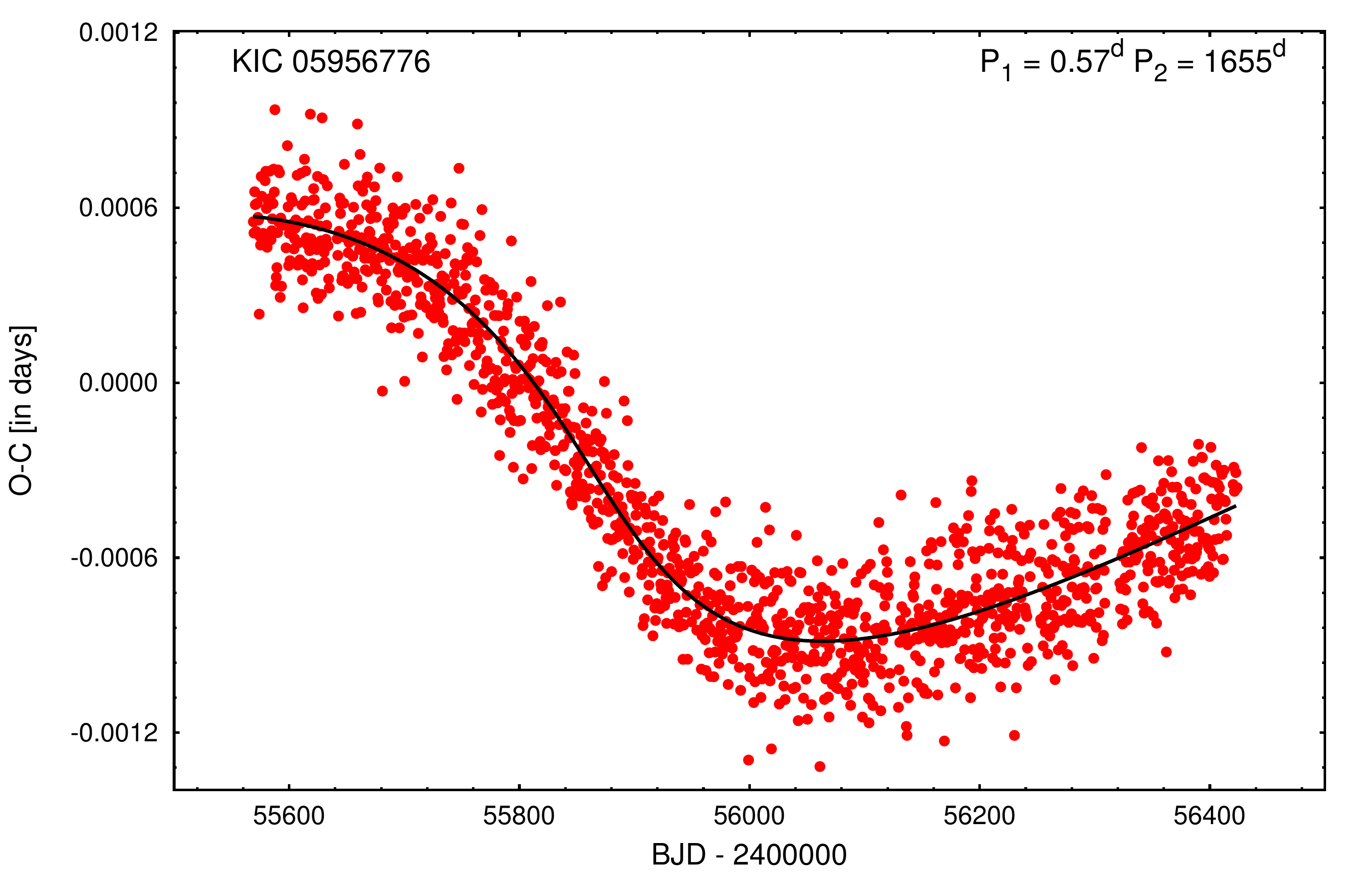}\includegraphics[width=60mm]{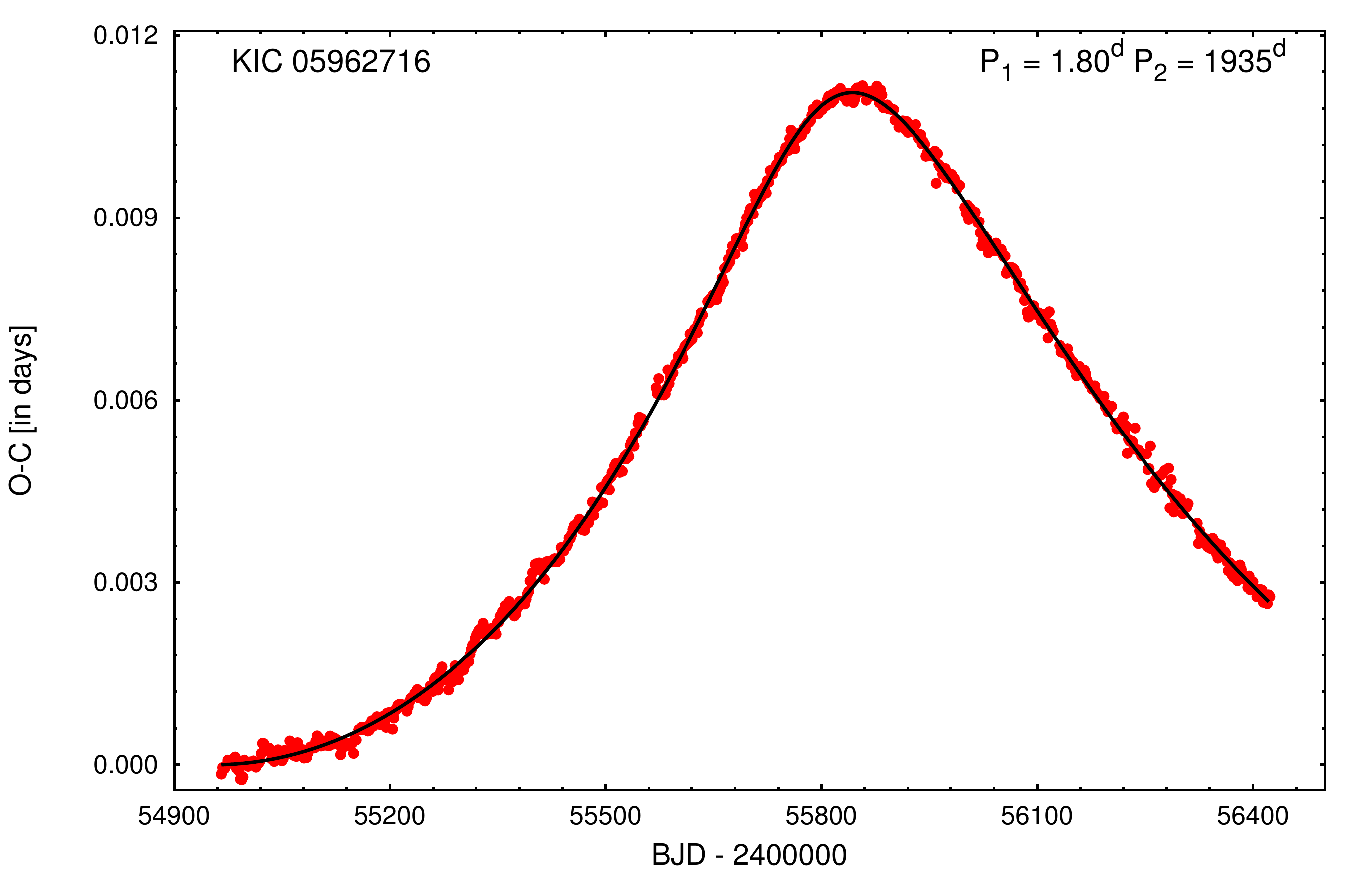}
\includegraphics[width=60mm]{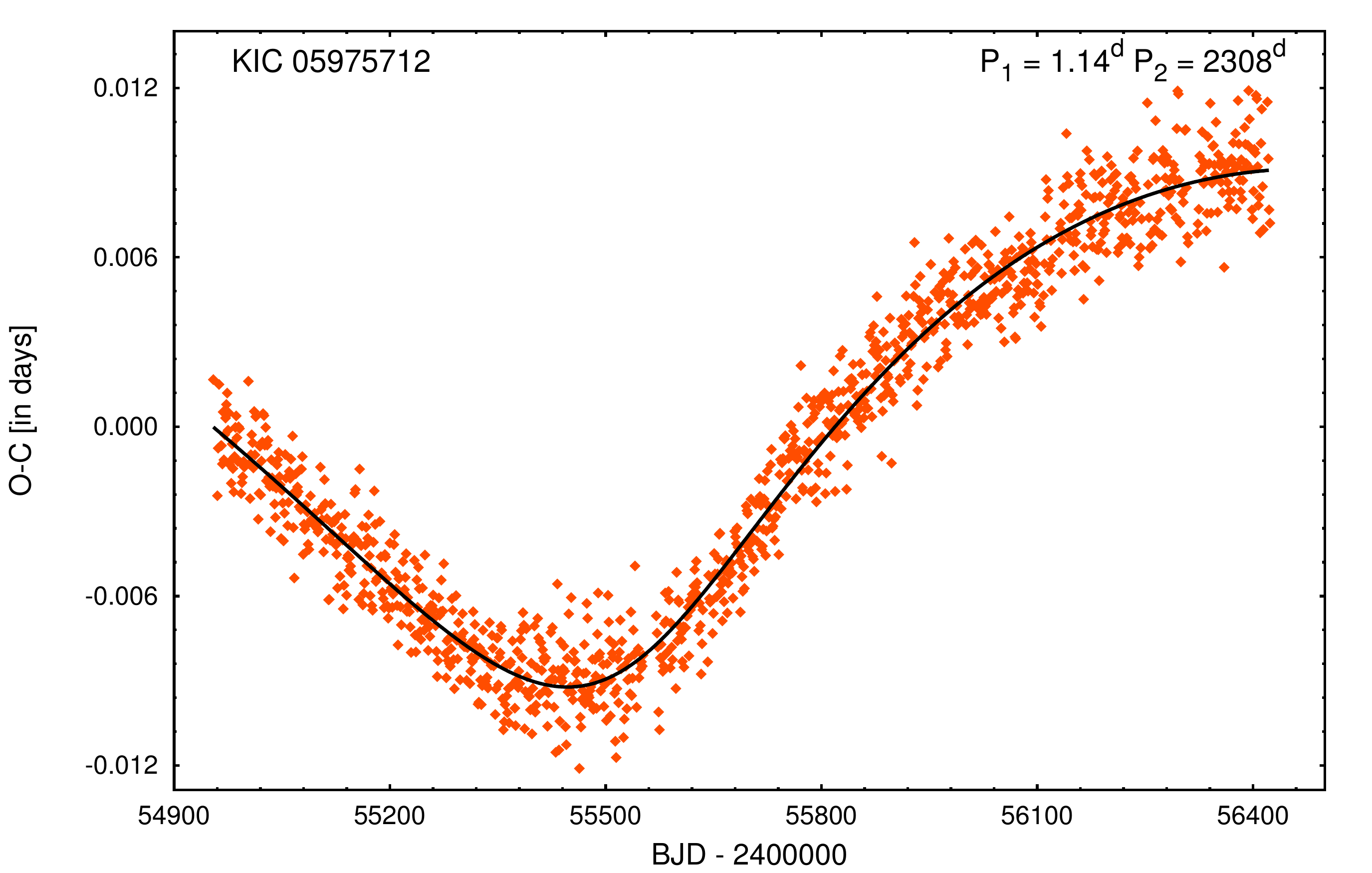}\includegraphics[width=60mm]{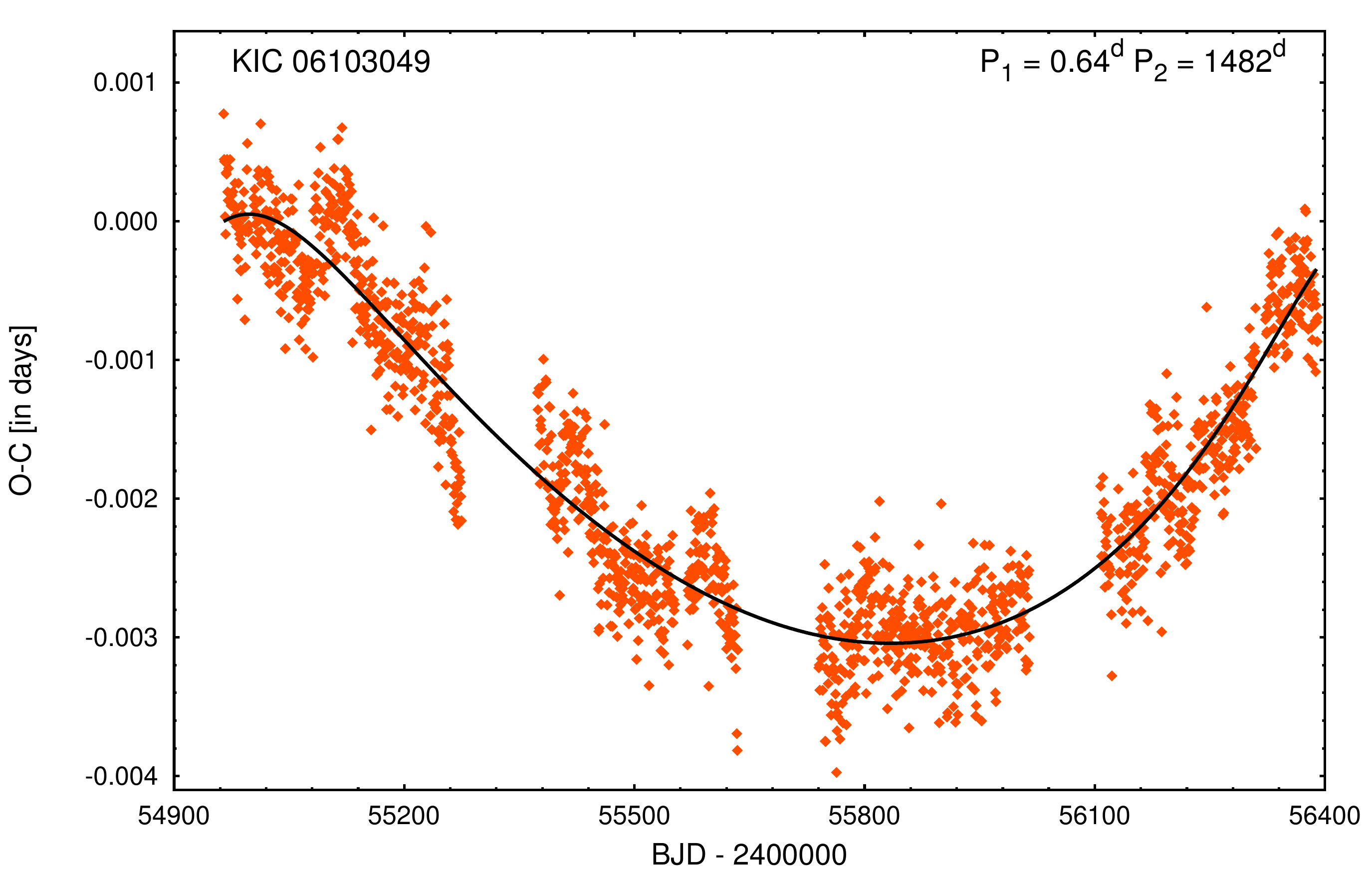}\includegraphics[width=60mm]{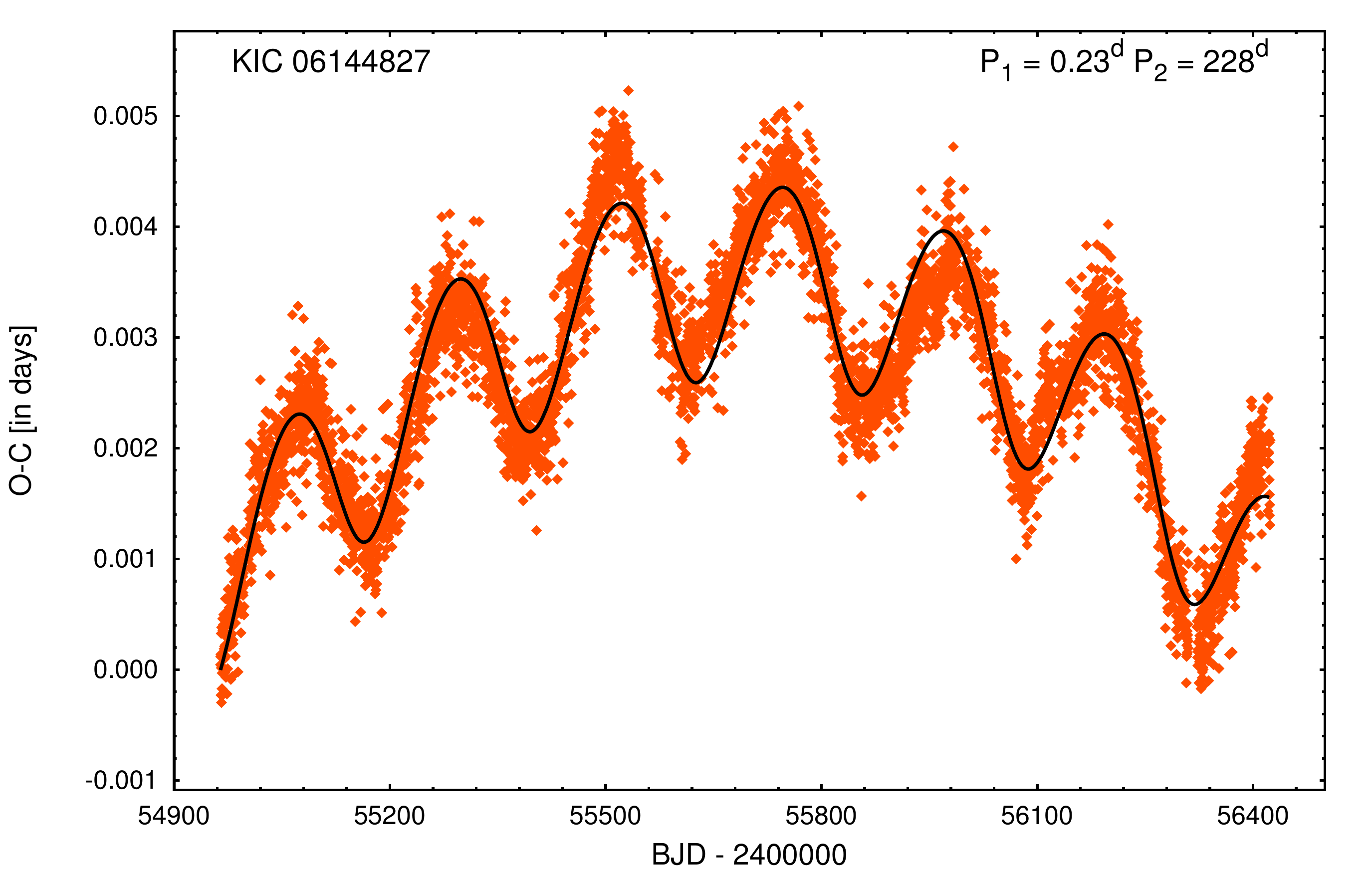}
\includegraphics[width=60mm]{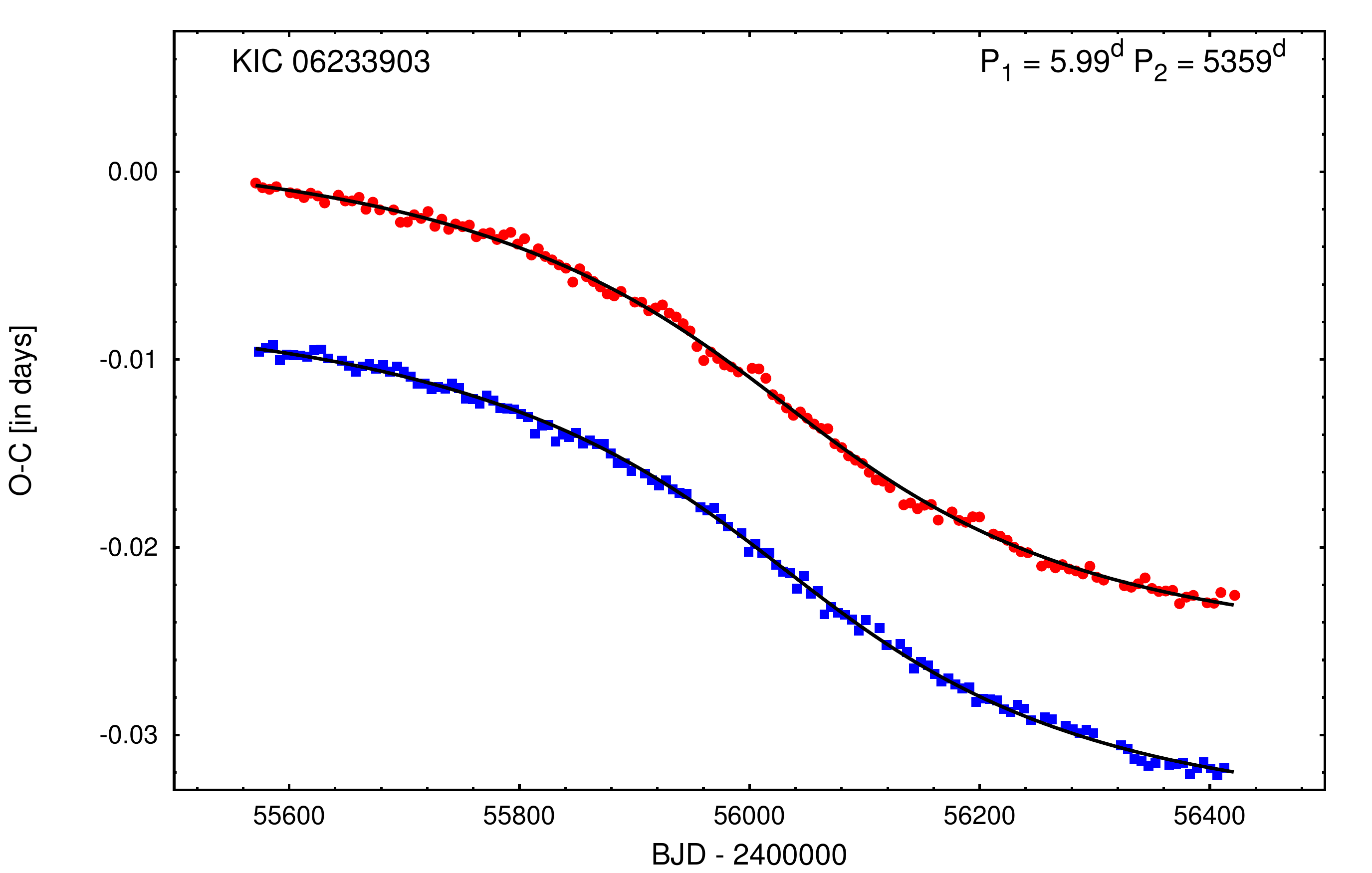}\includegraphics[width=60mm]{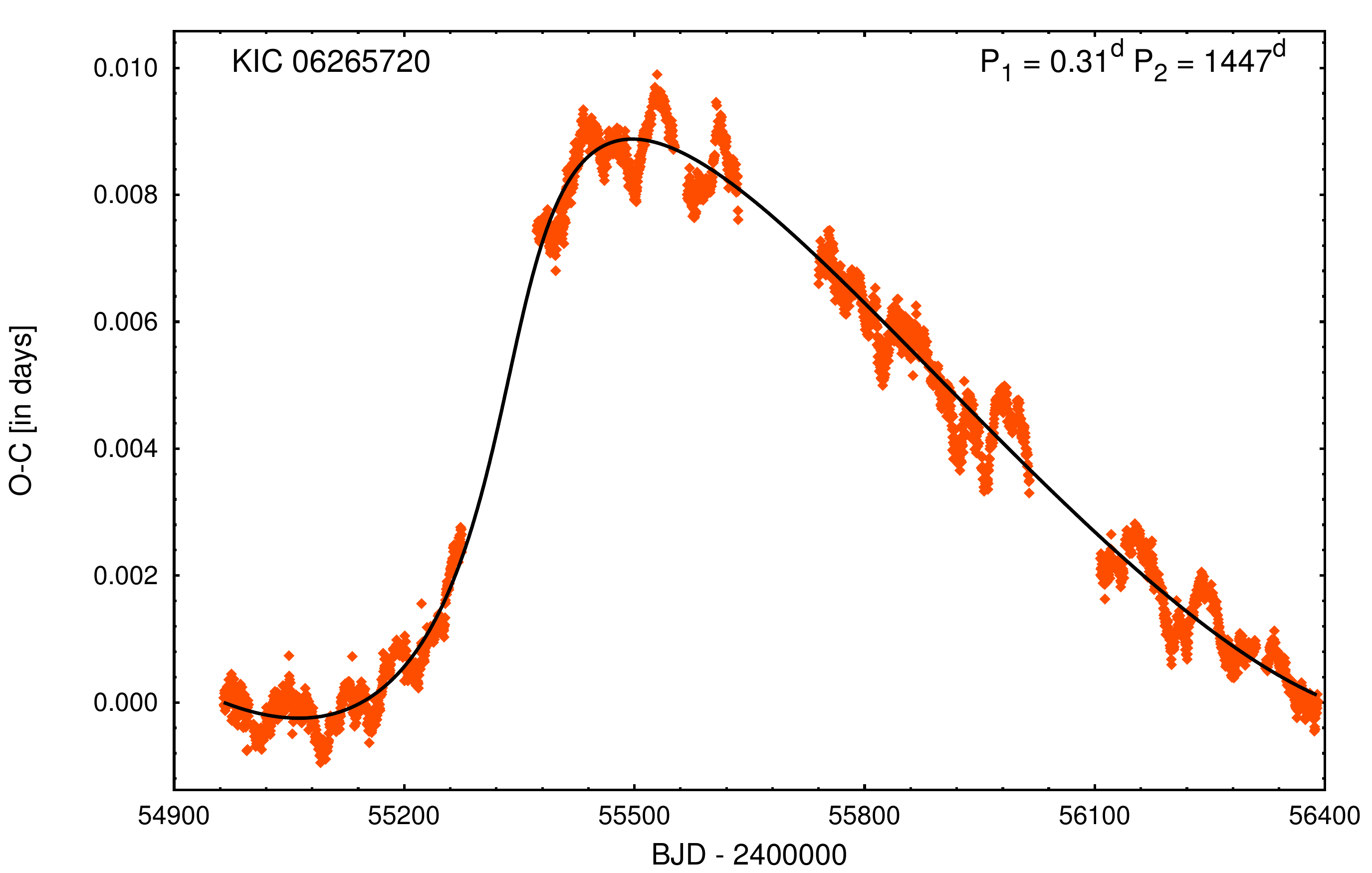}\includegraphics[width=60mm]{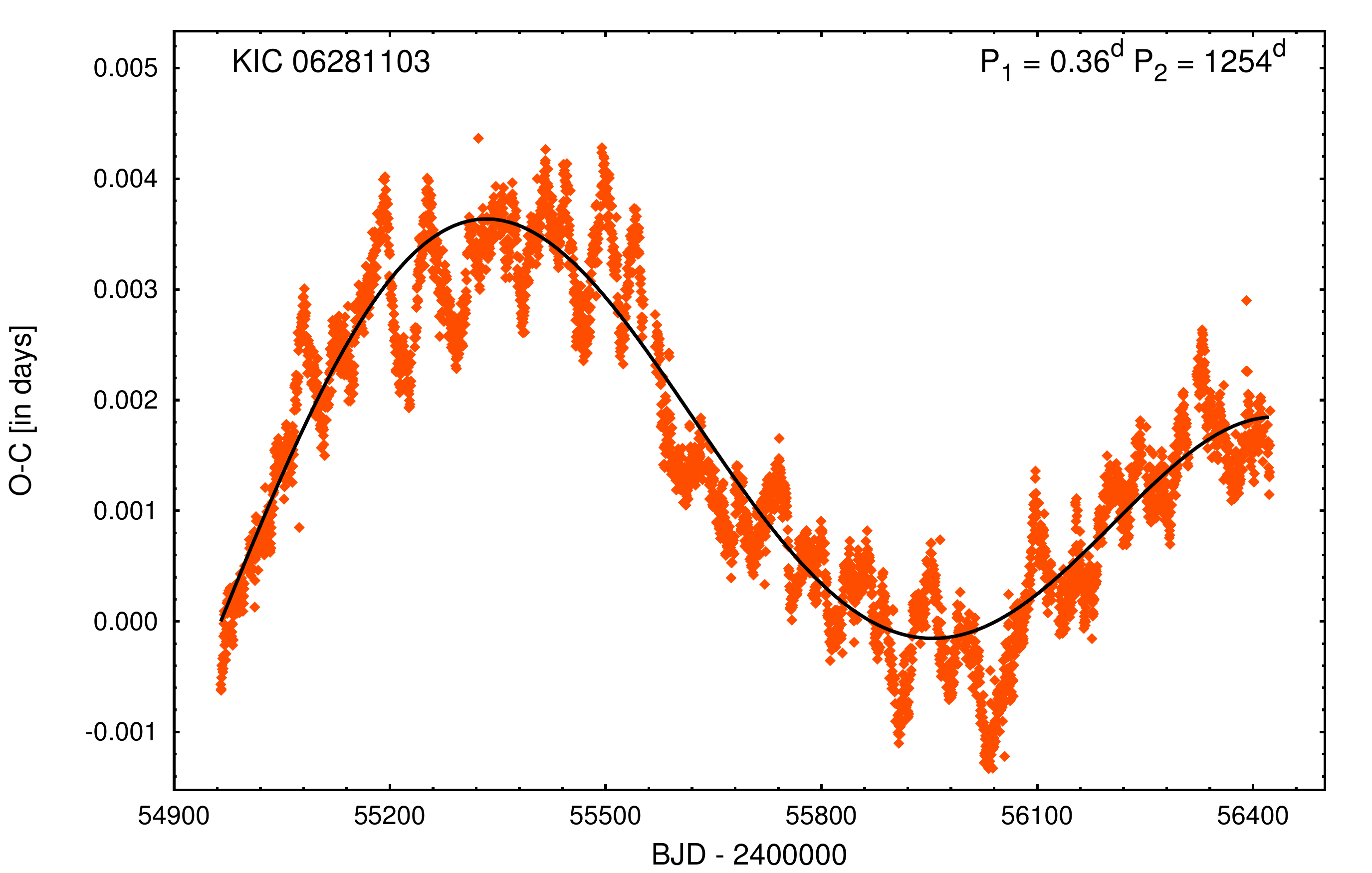}
\includegraphics[width=60mm]{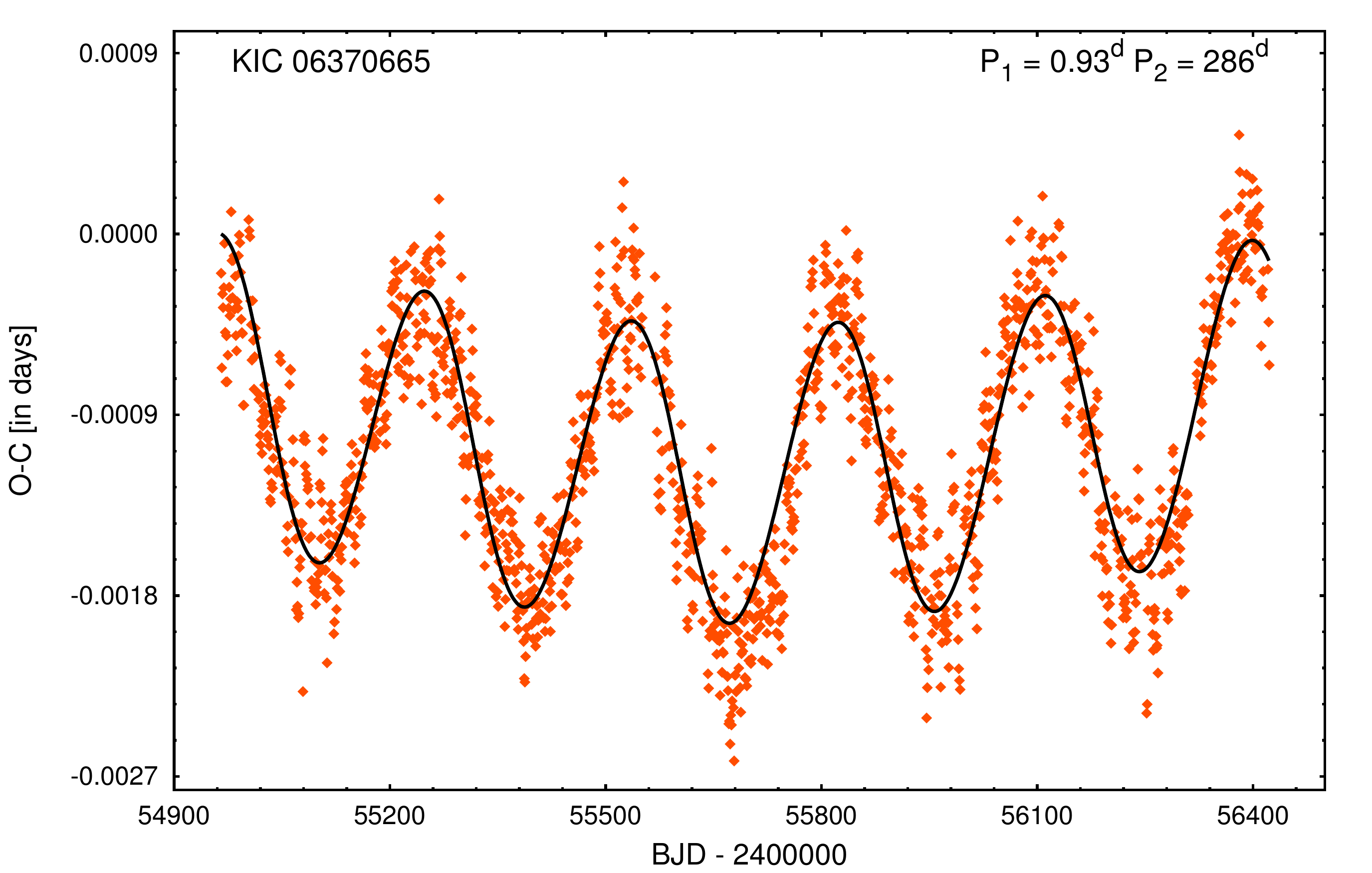}\includegraphics[width=60mm]{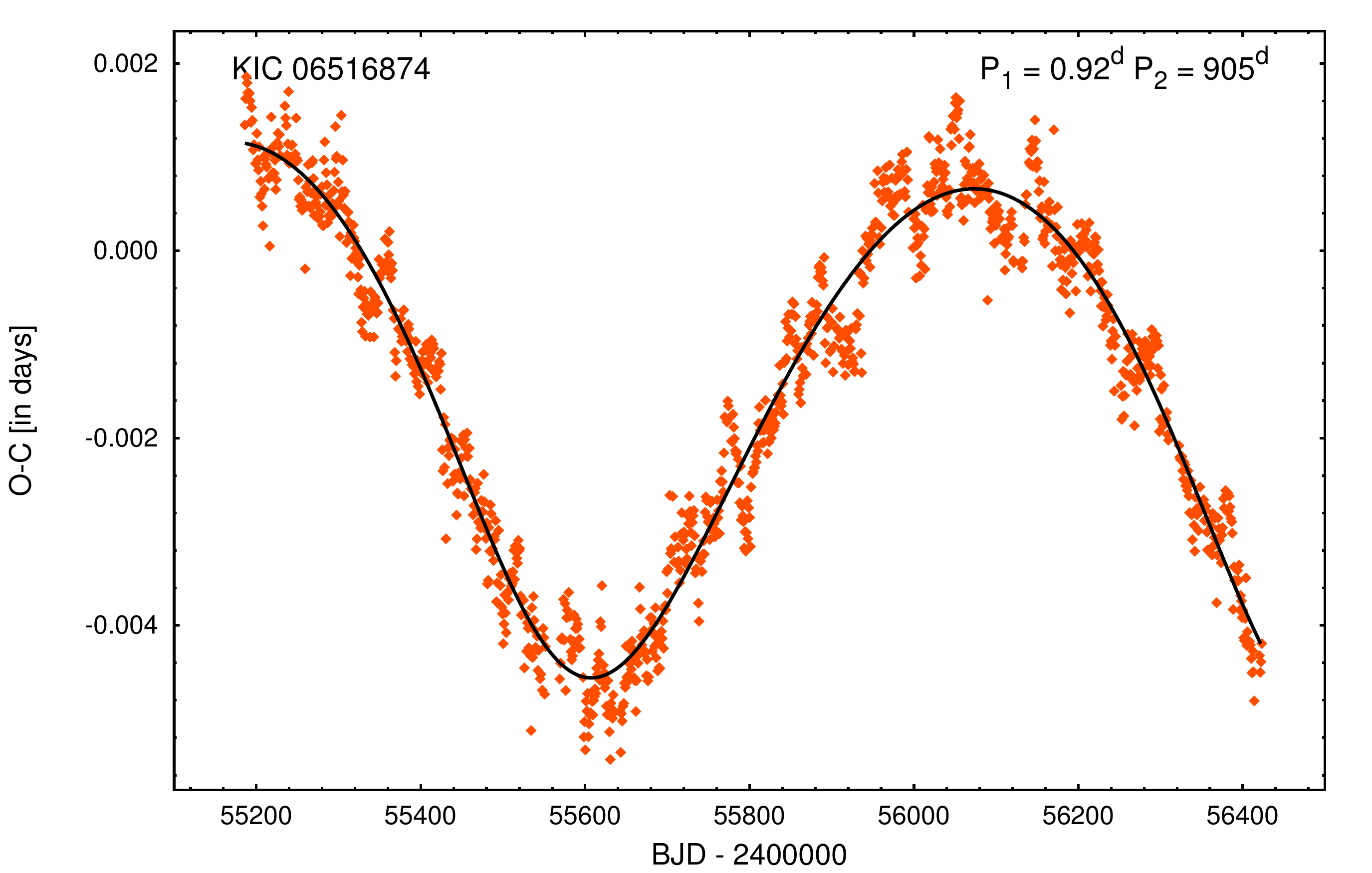}\includegraphics[width=60mm]{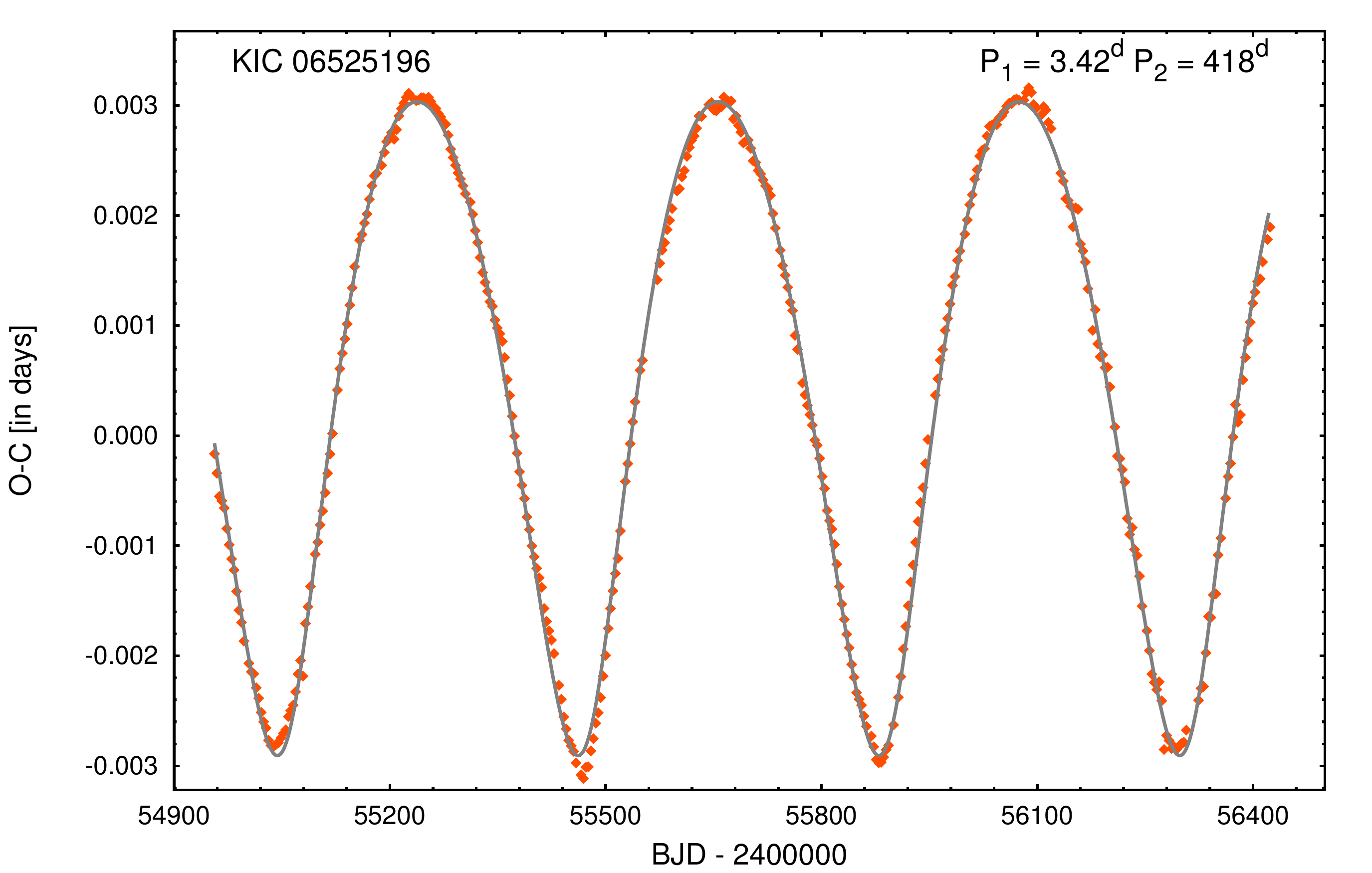}
\includegraphics[width=60mm]{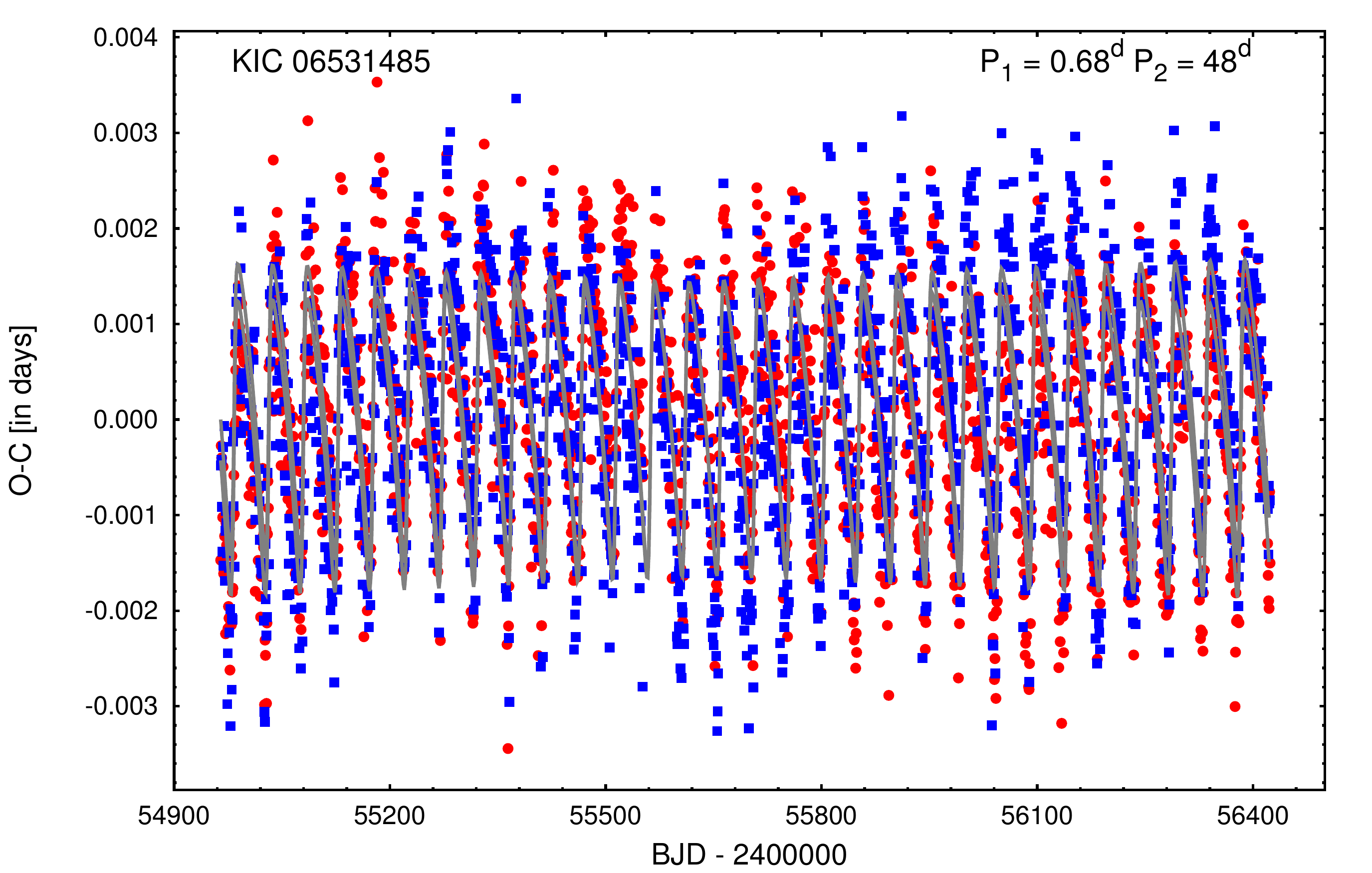}\includegraphics[width=60mm]{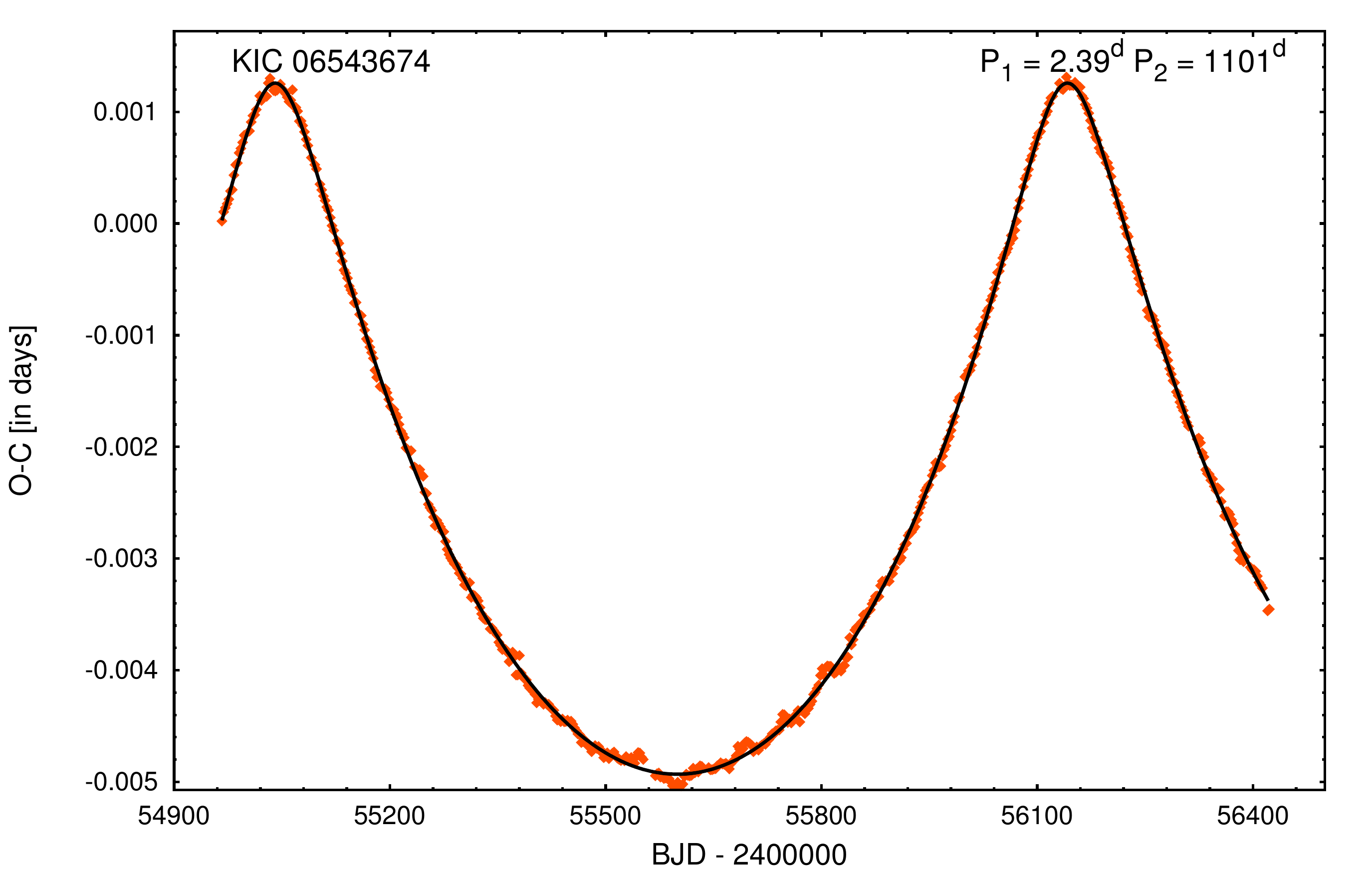}\includegraphics[width=60mm]{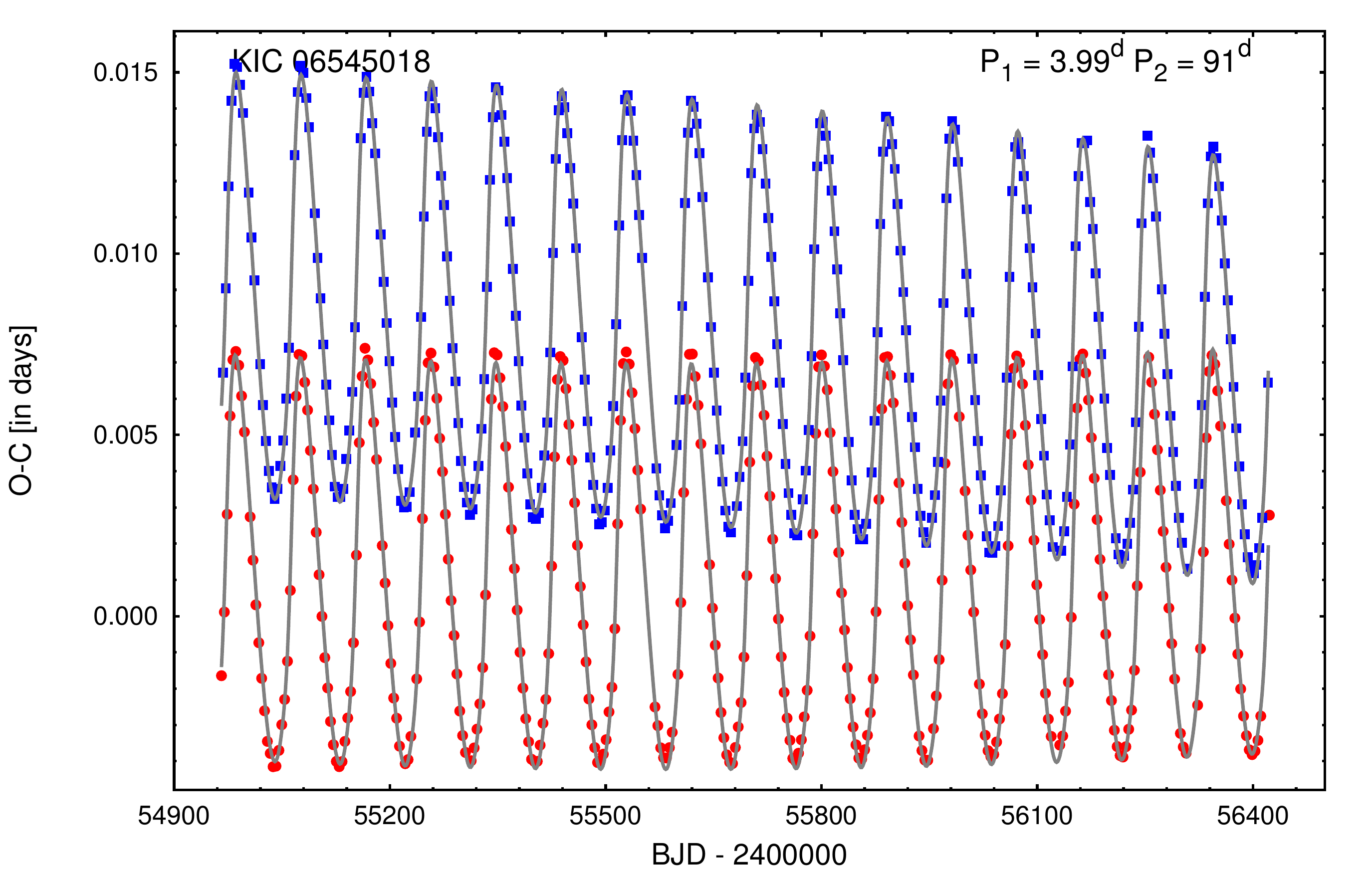}
\caption{(continued)}
\end{figure*}

\addtocounter{figure}{-1}

\begin{figure*}
\includegraphics[width=60mm]{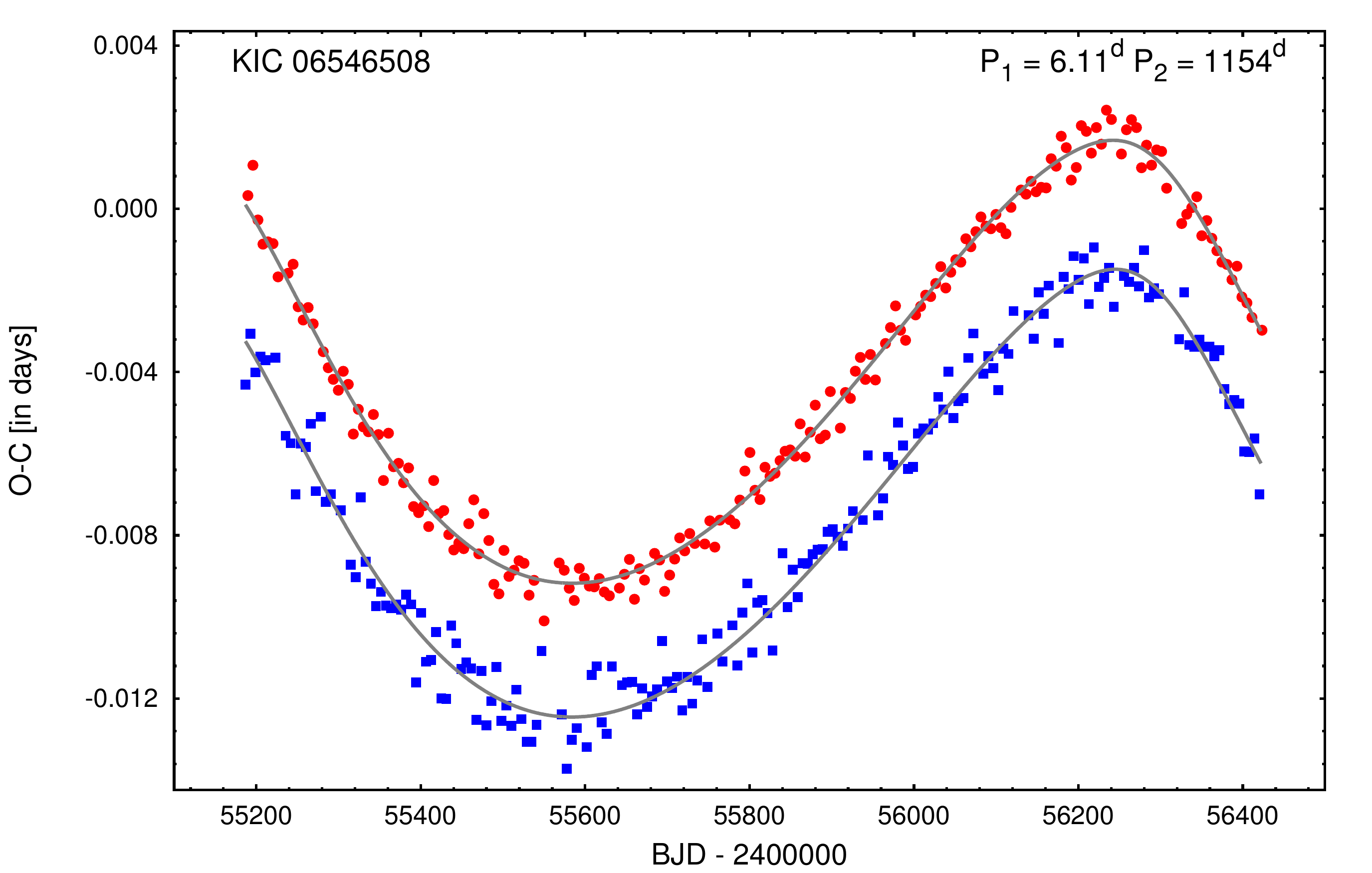}\includegraphics[width=60mm]{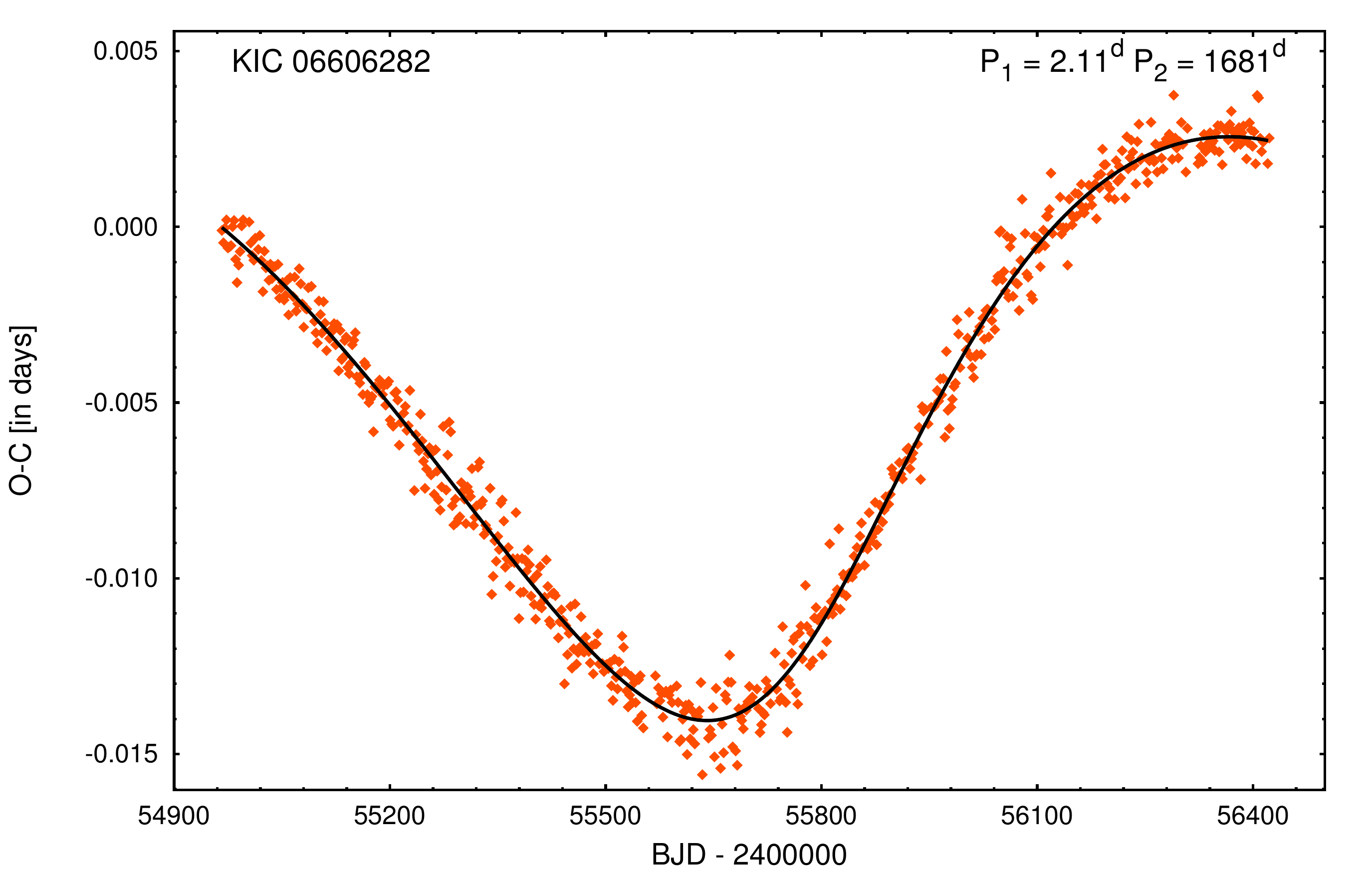}\includegraphics[width=60mm]{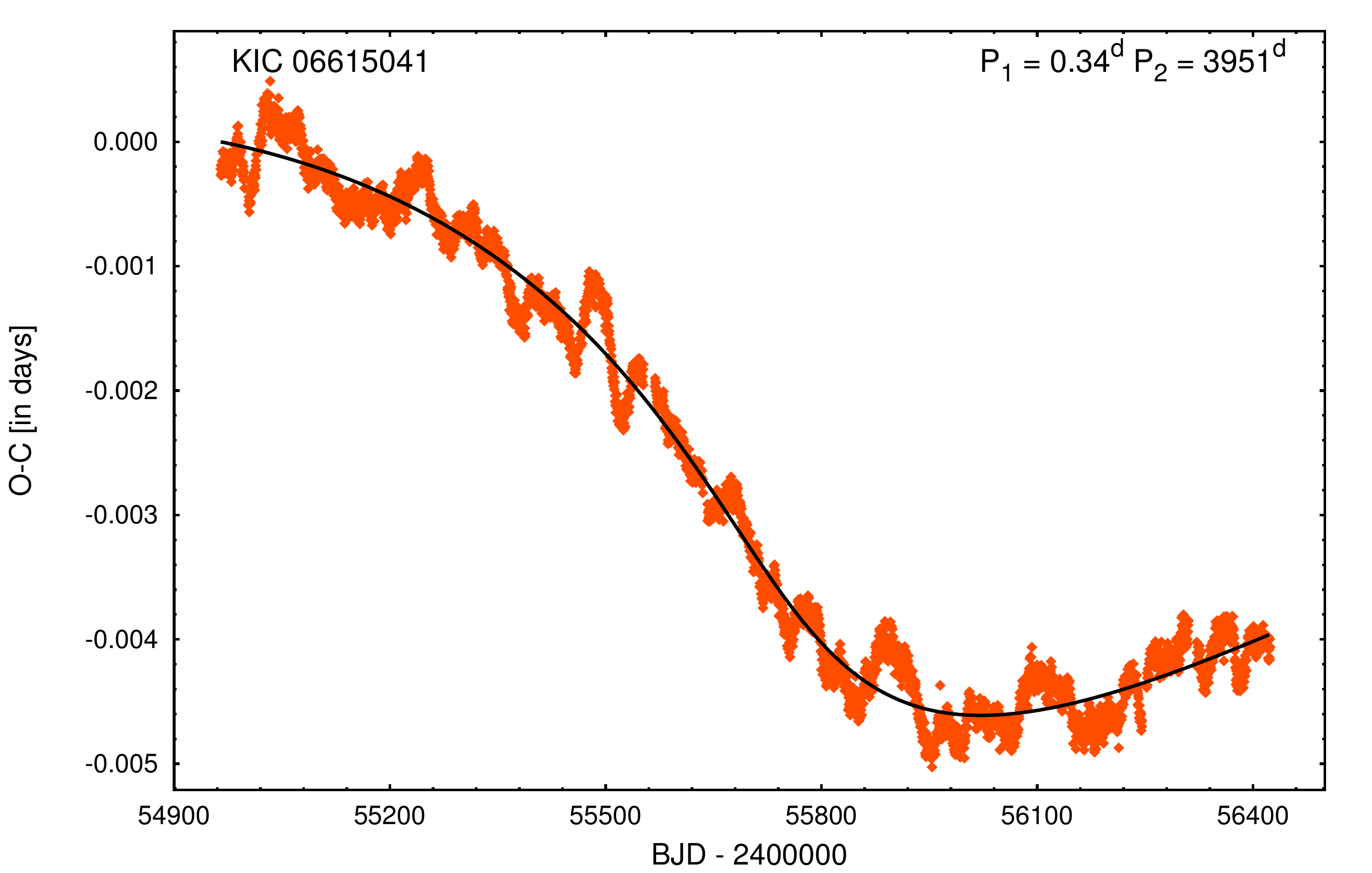}
\includegraphics[width=60mm]{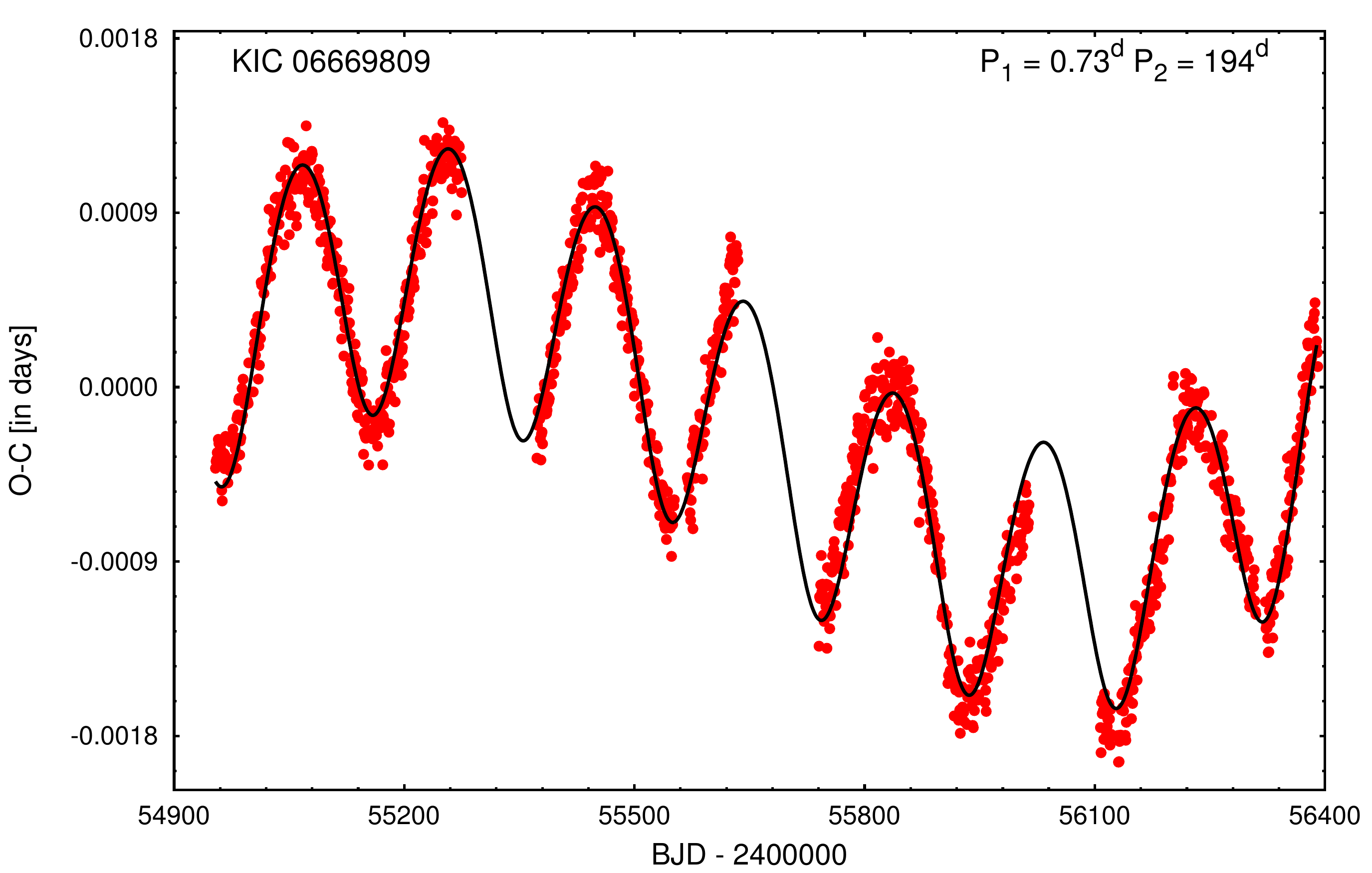}\includegraphics[width=60mm]{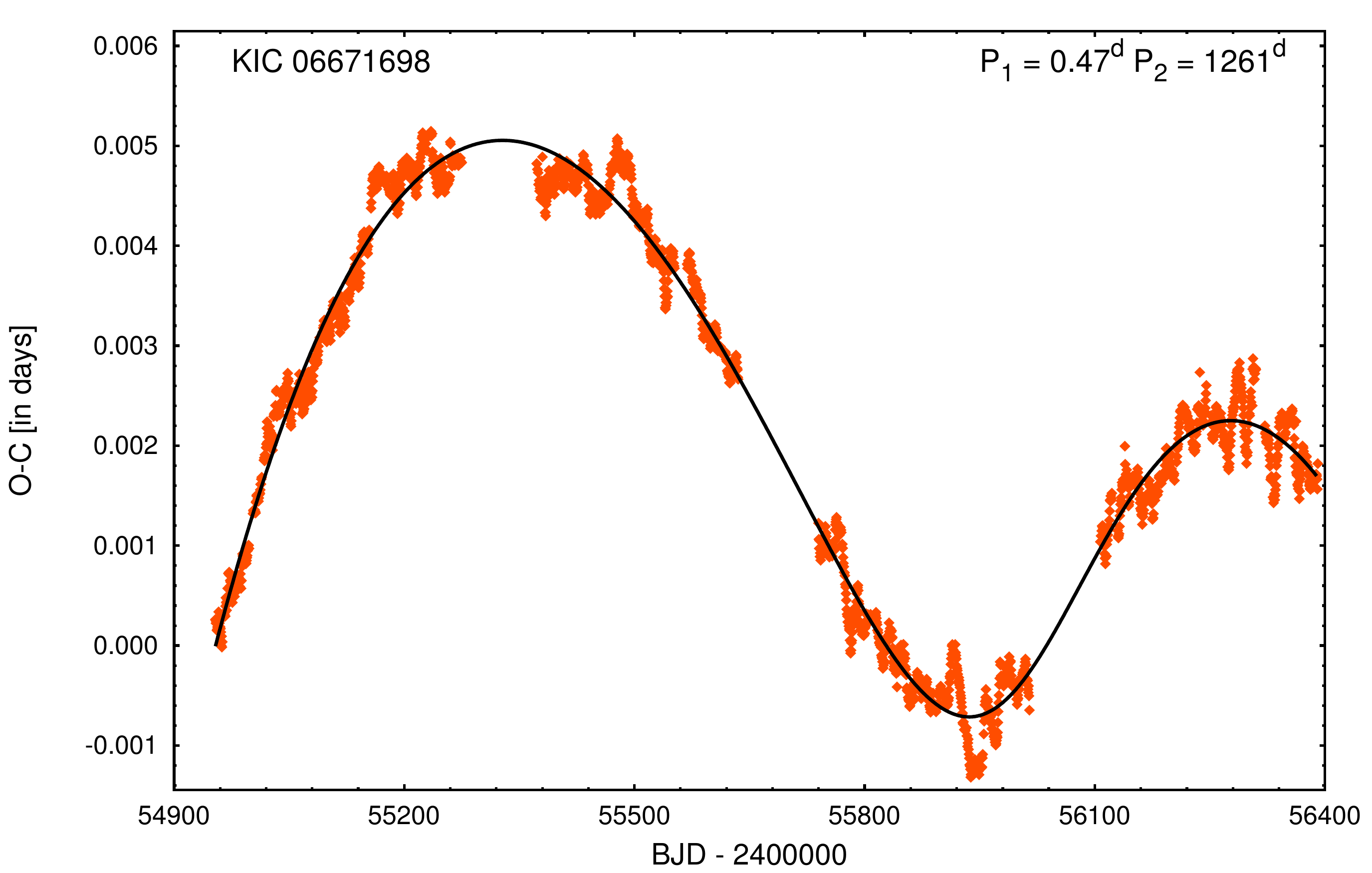}\includegraphics[width=60mm]{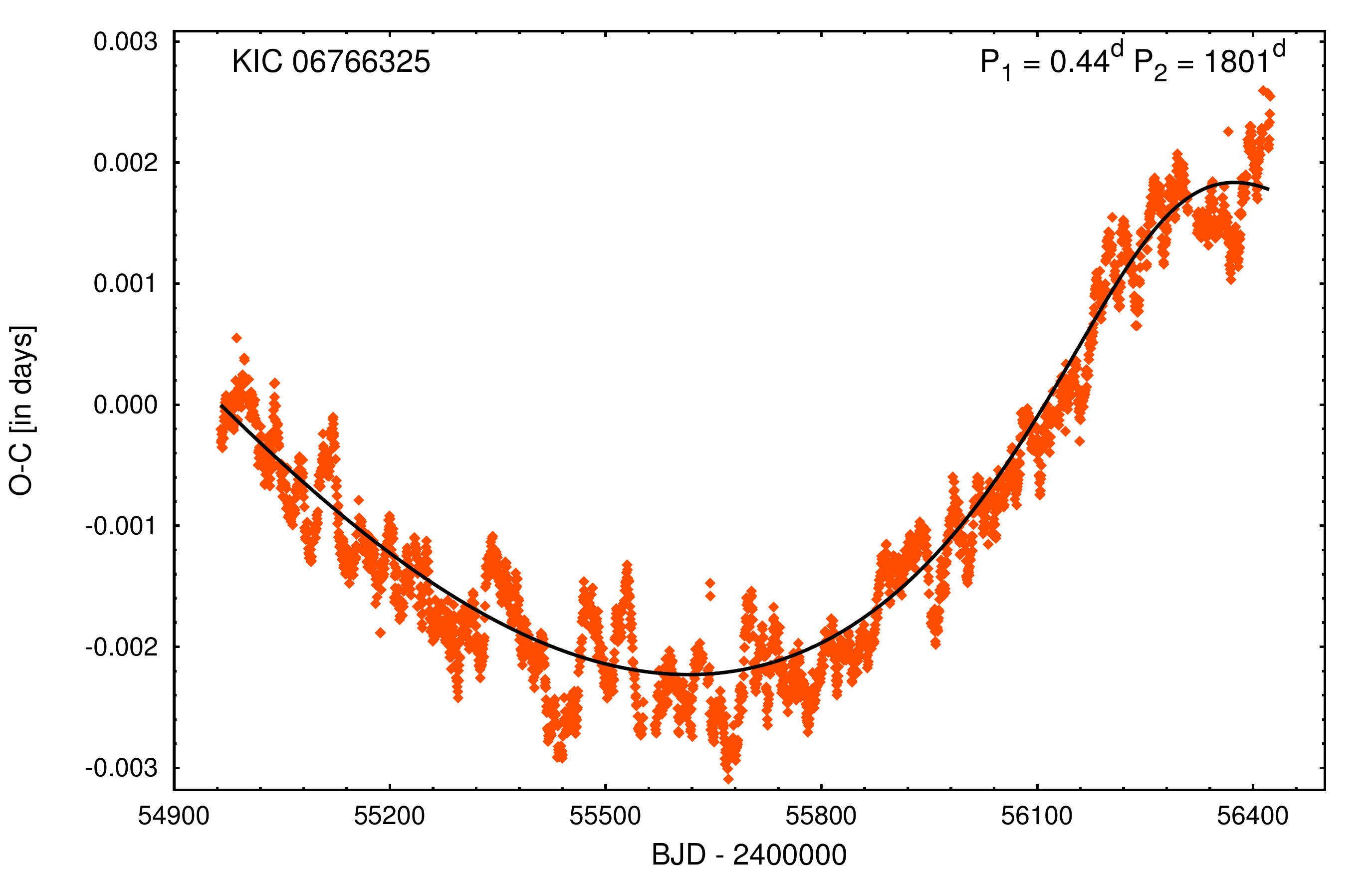}
\includegraphics[width=60mm]{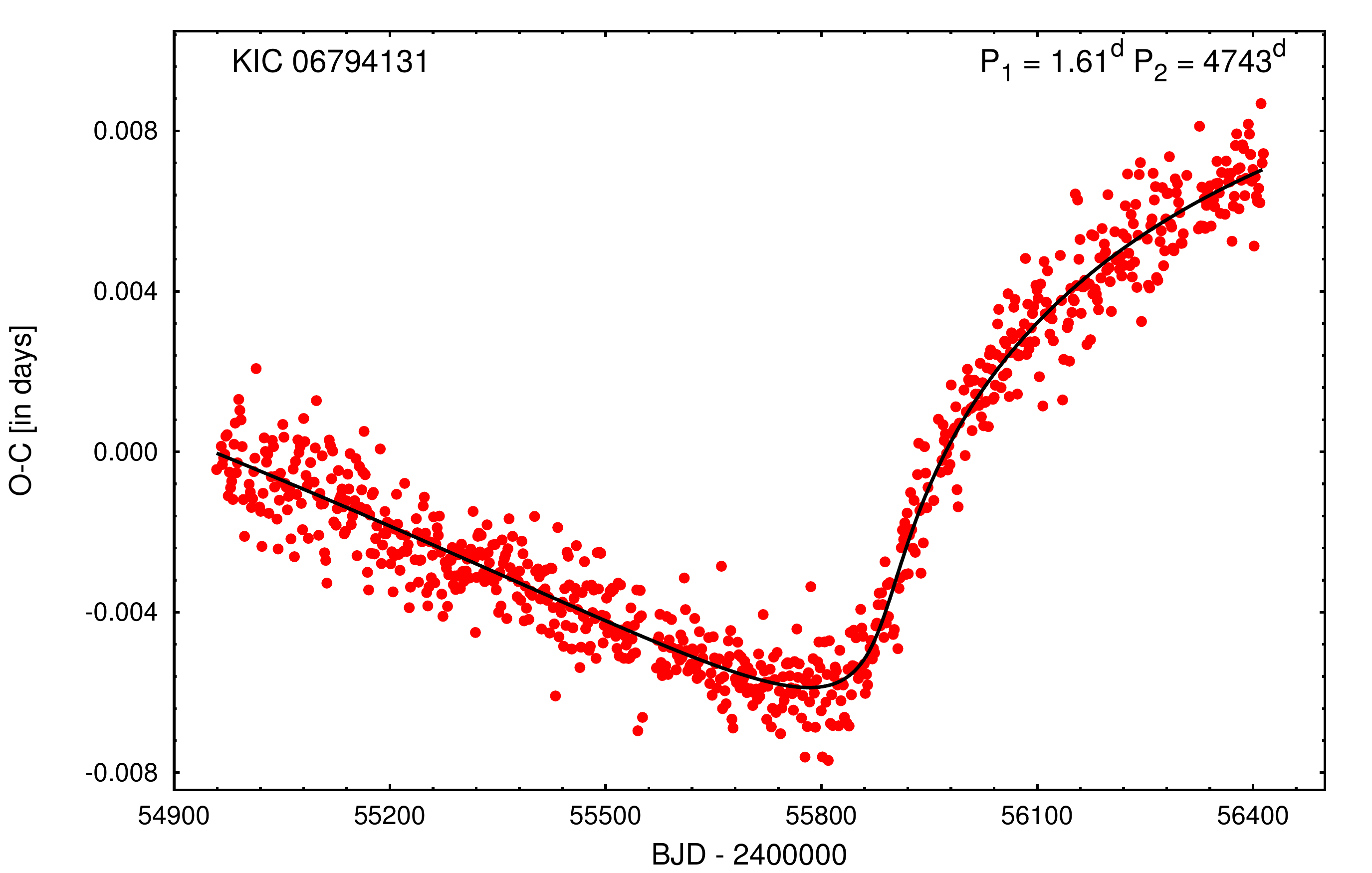}\includegraphics[width=60mm]{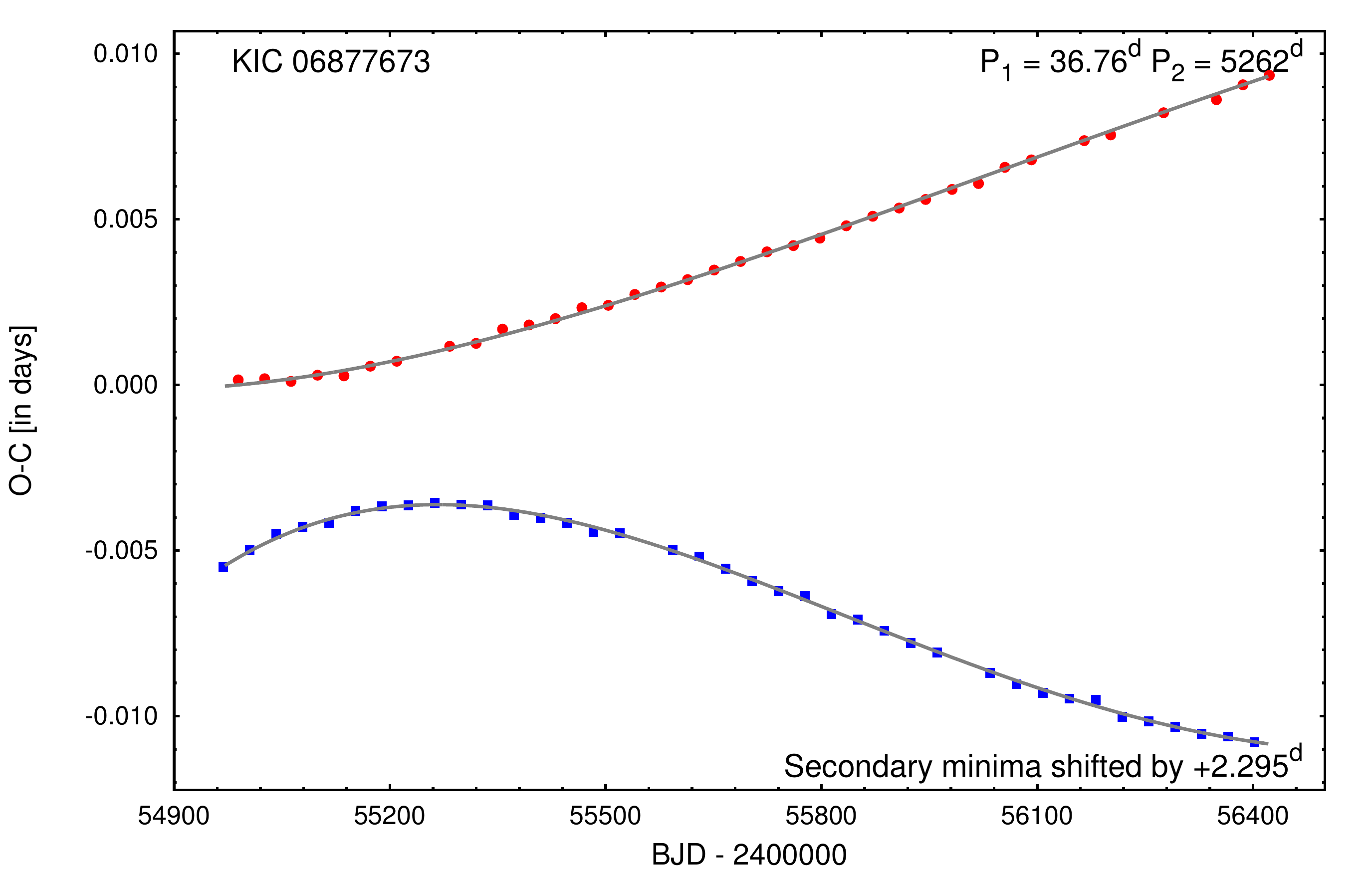}\includegraphics[width=60mm]{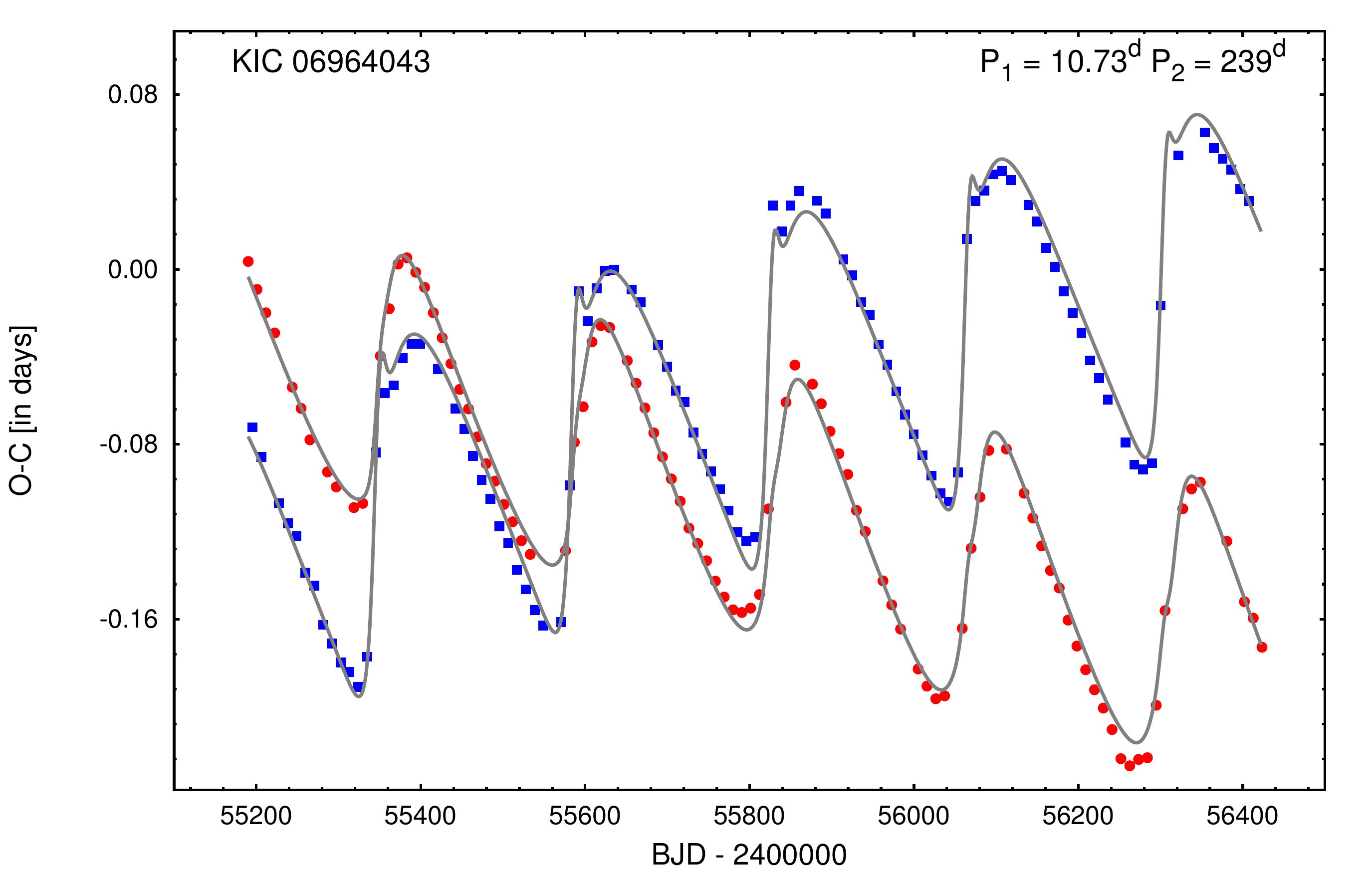}
\includegraphics[width=60mm]{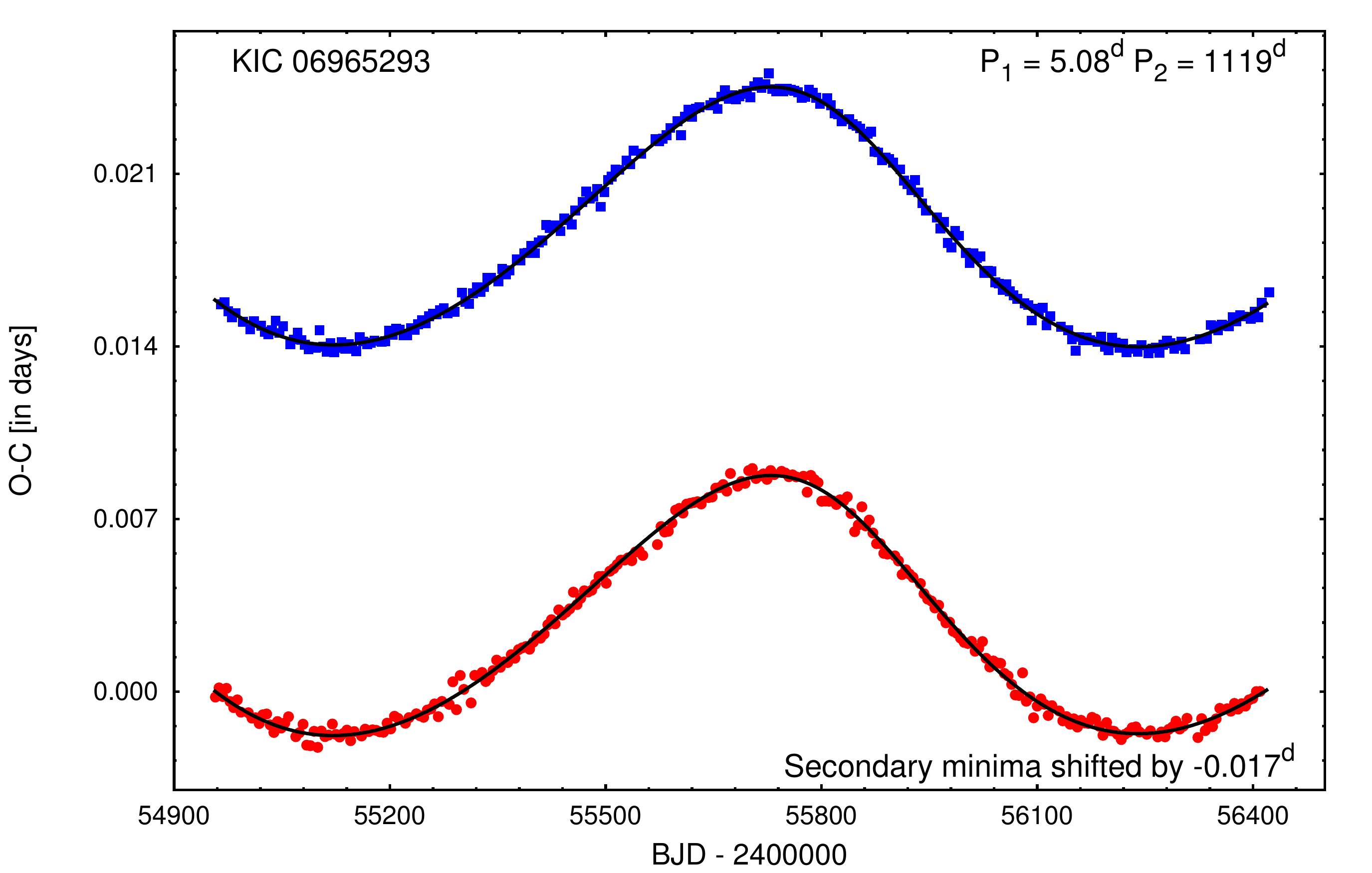}\includegraphics[width=60mm]{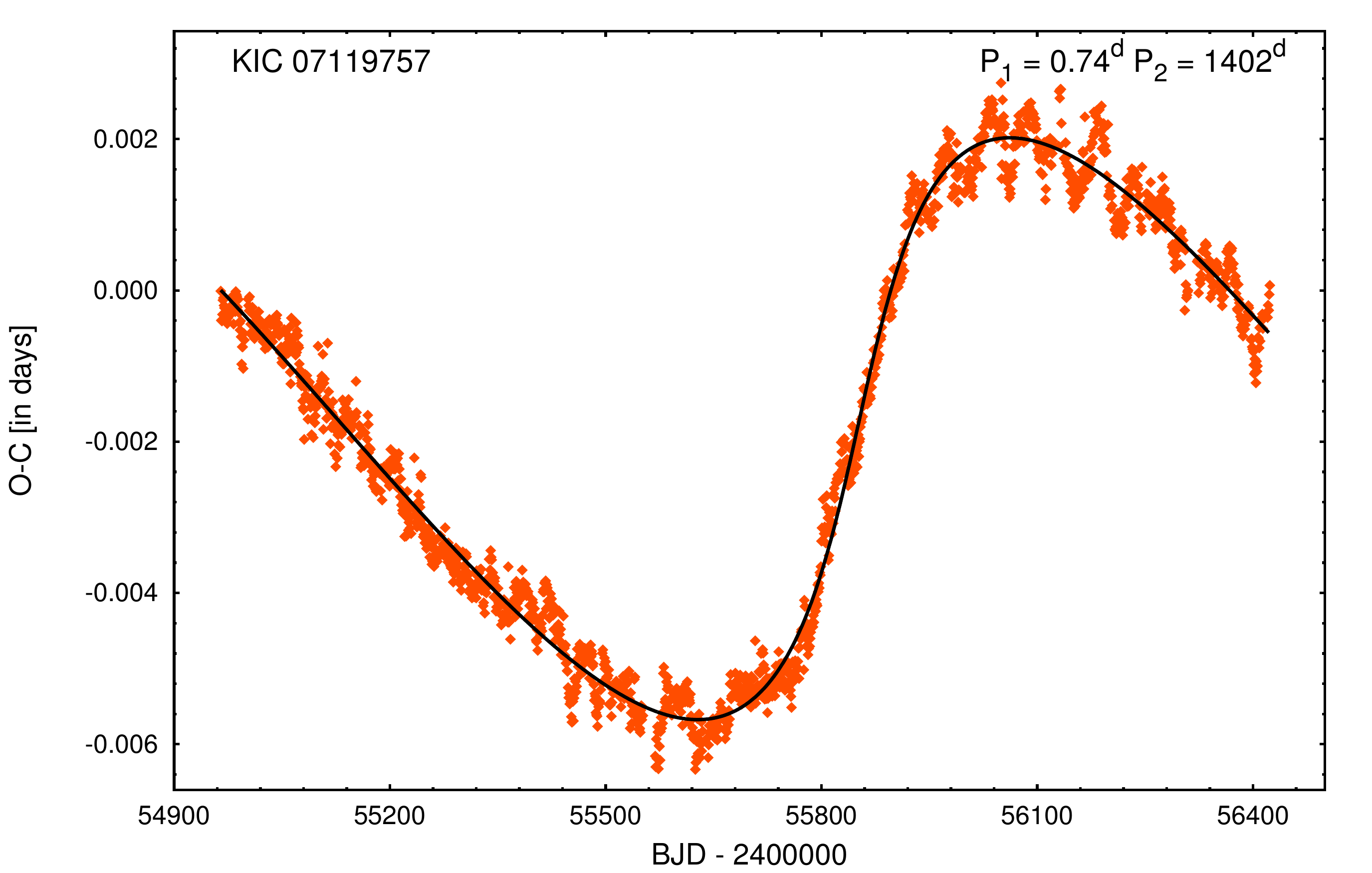}\includegraphics[width=60mm]{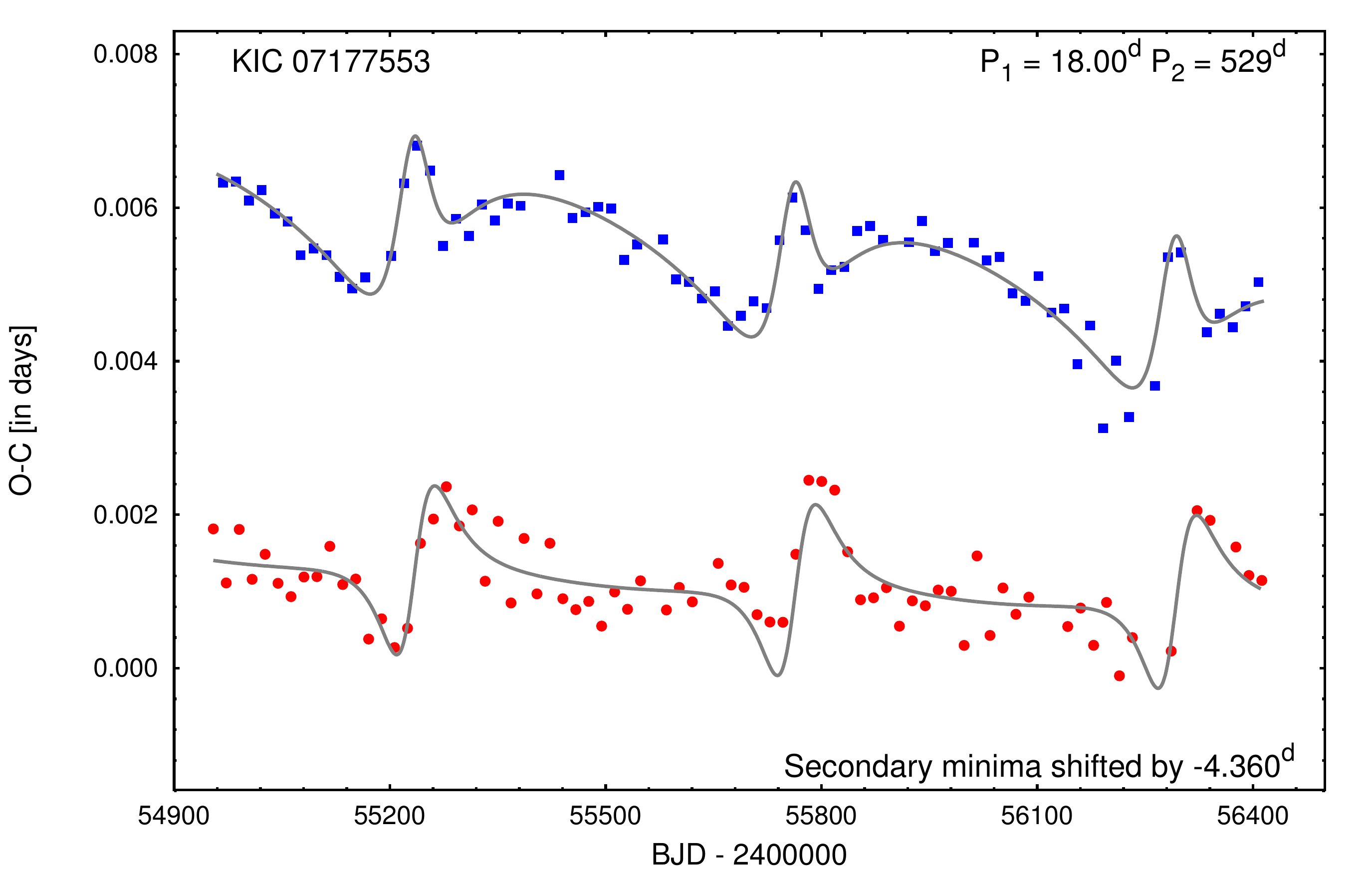}
\includegraphics[width=60mm]{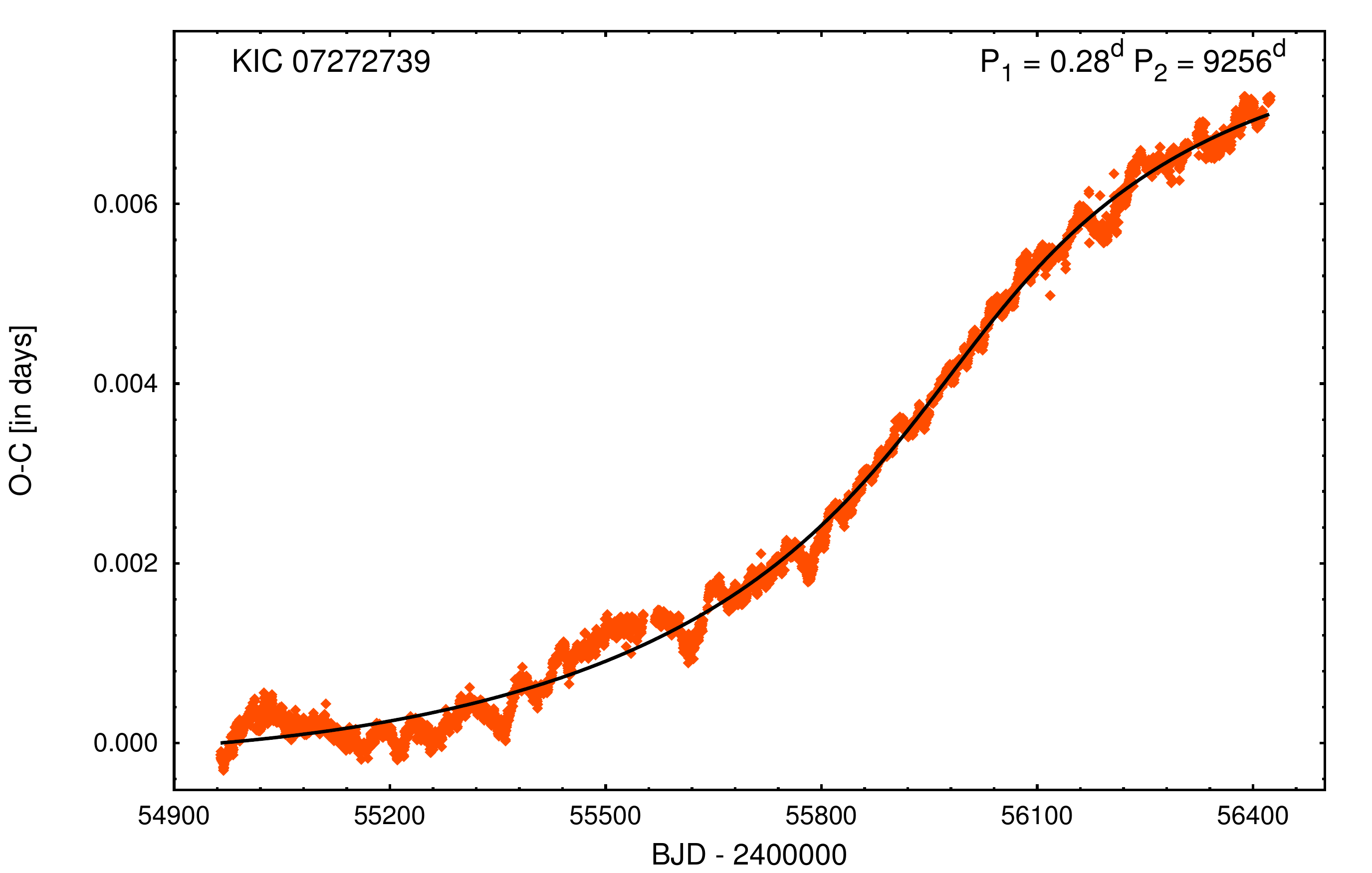}\includegraphics[width=60mm]{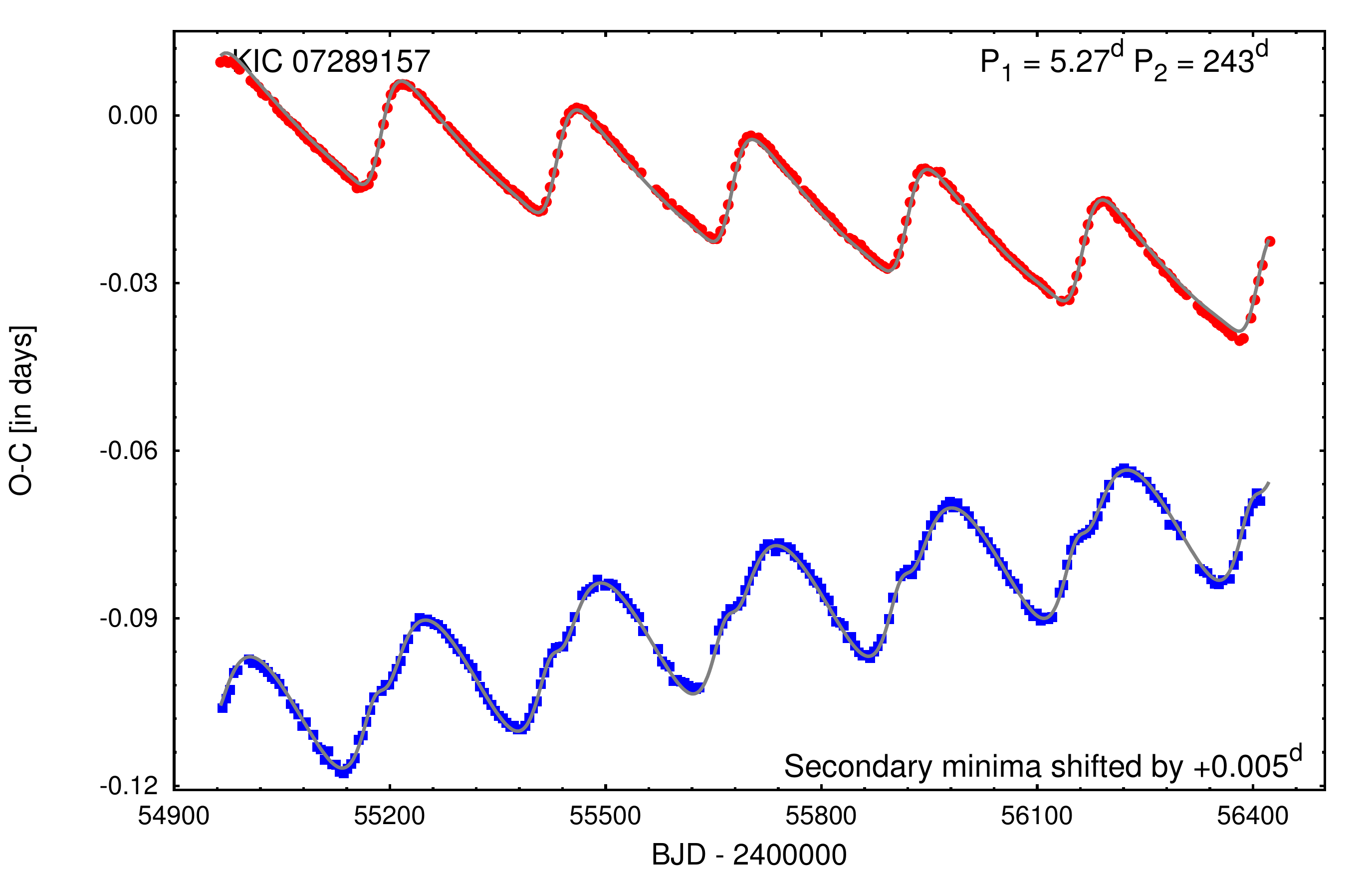}\includegraphics[width=60mm]{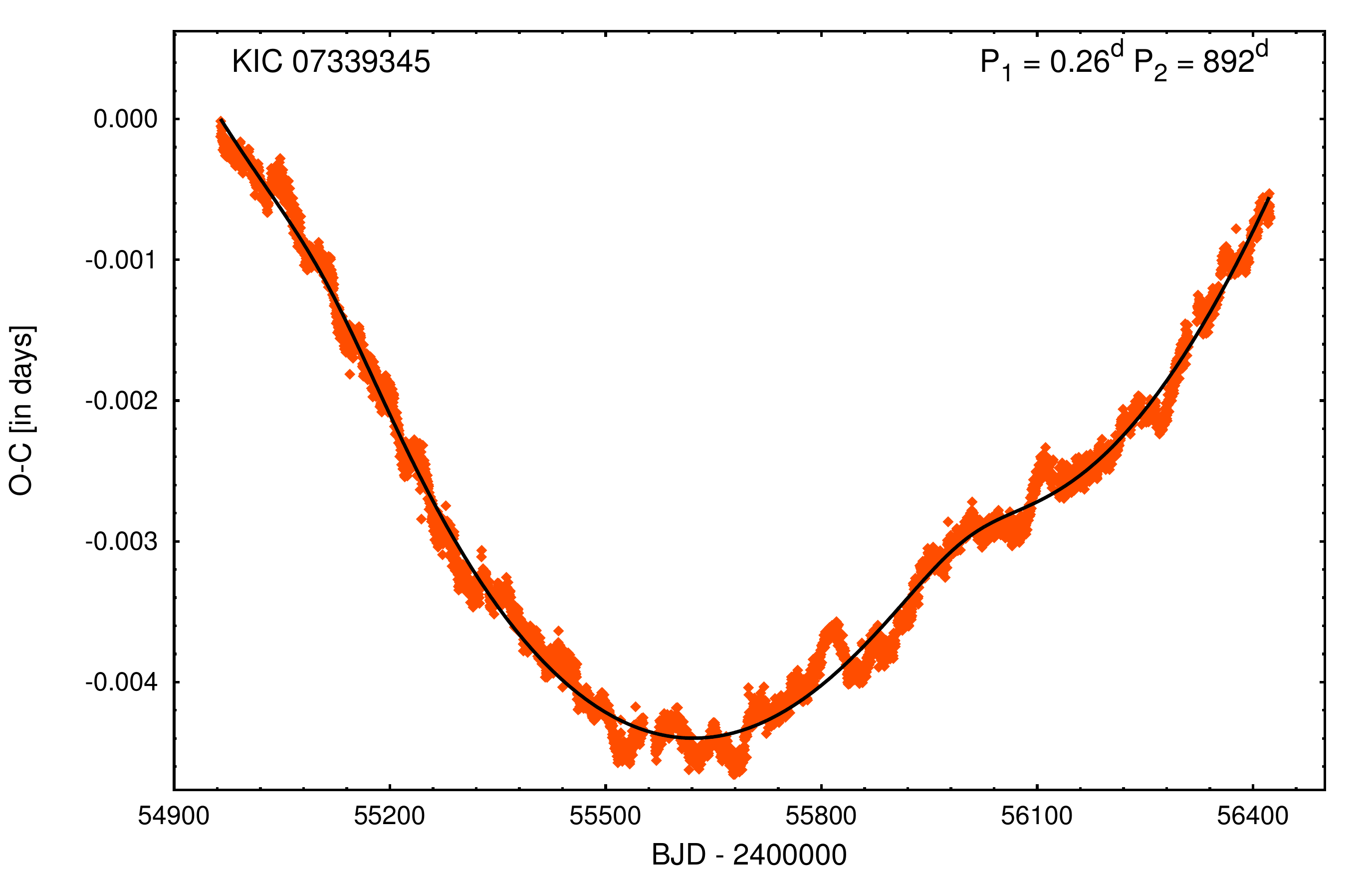}
\includegraphics[width=60mm]{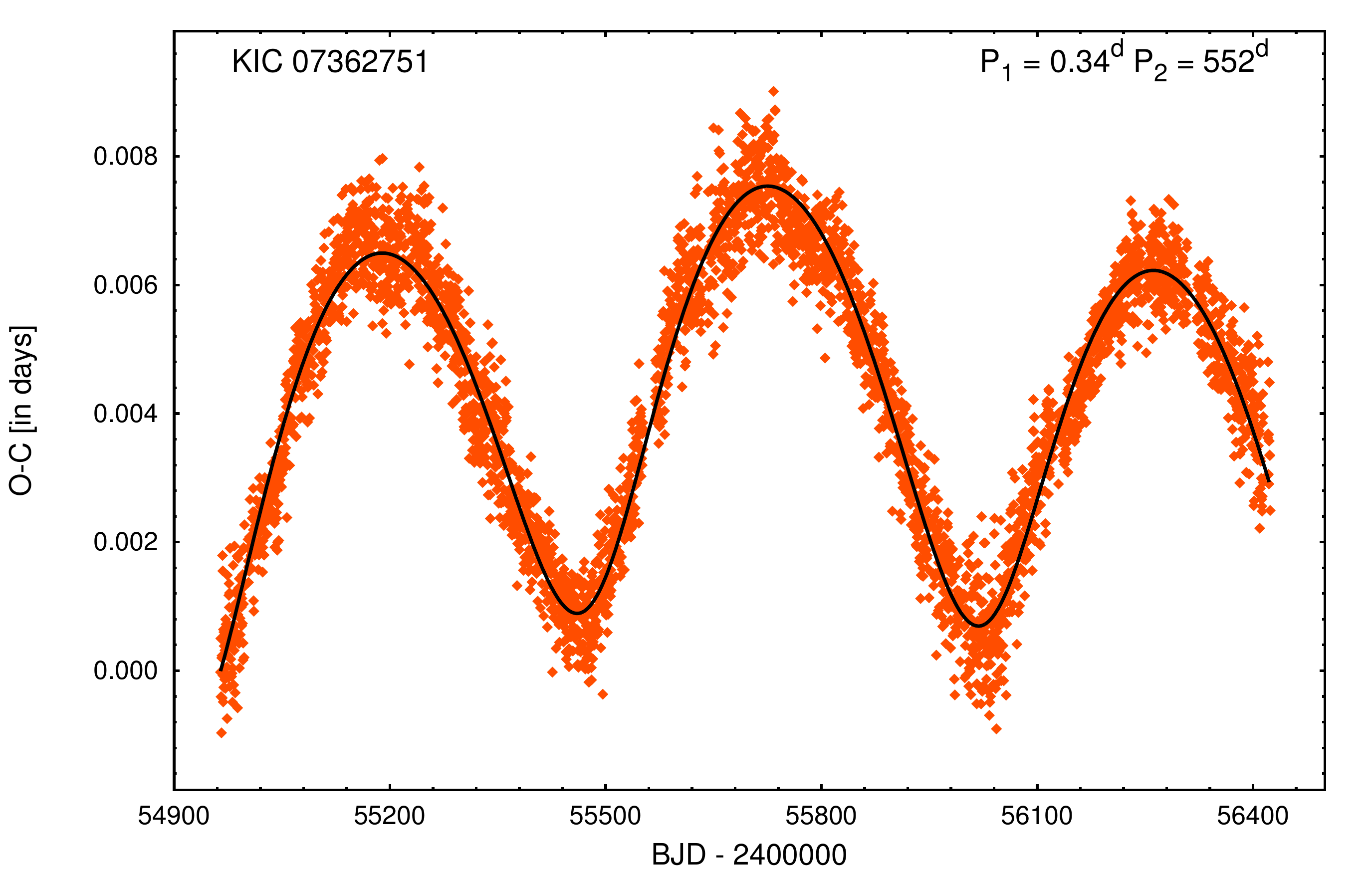}\includegraphics[width=60mm]{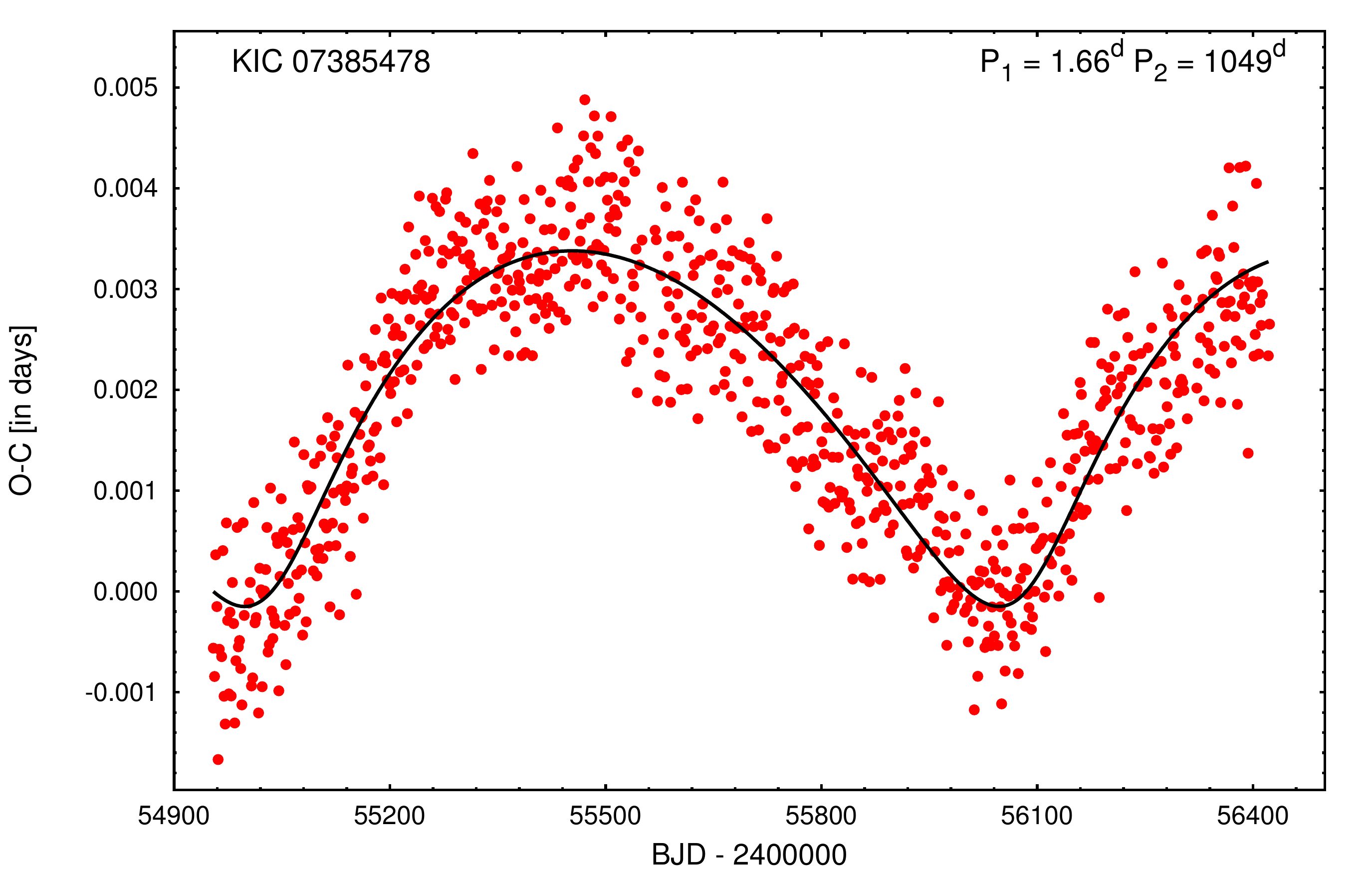}\includegraphics[width=60mm]{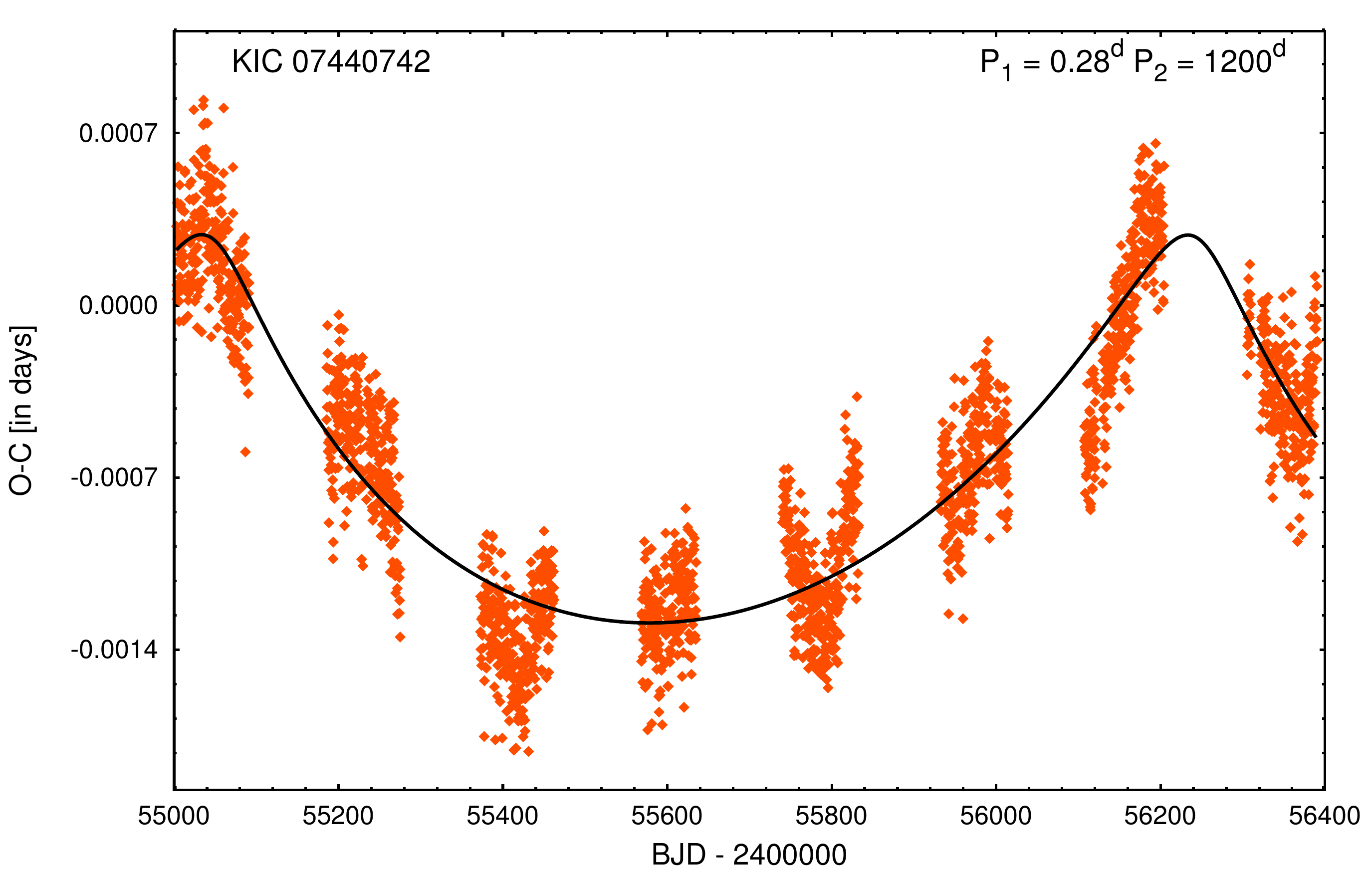}
\caption{(continued)}
\end{figure*}

\addtocounter{figure}{-1}

\begin{figure*}
\includegraphics[width=60mm]{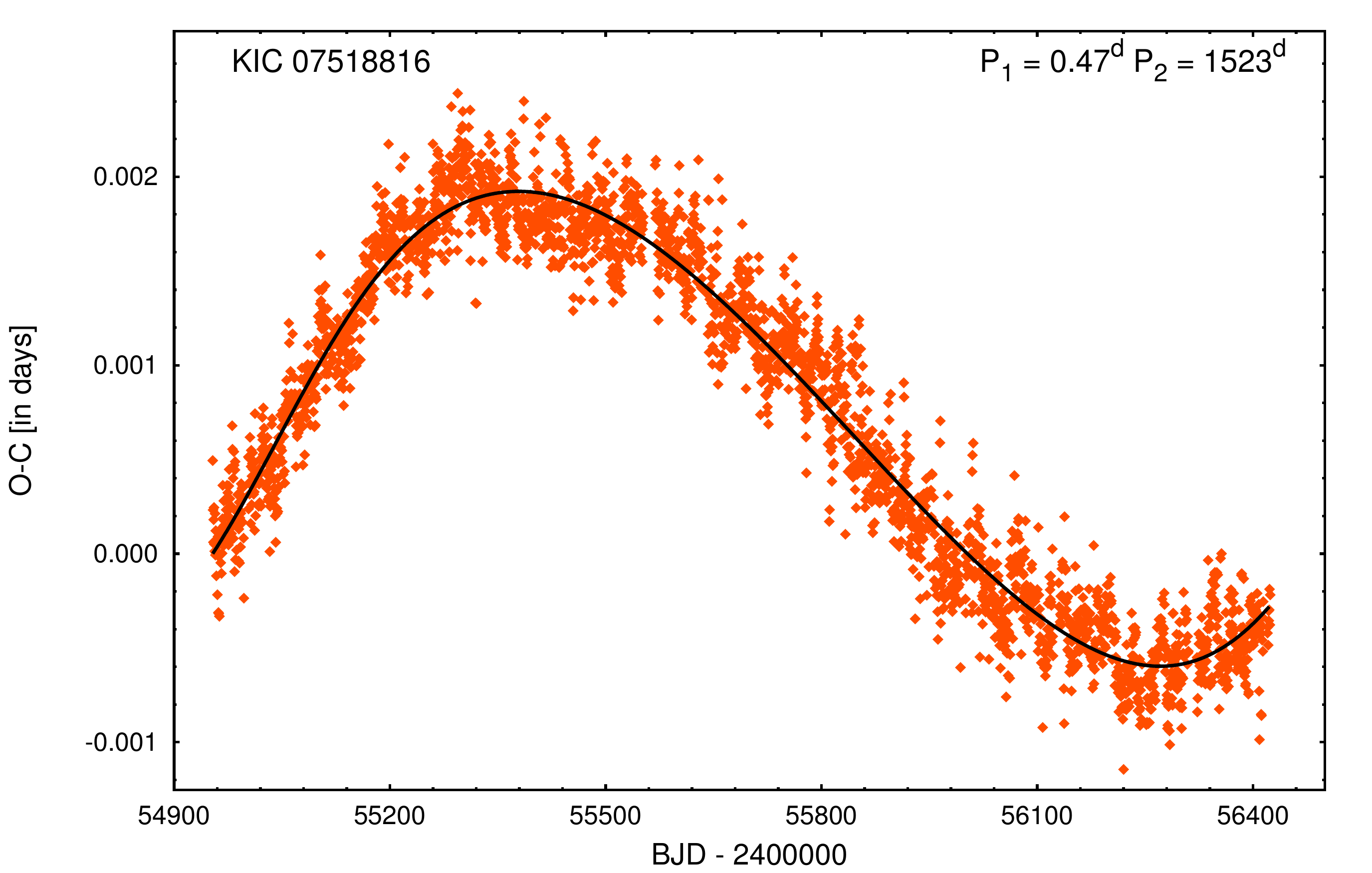}\includegraphics[width=60mm]{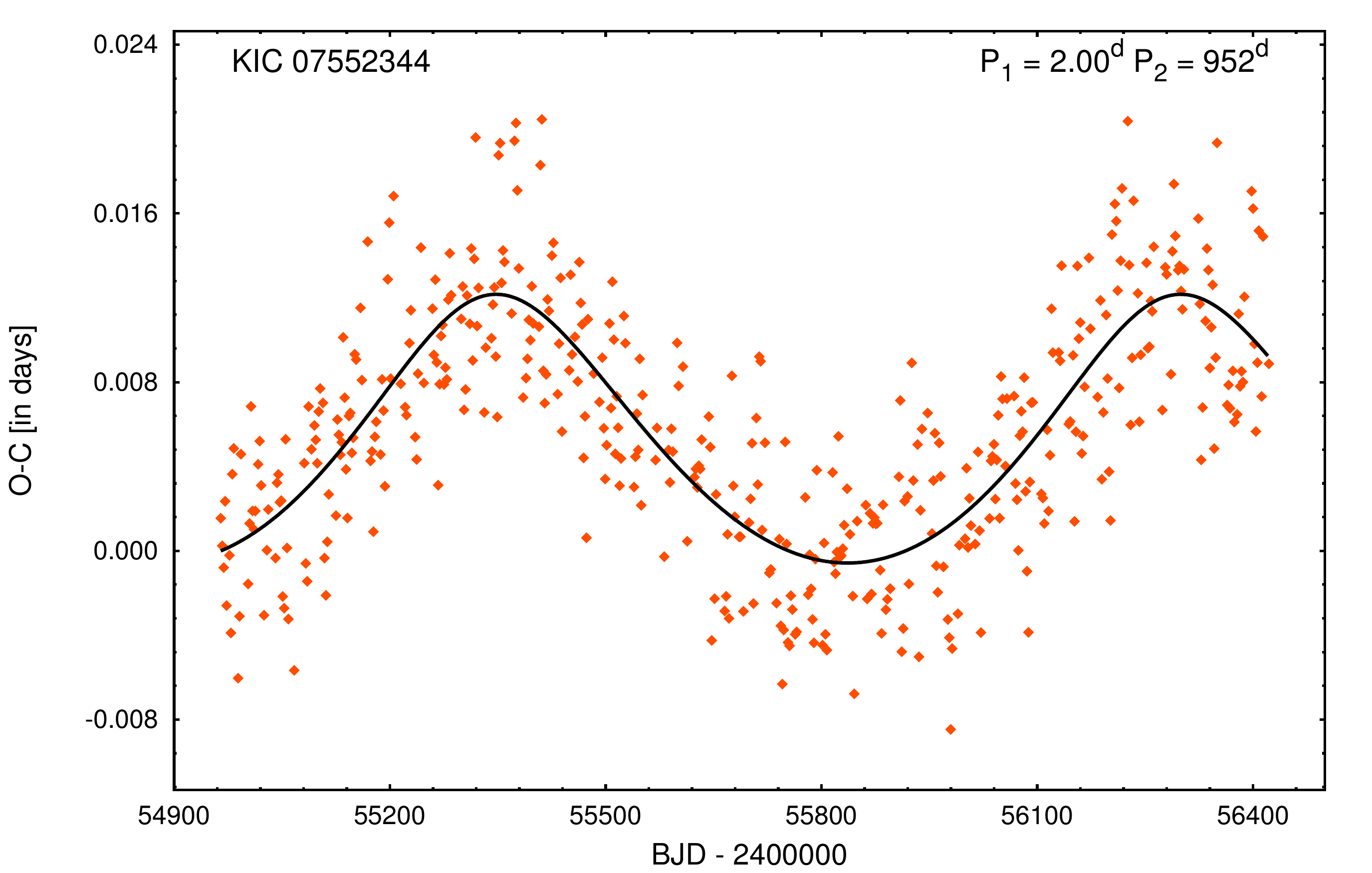}\includegraphics[width=60mm]{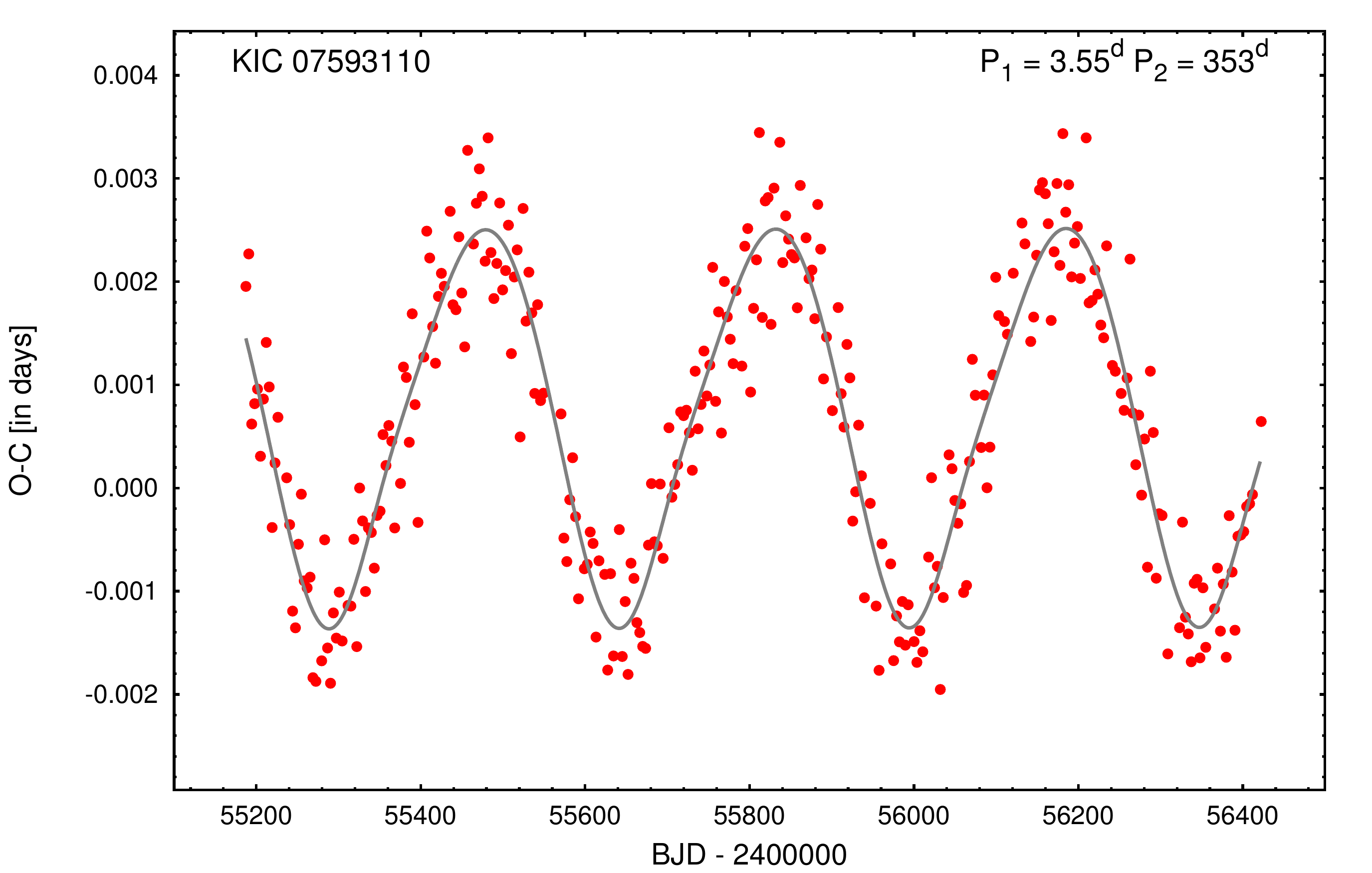}
\includegraphics[width=60mm]{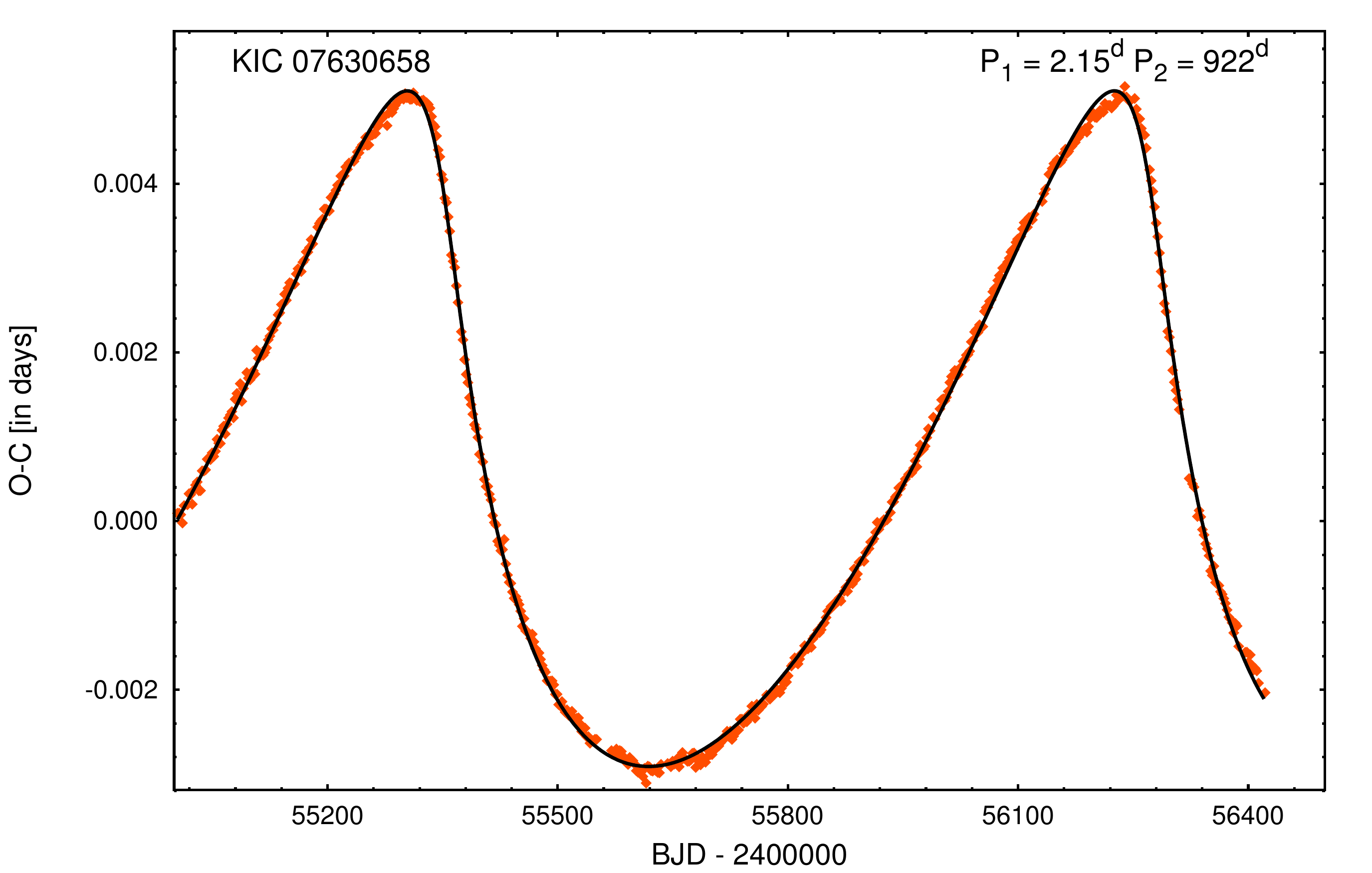}\includegraphics[width=60mm]{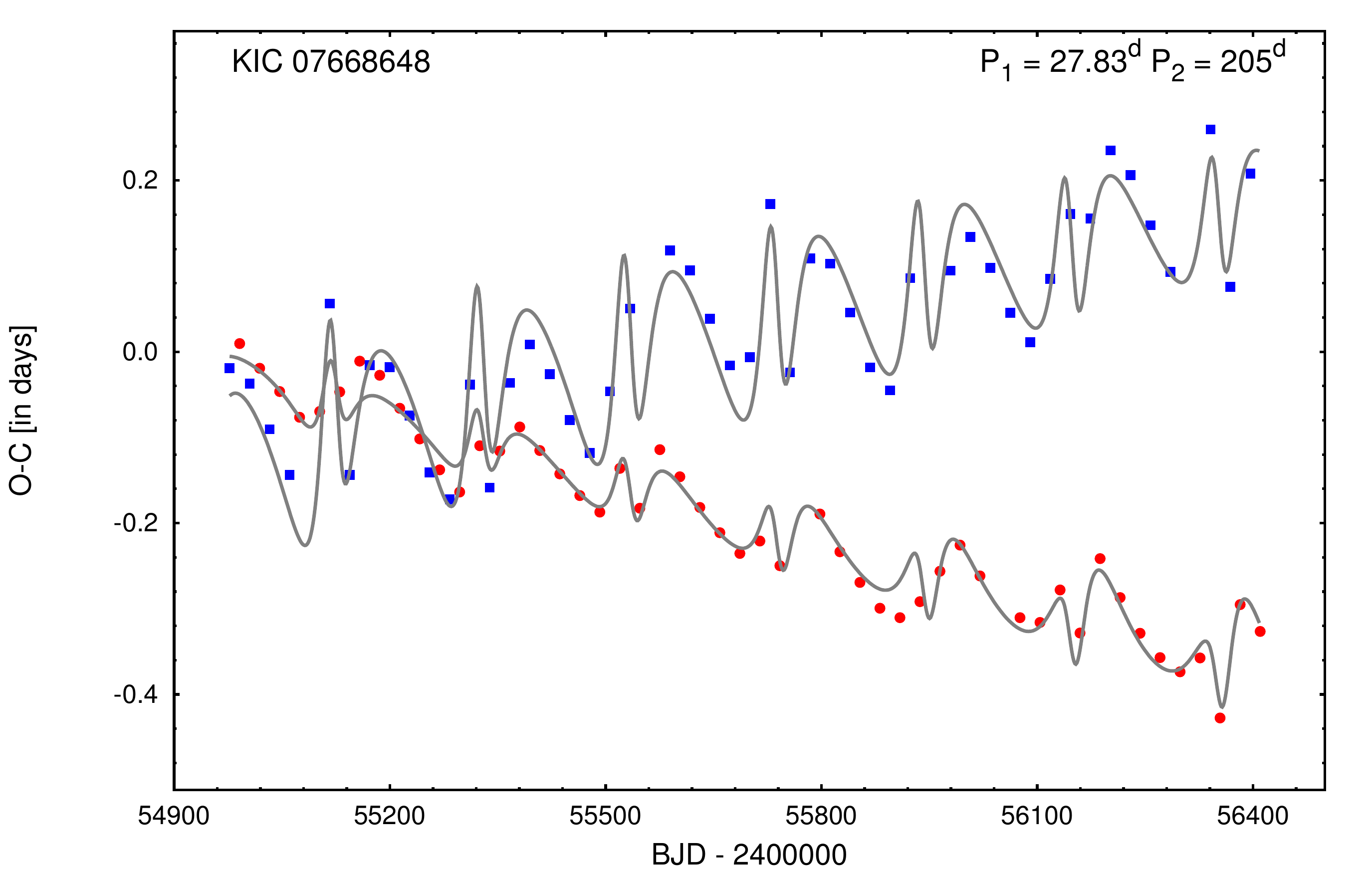}\includegraphics[width=60mm]{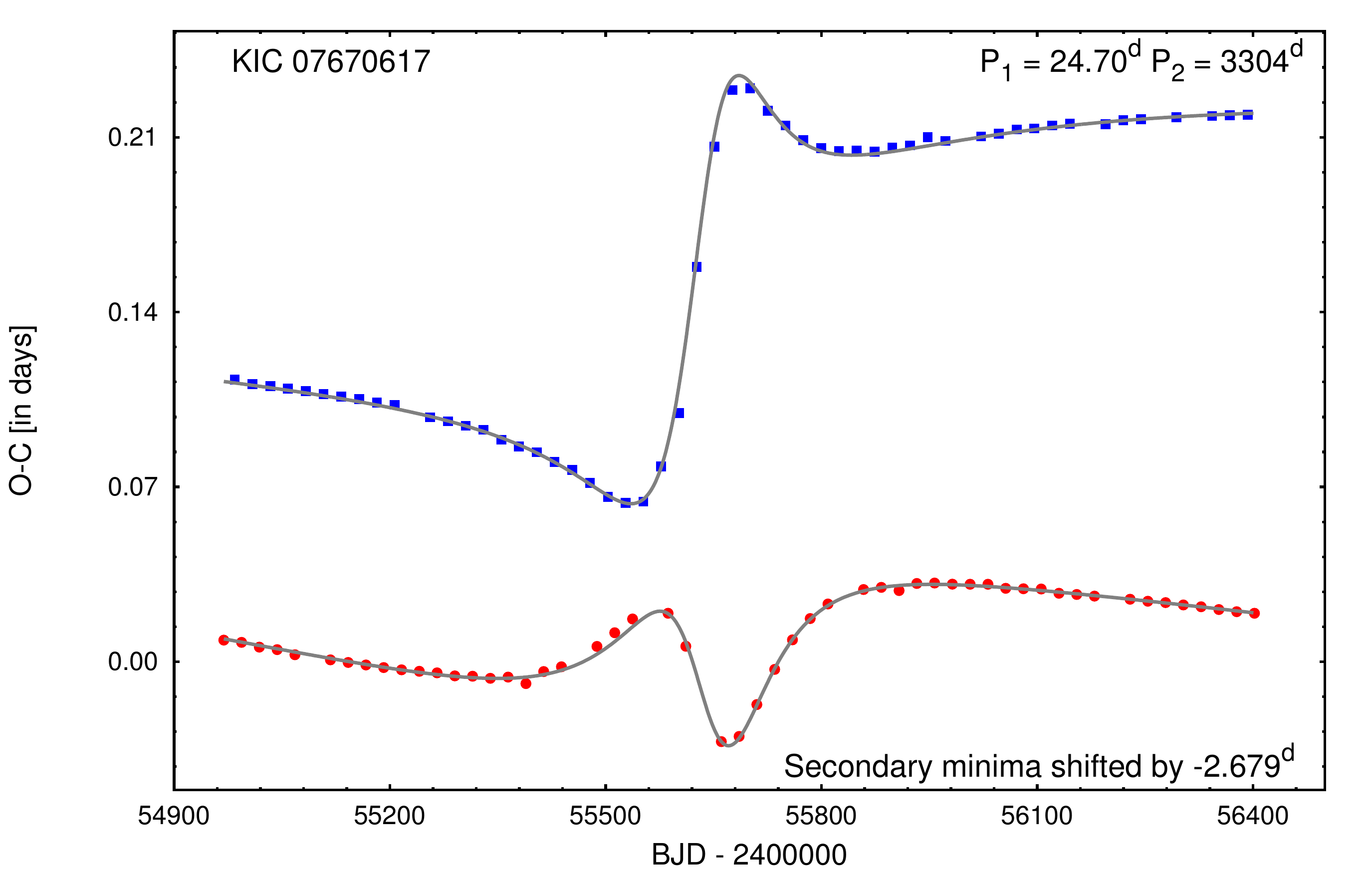}
\includegraphics[width=60mm]{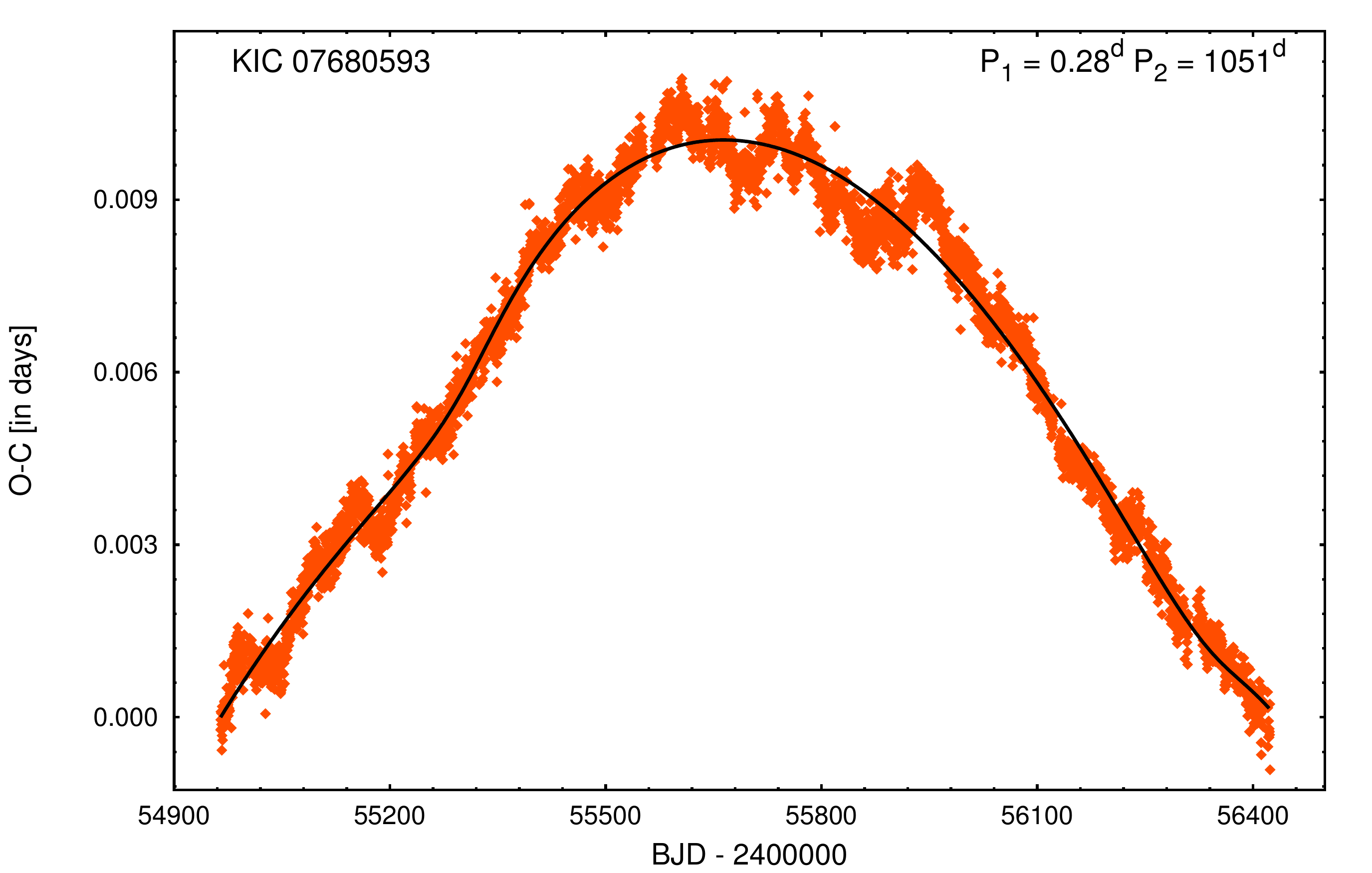}\includegraphics[width=60mm]{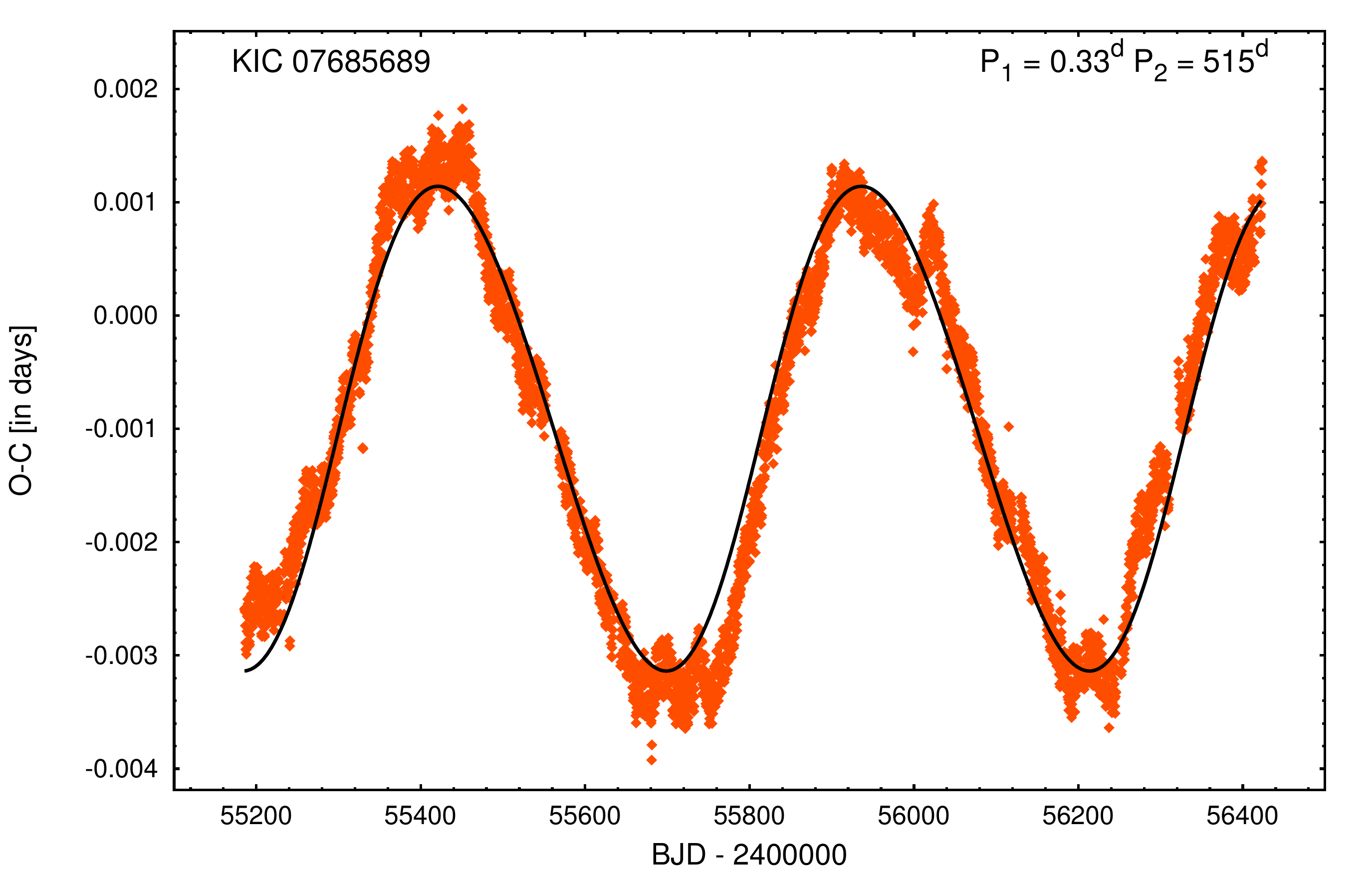}\includegraphics[width=60mm]{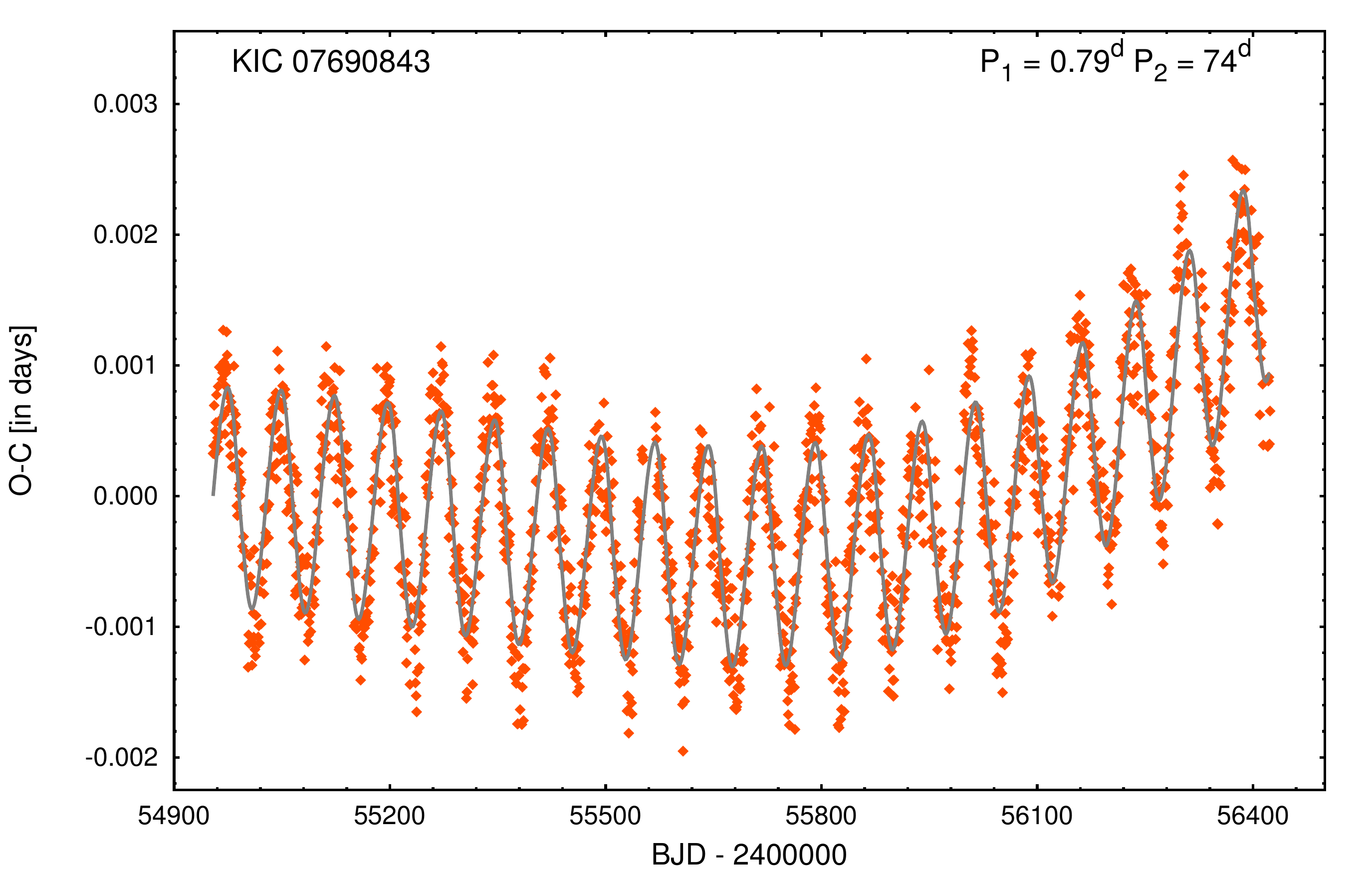}
\includegraphics[width=60mm]{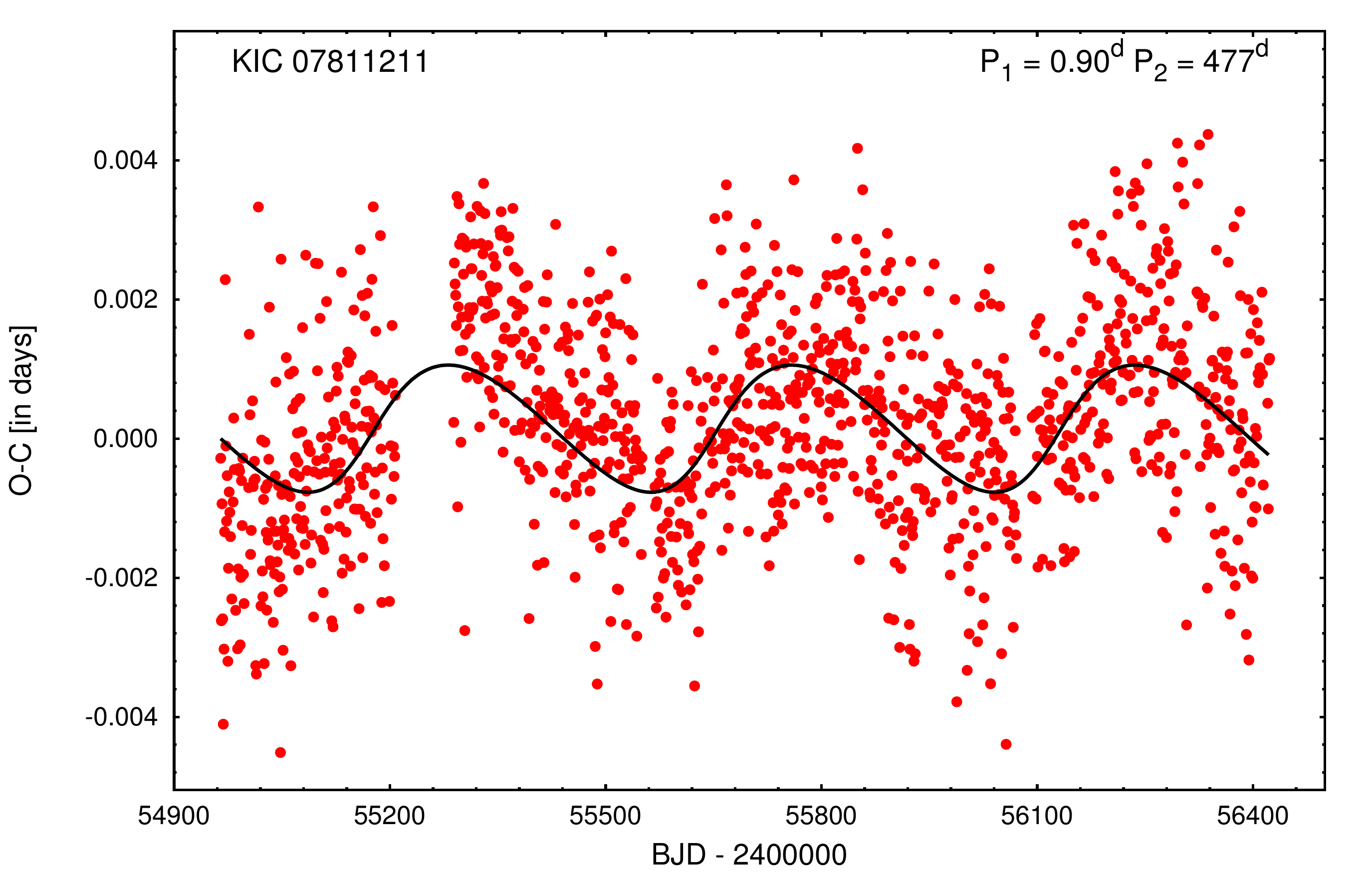}\includegraphics[width=60mm]{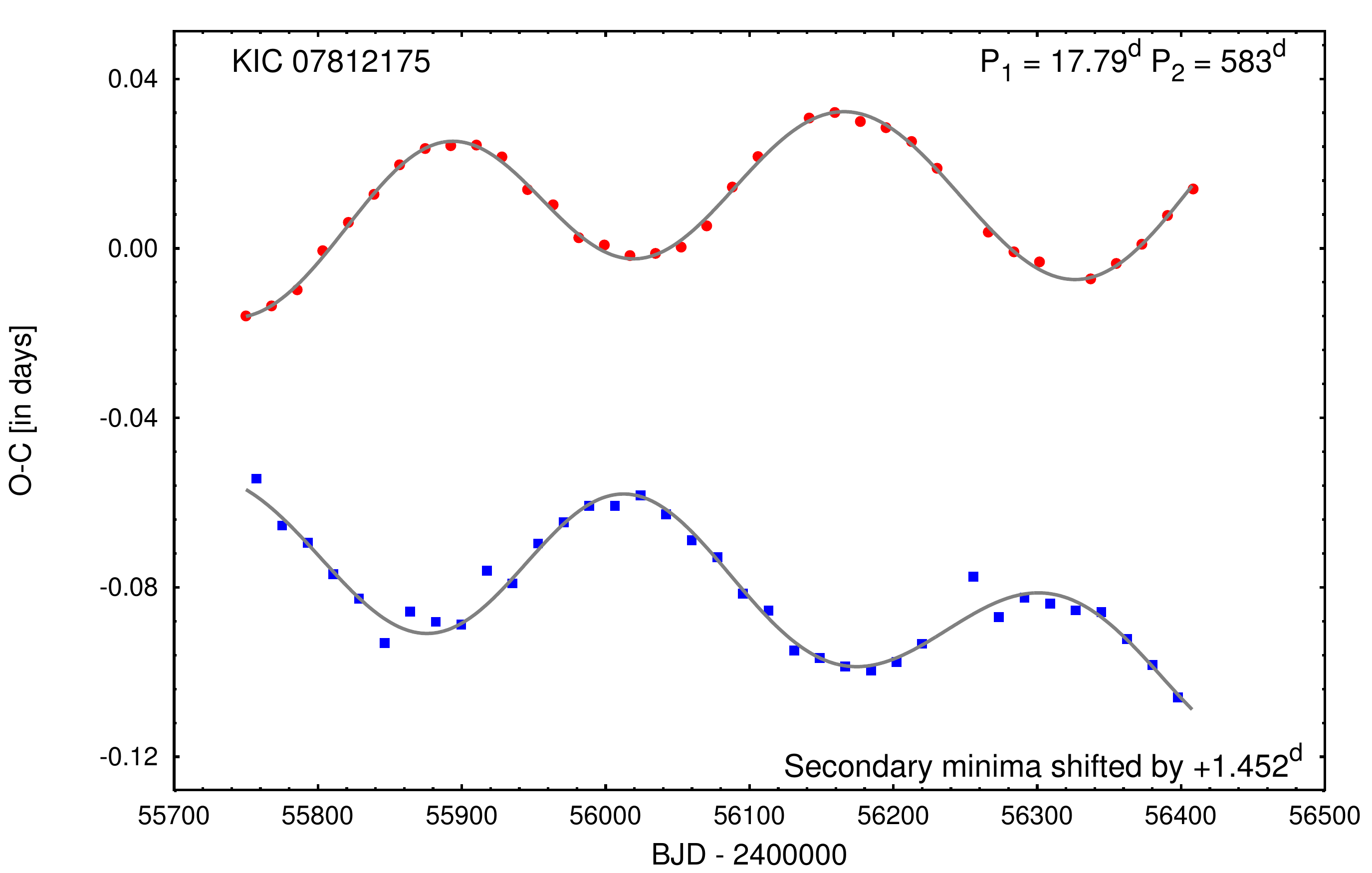}\includegraphics[width=60mm]{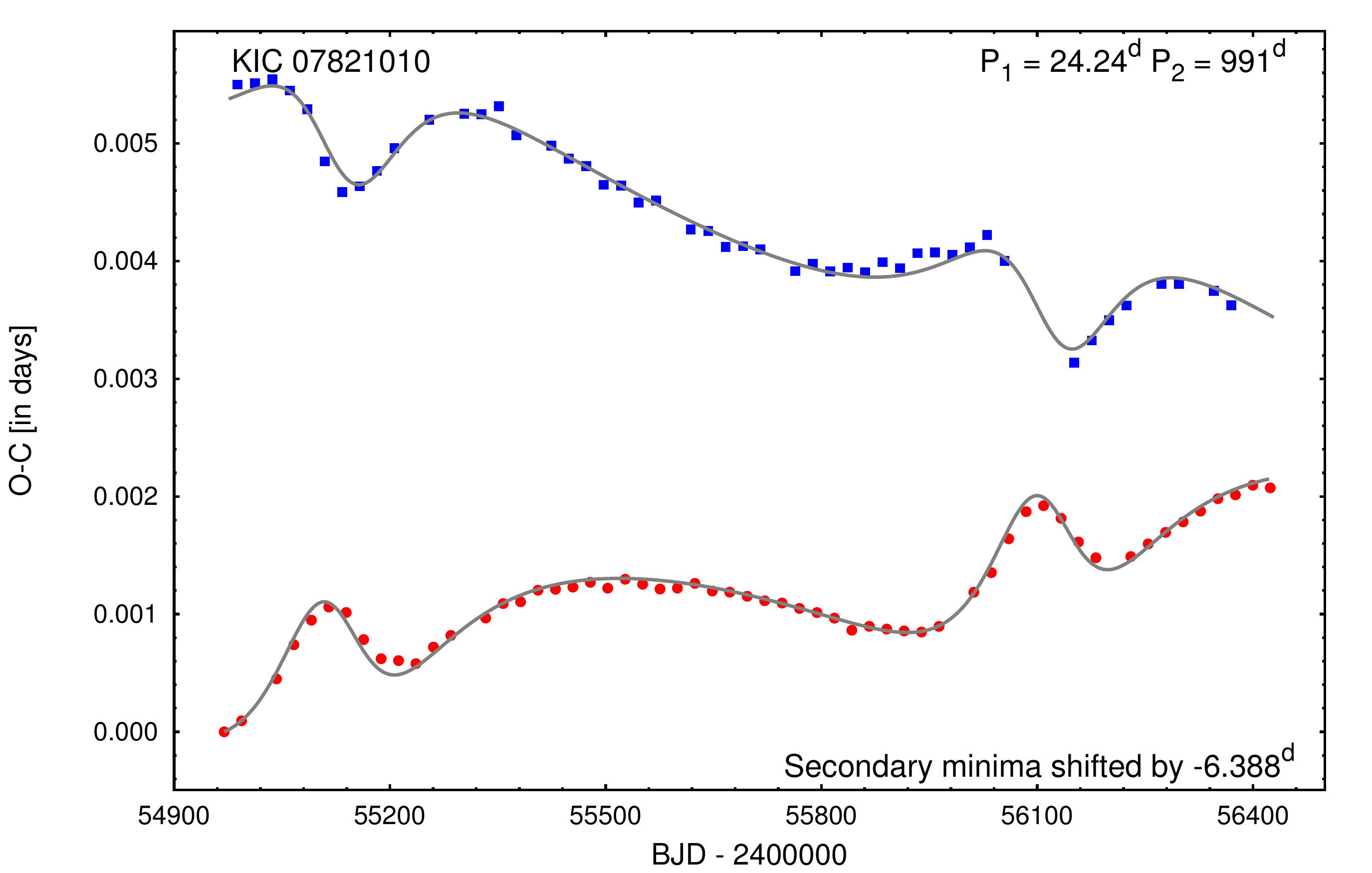}
\includegraphics[width=60mm]{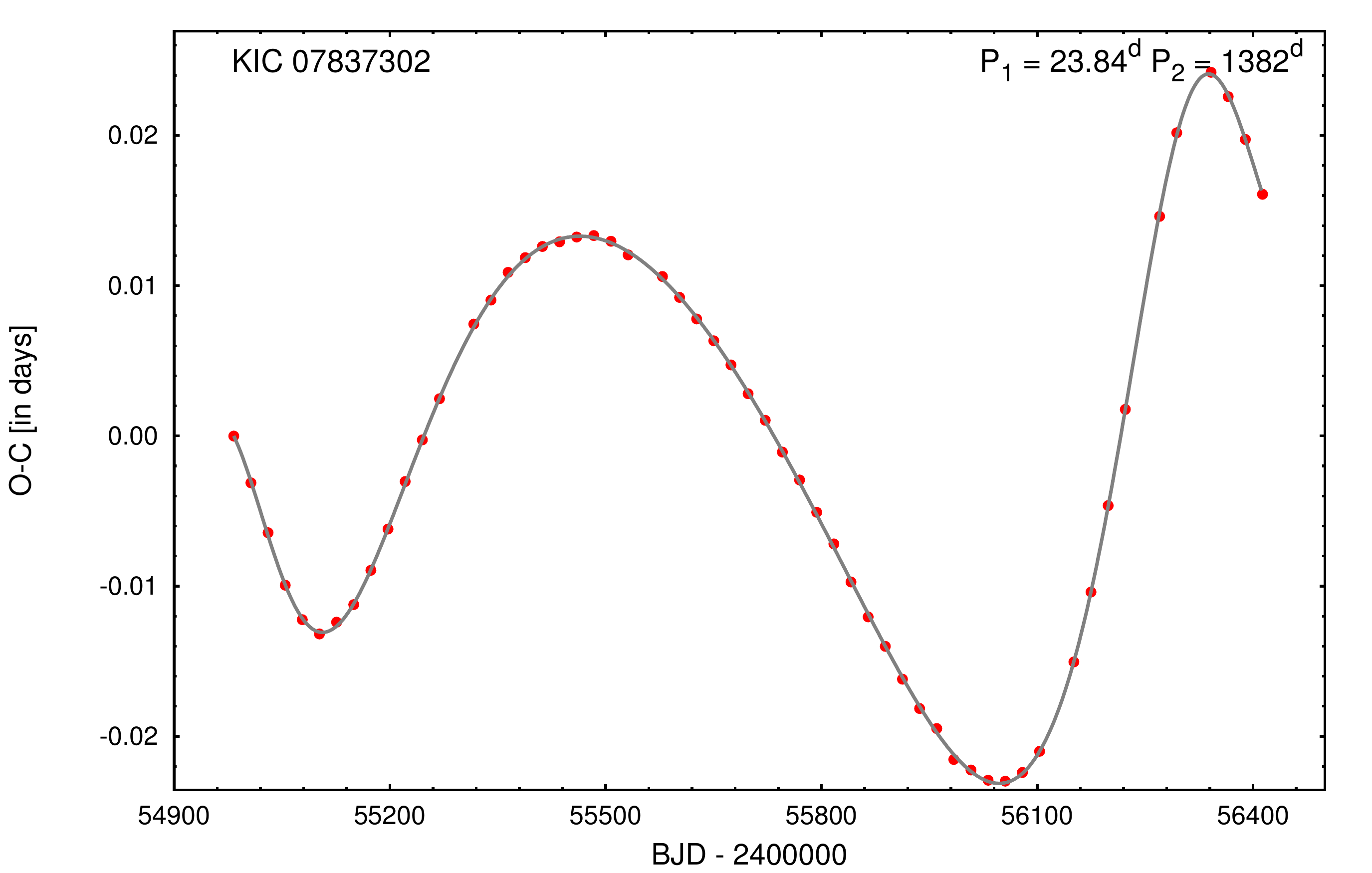}\includegraphics[width=60mm]{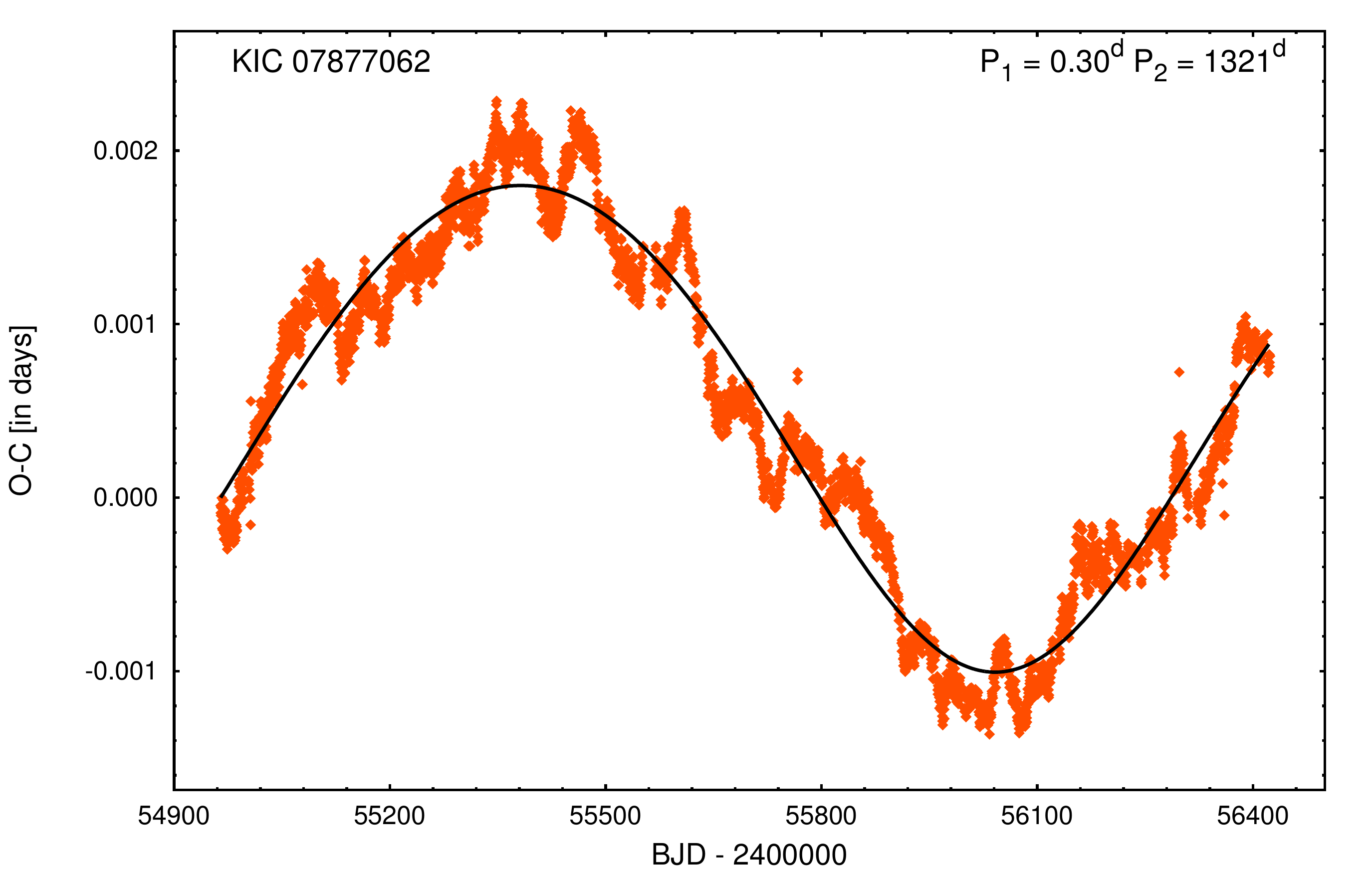}\includegraphics[width=60mm]{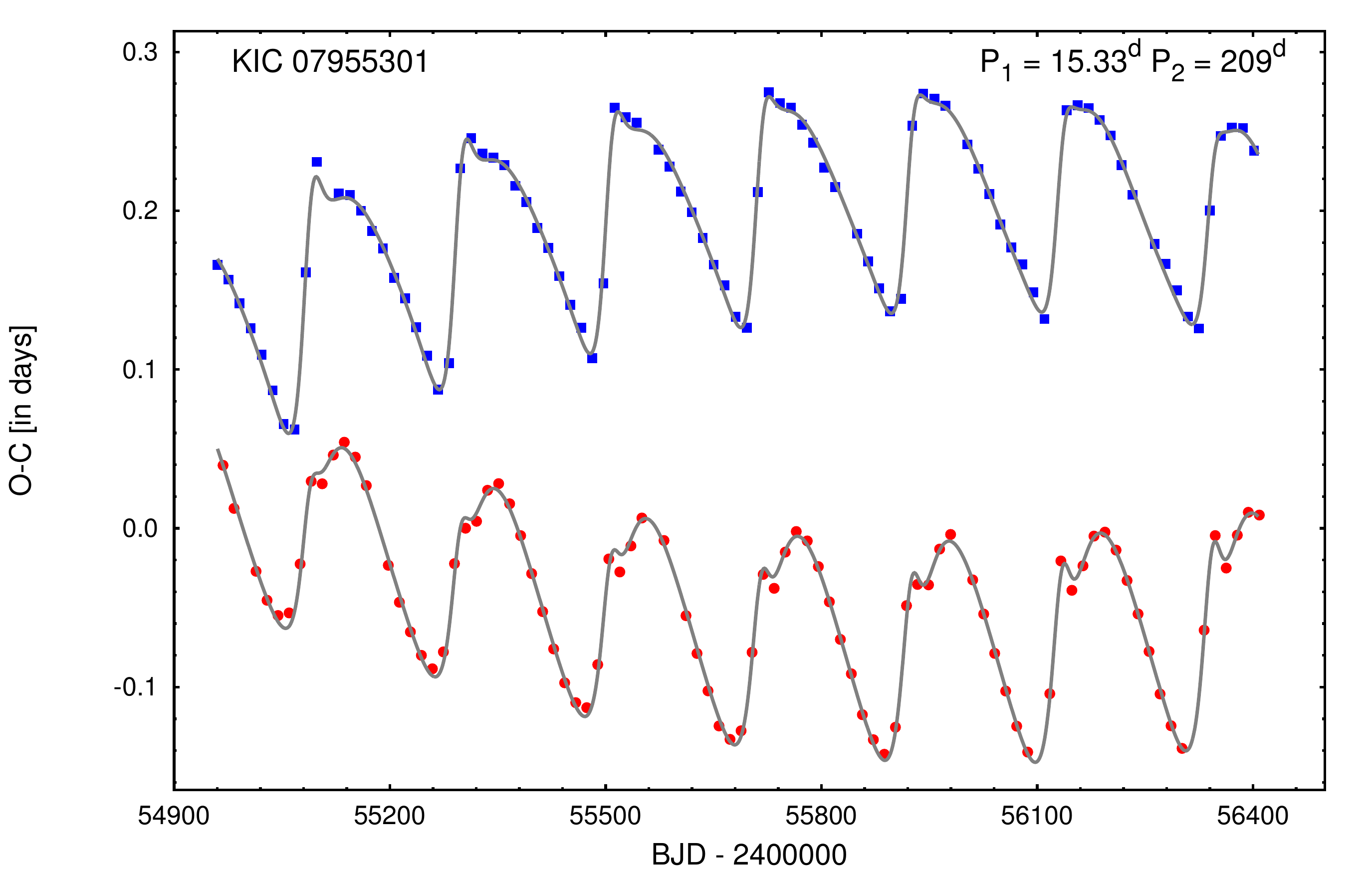}
\includegraphics[width=60mm]{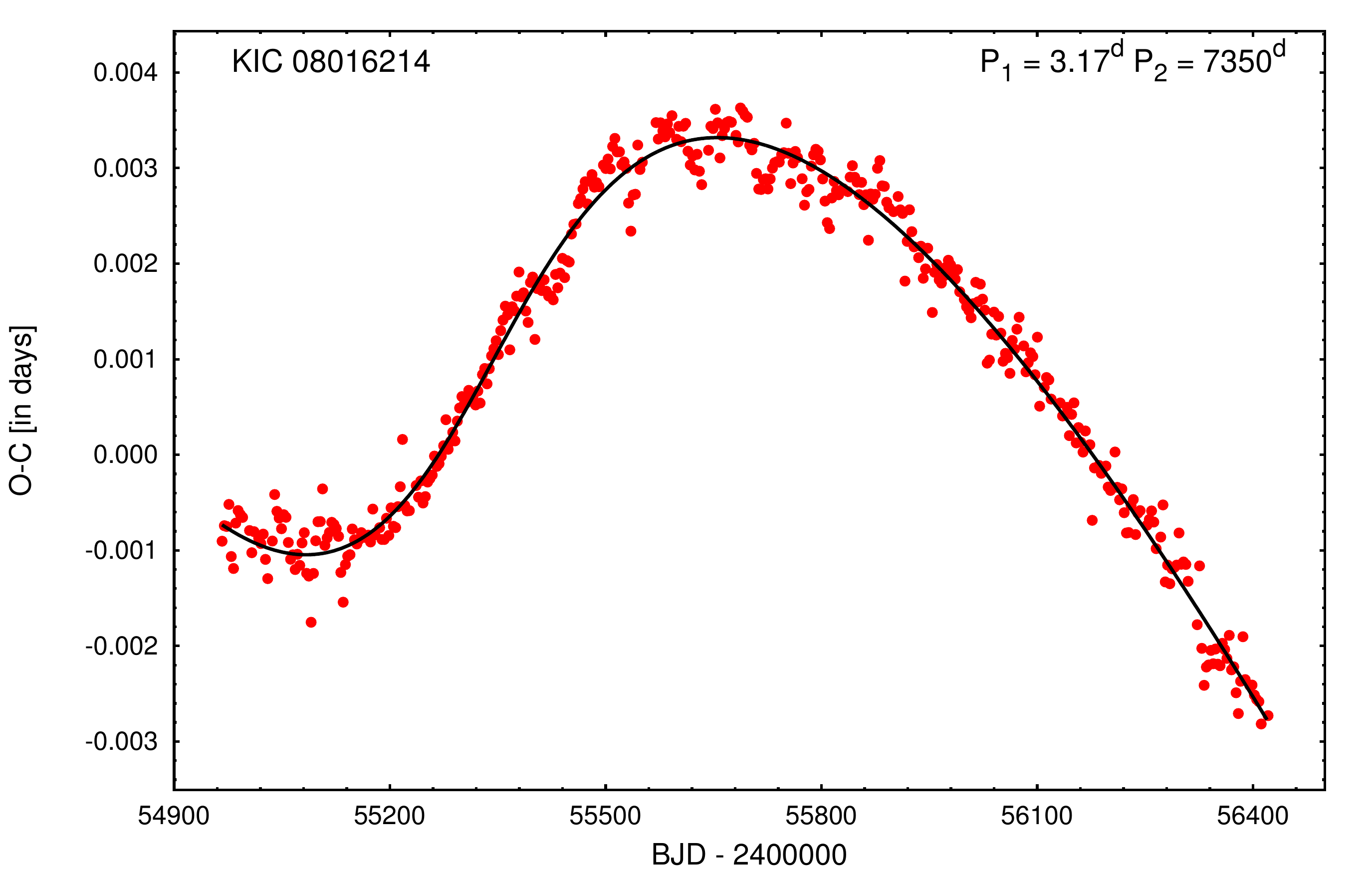}\includegraphics[width=60mm]{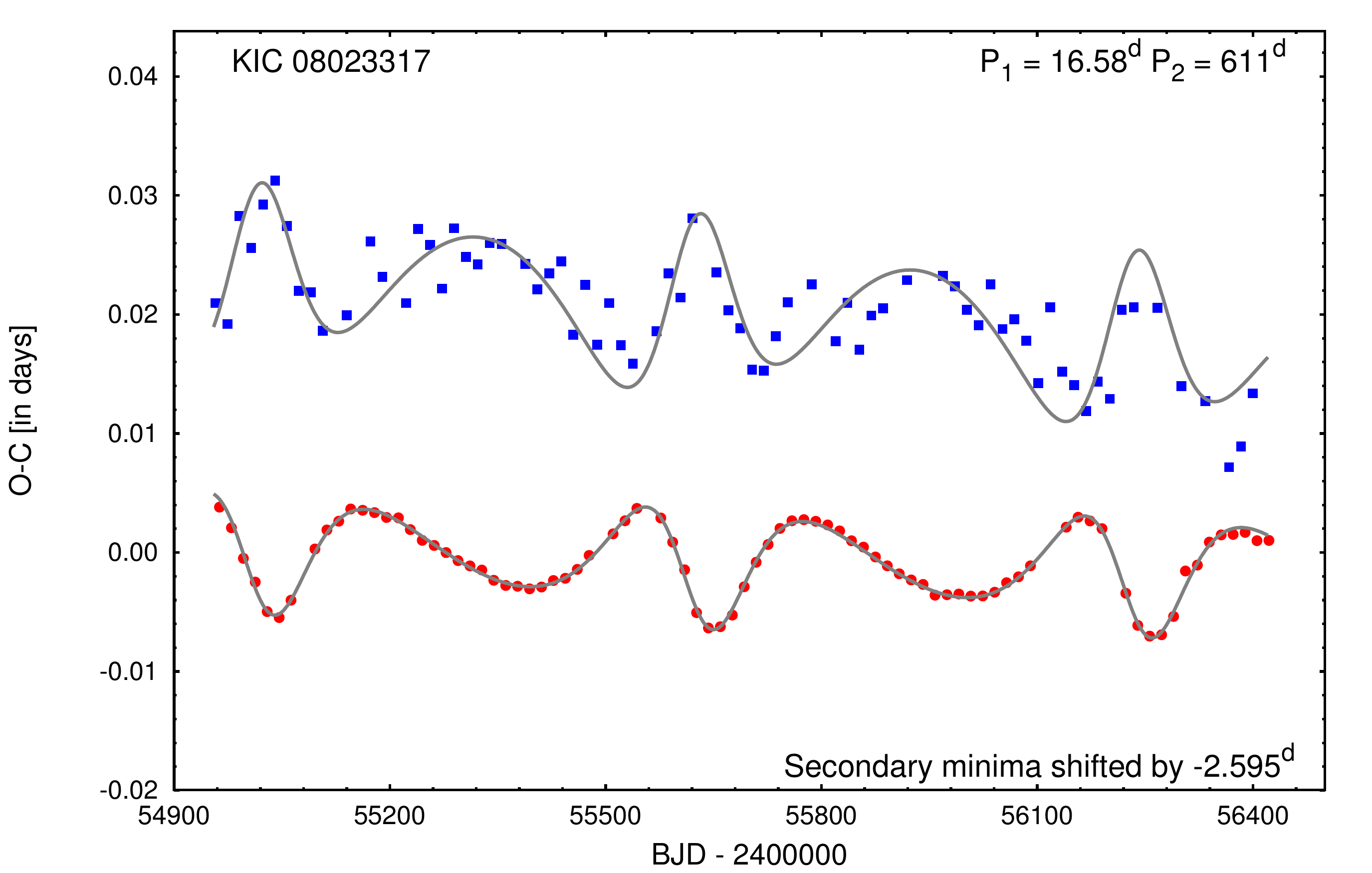}\includegraphics[width=60mm]{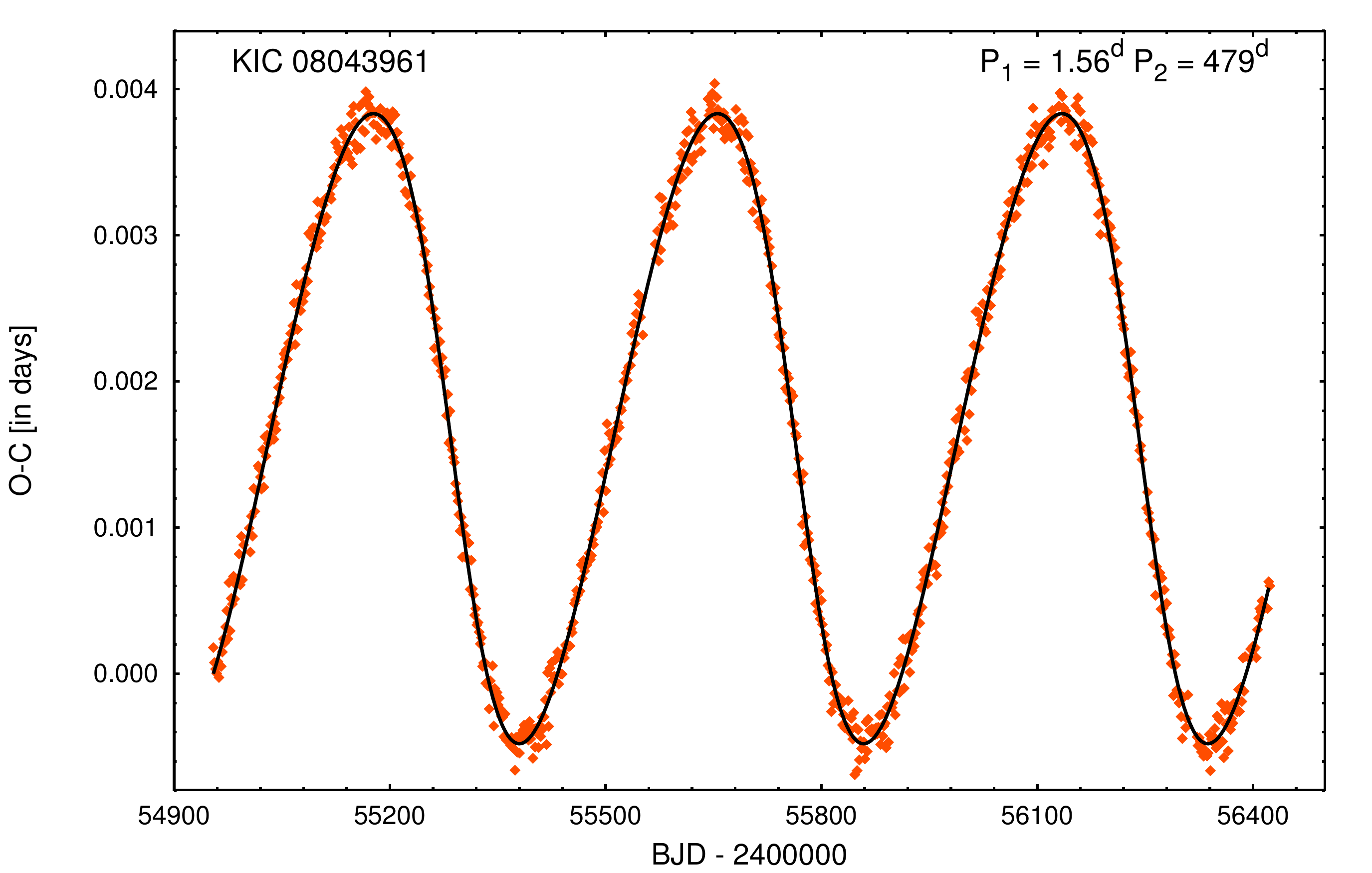}
\caption{(continued)}
\end{figure*}

\addtocounter{figure}{-1}

\begin{figure*}
\includegraphics[width=60mm]{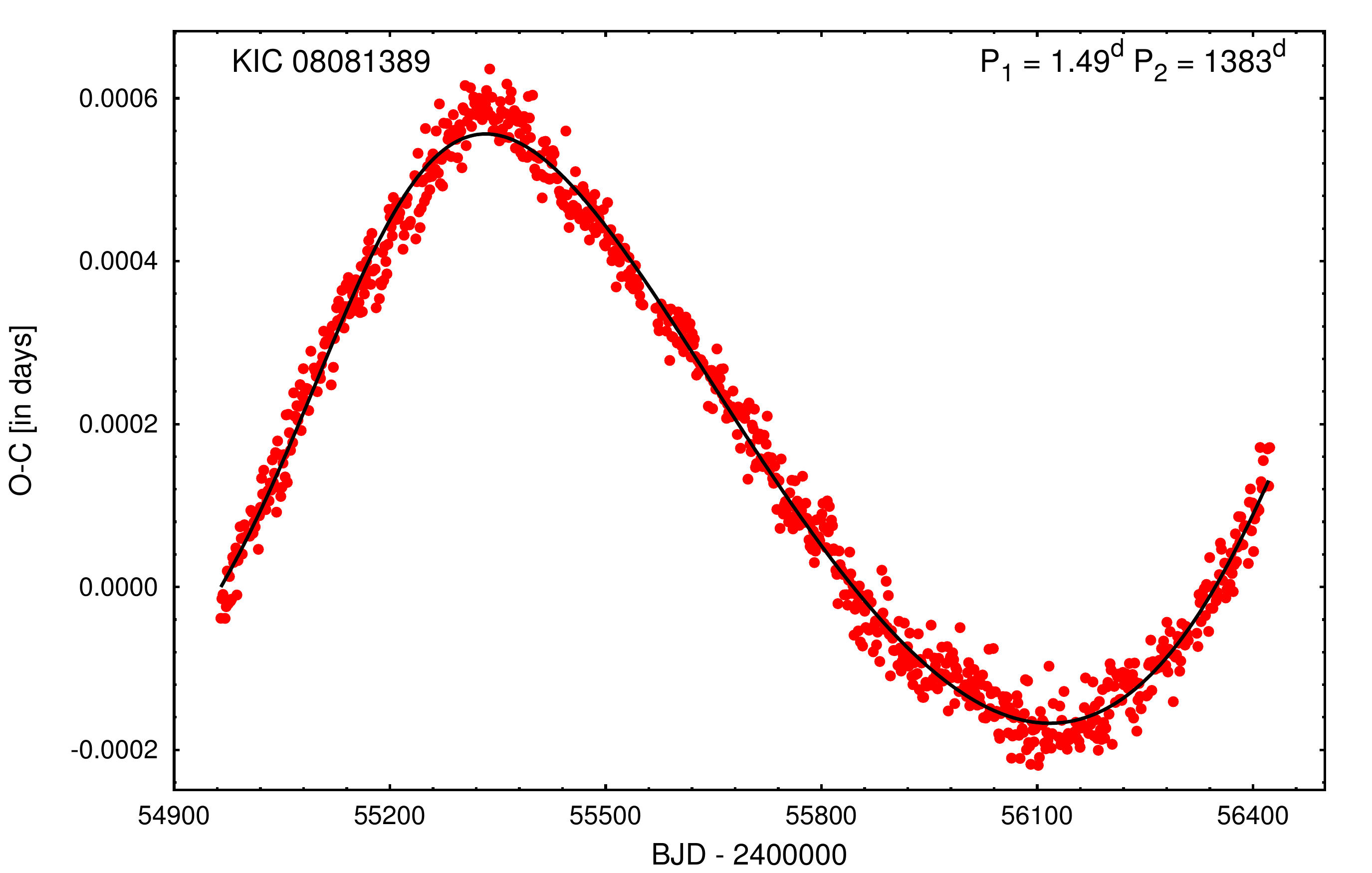}\includegraphics[width=60mm]{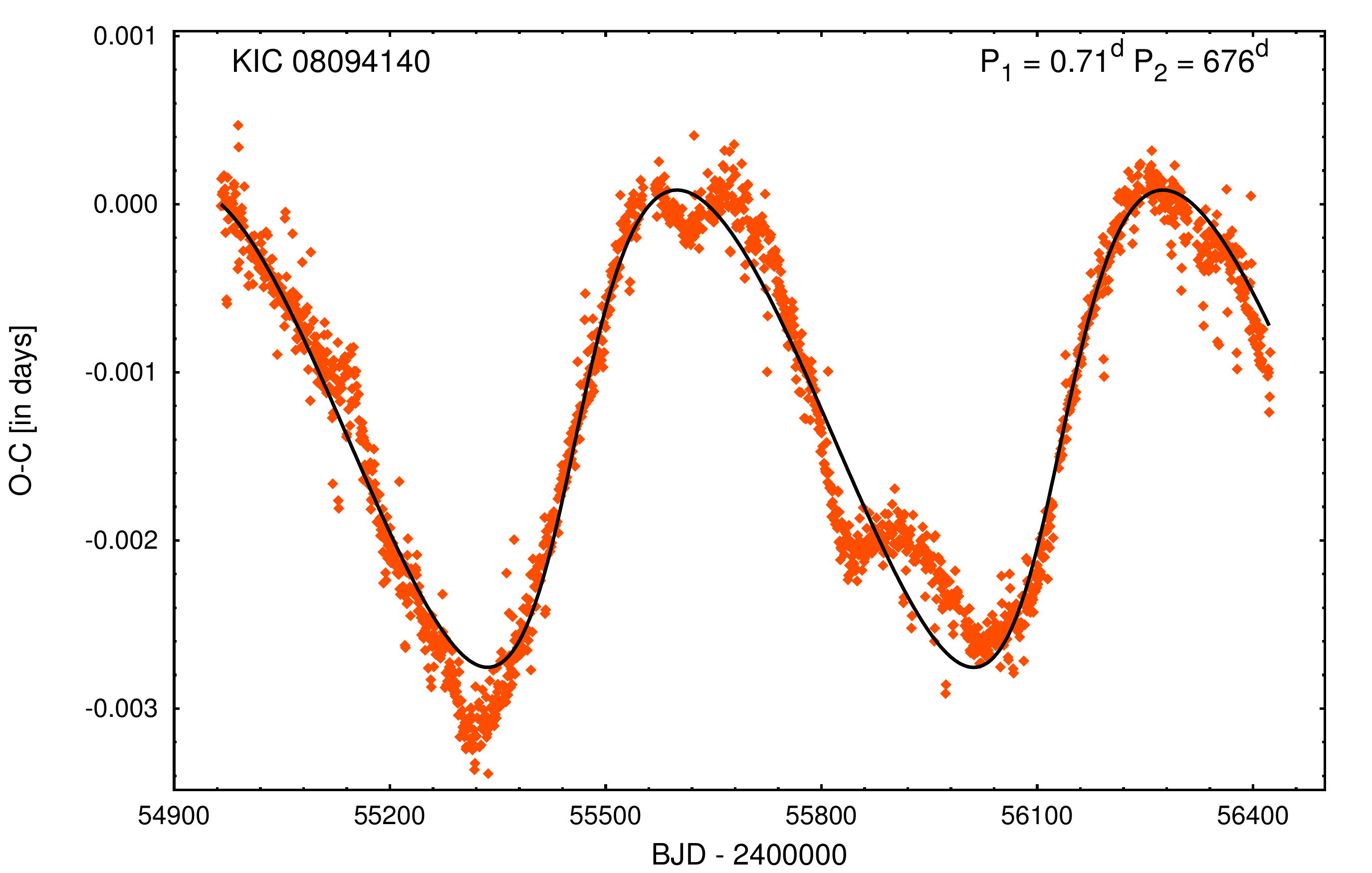}\includegraphics[width=60mm]{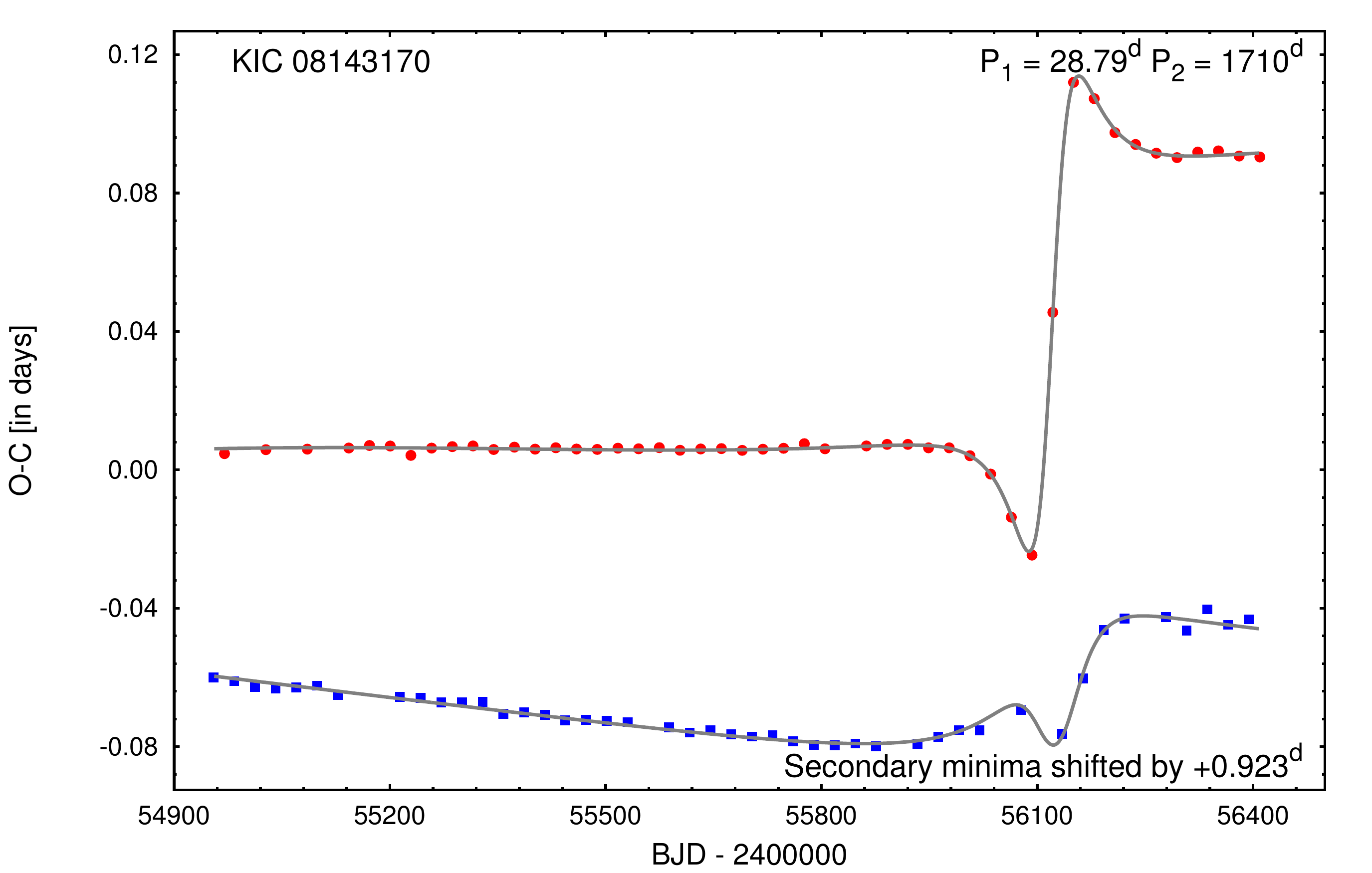}
\includegraphics[width=60mm]{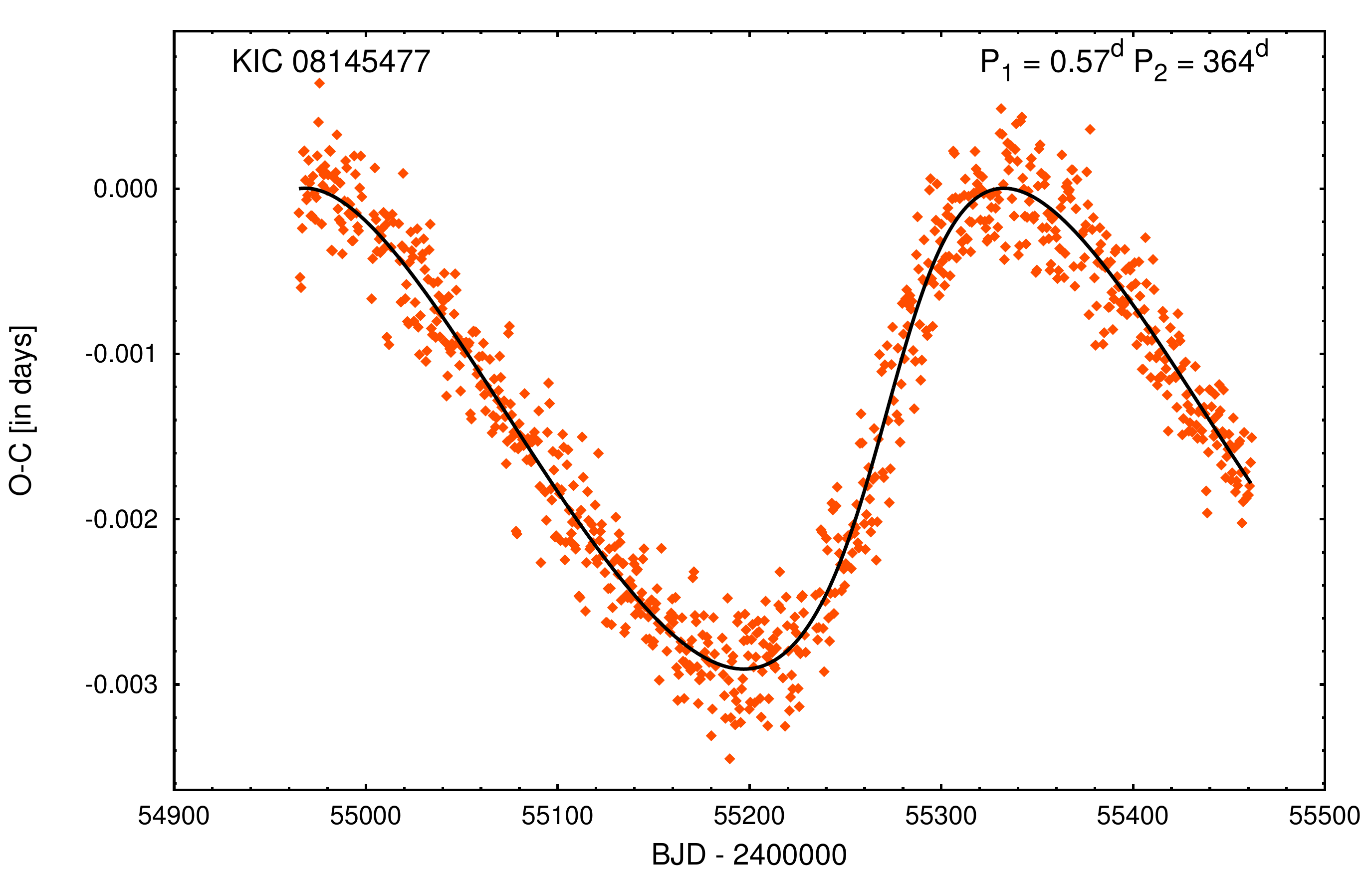}\includegraphics[width=60mm]{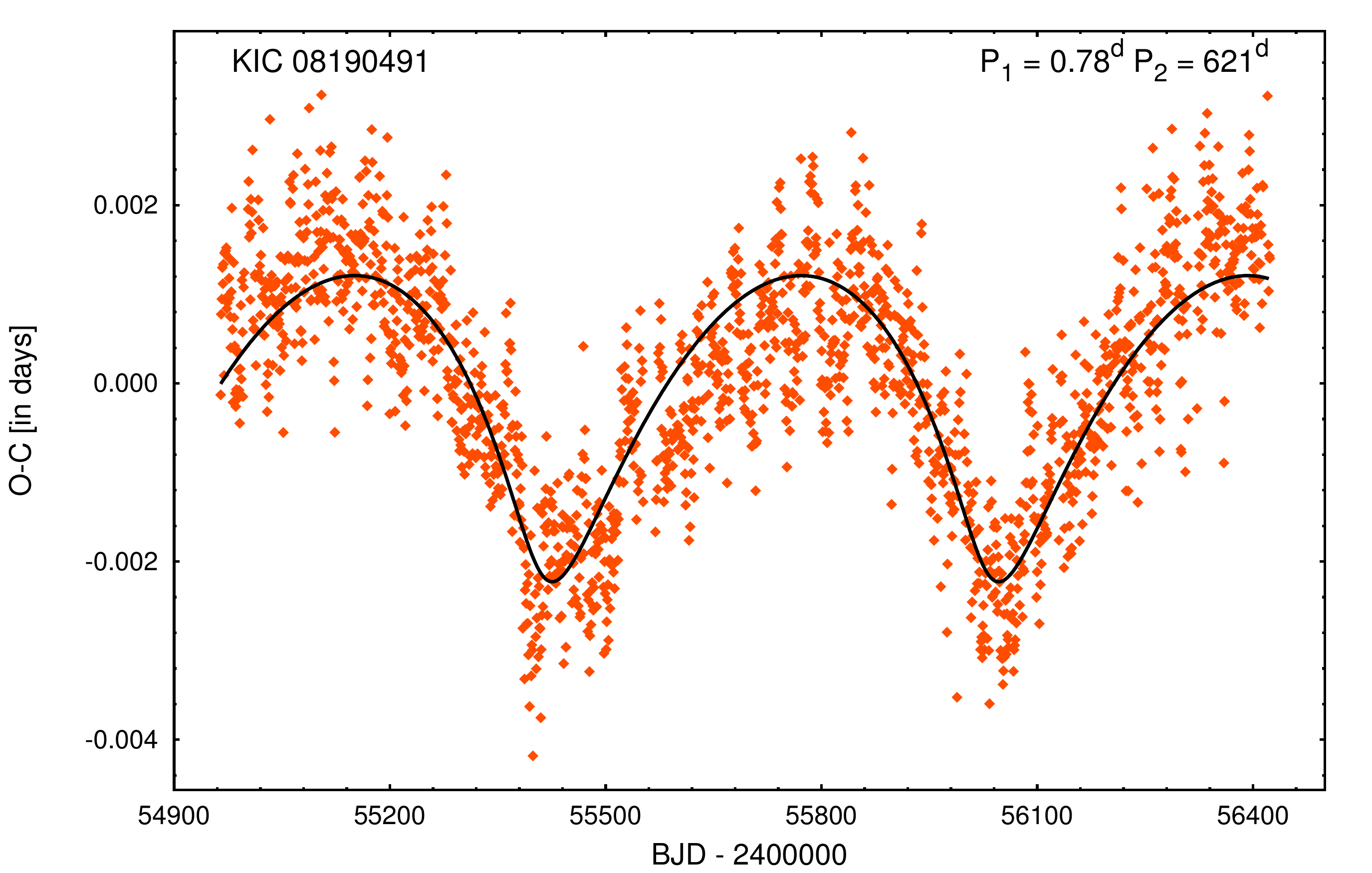}\includegraphics[width=60mm]{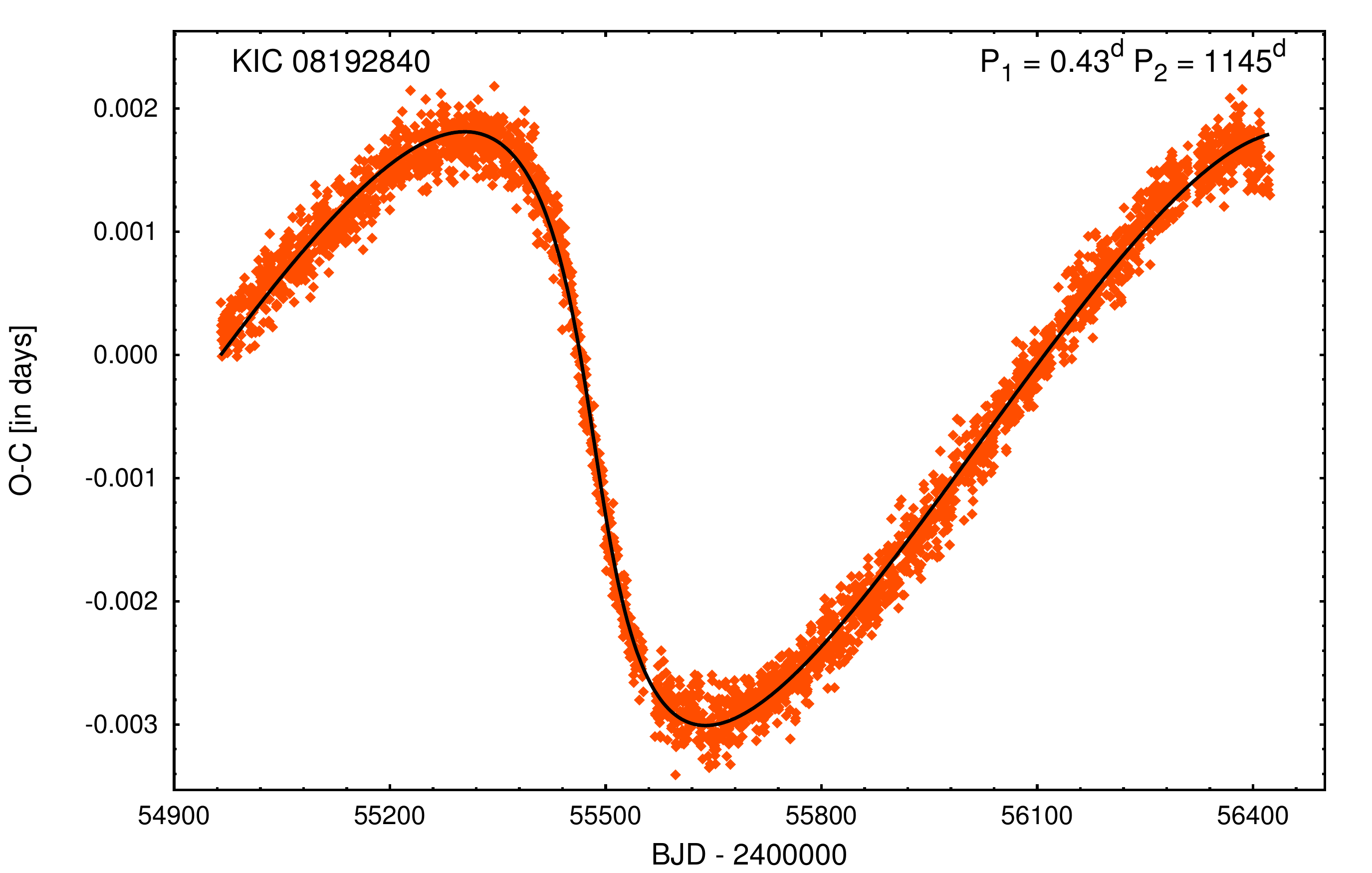}
\includegraphics[width=60mm]{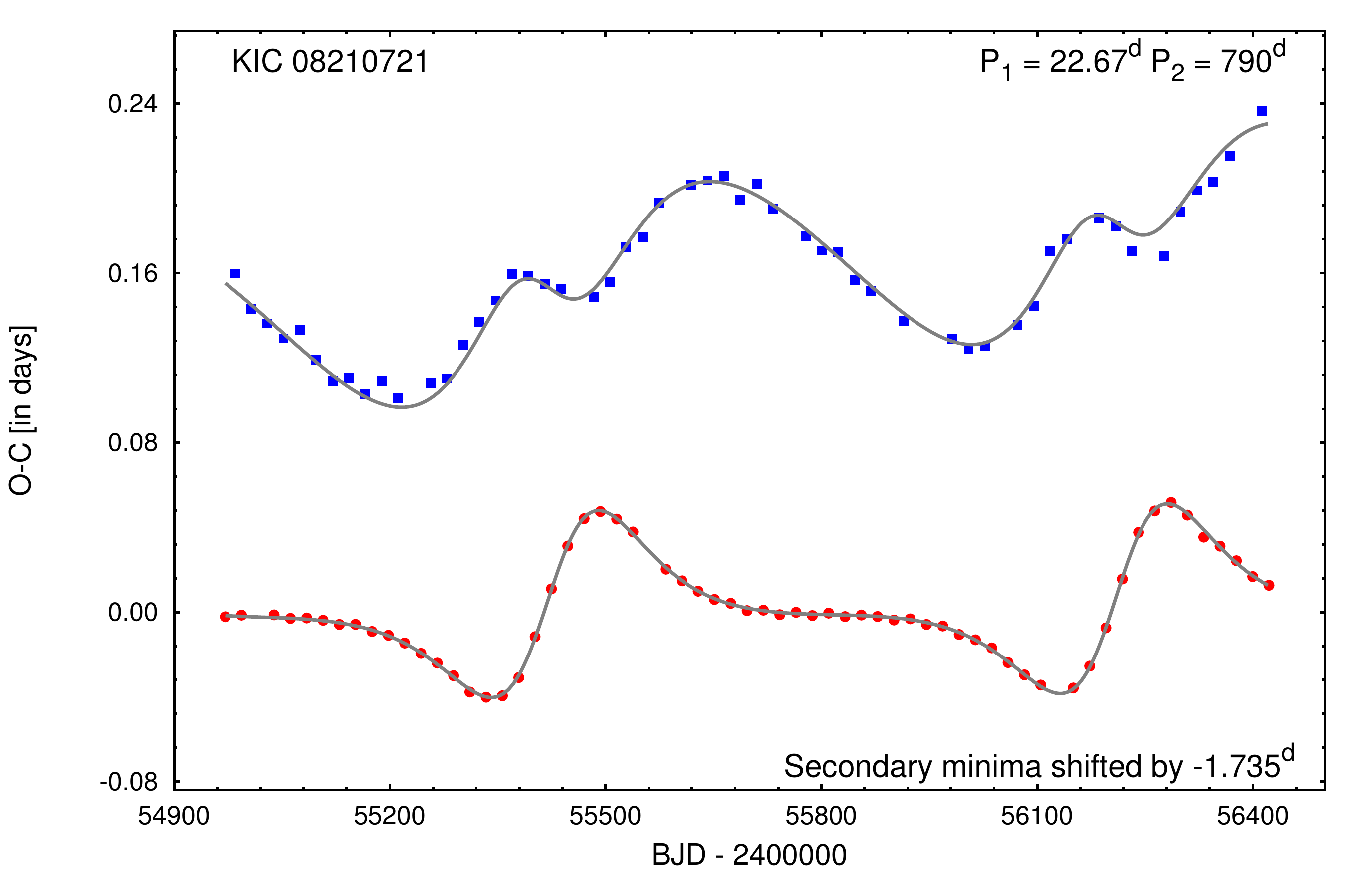}\includegraphics[width=60mm]{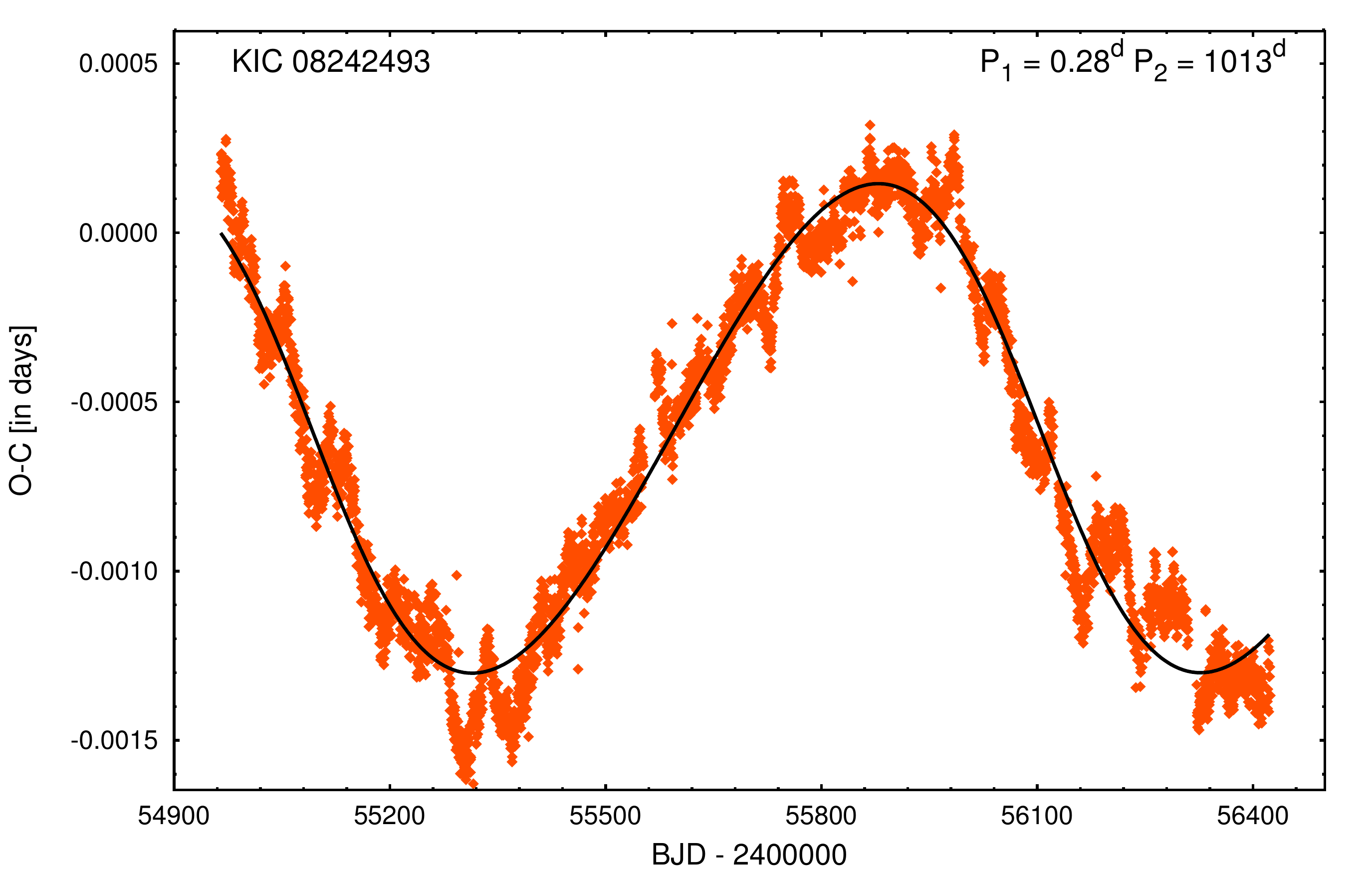}\includegraphics[width=60mm]{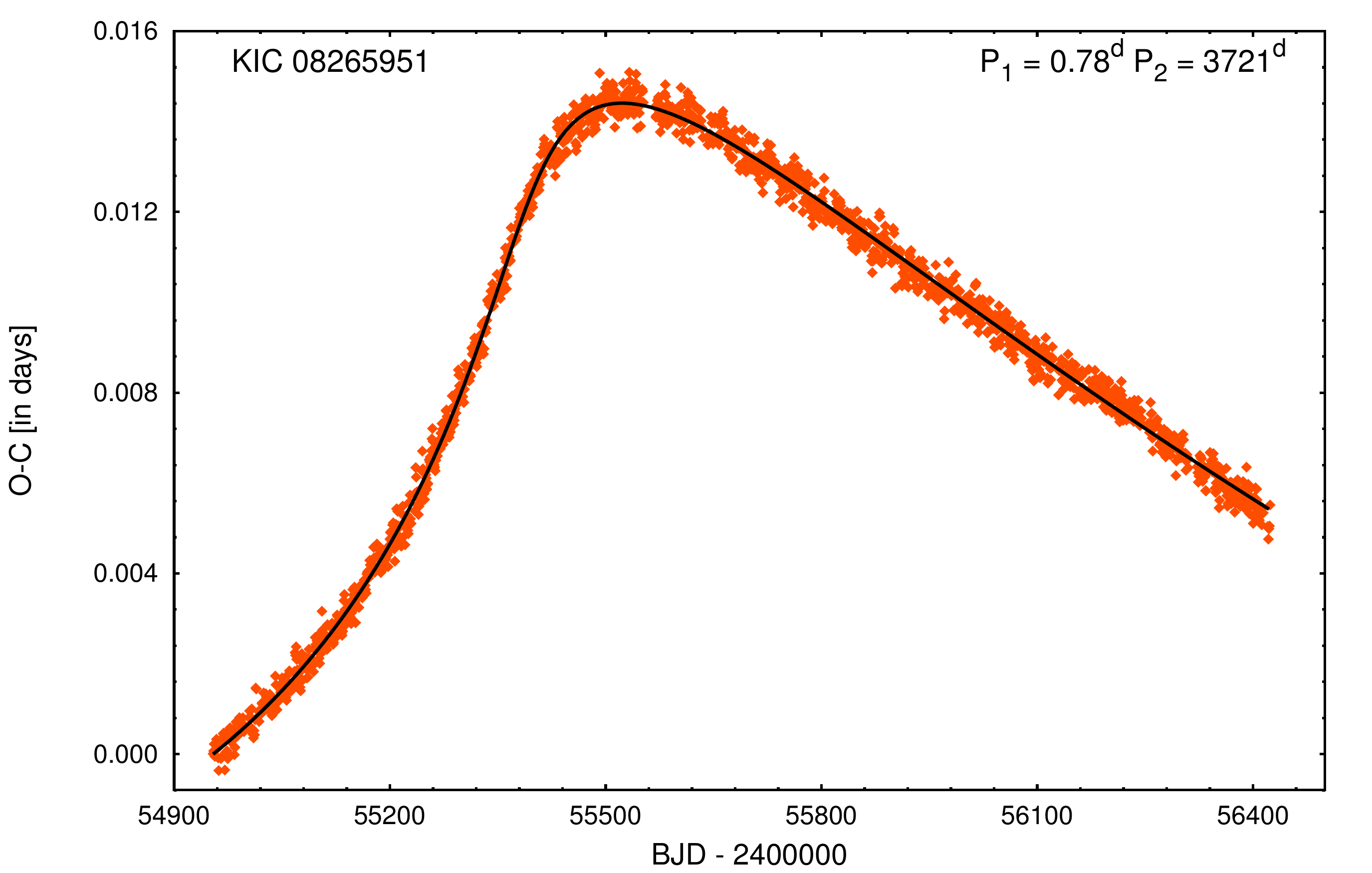}
\includegraphics[width=60mm]{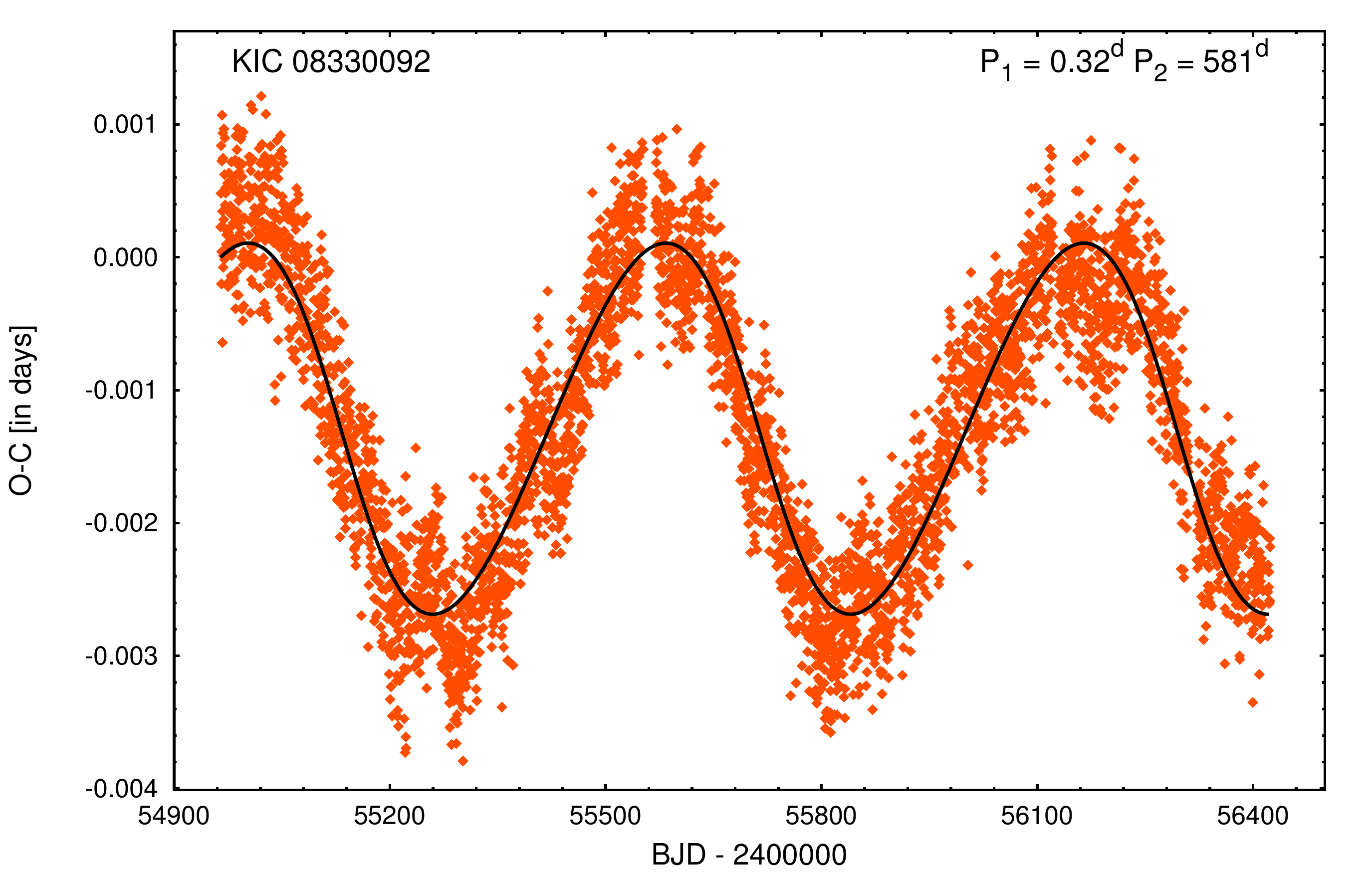}\includegraphics[width=60mm]{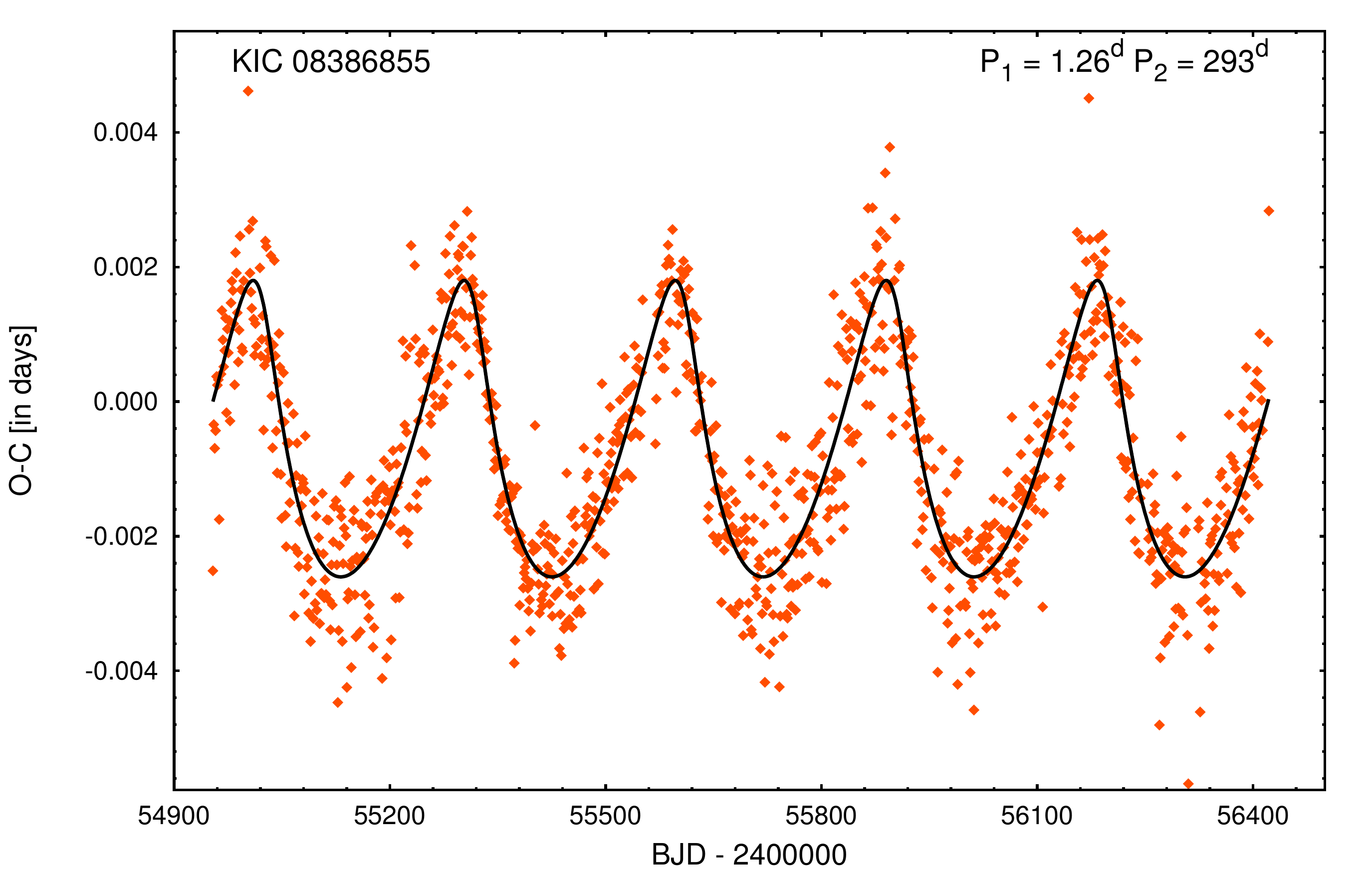}\includegraphics[width=60mm]{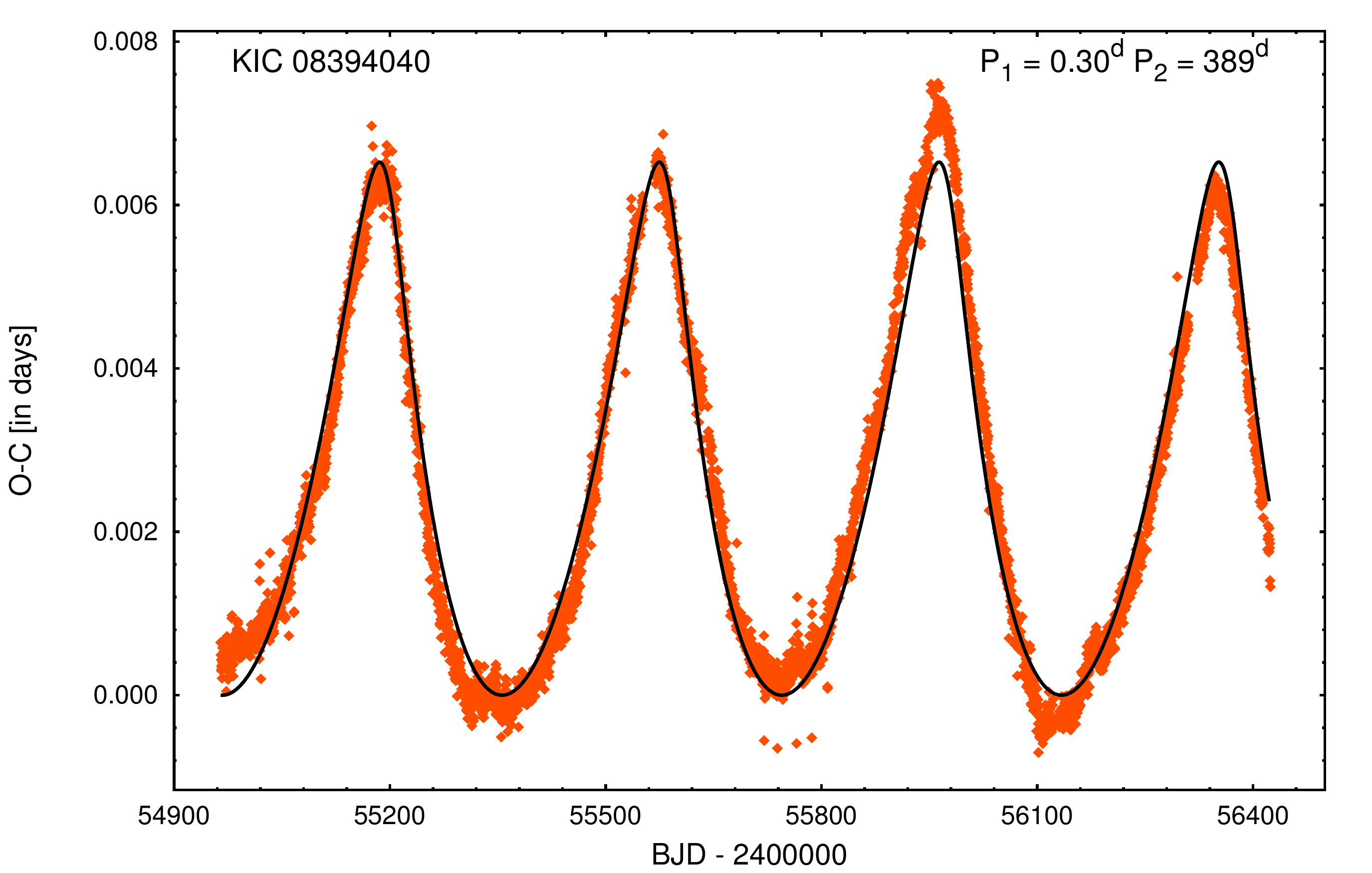}
\includegraphics[width=60mm]{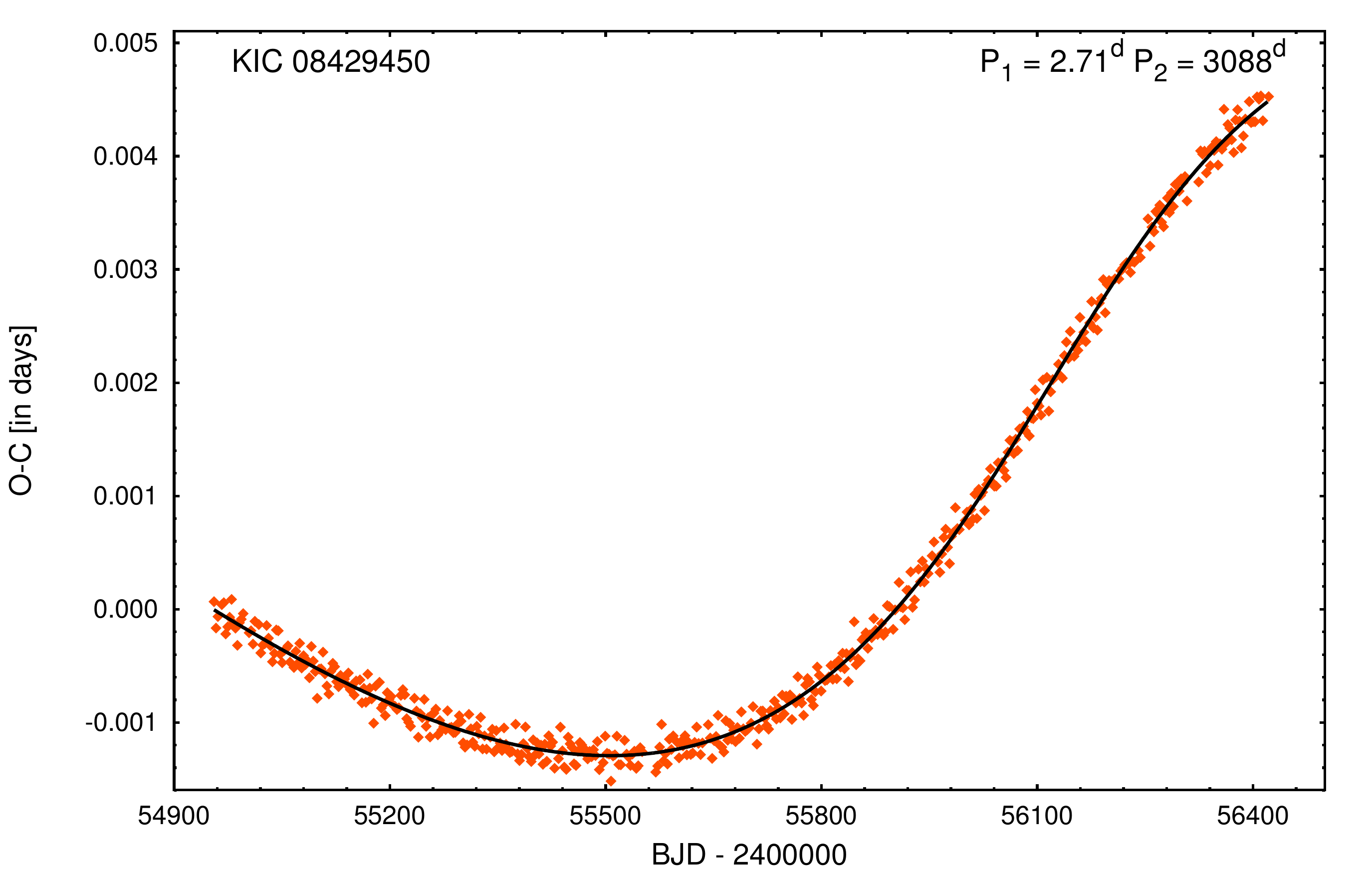}\includegraphics[width=60mm]{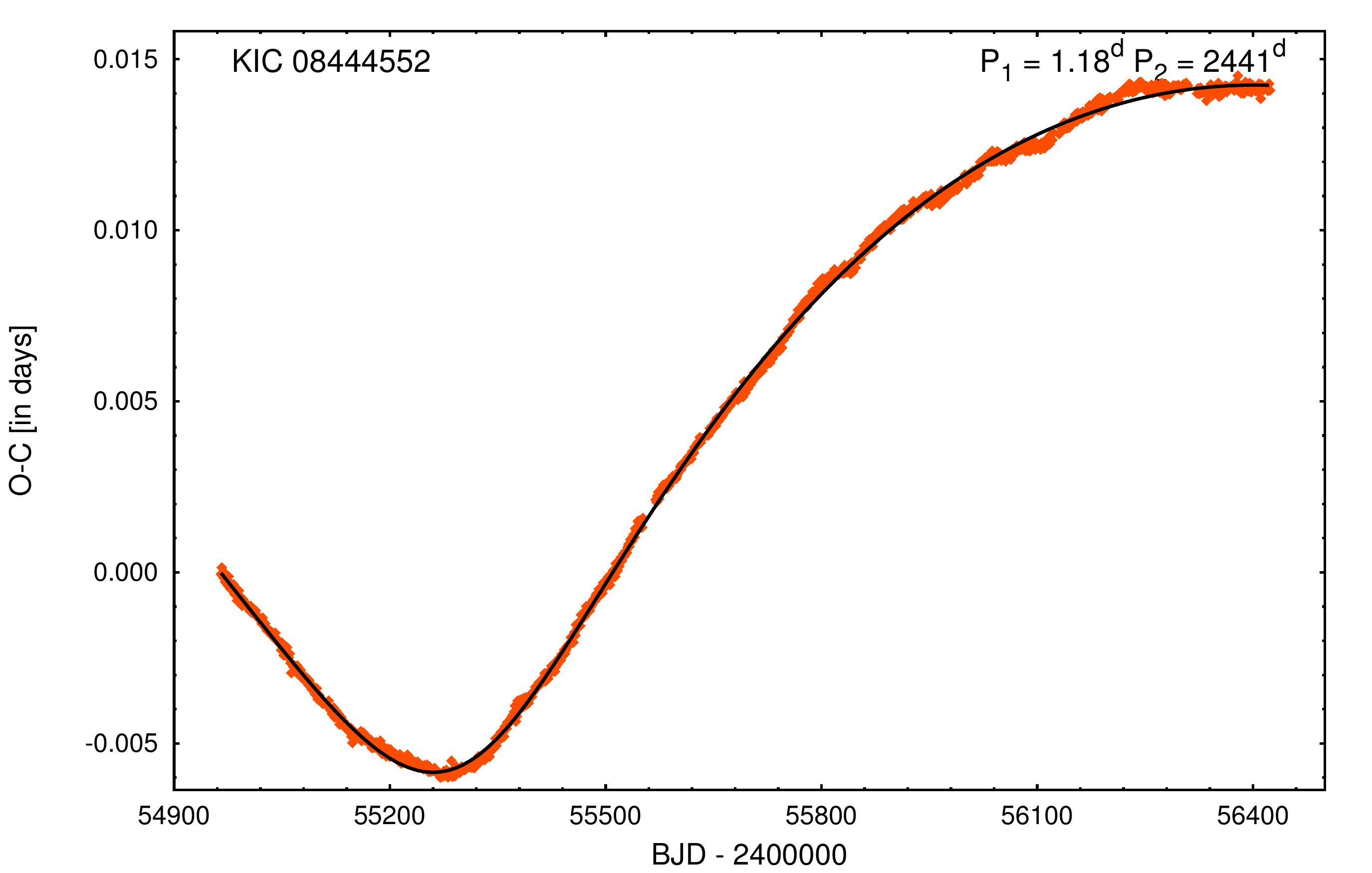}\includegraphics[width=60mm]{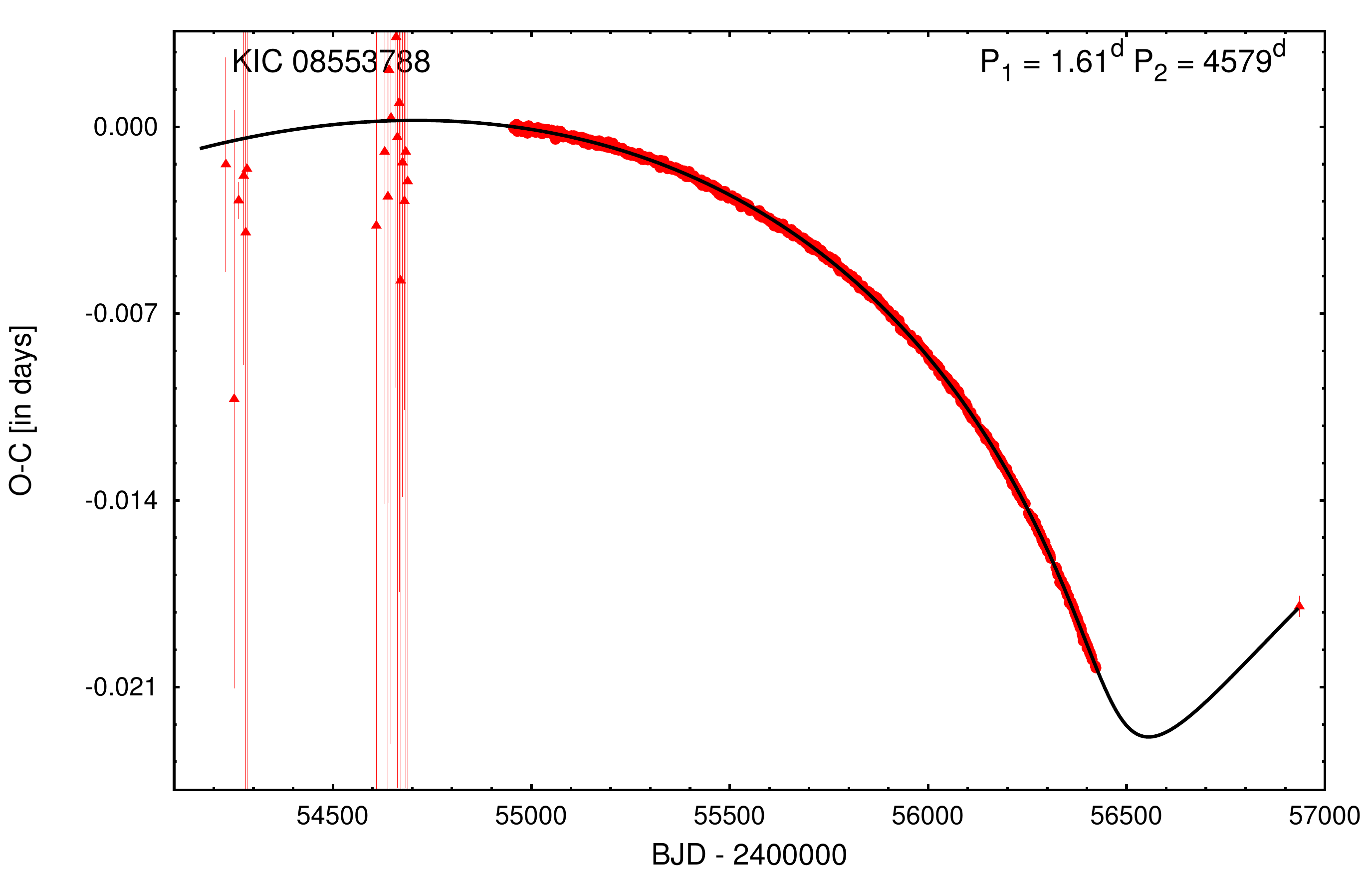}
\includegraphics[width=60mm]{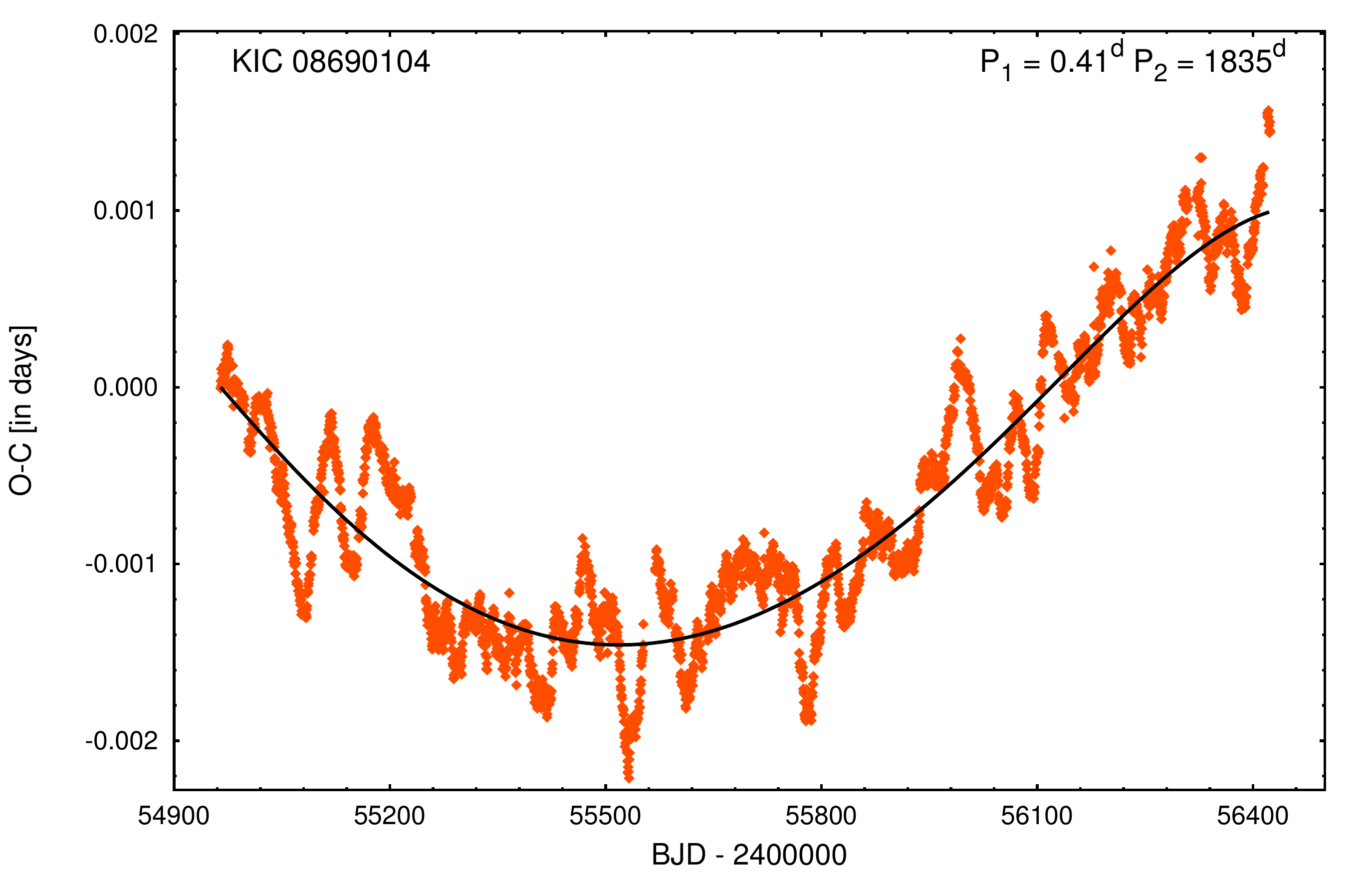}\includegraphics[width=60mm]{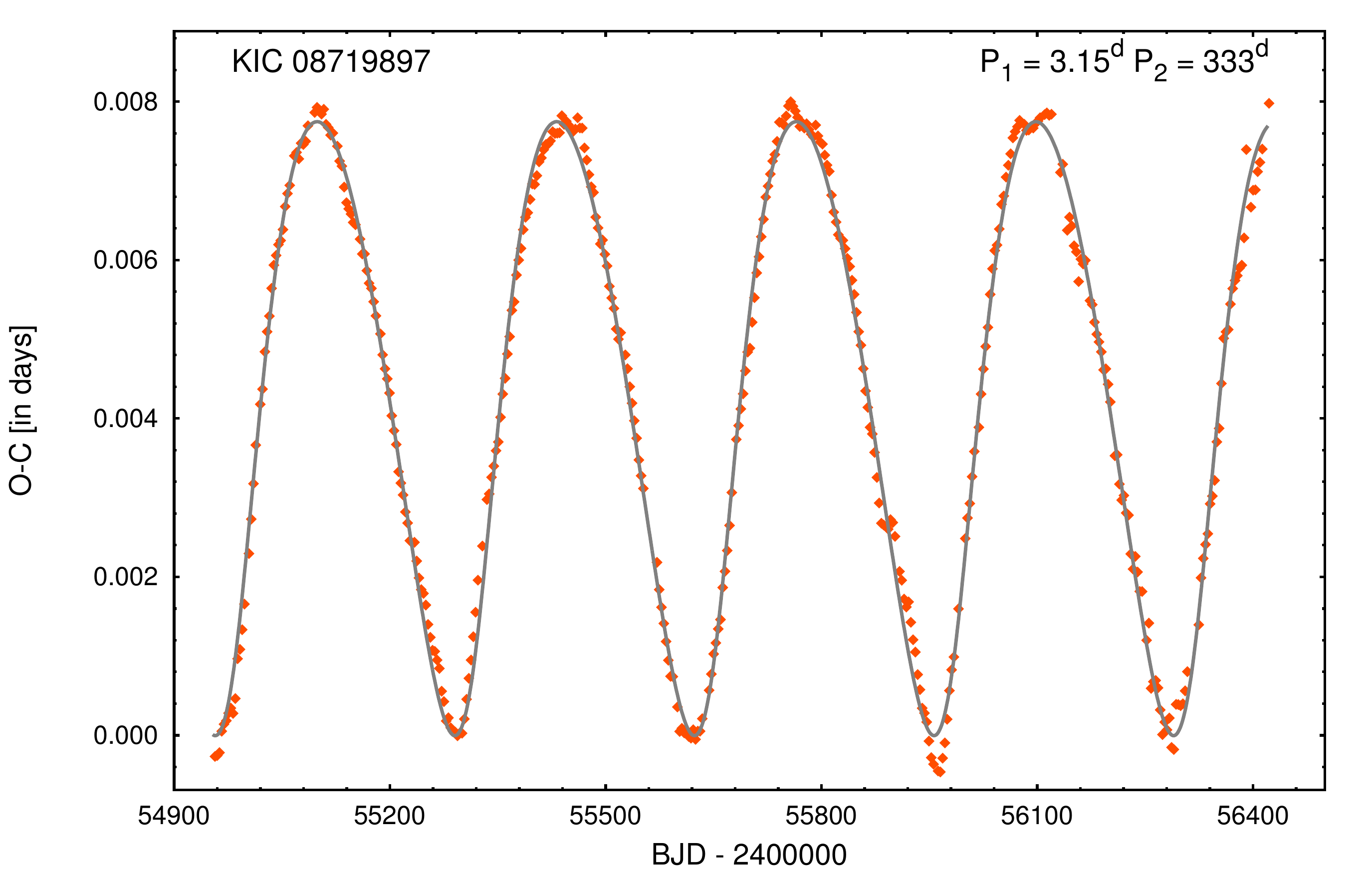}\includegraphics[width=60mm]{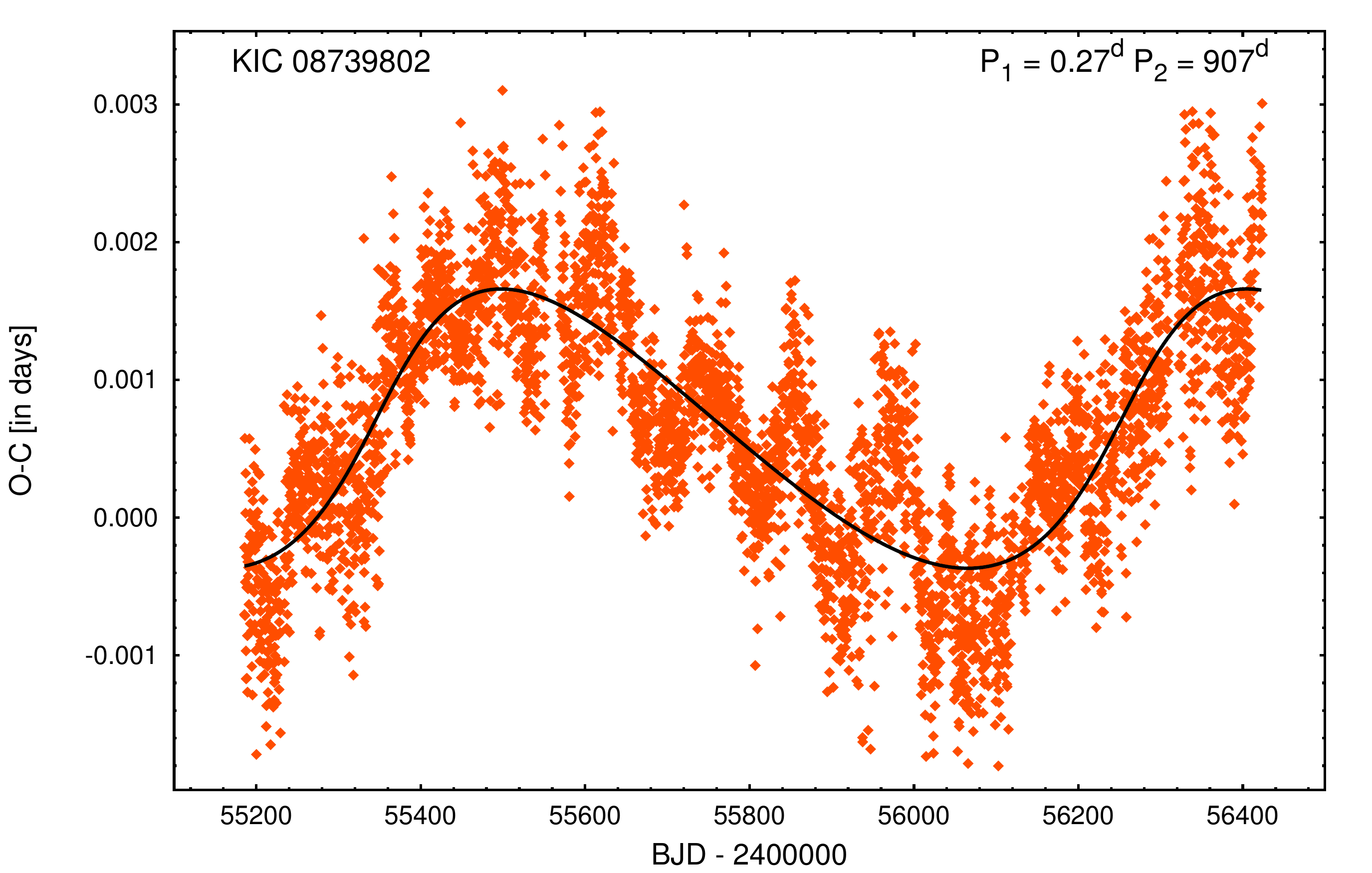}
\caption{(continued)}
\end{figure*}

\addtocounter{figure}{-1}

\begin{figure*}
\includegraphics[width=60mm]{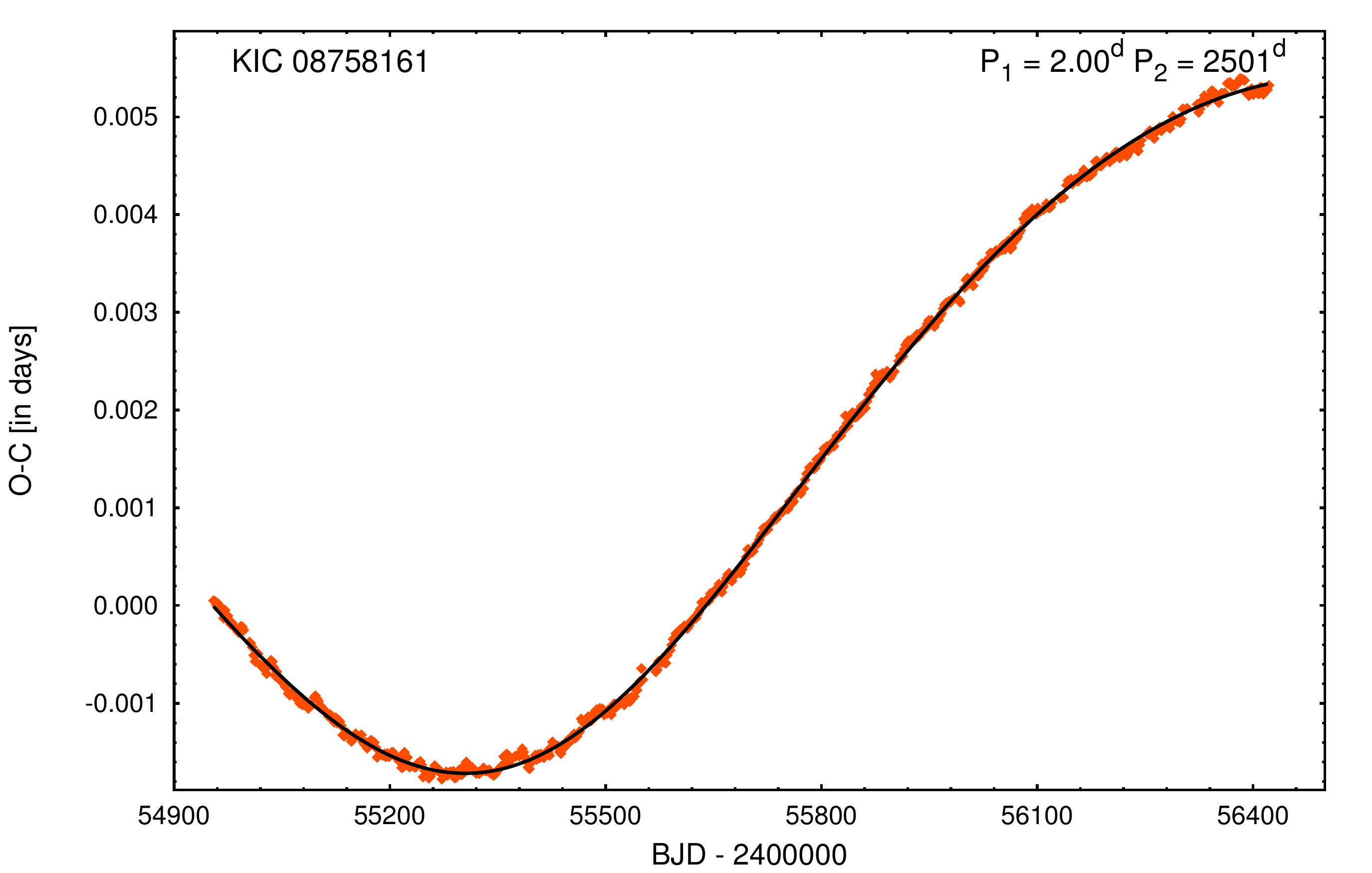}\includegraphics[width=60mm]{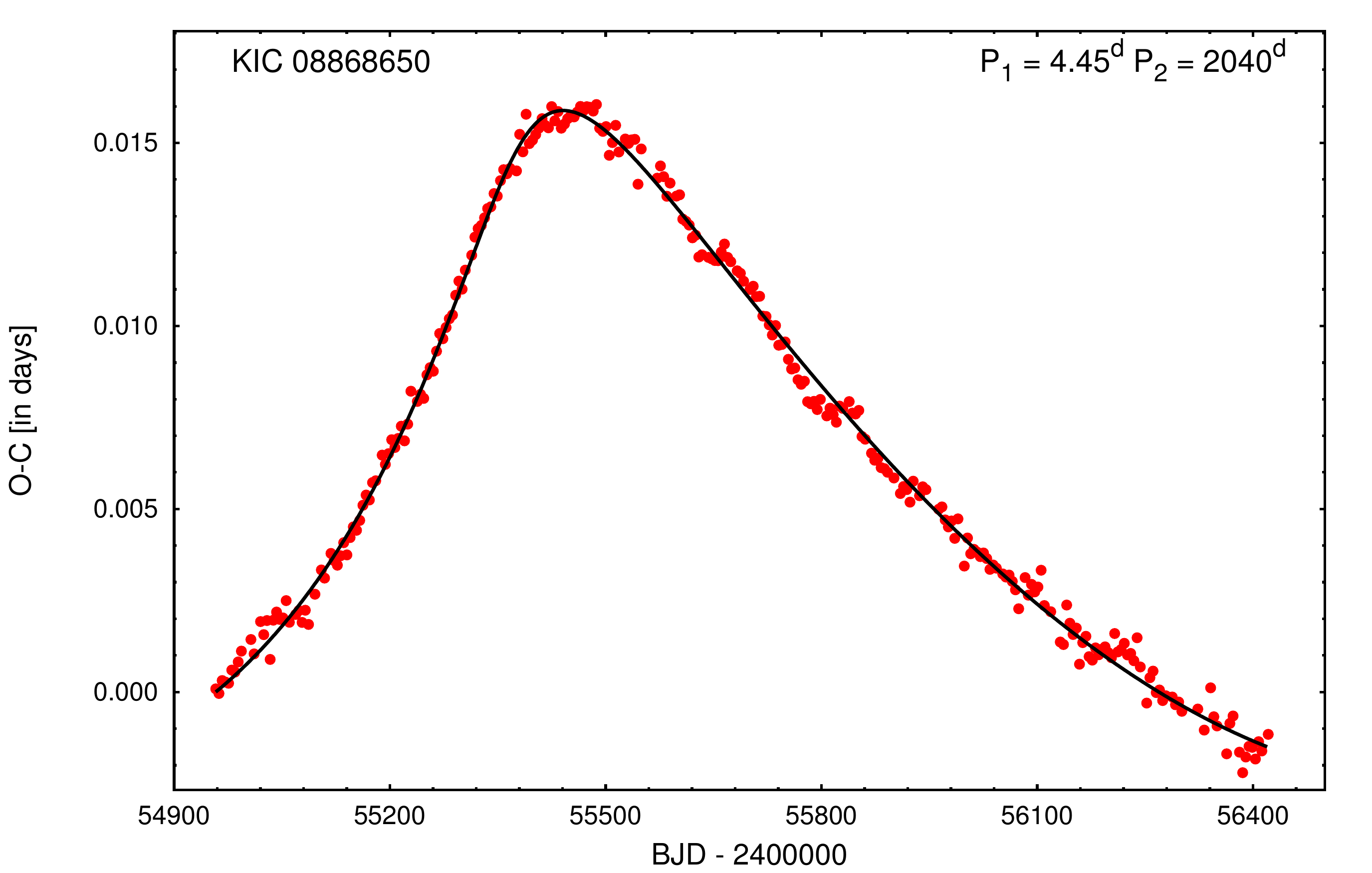}\includegraphics[width=60mm]{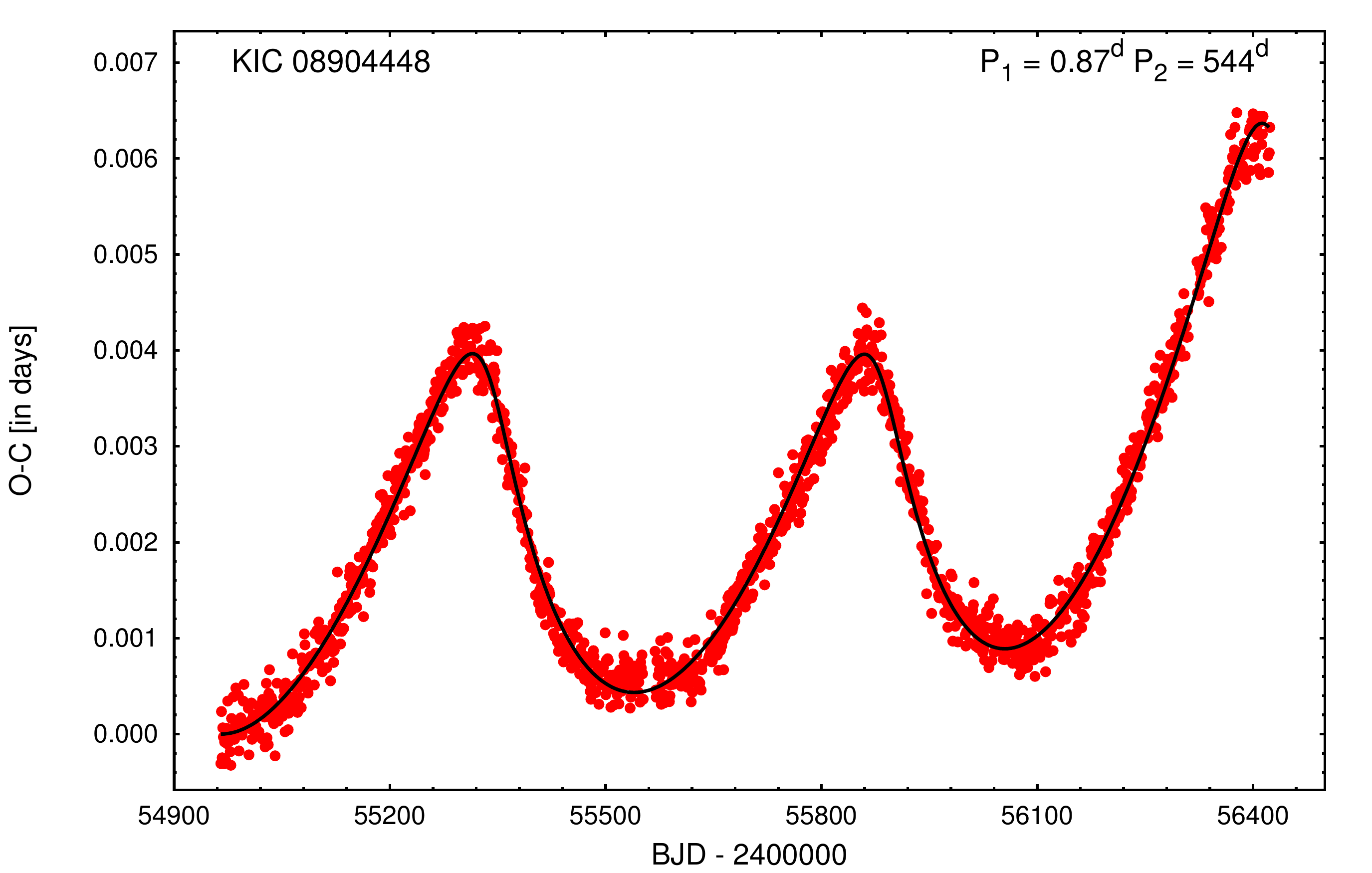}
\includegraphics[width=60mm]{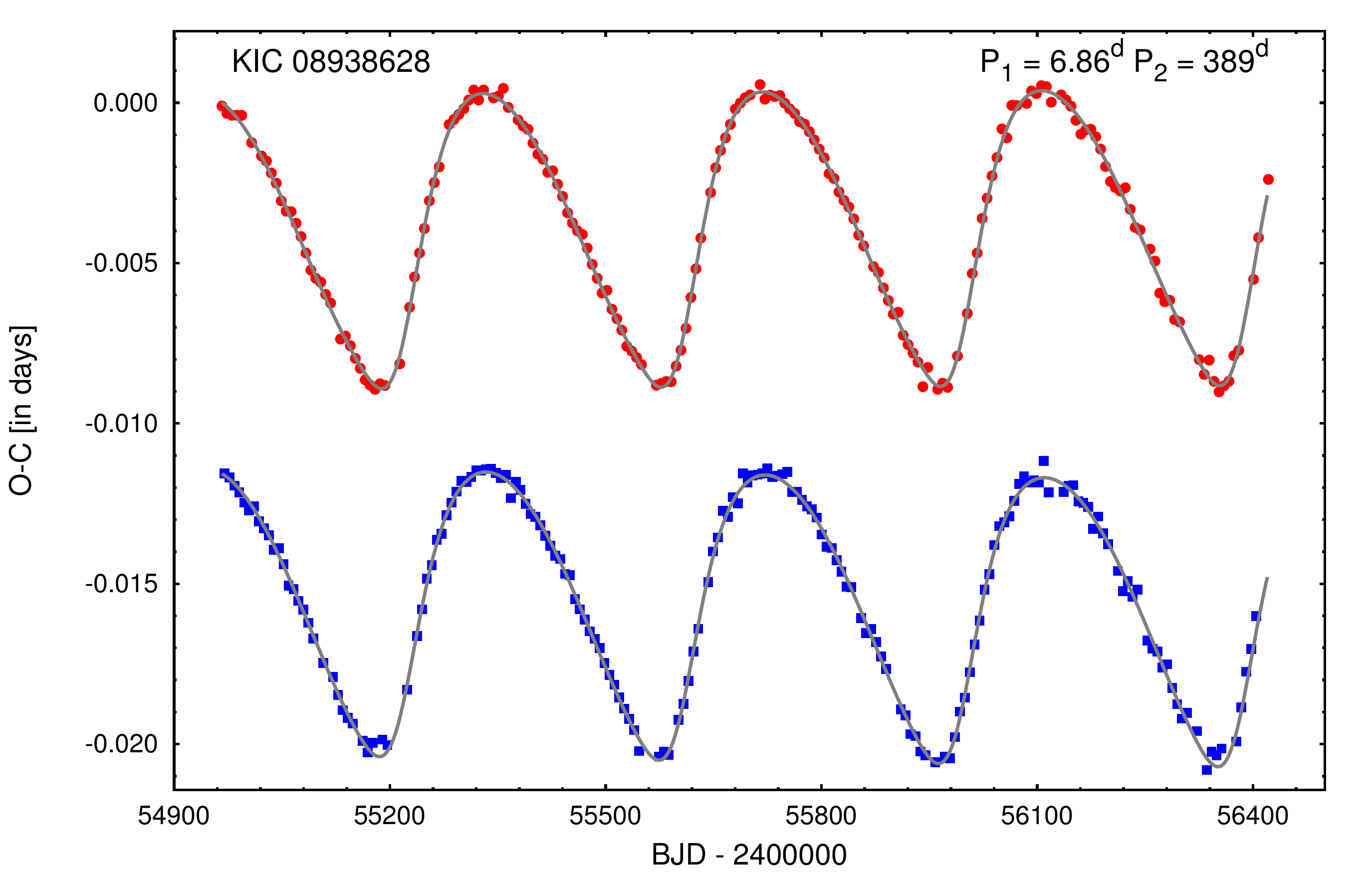}\includegraphics[width=60mm]{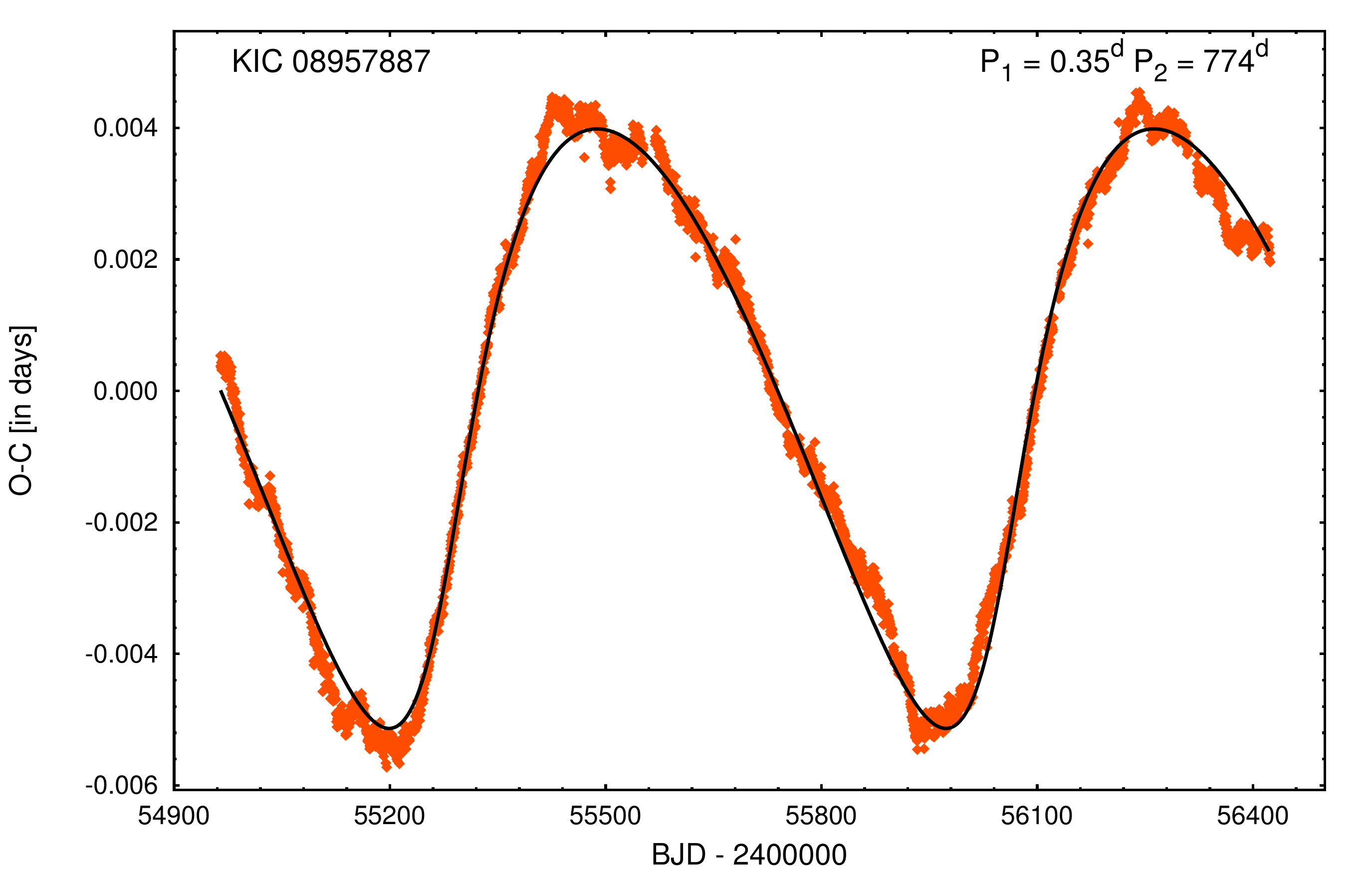}\includegraphics[width=60mm]{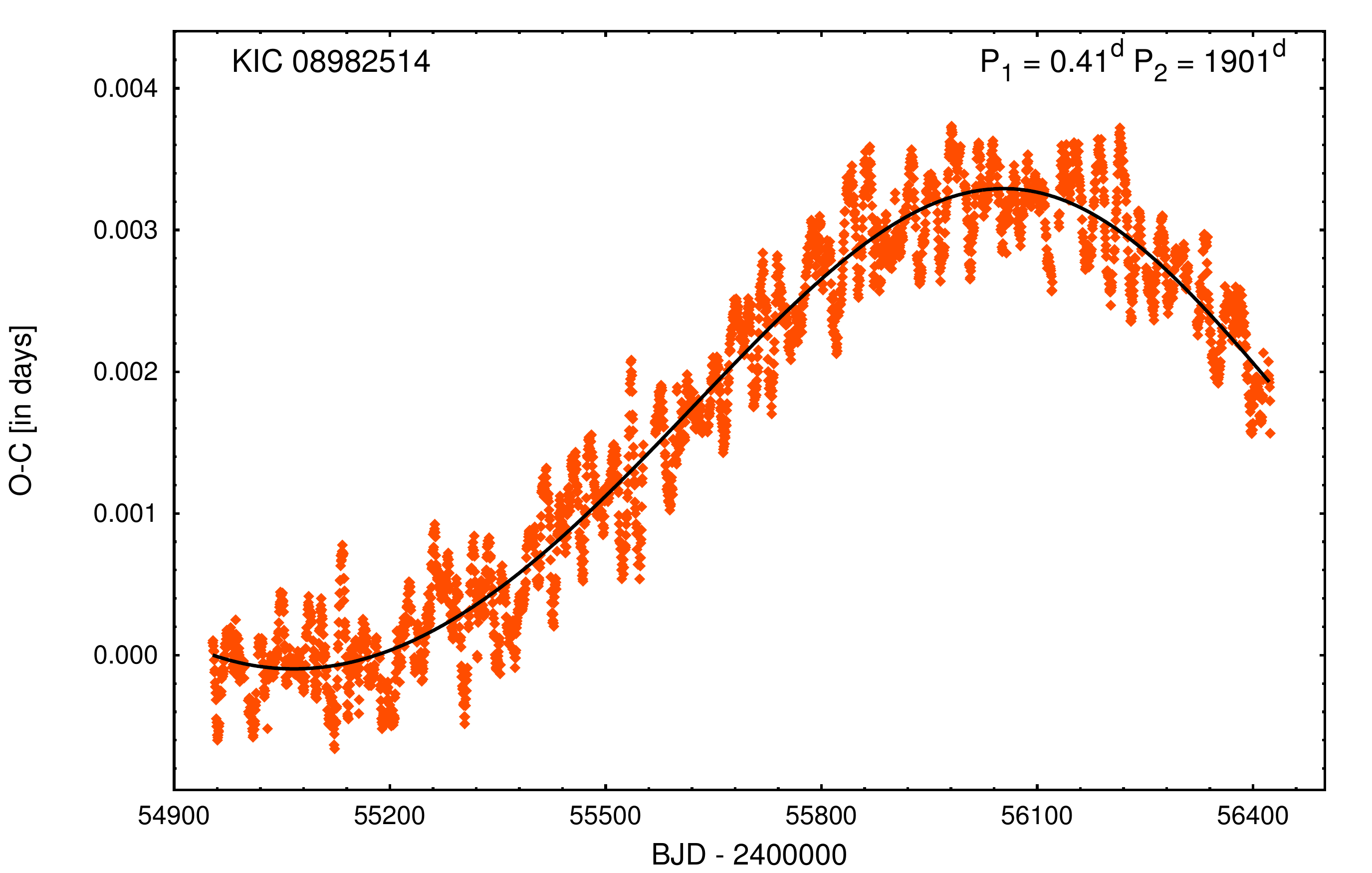}
\includegraphics[width=60mm]{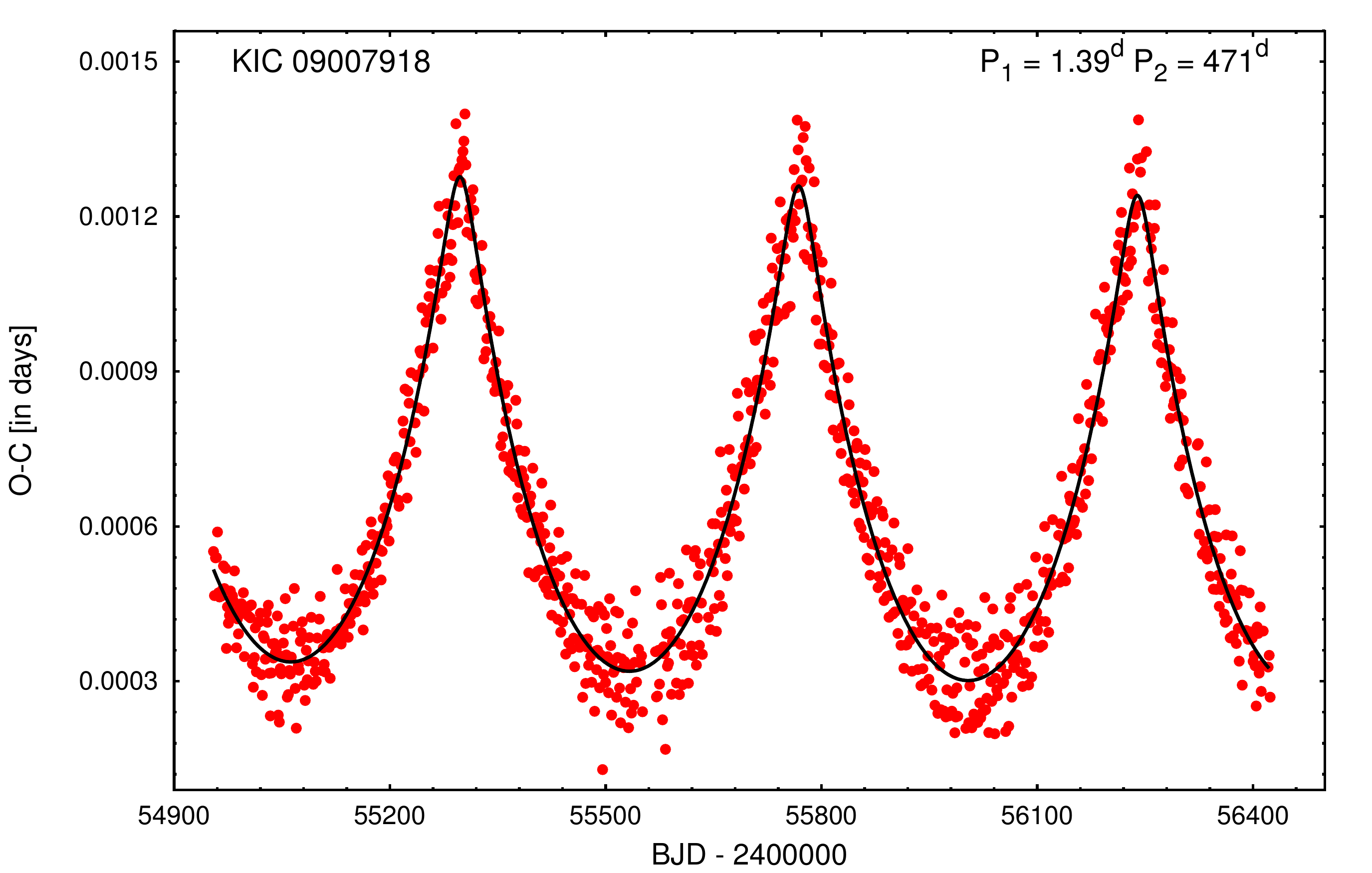}\includegraphics[width=60mm]{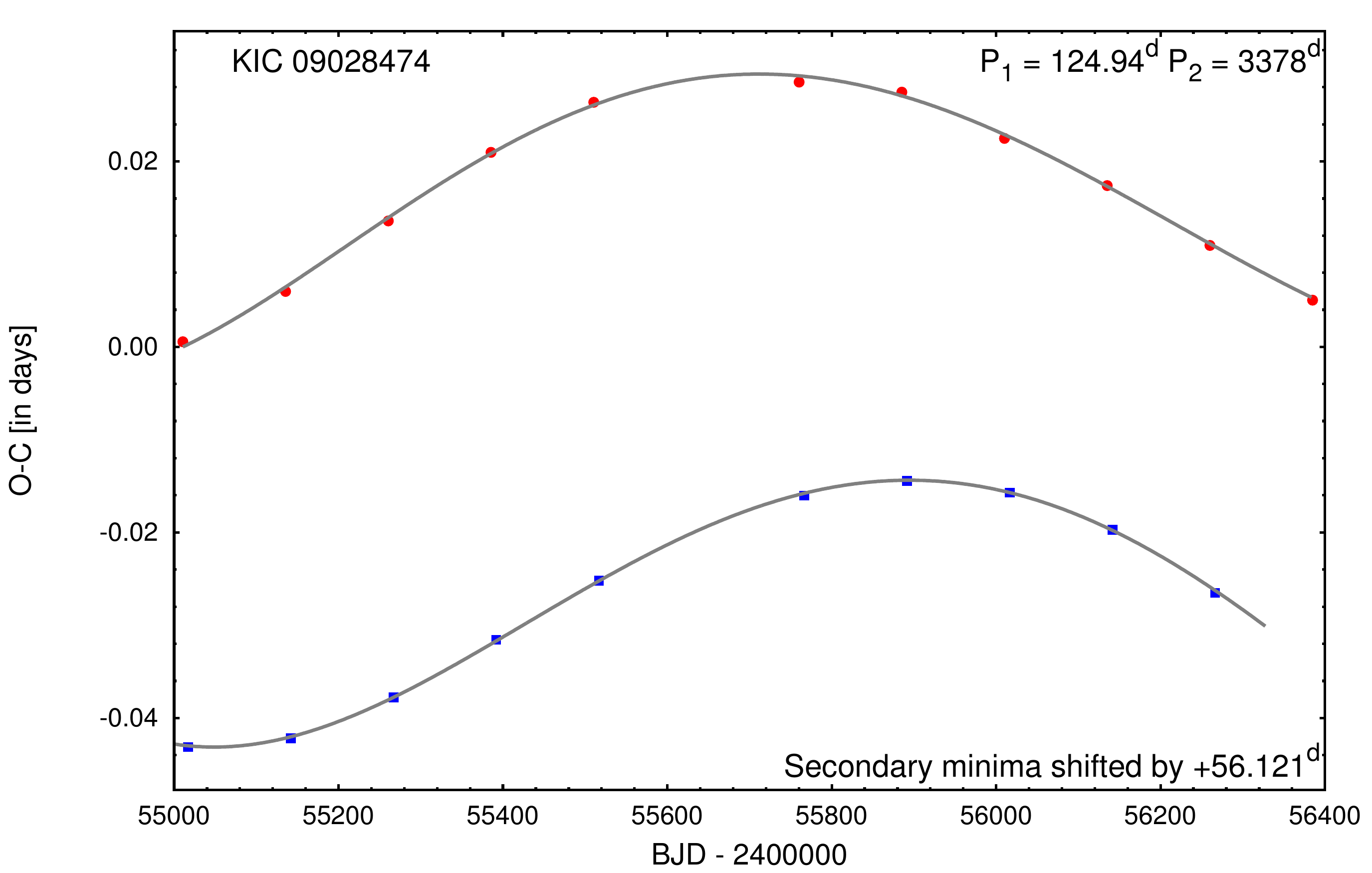}\includegraphics[width=60mm]{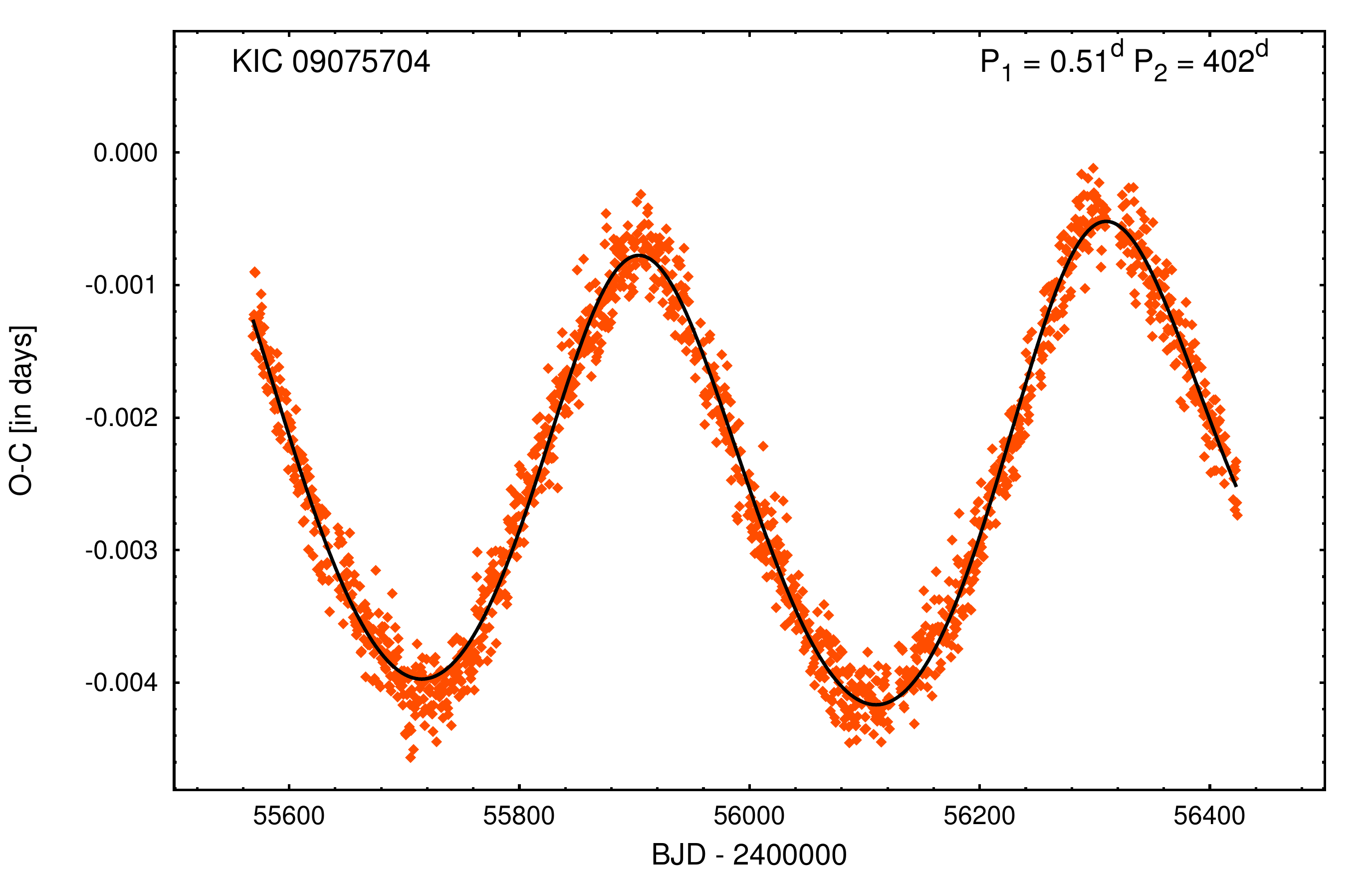}
\includegraphics[width=60mm]{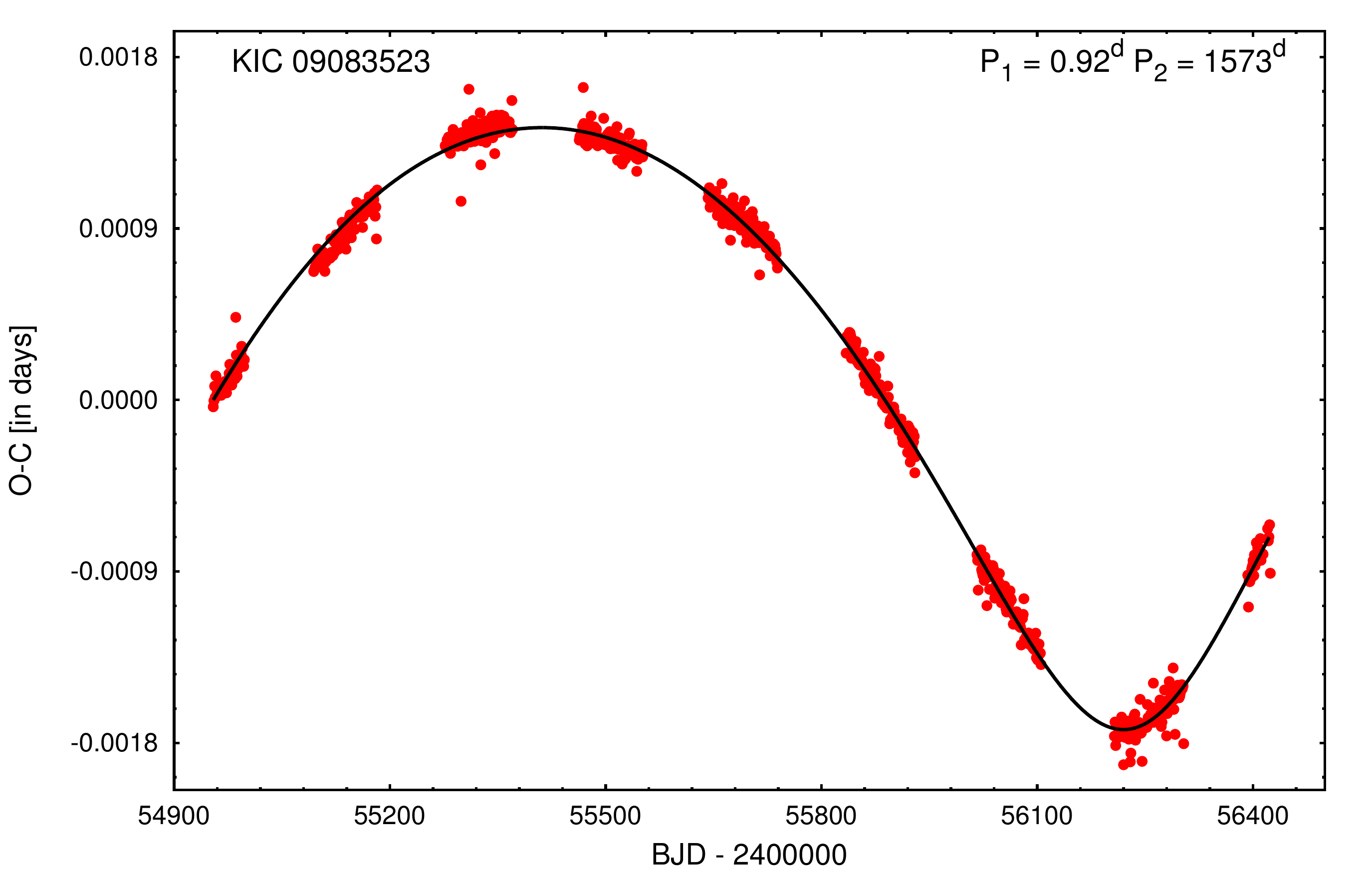}\includegraphics[width=60mm]{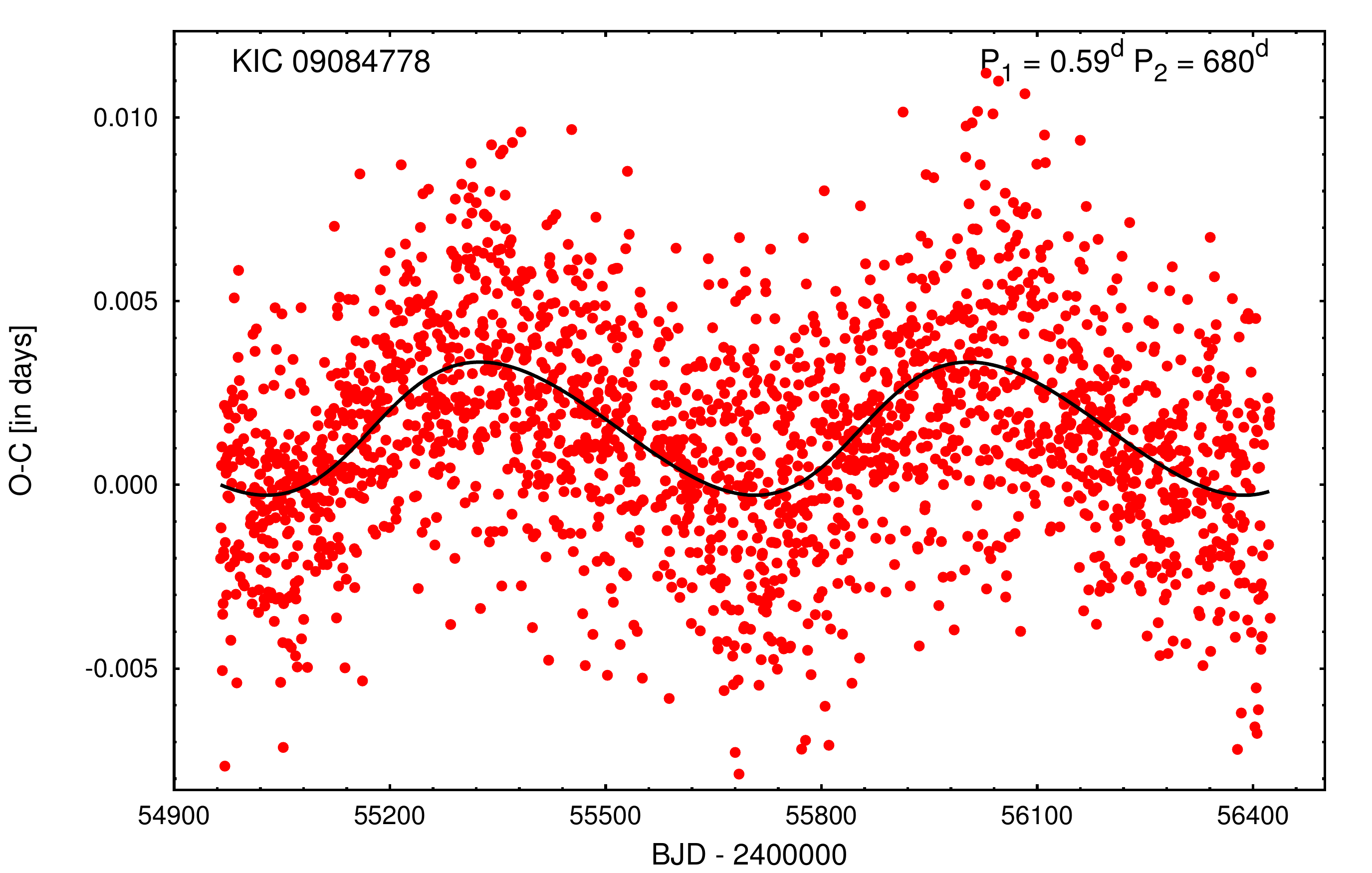}\includegraphics[width=60mm]{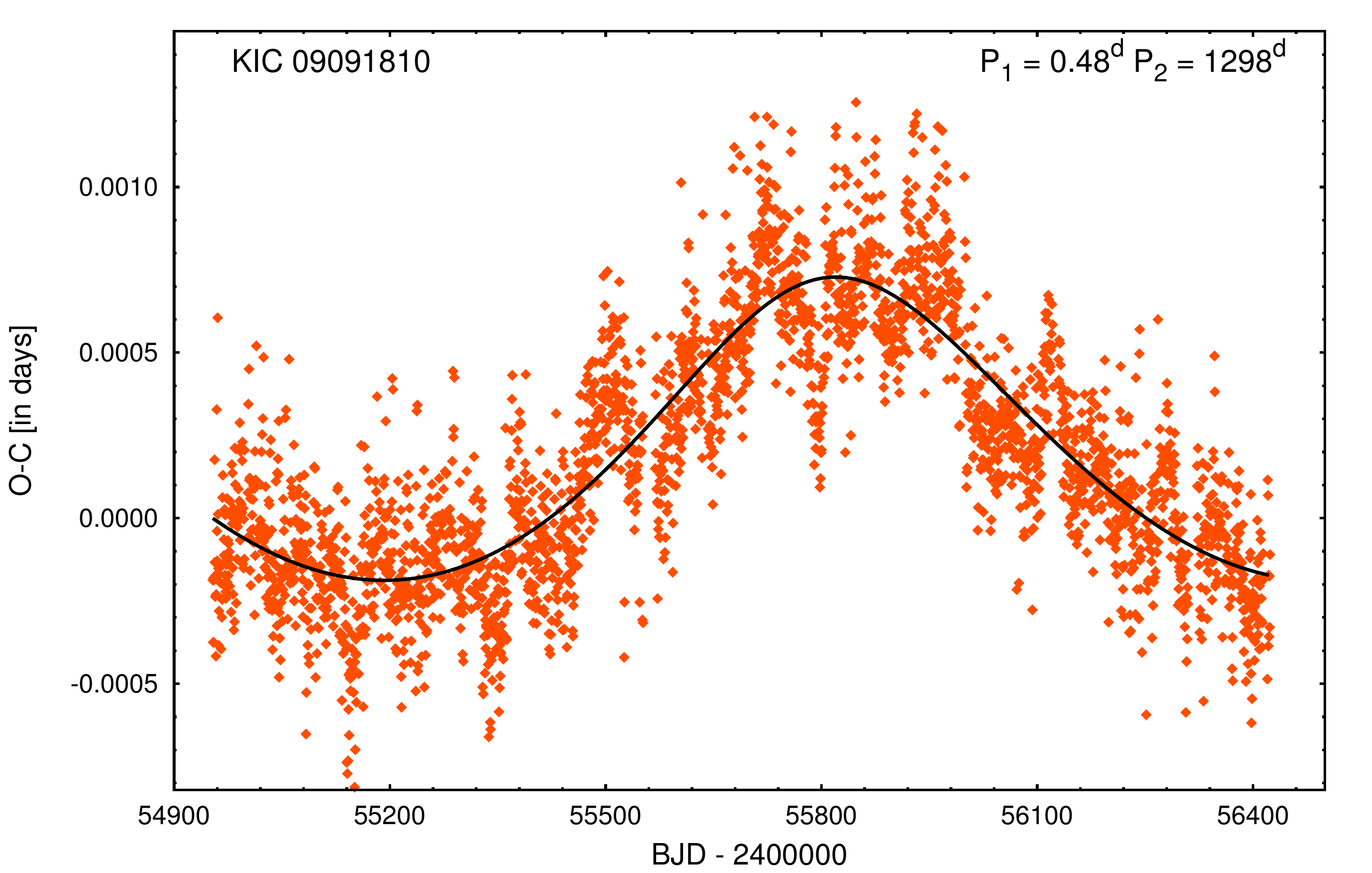}
\includegraphics[width=60mm]{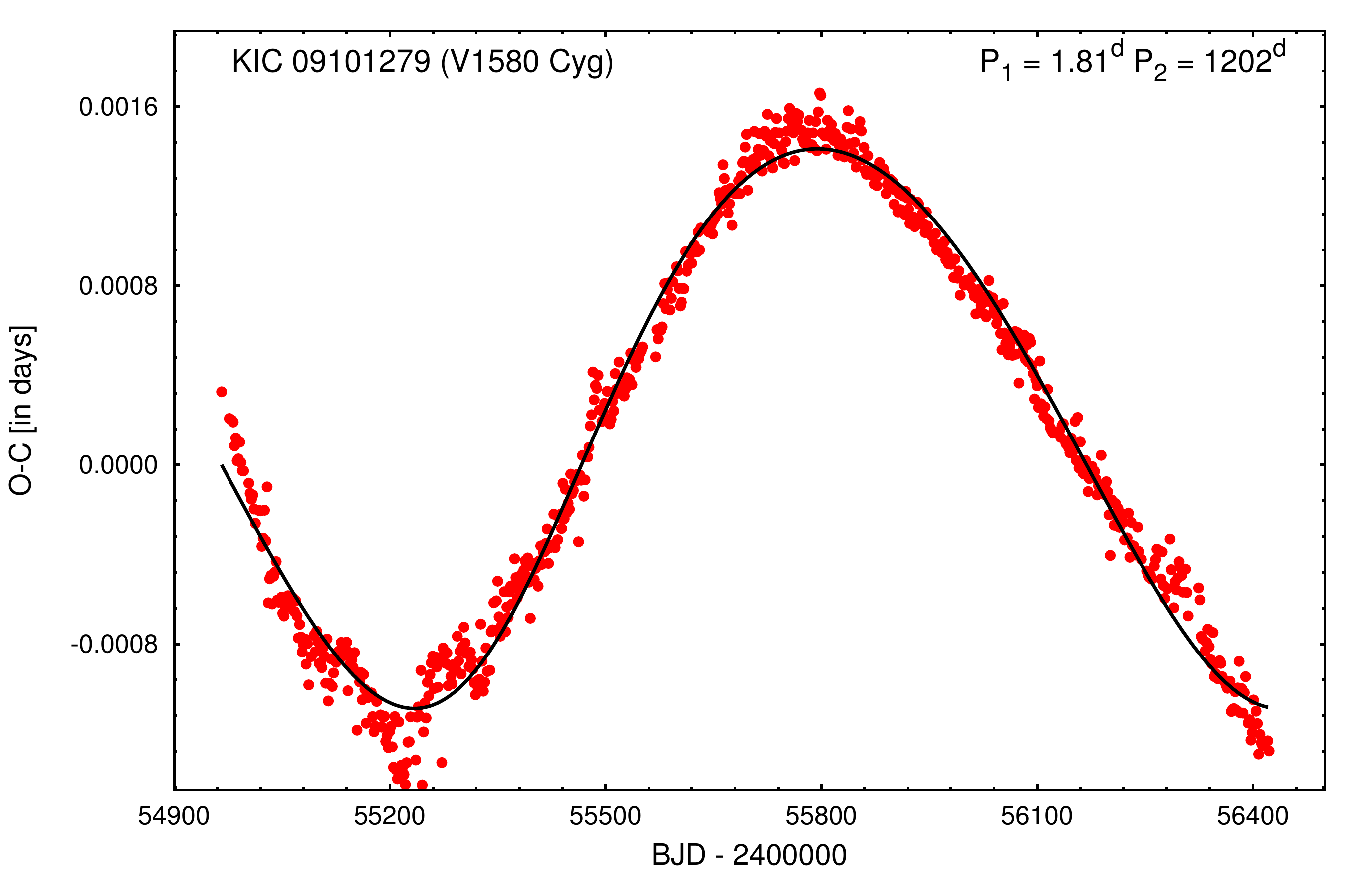}\includegraphics[width=60mm]{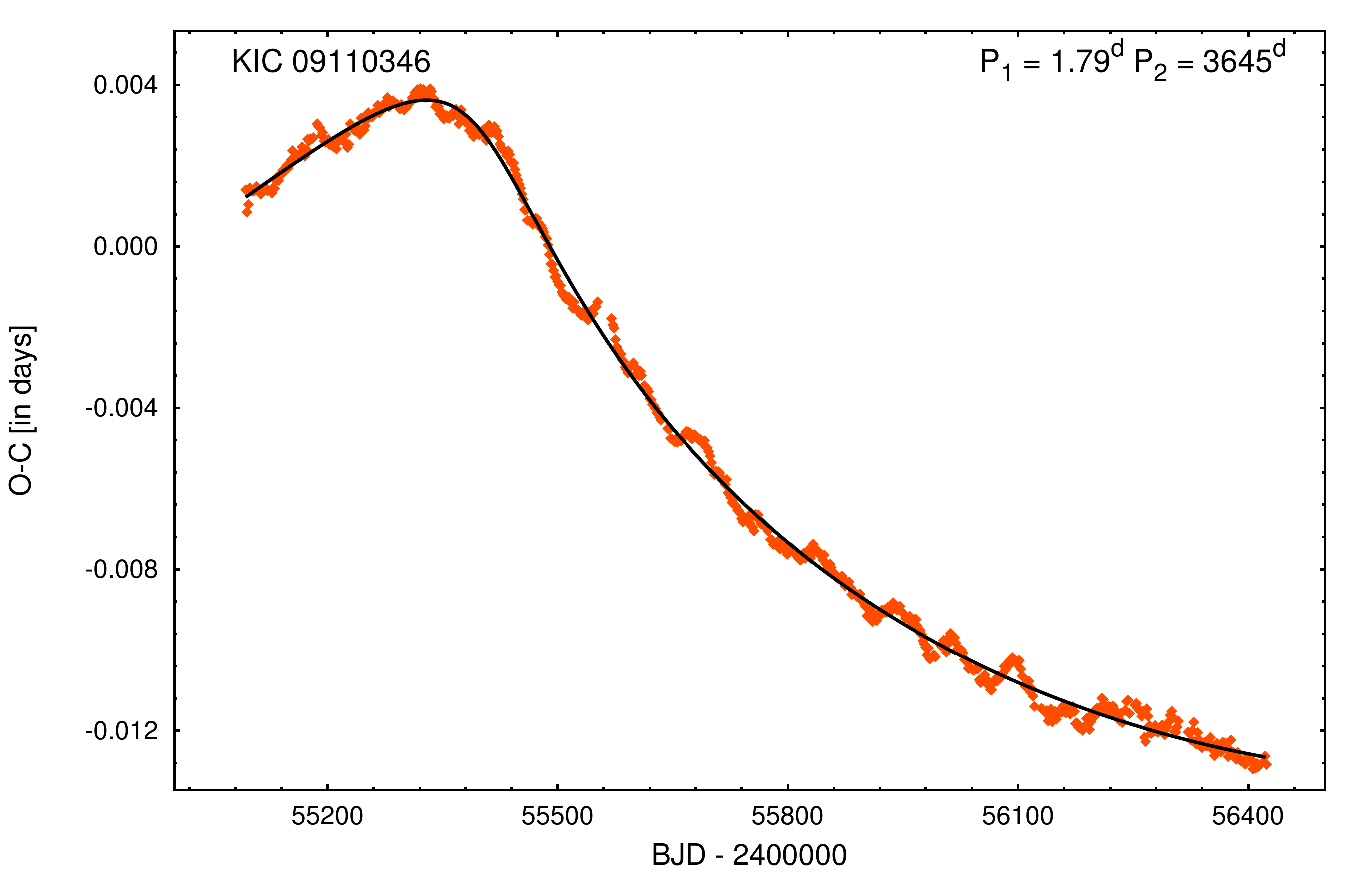}\includegraphics[width=60mm]{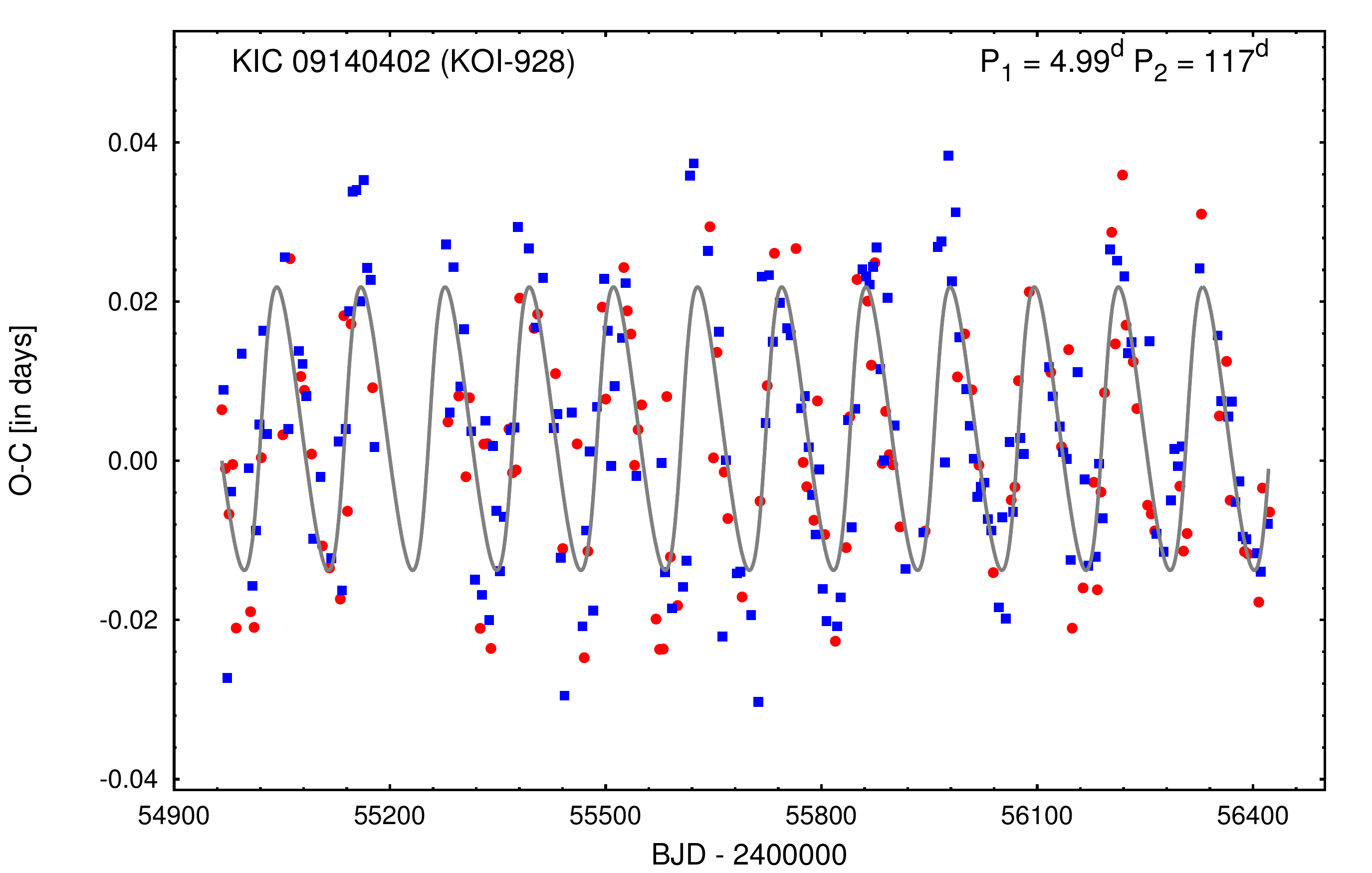}
\includegraphics[width=60mm]{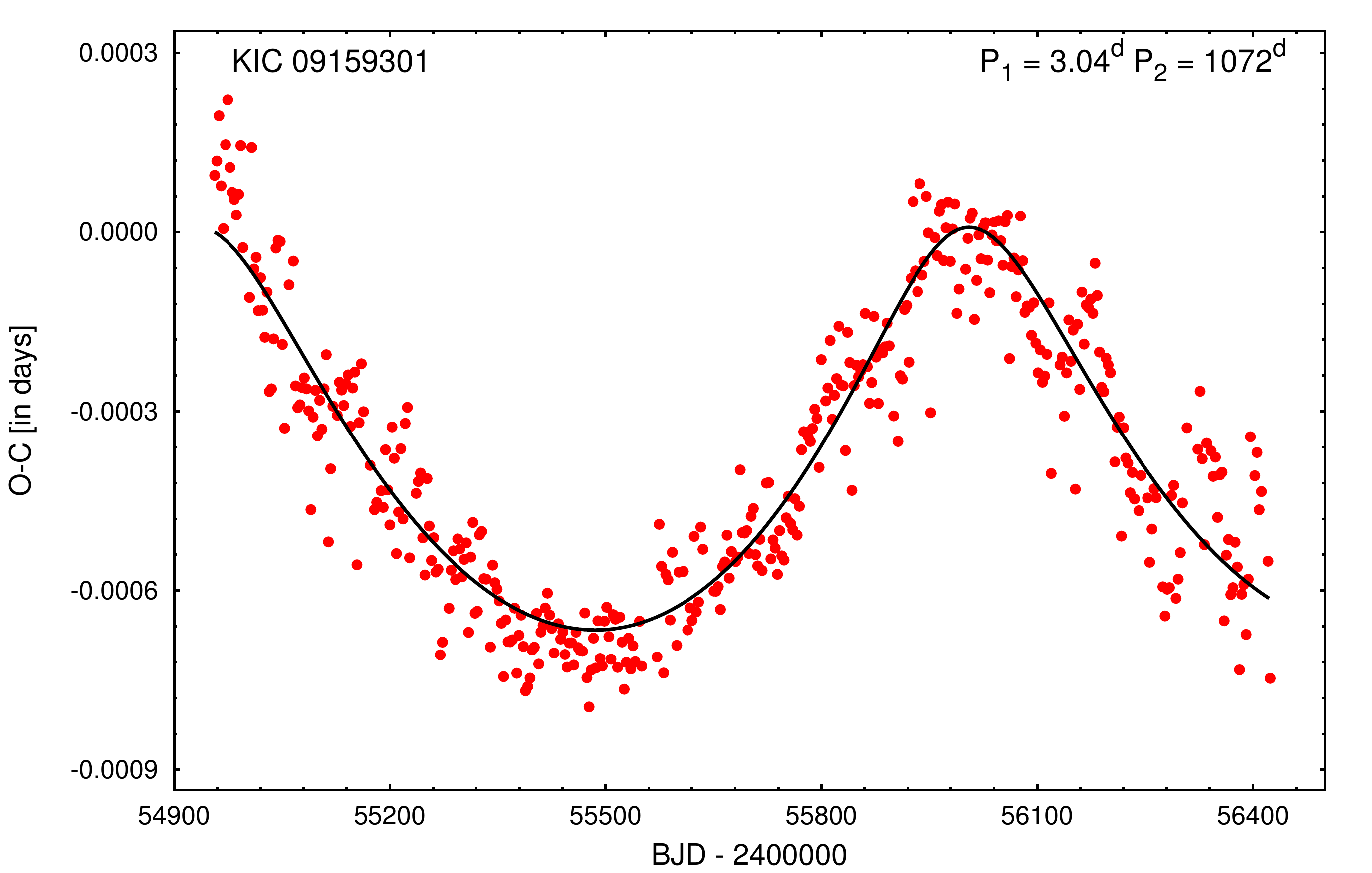}\includegraphics[width=60mm]{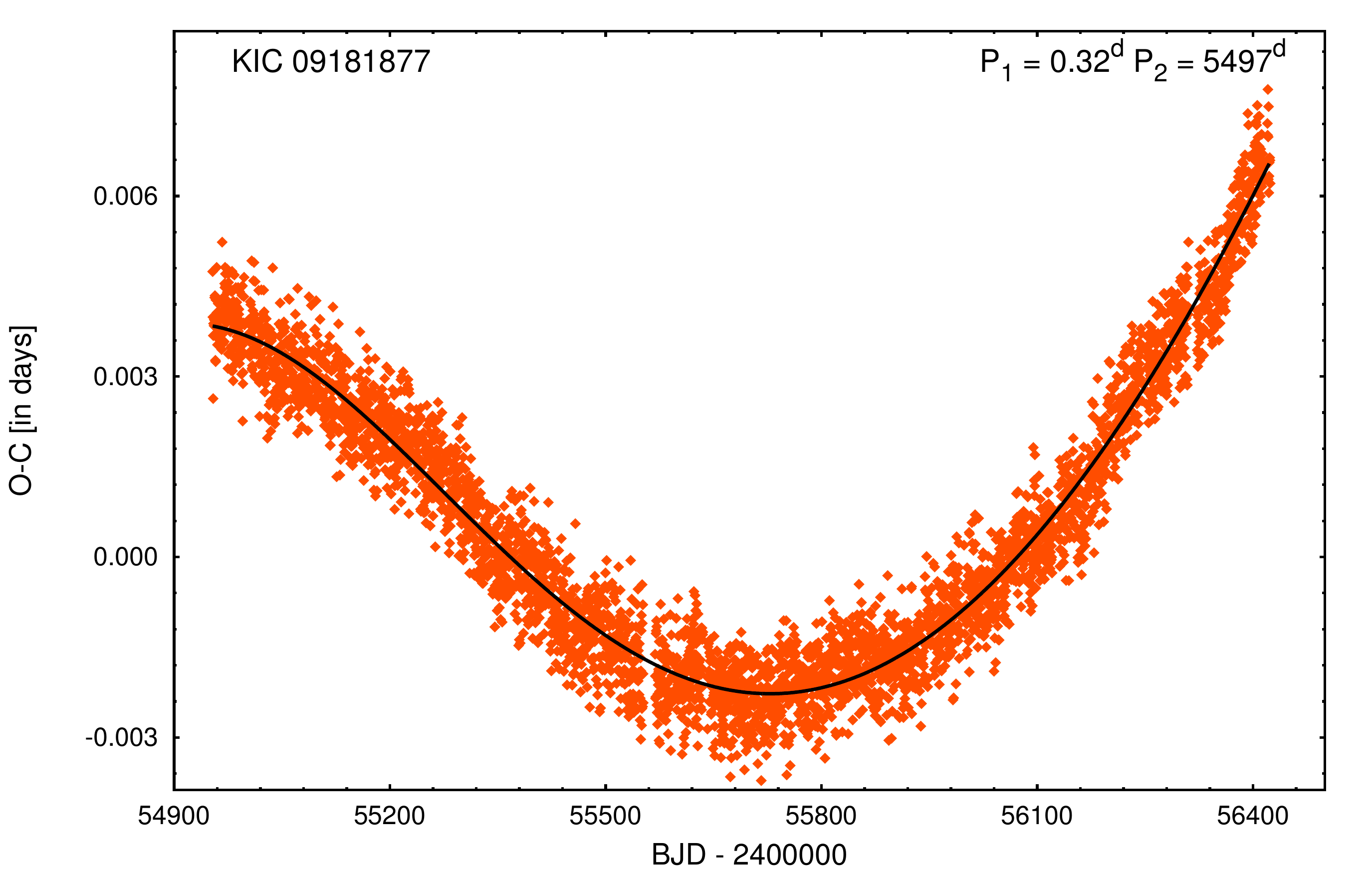}\includegraphics[width=60mm]{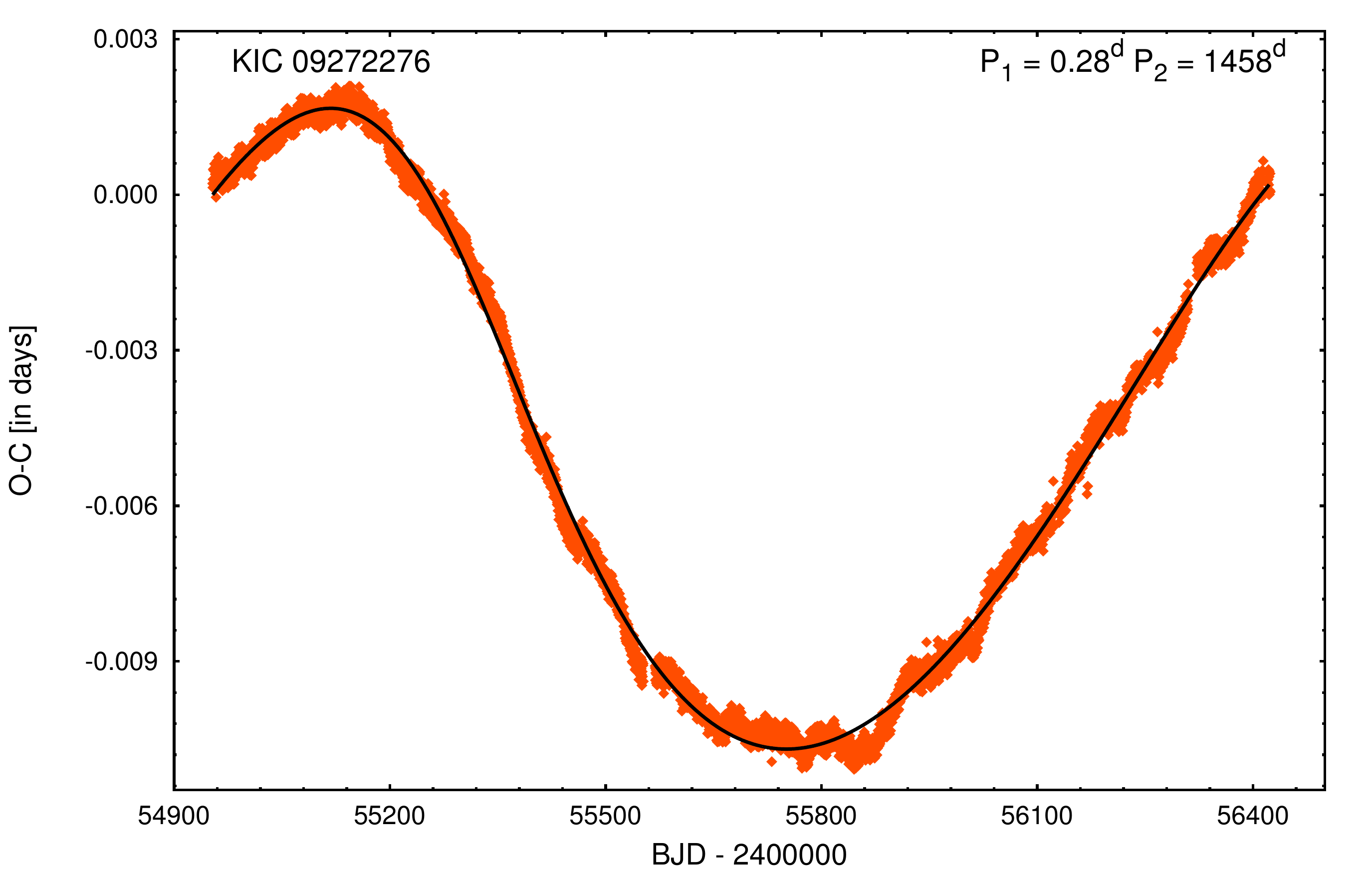}
\caption{(continued)}
\end{figure*}

\addtocounter{figure}{-1}

\begin{figure*}
\includegraphics[width=60mm]{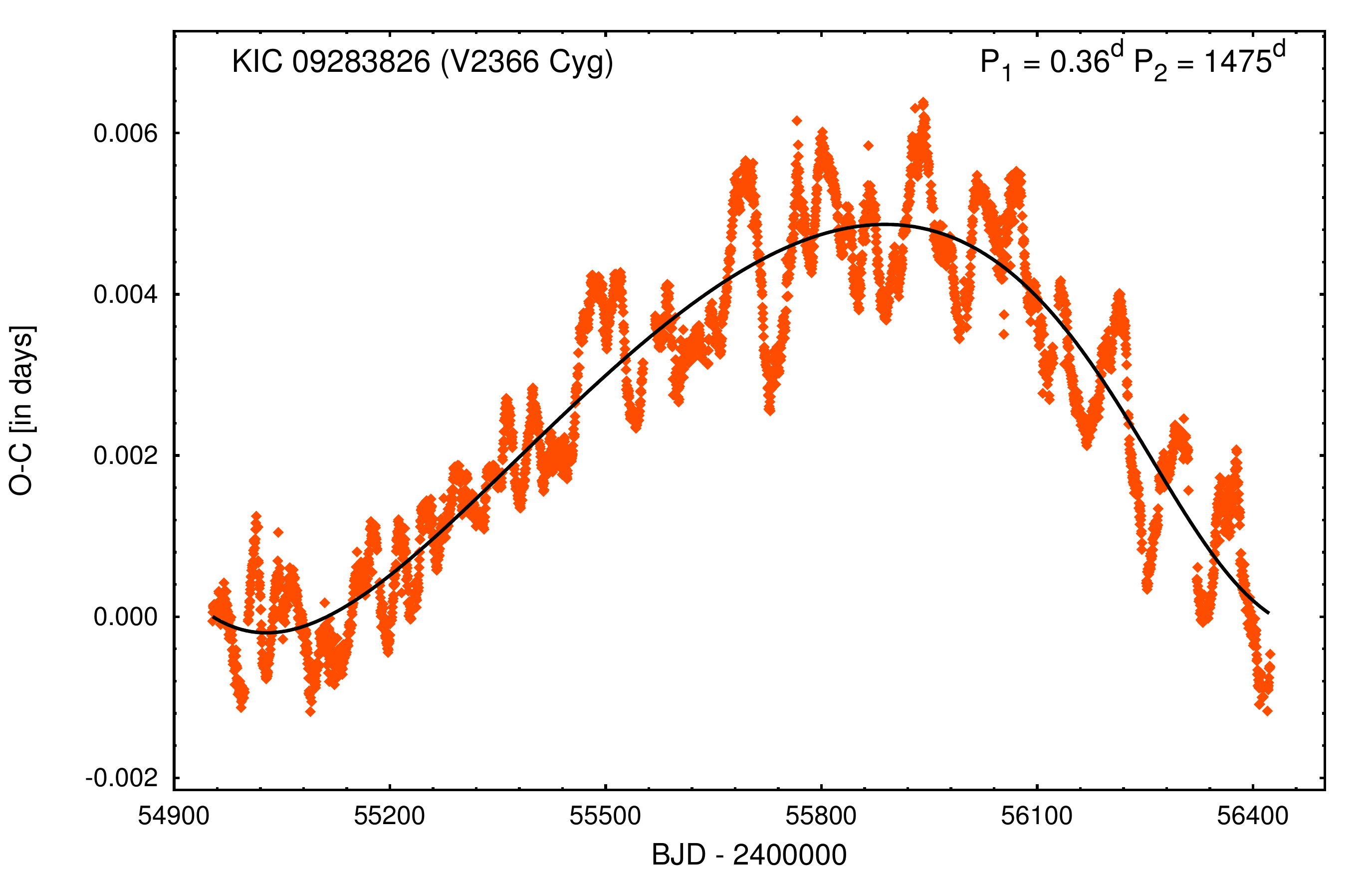}\includegraphics[width=60mm]{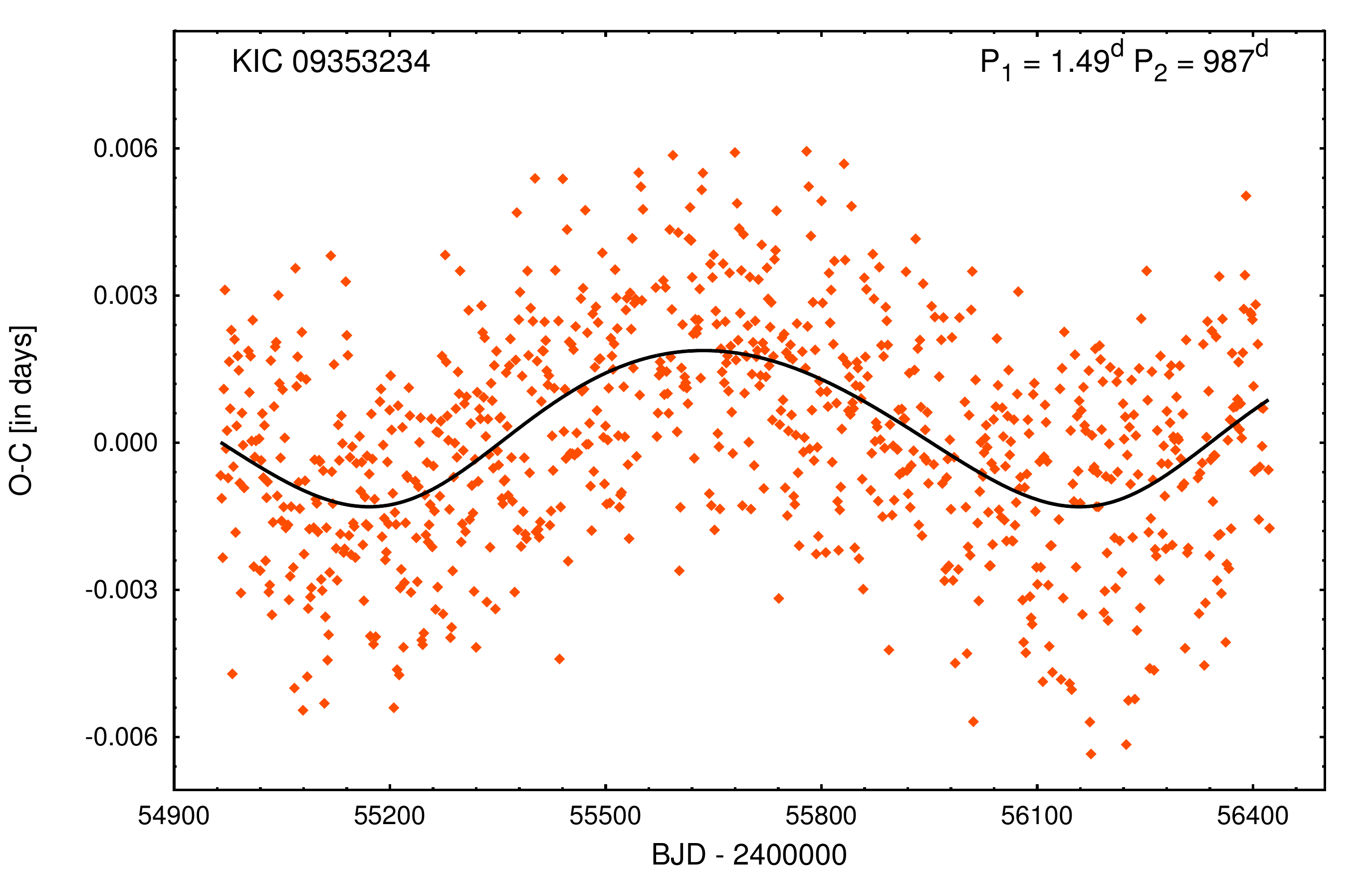}\includegraphics[width=60mm]{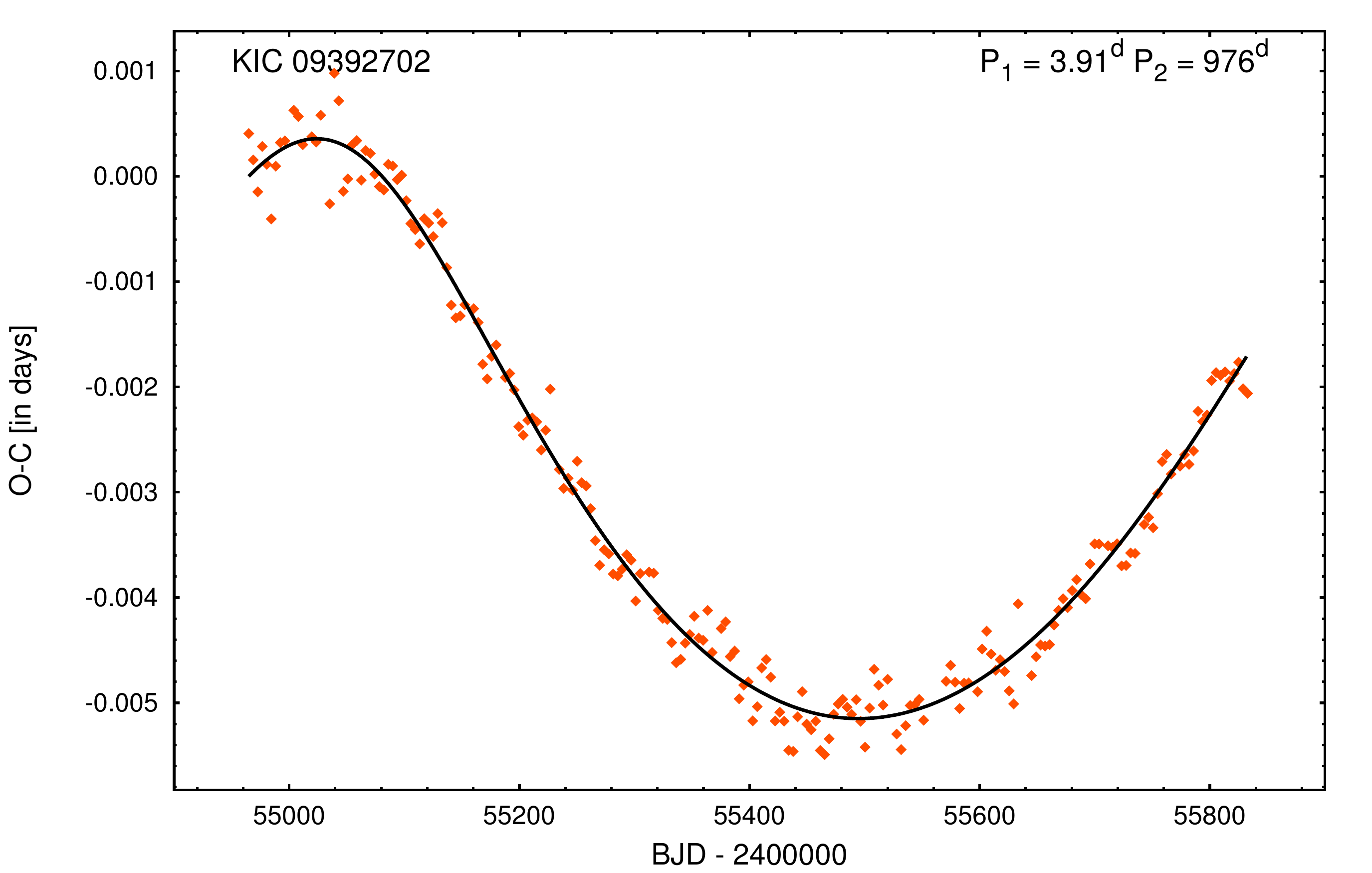}
\includegraphics[width=60mm]{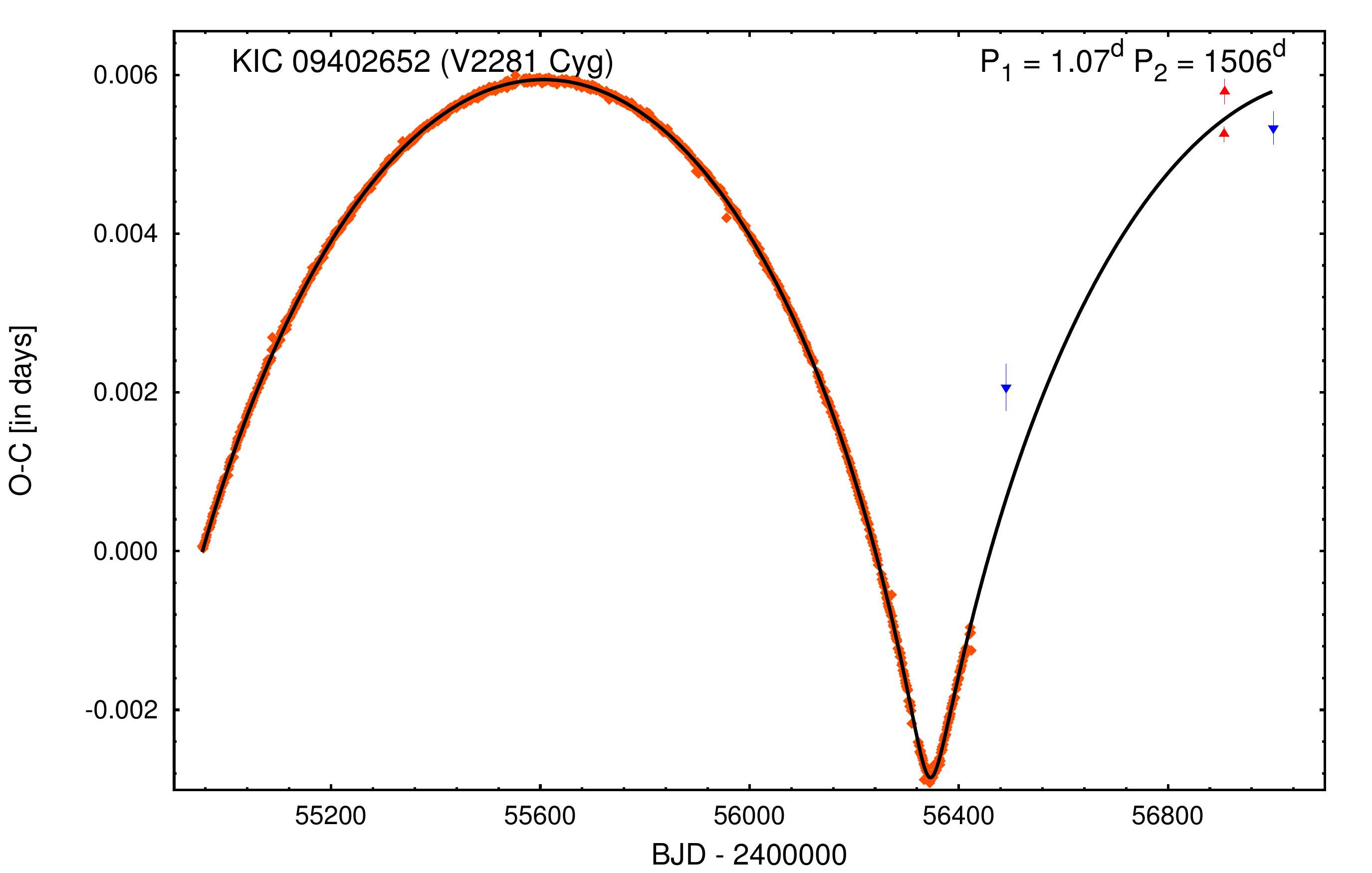}\includegraphics[width=60mm]{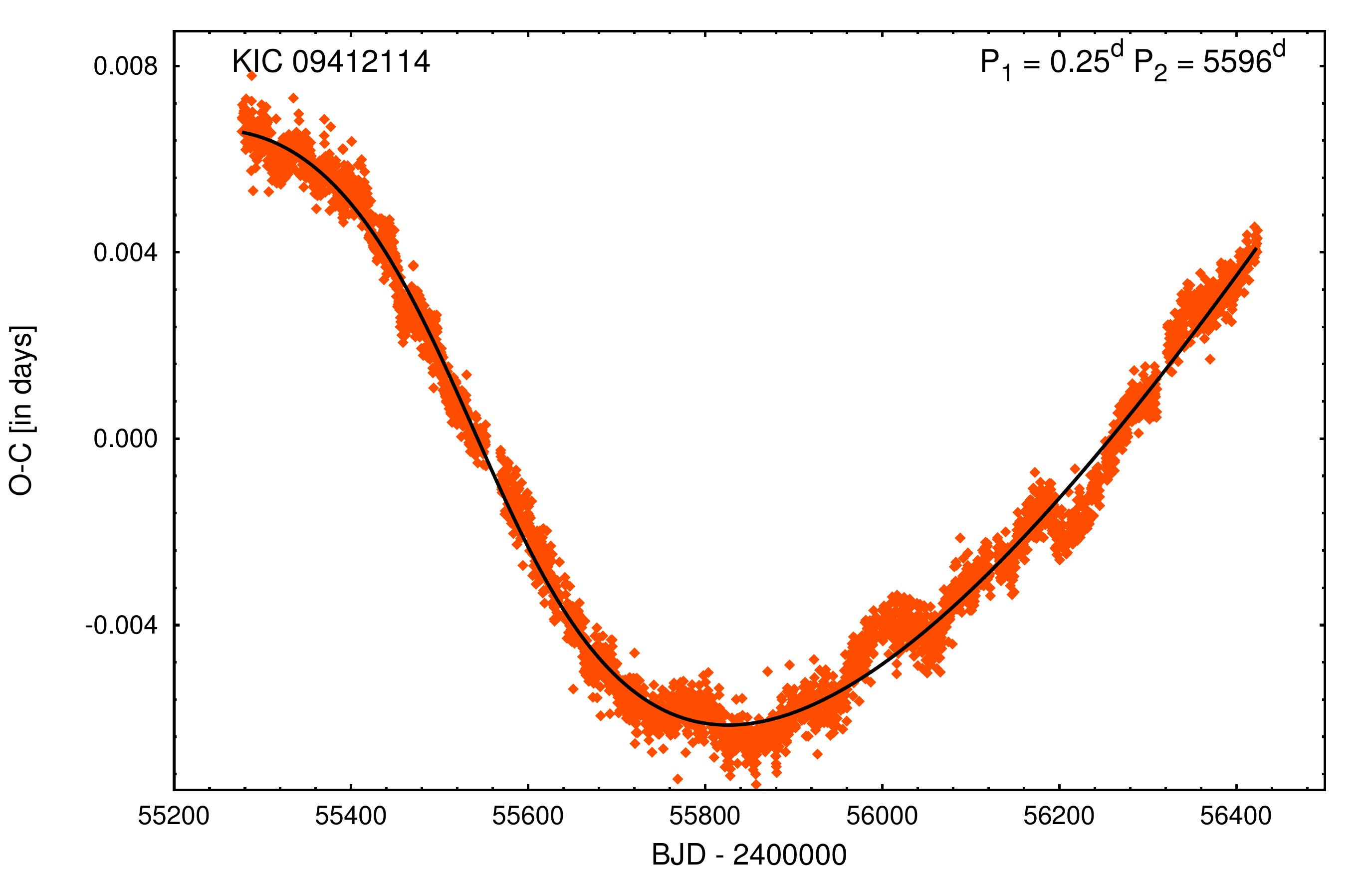}\includegraphics[width=60mm]{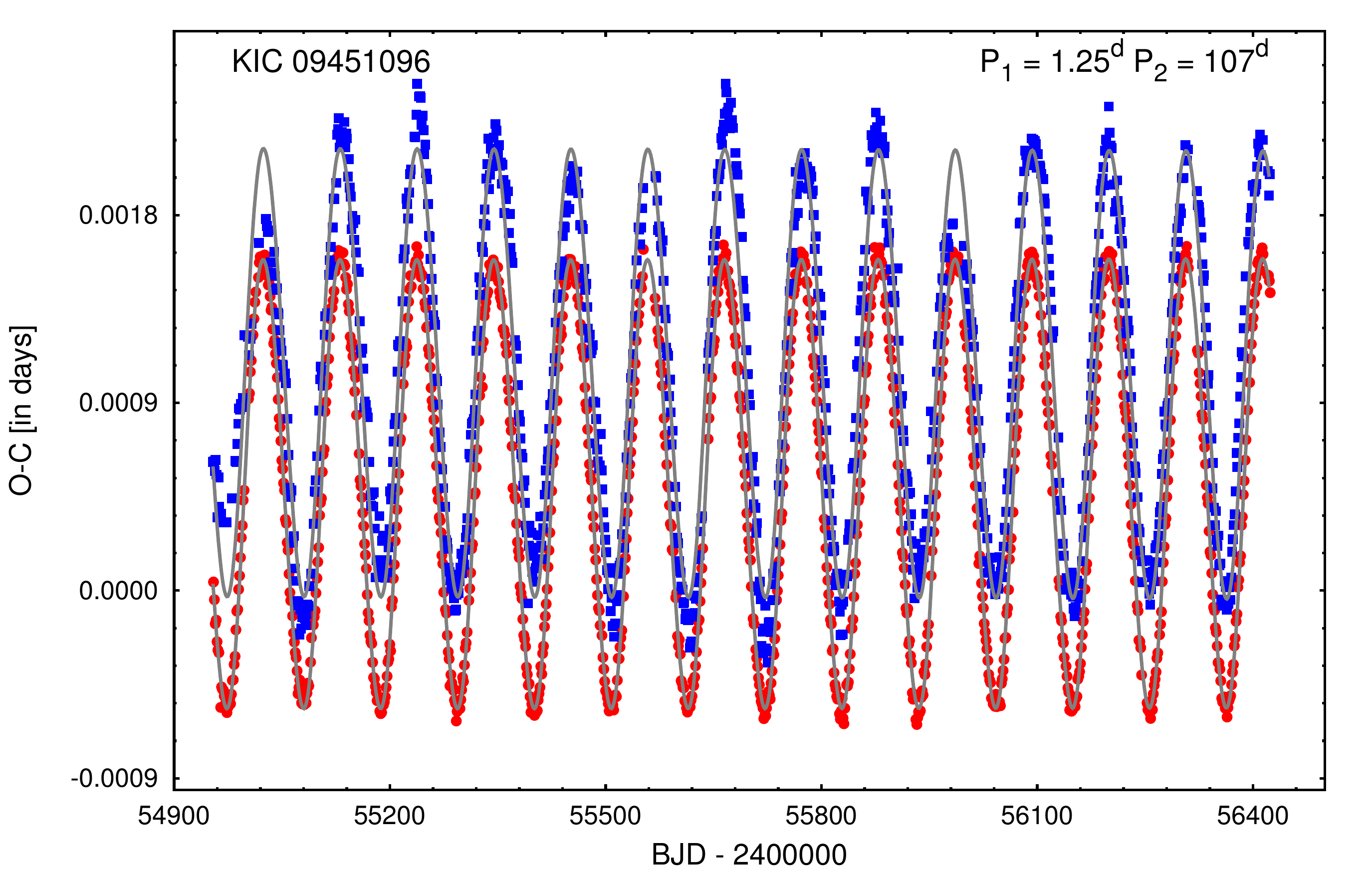}
\includegraphics[width=60mm]{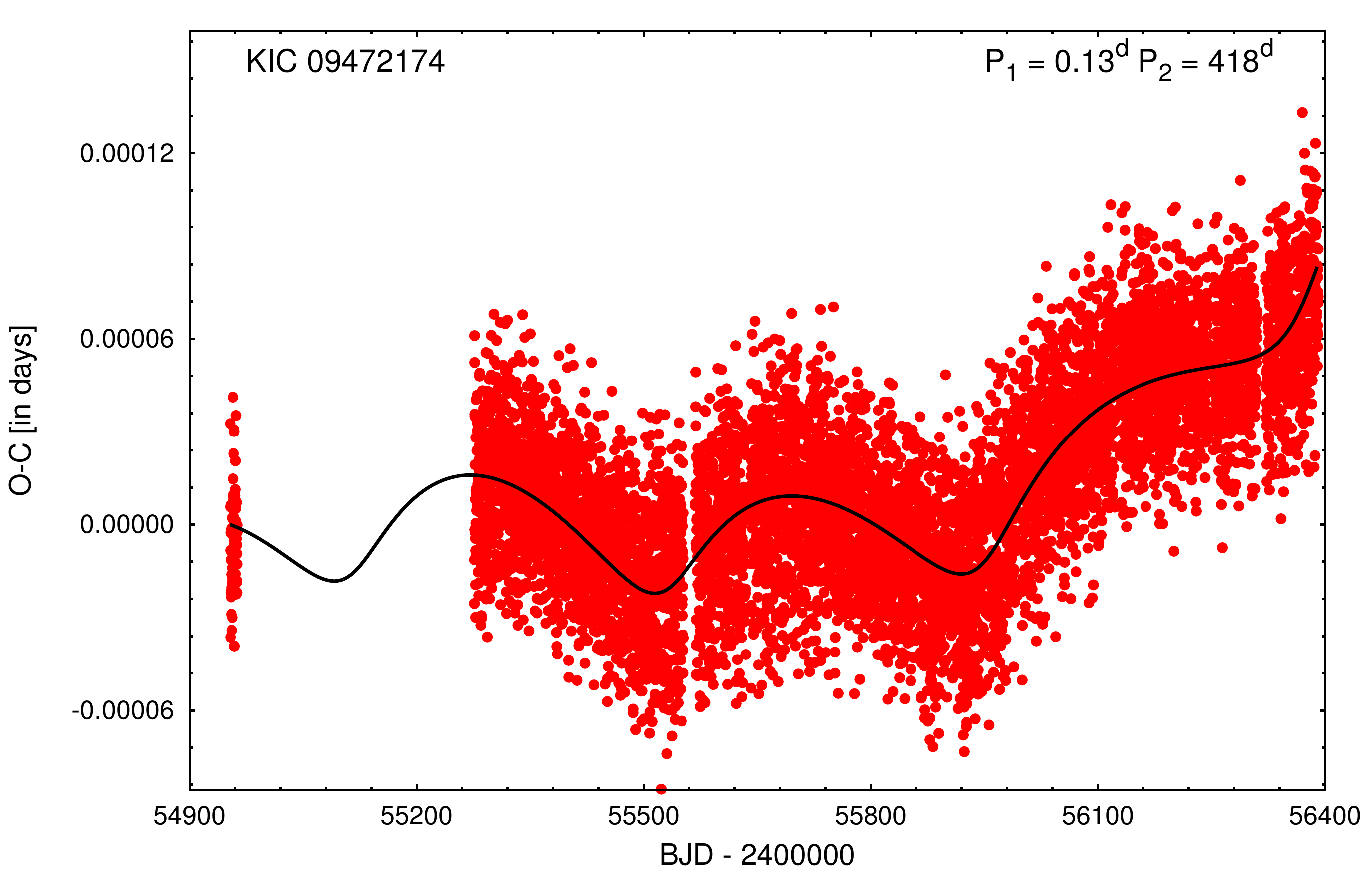}\includegraphics[width=60mm]{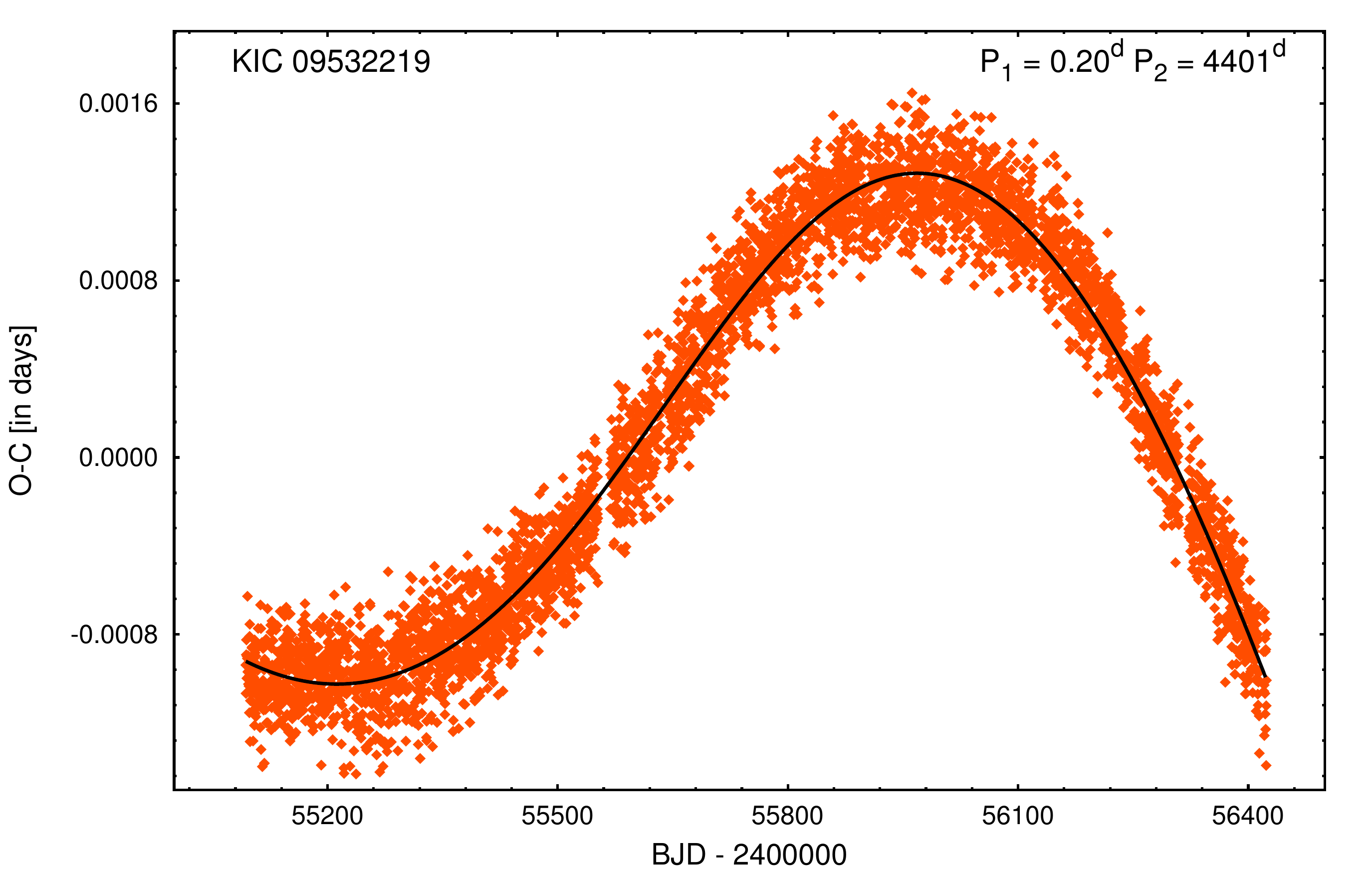}\includegraphics[width=60mm]{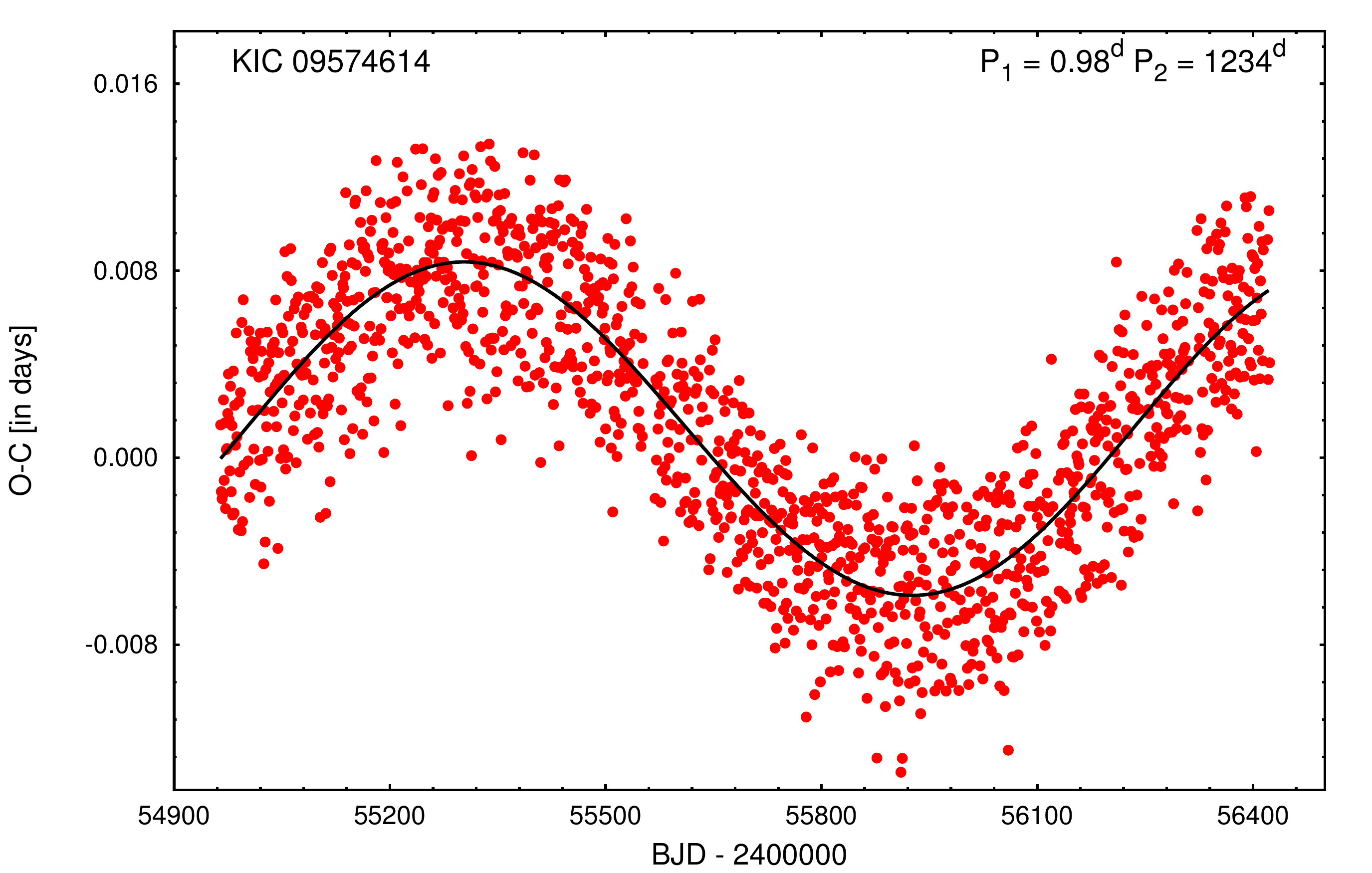}
\includegraphics[width=60mm]{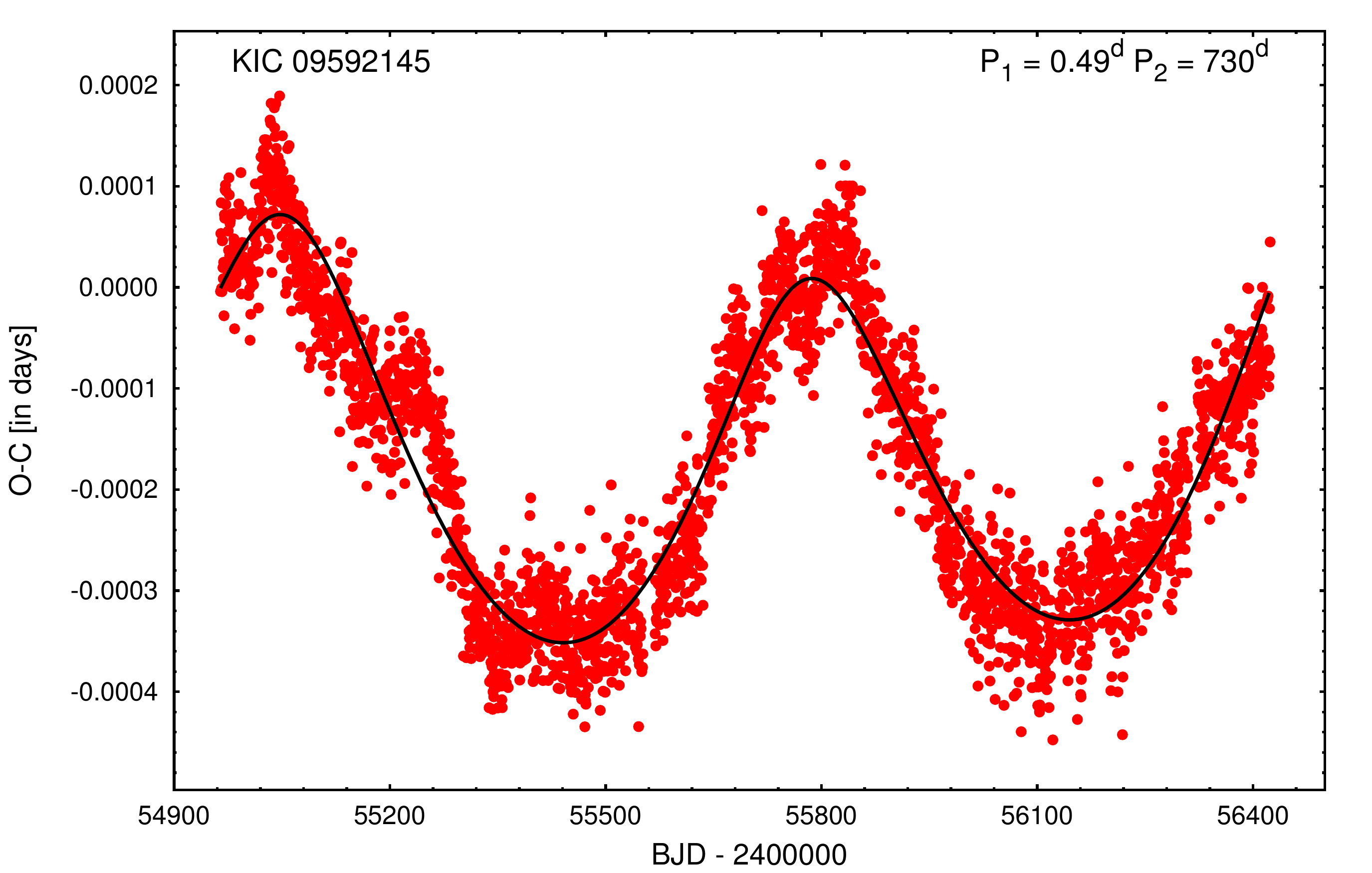}\includegraphics[width=60mm]{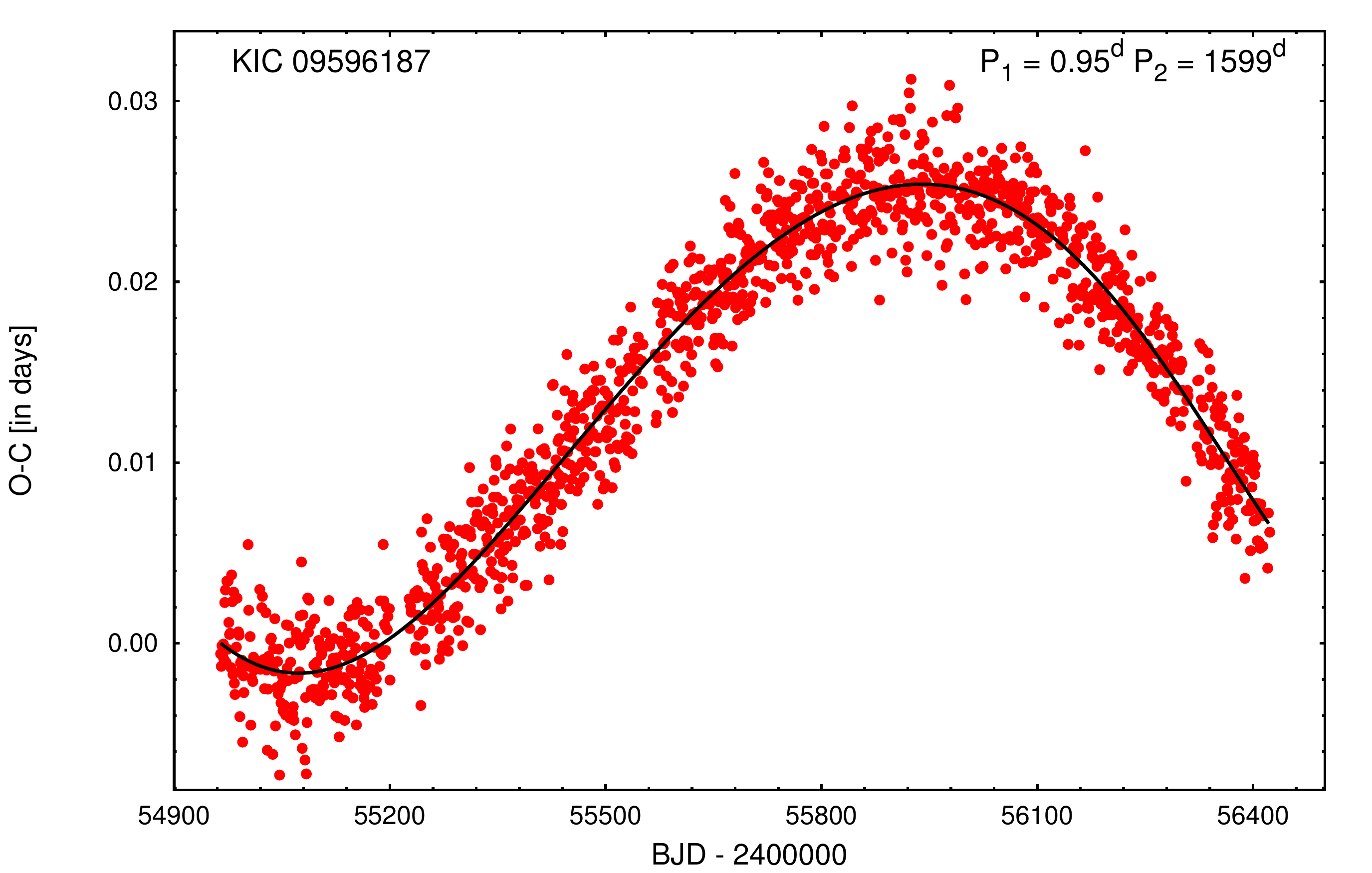}\includegraphics[width=60mm]{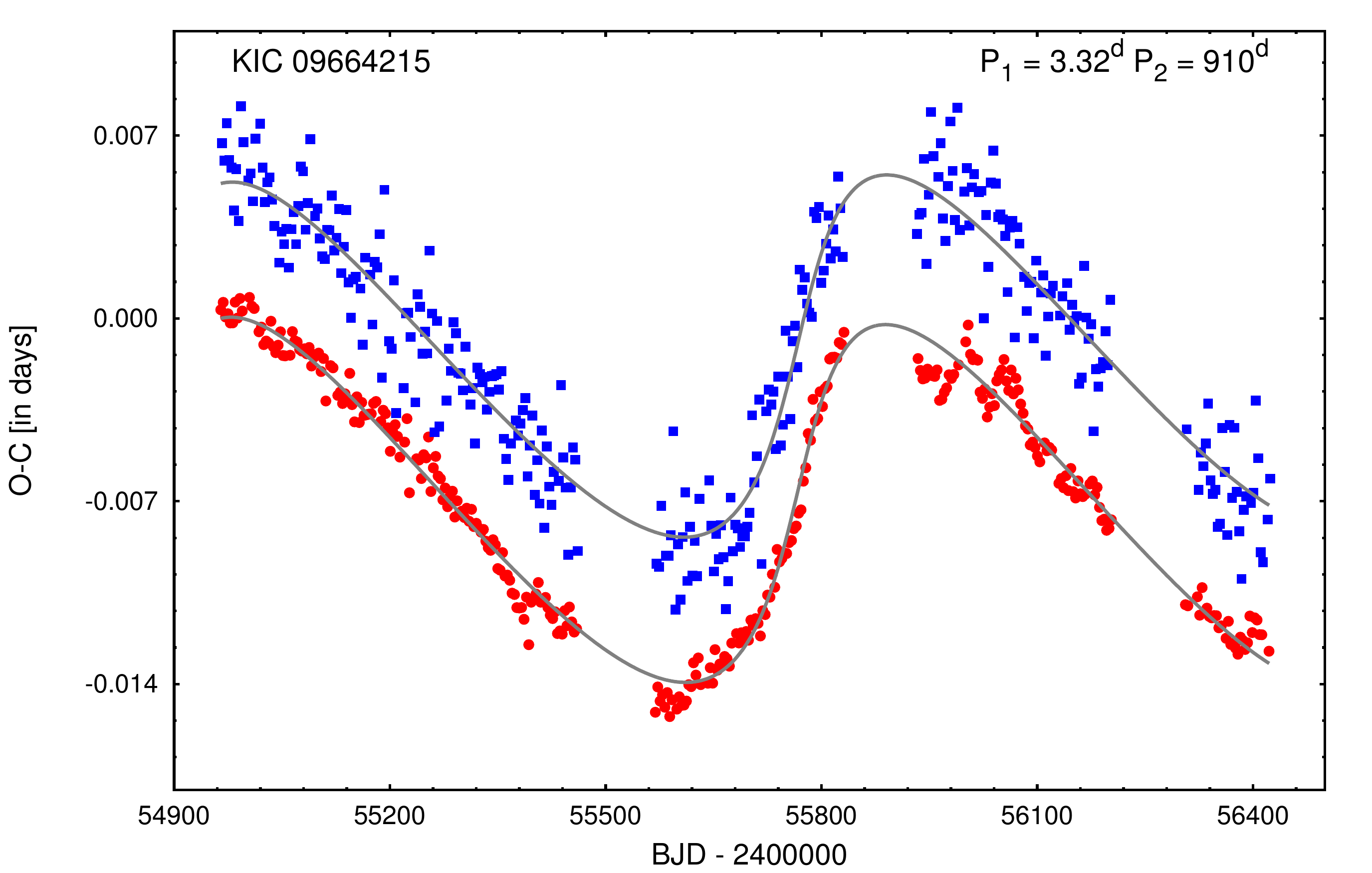}
\includegraphics[width=60mm]{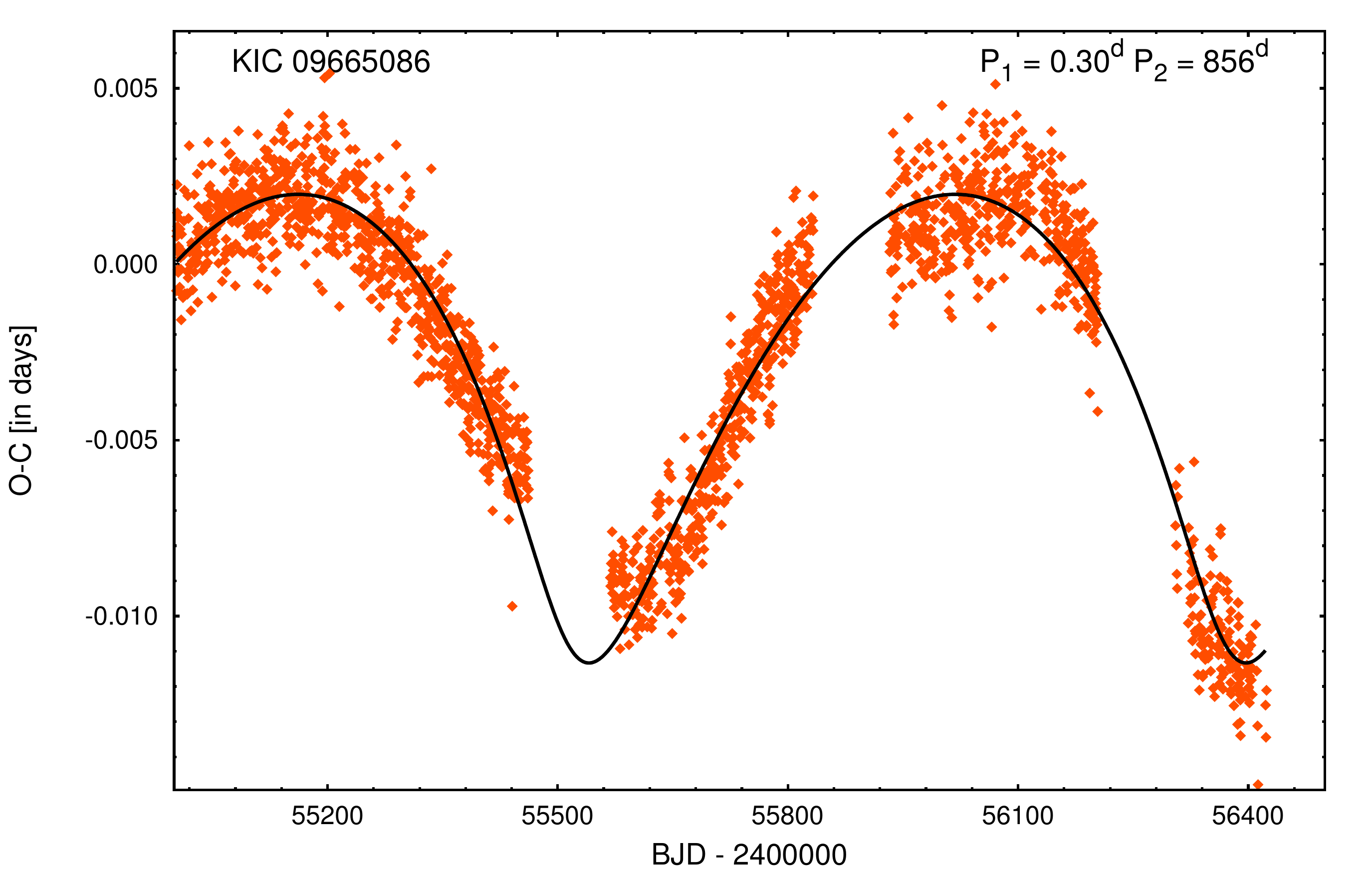}\includegraphics[width=60mm]{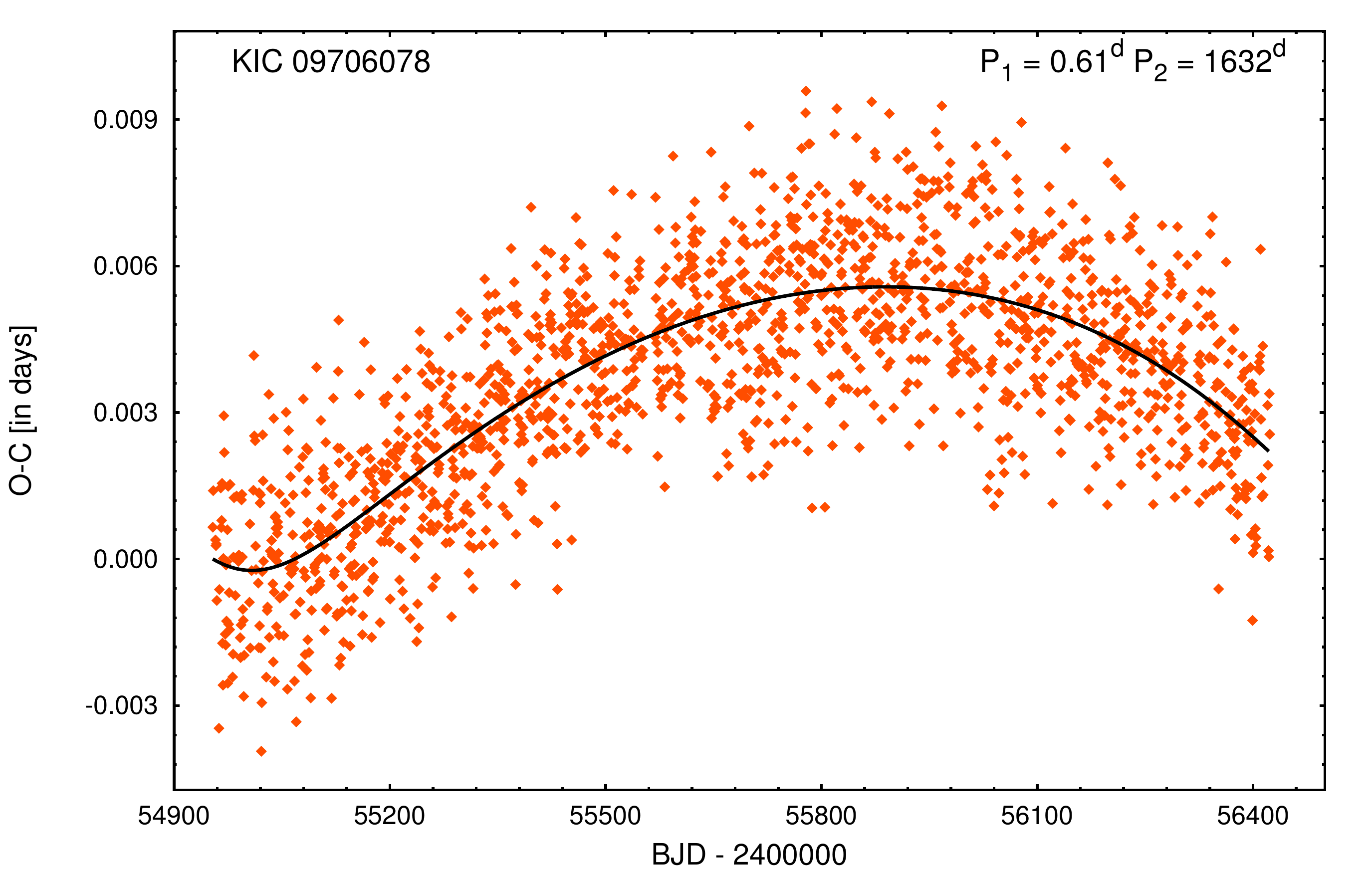}\includegraphics[width=60mm]{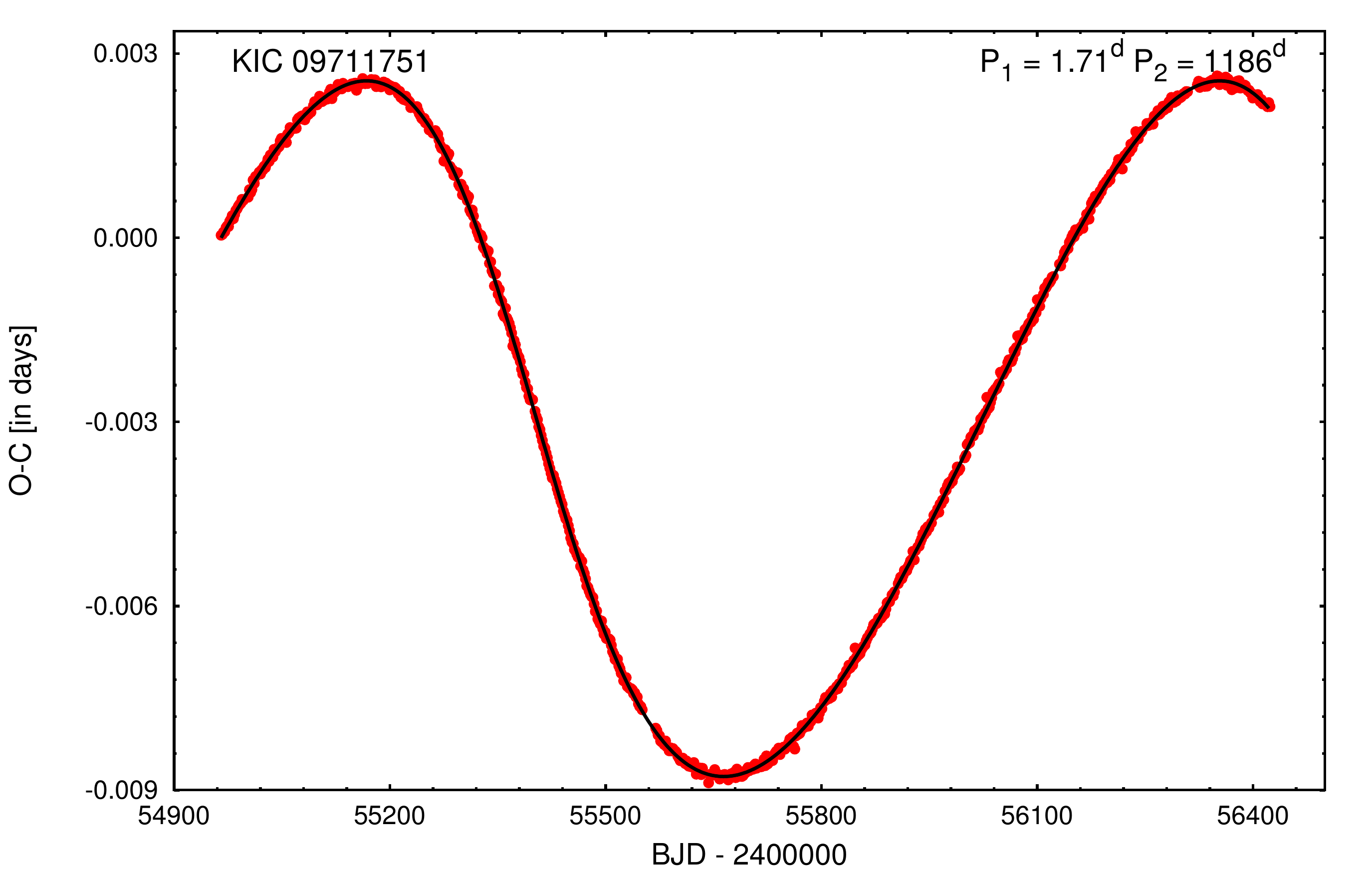}
\includegraphics[width=60mm]{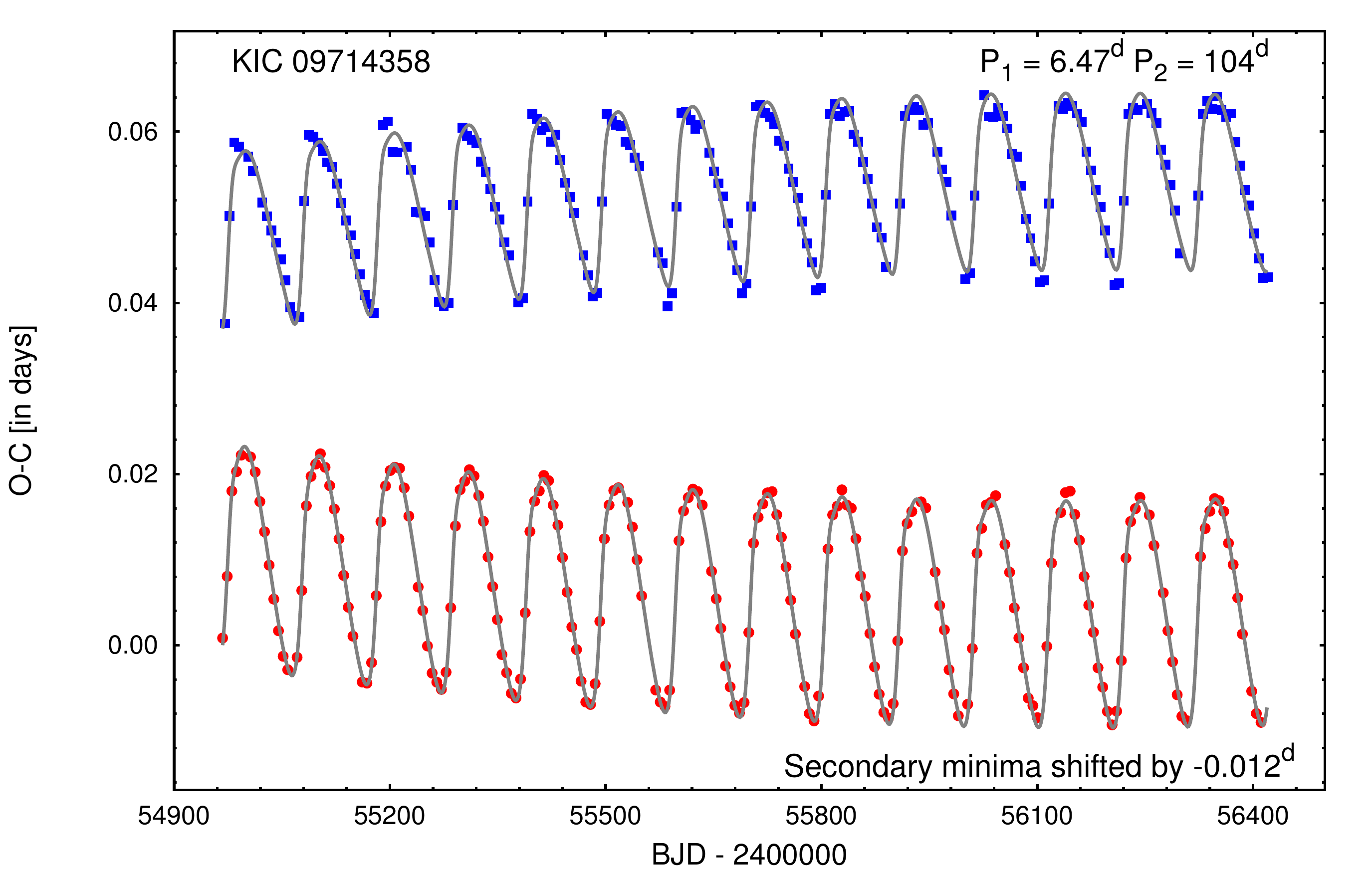}\includegraphics[width=60mm]{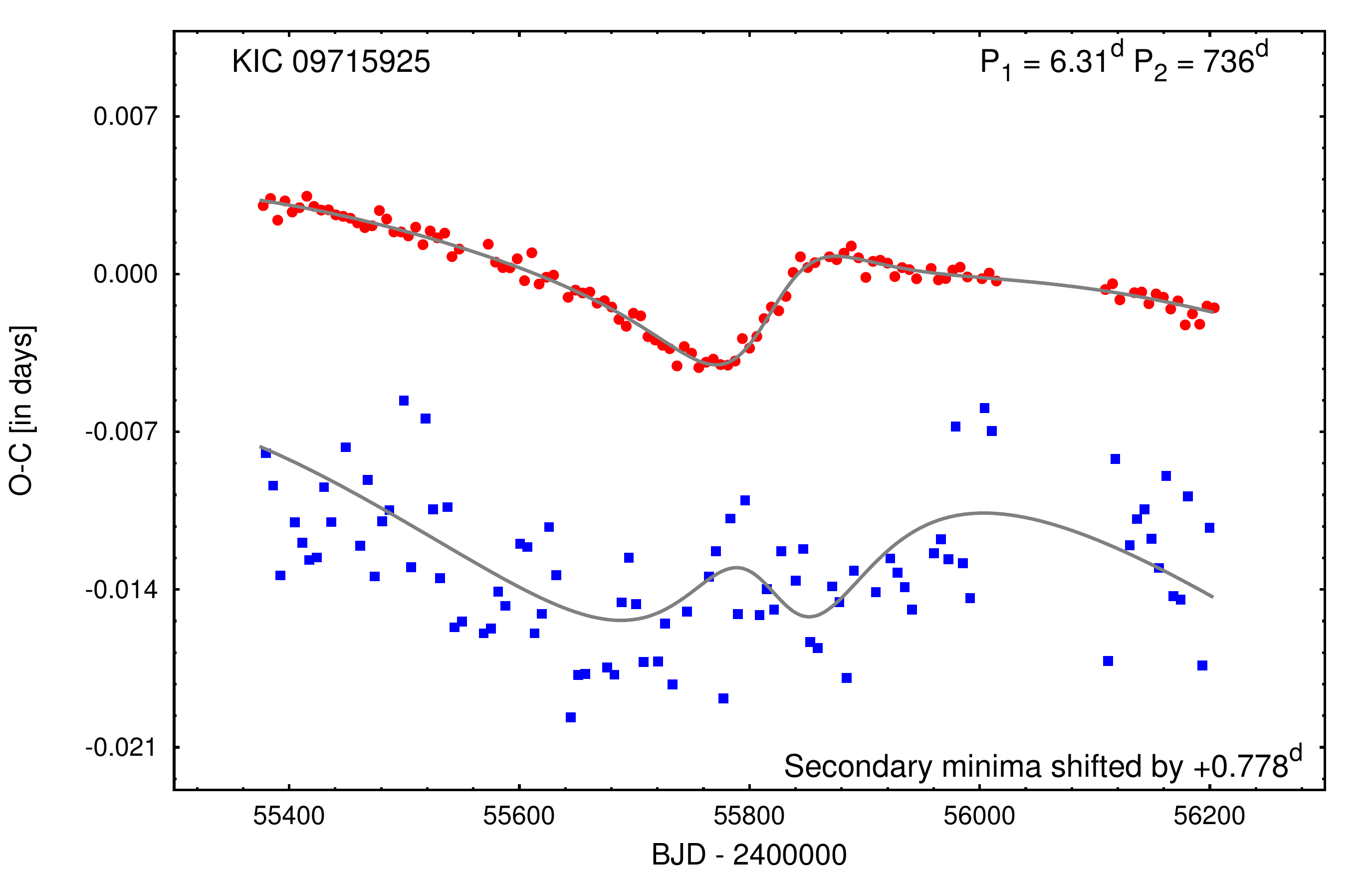}\includegraphics[width=60mm]{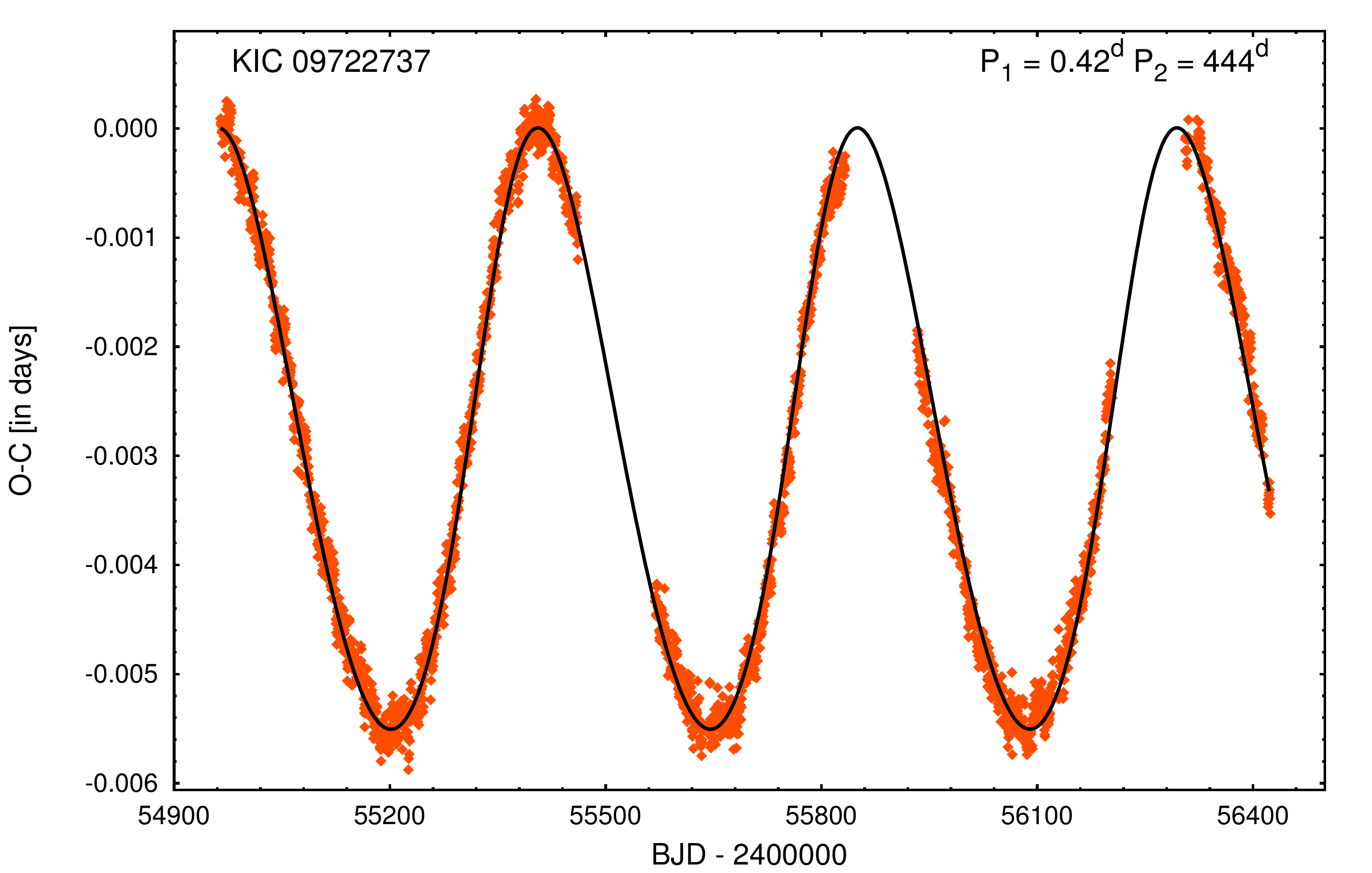}
\caption{(continued)}
\end{figure*}

\addtocounter{figure}{-1}

\begin{figure*}
\includegraphics[width=60mm]{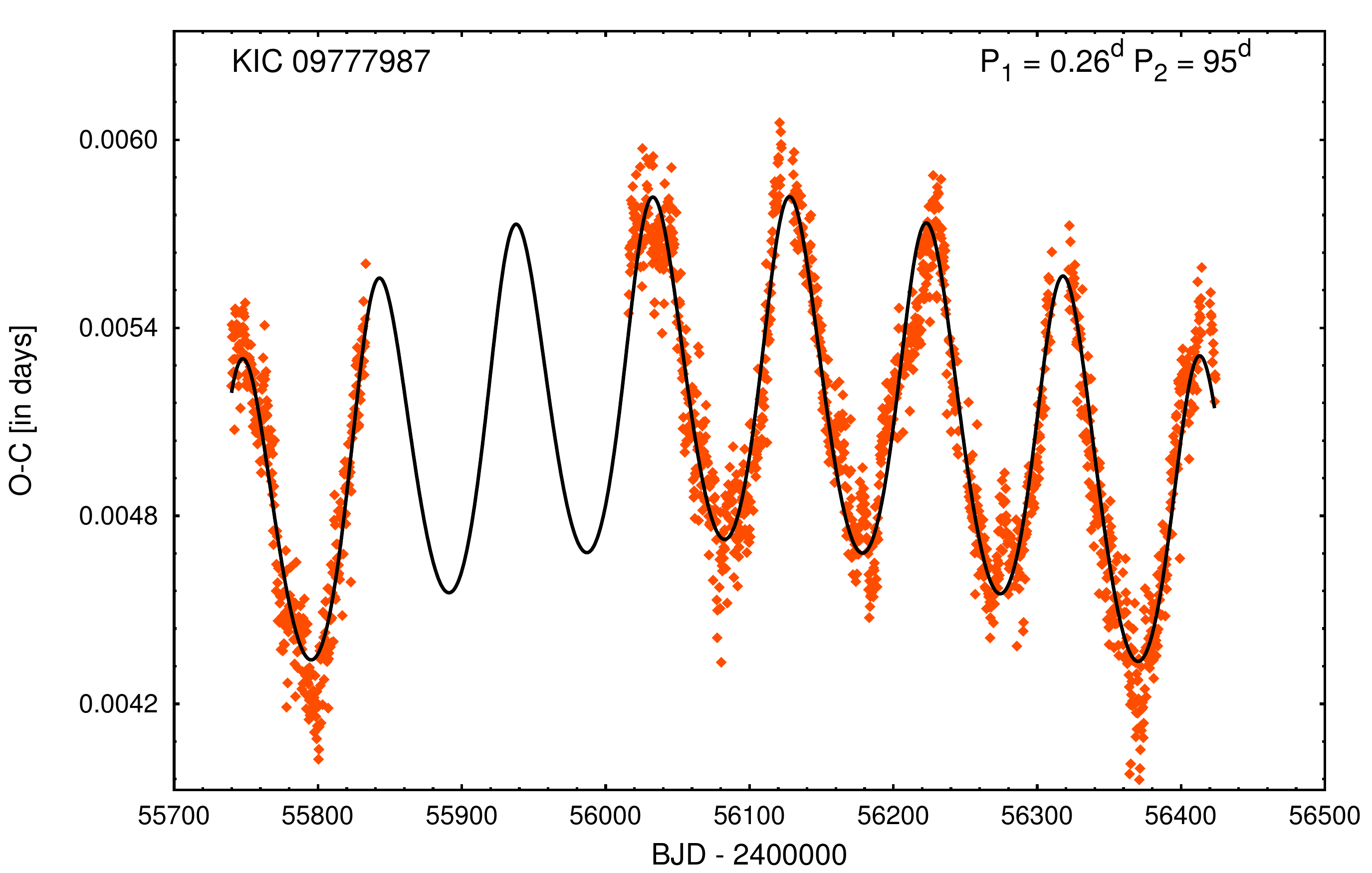}\includegraphics[width=60mm]{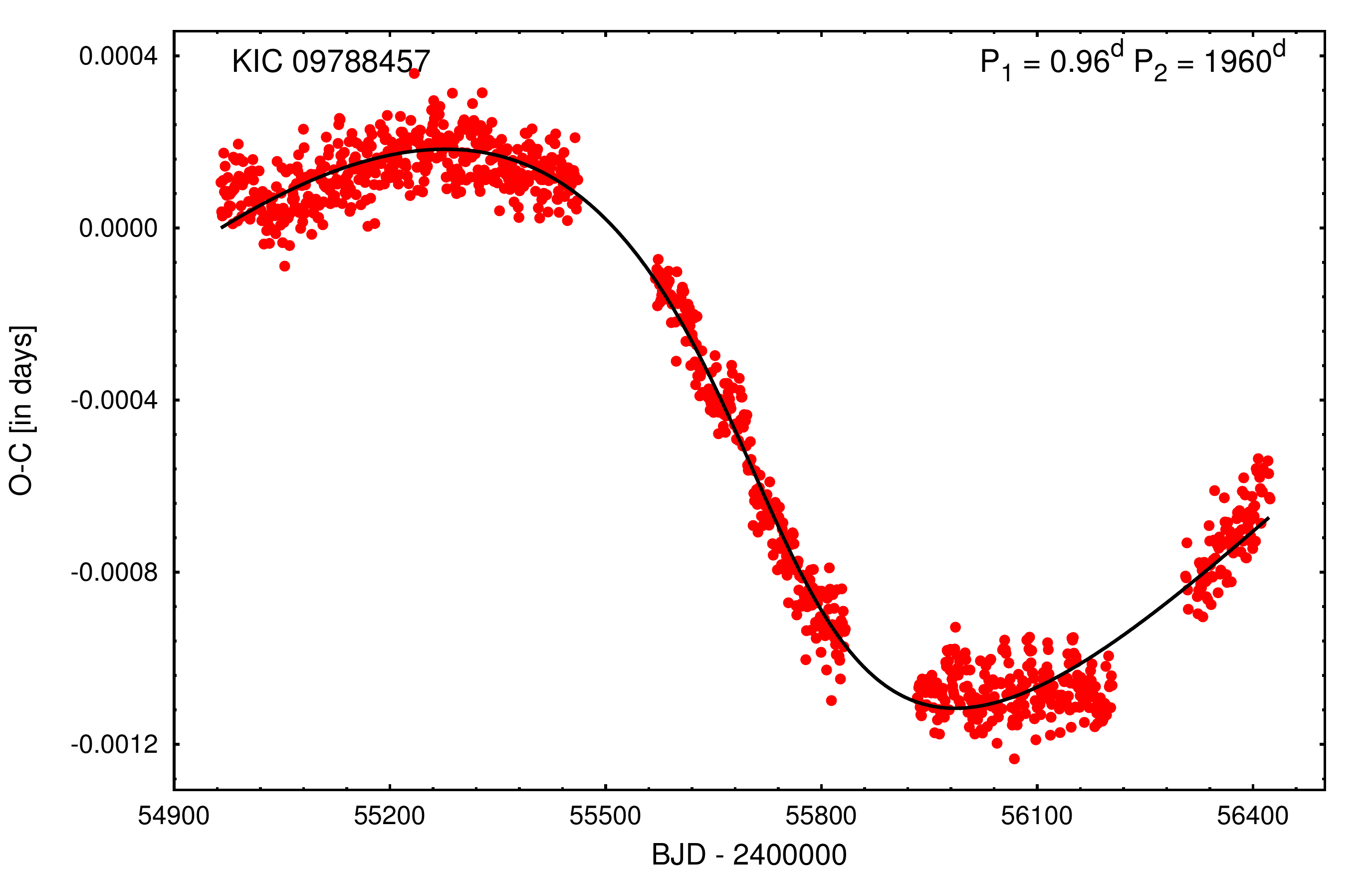}\includegraphics[width=60mm]{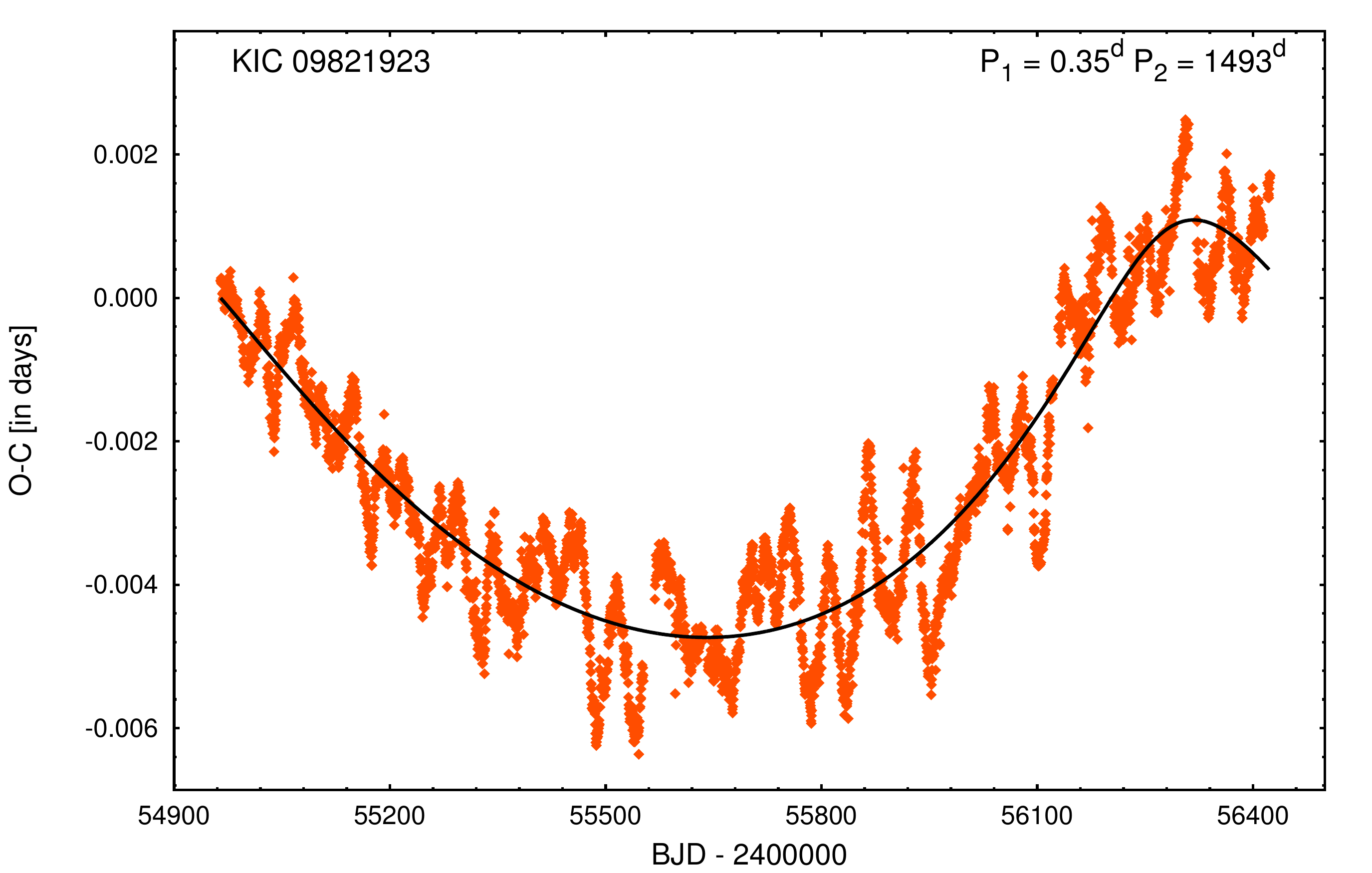}
\includegraphics[width=60mm]{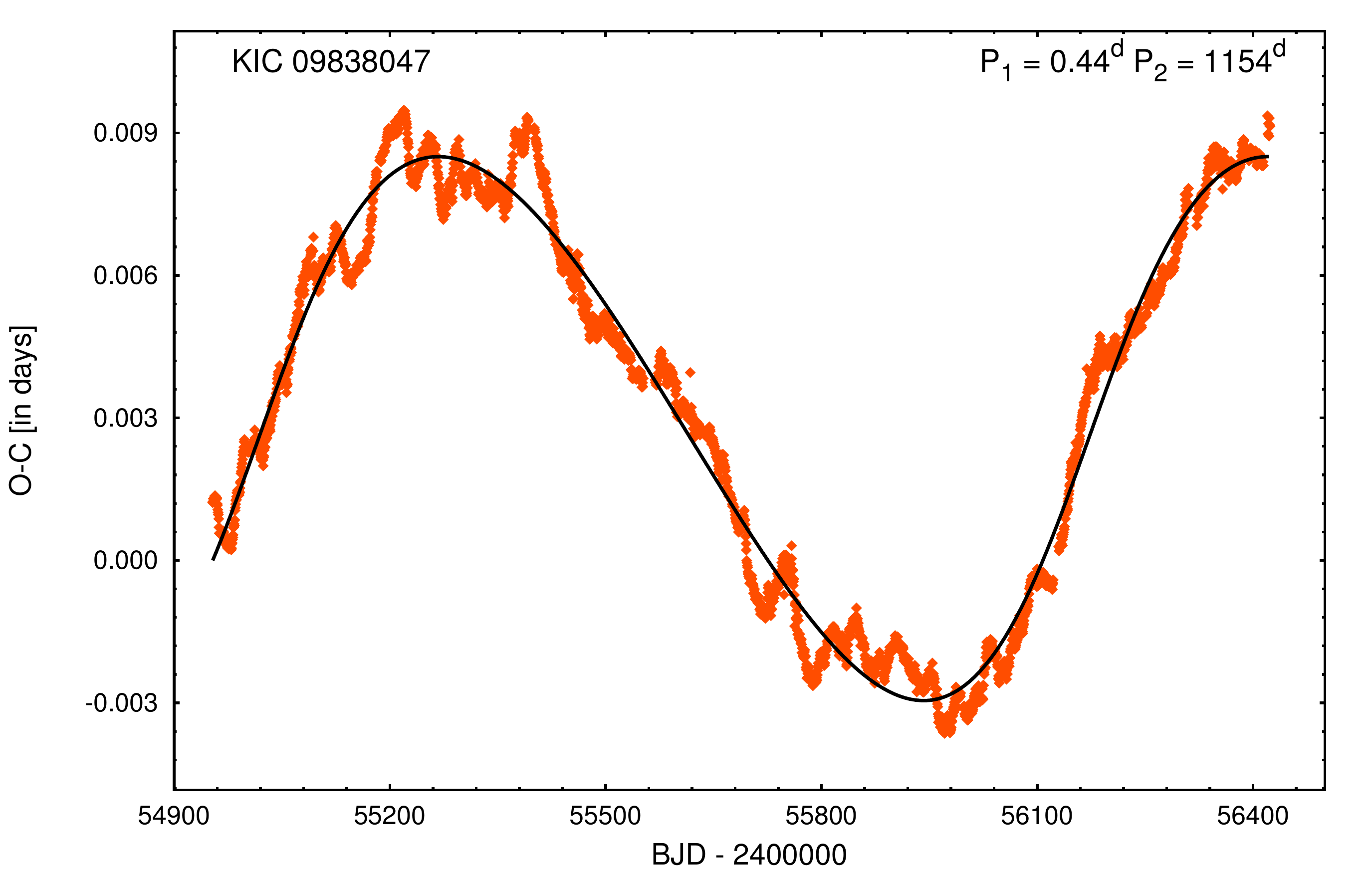}\includegraphics[width=60mm]{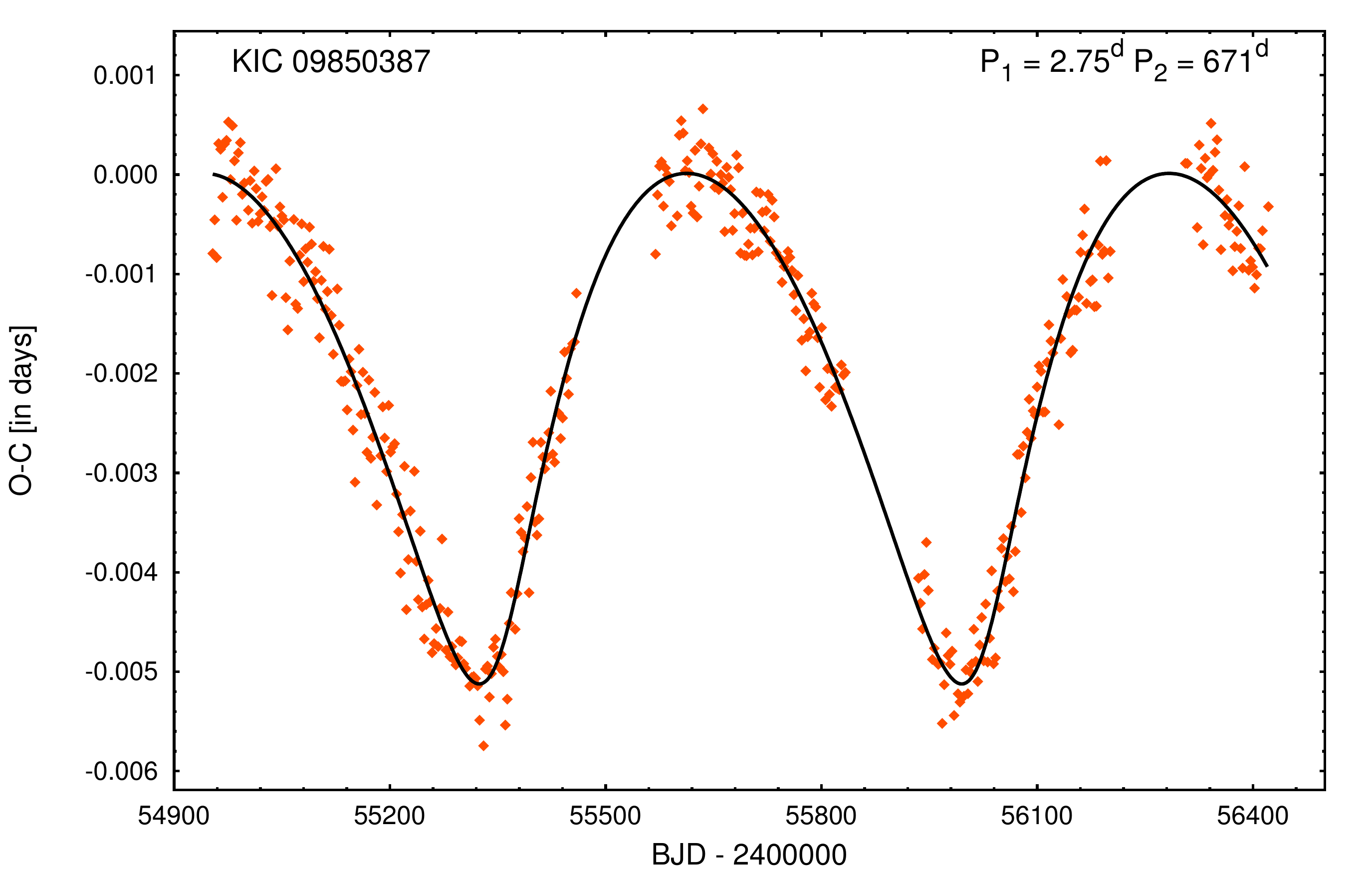}\includegraphics[width=60mm]{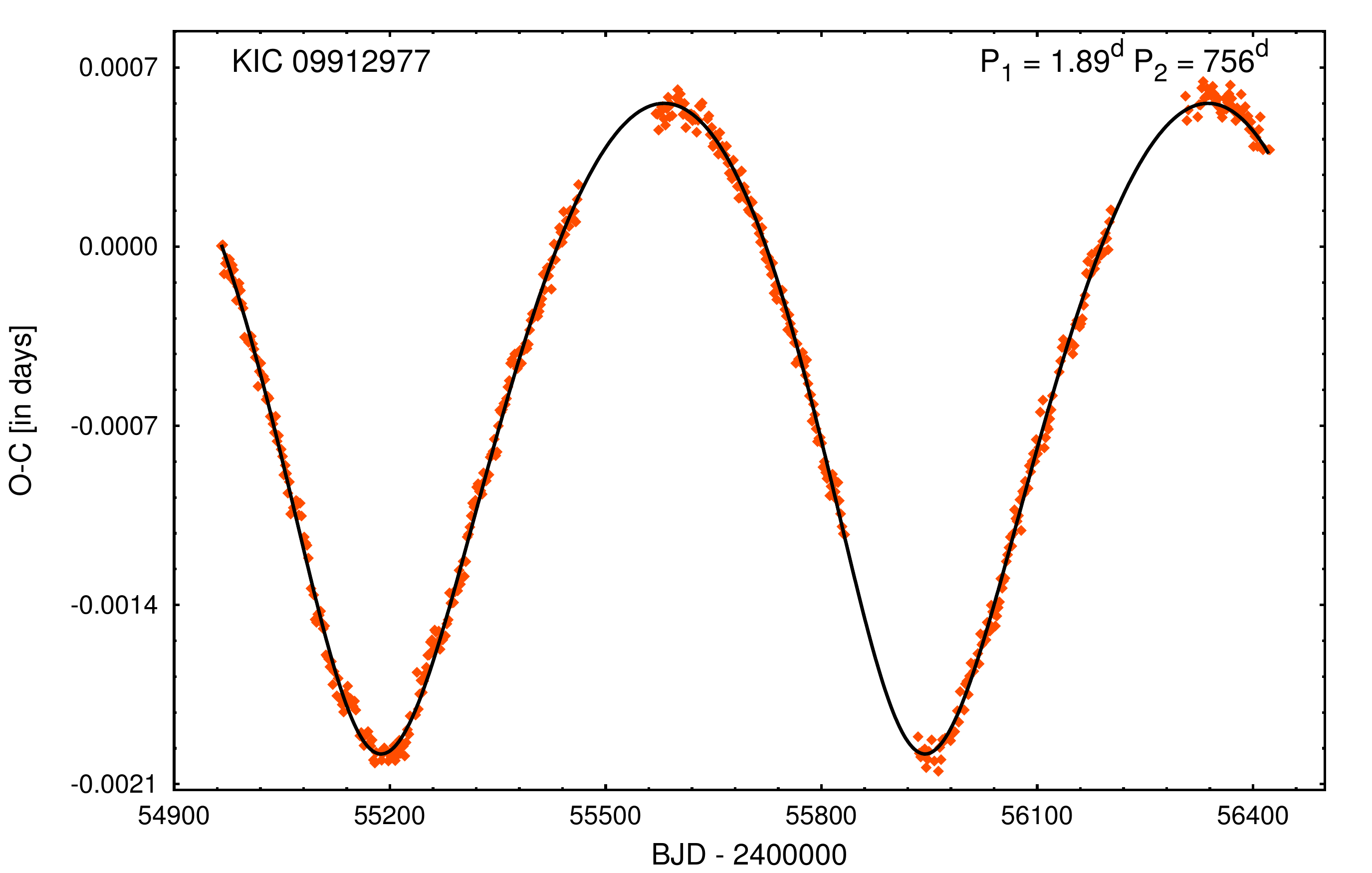}
\includegraphics[width=60mm]{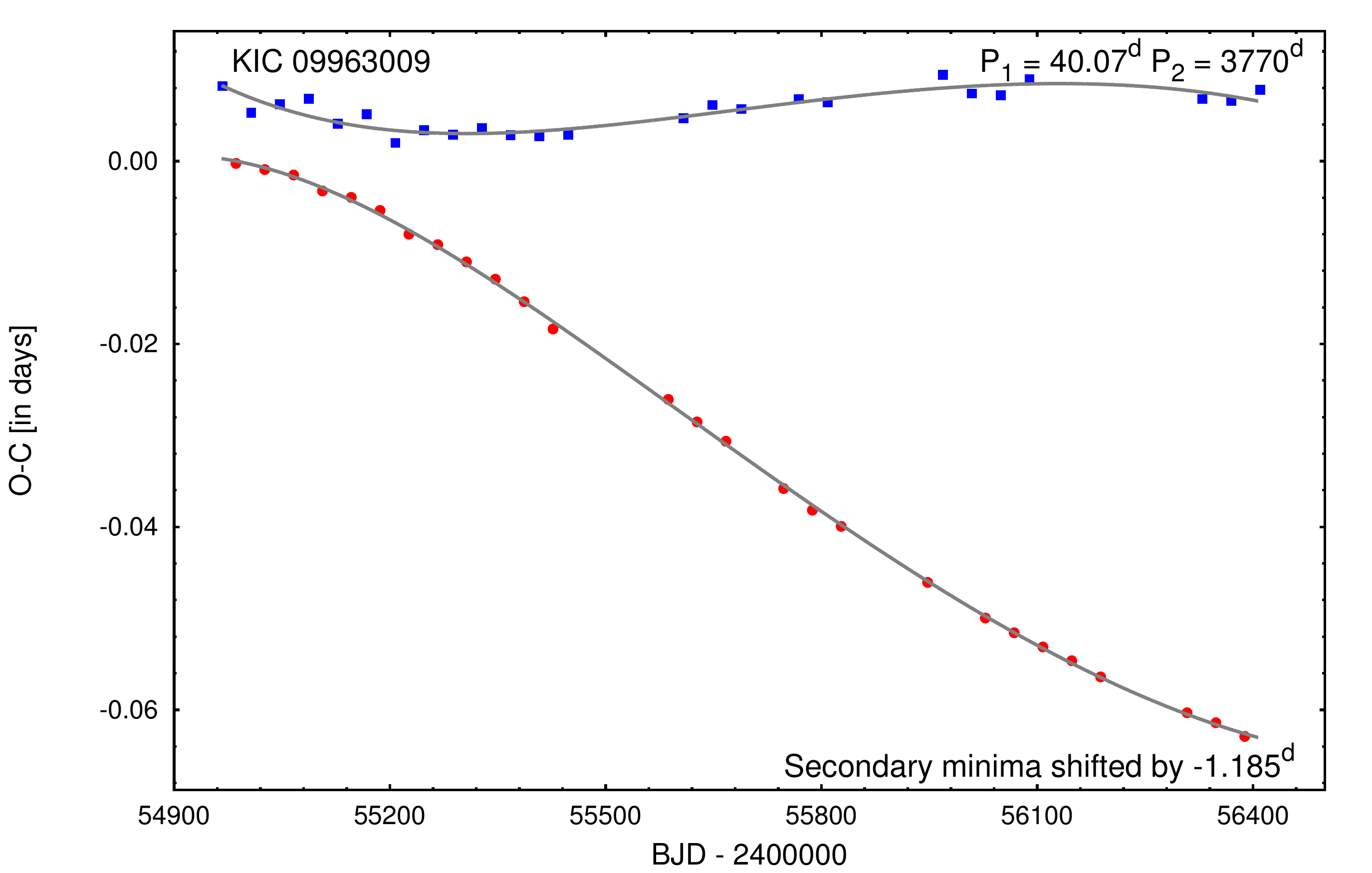}\includegraphics[width=60mm]{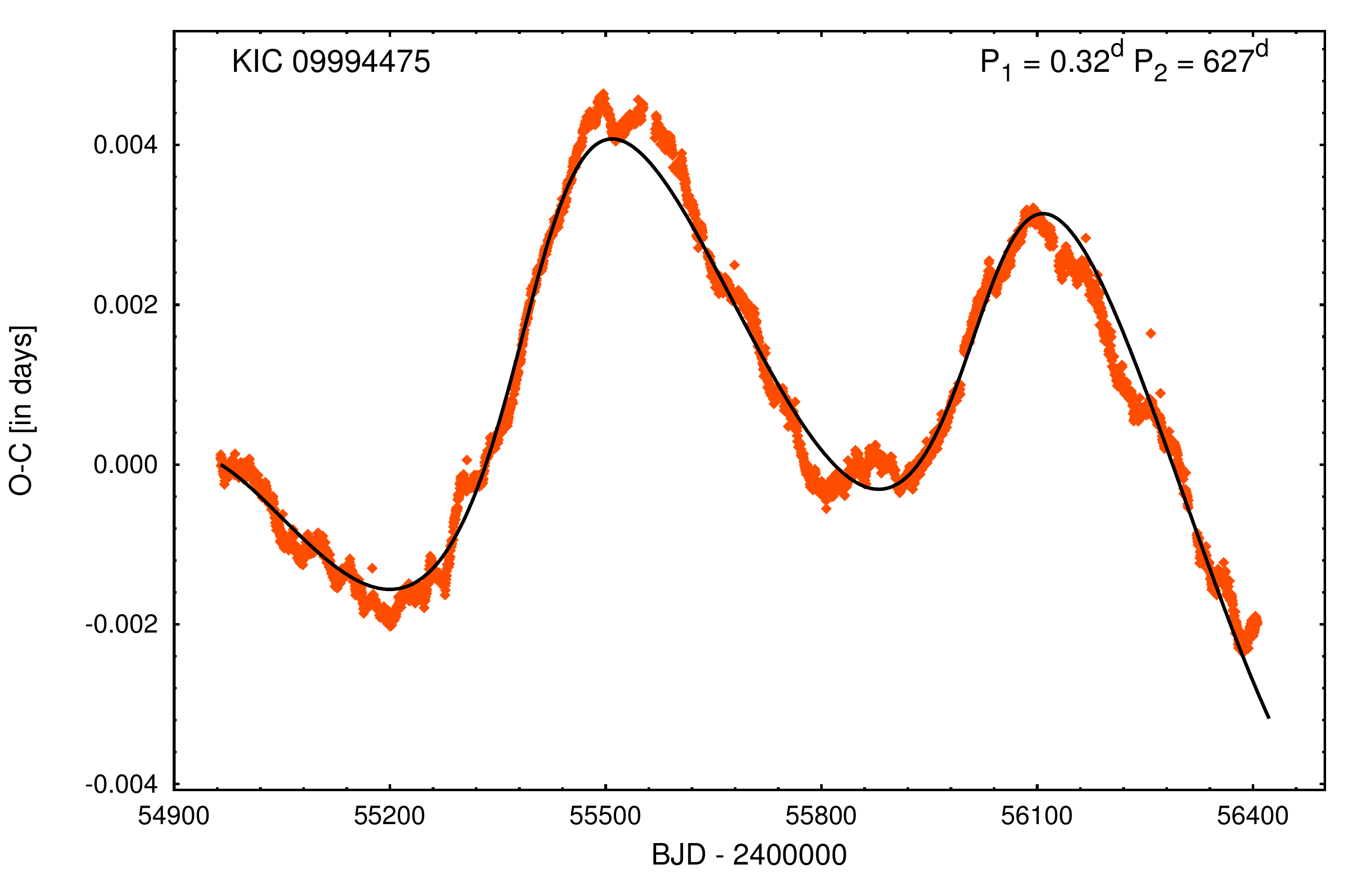}\includegraphics[width=60mm]{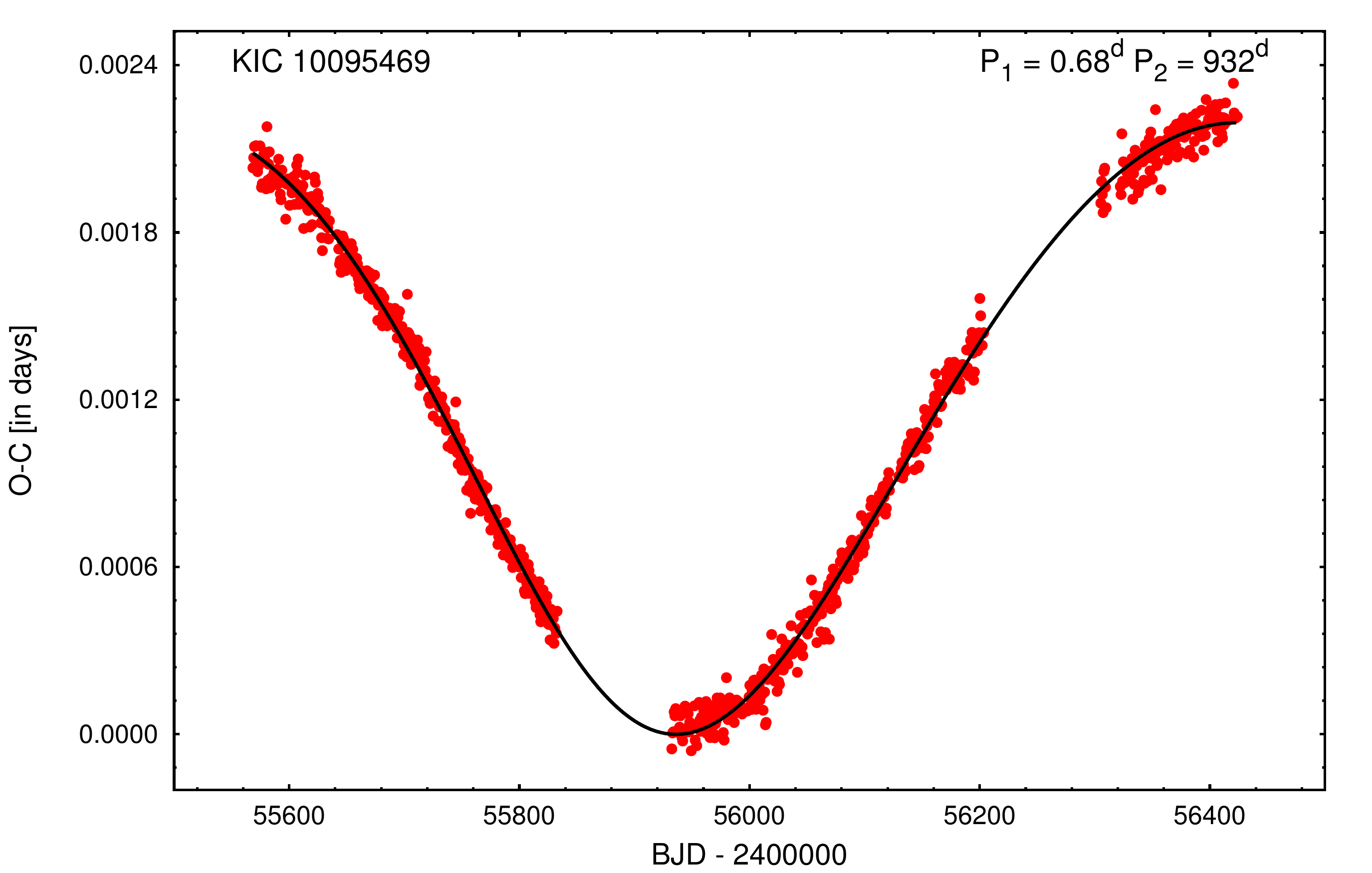}
\includegraphics[width=60mm]{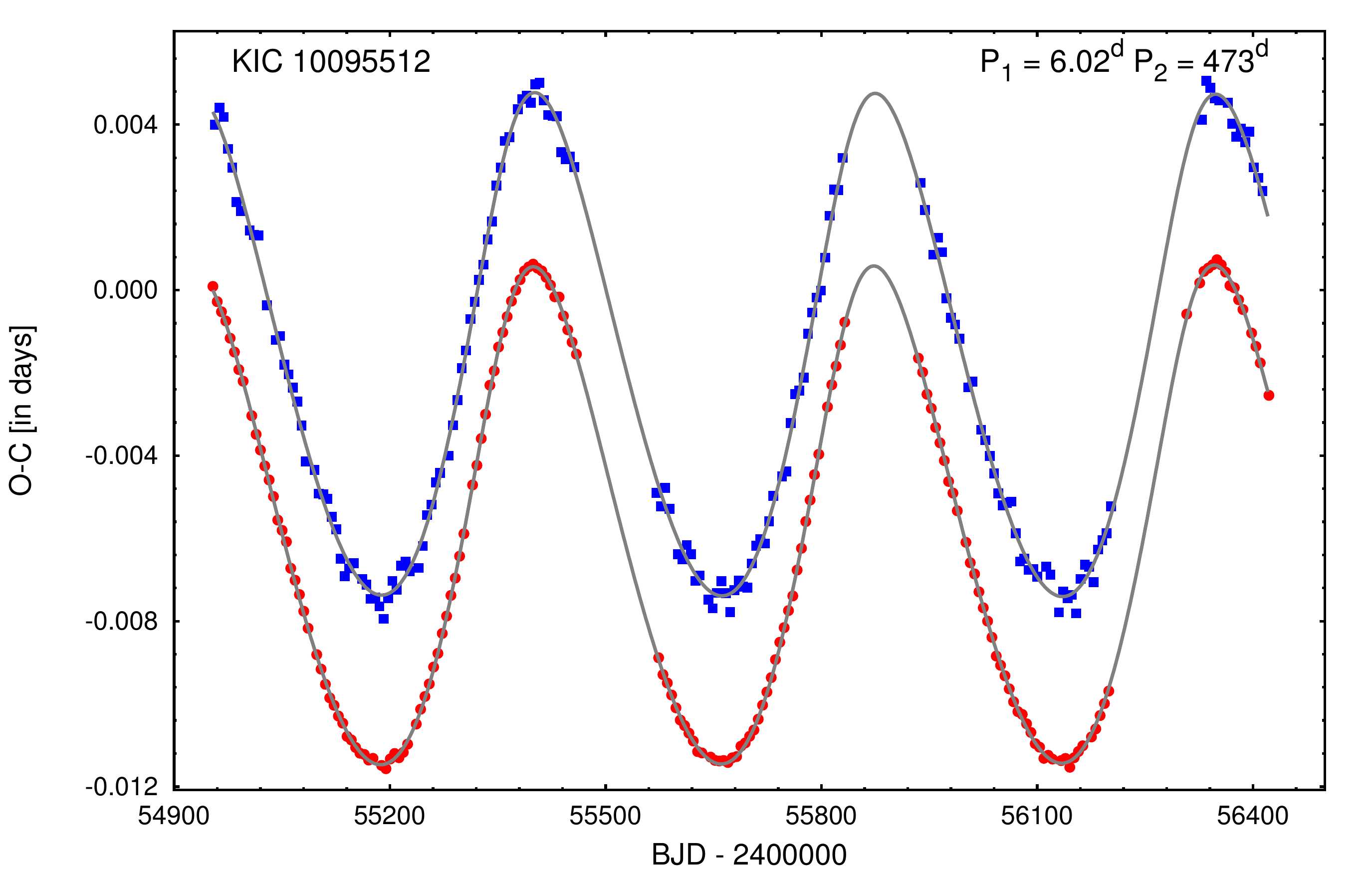}\includegraphics[width=60mm]{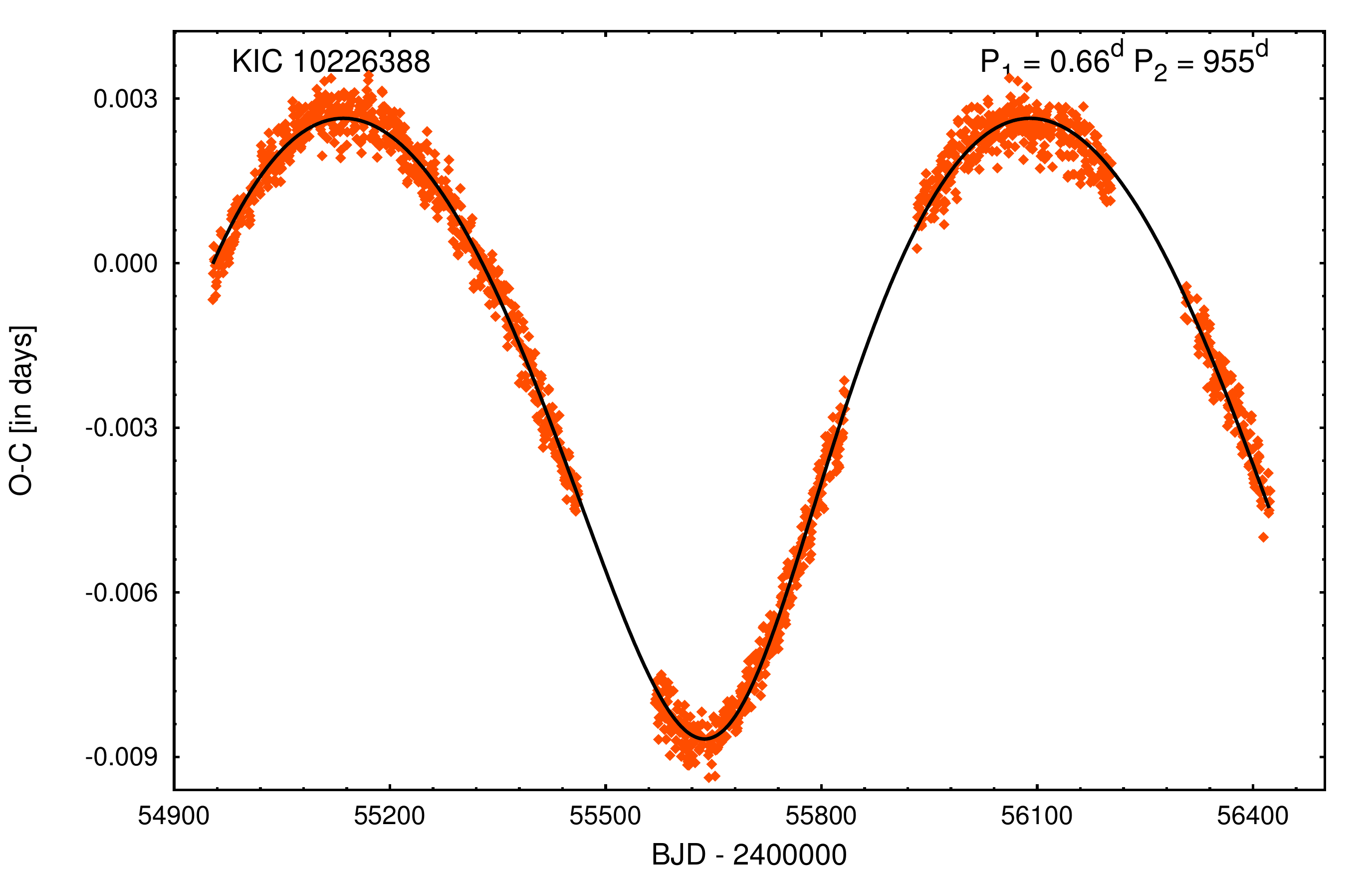}\includegraphics[width=60mm]{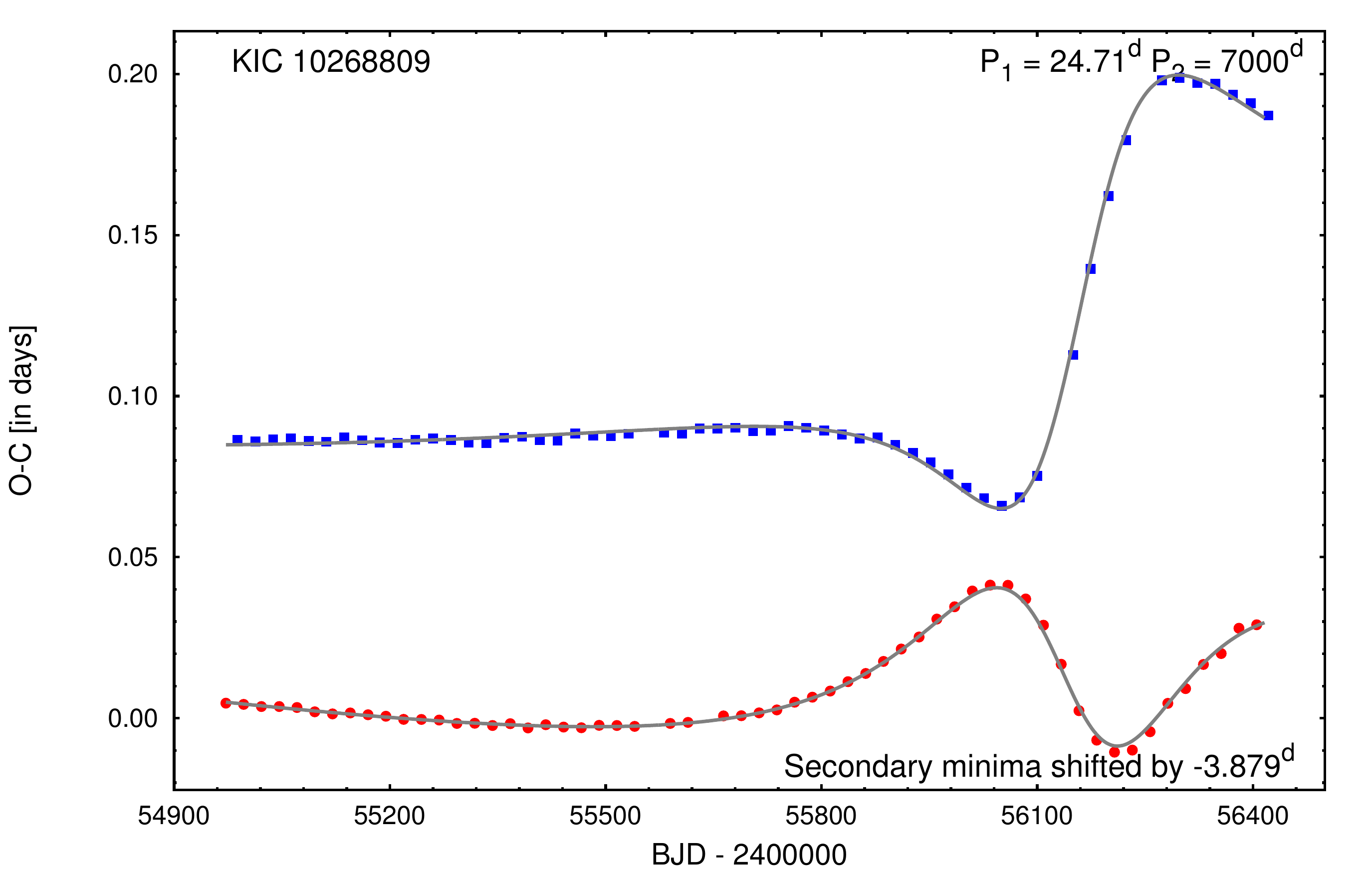}
\includegraphics[width=60mm]{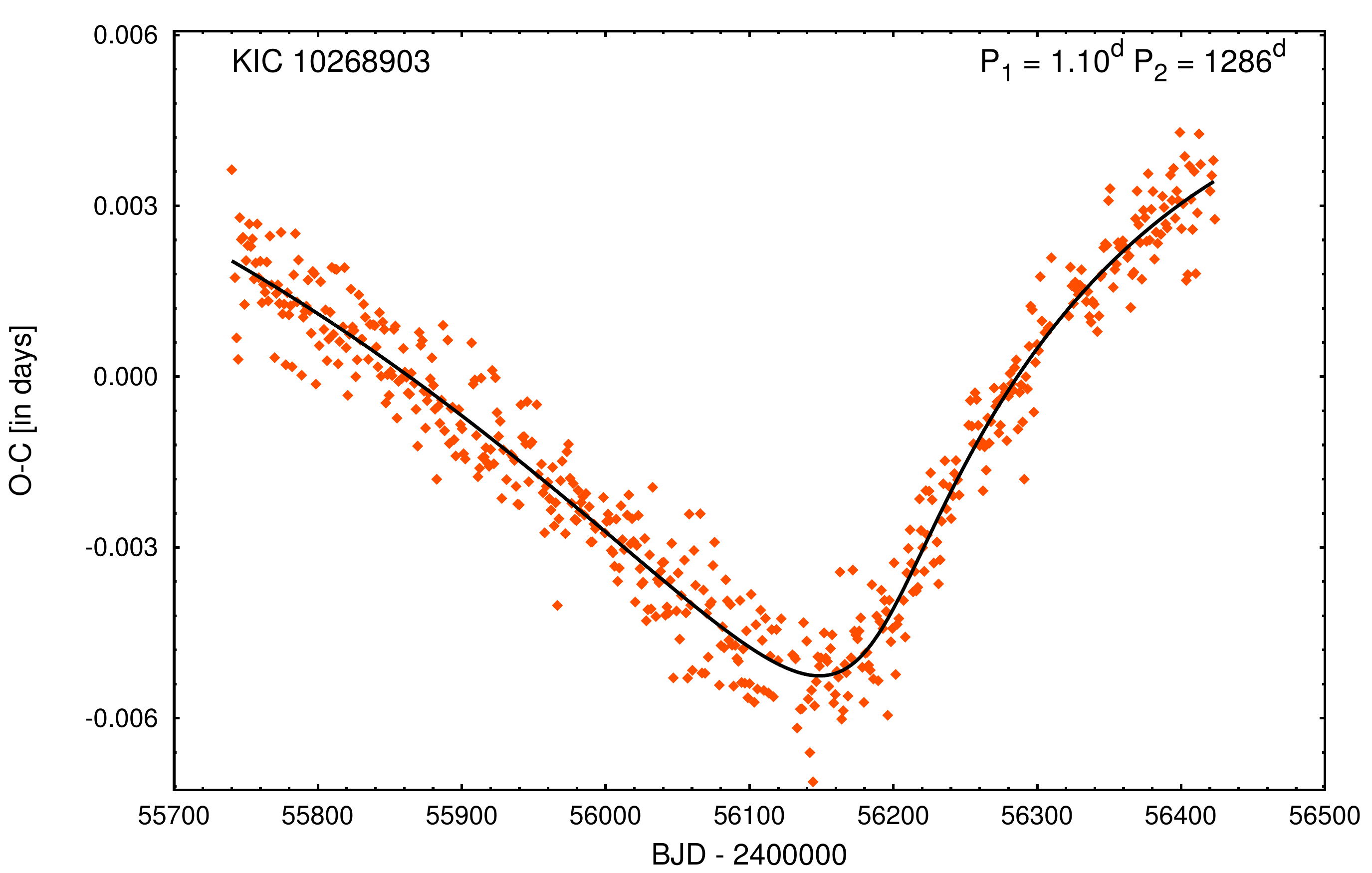}\includegraphics[width=60mm]{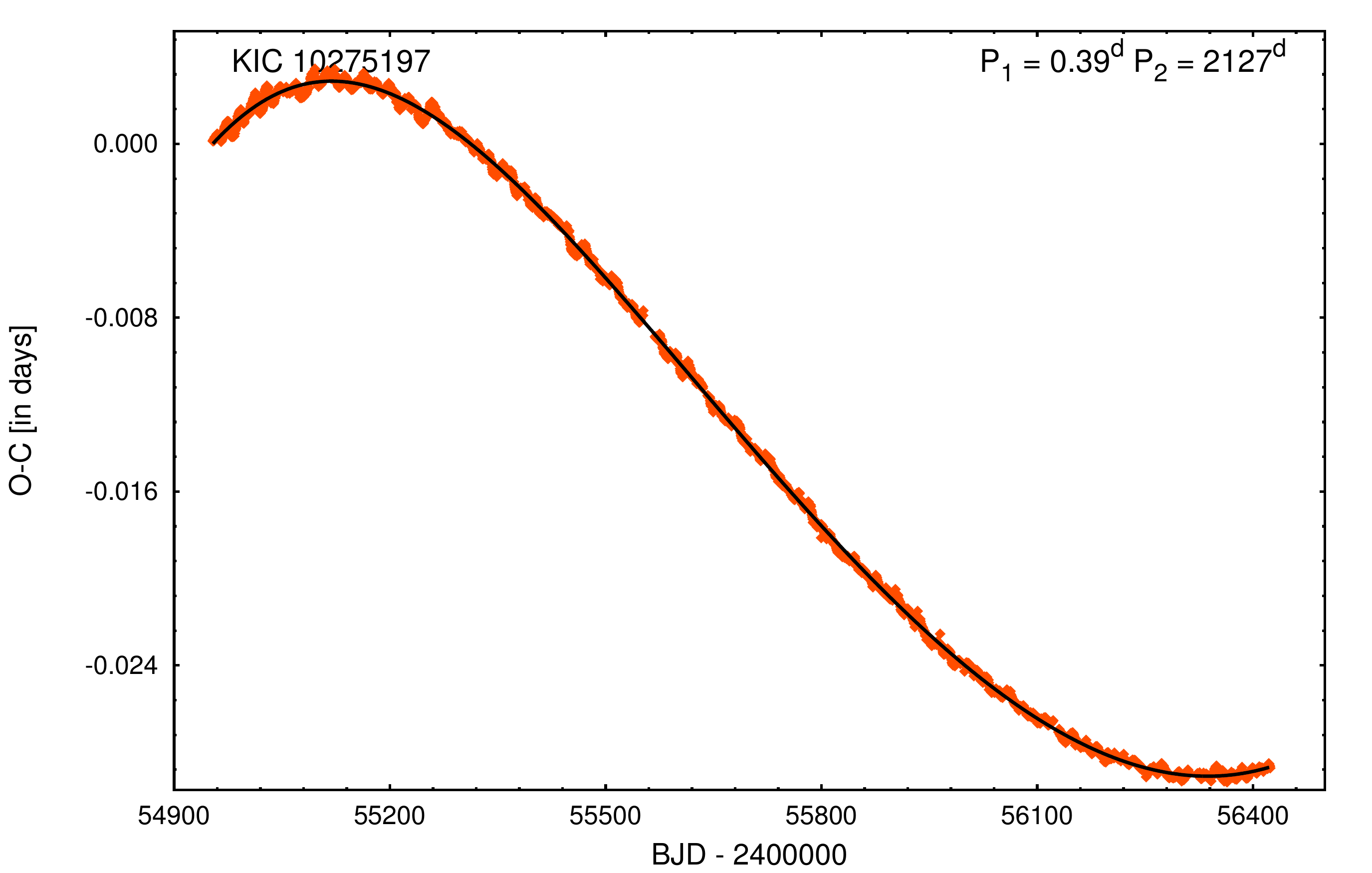}\includegraphics[width=60mm]{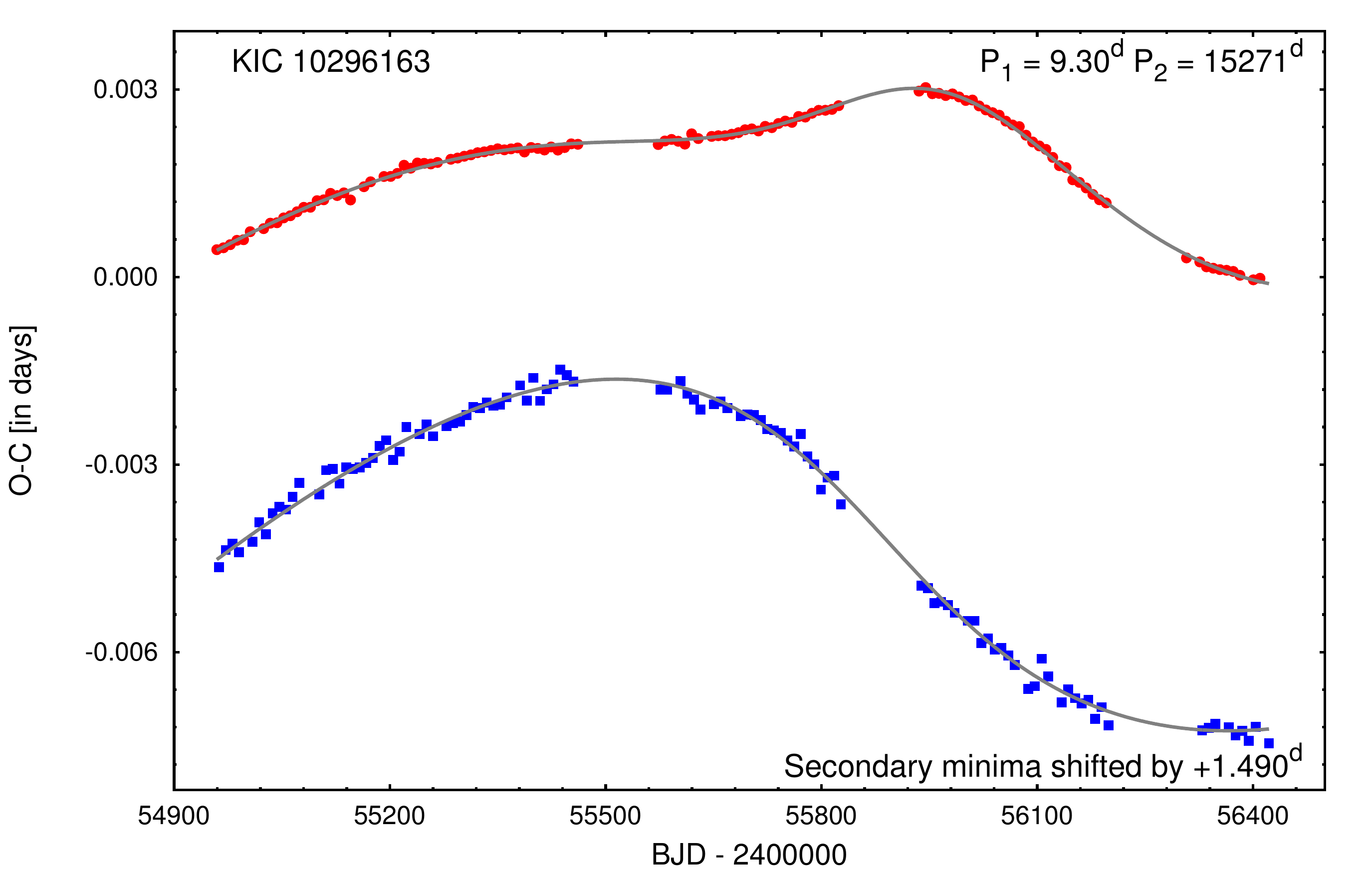}
\includegraphics[width=60mm]{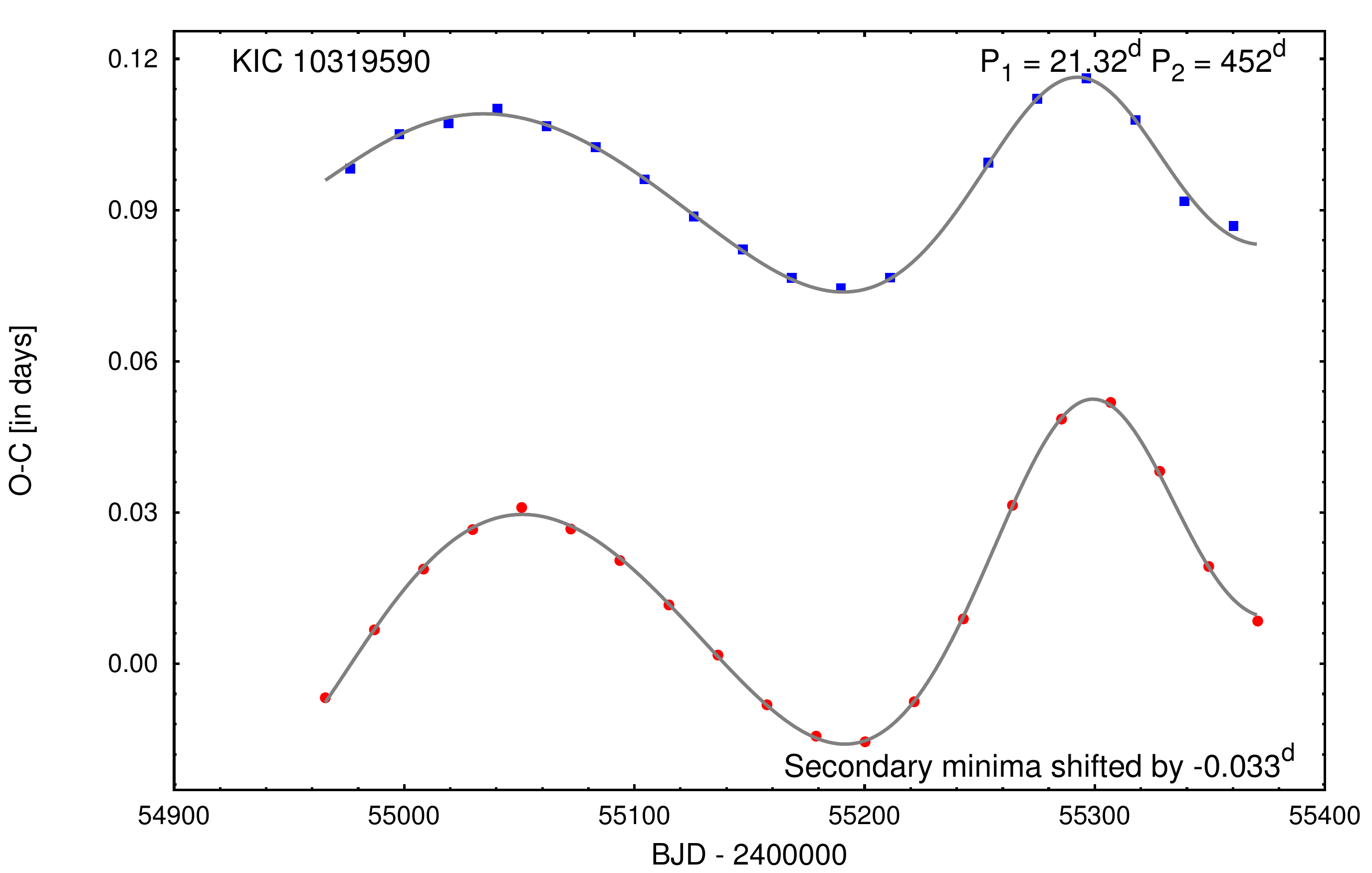}\includegraphics[width=60mm]{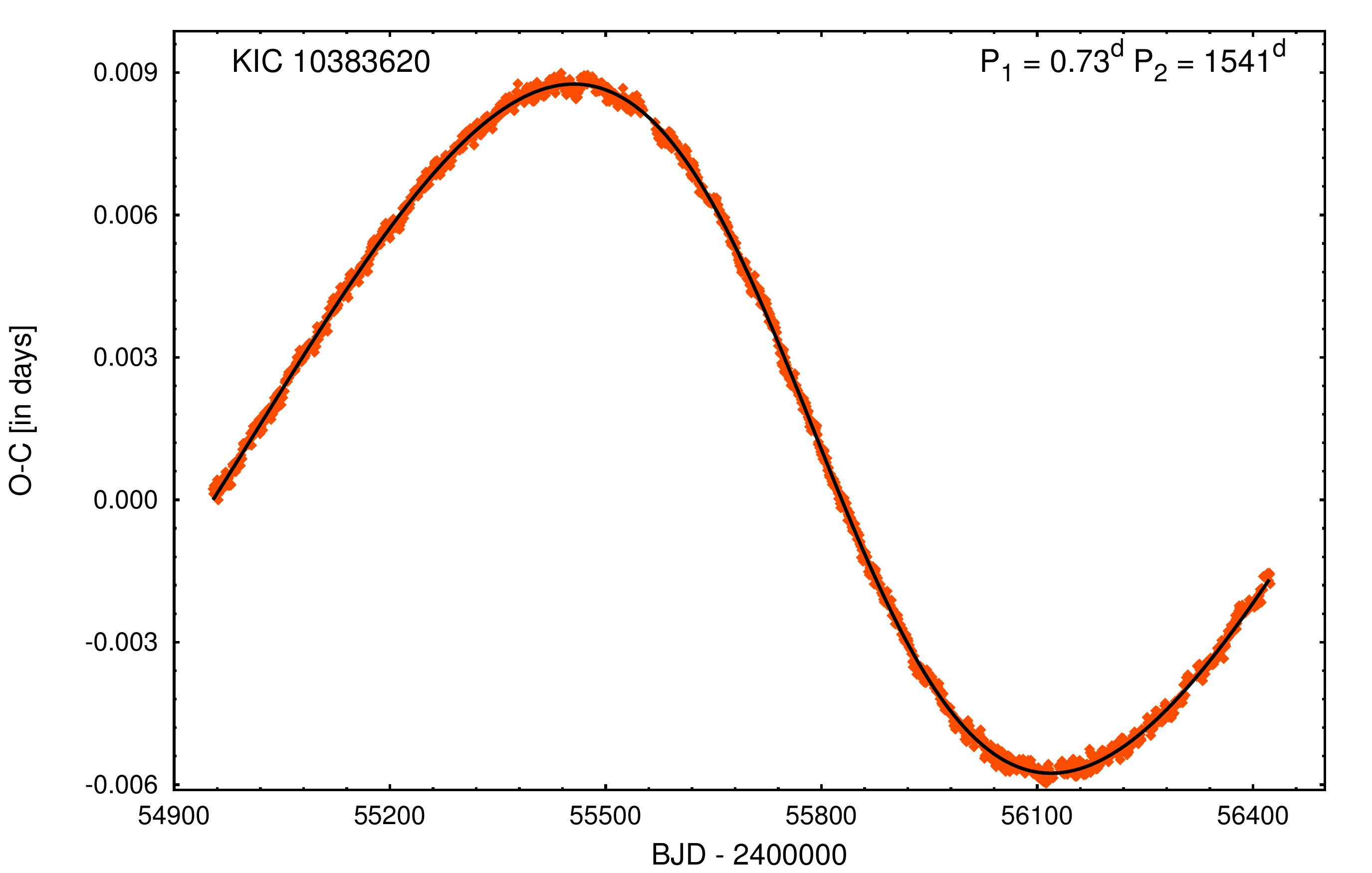}\includegraphics[width=60mm]{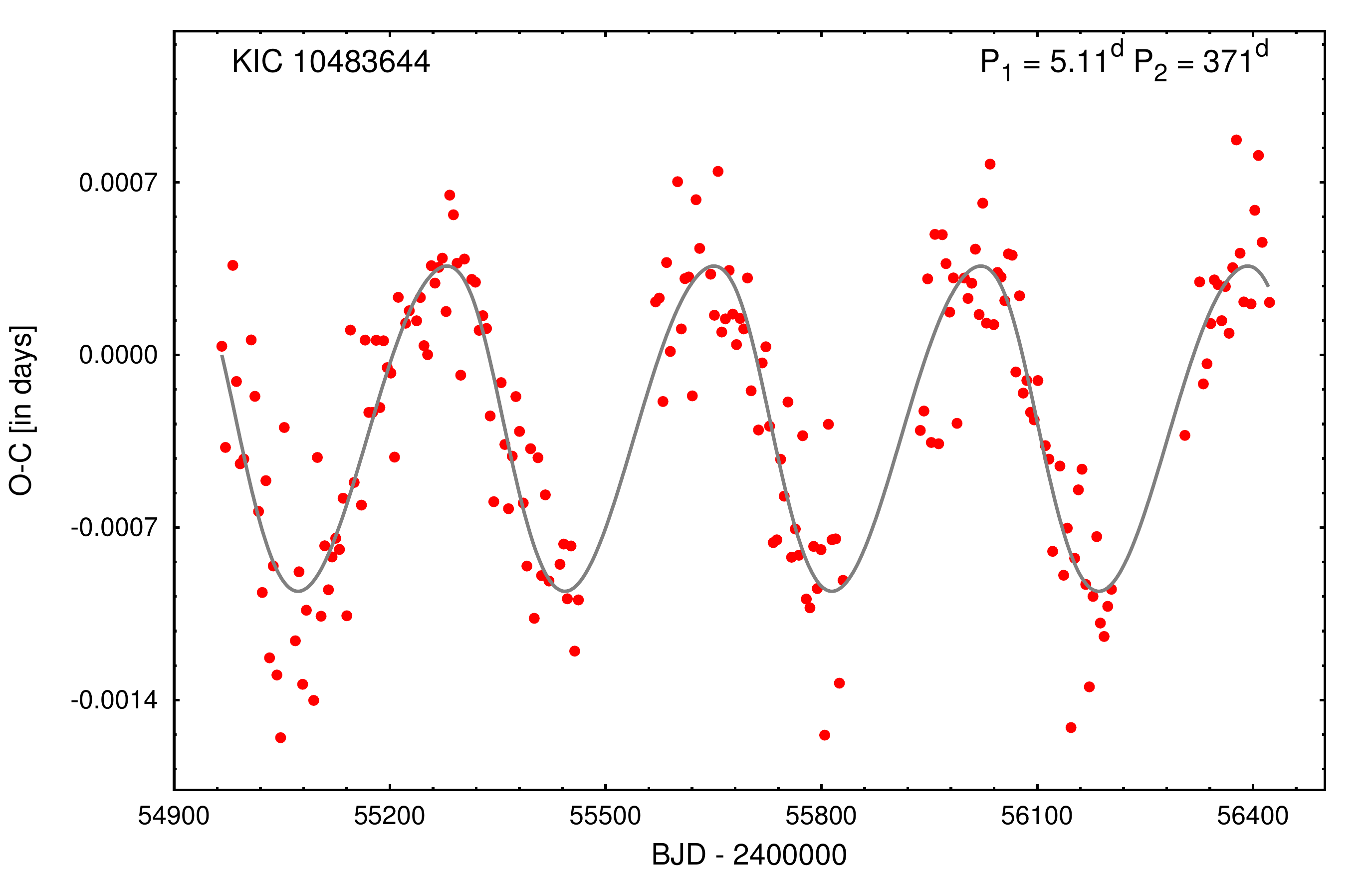}
\caption{(continued)}
\end{figure*}

\addtocounter{figure}{-1}

\begin{figure*}
\includegraphics[width=60mm]{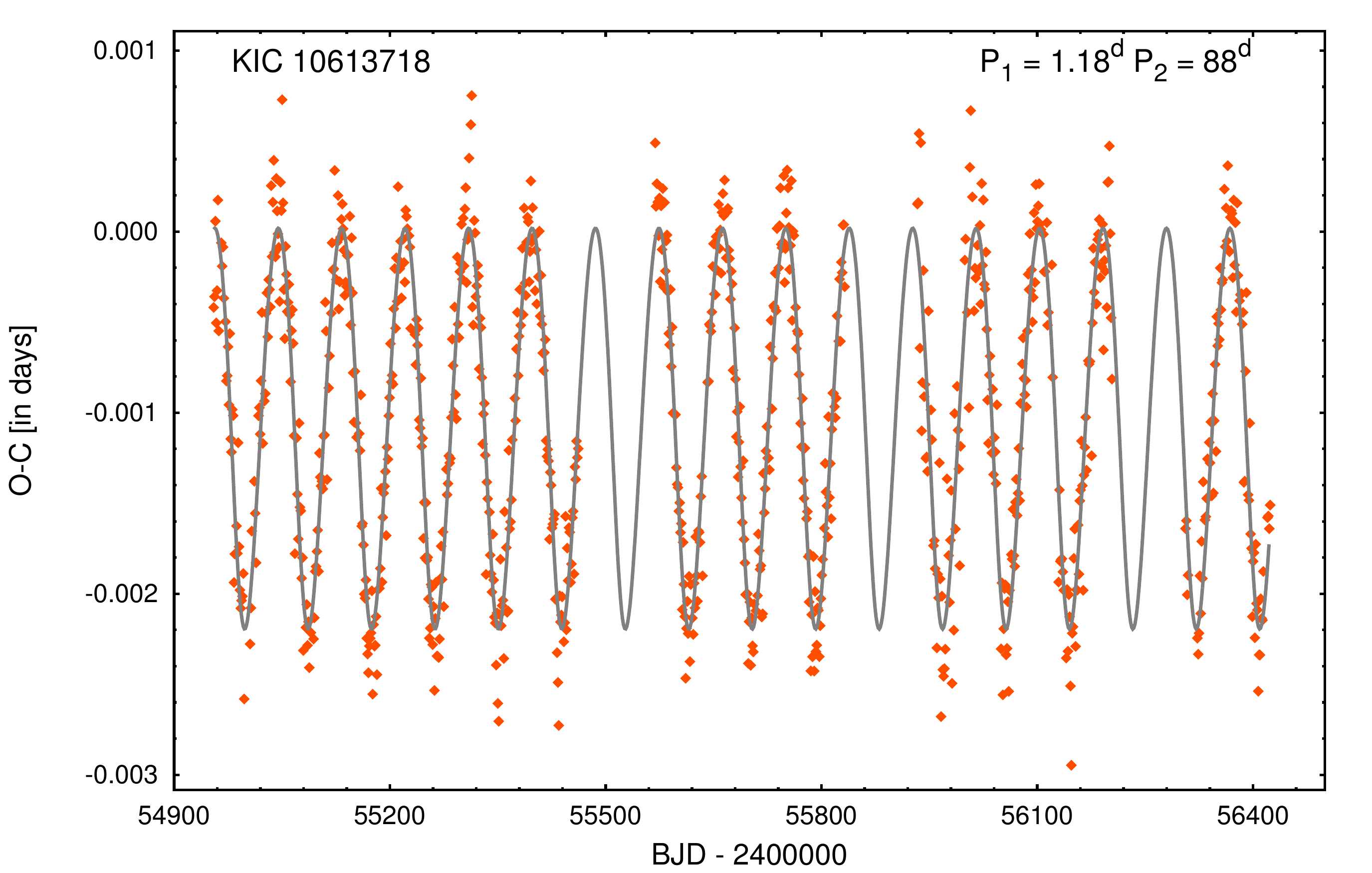}\includegraphics[width=60mm]{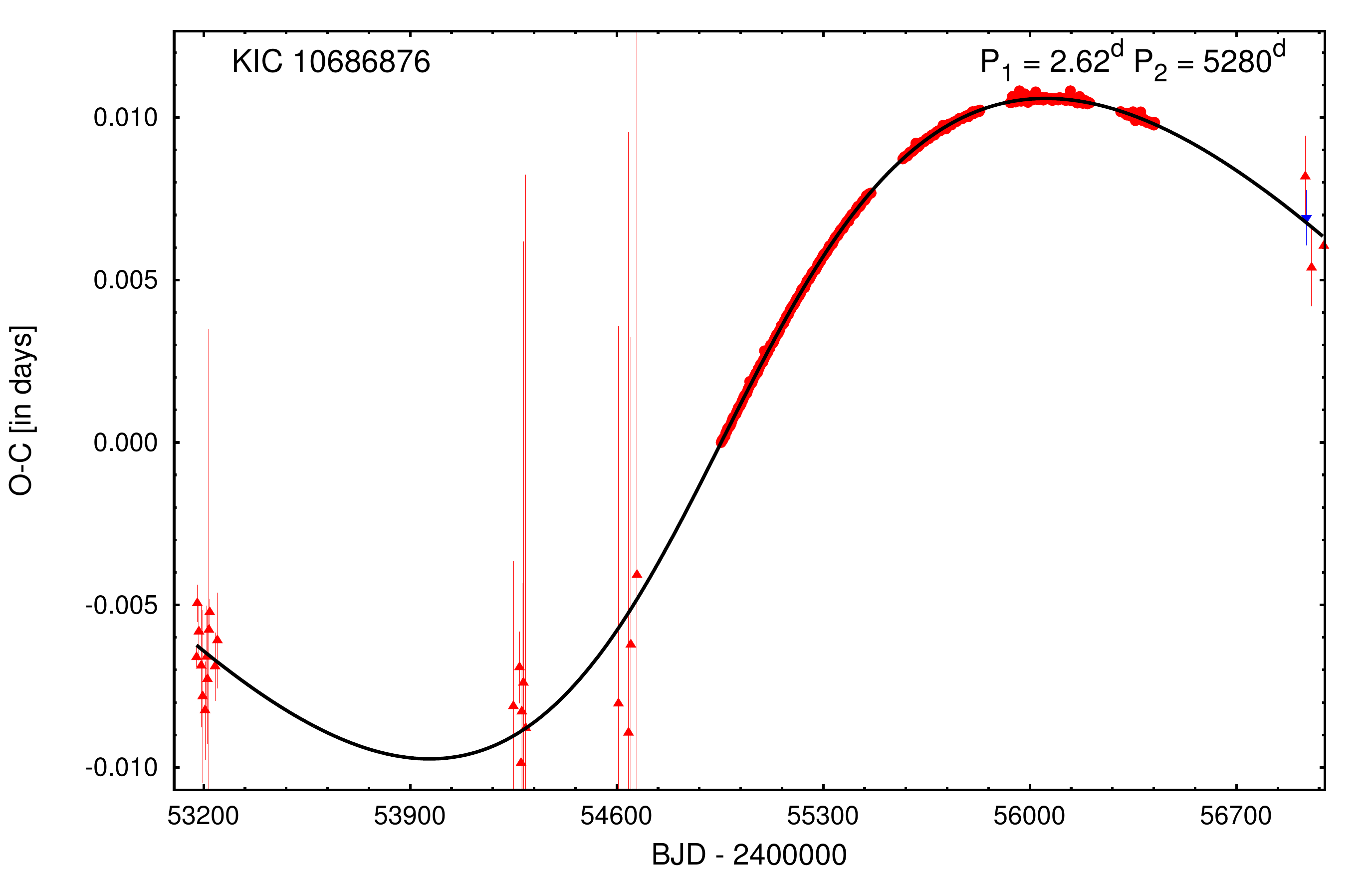}\includegraphics[width=60mm]{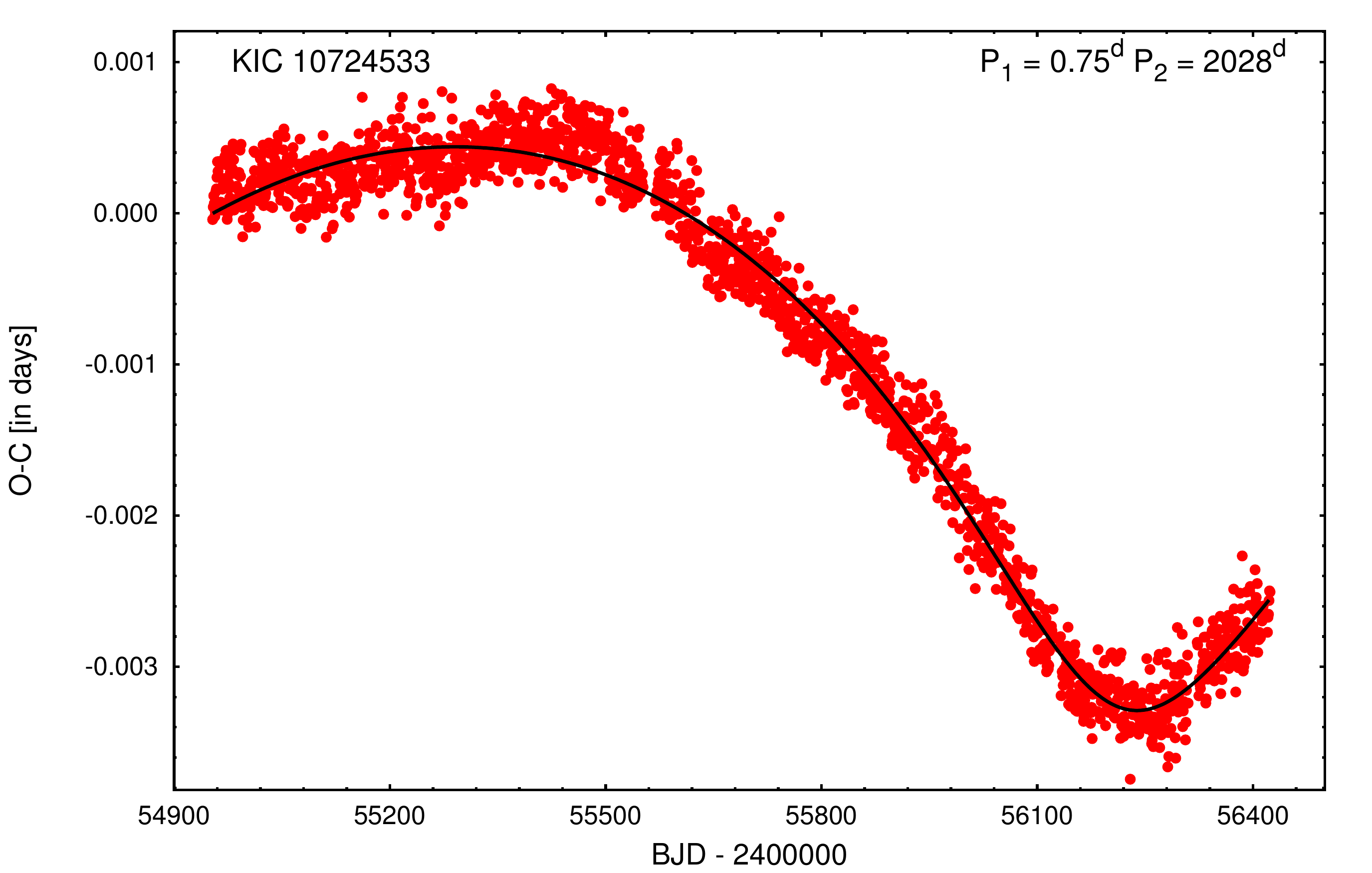}
\includegraphics[width=60mm]{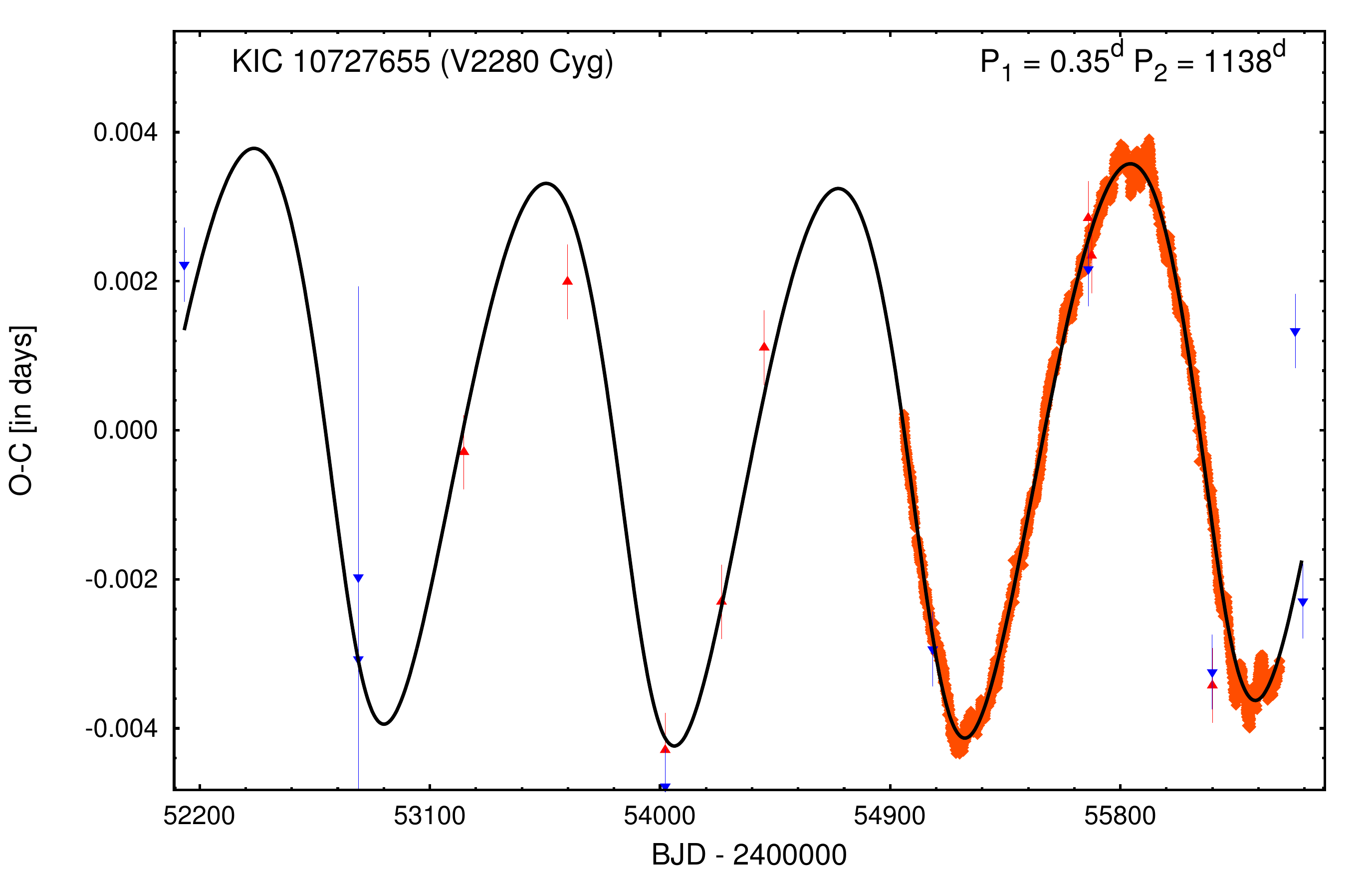}\includegraphics[width=60mm]{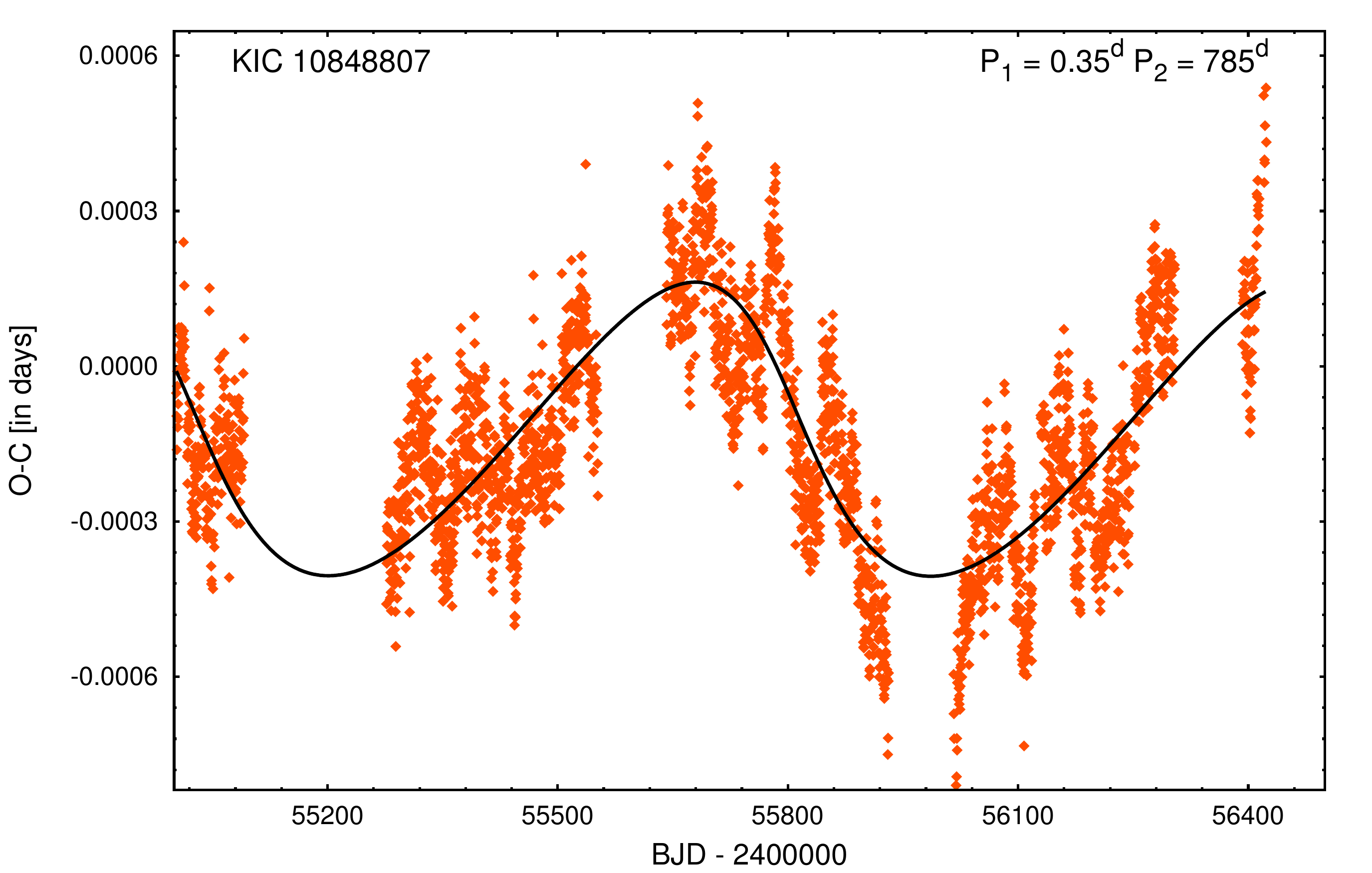}\includegraphics[width=60mm]{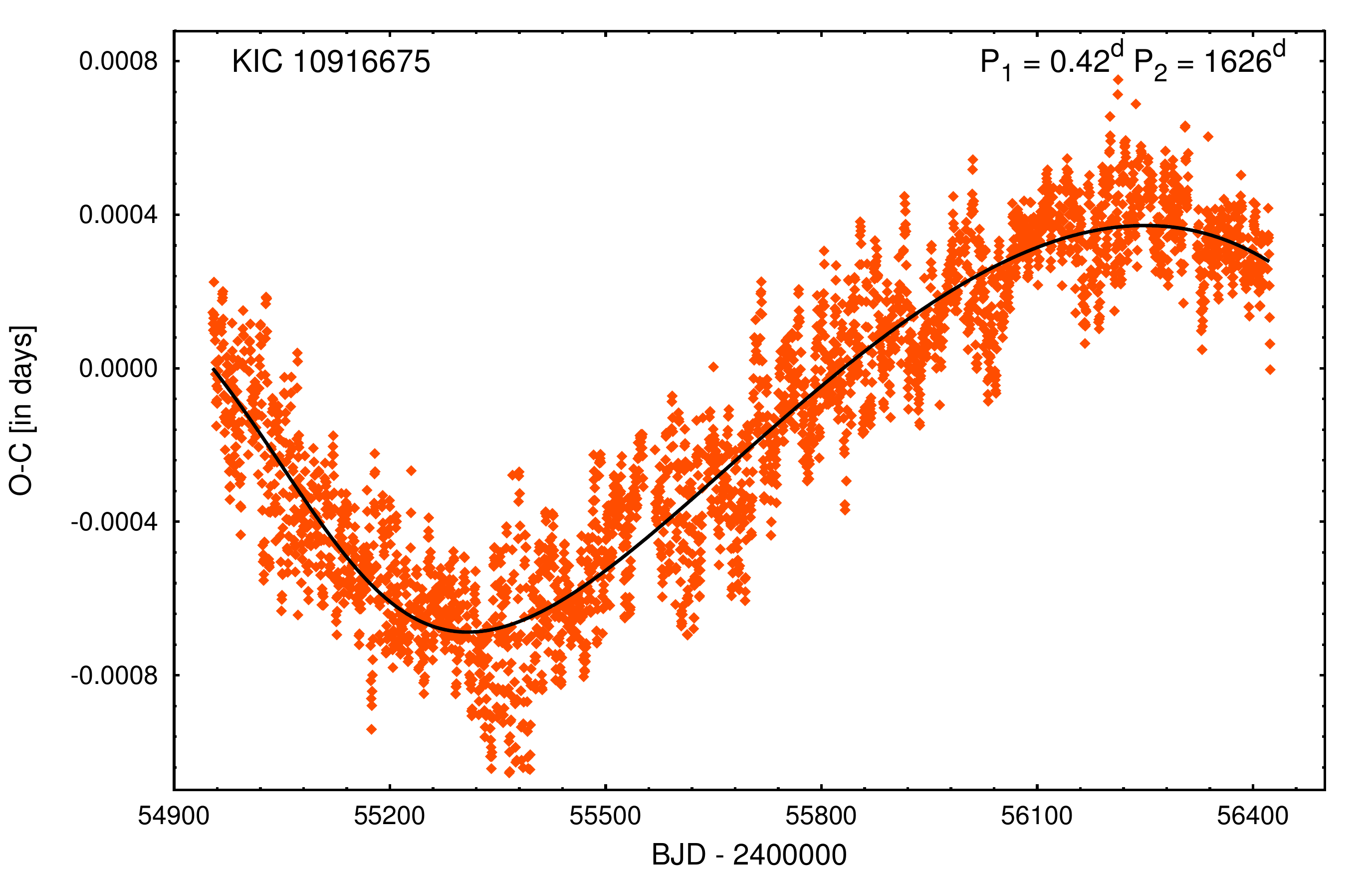}
\includegraphics[width=60mm]{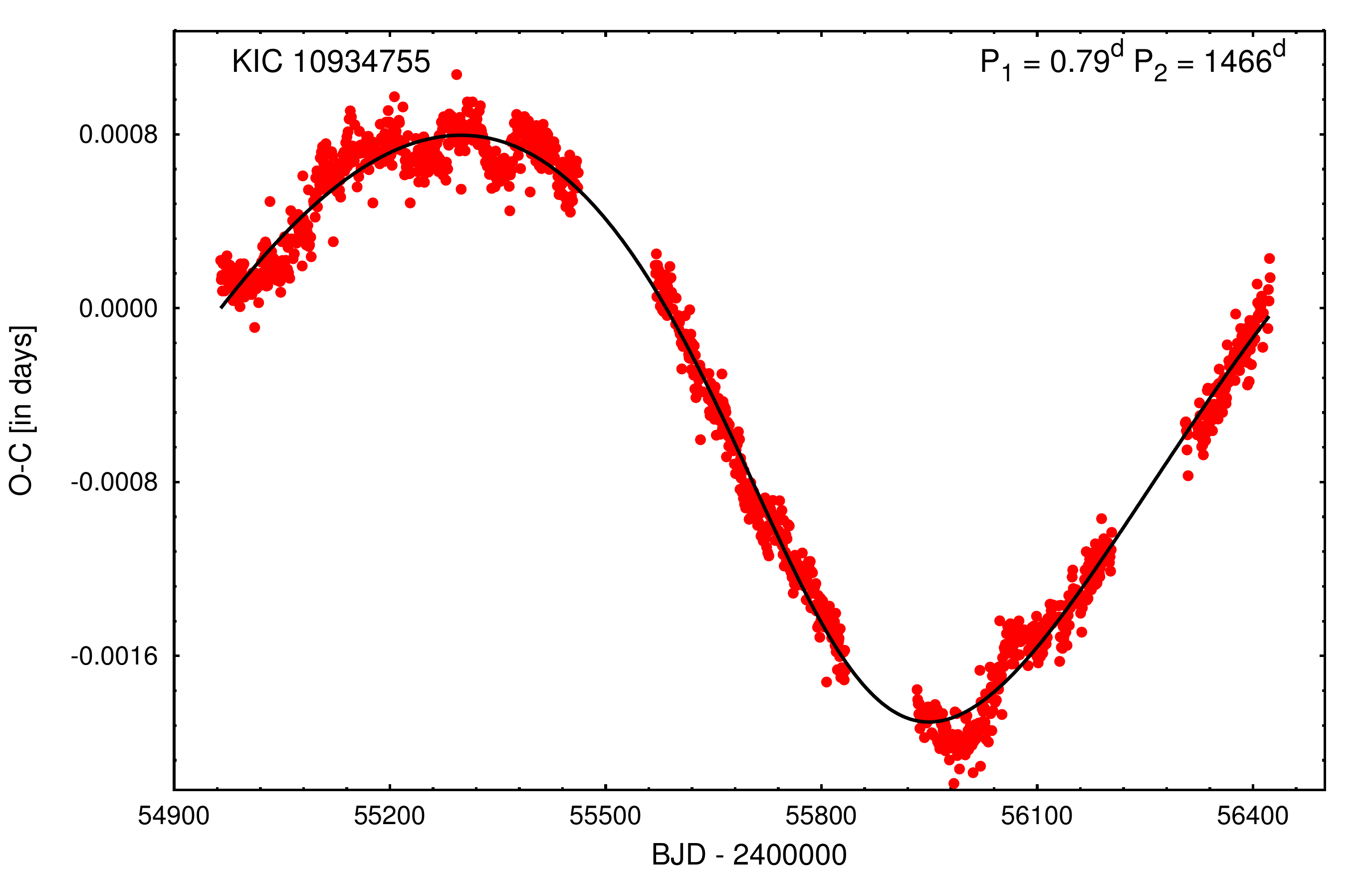}\includegraphics[width=60mm]{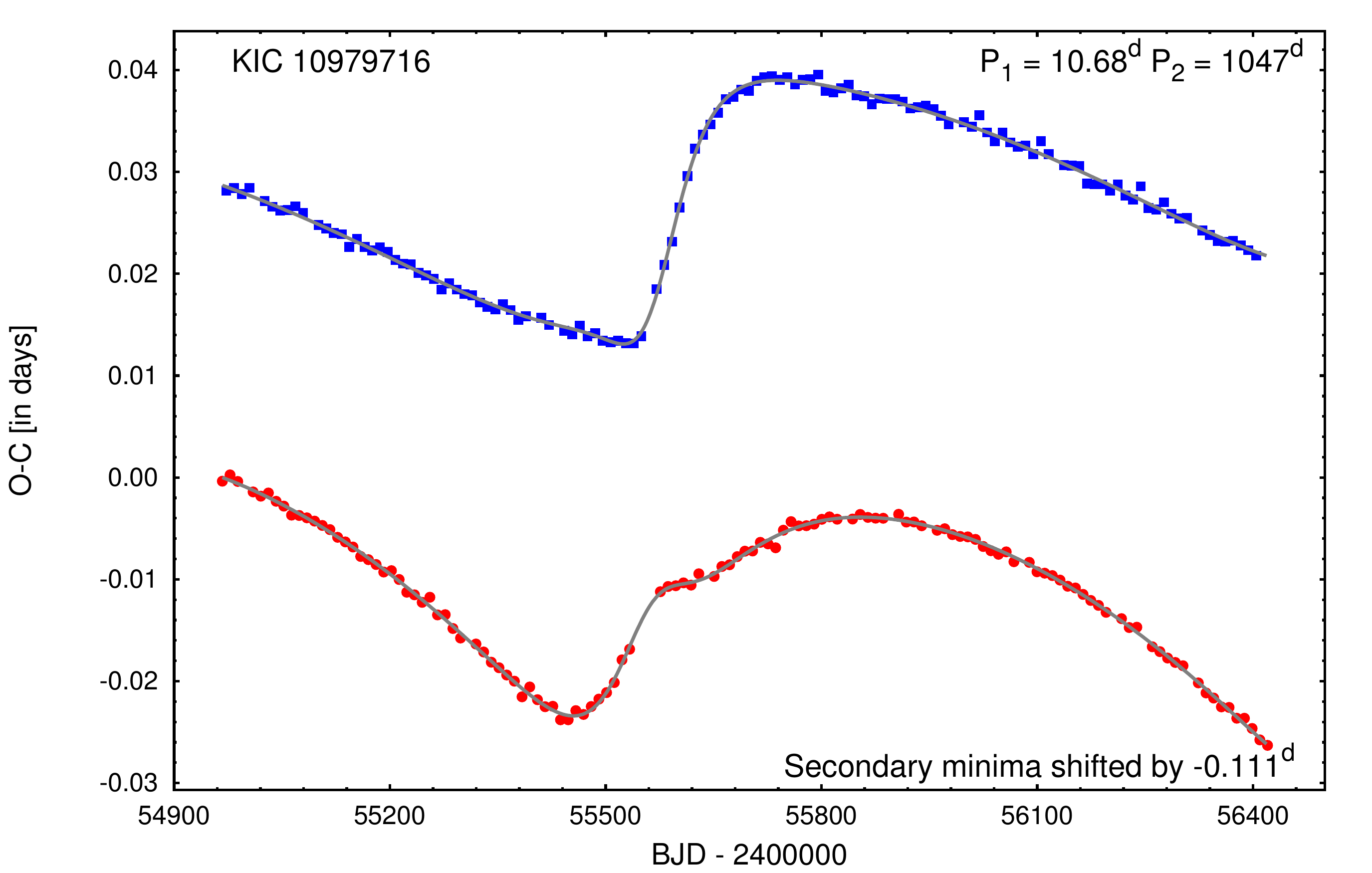}\includegraphics[width=60mm]{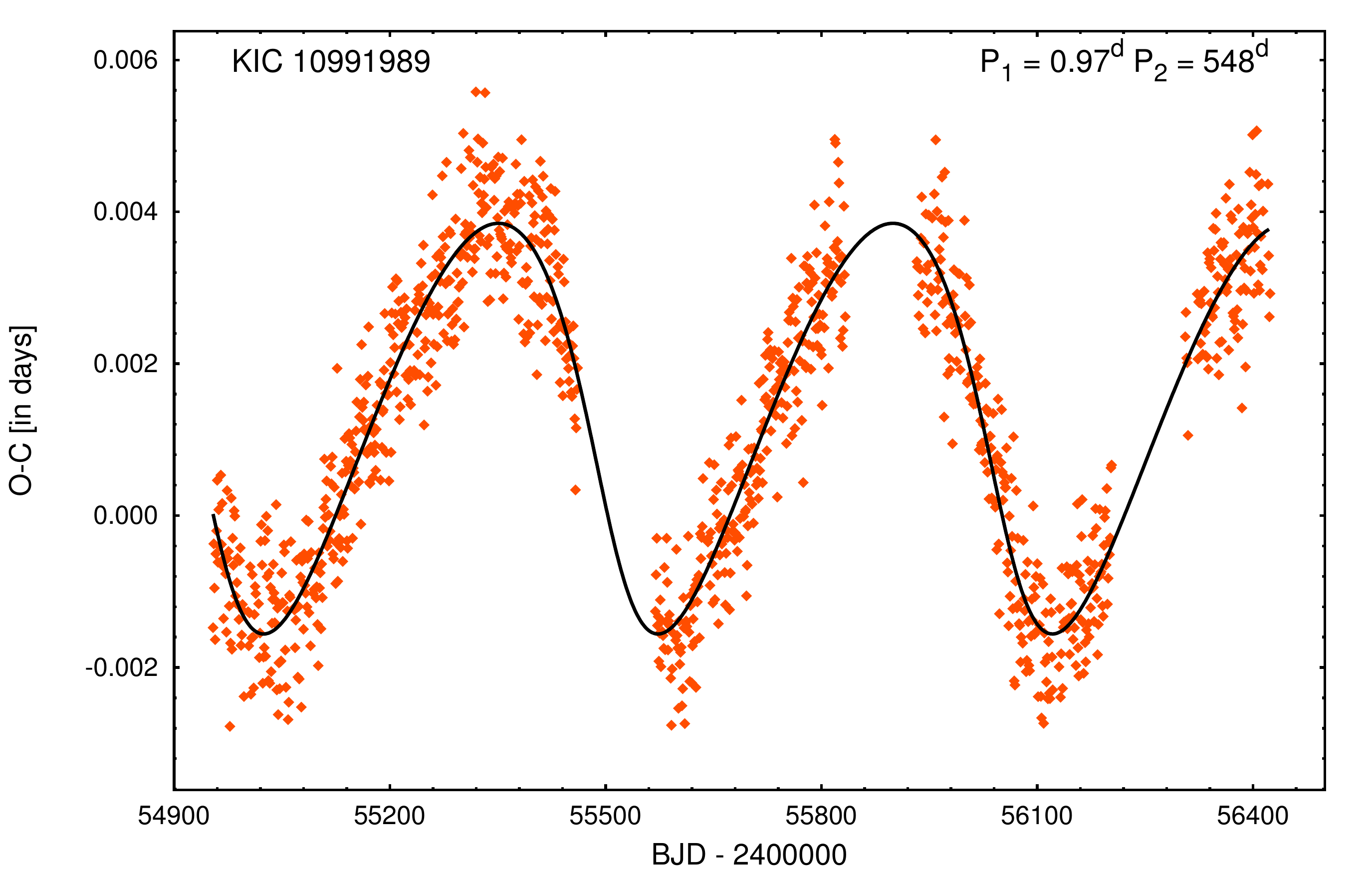}
\includegraphics[width=60mm]{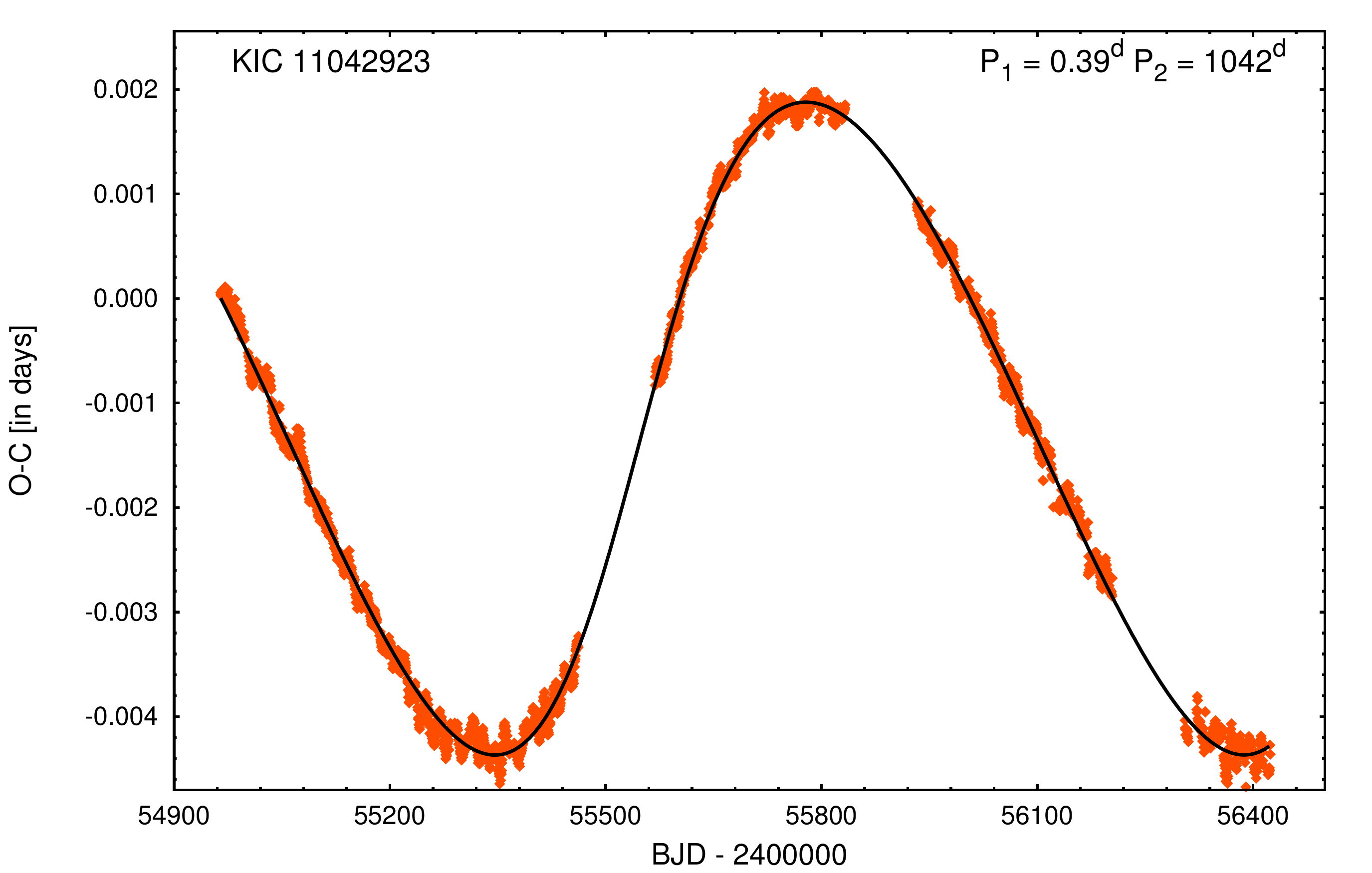}\includegraphics[width=60mm]{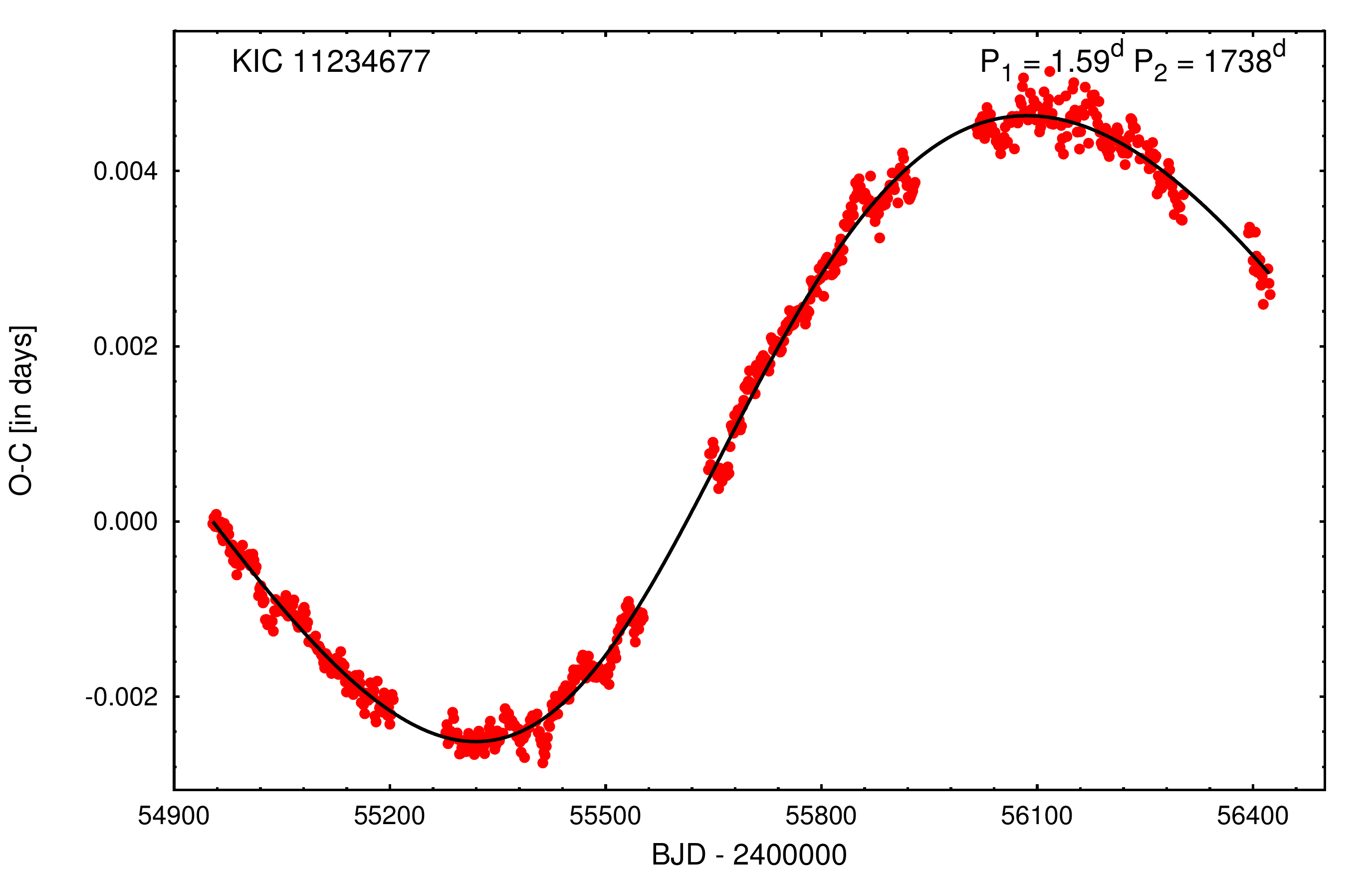}\includegraphics[width=60mm]{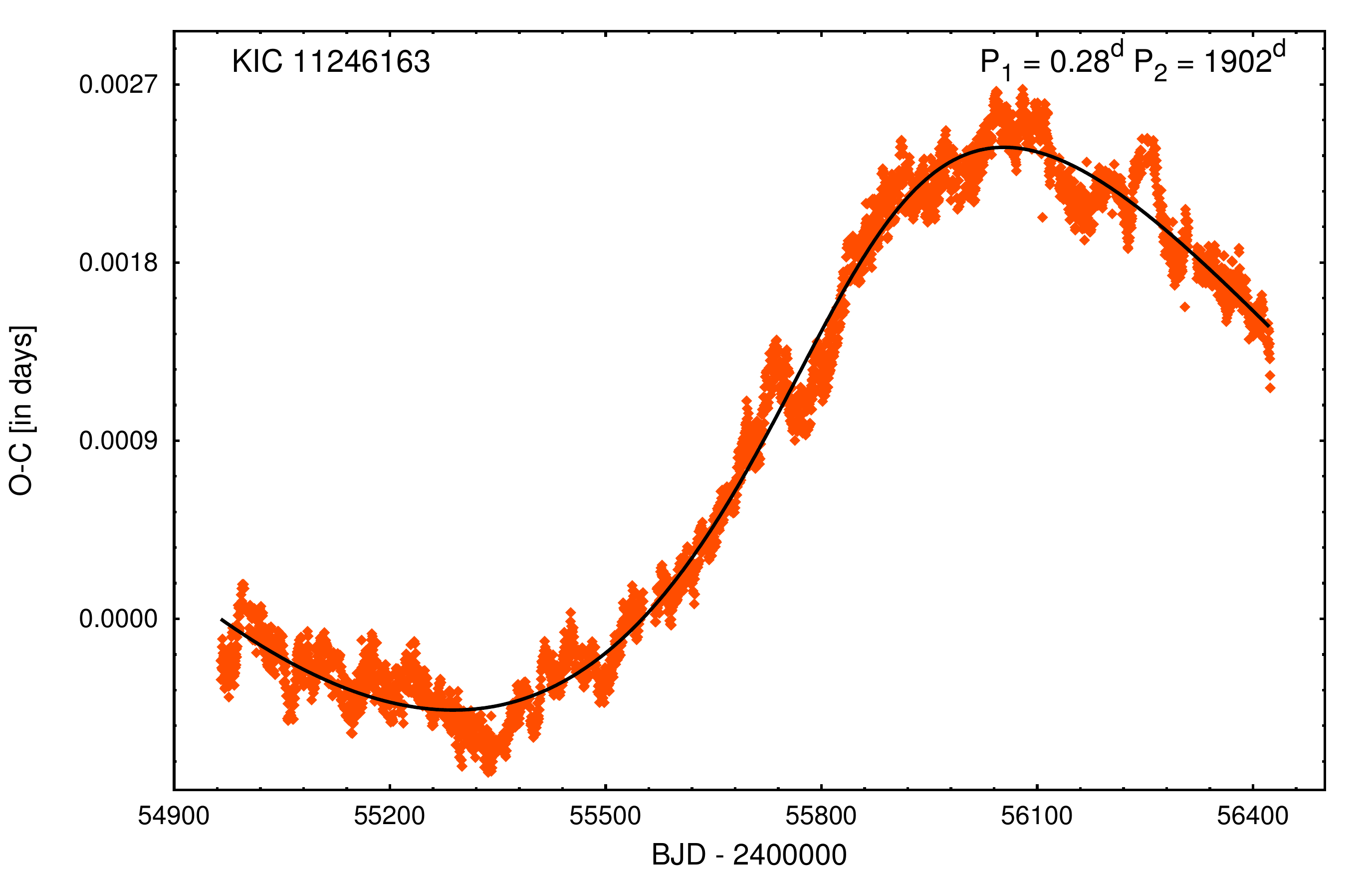}
\includegraphics[width=60mm]{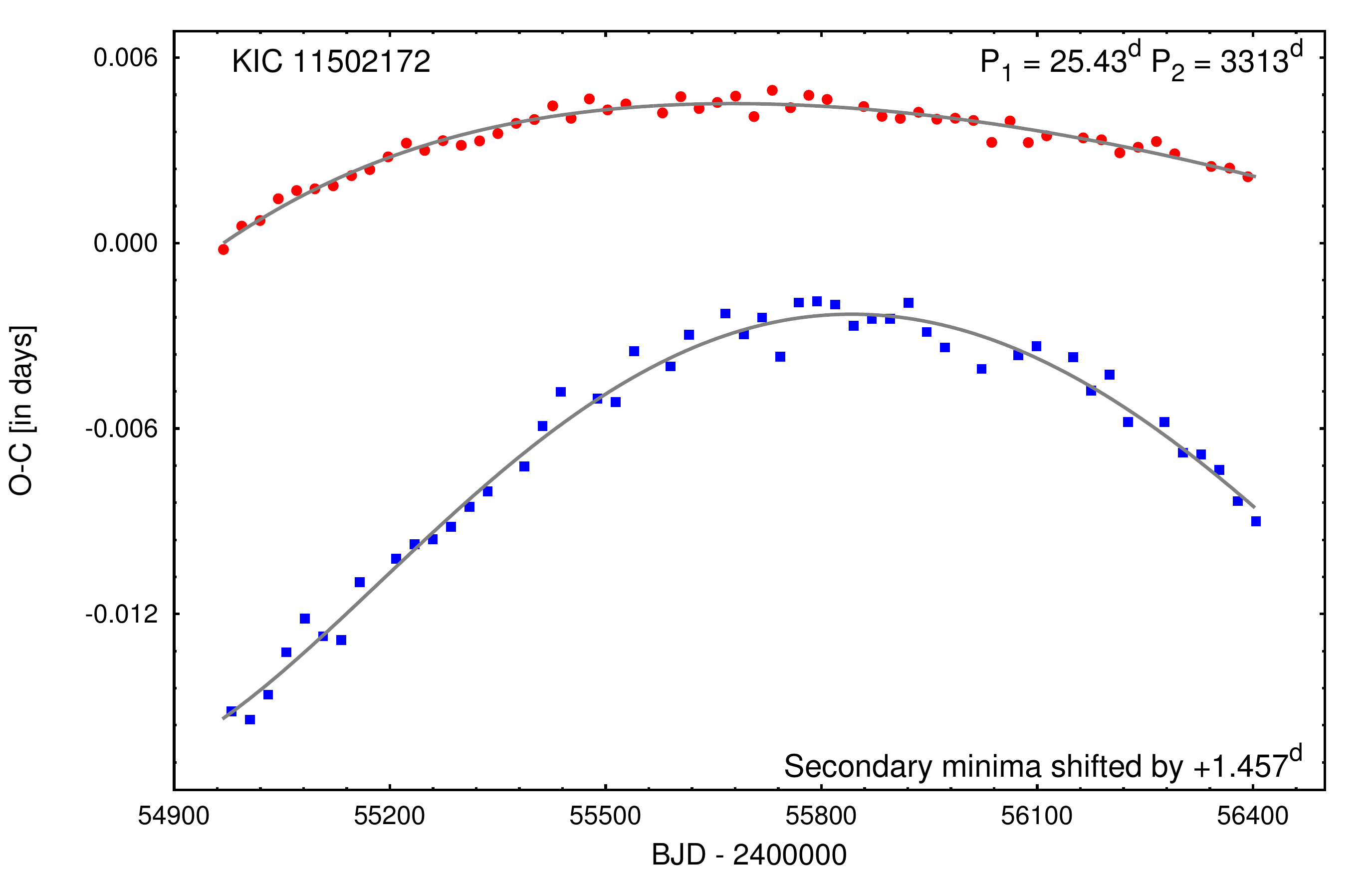}\includegraphics[width=60mm]{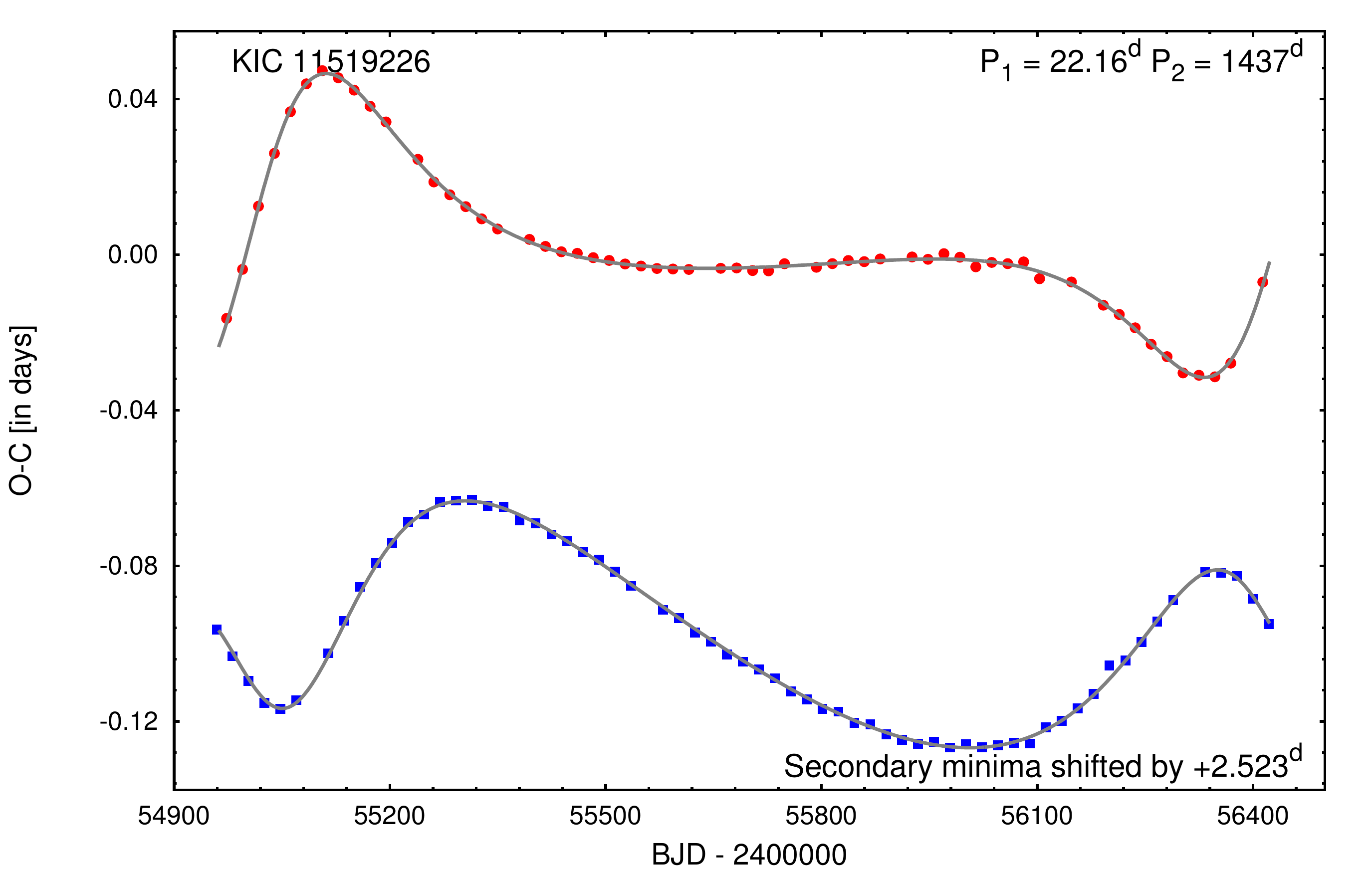}\includegraphics[width=60mm]{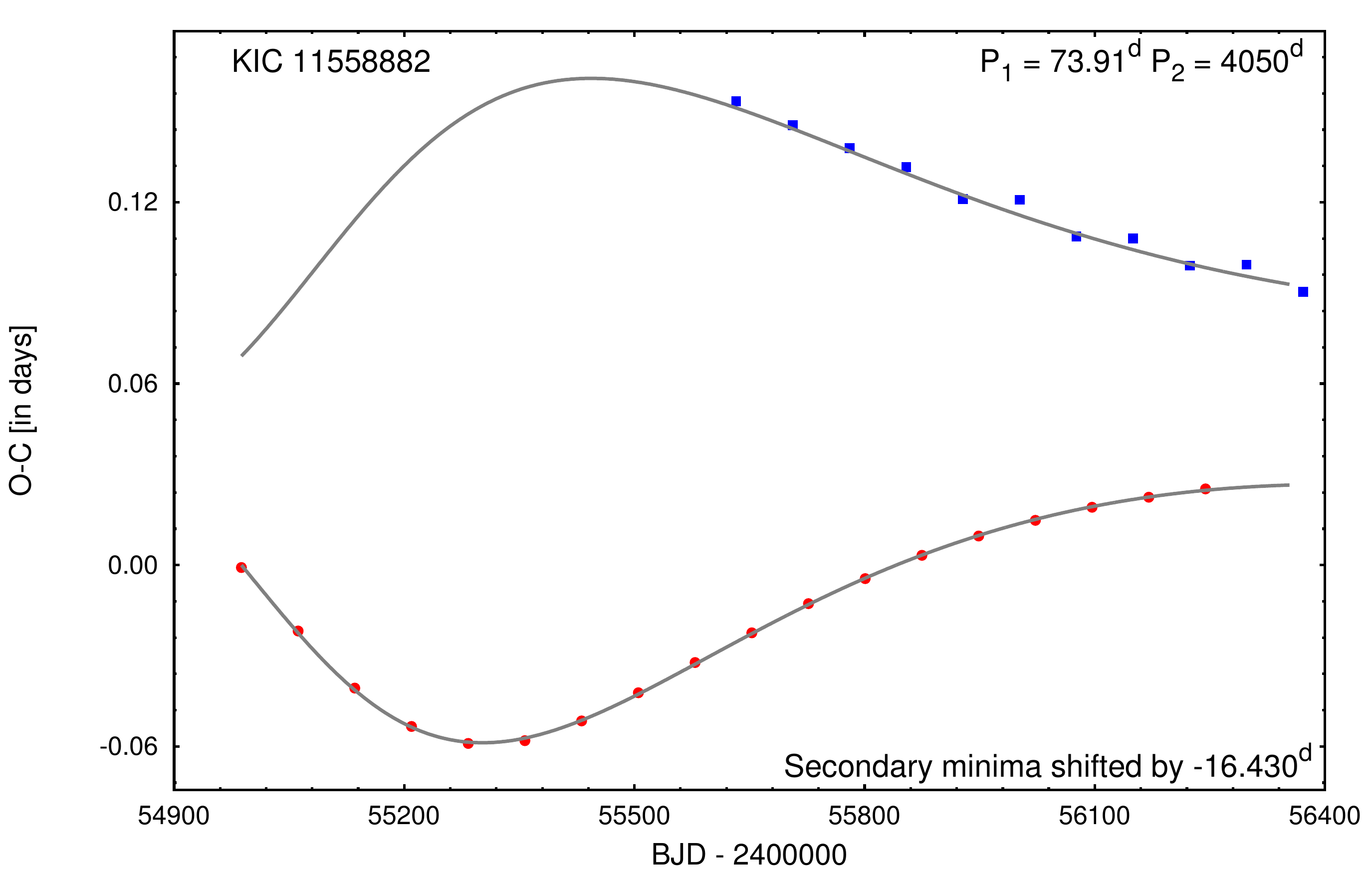}
\caption{(continued)}
\end{figure*}

\addtocounter{figure}{-1}

\begin{figure*}
\includegraphics[width=60mm]{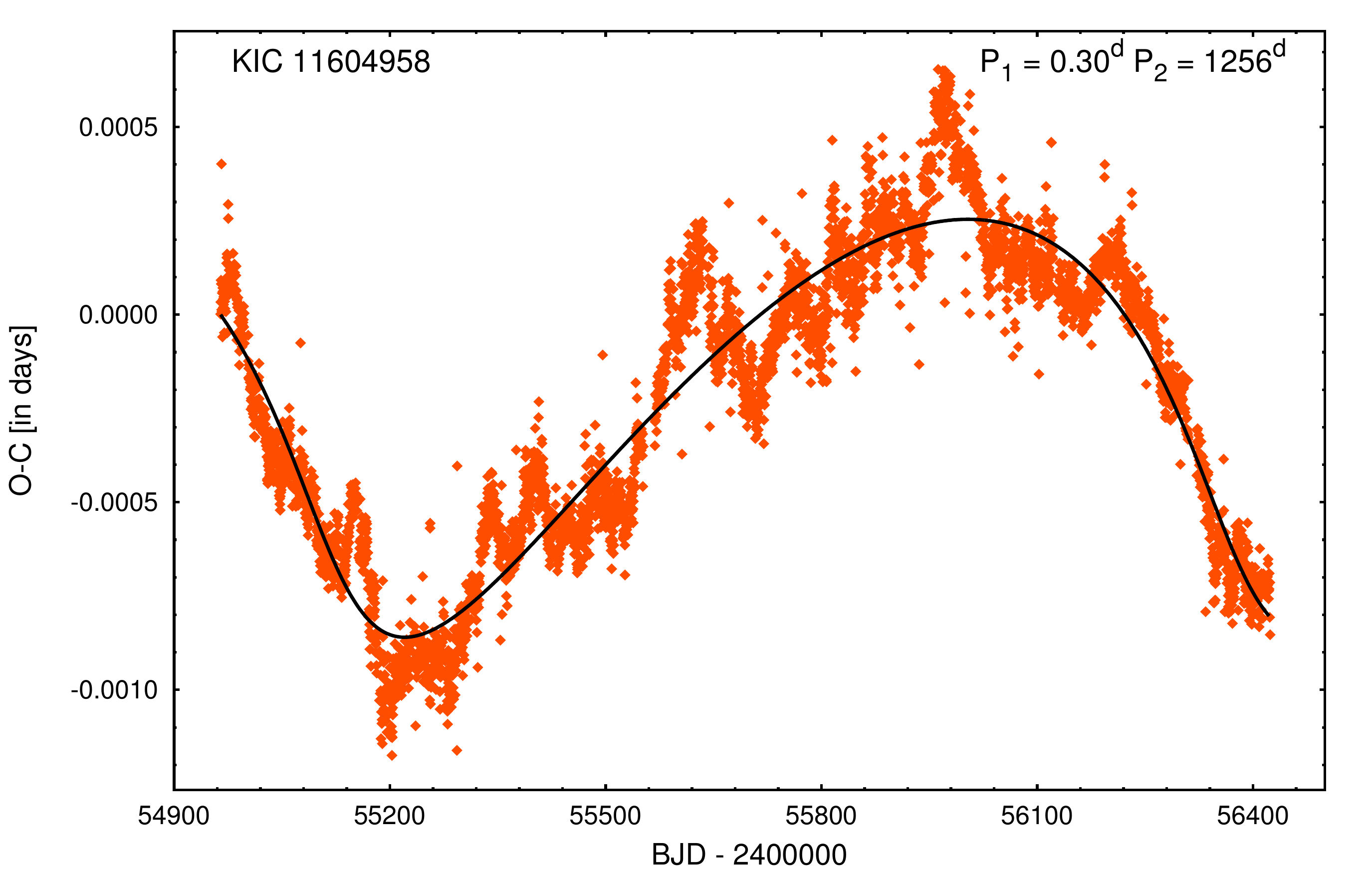}\includegraphics[width=60mm]{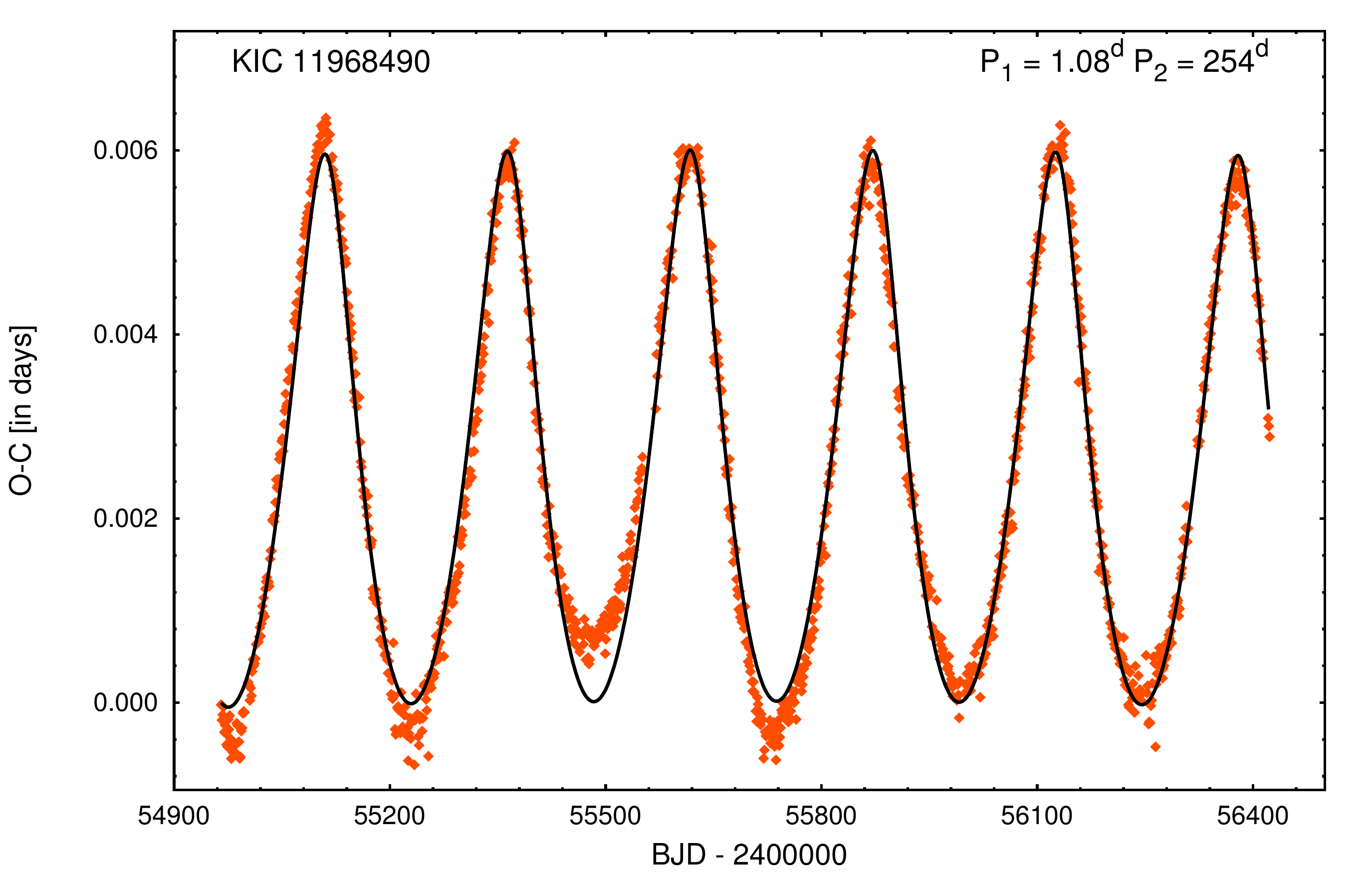}\includegraphics[width=60mm]{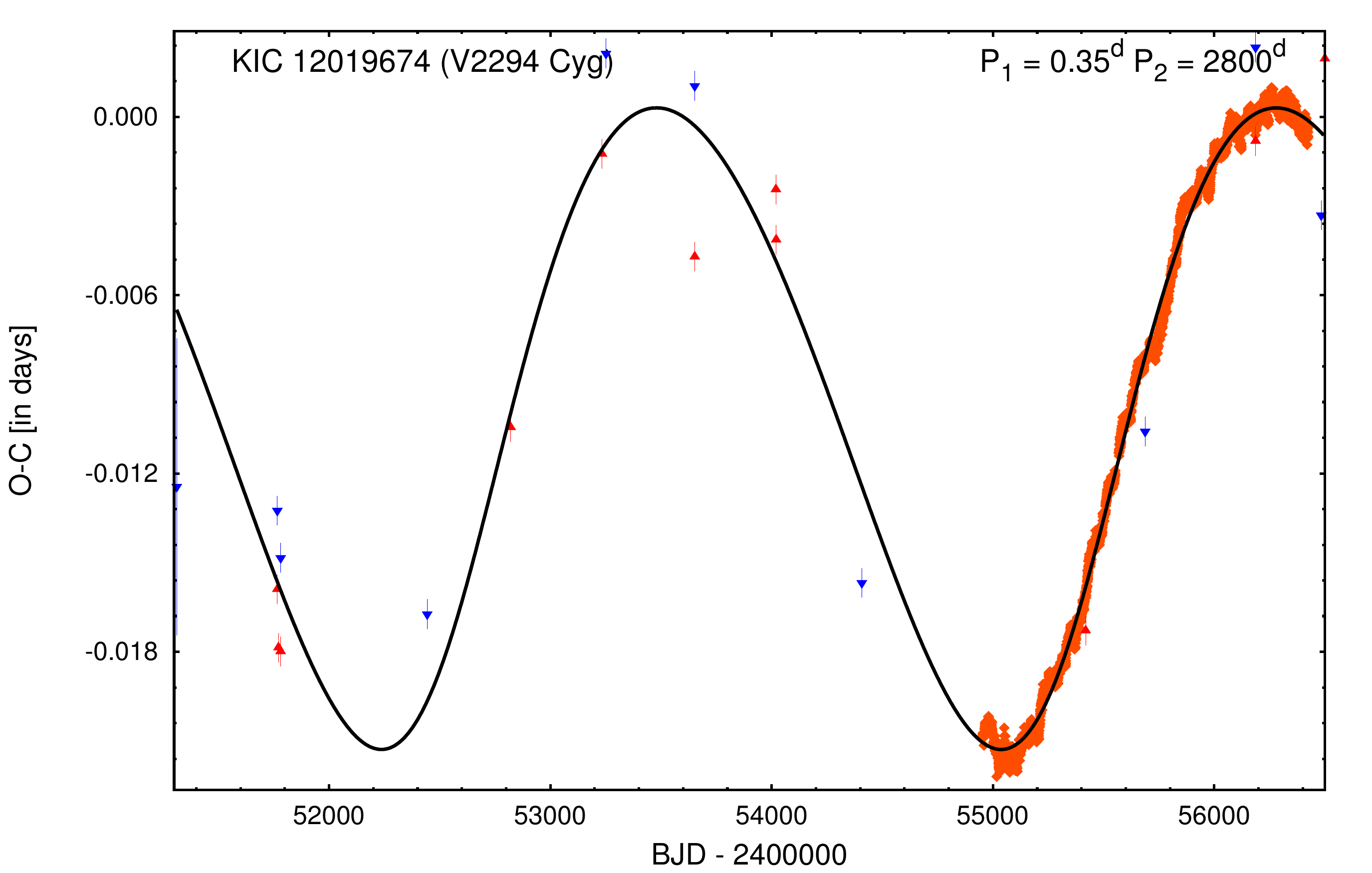}
\includegraphics[width=60mm]{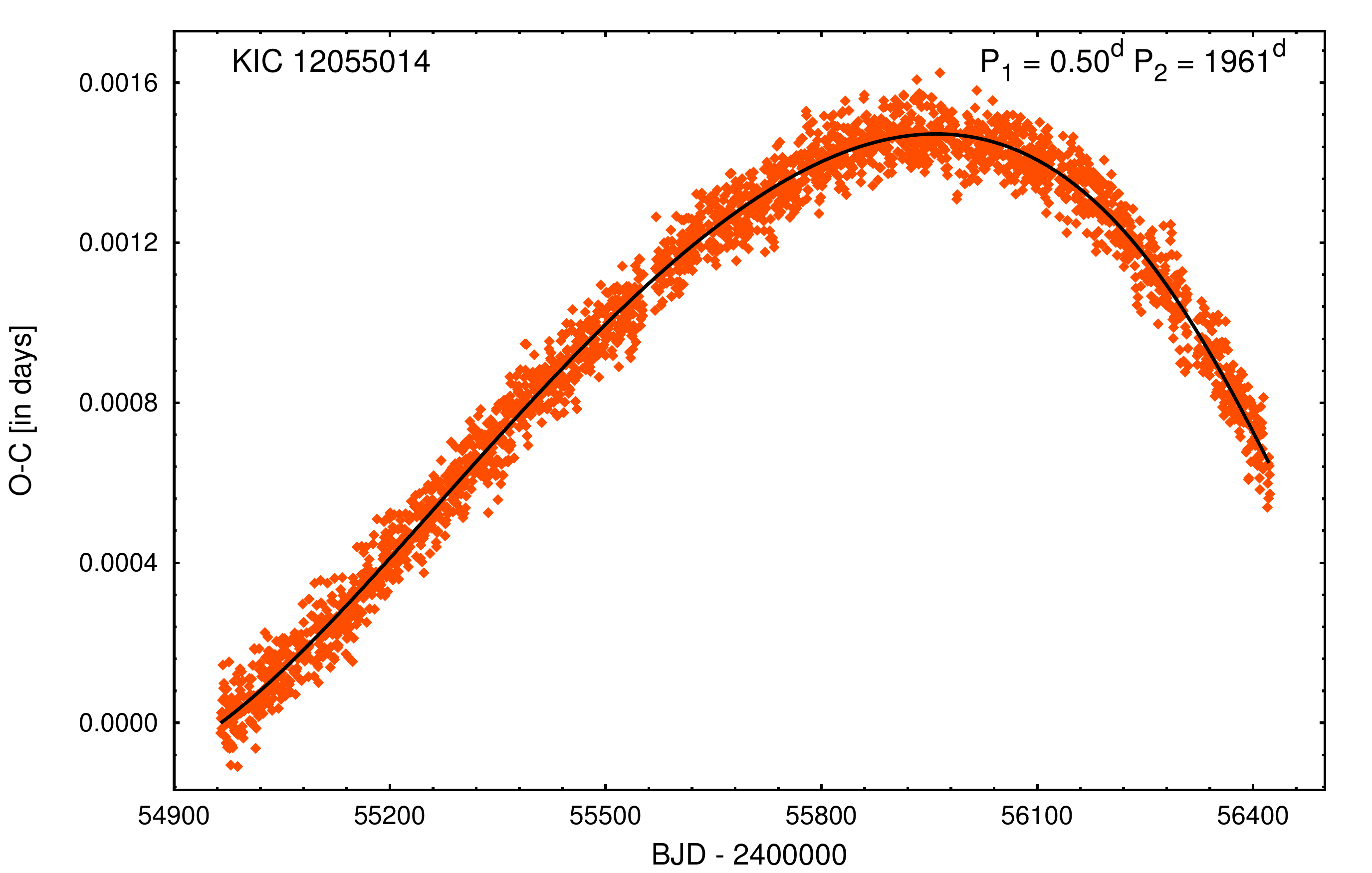}\includegraphics[width=60mm]{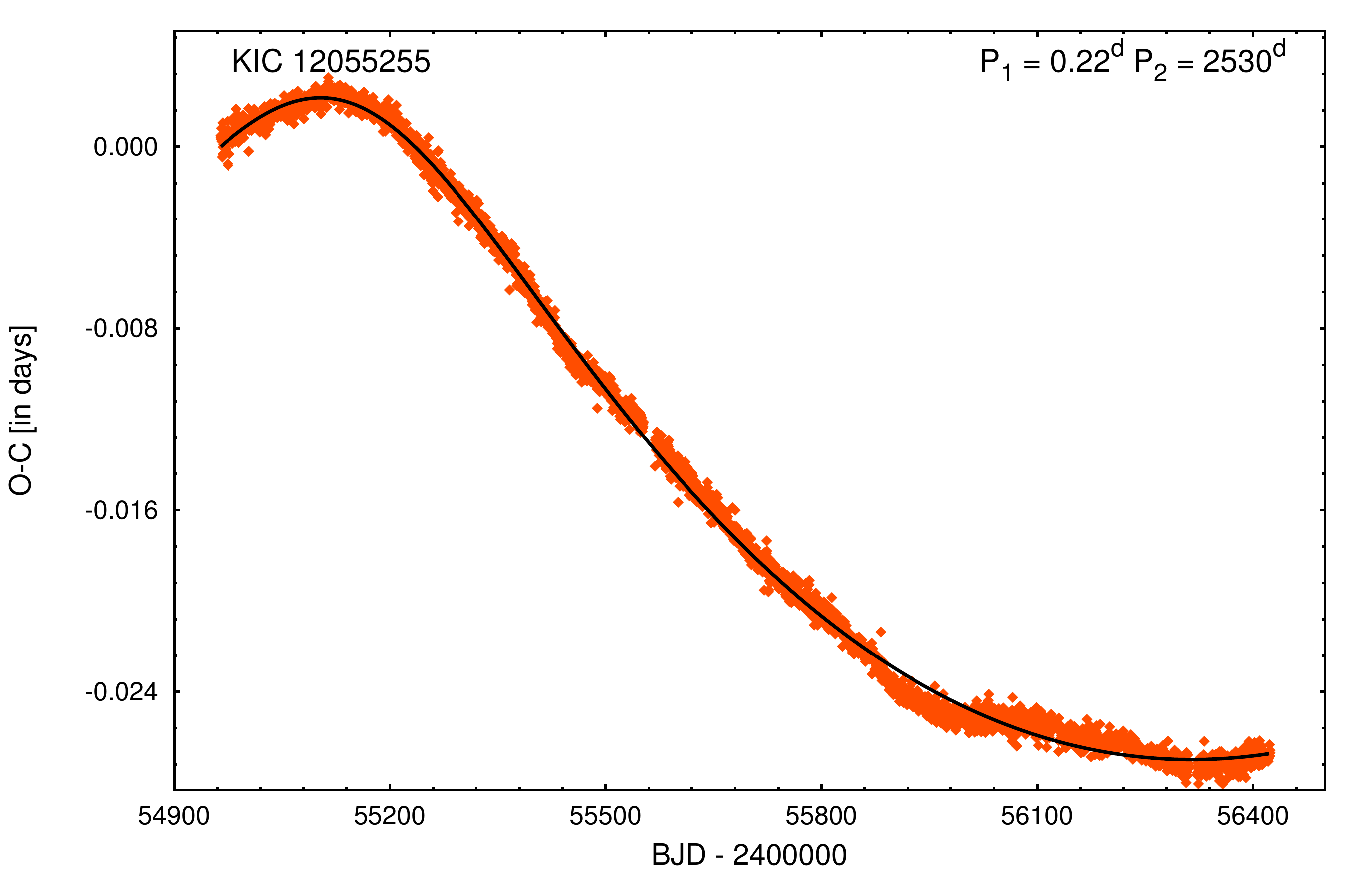}\includegraphics[width=60mm]{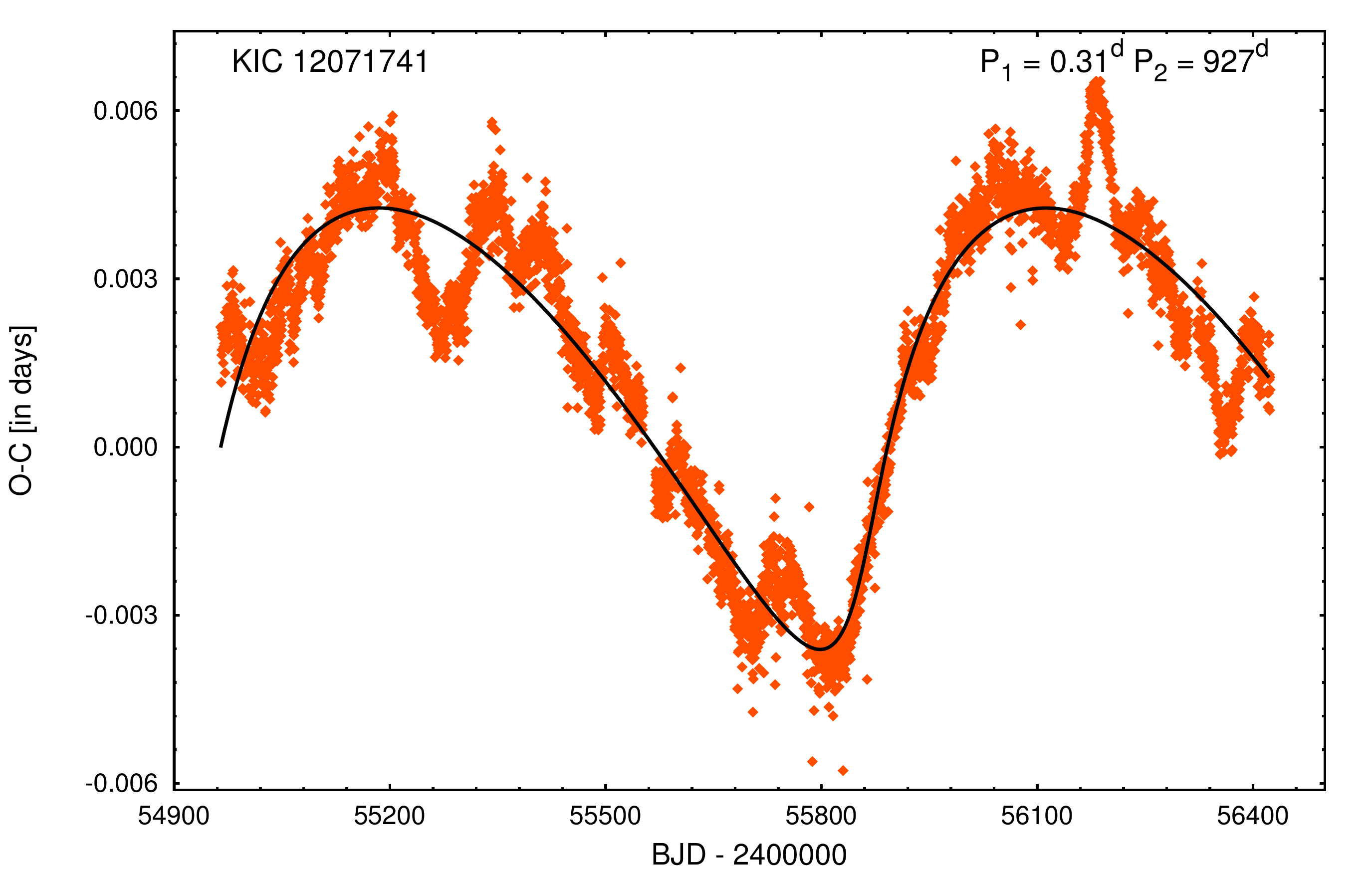}
\includegraphics[width=60mm]{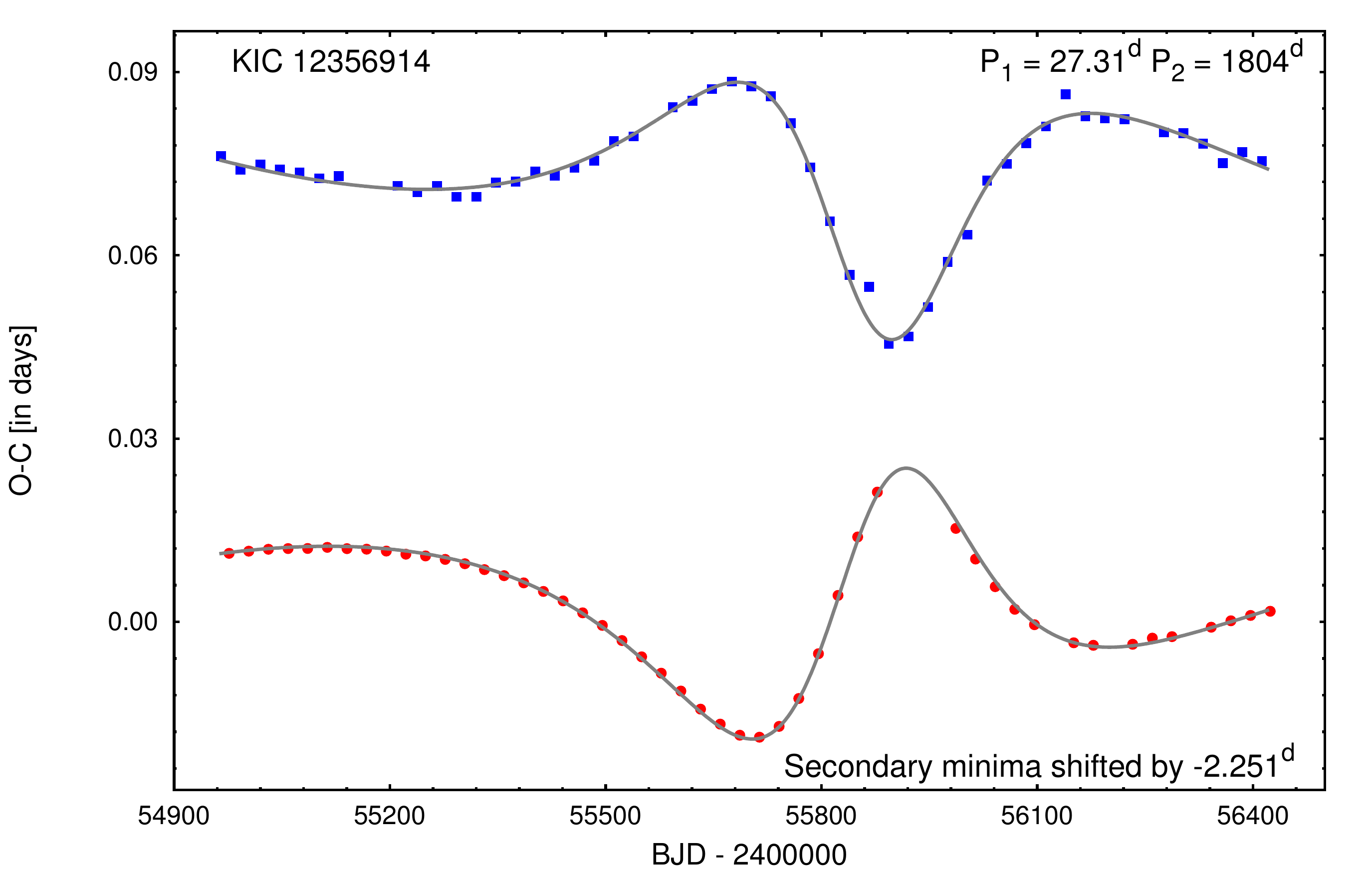}\includegraphics[width=60mm]{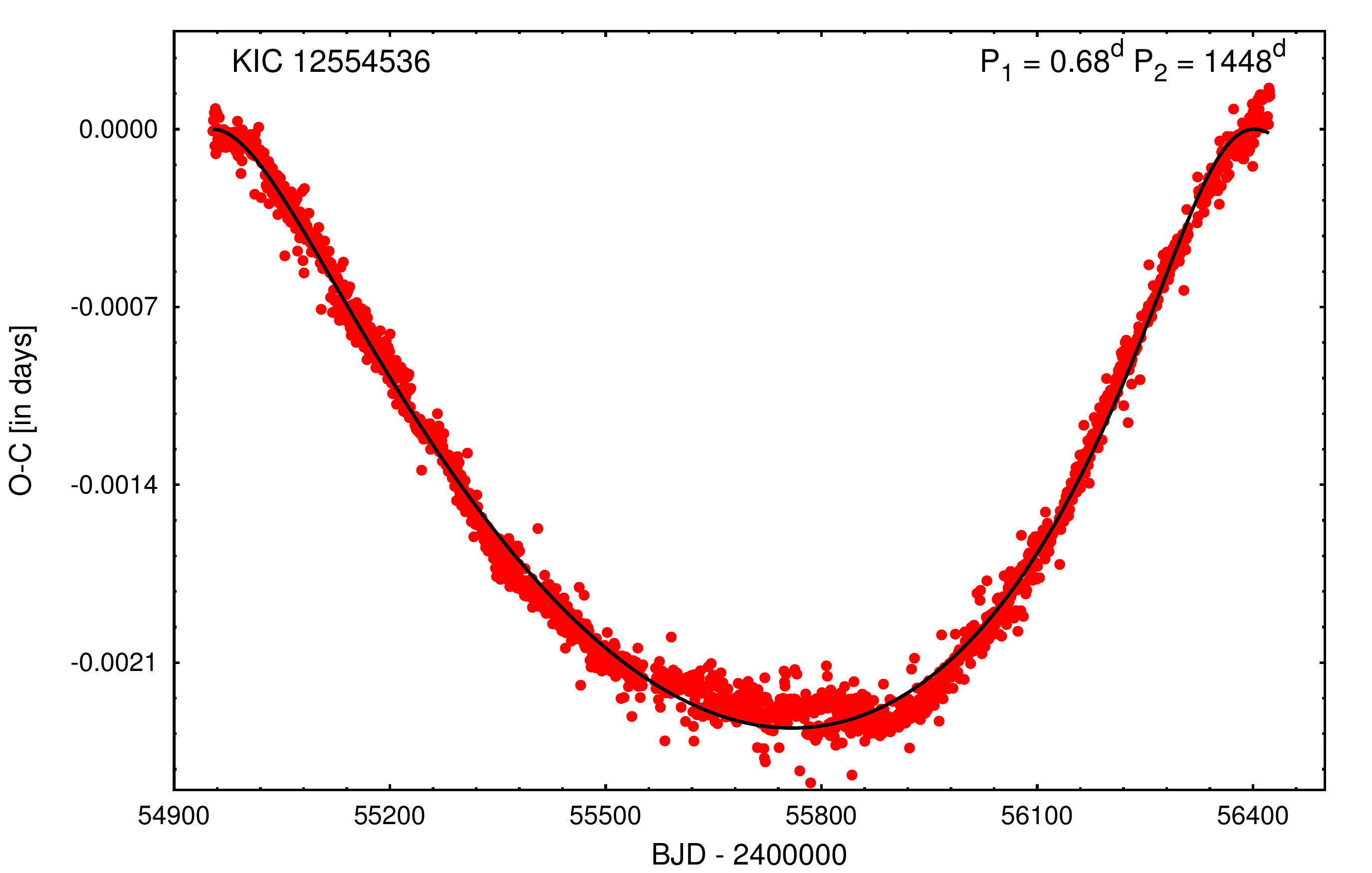}
\caption{(continued)}
\end{figure*}

\end{document}